\documentclass[11pt,a4paper]{article}
\usepackage{jheppub,amsmath,  amssymb,slashed,url,bm}
\usepackage{graphicx}
\usepackage{epstopdf}
\def\t{{ \sf t}} 
\def\kk{k}
\def\phif{f}
\def\dzzt{\D(\t z,z|\theta)}

\def\ggamma{\Gamma}
\def\Stt{\h S}
\def\Res{{\mathrm {R}}}

\def\ds{{\mathrm{ds}}}
\def\sD{\text{{\sf D}}}
\def\aalpha{\alpha}
\def\FF{{\eusm  F}}

\def\JJ{{\eusm J}}

\def\tt{{\mathfrak t}}
\def\ii{i}
\def\RR{{\mathcal R}}

\def\uu{u}
\def\piGSO{{\Pi_{\mathrm{GSO}}}}
\def\tpiGSO{{\t{\Pi}_{\mathrm{GSO}}}}
\def\UUU{\varPhi}
\def\SSigma{{\varTheta}}
\def\Stigma{{\varSigma}}
\def\btheta{{\bm \theta}}
\def\mm{{\bm m}}

\def\zizeta{\zeta}
\def\v{{\eurm v}}
\def\fD{ \eusm D}

\def\u{{\eurm u}}
\def\UU{U}
\def\p{{\eurm p}}

\def\VV{{\eusm V}}

\def\z{{\mathbf z}}

\def\w{{\eurm w}}

\def\g{\text{{\teneurm g}}}
\def\sg{\text{{\eighteurm g}}}
\def\ssg{\text{{\seveneurm g}}}
\def\be{\begin{equation}}
\def\ee{\end{equation}}

\def\XX{X}
\def\NS{{\mathrm {NS}}}
\def\Ra{{\mathrm R}}
\def\Ber{{\mathrm{Ber}}}
\def\BBer{{\textit{Ber}}}
\def\X{{\mathcal X}}

\def\hat{\widehat}

\def\tilde{\widetilde}

\def\h{\widehat}
\def\n{\text{{\teneurm n}}}
\def\sn{\text{{\eighteurm n}}}
\def\ssn{\text{{\seveneurm n}}}
\def\q{{\mathrm q}}
\def\y{{\mathrm y}}

\def\D{{\mathcal D}}
\def\S{{\mathcal S}}
\def\SIgma{\Sigma}
\def\YY{{\eusm Y}}
\def\V{{\mathcal V}}
\def\O{{\mathcal O}}

\def\Bbb{\mathbb}
\def\gh{{\mathrm{gh}}}
\def\red{{\mathrm{red}}}
\def\A{{\mathcal A}}
\def\d{{\mathrm d}}

\def\R{{\mathbb R}}
\def\C{{\mathbb C}}
\def\U{{\mathcal U}}
\def\D{{\mathcal D}}
\def\[{\bigl [}
\def\]{\bigr ]}
\def\J{{\mathcal J}}

\def\CP{{\mathbb{CP}}}
\def\N{{\mathcal N}}
\def\T{{\mathcal T}}

\def\Tr{{\mathrm {Tr}}}
\def\Z{{\mathbb Z}}
\def\ZZ{{\mathcal Z}}

\def\QQ{{\mathcal Q}}

\def\Tr{{\mathrm{Tr}}}

\def\L{{\mathcal  L}}
\def\t{\widetilde }
\def\h{\widehat}

\def\K{{\mathcal K}}
\def\V{{\mathcal V}}
\def\J{{\mathcal J}}
\def\I{{\mathcal I}}
\def\G{{\mathcal G}}
\def\B{{\mathcal B}}

\def\M{{\mathcal M}}

\def\MM{{\mathfrak M}}
\def\eusmm{{\MM_{\sg,\sn_\NS,\sn_\Ra}}}
\def\W{{\mathcal W}}
\def\P{{\mathcal P}}

\def\Ra{{\mathrm{R}}}

\def\lc{{\mathrm{lc}}}
\def\H{{\mathcal H}}

\def\st{{\mathrm{st}}}

\def\m{\cmmib m}
\def\tilde{\widetilde}

\def\bar{\overline}

\def\neg{\negthinspace}

\def\Ber{{\mathrm {Ber}}}
\def\Diff{{\eusm D}}
\def\Y{{\mathcal Y}}
\font\teneurm=eurm10 \font\seveneurm=eurm7 \font\eighteurm=eurm8 \font\fiveeurm=eurm5
\newfam\eurmfam
\textfont\eurmfam=\teneurm \scriptfont\eurmfam=\seveneurm
\scriptscriptfont\eurmfam=\fiveeurm
\def\eurm#1{{\fam\eurmfam\relax#1}}
\font\teneusm=eusm10 \font\seveneusm=eusm7 \font\fiveeusm=eusm5
\newfam\eusmfam
\textfont\eusmfam=\teneusm \scriptfont\eusmfam=\seveneusm
\scriptscriptfont\eusmfam=\fiveeusm
\def\eusm#1{{\fam\eusmfam\relax#1}}
\font\tencmmib=cmmib10 \skewchar\tencmmib='177
\font\sevencmmib=cmmib7 \skewchar\sevencmmib='177
\font\fivecmmib=cmmib5 \skewchar\fivecmmib='177
\newfam\cmmibfam
\textfont\cmmibfam=\tencmmib \scriptfont\cmmibfam=\sevencmmib
\scriptscriptfont\cmmibfam=\fivecmmib
\def\cmmib#1{{\fam\cmmibfam\relax#1}}

\title{Perturbative Superstring Theory Revisited}

 \author{Edward Witten }
\affiliation{School of Natural Sciences, Institute for Advanced Study,\\ 1 Einstein Drive, Princeton, NJ 08540 USA}
\abstract{Perturbative superstring theory  is revisited, with the goal of giving a simpler and more
direct  demonstration that multi-loop amplitudes are gauge-invariant (apart from known anomalies), 
satisfy space-time supersymmetry when expected, 
and have the expected infrared behavior.   The main technical tool is to make the whole analysis, including 
especially those arguments that involve
integration by parts, on supermoduli space, rather than after descending to ordinary moduli space. }
\begin{document} \maketitle

\section{Introduction}\label{intro}

Powerful covariant methods to compute superstring scattering amplitudes in the RNS formulation were introduced in the 1980's
\cite{FMS} and have been widely used ever since.\footnote{Parts of this construction were developed independently in
\cite{Knizhnik}.} 
 But some aspects of the formalism have remained slightly opaque.   
In the present paper, we will revisit this subject, aiming to give simpler and more direct demonstrations
that multi-loop superstring amplitudes are gauge-invariant (apart from known anomalies), satisfy space-time supersymmetry 
when expected, and have the infrared behavior that one would expect of a field theory with the same massless particles and
low energy interactions.  

These results are more transparent if superstring perturbation theory is formulated in terms of super Riemann surfaces
and supermoduli space, rather than aiming to reduce everything to ordinary Riemann surfaces and  moduli space.  This is not
a new idea and most of the pieces of the puzzle have been described long ago.  The literature is too vast to be properly summarized
here,\footnote{The history and development of string theory up to 1984 are
recounted in \cite{Birth}.}  but a few relevant points are as follows.  The measures on supermoduli space
that should be integrated to compute perturbative superstring scattering amplitudes were defined from several points
of view 
\cite{MNP,superoperator,RSV} in the 1980's.  The link between the picture-changing formalism of \cite{FMS} and
integration over fermionic moduli was made in \cite{EHVerl}.  The article
\cite{DPh} gives a thorough review of much of what was known in the late 1980's.  
Some of the ideas that will be important in the present
paper were introduced in the 1990's in work that has unfortunately remained little-known
\cite{Bel,Beltwo,Belthree}.   Finally, we mention a different kind of milestone; the first completely
consistent one-loop computations were performed in 
\cite{GSold,BGS}, and two-loop calculations were first performed in \cite{DPhgold}.

We start by describing the appropriate measures on moduli space and supermoduli space, and
the framework for understanding gauge-invariance of loop amplitudes.      We review the bosonic 
string in section  \ref{bosmeasure} in a way that straightforwardly generalizes to superstrings in section  \ref{measure}.
In sections \ref{vertex} and \ref{ramond}, 
we extend the superstring analysis to include external vertex operators, among
other things incorporating the relevant aspects of the covariant quantization of superstrings \cite{FMS}
and discussing the role of pictures.
In section \ref{propagator}, we use the general formalism to compute the string propagator, or in other words 
the integration measure for 
a string that propagates almost on-shell for a long proper time.   This is a good illustration,
and also  an important example, since it is the key to understanding the infrared region and explaining 
why superstring theory has precisely the same singularities as a field
theory with the same low energy content.  In section \ref{massren}, we analyze BRST anomalies and
introduce the massless tadpoles
that present the most subtle challenge for superstring perturbation theory.   We study loop-induced symmetry breaking 
in detail, even though it is fairly rare,  because it is a potential failure mode of superstring perturbation 
theory and it is important
to have criteria under which this failure mode does not occur.
In section \ref{tadpoles}, we analyze the spacetime supersymmetry of loop
amplitudes and the vanishing of massless tadpoles. The corresponding 
analysis for open and/or unoriented
superstrings involves some new ingredients related to spacetime anomalies and Ramond-Ramond tadpoles
and is treated in section \ref{anomalies}.  
Finally, in section \ref{betag}, we provide more details about the path integral of the bosonic ghost system
 in the presence of the delta function insertions that are ubiquitous in superstring perturbation theory.

We will generally attempt to minimize the demands on the reader's familiarity 
with supermanifolds and integration on them and with super Riemann surfaces and their
moduli.  More detail can be found, for example, in \cite{Wittenone,Wittentwo} and references cited there. 

It is tedious to always repeat everything for all of the open and closed, oriented and unoriented, bosonic and supersymmetric
string theories.  Our default examples are usually closed oriented bosonic strings and heterotic strings.  When this is simplest, 
we explain a point in the context of open strings. We only
run through the full roster of string theories when there are novelties involved, and otherwise leave generalizations to the reader.

We make   several restrictions on the scope of this paper. First and very important, 
we assume that the reader has already learned the basics about
superstring perturbation theory elsewhere, and we only review the basics here to the extent that it seems particularly useful.
 Second, as explained in section \ref{cvo}, we consider only the simplest
class of vertex operators that suffice for computing the $S$-matrix.  These are vertex 
operators that are conformal or superconformal
primaries with the simplest possible dependence on the ghost fields.  Techniques for 
computing with more general vertex operators
(arbitrary BRST-invariant vertex operators that are annihilated by $b_0-\t b_0$) are 
well-established \cite{OldPolch,PN}, but the necessary details would make
the present paper more complicated, possibly without interacting  in an interesting way with our main points.  

Accordingly, we omit certain questions, such as mass renormalization and an analysis of models in which a perturbative shift in the vacuum
is necessary to maintain spacetime supersymmetry, that 
are difficult to study in a conformally-invariant formalism.  Actually, an off-shell formalism appropriate for these
questions has been developed in a relatively recent series of papers (these papers 
were not available when the original version of the present paper was
submitted to the arXiv in 2012).  For a selection of these papers, see \cite{Sen1,Sen2,Sen3,Sen4,Sen5,Sen6}.    These papers
have been written in the language of picture-changing operators rather than in terms of super Riemann surfaces.  It would
certainly be possible to restate the results of those papers in the super Riemann surface language employed in the present paper,
basically by working on a super Riemann surface with a choice of local superconformal coordinates at each puncture.  However,
this has not yet been done.  

Likewise, it is difficult to incorporate expectation values of Ramond-Ramond fields in the superconformal framework,
so we do not 
 consider backgrounds with such expectation values.  Similarly, we do 
 not make contact with the pure spinor formalism \cite{PureSpinor}.

We will not discuss ultraviolet issues in this paper, since there are none.
Modular invariance, which was discovered over forty years ago \cite{Shapiro} following the recognition of the 
special role of 26 dimensions
\cite{Lovelace}, removes the ultraviolet region from superstring perturbation theory, 
though it took some time for this to be understood.

An informal account of some of the main ideas in this paper can be found in \cite{More}.

\section{A Measure On Moduli Space}\label{bosmeasure} 
\subsection{The Vacuum Amplitude}\label{vacamp}

We begin by recalling how the path integral of bosonic string theory generates a 
measure on the moduli space of Riemann surfaces.  
For standard explanations, see \cite{Alvarez,GM,operator} or section 5.3 of \cite{Polch} 
(whose conventions we generally follow).  
The explanation that follows is chosen to extend straightforwardly
to superstring theory. 

As usual, the worldsheet of a bosonic string is a Riemann surface $\Sigma$, endowed with a metric tensor $g$
and parametrized by local coordinates $\sigma^i$, $i=1,2$. 
We always assume Weyl invariance,
so we are really interested in the conformal structure, or equivalently the complex structure, defined by $g$.  
The worldsheet theory has matter fields $X$ and ghost fields $c^i$, $b_{ij}$; geometrically, $c$ is a vector field on $\Sigma$ and $b$ is
a symmetric traceless tensor.  The gauge-fixed action $I$
is the sum of a matter action $I_X$ and a ghost action $I_{\gh}$,
\begin{align}\label{zolk}\notag  I& = I_X+I_{\gh}\\
     I_{\gh} & = \frac{1}{2\pi}\int \d^2\sigma\sqrt g\, b_{ij}D^ic^j    .\end{align}
One defines an anomalous ghost number charge $N_\gh$ such that $c$ and $b$ have $N_\gh=1$ and $-1$ respectively. On a surface\footnote{\label{zondik} 
The function $F(g|\delta g)$ that we will define vanishes if $\Sigma$ has any conformal Killing vector fields, since in this case the ghost field $c$ has
zero-modes and the integral over those zero-modes will vanish.  Hence, until we introduce punctures in section \ref{vertop}, the following discussion
is non-trivial only if $\sg>1$.} $\Sigma$ of
genus
$\eurm g$, the ghost number anomaly violates $N_\gh$ by $-(6\g-6)$, meaning that a product of operators of definite ghost number
can have an expectation value only if their total ghost number is
$-(6\g-6)$.
     
The gauge-fixed theory has a BRST symmetry with generator $Q_B$.  Apart from the fact that $Q_B$ has ghost number 1 and obeys $Q_B^2=0$,
its most important property for our purposes is that
\begin{equation}\label{zonk} \{Q_B,b_{ij}\}=T_{ij}, \end{equation}
where $T_{ij}$ is the stress tensor, defined as the response of the action to a change in the metric:
\begin{equation}\label{wonk}\delta I= \frac{1}{4\pi}\int \d^2\sigma \sqrt g\, \delta g_{ij}\, T^{ij}. \end{equation}

In asserting that the action $I$ is BRST-invariant, one views the metric $g$ as a fixed, $c$-number quantity.  However, it will be convenient
to introduce a new fermionic variable $\delta g$, of $N_\gh=1$, and to extend the BRST symmetry by
\begin{equation}\label{zildoo}[Q_B,g_{ij}] =\delta g_{ij}, ~~~ \{Q_B,\delta g_{ij}\}=0. \end{equation}
We take $\delta g_{ij}$ to be a symmetric traceless tensor  (we define $\delta g$ to be traceless as we are only interested in $g$
up to Weyl transformations).
    
The formula (\ref{wonk}) shows that if we allow the metric to vary with an unspecified BRST variation $\delta g_{ij}$, then the action $I$
is not invariant.  But the fundamental formula (\ref{zonk}) shows that we can restore BRST symmetry with a simple addition to the action:
\begin{equation}\label{tolf}I\to \hat I=I+\frac{1}{4\pi}\int \d^2\sigma \sqrt g \delta g_{ij} b^{ij}. \end{equation}

Now we define a function\footnote{The vertical bar in $F(g|\delta g)$ is meant to remind us that $g$ is bosonic or even and $\delta g$ is fermionic or odd.
We often write  $F(t^1\dots t^p|\theta^1\dots \theta^q)$
for  a function $F$ that depends on even variables $t^1\dots t^p$ and odd variables $\theta^1\dots \theta^q$.}  of $g$ and $\delta g$ by integration over the fields $X,b,c$:
\begin{align}\label{olf} \notag F(g|\delta g)&= \int\D(X,b,c)\,\exp\left(-\hat I(X,b,c;g,\delta g)\right)\\ &=\int\D(X,b,c)\exp(-I)\exp\left(-\frac{1}{4\pi}\int_\Sigma
\d^2\sigma \sqrt g\,\delta g_{ij} \,b^{ij}\right). \end{align}
Since the action and the integration measure are BRST-invariant, integrating out the fields $X,b,c$ is a BRST-invariant operation, so $F(g|\delta g)$ is
BRST-invariant:
\begin{equation}\label{holf}[Q_B,F(g|\delta g)\}=0.\end{equation}

The integrand in (\ref{olf}) is an inhomogeneous function of $\delta g$.  However, the function $F(g|\delta g)$ is actually homogeneous in $\delta g$
with a definite weight.  To see this, we expand in powers of $\delta g$:
\begin{equation}\label{moof} \exp\left(-\frac{1}{4\pi}\int_\Sigma \d^2\sigma\sqrt g \,\delta g_{ij}b^{ij}\right) =\sum_{n=0}^\infty\frac{1}{n!}\left(-\frac{1}{4\pi}\int_\Sigma\d^2\sigma \sqrt g \,\delta g_{ij}b^{ij}\right)^n .\end{equation}
Each power of $\delta g$ accompanies a power of $b$, and because of the ghost number anomaly, a non-zero contribution comes only from $n=6\g - 6$.  So actually, $F(g|\delta g)$ is homogeneous in $\delta g$ of degree $6\g - 6$.  (We are limited here to $\g>1$ as we have assumed that $c$ has no zero-modes. In section \ref{vertop}, we will make vertex operator insertions and then there is no such restriction.)

Acting as in (\ref{holf}) on a function that depends only on $g$ and $\delta g$, $Q_B$  can be written in more detail as
\begin{equation}\label{oof}Q_B=\int_\Sigma \d^2\sigma\sum_{i,j=1,2} \sqrt g \delta g_{ij}\frac{\delta }{\delta g_{ij}}. \end{equation}
At this point, it is good to remember that on a manifold $M$ with local coordinates $x^i,~i=1,\dots , s=\mathrm{dim}\,M$, 
one introduces odd variables  $\d x^i$ (which are usually called one-forms)
and defines the exterior derivative operator
\begin{equation}\label{boof} \d =\sum_{i=1}^s \d x^i\frac{\partial}{\partial x^i},\end{equation}
which obeys $\d^2=0$.  A function $F(x^1\dots x^s|\d x^1\dots \d x^s)$ that is homogeneous of degree $p$ in the odd variables
has an expansion
\begin{equation}\label{toof} F(x^1\dots |\dots \d x^s) =\sum_{i_1<\dots < i_{p}}F_{i_1\dots i_p}(x^1\dots x^s) \d x^{i_1}\dots \d x^{i_p} \end{equation}
and is usually called a $p$-form.  So $\d$ maps $p$-forms to $(p+1)$-forms; a $p$-form $\omega$ is said to be closed if $\d\omega=0$
and to be exact if $\omega=\d\lambda$ for some $(p-1)$-form $\lambda$.   If we just identify $\delta g_{ij}$ with $\d g_{ij}$, we can
think of the operator on the left hand side of (\ref{oof}) as the exterior derivative on the space $\JJ$ of all
 of conformal structures on $\SIgma$ (that is, the space
of metrics on $\Sigma$ modulo Weyl transformations): the integral over $\Sigma$ and sum 
over $i,j=1,2$ in (\ref{oof}) is the analog of the sum over $i$ in
(\ref{boof}).   So in fact, we can view $F(g|\delta g)$ as a $(6\g-6)$-form on $\JJ$.  With this interpetation,  (\ref{holf}) says that this form is closed, $\d F=0$.

\subsection{Reducing To Moduli Space}\label{redmod}

To get farther, we must consider diffeomorphism invariance.  We want to interpret $F(g|\delta g)$ as a form not on  the 
infinite-dimensional space $\JJ$, but on the quotient $\M_\sg=\JJ/\Diff$, where $\Diff$ is the group of orientation-preserving diffeomorphisms of $\Sigma$. 
$\M_\sg$ is the moduli space of Riemann surfaces of genus $\g$.  To show that $F(g|\delta g)$ is the ``pullback'' from $\M_\sg$ of a differential
form on $\M_\sg$, we need to establish two things (see appendix \ref{pullback}):

(1) $F(g|\delta g)$ must be $\Diff$-invariant.  This is actually manifest from the diffeomorphism-invariance of the whole construction.

(2) $F(g|\delta g)$ must vanish if contracted with one of the vector fields on $\JJ$  that generates the action of $\Diff$.  Such a vector field is
\begin{equation}\label{tsolt} g_{ij}\to g_{ij}+\epsilon (D_iv_j+D_jv_i),\end{equation}
where $v^i$ is an ordinary vector field on $\Sigma$ and $\epsilon$ is an infinitesimal parameter. 
The operation of contraction with this vector
field transforms $\delta g_{ij}$ by
\begin{equation}\label{omigo}\delta g_{ij}\to \delta g_{ij}+\epsilon (D_iv_j+D_jv_i).\end{equation}
(It may be helpful to compare to the finite-dimensional formulas (\ref{onx}) and (\ref{tonx}).)
 For  $F(g|\delta g)$ to be invariant under this shift of $\delta g_{ij}$  
means precisely that
\begin{equation}\label{solt}\int_{\Sigma}\d^2\sigma \sqrt g\,(D_iv_j+D_jv_i) \frac{\delta}{\delta(\delta  g_{ij})} F(g|\delta g)=0.\end{equation}
Explicitly, given the definition (\ref{olf}) of $F(g|\delta g)$, the requirement is
\begin{equation}\label{poft}\int \D(X,b,c)\exp(-\hat I) \int_{\Sigma}\d^2\sigma \sqrt g (D_iv_j+D_jv_i) b^{ij}=0.\end{equation}
This is indeed true, as we learn upon integrating by parts and using the equation of motion
\begin{equation}\label{moft} D_ib^{ij}=0. \end{equation}

So in fact, $F(g|\delta g)$ is the pullback to $\JJ$ of a $(6\g-6)$-form on $\M_\sg$.  This form, moreover, is closed, since 
$F(g|\delta g)$ is BRST-invariant.
Actually, once we know that $F(g|\delta g)$ is a pullback from $\M_\sg$, we do not really need BRST symmetry to prove that $\d F=0$.
This is automatically true;  the dimension of $\M_\sg$ is $6\g-6$, so a $(6\g-6)$-form on $\M_\sg$ is automatically closed.  However, as we will
see in section \ref{vertop}, once we consider vertex operator insertions, the closedness of $F(g|\delta g)$ does give non-trivial information.

The equation (\ref{moft}) arises as the classical equation of motion $\delta \hat I/\delta c^j=0$.  The use of the classical equation of motion is
really a shorthand for the fact that the path integral is invariant under the change of variables
\begin{equation}\label{orby}c^i\to c^i+\epsilon v^i,\end{equation}
with $\epsilon$ a parameter.  Under this transformation, the measure is invariant and the action shifts by $\hat I\to \hat I-(\epsilon/2\pi)
\int_\Sigma \d^2\sigma \sqrt g \,b_{ij}D^iv^j.$  To first order in $\epsilon$,  the path integral transforms by
\begin{equation}\label{gorby}\int\D(X,b,c)\,\exp(-\hat I)\to \int \D(X,b,c)\,\exp(-\hat I)\,\left(1+\frac{\epsilon}{2\pi}\int_\Sigma\d^2\sigma\,\sqrt{g}
b_{ij}D^iv^j\right).\end{equation}
The fact that the right hand side is independent of $\epsilon$ gives (\ref{poft}).

In our derivation, we have not made use of the fact that $\JJ$ and $\M_\sg$ have complex structures.  In fact, picking
local holomorphic and antiholomorphic coordinates\footnote{We write $\t z$ for what is more commonly called $\bar z$.  Since all
correlation functions are real-analytic, we can analytically continue slightly away from $\t z =\bar z$ and that is why we adopt this more flexible notation.  The ability to make this analytic continuation is more important in superstring theory, in which  a condition $\t z=\bar z$ would not be invariant under superconformal transformations.}
 $z$ and $\t z$ on $\Sigma$, 
we can decompose $c^i$ and $b_{ij}$ into
holomorphic  components $c^z,\,b_{zz}$ and antiholomorphic components    $c^{\t z},\,b_{\t z \t z}$, which we 
often denote as $c,b$ and $\t c, \t b$, respectively.
One can define separate anomalous ghost number symmetries for $c,b$ and for $\t c,\t b$.  Each has an anomaly $-(3\g-3)$.  This means in practice
that both $b$ and $\t b$ have $3\g-3$ zero-modes on a surface $\Sigma$ of genus $\g$, so it requires $3\g-3$ insertions of $b$ and 
$3\g-3$ insertions of
$\t b$ to get a non-zero path integral.  From this, it follows that $F(g|\delta g)$ is a form of degree $(3\g-3,3\g-3)$, in other words, its holomorphic and antiholomorphic
degree are both $3\g-3$.

A few more comments will be helpful background for the generalization to superstrings.  By a zero-mode of $b$, we will mean 
simply a solution of the classical
equation $D_ib^{ij}=0$.  Let us expand $b_{ij}$ in $c$-number
zero-modes ${\eurm b}_\alpha,$ $\alpha=1,\dots, 6\g -6$, and non-zero
modes ${\eurm b}'_\lambda$:
\begin{equation}\label{zumbo}b_{ij}=\sum_{\alpha=1}^{6\sg-6} \u_\alpha {\eurm b}_{\alpha,ij}+\sum_\lambda \w_\lambda 
{\eurm b}'_{\lambda\,ij}.\end{equation}
Here $\u_\alpha$ and $\w_\lambda$ are anticommuting coefficients.
When we make this expansion, the coefficients $\u_\alpha$ do not appear in the original action $I(X,b,c)$, precisely 
because the corresponding modes are zero-modes.
They do appear in the extended action $\hat I$:
\begin{equation}\label{morko}\hat I=\dots +\frac{1}{4\pi}\sum_{\alpha=1}^{6\sg-6}\int_\Sigma\d^2\sigma \sqrt{g}\delta g_{ij} 
\u_\alpha {\eurm b}_{\alpha}^{ij}.\end{equation}
The part of the path integral that depends on the zero-mode coefficients $\u_\alpha$ is therefore particularly simple.  This is a factor
\begin{equation}\label{porko}\prod_{\alpha=1}^{6\sg-6}\int \d \u_\alpha\,\exp\left(\frac{\u_\alpha}{4\pi}
\int_{\Sigma}\d^2\sigma\sqrt g \delta g_{ij} {\eurm b}_\alpha^{ij}\right)
=\prod_{\alpha=1}^{6\sg-6}\frac{1}{4\pi}\int_\Sigma\d^2\sigma\sqrt g \delta g_{ij}{\eurm b}_\alpha^{ij}. \end{equation}  
We have used the fact that if $\u$ and $\v$ are odd variables, then $\int \d \u \exp(\u\v)=\v$.   
On the right hand side of (\ref{porko}), we see explicitly
that $F(g|\delta g)$ is proportional to $6\g-6$ factors of $\delta g$.  It is instructive to recall that if $\v$ is an odd variable, 
then $\v=\delta(\v)$.  So if we set
\begin{equation}\label{orko}\v_\alpha=\frac{1}{4\pi}\int_\Sigma\d^2\sigma\sqrt g \delta g_{ij}{\eurm b}_\alpha^{ij},\end{equation}
then $F(g|\delta g)$ is proportional to 
\begin{equation}\label{norko}\prod_{\alpha=1}^{6\sg-6}\delta(\v_\alpha).\end{equation}

Actually,  (\ref{norko}) gives the complete dependence of $F(g|\delta g)$ on $\delta g$.
To see this, we can assume that $c$ has no zero-modes, since otherwise $F(g|\delta g)=0$.  So the mode expansion of $c$ involves only
non-zero modes ${\eurm c}_\lambda$ with coefficients $\gamma_\lambda$:
\begin{equation}\label{hombe}c^i=\sum_\lambda \gamma_\lambda {\eurm c}_\lambda^i.\end{equation}
The part of the action $\hat I$ that depends on the  $\gamma_\lambda$ has the general form $-\sum_\lambda m_\lambda \w_\lambda \gamma_\lambda$
with non-zero constants $m_\lambda$, and the  integral over the $\gamma_\lambda$ is
\begin{equation}\label{ombe}\prod_\lambda \int D \gamma_\lambda\exp(m_\lambda \gamma_\lambda \w_\lambda)=\prod_{\lambda}m_\lambda\cdot \prod_\lambda \delta(\w_\lambda)\end{equation}  (here $\prod_\lambda m_\lambda$ is the determinant of the non-zero ghost and antighost modes; of course
it requires regularization).  The integral over $\w_\lambda$ can be performed with the aid of the $\delta$ functions in (\ref{ombe}), so
 even though in the definition (\ref{tolf}) of the extended action, the coefficients $\w_\lambda$  do couple to $\delta g$,
this coupling does not affect the evaluation of $F(g|\delta g)$.  Hence (\ref{porko}) or (\ref{norko})
does give the full dependence of $F(g|\delta g)$ on $\delta g$.

\subsection{Integration Over Moduli Space}\label{slice}

Having understood that $F(g|\delta g)$ is the pullback of a differential form of top degree 
on $\M_\sg$, which we denote by the same
name, we can formally define the genus $\g$ vacuum amplitude of string theory by integration over $\M_\sg$:
\begin{equation}\label{turn}Z_\sg=\int_{\M_\ssg}F(g|\delta g).\end{equation}
The only problem with this formal definition is that $\M_\sg$ is not compact.    This is an essential
fact in string theory, since
the region at infinity in $\M_\sg$ is the infrared region that generates singularities -- such as poles associated to on-shell
particles and cuts associated to unitarity -- that are essential to the physical 
interpretation of the theory.  But by the same token, the noncompactness of $\M_\sg$ makes
infrared divergences possible.  In bosonic string theory,
in attempting to evaluate the integral (\ref{turn}), one indeed runs into infrared divergences 
associated to tachyons and massless scalar tadpoles.
The real arena of application of a formula such as (\ref{turn}) is superstring theory, where in an appropriate class of
tachyon-free vacua with spacetime supersymmetry,  a similar formula arises (in a formalism
based on super Riemann surfaces rather than ordinary ones) without the infrared divergences.

In addition, it is a little artificial to focus on the vacuum amplitude (\ref{turn}).  What we really want to compute
is the $S$-matrix, which is obtained by a generalization of what we have so far explained to 
include external vertex operators. We discuss this in section \ref{vertop}.

Here we want to explain a key point that is largely unaffected by these generalizations.  We can
explicitly evaluate the differential form that has to be integrated over $\M_\sg$ using any local section of the
fibration $\JJ\to \M_\sg$, or equivalently any explicit family of metrics $g_{ij}(\sigma;m_1,\dots,m_\p)$ on
$\SIgma$ that depends on modular parameters $m_1,\dots,m_\p$.  Here the $m_s$, $s=1,\dots,\p=\mathrm{dim}\,
\M_\sg$ are local coordinates on $\M_\sg$ that we can interpret as moduli of $\Sigma$, and $g_{ij}(\sigma;m_1,\dots,m_\p)$
is a corresponding family of metrics on $\Sigma$.  (We call this ``a'' corresponding family of metrics, not ``the'' family,
since the metric corresponding to a given point in $\M_\sg$ is only determined up to diffeomorphism and Weyl transformation.)

The family of metrics $g_{ij}(\sigma;m_1,\dots,m_\p)$ defines a local slice transverse to the action of the diffeomorphism
group on $\JJ$, or equivalently a local section $\S$ of the fibration $\JJ\to\M_\sg$.  We do not need to worry about the
global properties of $\S$; the differential form on $\M_\sg$ that we are trying to evaluate is intrinsically defined,
and we can evaluate it  in a given region of $\M_\sg$ using any local slice that we choose.  

Once we specify the metric $g_{ij}$ in terms of the modular parameters $m_1,\dots,m_\p$, we can similarly
express
 $\delta g=\{Q_B,g\}$  in terms of $m_s$ and\footnote{Once
we get down to a finite-dimensional slice, we denote $\{Q_B,m_s\}$ as a one-form $\d m_s$, rather 
than using the notation $\delta m_s$,
which is suggestive of infinite dimensions.}  $\d m_s=\{Q_B,m_s\}$:
\begin{equation}\label{zoboo} \delta g_{ij}=\sum_{s=1}^\p\frac{\partial g_{ij}}{\partial m_s}\d m_s. \end{equation}

Accordingly, we can evaluate the term in the action that involves $\delta g$:
\begin{equation}\label{ormib}\frac{1}{4\pi}\int_\Sigma \d^2\sigma\sqrt g \,\delta g_{ij}b^{ij}=
\frac{1}{4\pi}\sum_{s=1}^\p\d m_s
\int_\Sigma \d^2\sigma \frac{\partial(\sqrt g g_{ij})}{\partial m_s}b^{ij}.\end{equation}

After expressing $F(g|\delta g)$ in terms of $m_s$ and $\d m_s$, $s=1,\dots,\p$, we get a 
differential form $F(m_1,\dots,m_\p|\d m_1,\dots ,\d m_\p)$ that
can be integrated over the variables $m_s$ and $\d m_s$.  From the definition of $F(g|\delta g)$, we have
\begin{equation}\label{logor}F(g|\delta g)=\int\D(X,b,c)\exp\left(-\hat I\right)
=\int\D(X,b,c)\exp\left(-I-\frac{1}{4\pi}\sum_{s=1}^\p\d m_s
\int_\Sigma \d^2\sigma \frac{\partial(\sqrt g g_{ij})}{\partial m_s}b^{ij}\right).\end{equation}
$F(g|\delta g)$ is an inhomogeneous differential form, rather than a form of definite
degree, since the right hand side is not of definite degree in $\d m_1,\dots,\d m_\p$.  The only
part that can be integrated over $\M_\sg$ is the part of top degree $\p$.  This is obtained by expanding the
exponent in (\ref{logor}) and picking out the term that is proportional to each of the odd variables 
$\d m_1,\dots, \d m_\p$:
\begin{equation}\label{ygor}F_{\mathrm{top}}(m_1\dots|\dots \d m_\p)=(-1)^{\p(\p+1)/2}\d m_1\dots \d m_\p
\int \D(X,b,c)e^{-I}\prod_{s=1}^\p \frac{1}{4\pi}\int_\Sigma \d^2\sigma \frac{\partial(\sqrt g g_{ij})}{\partial m_s}b^{ij}.
\end{equation}

In short, for each modulus $m_s$ over which one
wishes to integrate, one must insert in the path integral a factor of
\begin{equation}\label{gor}\Psi_s=\frac{1}{4\pi}\int_\Sigma \d^2\sigma \frac{\partial(\sqrt g g_{ij})}{\partial m_s}b^{ij}.
\end{equation}
This is a standard result (see \cite{GM} or eqn. (5.3.9) of \cite{Polch}).   The integral over $\M_\sg$ is
\begin{equation}\label{zor}(-1)^{\p(\p+1)/2}\int \D(m_1\dots|\dots \d m_\p) \d m_1\dots\d m_\p
\int\D(X,b,c)\exp(-I)\Psi_1\dots\Psi_\p.\end{equation}
Since the $\Psi_i$ are odd variables, we have $\Psi_i=\delta(\Psi_i)$, and we could equivalently write
\begin{equation}\label{zorm}(-1)^{\p(\p+1)/2}\int \D(m_1\dots|\dots \d m_\p) \d m_1\dots\d m_\p
\int\D(X,b,c)\exp(-I)\delta(\Psi_1)\dots\delta(\Psi_\p).\end{equation}  One
can perform the Berezin integral over the odd variables $\d m_s$, $s=1,\dots,\p$ and write an equivalent
formula 
\begin{equation}\label{por}\int \D(m_1,\dots,m_\p)\int \D(X,b,c)\exp(-I)\delta(\Psi_1)\dots\delta(\Psi_\p)\end{equation}
that involves integration over only the $m_s$ and not the $\d m_s$.  This is a more standard version of the formula.
Any of the formulas in this paragraph expresses the form that has to be integrated over $\M_\sg$ to compute the vacuum
amplitude of the bosonic string in terms of correlation functions on the Riemann surface $\Sigma$.    

We used a local
slice in writing the formula, but the formula is independent of the slice.  Concretely, the mechanism by which
this occurs is that the factors $\delta(\Psi_s)$ involve both zero-modes and non-zero modes of $b$, but
when we perform the integral over $c$, the coefficients of non-zero modes are set to zero, as in the discussion following eqn.
(\ref{norko}), and only the zero-modes remain. At that point the 
dependence on the slice drops out.  Hopefully, the reader can see 
that the derivation of (\ref{por}) and the other formulas of the last paragraph is
a close relative of the derivation of eqn. (\ref{norko}).  In each case, the key was the coupling of $\delta g$
to the zero-modes of $b$.  To get eqn. (\ref{norko}), we integrated over those zero-modes to get a function of 
$\delta g$, while to get (\ref{por}), we integrated over $\delta g$ to get a function of the zero-modes.

\subsection{Vertex Operator Insertions}\label{vertop} 

Typically, we want to calculate not the vacuum amplitude (\ref{olf}) but a more general amplitude with insertion of some
function $\Omega(X,b,c,g,\delta g)$:
\begin{equation}\label{zool} F_\Omega(g|\delta g)=\int \D(X,b,c) \,\exp\left(-\hat I(X,b,c,g,\delta g)\right)\,\Omega.\end{equation}
The right hand side of (\ref{zool}) is invariant if we transform all variables $\Phi=X,b,c,g,$ or $\delta g$ by
$\Phi\to (1+\epsilon Q_B)\Phi$, with a small parameter $\epsilon$.  In general, $\Omega$ may not be invariant under this transformation, so 
from this invariance we deduce not that $\d F_\Omega=0$, but that
\begin{equation}\label{ool} \d F_\Omega +F_{Q_B\Omega}=0.\end{equation}

The usual case is that  $\Omega$ is a product of vertex operators $\V_s,\,s=1,\dots,\n$ inserted at points $p_s\in\Sigma$:
\begin{equation}\label{obbo}\Omega=\prod_{s=1}^\sn\V_s(X,b,c;p_s).\end{equation}
The points $p_s$ are often called marked points or punctures.  In this case, we write 
$F_{\V_1,\dots,\V_\ssn}$ for $F_\Omega$.  

\subsubsection{Conformal Vertex Operators}\label{cvo}

We will consider only the simplest class of operators that suffices for computing the $S$-matrix.   These
are operators of the form $\V_s=\t c c V_s$, where $V_s$ is a  (1,1) primary
field\footnote{\label{bozzoc} By an $(n,m)$ primary field, we mean a primary field of holomorphic and antiholomorphic
conformal dimensions $n$ and $m$.}
 constructed from the matter fields only.  The reason that these are the simplest operators to use is as follows.

To express $S$-matrix elements in terms of integrals of naturally-defined differential forms on the appropriate finite-dimensional
moduli space, we will need
the vertex operators to depend only on $c$ and $\t c$ but not their derivatives.  (See section \ref{intmod}.)
An equivalent condition\footnote{This condition may have first been stated in eqn. (5.18) of \cite{operator}.  See also \cite{PN}.}  is that  \begin{equation}\label{morg}b_n\V_s=\t b_n\V_s=0,~~n\geq 0,\end{equation}
 where $b_n$ and $\t b_n$ are the antighost modes
\begin{equation}\label{dobbo}b_n=\frac{1}{2\pi i}\oint \d z\,z^{n+1}b_{zz},~~\t b_n=\frac{1}{2\pi i}\oint \d\t z \,\t z^{n+1} b_{\t z\,\t z}. \end{equation}
If $Q_B\V_s=0$ and $\V_s$ also obeys (\ref{morg}), then $\V_s$ is annihilated for $n\geq 0$
by $L_n=\{Q_B,b_n\}$ and by $\t L_n=\{Q_B,\t b_n\}$, $n\geq 0$.
So $\V_s$ will be a conformal primary of dimension $(0,0)$.

The most simple formalism also represents the external states by vertex operators of ghost number 2; the
usual physical states appear at this ghost number, and this value will lead to differential forms on moduli
space of the right degree to be integrated, as we will see. 
Even without assuming BRST invariance, 
the most general vertex operator of ghost number 2 that does not involve derivatives of
$c$ and $\t c$ is $\V_s=\t c c V_s$, where $V_s$ is constructed from matter fields only.  
Such a $\V_s$ is BRST-invariant if and only if
$V_s$ is a conformal primary field of dimension $(1,1)$, and in this case $\V_s$ is indeed 
a conformal primary of dimension $(0,0)$.  So these are the simplest vertex operators to use and we will
call them conformal vertex operators.

Every physical state of string theory with non-zero momentum\footnote{\label{miss}This assertion 
is part of the BRST version of the no-ghost theorem; see \cite{KO,FGZ,FO,Th}, or, for example,  section 4.2 of \cite{Polch}. (For the original
no-ghost theorem, see \cite{GT,Brower}.)  The usual
proof holds in compactification to $d\geq 2$ dimensions -- or more precisely if the matter
conformal field theory has at least 2 free fields.    In typical bosonic
string compactifications, what we miss by considering only
conformal vertex operators is the ability to separate the zero momentum dilaton from the trace
of the metric.    (See section \ref{remark}.) This  does
 not affect the ability to compute the $S$-matrix.}
can be represented by a conformal vertex operator, so vertex operators of this type suffice for 
computing the $S$-matrix.  However,
the representation of a physical state by a conformal vertex operator is in general not unique.  
A matter primary $V_s$ of
dimension $(1,1)$ may be a null vector, in which case the corresponding vertex operator $\V_s$ is a BRST commutator, $\V_s=\{Q_B,\W_s\}$.
We will have to demonstrate gauge-invariance of the $S$-matrix, by which we mean that a conformal vertex
operator of the form $\V_s=\{Q_B,\W_s\}$ decouples from the $S$-matrix.  In doing so, an important fact, explained in appendix
\ref{really}, will be that
if a conformal vertex operator $\V_s$ is of the form $\{Q_B,\W_s\}$, then $\W_s$ can be chosen to obey the same
conditions as $\V_s$:
\begin{equation}\label{obbox} b_n\W_s=\t b_n\W_s = 0,~~n\geq 0.\end{equation}
This is equivalent to saying that just like $\V_s$,  $\W_s$ is constructed only from $c$ and $\t c$ and not their derivatives.
Given this,  $\W_s$ is a conformal primary of dimension $(0,0)$, just like $\V_s$.
Indeed, we have $L_n\W_s=\{Q_B,b_n\}\W_s=Q_Bb_n\W_s+b_n\V_s$, and this vanishes for $n\geq 0$, since $b_n\V_s=b_n\W_s=0$.

For insertions of conformal vertex operators,  (\ref{ool}) reduces to 
$\d F_{\V_1,\dots,\V_\ssn}=0$, since conformal vertex operators are $Q_B$-invariant not just in the usual sense  but also
in the extended sense.  The usual $Q_B$ transformations for an operator
 constructed just from matter fields $X$, ghosts $c$, and the world sheet metric $g$ -- such as a conformal
 vertex operator $\V_s$ -- are
 \begin{align}\label{ustrans} \delta c^i & = c^j\partial_j c^i \cr
                                               \delta X & = c^j\partial_j X \cr
                                                \delta g_{ij}&=D_ic_j+D_jc_i-g_{ij}D_k c^k. \end{align}                                             
A conformal vertex operator $\V_s(p)$  is invariant under this transformation.  In the extended formalism
that we use here, only one thing is different:  instead of $\delta g_{ij}$ being defined as in (\ref{ustrans}), 
it is treated, off-shell, as an arbitrary
symmetric traceless tensor.  However, on-shell the formula $\delta g_{ij}=D_ic_j+D_jc_i-g_{ij}D_kc^k$ is valid since this
is the equation of motion for the antighost field $b$ derived from the extended action $\h I$ of eqn. (\ref{tolf}).  In general, the equations
of motion may be used in proving the invariance of an operator under a symmetry.  One may
be slightly surprised at the need here to use the antighost equation of motion; however, in standard approaches,
this is necessary after gauge-fixing.  Customarily, after gauge-fixing, $g_{ij}$ is  treated as a 
$c$-number and so $\delta g_{ij}$ is set to 0; after doing so, $Q_B$-invariance of conformal vertex operators
relies on the fact that the equation of motion for $b$ derived from
the ordinary action $I$ is $D_ic_j+D_jc_i-g_{ij}D_kc^k=0$.

\subsubsection{Integration On Moduli Space}\label{intmod}

As we have defined it so far, $F_{\V_1,\dots,\V_\ssn}$ is a differential form  
on the space $\JJ$ of conformal structures on $\Sigma$.
However,  as in section \ref{vacamp}, we can  interpret $F_{\V_1,\dots,\V_\ssn}$ as the pullback of
a differential form on the appropriate moduli space, which in the present case 
is the moduli space $\M_{\sg,\sn}$ that parametrizes a Riemann surface $\Sigma$ of
genus $\g$ with $\n$ punctures $p_1,\dots,p_\sn$.  Let $\Diff_{p_1,\dots,p_\ssn}$ be the group of 
orientation-preserving diffeomorphisms of $\Sigma$
that are trivial at $p_1,\dots,p_\sn$. This group is generated by vector fields 
$v$ on $\Sigma$ that vanish at $p_1,\dots,p_\sn$:
\begin{equation}\label{yort}v^i(p_1)=\dots = v^i(p_\sn)=0.\end{equation}
 $\M_{\sg,\sn}$ is a quotient:
\begin{equation}\label{otonf}\M_{\sg,\sn}=\JJ/\Diff_{p_1,\dots,p_\ssn}.\end{equation}
Just as in section \ref{vacamp}, for $F_{\V_1,\dots,\V_\ssn}(g|\delta g)$ to be a pullback from $\M_{\sg,\sn}$, 
we need two properties:

(1)  $F_{\V_1,\dots,\V_\ssn}(g|\delta g)$ must be $\Diff_{p_1,\dots,p_\ssn}$-invariant.  
This is manifest from the definition.

(2)  $F_{\V_1,\dots,\V_\ssn}(g|\delta g)$ must also be annihilated by contraction with any of the vector 
fields generating $\Diff_{p_1,\dots,p_\ssn}$.
The condition is as in eqn. (\ref{poft}) but now with the external vertex operators included:
\begin{equation}\label{pofit}\int \D(X,b,c)\exp(-\hat I)\prod_{s=1}^\sn\V_s(X,b,c;p_s) 
\int_{\Sigma}\d^2\sigma \sqrt g (D_iv_j+D_jv_i) b^{ij}=0.\end{equation}
To prove this using the change of variables (\ref{orby}), we need to know that the vertex operators 
$\V_s(p_s)$ are invariant under
$c^i\to c^i+\epsilon v^i$, so that including these vertex operators does not modify the previous argument.  
This is true because
$\V_s(p_s)$ is proportional to $c(p_s)$ and $\t c(p_s)$ but not their derivatives,
and moreover $v^i(p_s)=0$ as in (\ref{yort}).

Having understood that $F_{\V_1,\dots,\V_\ssn}(g|\delta g)$ is the pullback of a differential form on $\M_{\sg,\sn}$ -- 
which we denote by the same
name -- we can give  a formal recipe for computing $S$-matrix elements.  
If the vertex operators $\V_s$ all have ghost number 2, then the number of antighost 
insertions needed to get a non-zero path
integral is $6\g -6+2\n$, so that is
the degree of $F_{\V_1,\dots\V_\ssn}$.  But this number is the same
as the real dimension of $\M_{\sg,\sn}$, so formally we can integrate $F_{\V_1,\dots,\V_\ssn}$ over $\M_{\sg,\sn}$:
\begin{equation}\label{bozzo}\langle \V_1\dots \V_\sn\rangle_\sg 
=\int_{\M_{\ssg,\ssn}}F_{\V_1,\dots,\V_\ssn}. \end{equation}
This integral is supposed to give the genus $\g$ contribution to the scattering amplitudes of the string states 
corresponding to $\V_1,\dots,\V_\sn$.

Just as in section \ref{slice}, the  only problem is that $\M_{\sg,\sn}$ is not compact.
The integrals (\ref{bozzo}) are infrared-divergent because 
of tachyons and massless tadpoles, and hence in bosonic string theory the definition of the $S$-matrix is purely
formal. The real arena of application of the ideas that we have explained is superstring theory, 
which admits a similar
formalism,  but with a sensible infrared behavior.  At the most basic
level, the infrared behavior of superstring theory is sensible because supersymmetric theories 
are tachyon-free; more delicate
questions of massless tadpoles will be discussed in section \ref{tadpoles}.

\subsubsection{Gauge Invariance}\label{ginv}

Now let us discuss the gauge-invariance of the formalism.  Gauge-invariance 
of the $S$-matrix means that the integral (\ref{bozzo}) must vanish if we replace one of the vertex operators, say
$\V_1$, with $\{Q_B,\W_1\}$:
\begin{equation}\label{armob}\int_{\M_{\ssg,\ssn}}F_{\{Q_B,\W_1\},\V_2,\dots,\V_\ssn}=0.\end{equation}
To explore this question, we can invoke (\ref{ool}):
\begin{equation}\label{ondo} \d F_{\W_1,\V_2,\dots,\V_\ssn}+F_{\{Q_B,\W_1\},\V_2,\dots,\V_\ssn}=0.\end{equation}
As explained in appendix \ref{really}, we can assume here that $\W_1$ is annihilated by 
$b_n$ and $\t b_n$, $n\geq 0$, or equivalently
that it is constructed from $c$ and $\t c$ but not their derivatives.  This ensures that $\W_1(p_1)$ is invariant
under $c^i\to c^i+\epsilon v^i$, where $v^i$ vanishes at the points $p_s$. So the same 
argument that we have already given
shows that $F_{\W_1,\V_2,\V_3,\dots,\V_\ssn}$ is the pullback of
a differential form on $\M_{\sg,\sn}$, in this case a form of degree $6\g-7+2\n=\mathrm{dim}\,\M_{\sg,\sn}-1$.   
So we can
understand eqn. (\ref{ondo}) as a relation between differential forms on $\M_{\sg,\sn}$, and in particular we have
\begin{equation}\label{tarmob}\int_{\M_{\ssg,\ssn}}F_{\{Q_B,\W_1\},\V_2,\dots,\V_\ssn} 
=-\int_{\M_{\ssg,\ssn}}\d F_{\W_1,\V_2,\dots,\V_\ssn}. \end{equation}

If $\M_{\sg,\sn}$ were compact, we would now invoke Stokes's theorem to 
deduce the vanishing of the right hand side of eqn. (\ref{tarmob}),
and this would establish gauge-invariance. Since actually $\M_{\sg,\sn}$ is non-compact, 
to demonstrate the vanishing of the right hand side of eqn. (\ref{tarmob}), we have to integrate 
by parts and show that
there are no surface terms arising in the infrared region at infinity.  
Because  of the unphysical infrared singularities
of bosonic string theory, the proper framework for this discussion is really superstring theory.
We return to this question in section \ref{massren}, after describing the appropriate superstring measure
and gaining some information about its behavior in the infrared region.

\subsubsection{More General Vertex Operators}\label{morge}

Had we represented our external states by arbitrary $Q_B$-invariant vertex operators,\footnote{Actually, even in
a very general formalism, the vertex operators
should be constrained to be annihilated by $b_0-\t b_0$, for a reason that will appear in the next paragraph. It follows that they are 
also annihilated
by $L_0-\t L_0=\{Q_B,b_0-\t b_0\}$.} 
rather than the special ones we actually used, 
we would have needed a more elaborate formalism \cite{OldPolch,PN}. This does not 
seem necessary for the purposes of the present
paper, so we will just give a few hints of what is  involved.    
If the $\V_s$ (or the gauge parameters $\W_s$ in the above analysis) are $Q_B$-invariant and primary, but 
depend on derivatives of $c$ and $\t c$, we can define $F_{\V_1,\dots,\V_\ssn}$ (or $F_{\W_1,\V_2,\dots,\V_n}$) in 
a conformally invariant fashion as a closed form on the space $\JJ$ of conformal 
structures, but it is not a pullback from $\JJ/\Diff_{p_1,\dots,p_\ssn}$.    However, at a given mass level, the
$\V_s$ and $\W_s$ can only depend on derivatives of $c$ and $\t c$ up to some finite order $N$. So  $F_{\V_1,\dots,\V_\ssn}$ (or $F_{\W_1,\V_2,\dots,\V_\ssn}$) is always a pullback from 
$\JJ/\Diff^{(N+1)}_{p_1,\dots,p_\ssn}$, 
where $\Diff^{(N+1)}_{p_1,\dots,p_\ssn}$ consists of orientation-preserving diffeomorphisms of $\Sigma$ that are
trivial up to order $N+1$ near the $p_s$.  If the $\V_s$ or $\W_s$ are not primary, one must drop the assumption
of conformal invariance and endow $\Sigma$ with a metric (or a suitable system of local parameters) near the punctures
$p_s$.

 In either or both of these cases, one can define $F_{\V_1,\dots,\V_\ssn}$ (or $F_{\W_1,\V_2,\dots,\V_\ssn}$)
as a closed form of the appropriate degree on a space ${\mathcal Z}_{\sg,\sn}$ that is a fiber bundle 
over $\M_{\sg,\sn}$.  Provided that the vertex operators are all annihilated by $b_0-\t b_0$, one can define $\mathcal Z_{\sg,\sn}$ so
that its fibers are contractible (see \cite{PN}).  This ensures that it is possible  to  choose a section $\Upsilon$ of the fibration $\mathcal Z_{\sg,\sn}\to\M_{\sg,\sn}$  and that all such sections are homologous.
One defines the scattering amplitudes by integrating 
$F_{\V_1,\dots,\V_\sn}$ over $\Upsilon$.
This procedure would give a unique and gauge-invariant result, independent of the choice of $\Upsilon$,
if $\M_{\sg,\sn}$ were compact.  So as usual the only real subtleties -- beyond the need for a somewhat intricate formalism -- involve
the behavior in the infrared region at infinity.

\subsubsection{Open And/Or Unoriented Strings}\label{openor}

So far we have considered oriented closed strings only.  The generalizations to open and/or unoriented
strings  do not change much of what we have said
so far.

In a theory of oriented open and closed strings, the string worldsheet $\Sigma$ is 
an oriented two-manifold that may have a boundary.
Open-string vertex operators are inserted at boundary points.  The simplest open-string vertex 
operators are of the form $\U=cU$,
where $U$ is a primary field of dimension 1 constructed from matter fields only.  The worldsheet path
integral may be analyzed along the above lines and -- in the presence of both closed- and open-string vertex operators -- gives rise to a naturally defined differential form on an appropriate 
moduli space of Riemann surfaces with boundary
and in general with both bulk and boundary punctures.  To get a result of this form, one needs the fact that
$b_n\U=0$, $n\geq 0$; this is the case for $\U=cU$.

In a theory of unoriented strings, the only change is that the worldsheet $\Sigma$ is unoriented, and the vertex operators
must be defined in a way that does not depend on a choice of orientation; they must 
be invariant under exchange of local holomorphic
and antiholomorphic variables.   

We will say much more about open and/or unoriented strings in section \ref{anomalies}.
Until that point, we illustrate most ideas with closed oriented strings; we consider open strings
when this is illuminating or there is something distinctive to say.

\subsection{Integrated Vertex Operators}\label{integrated}

\subsubsection{Two Types Of Vertex Operator}\label{persp}

As an important illustration of some of these ideas, we will explain the relation between unintegrated and integrated
vertex operators.  See for example pp. 163-4 of \cite{Polch}.   

Consider a closed-string vertex operator of the form $\V=\t c c V$, where $V$ is a primary of dimension $(1,1)$.  
A primary of dimension $(1,1)$ is a two-form or measure that can be integrated over the worldsheet in a natural
way. So instead of inserting $\V$ at a point in $\Sigma$, we could consider inserting a factor of $\int_\Sigma V$,
which one might prefer to write as $\int_\Sigma\d z\d\t z V_{z\t z}$.  The operator $\V$, inserted at a point in $\Sigma$,
is called the unintegrated form of the vertex operator, while the insertion $\int_\Sigma V$ (or sometimes just the operator
$V$) is called the integrated form.  

Roughly speaking, it is equivalent to insert $\V$ at a point in $\SIgma$ -- the choice of this point is then one of the
moduli over which we integrate in computing the scattering amplitude -- or to insert a factor of $\int_\SIgma V$.

Similarly, for open strings, it is, roughly speaking, equivalent to insert an unintegrated vertex operator $\U=cU$
or an integrated vertex operator $\oint_{\partial \Sigma}U$.  Here $U$ is integrated over the boundary of $\Sigma$
or a component thereof (if several open-string vertex operators are inserted on the same boundary component,
then as usual one can fix their cyclic order).  

  The formalism based on unintegrated vertex operators is always correct when used in conjunction
with the Deligne-Mumford compactification of the moduli space of Riemann surfaces or super Riemann surfaces\footnote{The
Deligne-Mumford compactification for super Riemann surfaces was constructed by Deligne in the 1980's \cite{Deligne}.  This
work is unfortunately unpublished, but has
been presented in a series of lectures \cite{DeligneLectures}.  An introduction to the Deligne-Mumford compactification
can be found, for example, in section  6 of \cite{Wittentwo}.}
and with an appropriate 
infrared regulator.  To make this clear is one of the main goals of the present paper.

\begin{figure}
 \begin{center}
   \includegraphics[width=2.5in]{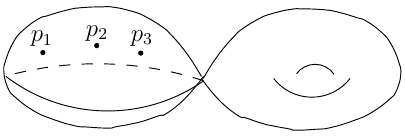}
 \end{center}
\caption{\small On the locus in moduli space at which  naively speaking several punctures in a Riemann surface $\Sigma$ collide, what actually happens
in the Deligne-Mumford compactification is that $\Sigma$ splits off a genus zero component that contains the punctures.
This is illustrated here for the case of three coinciding punctures, labeled $p_1,p_2,p_3$.  The genus zero component in
this example has four distinguished points -- the $p_i$ and the ``node'' at which it meets the other component of $\Sigma$ -- so
it has one complex modulus.  This modulus is lost if one represents the vertex operator insertions at the $p_i$ (or even 2 of them)
in integrated form.  Because of this, the use of integrated vertex operators does not treat correctly questions for which
this region of the moduli space is important.  The difficulties  arise when the total momentum flowing
through the node is on-shell.}
 \label{failint}
\end{figure}

By contrast, relying on integrated vertex operators, though extremely useful in practice, will run into trouble for
certain questions involving behavior at exceptional momenta, including but not limited to
questions about mass renormalization and  massless tadpoles.
(For some examples, see \cite{DIS,ADS,GrSe}; the relevant models were studied at the level of low energy
effective field theory in \cite{DSW}.)  The basic reason for this is illustrated in fig. \ref{failint}: the Deligne-Mumford compactification
is constructed in such a way that  external vertex operators never collide, while collisions of vertex operators
occur  if one represents an external
string state by an integrated insertion $\int_\Sigma V$ (the integral is over all of $\Sigma$ including the locations of
other vertex operators).  Accordingly, the map from integrated to unintegrated vertex operators,
which we derive in section \ref{horsy}, works only over the uncompactified moduli space $\M_{\sg,\sn}$, where all punctures are distinct, and does not
give a good treatment of the compactification. 

Still, the formalism with integrated vertex operators is important as it is usually the simplest approach
 when it is applicable and it is the basis for many
practical calculations.  A sufficient criterion that, with minor extensions, accounts for most successful
uses of integrated vertex operators is that 
one may use integrated vertex operators if the external
momenta are sufficiently generic that the sum of the momenta in any set of colliding vertex operators is off-shell.\footnote{For example, in computing a genus 0 scattering amplitude, three vertex operators are fixed and one usually uses
the integrated form of all other vertex operators.  If the three fixed vertex operators have reasonably generic momenta,
then the criterion just stated is satisfied. This justifies the usual computation of tree-level scattering
amplitudes.}
This ensures that one can  compute by analytic continuation from a region of external momenta in which
the subtleties of the compactification are irrelevant.

\subsubsection{The Derivation}\label{horsy}

Let $z$ and $\t z$  be  local holomorphic and antiholomorphic coordinates on $\Sigma$. We could insert a vertex operator $\V=c\t c V$ at any point 
\begin{equation}\label{tello}z=\z,~~~~\t z=\t \z.\end{equation}
We are going to evaluate the formula (\ref{ormib}) -- or any of the related formulas of section \ref{slice} -- with two of the moduli $m_s$ of $\Sigma$
being the choice of $\z, \,\t \z$.

In the formalism of section \ref{vertop}, we are instructed to keep fixed the point $p$ at which $\V$ is inserted   
and instead vary the metric of $\Sigma$.  How then do we vary $\z$ and $\t \z$ keeping everything else fixed?
We do this by letting the metric $g_{ij}$ of $\Sigma$ depend on $\z$ and $\t\z$ in a way that can be 
removed by a diffeomorphism,
but not by a diffeomorphism that leaves fixed the point $p$.  

We will make a diffeomorphism between two different parametrizations of $\Sigma$, which we will call {\it (i)} and {\it (ii)}.  In
{\it (i)}, we describe $\Sigma$ by local complex coordinates $w, \,\t w$ (which for
what follows may as well be complex conjugates).   We suppose that a vertex operator $\V=\t c c V$ is
inserted at the point $w=\t w=0$, which we call $p$.     We assume also that the parameter $w$ is well-defined in the disc $|w|\leq 1$,
and that no vertex operators other than $\V$ are inserted in this disc.  Here $|w|=\sqrt{w\t w}$.  In general, $w$ is a complex parameter, not necessarily holomorphic,
but it will be a holomorphic parameter near $w=0$.

In description {\it (ii)}, we use holomorphic and antiholomorphic local coordinates $z$ and $\t z$ (which may as well be complex conjugates).
 The two descriptions will be related by a diffeomorphism that depends
on a parameter $\z$ -- which will be one of the moduli of $\Sigma$ -- and which we will restrict to the region $|\z|\leq \epsilon$ for some
small $\epsilon$.  We want
\begin{equation}\label{onfov} z=\begin{cases}  w+\z   & \mbox{if}  ~ |w|<\epsilon \\ w &\mbox{if}~ |w|>1-\epsilon. \end{cases}\end{equation} 
For example, we can have
\begin{equation}\label{tonfo} z= w +f(|w|)\z ,\end{equation}
where $f(|w|)$ is a monotonic function that equals 1 for $|w|<\epsilon$ and 0 for $|w|>1-\epsilon$.  
We similarly express $\t z$ in terms of $w$ and $\t w$ and also $\t \z$:
\begin{equation}\label{itonfo} \t z= \t w +f(|w|)\t \z .\end{equation}
The coordinate transformations (\ref{tonfo}) and (\ref{itonfo}) are not holomorphic in general, though they are holomorphic when restricted
to the region $|w|\leq \epsilon$.  So although we assume $z$ to be a holomorphic parameter, $w$ is only holomorphic for $|w|<\epsilon$.

In description {\it (ii)}, the vertex operator $\V$ is inserted at a $\z$-dependent position, namely $z=\z$, $\t z=\t \z$.  But we 
assume that in description {\it (ii)}, the metric
 of $\Sigma$ is independent of $\z$.  In description {\it (i)}, $\V$ is inserted at $w=\t w=0$, independent of $\z$, but
  the metric of $\Sigma$ will then have to depend on $\z$ and $\t \z$.  Since this $\z$ and $\t \z$-dependence of the metric could be removed by using
  the coordinates $z,\t z$ instead of $w,\t w$, it is of the ``pure gauge'' form
  \begin{align}\label{poffo}\partial_{\z}g_{IJ} & = D_I v_J+D_Jv_I \cr
      \partial_{\t \z}g_{IJ}& = D_I\t v_J+D_J\t v_I,\end{align}
                                         for  some vector fields $v^I$ and $\t v^I$.  
                                         In these formulas, one can consider the indices $I,J$ to take the values
                                         $w$, $\t w$.  We can describe 
                                         $v$ by its components $(v^w,v^{\t w})$ in this basis.  
                                         Up to sign,\footnote{To understand the sign, observe that a function $h$ that depends only on $z$ and $\t z$ is annihilated by $\partial_\z+
                                         \partial_\z w|_{z,\t z}\partial_w +\partial_\z \t w|_{z,\t z}\partial_{\t w}$, and so obeys $\partial_\z h=-
                                         ( \partial_\z w|_{z,\t z}\partial_w +\partial_\z \t w|_{z,\t z}\partial_{\t w})h$, with a minus sign.  In (\ref{imx}),
                                         we  are applying
                                         the same idea for a metric that depends only on $z$ and $\t z$.}
                                          these components
                                         are just the derivatives of $(w,\t w)$ with respect to $\z$ at fixed $z,\t z$, since those derivatives determine
                                         how $w,\t w$ must vary with $\z$ to keep $z,\t z$ fixed:
          \begin{equation}\label{imx}(v^w,v^{\t w})=-\left(\left.\partial_\z\right|_{z,\t z}w,\left.\partial_\z\right|_{z,\t z}\t w\right).\end{equation}
To evaluate these derivatives in general, we would need to differentiate the function $f(|w|)$ that appears in eqns. (\ref{tonfo}) and (\ref{itonfo}).
But we are only interested in what happens near $w=0$.  For $|w|<\epsilon$, we have $z=w+\z$, $\t z=\t w+\t \z$, so the evaluation
of the derivatives in (\ref{imx}) just gives
\begin{equation}\label{jimx} (v^w,v^{\t w})=(1,0).\end{equation}
Similarly, near $w=0$, we have
\begin{equation}\label{wimx}(\t v^w,\t v^{\t w})=(0,1).\end{equation}

Now we want to evaluate the measure of integration over the modular parameters $\z$ and $\t \z$. For this, we use description {\it (i)},
since it satisfies the assumptions of our 
 general formalism that the metric of $\SIgma$ varies but the insertion points of vertex operators are held fixed.
The measure for integration over $\z$ and $\t \z$ is
\begin{equation}\label{uryt}\d \z\,\d \t\z \,\,\Psi_\z\,\Psi_{\t\z}, \end{equation}
where we evaluate (\ref{gor}) to get
\begin{equation}\label{zorr}\Psi_\z=\frac{1}{2\pi}\int_\Sigma \d w\,\d\t w\sqrt g b^{ij}D_iv_j,~~\Psi_{\t\z}=\frac{1}{2\pi}\int_\Sigma \d w\,\d\t w\sqrt g b^{ij}D_i\t v_j.\end{equation}

We have to evaluate the product
\begin{equation}\label{buryt}\Psi_\z\Psi_{\t \z} \,\t c c V(0).\end{equation}
with an insertion of $\V=\t c c V$ at $w=\t w=0$.
If we had no insertion $\t cc V(0)$, we would simply say that $\Psi_\z=\Psi_{\t\z}=0$ after integrating by parts
and using the equation of motion $D_Ib^{IJ}=0$.  However, with an insertion of $c $ or $\t c$, we actually
run into a delta function contribution\footnote{The statement that there is such a delta function is
a variant of the explanation in eqn. (\ref{orby}).} $(\sqrt g/2\pi)D_i b^{ij}(p')c^k(p)=\delta^{jk}\delta(p'-p)$.  Evaluating $\Psi_\z$ and
$\Psi_{\t\z}$ with the help of these delta functions, we find that we can replace $\Psi_\z\Psi_{\t \z}\t c c V(p)$
with $v^w\t v^{\t w}V(p)=V(p)$, where in the last step we used (\ref{jimx}) and (\ref{wimx}).  So we can replace $\d \z\, \d \t\z \Psi_\z
\Psi_{\t\z}\t c c V(p)$ by $\d\z\, \d\t \z V(p)$.  (There are no such contributions for other vertex operators, since by hypothesis there are no other
vertex operators inserted in the region where the metric in description {\it (i)} depends on $\z$ and $\t \z$.)

In description {\it (i)}, the integral over $\z $ and $\t \z$ will not be very transparent because the metric of $\Sigma$
depends on $\z$ and $\t \z$.  However, at this point, we can go over to description {\it (ii)}, in which the metric of $\Sigma$ does
not depend on $\z$ and $\t \z$.  In that description, the point $w=\t w=0$ is mapped to $(z,\t z)=(\z,\t\z),$ so
the vertex operator insertion is $V(\z,\t\z)$
and the dependence on $\z$ and $\t \z$ is entirely contained in the expression $\d\z\,\d\t\z\,V(\z,\t\z)$.  So the dependence
of the scattering amplitude on the insertion of the vertex operator $\V$ that we have been analyzing comes entirely
from a factor
\begin{equation}\label{obo}\int_\Sigma \d \z\,\d\t \z\,V(\z,\t \z). \end{equation}
We have to include this factor in addition to whatever other vertex operator insertions there may be, and in addition
to whatever other moduli $\Sigma$ may have.  We have arrived at the description of the $S$-matrix in terms
of an integrated vertex operator.

In this derivation, we chose local coordinates and considered only a small region $|\z|<\epsilon$.  
To justify the statement that the integral in (\ref{obo}) should extend over all of $\Sigma$, we simply cover $\Sigma$ with small open
sets in each of which we can proceed in the fashion just described, using convenient local parameters.  (Of course, there is a potential
problem when the vertex operator $\V$ meets another vertex operator.  A sufficient criterion for avoiding trouble was stated at the
end of section \ref{persp}.)

Obviously, in this derivation we assumed that the position at which $\V$ is inserted on 
$\Sigma$ is one of the moduli of the problem.
There are a few exceptional cases in which this is not true -- the three-point function in genus 0, and the one-point function in
genus 1.  Those examples are simple enough to be treated by hand and lead to natural formulas involving correlation
functions of the integrated 
vertex operators, even though the above derivation does not apply.  For example, for the three-point
function in genus 0, there are no moduli at all; one simply has to evaluate the three-point function $\langle \t ccV_1(z_1,\t z_1)
\t c c V_2(z_2,\t z_2)\t c cV_3(z_3,\t z_3)\rangle$, which factors as $ \langle \t c c(z_1,\t z_1)
\t cc(z_2,\t z_2)\t c c(z_3,\t z_3)\rangle
\langle V_1(z_1,\t z_1)
V_2(z_2,\t z_2)V_3(z_3,\t z_3)\rangle$.  So the result can be conveniently expressed in terms of the three-point functions
of the ``integrated'' vertex operators $V_i$, even though in this example those operators are not integrated.

\subsubsection{More On Gauge Invariance}\label{moregauge}

Let us reconsider gauge invariance -- that is, the decoupling of BRST-trivial states -- from the point of 
view of integrated vertex operators.
The integrated vertex operator $V$ is a matter primary field.  We expect it to decouple if $V$ is a null vector.  

The cases of massless and massive particles are rather different, as we will see momentarily.  A null vector that represents a pure gauge mode of a massless particle is necessarily of the form\footnote{$\Phi_{n,m}$ will denote a matter primary field of antiholomorphic and holomorphic
conformal dimensions $n$ and $m$. The
simplest null vectors are treated in appendix  \ref{really}, with the antiholomorphic modes suppressed.}
 $V=L_{-1}\Phi_{1,0}$ or
$V=\t L_{-1}\Phi_{0,1}$.  We have $L_{-1}\Phi_{1,0}=\partial_z\Phi_{1,0}$, $\t L_{-1}\Phi_{0,1}=\partial_{\t z}\Phi_{0,1}$, leading to
 $\int_\Sigma V=\int_\Sigma \partial_z\Phi_{1,0}$ or $\int_\SIgma V=\int_\Sigma\partial_{\t z}\Phi_{0,1}$.
In each case, assuming for the moment that there are no surface terms,
$\int_\Sigma V=0$ vanishes since $V$ is a total derivative, or equivalently an exact form.  

For massive particles, the story is less simple.  At a massive level, a null vector can be, for example, of the form
$V=(L_{-2}+(3/2)L_{-1}^2)\Phi_{1,-1}$.  This is not a total derivative on the Riemann surface $\Sigma$, so it is not
 true that $\int_\Sigma V$ vanishes.  Rather, the null vector decouples upon performing the full integral over
the moduli space $\M_{\sg,\sn}$ of Riemann surfaces of genus $\g$ with $\n$ punctures.  However, it is clumsy to make
this argument in terms of integrated vertex operators.  It is much more effective to go back to the unintegrated vertex
operator $\V=\t c cV$, which is of the form $\{Q_B,\W\}$ if $V$ is a null vector.  Then as in section \ref{ginv}, the BRST
machinery shows that the coupling of $\{Q_B,\W\}$ involves an exact form on $\M_{\sg,\sn}$, and the decoupling of $\{Q_B,\W\}$
follows by integration by parts on $\M_{\sg,\sn}$.  

In each of these cases, one needs to analyze possible surface terms at infinity in the integral over $\Sigma$
or over $\M_{\sg,\sn}$.  In closed-string theory,\footnote{The restriction
to closed-string theories is important here because if $\Sigma$ has a boundary, then
integration by parts on $\Sigma$ can certainly produce a boundary term, giving a simple mechanism for gauge symmetry
breaking  (see section \ref{burgo}).  This is not a surface term at infinity in the everyday sense, though it has
that interpretation in the Deligne-Mumford compactification (see section \ref{strapple}). The argument given in the text
does apply to open-string gauge invariances in theories of open and closed strings, since the boundary of a string worldsheet
itself has no boundary.}
 in
 the integral over $\Sigma$, by ``infinity'' one means points at which $V$ meets another
vertex operator.  A simple criterion for ensuring that there can be no difficulty is that the external momenta are sufficiently
generic in a sense stated in the last paragraph of section \ref{persp}.   This case suffices for computing the
$S$-matrix, so  for massless modes of the closed
bosonic
string, there are no anomalies in BRST-symmetry.
Another way to reach this conclusion is to observe that a collision of two vertex
operators can already occur in genus zero, so whatever effects it produces are already included in the definition of
the tree-level BRST symmetry.  A related statement from a spacetime point of view is that in closed-string theories, loop effects do not trigger spontaneous
breaking of gauge symmetries that can be proved by integration by parts on $\Sigma$.  We elaborate on this  in sections
\ref{massren} and \ref{tadpoles}.  

Matters are different for massive modes. Gauge-invariance for massive modes has to be proved by integration by
parts on  $\M_{\sg,\sn}$.  In that context,
``infinity'' is associated to all possible degenerations of $\Sigma$ that occur in the Deligne-Mumford compactification
of $\M_{\sg,\sn}$ (see for example section 6 of \cite{Wittentwo} for an introduction). In this case, a full treatment involves
a variety of issues that we will attempt to elucidate in section \ref{massren}.

The distinction between the two cases of massless and massive null vectors can be phrased as follows.  There is a forgetful
map  $\pi:\M_{\sg,\sn}\to\M_{\sg,\sn-1}$ that forgets one of the punctures.  This map is a fibration; 
the fiber is a copy of $\Sigma$, parametrizing the position of the puncture
that one is forgetting:
\begin{equation}\label{hobbo}\begin{matrix} \Sigma& \to & \M_{\sg,\sn}\cr && 
\downarrow \pi\cr & & \M_{\sg,\sn-1}.\end{matrix}\end{equation}
The use of the integrated vertex operator amounts to performing an integral over $\M_{\sg,\sn}$ by integrating first
over the fibers of this fibration.  The integral over $\M_{\sg,\sn}$ that represents the coupling of a massless null vector
vanishes upon integration over the fibers of the fibration (\ref{hobbo}), because in this case $V$ is a 
total derivative on $\Sigma$.
By contrast, integration over the fibers of $\pi$ does not help much in understanding the coupling of a massive null vector;
in that case, the only simplicity arises upon performing the full integral over $\M_{\sg,\sn}$.

What we have just described is a close analog of a phenomenon that occurs in superstring theory and accounts for
some of the subtlety of that subject.  In superstring theory, $\Sigma$ becomes a super Riemann surface.  As long
as one considers only Neveu-Schwarz vertex operators, there is a fibration like (\ref{hobbo}).  Integration over the fibers
of this fibration suffices to establish the decoupling of massless null vectors in the NS sector.  For Ramond vertex operators,
however, there is no fibration analogous to (\ref{hobbo}); this is because a Ramond vertex operator is inserted at a singularity
of the superconformal structure of $\Sigma$, and it does not make sense to move such a vertex operator while keeping fixed
the moduli of $\SIgma$.  Accordingly, in the Ramond sector, even for massless states -- gravitinos -- 
gauge-invariance can only be established by  a full integral over the supermoduli space, not just an integral over 
the worldsheet.  This fact was one source of difficulty in the literature of the  1980's.  

An important point is that as long as one is only forgetting one puncture, the fibration (\ref{hobbo}) extends over the 
Deligne-Mumford compactifications of
$\M_{\sg,\sn}$ and $\M_{\sg,\sn-1}$.  For an analogous fibration that forgets two or more punctures, this would not be true.
A locus in which three or more punctures collide needs to be blown up to get the Deligne-Mumford compactification
(this was depicted in fig. \ref{failint}).  The three collliding vertex operators can be, for example,
two that are expressed in integrated form and a third that is either integrated or unintegrated.
  Accordingly, the
use of integrated vertex operators  leads to difficulty in some calculations, as already remarked in section \ref{persp}.

\subsection{Some Variations On The Theme}\label{another}

Since an odd variable is its own delta function, instead of writing the vertex operator as $\V=\t c cV$, we can write
\begin{equation}\label{moby}\V=\delta(\t c)\delta(c) V. \end{equation}
This will be a good starting point for understanding superstring vertex operators.

The delta functions in (\ref{moby}) have an intuitive explanation.  The ghost field $c^i$ is a vector field on $\Sigma$ -- a generator of the diffeomorphism
group of $\Sigma$.  In the presence of an unintegrated vertex operator insertion at a point $p\in\Sigma$, we want to consider as symmetries only those diffeomorphisms
that leave $p$ fixed.  They are generated by vector fields that vanish at $p$.
So we want to set the ghost fields $\t c,\, c$ to zero at $p$.  This is accomplished by the delta functions.

Another useful variation on what we have explained is as follows.
A metric $g$ on an oriented two-manifold $\Sigma$ determines a complex structure $J$.  Instead of using 
the path integral to define a differential
form $F(g|\delta g)$ on the space $\JJ$ of all metrics on $\Sigma$, we could have used it to define a differential form 
$F(J|\delta J)$ on the space $\ZZ$ of all complex
structures on $\Sigma$.  In fact, this has several advantages, including manifest conformal invariance.

If we work with variations of complex structures rather than  metrics, then eqn. (\ref{wonk}) for the variation of the action
becomes, in local complex coordinates,
\begin{equation}\label{ofox} \delta I=\frac{-i}{4\pi}\int_\Sigma \d\t z \,\d z\left( \delta J_{\t z}^zT_{zz}-\delta J_{z}^{\t z}T_{\t z \t z}\right),
\end{equation}
and similarly eqn. (\ref{tolf}) for the extended action becomes
\begin{equation}\label{bolf}\hat I=I+\frac{-i}{4\pi}\int_\Sigma \d\t z\,\d z\left(\delta J_{\t z}^zb_{zz}-\delta J_z^{\t z}b_{\t z\t z}\right).
\end{equation}
From this starting point, the  analysis proceeds in an obvious way.  
The complex structure of $\M_{\sg,\sn}$ is more evident in this description.
$\delta J_{\t z}^z$ is a form of type $(1,0)$ on the space of complex structures and $\delta J_z^{\t z}$ is a form of type $(0,1)$.
After dividing by diffeomorphisms (which act holomorphically on the space of complex structures), $\delta J_{\t z}^z$
and $\delta J_z^{\t z}$ descend, respectively, to $(1,0)$- and $(0,1)$-forms on $\M_{\sg,\sn}$.

For bosonic strings, it does not much matter if one works with metrics or with complex structures.  In the generalization to
superstrings, the description in terms of complex structures is more efficient.

\section{A Measure On Supermoduli Space}\label{measure}

In this section, we will adapt the derivation of section \ref{bosmeasure} to superstring theory.  
For this, one simply replaces Riemann surfaces by super Riemann surfaces, 
and  extends the  stress tensors, ghosts, and antighosts
of section \ref{bosmeasure}  to supermultiplets.  All of the bosonic formulas have natural superanalogs as long
as one has the courage to write them down.

To appreciate the construction of the superstring measure, one has to become comfortable with interpreting the integral over
certain even variables
(the commuting ghosts and the differentials of odd moduli)
 as an algebraic operation -- a slightly generalized Gaussian integral -- rather than interpreting it
literally as an integral.   This  is natural in the general theory of integration over supermanifolds;
see for example section 3.3.2 of \cite{Wittenone}.   One also has to become comfortable with certain delta function
operators (aspects of which were introduced in \cite{EHVerl} and \cite{superoperator})
 that may be slightly off-putting at first sight, but that on further reflection make perfect sense.
In the present section, we make only the minimum necessary remarks on these matters, deferring more detail on
the commuting ghost system to section \ref{betag}.  

One difference between superstrings and bosonic strings is that technically in the superstring case, it seems simpler to
express the construction in terms of complex structures (as in section \ref{another}) rather 
than in terms of metrics (as in the rest of section
\ref{bosmeasure}).

For brevity, we concentrate here on the heterotic string, in which only the 
right-moving modes have superconformal
symmetry. The generalization to superstring constructions in which both left- and right-movers have superconformal 
symmetry should be clear and occasionally we make a few remarks.

\subsection{A Short Account Of Super Riemann Surfaces}\label{introstring}

We begin with a bare minimum background from super Riemann surface theory.  Many topics are treated
much more fully in  \cite{Wittentwo}.  

\subsubsection{Worldsheets}\label{worldsheets}

The worldsheet $\Sigma$ of a heterotic string has a local antiholomorphic coordinate $\t z$ and local holomorphic coordinates 
$z|\theta$.
From a holomorphic point of view, $\Sigma$ is endowed with a superconformal structure, which in suitable coordinates -- called
superconformal coordinates -- is determined by the operator
\begin{equation}\label{izzop}D_\theta=\frac{\partial}{\partial \theta}+\theta\frac{\partial}{\partial z}.\end{equation} 
We are really not interested in $D_\theta$ but in the line bundle  $\D$ whose holomorphic sections are of the form $f(z|\theta)D_\theta$;
this is a holomorphic subbundle of what we will call $T_R\Sigma$, the holomorphic tangent bundle of $\Sigma$.
$T_R\SIgma$ 
is generated by $\partial_z$ and $\partial_\theta$.  Similarly, we write $T_L\SIgma$ for the antiholomorphic tangent bundle,
which is generated by $\partial_{\t z}$;  $T_L^*\SIgma$ for the antiholomorphic cotangent bundle, generated by $\d\t z$;
and $T_R^*\SIgma$ for the holomorphic cotangent bundle, generated by $\d z$ and $\d\theta$.  We also call $T_L^*\Sigma$
and $T_R^*\Sigma$ the spaces of $(0,1)$-forms and $(1,0)$-forms on $\Sigma$.  

The intuition that holomorphic and antiholomorphic variables on $\Sigma$ are independent of each other can be
captured by thinking of 
$\SIgma$ as a smooth submanifold of a complex supermanifold $\Sigma_L\times \SIgma_R$, where $\Sigma_L$ is
an ordinary Riemann surface, $\Sigma_R $ is a super Riemann surface, and $\Sigma_L$ is very close to the complex
conjugate of the reduced space $\SIgma_{R,\red}$ of $\SIgma_R$.  (See section 5 of \cite{Wittenone}.) 
From that point of view, holomorphic and antiholomorphic functions on $\Sigma$
are restrictions to $\Sigma$ of holomorphic functions on $\Sigma_R $ and $\SIgma_L$; similarly $T_R\Sigma$ and $T_L\Sigma$
are restrictions to $\Sigma$ of the holomorphic tangent bundles $T\Sigma_R$ and $T\SIgma_L$, respectively.  This gives
the most precise approach to a heterotic string worldsheet and we will adopt this point of view  in the present
paper when needed.

The simplest type of holomorphic field on $\Sigma$ is a field  $\Phi^{[n]}$  that is a section of $\D^n$.   In 
superconformal coordinates, such a field has an expansion
\begin{equation}\label{polko}\Phi^{[n]}=u+\theta v,\end{equation}
where $u$ has conformal dimension $-n/2$ and $v$ has conformal dimension $-n/2+1/2$.

A coordinate transformation is called superconformal if it  multiplies $D_\theta$
by a scalar function (and hence preserves the line bundle $\D$ generated by $D_\theta$).  Superconformal coordinates
are only unique up to a superconformal transformation.  A vector field generates
a superconformal transformation -- and we call it a superconformal vector field -- if its commutator with $D_\theta$ is
a ($z|\theta$-dependent) multiple of $D_\theta$.   
Odd and even superconformal vector fields take the form
\begin{align}
\label{durov} \nu_f&=f(z)\left(\partial_\theta-\theta\partial_z\right)\cr V_g&=g(z)\partial_z+\frac{g'(z)}{2}\theta\partial_\theta,  \end{align}
with holomorphic functions $f(z),$ $g(z)$.  These functions can be combined to a superfield $\VV^{[2]}(z|\theta)=g(z)+2\theta f(z)$, which
is a section of $\D^2$.  

The following are a basis of superconformal vector fields that are regular except possibly for a pole at $z=0$:
\begin{align}\label{urv} G_r& = z^{r+1/2}\left(\partial_\theta-\theta\partial_z\right),~~~~~~~~~~~~~    r\in \Z+1/2 \cr
                                     L_n&= -z^{n+1}\partial_z-\frac{1}{2}(n+1)z^n\theta\partial_\theta, ~~~n\in \Z.\end{align}   
The pole is absent if
\begin{equation}\label{oggo} r\geq -1/2,~~~n\geq -1.\end{equation}                                                                       
A short calculation shows that the vector fields (\ref{urv}) obey the super Virasoro algebra (with zero central charge)
in the Neveu-Schwarz (NS) sector:
\begin{align}\label{rov} [L_m,L_n]& =(m-n)L_{m+n}\cr \{G_r,G_s\}&=2L_{r+s} \cr [L_m,G_r]&=\left(\frac{m}{2}-r\right) G_{m+r}.\end{align}
     
If $z|\theta$ are local superconformal coordinates and $\t z$ is an antiholomorphic coordinate that is sufficiently close
to the complex conjugate of $z$, we call $\t z;\neg z|\theta$ a system of standard local coordinates.  (We do not
ask for $\t z$ to equal the complex conjugate of $z$ as this condition is not invariant under superconformal transformations.
See section 5 of \cite{Wittenone} for more discussion.)

We will use the term supercomplex structure to refer to the complex structure of $\Sigma$ plus its holomorphic superconformal
structure (the choice of the line bundle $\D$ generated by $D_\theta$).  Just as on an ordinary complex manifold,
the complex structure can be defined by a linear transformation $\J$ of the cotangent bundle of $\Sigma$
that obeys $\J(\d z)=i\d z$, $\J(\d\theta)=i\d\theta$, $\J(\d\t z)=-i\d\t z$.  These conditions can be described by saying
that $\d z$ and $\d \theta$ are one-forms of type $(1,0)$, and furnish a basis of the holomorphic cotangent bundle $T_R^*\Sigma$,
while $\d\t z$ is of type $(0,1)$, and furnishes a basis of the antiholomorphic cotangent bundle $T_L^*\SIgma$.  
$\J$ obeys $\J^2=-1$, just as on an ordinary complex manifold, and also is compatible with the existence of a holomorphic
superconformal structure, as described above.  The space of all such $\J$'s is an infinite-dimensional complex supermanifold
$\JJ$.

\subsubsection{Deformations}\label{deformational}

To construct
superstring perturbation theory, we have to study deformations of $\SIgma$.  {\it A priori}, we have to consider
both deformations of the complex structure $\J$ of $\Sigma$ and deformations of its superconformal structure, that is
deformations of
the embedding of $\D$ in $T_R\Sigma$.  We will see shortly that there are no nontrivial deformations of the
superconformal structure without changing the complex structure, so we really only need to consider deformations of $\J$.
Of course, we only
 allow deformations of $\J$ that preserve
the existence of a holomorphic superconformal structure. 
For our purposes in this paper, what we need to know can be summarized as follows (see for example sections 3.5.3-4 of
\cite{Wittentwo}).  The allowed
variations of   $\J$ are determined\footnote{\label{blogo} The constraint $\J^2=-1$ implies the vanishing
of $\delta\J^\theta_\theta$, $\delta\J^\theta_z$, $\delta\J^z_\theta$, $\delta\J^z_z$, and $\delta\J^{\t z}_{\t z}$.  The superconformal
structure determines $\delta\J_{\t z}^\theta$ in terms of $\delta\J_{\t z}^z$ and $\delta\J_z^{\t z}$ in terms of $\delta\J_\theta^{\t z}$.} by the components $\delta \J_{\t z}^z$ and $\delta\J_\theta^{\t z}$ which respectively
represent deformations of the holomorphic and antiholomorphic structure of $\Sigma$.  
Their $\theta$ expansions read
\begin{align}\label{hosetof}\delta\J_{\t z}^z&=h_{\t z}^z+\theta \chi_{\t z}^\theta \cr \delta\J_\theta^{\t z}&= e_\theta^{\t z}+\theta h_z^{\t z}.\end{align}
In the first equation, in conventional language, $h_{\t z}^z$ is a metric perturbation and $\chi_{\t z}^\theta$ is the gravitino field.  In
the second equation, $h_z^{\t z}$ is a metric perturbation while $e_\theta^{\t z}$ can be set to zero as a gauge condition
(using the $\theta$-dependent part of $q^{\t z}$ in eqn. (\ref{itzo}) below).
Geometrically, $\delta\J_{\t z}^z$ is a $(0,1)$-form on $\Sigma$
with values in $\D^{2}$, and $\delta\J_\theta^{\t z}$ is a section of $\D^{-1}$ with values in $T_L\Sigma$.              
 For any point $p\in\Sigma$, we can interpret\footnote{This is precisely analogous to what happens on an ordinary Riemann
 surface with complex structure $J$: the variation $\delta J_{\t z}^z$ is a $(1,0)$-form on the space of complex structures,
 while $\delta J_z^{\t z}$ is a $(0,1)$-form.}   $\delta\J_{\t z}^{z}(p)$  as a $(1,0)$-form on $\JJ$, the space of supercomplex
 structures, 
and $\delta\J_\theta^{\t z}(p)$ as a $(0,1)$-form on $\JJ$.  

Now we have to take account of diffeomorphisms.
Deformations of $\SIgma$ that are generated by a vector field on $\Sigma$ -- the generator of an infinitesimal diffeomorphism of $\Sigma$
-- are uninteresting in a diffeomorphism-invariant worldsheet theory.  
A general vector field takes the form
\begin{equation}\label{vectorfields} q^{\t z}\partial_{\t z}+\left(q^z\partial_z+\frac{1}{2}D_\theta q^z
D_\theta\right)+q^\theta D_\theta.\end{equation}
(The advantage of writing the expansion this way is explained in \cite{Wittentwo}.)
The $q^\theta D_\theta$ term is used to eliminate deformations of $\SIgma$ in which the embedding of 
$\D$ in $T_R\Sigma$ is  changed.  This works as follows.  The change in $D_\theta$ generated by $q^\theta D_\theta$ is
\begin{equation}\label{duffo}\delta D_\theta =[D_\theta, q^\theta D_\theta]=(D_\theta q^\theta) D_\theta -2q^\theta \partial_z.\end{equation}  On the right hand side, the first term proportional to $D_\theta$ does not change the line bundle $\D$
generated by $D_\theta$, but the second term $-2q^\theta \partial_z$, since $q^\theta$ is an arbitrary function, 
represents an arbitrary change in the embedding of
$\D$ in $T_R\Sigma$.  ($T_R\Sigma$ is generated by $D_\theta$ and $\partial_z$, while $\D$ is generated by $D_\theta$,
so to change the embedding of $\D$ in $T_R\Sigma$, one must shift  $D_\theta$ by a multiple of $\partial_z$.)  Thus, we need only consider deformations of the complex structure $\J$ of $\Sigma$.

Given this, the nontrivial deformations of $\Sigma$ are deformations of $\J$ modulo deformations generated by  
$q^{\t z}$ and $q^z$.
Such deformations take the form 
\begin{align}\label{itzo}\delta\J_{\t z}^z &= \partial_{\t z}q^z\cr
                                          \delta\J_\theta^{\t z}&= D_\theta q^{\t z}.\end{align}
               The moduli of $\Sigma$ over which one ultimately has to integrate in order to compute scattering amplitudes
correspond to deformations of $\J$ modulo these trivial ones.                           

\subsubsection{Action And Ghosts}\label{actionghosts}

Now consider a superconformal field theory on $\Sigma$ with action $I$.  In superstring theory, $I$ is the sum of a matter
action that we call $I_\XX$, where $\XX$ refers generically to all matter fields on the worldsheet, and a ghost action
$I_\gh$.  When we vary the supercomplex structure of $\Sigma$, $I$ changes by an amount proportional to the
supercurrent and stress tensor of $\Sigma$.  In heterotic string theory, the formula reads
\begin{equation}\label{bitzo} \delta I=  \frac{1}{2\pi}\int_\Sigma\dzzt\left( \delta \J_{\t z}^z \S_{z\theta } +\delta\J_\theta^{\t z}
T_{\t z\t z}\right)       . \end{equation}
This is the analog of the bosonic formula eqn. (\ref{wonk}).  As in section 3.3 of \cite{Wittentwo}, we make
the convenient abbreviation
\begin{equation}\label{conabb}\dzzt=-i[\d\t z;\neg\d z|\d\theta],\end{equation}
to get more natural-looking formulas.\footnote{The analog on an ordinary Riemann surface is
as follows.  If $z=x+iy$, then $\d \bar z\wedge \d z=2i\,\d x\wedge \d y$ is imaginary, so it is
sometimes convenient to introduce the real two-form $\d^2z=-i \d\bar z\wedge \d z$, which is
a bosonic analog of $\dzzt$.}
In eqn. (\ref{bitzo}),  $\S_{z\theta }$ is the holomorphic superfield that contains the holomorphic supercurrent and stress tensor,
\begin{equation}\label{ongo}\S_{z\theta }= S_{z\theta}+\theta T_{zz} ,\end{equation}
 and $T_{\t z\t z}$ is
the antiholomorphic stress tensor.  
These fields obey \begin{equation}\label{thoby}\partial_{\t z}\S_{z\theta}=0=D_\theta T_{\t z\t z}=\partial_z T_{\t z\t z},\end{equation}
ensuring that $\delta I$ vanishes if $\delta \J$ is generated by an infinitesimal diffeomorphism 
(in other words if $\delta\J$ has the form of eqn. (\ref{itzo})).  
Geometrically, $S_{z\theta }$ is a section of $\D^{-3}$ and $T_{\t z\t z}$ is a section of $(T^*_L\Sigma)^2$.  These
facts ensure that  the integrand in (\ref{bitzo})  is a $(0,1)$-form on $\Sigma$ with values in $\D^{-1}$, so that the integral 
 does not depend on
the choice of coordinates.  

Acting on possible operator insertions at $z=0$, $\S_{z\theta}$ has a mode expansion
\begin{equation}\label{domely}\S_{z\theta}(z|\theta)=\frac{1}{2}\sum_{r\in\Z+1/2}z^{-r-3/2}G_r+\theta\sum_{n\in\Z}z^{-n-2}L_n.\end{equation}  The coefficients are quantum operators corresponding to the superconformal vector fields in eqn. (\ref{urv}).
Similarly, the modes of  $T_{\t z\t z}$ are the left-moving Virasoro generators:
\begin{equation}\label{iltzo}T_{\t z\t z}=\sum_{n\in \Z}\t z^{-n-2}\t L_n.\end{equation}

The last characters that we have to introduce are the supersymmetric ghosts and antighosts, or superghosts for short. The holomorphic ghosts 
are a superfield\footnote{Though we will display them for the moment, the
 traditional subscripts and superscripts in $C^z$ and $B_{z\theta}$
 are natural only when the super Riemann surface $\SIgma$ is split.  In general, it is best to simply think of $C$ and $B$
as sections of $\D^2$ and $\D^{-3}$, respectively.  The same goes for related objects such as the supercurrent $\S$.} $C^z$ of $N_\gh=1$.  $C^z$ is an odd  section of $\D^2$
and its theta expansion reads
\begin{equation}\label{turko} C^z=c^z+\theta\gamma^\theta  \end{equation}
where $c^z$ and $\gamma^\theta$ have conformal dimensions $-1$ and $-1/2$, respectively. Similarly, the holomorphic antighosts 
are a superfield $B_{z\theta}$ of $N_\gh=-1$.  $B$ is a section of $\D^{-3}$, with theta expansion
\begin{equation}\label{murko}B_{z\theta}=\beta_{z\theta}+\theta b_{zz},\end{equation}
where $\beta$ and $b$ have conformal dimensions $3/2$ and $2$, respectively. 
 The fields $b_{zz} ,\,c^z$ are the 
anticommuting Virasoro ghosts that are familiar from
the bosonic string, while $\beta_{z\theta},\,\gamma^\theta$ are commuting ghost fields that are associated to the odd 
generators of the super-Virasoro algebra.  The $\beta\gamma$
system has unusual properties that have long played a central role in quantization of superstrings \cite{FMS}.  
The antiholomorphic ghosts and antighosts
are anticommuting fields $\t C^{\t z}$ of $N_\gh=1$ and $\t B_{\t z\t z}$ of $N_\gh=-1$ that are 
sections respectively of $T_L\SIgma$ and $(T^*_L\Sigma)^2$.  Their theta expansions
read
\begin{equation}\label{lurko} \t B_{\t z\t z}=\t b_{\t z \t z}+\theta \t f_{\t z\t z \theta},~~~ \t C^{\t z}=\t c^{\t z}+\theta\t g^{\t z}_\theta,\end{equation}
where $\t c$ and $\t b$ play the role of antiholomorphic Virasoro ghosts that are familiar in the 
bosonic string, while $\t f$ and $\t g$ are auxiliary fields that vanish on-shell. 
The ghost and antighost fields are governed by the action
\begin{equation}\label{tomgor}I_{\gh}=\frac{1}{2\pi}\int\dzzt\, \left(B_{z\theta}\partial_{\t z}C^{z}+\t B_{\t z\t z} D_\theta \t C^{\t z}\right),\end{equation}
and in particular the equations of motion for the antighosts read
\begin{equation}\label{omgor}\partial_{\t z}B_{z\theta}=0=D_\theta \t B_{\t z\t z}. \end{equation}
The subscripts and superscripts carried by the fields are intended as reminders of how they transform under reparametrizations;
we often omit them to reduce clutter.

For the present paper, one of the most important properties of the ghost and antighost fields is the BRST transformation law
of the antighosts:
\begin{align}\label{formox}[Q_B,B_{z\theta}]&=\S_{z\theta} \cr \{Q_B,\t B_{\t z\t z}\}&=T_{\t z\t z}.\end{align}

\subsection{The Extended Action And The Integration Measure}\label{exform}

The matter plus ghost action $I$ of the heterotic string is BRST-invariant if  $\J$ 
is regarded as a fixed quantity.
Suppose, however, that we let the BRST charge $Q_B$ act on $\J$, producing variations 
$\delta\J_{\t z}^z$ and $\delta\J_\theta^{\t z}$. 
We understand these as fields of  $N_\gh=1$;
$\delta\J_{\t z}^z$ is odd and $\delta\J_\theta^{\t z}$ is even.  
To preserve $Q_B^2=0$,
we take 
\begin{equation}\label{mog}\{Q_B,\delta\J_{\t z}^z\}=[Q_B,\delta\J_\theta^{\t z}]=0.\end{equation}

If we extend the action of $Q_B$ in this way, the action is no longer $Q_B$-invariant.  But from our study of the bosonic string, we know what
to do.  We restore the $Q_B$-invariance by extending the action, adding a coupling of $\delta\J$ to the antighosts:
\begin{equation}\label{extac}I\to \hat I=I+\frac{1}{2\pi}\int_\Sigma\dzzt \left( \delta\J_{\t z}^z B_{z\theta}-\delta\J_\theta^{\t z}
\t B_{\t z\t z}\right). \end{equation}   The extended action $\h I$ is $Q_B$-invariant.

From here we proceed as in our study of the bosonic string.
We integrate out all other variables -- both matter fields $\XX$ and superghosts -- to define a function of $\J$ and $\delta \J$ only:
\begin{equation}\label{texac}F(\J,\delta\J)=\int\D(\XX,B,C,\t B,\t C)\,\exp(-\h I).\end{equation}
Integrating over the BRST multiplets $\XX,B,C,\t B,\t C$ is a $Q_B$-invariant operation, so $F(\J,\delta\J)$ is $Q_B$-invariant:
\begin{equation}\label{exac}[Q_B,F(\J,\delta\J)\}=0. \end{equation}

From here, we can, up to a certain point, reason as we did in the case of the bosonic string.  Acting on functions of $\J$ and $\delta\J$ only,
$Q_B$ can be regarded as the exterior derivative on the space $\JJ$ of all supercomplex structures.  Here, for any
point $p\in\Sigma$, we interpret the matrix elements of $\J$  acting on the cotangent space to $\Sigma$ at $p$ as functions on $\JJ$, while   $\delta\J_{\t z}^z(p)$ and  
$\delta\J_\theta^{\t z}(p)$  are respectively $(1,0)$-forms and $(0,1)$-forms on $\JJ$.  With this interpretation, we can understand
$F(\J,\delta\J)$ as a form on $\JJ$; eqn. (\ref{exac}) says that this form is closed:
\begin{equation}\label{exacto}  \d F(\J,\delta\J)=0.  \end{equation}

\subsubsection{A Lightning Review Of Integration On Supermanifolds}\label{lightning}

We did not call $F(\J,\delta\J)$ a differential form; in fact, technically it is better called a pseudoform.  The reader may wish
to consult an introduction to integration theory on supermanifolds (such as \cite{Wittenone}), but here is a very brief summary.  Let
$M$ be a supermanifold with even coordinates $t^1\dots t^m$ and odd coordinates $\theta^1\dots\theta^n$.  For each even or odd
coordinate $t^i$ or $\theta^j$, we introduce  corresponding variables $\d t^i$ or $\d\theta^j$ with the opposite statistics (so $\d t^i$ is
odd and $\d\theta^j$ is even).  We write generically $x$ for all variables $t^1\dots|\dots\theta^n$ and $\d x$ for the corresponding
differentials.  We say that a function $F(x,\d x)$ has degree $s$ if
\begin{equation}\label{degor} F(x,\lambda\d x)=\lambda^s F(x,\d x).  \end{equation}
We define the exterior derivative, mapping functions of degree $s$ to functions of degree $s+1$:
\begin{equation}\label{cormo}\d=\sum_I\d x^I\frac{\partial}{\partial x^I},~~\d^2=0.\end{equation}
Whether or not the original set of variables $t^1\dots|\dots\theta^n$ had a natural measure, there is always a natural measure
for the extended set of variables $x,\d x$, because of the way the variables come in pairs with opposite statistics.  The integral of a function
$F(x,\d x)$ is defined as a Berezin integral over all variables $x$ and $\d x$, whenever this makes sense:
\begin{equation}\label{umby}\int \D(x,\d x)\,F(x,\d x). \end{equation}
Here, however, we meet the main difference between ordinary integration theory and integration theory on supermanifolds.
On an ordinary manifold  with only even coordinates $t^1\dots t^m$ and therefore only odd differentials $\d t^1\dots\d t^m$,
a function $F(x,\d x)$ is inevitably polynomial in the differentials.  But on a supermanifold,  there are also odd coordinates $\theta^1\dots
\theta^n$ and therefore even differentials $\d\theta^1\dots\d\theta^n$, so $F(x,\d x)$ is not necessarily polynomial in the $\d x$'s.  If
$F(x,\d x)$ has polynomial dependence on the $\d x$'s, we call it a differential form; otherwise we call it a pseudoform or just a form.

In fact, if $F(x,\d x)$ is polynomial in the even variables $\d\theta^1\dots\d\theta^n$, then the integral over those variables diverges.
A typical example of a form that can be integrated over a supermanifold of dimension $m|n$ is 
\begin{equation}\label{mondek} F(x,\d x)=f(t^1\dots|\dots \theta^n) \d t^1\dots \d t^m \delta(\d\theta^1)\dots \delta(\d\theta^n).\end{equation}
The integrals over the $\d\theta$'s can be done with the aid of the delta functions.  
Notice that $F(x,\d x)$ has degree $m-n$.  It also has another important ``quantum number'' which has no analog in integration theory
on ordinary manifolds.  This is the ``picture number,'' defined as minus\footnote{The minus sign is included in the definition to agree with
the choice in \cite{FMS}.}  the number of even differentials with respect to which $F(x,\d x)$
has delta function support.  So the form $F(x,\d x)$ defined in (\ref{mondek}) has picture number $-n$.
By a form of superdegree $m|n$, we mean a form of degree $m-n$ and picture number $-n$.  A form of superdegree $m|n$ is also
called an $m|n$-form.   An $m|n$-form, such as $F(x,\d x)$
as defined above, can  be integrated
over a supermanifold of dimension $m|n$.

The typical example (\ref{mondek}) of a form that can be integrated may look unappealing, because the odd differentials $\d t^i$ and 
even differentials $\d\theta^j$ are treated differently.  We can write the definition in a possibly more pleasing way if we recall that
an odd variable is its own delta function, so that we can replace $\d t^i$ with  $\delta(\d t^i)$.  Thus, we can 
rewrite (\ref{mondek}) as follows:
\begin{align}\label{ondek} F(x,\d x)=&f(t^1\dots|\dots\theta^n) \delta(\d t^1)\dots \delta(\d\theta^n)\cr =&
f(t^1\dots|\dots\theta^n) \delta^{m|n}(\d t^1\dots |\dots\d\theta^n).\end{align}

In general, we want to consider forms that have delta function localization at $\d\theta=0$ for  some of the $\d\theta$'s and polynomial
dependence on the others.  By ``delta function localization,'' we refer to a function with distributional support at $\d\theta=0$, either
 a delta function or a (possibly repeated) derivative of a delta function. 
For example, for a single odd variable $\theta$,
the form 
\begin{equation}\label{ipso} \frac{\partial^r}{\partial (\d\theta)^r}\delta(\d\theta) \end{equation}
has degree $-r-1$ and picture number $-1$. So it is a form of superdegree $-r|1$.  We allow derivatives of a
delta function, not just delta functions, because this is necessary  to get a  theory in which all important
operations can be defined (a key example is contraction with a vector field, discussed in appendix \ref{pullback},
which can involve the derivative with respect to an even differential $\d\theta$).
 
 On a supermanifold $M$ of any dimension, an $m|n$-form can potentially be integrated on a submanifold $N\subset M$
of dimension $m|n$.  We say ``potentially'' because just as in integration on a bosonic manifold, sometimes the integral may diverge.

We conclude this lightning introduction to integration theory on supermanifolds with a simple example on $\R^{1|2}$ with coordinates
$t|\theta^1\theta^2$.  One purpose of the example is to show that it is useful to consider objects like $\delta'(\d\theta)$.
 Consider the form
\begin{equation}\label{condo} F(x,\d x)= \theta^2\, \d t\,\delta'(\d\theta^1).\end{equation}
This form has degree $-1$ and picture number $-1$, so it is a $0|1$-form and can potentially be integrated on a submanifold
$N\subset M$ of that dimension.
We define $N$ by $t=\alpha\theta^1$, $\theta^2=\lambda\theta^1$, where $\alpha$ is an odd parameter and $\lambda$ is an even one. 
So $N$ is parametrized by $\theta^1$.  
Restricted to $N$, we have $F(x,\d x)=-\alpha\lambda \theta^1\d\theta^1\delta'(d\theta^1)=-\theta^1\alpha\lambda\delta(\d\theta^1)$
(where we used the fact that for an even variable $y$, $y\delta'(y)=-\delta(y)$).  Performing the Berezin integral over $\theta^1$ and
$\d\theta^1$, we get finally $\int_N\D(\theta^1,\d\theta^1)\,F=-\alpha\lambda$.  

\subsubsection{Bosonic Integration As An Algebraic Operation}\label{bosal}

An unusual feature of superstring quantization  is that \cite{FMS} the path integral of the $\beta\gamma$ commuting ghost system
is not really an integral in the usual sense.  That is because the fields $\beta,\gamma$ obey no reality condition and the action
\begin{equation}\label{thac}I_{\beta\gamma}=\frac{1}{\pi}\int_\Sigma \d^2z \,\beta\partial_{\t z}\gamma \end{equation}
has no reality or 
positivity property.  As a result, the ``path integral'' of the $\beta\gamma$ system has to be understood as a formal algebraic operation,
somewhat like the Berezin integral for fermions.

Something similar happens in general integration theory over supermanifolds when the odd 
variables $\theta^i$ do not have a real structure.\footnote{See for example section 3.3.2 of \cite{Wittentwo}.}  
One has to postulate formulas like
\begin{equation}\label{tofoto}\int \D(\d\theta)\frac{\partial^r}{\partial(\d\theta)^r}\delta(\d\theta)=\delta_{r,0},\end{equation}
which have to be understood in a formal sense if $\d\theta$ is a complex variable.
We are in that situation here, because $\delta\J$ contains even as well as odd modes
and these even modes have no real structure. (Both
$\delta\J_\theta^{\t z}$ and $\delta\J_{\t z}^z$ have even modes, but the even modes in $\delta\J_\theta^{\t z}$ play little role
because they couple to auxiliary fields in $\t B$ that vanish by their equation of motion.  The important even modes are the modes
in $\delta\J_{\t z}^z$ that couple to $\beta$.) 
 So the coupling $\int \delta \J B$ in the extended action $\h I$ has the same basic
property as the original action $I_{\beta\gamma}$: it contains a bilinear coupling of complex bosons, with no reality or positivity property.
By analogy with what we did for the bosonic string in section \ref{bosmeasure}, 
we eventually will want to integrate over some modes of $\delta \J$ to construct superstring scattering amplitudes.  Because of the lack
of any reality or positivity property in the $\int\delta\J B$ coupling, this integral will pose exactly the same type of question as is posed
by the original $\beta\gamma$ path integral.  

In section \ref{betag}, after gaining more experience with what sort of
operator insertions are important in the $\beta\gamma$ path integral, we will give a more extensive discussion of such matters (see
also \cite{EHVerl,EHV,OL}).
For now, we just explain the minimum so that we can proceed.  The troublesome integrals over complex bosons will always be Gaussian
integrals or slight generalizations thereof.  Let $\u_i,\v_j$, $i,j=1,\dots,n$ be bosonic variables and let $m$ be an $n\times n$ complex
matrix that is nondegenerate but obeys no reality or positivity condition.  We define the basic Gaussian integral
\begin{equation}\label{irto}\int \D(\u,\v)\exp\left(-\sum_{i,j}\u_i m_{ij}\v_j\right)=\frac{1}{\det m}. \end{equation}
If $m$ is positive define, $\u_j$ is understood as the complex conjugate of $\v_j$, and the integration measure $\D(\u,\v)$ is suitably
normalized (with a factor of $1/2\pi i$ for each pair of variables), then this formula is actually a true theorem about ordinary integration.
In general, we just take it as a definition of what we mean by Gaussian ``integration'' for complex bosons, somewhat like the formal
definition of the Berezin integral for fermions.  As usual, we also extend the Gaussian integral to allow linear ($c$-number) sources $r$ and $s$
in the exponent:
\begin{equation}\label{pirto}\int\D(\u,\v)\exp\left(-\sum_{i,j}\u_i m_{ij}\v_j-\sum_k (r_k\u_k+s_k\v_k)\right)=\frac{\exp(\sum_{i,j} s_i (m^{-1})_{ij} r_j)}
{\det m}.\end{equation}
We also need the conventional formulas for the integral of a Gaussian  times a polynomial, which involve Wick contractions
using $m^{-1}$ as a ``propagator.''   In a standard way, these formulas
can be obtained by differentiating (\ref{pirto}) with respect to $r$ and $s$ a desired number of times and then setting $r=s=0$.
And we want to allow an operation of integrating over some components of $r$ and $s$.   For now, we just observe that since the
exponent in (\ref{pirto}) is quadratic in the combined set of variables $\u,\v,r,s$, an extension of (\ref{pirto}) in which we want to integrate
over some components of $r$ and $s$ is just a Gaussian integral with more integration variables, so it is covered by the same definitions.
We return to this in section \ref{betag}.

If we set $r=0$, we get a special case of (\ref{pirto}):
\begin{equation}\label{zpirto}\int\D(\u,\v)\exp\left(-\sum_{i,j}\u_i m_{ij}\v_j-\sum_k s_k\v_k\right)=\frac{1}
{\det m},\end{equation}
independent of $s$.  This formula is useful for understanding how we should define the following important integral:
\begin{equation}\label{zupiro} G(\v)=\int\D(\u)\exp(-(\u,m\v)).\end{equation}
If it is going to be possible to integrate over $\u$ and $\v$ by integrating first over $\u$ and then over $\v$, we should
have
\begin{equation}\label{upiro}\int \D(\v) A(\v)G(\v) =\int\D(\u,\v)A(\v)\exp(-(\u,m\v)),\end{equation}
for any allowed function $A(\v)$.  Taking $A(\v)=\exp(-(s,\v))$ and using (\ref{zpirto}), we see that we need
\begin{equation}\label{mupiro} \int\D(\v) \exp(-(s,\v)) G(\v)=\frac{1}{\det m},\end{equation}
independent of $s$.  This motivates the definition $G(\v)=\delta(m\v)$, that is,
\begin{equation}\label{tofo}\int\D(\u)\exp(-(\u,m\v))=\delta(m\v),\end{equation}
along with
\begin{equation}\label{yofo}\int\D(\v)\delta(m\v)=\frac{1}{\det m}. \end{equation}

All of these formulas have analogs in the Berezin integral for fermions -- the difference being that the factors of $\det m$ would
appear in the numerator rather than the denominator -- and like the Berezin integral,
they should be understood as a convenient algebraic machinery
for manipulating certain expressions.

\subsubsection{The Zero-Mode Coupling}\label{zerom}

Given the formal calculus that we have just described, it is straightforward to generalize  bosonic 
formulas such as  (\ref{porko}) or (\ref{norko})  and determine the exact dependence of $F(\J,\delta\J)$ on $\delta\J$.  We assume that
$\Sigma$ has genus $\g\geq 2$ so that the ghost fields $C$, $\t C$ have no zero-modes.  However,
the antighost fields $\t B$ and $B$ do have zero-modes.   In the following derivation, the holomorphic
and antiholomorphic ghosts are decoupled from each other and can be treated separately.  We consider
first the $B,C$ system, whose coupling to $\delta\J$ is given in eqn. (\ref{extac}):
\begin{equation}\label{zonick}\t I_{\delta\J B}=\frac{1}{2\pi}\int \dzzt\,\delta \J_{\t z}^zB_{z\theta}.\end{equation}

The number of zero-modes of the field $B_{z\theta}$ is $3\g-3|2\g-2$ (that is, $3\g-3$ even zero-modes
and $2\g-2$ odd ones).  We will denote the zero-modes as $\eurm B_\alpha$, $\alpha=1\dots 3\g-3|1\dots 2\g-2$; 
that is, the $\eurm B_\alpha$ include all even and odd zero-modes of
$B$.  

Rather as in (\ref{zumbo}), we expand $B$ as a sum of zero-modes $\eurm B_\alpha$ and non-zero modes
$\eurm B'_\lambda$ with coefficients $\u_\alpha$ and $\w_\lambda$:
\begin{equation}\label{mixto} B=\sum_{\alpha=1\dots|\dots 2\sg-2} \u_\alpha \eurm B_\alpha
+\sum_\lambda \w_\lambda \eurm B'_\lambda.   \end{equation}
$C$ has an analogous expansion
\begin{equation}\label{turx}C=\sum_\lambda \gamma_\lambda {\eurm C}_\lambda, \end{equation}
now with only non-zero modes ${\eurm C}_\lambda$ with coefficients $\gamma_\lambda$.  The ghost action (\ref{tomgor}) pairs the $\w_\lambda$ and the $\gamma_\lambda$
by a nondegenerate bilinear pairing $\sum_\lambda m_\lambda \w_\lambda \gamma_\lambda$, $m_\lambda\not=0$.
When we integrate over $\gamma_\lambda$, we get delta functions setting the $\w_\lambda$ to zero.  This results from the
formula
\begin{equation}\label{pixto}\int\d\gamma\exp(-m\w\gamma)=\delta(m\w),\end{equation}
If $\w$ and $\gamma$ are odd variables, this formula is a consequence of the Berezin integral (and was used
in our study of bosonic strings), while if they are even variables, the formula has been explained
in eqn. (\ref{tofo}).  
Once we set the $\w_\lambda$ to zero, the coupling of $B$ to $\delta \J$  reduces to
\begin{equation}\label{izzo}\t I_{\delta\J B}=\sum_{\alpha=1\dots|\dots 2\sg-2} \u_\alpha \Psi_\alpha ,\end{equation}
with
\begin{equation}\label{pizzo}\Psi_\alpha=\frac{1}{2\pi}\int_{\Sigma}\dzzt\,\delta\J_{\t z}^z \eurm B_{z\theta\,\alpha}
.\end{equation} 
The integral over the zero-mode coefficients $\u_\alpha$ therefore becomes
\begin{equation}\prod_{\alpha=1\dots|\dots 2\sg-2} \int \d\u_\alpha\exp(\u_\alpha\Psi_\alpha)=\prod_{\alpha=1\dots|
\dots 2\sg-2}\delta(\Psi_\alpha)=\delta^{3\sg-3|2\sg-2}(\Psi'_1\dots|\dots \Psi''_{2\sg-2}).\end{equation}
(In the last version of the formula, we write $\Psi_\alpha$ as $\Psi_\alpha'$ or $\Psi_\alpha''$ according to whether it is even or odd.)

We can do a similar calculation for the antiholomorphic ghosts $\t B$ and their coupling
\begin{equation}\label{yotiz}\t I_{\delta\J\t B}=\frac{1}{2\pi}\int_\SIgma\dzzt\,\delta\J_\theta^{\t z}\t B_{\t z \t z}
\end{equation}
to $\delta\J$.  The number of zero-modes of $\t B_{\t z \t z}$ is $3\g-3$. We write $\t B=\sum_{\alpha=1\dots 3\sg-3}
\t \u_\alpha\t {\eurm B}_\alpha+\sum_\lambda\t\w_\lambda\t{\eurm B}_\lambda'$ where $\t{\eurm B}_\alpha$
are zero-modes and $\t{\eurm B}'_\lambda$ are non-zero modes.  The integral over $\t C$ sets the coefficients 
$\t\w_\lambda$
of non-zero modes to zero, whereupon we get
\begin{equation}\label{otz}\t I_{\delta\J\t B}=\sum_{\alpha=1}^{3\sg-3}\t\u_\alpha\t\Psi_\alpha,\end{equation}
with
\begin{equation}\label{botz}
\t\Psi_\alpha=\frac{1}{2\pi}\int_\Sigma[\d\t z;\neg\d z|\d\theta]\,\delta\J_\theta^{\t z}\,\t {\eurm B}_{\t z\t z\,\alpha}.\end{equation}
The integral over the $\t u_\alpha$
gives
\begin{equation}\prod_{\alpha=1}^{3\sg-3}\int\d\t \u_{\alpha}\exp(\t \u_\alpha\t\Psi_\alpha)=\delta^{3g-3}(\t\Psi_1,\dots,
\t\Psi_{3\sg-3}). \end{equation}

We have determined the exact dependence of $F(\J,\delta\J)$ on $\delta \J$: it
is the product of a function
that depends only on $\J$ and not on $\delta\J$ times
\begin{equation}\label{reddo}\delta^{3\sg-3}(\t\Psi_1\dots\t\Psi_{3\sg-3})\delta^{3\sg-3|2\sg-2}(\Psi'_1\dots|\dots\Psi''_{2\sg-2}).
\end{equation}
This formula (cousins of which can be found in \cite{superoperator,Bel}) has an obvious
resemblance to the formula (\ref{ondek}) giving the prototype of a form of degree $m|n$.  Clearly, we have
learned that $F(\J,\delta\J)$ is a form on $\JJ$ of degree $6\g-6|2\g-2$.  More specifically, this form
has holomorphic degree $3\g-3|2\g-2$ and antiholomorphic degree $3\g-3|0$.  It is hopefully clear that a similar derivation for Type II superstrings would proceed in essentially the same way and 
give a form whose holomorphic and antiholomorphic degree would both be $3\g-3|2\g-2$.

\subsection{Where To Integrate}\label{intor} 

\subsubsection{Reduction To Supermoduli Space}\label{zonno}

The superdegree of $F(\J,\delta\J)$ is appropriate for a form that would be integrated over a supermanifold
of dimension $6\g-6|2\g-2$.  The reader who has gotten this far will undoubtedly anticipate that the supermanifold
in question will be, in some sense, the moduli space of super Riemann surfaces.  

The basic idea for reducing from $\JJ$ to the moduli space of super Riemann surfaces is the same as it was
in section \ref{redmod}.  We introduce the group $\Diff$ of superdiffeomorphisms of $\Sigma$.  
We want to show that $F(\J,\delta\J)$ is a pullback from, roughly speaking, the quotient $\JJ/\Diff$.
For this, we need the same two facts as before: (1) $F(\J,\delta\J)$ should be $\Diff$-invariant; (2) $F(\J,\delta\J)$
should vanish if contracted with one of the vector fields that generates the action of $\Diff$.  

Property (1) is manifest from the definition of $F$, and property (2) is true by virtue of the same reasoning as
in section \ref{redmod}.  In this case, we have to show that $F(\J,\delta\J)$ is invariant if we shift $\delta\J$
in the fashion analogous to (\ref{omigo}), or in other words, in view of eqn. (\ref{itzo}), by
\begin{equation}\label{helmo}\delta\J_{\t z}^z\to \delta\J_{\t z}^z+\partial_{\t z}q^z,~~\delta\J_\theta^{\t z}
\to \delta\J_\theta^{\t z}+D_\theta q^{\t z}. \end{equation}
To show that $F(\J,\delta\J)$ has this symmetry, we simply observe that the extended action $\h I$ of
eqn. (\ref{extac}) has this symmetry, as one sees by integration by parts and using the classical
equations of motion for $B$ and $\t B$ that can be deduced from the ghost action (\ref{tomgor}).
As we discussed in relation to eqn. (\ref{orby}), instead of using the classical equations of motion for
$B$ and $\t B$, it is better to accompany the transformation (\ref{helmo}) of $\delta\J$ by a shift in $C$ and
$\t C$ such that $\hat I$ is invariant.  Since we integrate out $C$ and $\t C$ in defining $F(\J,\delta\J)$,
the invariance of the extended action $\h I$ under the shift (\ref{helmo}) of $\delta\J$ together with a suitable shift
in $C$ and $\t C$ implies that $F(\J,\delta\J)$ is invariant under the shift in $\delta\J$.  (After introducing
vertex operators, we spell out some details of this argument in section \ref{scattering}.)

Thus, roughly speaking, $F(\J,\delta\J)$ is a pullback from the moduli space $\MM_\sg$ of super Riemann surfaces
of genus $\g$.
Moreover, counting both holomorphic and antiholomorphic variables, $F(\J,\delta\J)$  has the right
superdegree to be  integrated  over 
$\MM_\sg$.  So we can define the integral of $F(\J,\delta\J)$ over this moduli
space, and this will give the genus $\g$ contribution to the vacuum amplitude of the heterotic string.  

As usual, some
subtleties arise in the infrared region because $\MM_\sg$ is not compact; these will occupy our attention later.  But as a preliminary, one has to grapple with an important detail.

\subsubsection{A Hard-To-Avoid Detail}\label{harder}

 To explain the relevant point
without excess clutter,
let us simplify and suppose that the moduli space $\MM$ of super Riemann surfaces
has dimension $1|2$ rather than $3\g-3|2\g-2$.  Then locally we could pick holomorphic
coordinates $m|\eta^1,\eta^2$ on
$\MM$.  Any such coordinate system would be valid only locally in $\MM$.  In another region of $\MM$,
we might use another holomorphic
coordinate system $m'|\eta'{}^{1},\eta'{}^{2}$, related to the first by 
\begin{align}\label{delco} m'& = f(m|\eta^1,\eta^2) \cr
                                         \eta'{}^{i}&=\psi^i(m|\eta^1,\eta^2),~~i=1,2,\end{align}
                                         with some holomorphic functions $f|\psi^1,\psi^2$.
$\MM$ is said to be holomorphically projected if it is possible to choose the coordinates so that $f(m|\eta^1\eta^2)$
is a function of $m$ only, and not of the $\eta$'s.  The more general possibility is that the relation is
\begin{equation}\label{elco}m'=f_0(m)+\eta^1\eta^2f_2(m),\end{equation}
with functions $f_0$ and $f_2$ that depend on $m$ only.   $\MM$ is said to be holomorphically split if it is possible
to choose the coordinates so that $f$ is independent of the $\eta$'s and  the $\psi^i$ are linear in the $\eta$'s.  (With only two  odd
variables $\eta^i$,
the second condition is trivial since the $\psi^i$ are odd functions.)  

It is rather special for a complex supermanifold to be holomorphically projected or split.  There has been no reason to believe
that the moduli space of super Riemann surfaces has this property, and recently it has been shown that this
is not the case in general \cite{DonW}.   (Supermoduli space is holomorphically projected in low
orders and this fact has been exploited in explicit computations, especially in \cite{DPhgold}.)

Now let us consider the antiholomorphic
degrees of freedom of the heterotic string.  If supermoduli space has dimension $1|2$, then the ordinary
moduli space of bosonic Riemann surfaces has dimension $1|0$.  We therefore describe the antiholomorphic
structure of $\Sigma$ by a local coordinate $\t m$.  In another region of the moduli space, we would use another
local coordinate $\t m'$ related to the first by a holomorphic change of coordinates
\begin{equation}\label{inty}\t m'=\t f(\t m).\end{equation}

What relation can we assume between $\t m$ and $m$?  Naively one would like to claim that the coordinates
can be chosen so that  $\t m$ is the complex
conjugate of $m$ and $\t m'$ is the complex conjugate of $m'$.  
However, if $\MM$ is not holomorphically projected, these assertions do not make sense.  
We cannot consistently claim that $\t m=\overline m$ and  $\t m'=\overline m'$ 
if the relation between $m'$ and $m$ depends on the $\eta$'s while the relation between $\t m$ and $\t m'$ does not.  

An elegant way to proceed\footnote{See p. 95 of \cite{DEF} for
a very brief explanation, and section 5 of \cite{Wittenone} for a more leisurely account, with
discussion of some alternative approaches.}  
is as follows.   We make use
of the fact that the partition functions and correlation functions of the worldsheet conformal field theories that are
relevant in string theory are all real-analytic.  So we can analytically continue away from $\t m=\overline m$
as long as we do not go too far away.  Similarly, if we insert a vertex operator at a point $\t z;\neg z|\theta$;
we can analytically continue away from $\t z =\bar z$, as long as we do not go too far.  When we make
this sort of analytic continuation, we can treat the moduli of left and right-movers on the worldsheet
as independent complex variables.  

If we proceed in this way, the arguments of section \ref{zonno} show that $F(\J,\delta\J)$ is
the pullback of a holomorphic form (which we denote by the same name)
on a product $\M_L\times \M_R$, where $\M_L$ and $\M_R$ parametrize
respectively the complex structures ``seen'' by the left-movers and right-movers.\footnote{To be more exact, $F(\J,\delta\J)$ is defined
and holomorphic in an open set of $\M_L\times \M_R$ that includes the region near $\t m=\bar m$ that
will be used in the following construction.}   Any of the oriented closed-string theories can be described in this way.
For the heterotic string, $\M_R$ is a copy of the moduli space of super Riemann
surfaces, parametrizing holomorphic structures on $\Sigma$, and $\M_L$ is a copy of the moduli space
of ordinary Riemann surfaces, parametrizing antiholomorphic structures on $\Sigma$.  Analytic continuation
does not affect the statement that  $F(\J,\delta\J)$  is a form of superdegree $6\g-6|2\g-2$, which is the same as the 
complex dimension of $\M_L\times \M_R$, or the fact that $F$ is closed, $\d F(\J,\delta\J)=0$.

In general, suppose that $M$ is a complex supermanifold of dimension $p|q$, and $F(x,\d x)$ 
is a closed holomorphic $p|q$-form
on $M$.  Then $F$ can be integrated on a subsupermanifold $\varGamma\subset M$ of real codimension $p|0$ (and thus real
dimension $p|q$), with
a result that only depends on the homology class of $\varGamma$.
A natural source of such $\varGamma$'s is as follows. Let $M_\red$ be the reduced space of $M$, defined 
by setting the odd coordinates to zero.  So $M_\red$ is an ordinary complex manifold of complex dimension $p$.  
Let $\Gamma$ be
a middle-dimensional cycle in $M_\red$, or in other words a cycle of real codimension $p$.  Then by relaxing the
condition that the odd variables should vanish, $\Gamma\subset M_\red$ can 
be thickened to a cycle $\varGamma\subset M$ of the same codimension
in a way that is unique up to homology.\footnote{The idea is very simple:
whatever equations define $\Gamma$, one includes the odd variables in those equations in an arbitrary way to get $\varGamma$.
Because the odd variables are infinitesimal, there is no topology in whatever choices have to be made.
See appendix
\ref{leisurely} for an example of choosing $\varGamma$, and eqn.
\ref{zrobo} for an explanation of the fact that in the example, the possible choices are homologous.  }  So we can define the integral
$\int_{\varGamma}\D(x,\d x) \,F(x,\d x)$, and (if $\Gamma$ is compact) it only depends on the homology class of $\Gamma$.

The reduced space  $(\M_L\times\M_R)_\red\subset\M_L\times \M_R$ is defined by setting to zero all of the 
odd coordinates.  In the toy model above, the reduced space is defined by $\eta_1=\eta_2=0$.  The naive
conditions $\t m=\bar m$, $\t m'=\bar m'$ make sense once we set the odd variables to zero, so they should
be understood as defining a middle-dimensional cycle $\Gamma\subset (\M_L\times\M_R)_\red$.
Then we thicken $\Gamma$ to a cycle $\varGamma\subset\M_L\times \M_R$.  Though the thickening from 
$\Gamma$ to $\varGamma$ is not completely canonical, the integral 
\begin{equation}\label{yred} Z_\sg=\int_\varGamma \D(\J,\delta\J) \,F(\J,\delta\J)  \end{equation}
would be independent of  the choices involved if $\Gamma$ were compact.  As it is, what
we have said about the definition of $\varGamma$ needs to be supplemented with an explanation of how $\varGamma$
should behave at infinity.  We postpone this until we have acquired in section \ref{propagator} a basic understanding
of the behavior  of the superstring measure in the infrared region.

$\varGamma$ (and its generalizations with vertex operator insertions) 
is the appropriate integration cycle for heterotic string perturbation theory. The integral (\ref{yred})
is the $\g$-loop contribution to the vacuum amplitude of the heterotic string. 

One might have hoped to define a natural moduli space of heterotic string worldsheets over which one
would integrate to compute scattering amplitudes.  This  appears to be too optimistic, for the reasons
sketched above.  The  integration cycle $\varGamma$ is what exists instead.
 
\subsection{Integration Over A Slice}\label{yttro}

In the analytic continuation of section \ref{harder}, $F(\J,\delta\J)$ is understood as a holomorphic form on $\M_L\times \M_R$
(which is supposed to be integrated on a cycle $\varGamma$ that is close to the ``diagonal'').   To make
the following discussion concrete, it is useful to pick
local holomorphic coordinates $m_1\dots m_{3\sg-3}|\eta_1\dots\eta_{2\sg-2}$ on $\M_R$; we often denote them collectively
as $\mm_\alpha$, $\alpha=1\dots|\dots 2\g-2$.  And we pick local holomorphic coordinates $\t m_1\dots\t m_{3\sg-3}$
on $\M_L$.  We make no assumption about all these parameters except that they give a good parametrization of a region
of $\M_L\times \M_R$ (sufficiently close to the ``diagonal'' $\t m_\alpha=\bar m_\alpha$) in which we want to get a better
understanding of the heterotic string path integral.

The modular parameters $\t m_\beta$ and $\m_\alpha$ determine the supercomplex structure $\J$ of $\SIgma$
up to diffeomorphism.
Just as in our study of the bosonic string in section \ref{slice}, we can make the formula for the vacuum amplitude 
$Z_\sg$ more concrete
by picking a local slice transverse to the action of the diffeomorphism group on $\JJ$.   
   Picking a slice means that we pick a definite family of $\J$'s parametrized by the $\t m_\beta$ and $\m_\alpha$.
We then regard $\J$ as a function of the $\t m_\beta$ and $\m_\alpha$, and similarly we express $\delta\J$ in
terms of the corresponding differentials  $\d\t m_\beta$ and $\d \m_\alpha$:  
\begin{align}\label{rfo}\delta\J_{\t z}^z=&\sum_{\alpha=1\dots|\dots 2\sg-2}\frac{\partial\J_{\t z}^z}{\partial 
\bm m_\alpha}\d\mm_{\alpha} \cr
\delta\J_\theta^{\t z}=&\sum_{\beta=1\dots 3\sg-3}\frac{\partial \J_\theta^{\t z}}{\partial \t m_\beta}\d\t m_\beta.\end{align}

The worldsheet action of the heterotic string will in general have a complicated dependence on the modular parameters
$\mm_\alpha$ and $\t m_\beta$.  But it is always linear in the $\d\mm$'s and $\d\t m$'s, since it is linear in $\delta\J$.   
So -- as in section \ref{slice} -- it is possible to ``integrate out'' the differentials $\d\mm_\alpha$ and $\d\t m_\beta$, to get a reduced 
description with fewer variables.  

Going back to (\ref{extac}), we see that the part of the action that is linear in the differentials is
\begin{equation}\label{mixo}I_{\d\mm,\d\t m}=\sum_{\alpha=1\dots|\dots 2\sg-2}\d \mm_\alpha\,B^{(\alpha)}+\sum_{\beta=1\dots 3\sg-3}\d\t m_\beta\t B^{(\beta)},\end{equation}
with
\begin{align}\label{happy}B^{(\alpha)}&=\frac{1}{2\pi}\int_\SIgma\dzzt\,\frac{\partial \J_{\t z}^z}{\partial\mm_\alpha}B_{z\theta}\cr
   \t B^{(\beta)}&=\frac{1}{2\pi}\int_\Sigma\dzzt\,\frac{\partial \J_\theta^{\t z}}{\partial\t m_\beta}\t B_{\t z\t z}.  \end{align}
We can think of $B^{(\alpha)}$ and $\t B^{(\beta)}$ as the antighost modes that are 
conjugate to the moduli $\m_\alpha$ and $\t m_\beta$.
From (\ref{mixo}), the integral over the differentials gives
\begin{align}\label{zolod}\prod_{\beta=1\dots 3\sg-3}\int \D(\d\t m_\beta)\exp\left(-\d \t m_\beta \t B^{(\beta)}\right)&\cdot
\prod_{\alpha=1\dots |\dots 2\sg-2}\D(\d \m_\alpha)\exp\left(-\d\m_\alpha B^{(\alpha)}\right) \cr
&=\prod_{\beta=1\dots 3\sg-3}\delta(\t B^{(\beta)}) \cdot \prod_{\alpha=1\dots|\dots 2\sg-2}\delta(B^{(\alpha)}).\end{align}
In short, we can integrate out the differentials in favor of a delta function for each conjugate 
antighost mode.  We abbreviate
the product of delta functions that we have just obtained as
\begin{equation}\label{doofus}\delta^{3\sg-3}(\t B^{(\beta)})\delta^{3\sg-3|2\sg-2}(B^{(\alpha)}).\end{equation}

We now define a function $\Lambda(\t m,\m) $ of the moduli $\t m_\beta$ and $\m_\alpha$ by 
performing a path integral over all worldsheet
fields of the heterotic string (both matter and superghosts) with the above delta functions included 
and the moduli held fixed:\begin{equation}\label{ofus}\Lambda(\t m,\m)= \int \D(\XX,B,C, \t B, \t C)
\exp(-I)\,\delta^{3\sg-3}(\t B^{(\beta)})\delta^{3\sg-3|2\sg-2}(B^{(\alpha)}).\end{equation}
($\Lambda$  was first introduced in \cite{superoperator} and used
to define what we will shortly call $\Xi$.)  The argument given in the last paragraph of  section \ref{slice}
applies here to show that $\Lambda(\t m,\m)$ does not depend on the choice of slice that was used in the 
computation.  The delta functions depend on the slice, but the dependence disappears when we integrate over
the fields $C$ and $\t C$.

The delta function insertions in (\ref{ofus})
are actually needed in order for the worldsheet
path integral to give a sensible and  non-zero result.  An antighost zero-mode that is not 
removed by a delta function would cause the worldsheet
path integral to vanish (in the case of a fermionic zero-mode) or to diverge (in the case of a bosonic one).  
As usual, when studying vacuum amplitudes,
 we restrict to $\g\geq 2$ so that there are no $C$ or $\t C$ zero-modes and hence delta functions for
 antighost zero-modes suffice to give a sensible path integral.  Later,
when we include external vertex operators, we will always, for all $\g\geq 0$,  have a suitable set of 
delta functions either present in the vertex operators or coming from integrating out the $\d \m$'s and $\d \t m$'s
 so that the path
integral is sensible and not trivially zero.

Let us use an even more condensed notation in which $x$ represents all moduli 
$\m_\alpha$ and $\t m_\beta$ together.  With
 the differentials $\d x$ included, there is a natural measure $\D(x,\d x)$ for the combined 
 system of moduli and their differentials.  
(Here ``natural'' means that $\D(x,\d x)$ is invariant under any reparametrization of the $x$'s together with the induced
transformation of the $\d x$'s.) 
As usual, this is so because of the way the variables come in pairs with opposite statistics.  
Although $\D(x,\d x)$ is completely natural, if we factor it as the product of a measure for the $x$'s,
\begin{equation}\label{kinco}\D(x)=[\d \t m_1\dots \d \t m_{3\sg-3};\neg \d m_1\dots\d m_{3\sg-3}|
\d\eta_1\dots\d\eta_{2\sg-2}] \end{equation}
and the measure for the $\d x$'s that we used in integrating them out in eqn. (\ref{zolod}), 
then neither factor is natural by itself.
One way to get a natural formula is to multiply   $\D(x)$ by the function $\Lambda(x)$  that we generated 
by integrating over the $\d x$'s as well as other variables:
\begin{equation}\label{incozz}\Xi(x)=\D(x)\,\Lambda(x). \end{equation}
  Concretely, if we transform to a new set of holomorphic 
coordinates on $\M_L\times \M_R$, then $\D(x)$
is multiplied  by the Berezinian of the change of coordinates (this Berezinian is the 
superanalog of the Jacobian for a change of coordinates), 
while the inverse of this factor multiplies the product of delta functions in (\ref{doofus}), and likewise 
therefore the function $\Lambda(x)$. 

Geometrically, $\Xi(x)$ is a holomorphic section of the Berezinian line bundle of $\M_L\times \M_R$.  
This is just a fancy way to say that $\Xi(x)$  is a well-defined measure on $\M_L\times \M_R$, in the holomorphic
sense.  The relation between $F(\J,\delta\J)$ and $\Xi(x)$ is just that
\begin{equation}\label{urbo}\Xi(x)=\int \D(\d x)\,F(\J,\delta\J). \end{equation}
Concretely, on the right hand side, $F(\J,\delta\J)$ was defined by an integral over heterotic string
matter fields and superghosts, while on the left hand side $\Xi(x)$ was defined by integrating over
the $\d x$'s as well as the matter fields and superghosts.   So we can map $F$ to $\Xi$ by integrating
out the $\d x$'s.  
In general  (see for example section 3.3.3 of \cite{Wittenone}), 
integrating out the differentials in this fashion gives a natural map from a pseudoform, in our case $F(\J,\delta\J)$, to
a section of the Berezinian, here $\Xi(x)$.  

On a complex supermanifold $M$
of complex dimension $p|q$, in general a holomorphic section of the Berezinian  
can be integrated on a cycle of real codimension $p|0$; the integral only depends on the homology class
of the cycle.  
In our case, the cycle on which we wish to integrate $\Xi(x)$ is the integration cycle 
$\varGamma$ of the heterotic string.
This integral gives the genus $\g$ contribution to the heterotic string vacuum amplitude:
\begin{equation}\label{pyp}Z_\sg=\int_{\varGamma}\Xi(x).                  \end{equation}

The formula (\ref{pyp}) is completely equivalent to our earlier formula (\ref{yred}) for $Z_\sg$.  
The two formulas arise by performing
the same integrals in different orders.  Eqn. (\ref{yred}) arises in a procedure in which  
we compute $Z_\sg$ by integrating first over matter fields and superghosts
to compute $F(\J,\delta\J)$, after which we integrate over $x $ and $\d x$ to get $Z_\sg$.  
Eqn.  (\ref{pyp}) corresponds to a procedure
in which we first integrate over $\d x$ to get the product of delta functions in (\ref{doofus}), 
then integrate over matter fields and superghosts
to compute $\Xi(x)$, and finally integrate over $x$.   

The advantage of 
eqn. (\ref{yred}) it that it enables us to
maintain manifest BRST symmetry at all stages.   Eqn. (\ref{pyp}) has enabled us to make contact with
the literature, and has other applications  as well.

\subsubsection{More On Changes Of Coordinates}\label{yzzo}

By now we understand that the vacuum amplitude of the heterotic string can be computed by integrating
the naturally-defined section $\Xi(x)$ of  $\BBer(\M_L\times \M_R)$ or equivalently the pseudoform $F(\J,\delta\J)$
over the integration cycle $\varGamma$.  Moreover, once we introduce vertex operators, the same machinery
can be used to compute scattering amplitudes.   The only possible problems have to do with the behavior in the infrared
region at infinity.

Here we will describe an important subtlety\footnote{See \cite{ARS} for an early treatment of some of these issues.}  
that arises when one tries to choose coordinates and evaluate the
integral over $\varGamma$.  This subtlety is related to what was described in section \ref{harder},
so we will return to the simplified model in which we suppose that $\M_R$ has dimension $1|2$,
with local coordinates $m|\eta^1,\eta^2$, and $\M_L$ has dimension $1|0$ with a local coordinate $\t m$.

With only two odd coordinates, the general possible form of an even section $\Xi$ of $\BBer(\M_L\times \M_R)$ is
\begin{equation}\label{zyrto} \Xi=[\d\t m;\neg\d m|\d\eta^1,\d\eta^2]\left(\Upsilon_0(\t m,m)+
\Upsilon_2(\t m, m)\eta^1\eta^2\right). \end{equation} 
Naively speaking, since we are planning to perform a Berezin integral over all variables including
$\eta^1$ and $\eta^2$, it looks like we only care about $\Upsilon_2$; one would think that  $\Upsilon_0$
is projected out when we integrate over $\eta^1$ and $\eta^2$.

The trouble with this way of thinking is that although $\Xi$ is a naturally-defined object, its
separation into $\Upsilon_0$ and  $\Upsilon_2\eta^1\eta^2$
is not natural.  Consider the change of coordinates from $m|\eta^1,\eta^2$ to
\begin{align}\label{pyrto}m'&= m+a(m)\eta^1\eta^2 \cr
                                        \eta'{}^1&=\eta^1 \cr
                                         \eta'{}^2& =\eta^2.\end{align}
In the new coordinate system,
\begin{equation}\label{zyrtox} \Xi=[\d\t m;\neg\d m'|\d\eta'{}^1,\d\eta'{}^2]\left(\Upsilon'_0(\t m,m')+
\Upsilon_2'(\t m,m')\eta'{}^1\eta'{}^2\right),\end{equation}
with
\begin{equation}\label{troyo}
\Upsilon_2'(\t m ,m')=
\Upsilon_2(\t m, m')-\partial_{m'}\left(a(m')\Upsilon_0(\t m;m')\right). \end{equation}
We see that we need to know $\Upsilon_0$ in the old coordinate system in order to compute $\Upsilon_2'$
in the new coordinate system.

It is true that the term in $\Upsilon_2'$ that depends on $\Upsilon_0$ is a total derivative, so one might
hope that it would integrate to zero.
However, such arguments are not
as useful in practice as one might think.  Any coordinate system is only valid locally, and locally anything is
a total derivative.  It is only if $\M_R$ is holomorphically projected, so that we can avoid ever making changes
of coordinates  in which an even coordinate is shifted by a fermion bilinear, 
that one can in a fairly natural way forget $\Upsilon_0$ and study only $\Upsilon_2$.  

The moduli space of super Riemann surfaces is actually not holomorphically
projected in general \cite{DonW}.  
Even if it were, relying on such a projection would possibly make it difficult to understand the gauge
invariance and spacetime supersymmetry of superstring scattering amplitudes.  Finally, it turns out
that even in cases in which $\M_R$ is holomorphically projected, the proper 
definition of the integration cycle $\varGamma$
(in the infrared region at infinity that we will study in section \ref{propagator}) is sometimes such 
that a naive computation
using the holomorphic projection and throwing away $\Upsilon_0$ gives the wrong answer.  For an example,
see appendix \ref{leisurely}.

In short, to understand superstring theory without generating multiple complications, one has to work
with the naturally-defined objects  $\Xi(x)$ or $F(\J,\delta \J)$, rather than projections of them that
depend on coordinate choices.      

In the superstring literature of the 1980's, one typically integrated over the odd variables
to get a measure depending only on the even moduli  $\t m_\beta$ and $m_\alpha$.     This gave formulas that were only
locally defined and depended on a specific procedure for  integrating over the odd variables. The formulas would change 
by a total derivative, as in eqn. (\ref{troyo}) above, if one changed the procedure.  It was necessary to keep track
of these total derivatives.  
All this greatly added to the complexity of the subject.   Working on supermoduli space
with the natural objects $\Xi(x)$ and $F(\J,\delta\J)$ may or may not help with practical calculations, but it does make
the conceptual framework more simple. 
                                
\subsection{A Basis Of Odd Moduli}\label{basodd}

Now we will describe the parametrizations of $\M_L\times \M_R$  that are most usual in practice.  
One begins with a purely bosonic Riemann surface $\Sigma_\red$ with local moduli $\t m_\beta$ and $m_\alpha$,
and with a choice of spin structure.  The spin structure
determines a square root of the canonical bundle of $\Sigma_\red$ that we call $K^{1/2}$; its inverse we
call $T^{1/2}$.
 From $\Sigma_\red$,
we build a super Riemann surface $\Sigma$ by adding an odd variable $\theta$ that is valued in $T^{1/2}$.

We want to deform $\Sigma$
by turning on a gravitino field $\chi_{\t z}^\theta$.  This field enters
the general formula eqn. (\ref{hosetof}) for the $(1,0)$ part of the variation of $\J$:
\begin{equation}\label{nosetof} \delta\J_{\t z}^z=h_{\t z}^z+\theta \chi_{\t z}^\theta.       \end{equation}
Our main interest here is not the bosonic field $h_{\t z}^z$  that describes the bosonic moduli $m_\alpha$, but the
gravitino field $\chi_{\t z}^\theta$ that describes the odd moduli.  
At the linearized level, it is subject to the gauge equivalence
\begin{equation}\label{broof}\chi_{\t z}^\theta\to \chi_{\t z}^\theta+\partial_{\t z}y^\theta.\end{equation}
(This equivalence follows from the action (\ref{itzo}) of a vector field $q^z\partial_z$ on $\delta\J_{\t z}^z$;
one expands $q^z(\t z;\neg z|\theta)=s^z(\t z;\neg z)+\theta y^\theta(\t z;\neg z)$ and then (\ref{broof}) is the
symmetry generated by $y^\theta$.)    Modulo this gauge equivalence, $\chi_{\t z}^\theta$ takes
values in the sheaf cohomology group $H^1(\Sigma_\red,T^{1/2})$, whose dimension is $2\g-2$.  

So we can specify  a family of fermionic deformations of $\Sigma$  by picking a set of $2\g-2$  $c$-number 
$\chi_{\t z}^\theta$ fields, representing a basis of $H^1(\SIgma_\red,T^{1/2})$.  We call these fields 
$\chi^{(\sigma)\theta}_{\t z}$,
$\sigma =1 ,\dots, 2\g-2$.  Any generic
set  of $\chi^{(\sigma)}$'s will do, since the condition that they project to linearly independent elements
of $H^1(\SIgma_\red,T^{1/2})$ is satisfied generically. Then
we expand the gravitino field as 
\begin{equation}\label{zingo}\chi_{\t z}^\theta=\sum_{\sigma=1}^{2\sg-2}\eta_\sigma \chi_{\t z}^{(\sigma)\theta},\end{equation}
with odd coefficients $\eta_\sigma$.  The $\eta_\sigma$ can serve as the odd moduli of $\M_R$, in a suitable region.

This is actually a convenient way to give a local parametrization of the odd directions in supermoduli space.  
However, there are two
very important pitfalls.  The first is simply that we  have to make sure that the $\chi^{(\sigma)}$'s do give
a basis for $H^1(\Sigma_\red,T^{1/2})$.  This is { untrue} precisely if a linear combination of
the $\chi^{(\sigma)}$'s, with $c$-number coefficients $e_\sigma$ that are not all zero, can be gauged away by some $y^\theta$:
\begin{equation}\label{humby} \partial_{\t z}y^\theta =\sum_\sigma e_\sigma \chi^{(\sigma)\theta}_{\t z}.   \end{equation}
The cokernel of the operator $\partial_{\t z}$, mapping sections of $T^{1/2}$ to $(0,1)$-forms with values in $T^{1/2}$,
is $H^1(\Sigma_\red,T^{1/2})$, which has dimension $2\g-2$,  and generically
to get a non-zero solution of an equation such as (\ref{humby}) with a non-zero source on the right hand side,
we need to adjust $2\g-1$ parameters.\footnote{One of these parameters is an overall scaling of the right hand
side of the equation, or
in other words of the $e_\sigma$;
this does not affect whether the equation has a solution.  Modulo this scaling, we need to adjust $2{\sg}-2$
parameters to get a solution, the same number as the dimension of the cohomology group.}
The $2{\g}-2$ parameters $e_\sigma$ are not quite enough, but if we also vary one of the bosonic moduli 
$m_\alpha$, then we have enough parameters so that it is natural for eqn. (\ref{humby}) to have a solution
for some isolated values of $m_\alpha$. The fermionic gauge-fixing that is defined by the choice of slice (\ref{zingo})
is then wrong at such a value of $m_\alpha$, and the measure on supermoduli space, computed as in section 
\ref{pco} using this slice, will develop a pole.  (Explicitly, the pole arises because the product of delta functions in (\ref{pinzor})
does not remove all of the zero-modes of the field $\beta_{z\theta}$.)

Except in very 
low genus, this phenomenon is practically unavoidable and in the literature there is a technical name for these poles -- they
are called spurious singularities. They are spurious because they result from an invalid gauge-fixing. 
To compute correctly in a formalism with spurious singularities can become complicated in general, as one has to keep
track of total derivatives supported on the locus of spurious singularities.  We will not try to explain how to do
that here.  Rather, our goal is to understand how to calculate in a framework 
in which all singularities are physically sensible and represent the effects of
on-shell particles.  So we want to avoid spurious singularities.  

That does not mean avoiding the
use of the slices we have described.  It means that such a slice should be used in the following
sense.  We pick a suitable set of $\chi^{(\sigma)}$'s and a sufficiently small open set $\U$ in $\M_L\times \M_R$ so that eqn. (\ref{humby}) 
has no solution when
$\Sigma_\red$ is parametrized by $\U$. 
Then, in that region, we use eqn. (\ref{zingo}) to define fermionic moduli $\eta_\sigma$.  Together 
with the moduli $m_\alpha$ and $\t m_\beta$ of $\Sigma_\red$, 
this gives a local coordinate system on $\M_L\times \M_R$.
 In this coordinate system, we compute the invariant objects $F(\J,\delta\J)$ or $\Xi(x)$ on $\M_L\times \M_R$
 that have to be integrated
 to compute the heterotic string amplitude.  Of course, with a particular basis of gravitino modes, this procedure 
works only in a suitable open set $\U$.  But we can cover
$\M_L\times \M_R$ with open sets $\U^{(\zeta)}$ in each of which we can pick  an appropriate 
 basis  $\chi^{(\sigma;\zeta)}$
 of gravitino fields such that eqn. (\ref{humby}) nowhere has a non-zero solution.  This gives
 us in each $\U^{(\zeta)}$ 
 a good set of odd moduli 
$\eta_{\sigma;\zeta}$.  Using these local slices,
 we compute $F(\J,\delta\J)$ or $\Xi(x)$ in each $\U^{(\zeta)}$; the results automatically agree on intersections
 $\U^{(\zeta)}\cap \U^{(\zeta')}$ since 
 $F(\J,\delta\J)$ and $\Xi(x)$ are independent of the coordinate choices.
 
After covering $\M_L\times \M_R$ by the open sets $\U^{(\zeta)}$ and defining a basis of
 odd moduli $\eta_{\sigma;\zeta}, \,\sigma=1,\dots,2\g-2$ in each open set
as in the last paragraph, what we do {\it not} want to do is to
 integrate over the $\eta_{\sigma;\zeta}$ in each $\U^{(\zeta)}$ to get a measure on the reduced space parametrized
 by the bosonic moduli $m_\alpha$ and $\t m_\beta$ only.  
 These computations would not fit together in a simple
 way on intersections $\U^{(\zeta)}\cap \U^{(\zeta')}$, for the following reason.  As we will see shortly, 
 when we change basis from $\chi^{(\sigma;\zeta)}$ to $\chi^{(\sigma;\zeta')}$ in the intersection
 $\U^{(\zeta)}\cap \U^{(\zeta')}$, the moduli $m_\alpha$ undergo the sort of transformation that we studied in eqn.
(\ref{pyrto}).   In the language of that discussion, integrating over the odd moduli 
$\eta_{\sigma;\zeta}$ in each $\U^{(\zeta)}$ would
entail  dropping $\Upsilon_0$ and keeping $\Upsilon_2$.  
A correct computation has to keep track of $\Upsilon_0$.  Conceptually, the simplest  correct procedure
 is to use the local slices to compute the invariant objects $F(\J,\delta\J)$ or $\Xi(x)$  
 (or their generalizations with external vertex operators), and then evaluate
 the amplitude using (\ref{yred}) or (\ref{pyp}).  In that way,
 we extract only invariant information from the local slices and we never meet spurious singularities.   It is also possible but cumbersome
 to describe a correct procedure without the geometric interpretation that we have just explained.  See for example \cite{SenWitten}.

We still have to explain why, when we change  basis for the gravitino modes from $\chi^{(\sigma;\zeta)}$ to $\chi^{(\sigma;\zeta')}$, the bosonic moduli undergo a transformation such 
as  (\ref{pyrto}).  Any change of basis for the $\chi^{(\sigma)}$ can be made by replacing them with linear
combinations of themselves and modifying them by gauge transformations.
 The subtlety entirely comes
from the gauge transformations.\footnote{The reason for this is that replacing the $\chi^{(\sigma)}$ by linear combinations
of themselves does not change the space of gravitino modes over which we are integrating, only its parametrization by the $\eta$'s.
See the discussion of eqn. (\ref{underw}).}   To implement a general change of basis, we need to make
 a gauge transformation on each
$\chi^{(\sigma)}$ separately
\begin{equation}\label{hundo}\chi^{(\sigma)\theta}_{\t z}\to   \chi^{(\sigma)\theta}_{\t z}+{\partial_{\t z}}y^{(\sigma)\theta}.
\end{equation}
This amounts to the gauge transformation
\begin{equation}\label{bundo}\chi\to\chi+\partial_{\t z}y,\end{equation}
with $\chi$ as in (\ref{zingo}) and
\begin{equation}\label{dokey} y=\sum_{\sigma=1}^{2\sg-2}\eta_\sigma y^{(\sigma)}.\end{equation}

The problem is that although in leading order in odd variables, a gauge transformation with parameter $y$ acts on $\chi$
as in (\ref{broof}) while leaving $h$ invariant, in the next order, $h_{\t z}^z$ is transformed by
\begin{equation}\label{yundo}h_{\t z}^z\to h_{\t z}^z+y^\theta\chi_{\t z}^\theta. \end{equation}
In the present context, this is
\begin{equation}\label{brudo}h_{\t z}^z\to h_{\t z}^z+\sum_{\sigma,\sigma'}\eta_{\sigma}\eta_{\sigma'}y^{(\sigma)\theta}
\chi_{\t z}^{(\sigma')\theta}.\end{equation}
This deformation of $h_{\t z}^z$ arises as soon as there are two or more odd moduli.
Importantly, even though it is induced by a gauge transformation of $\chi$,
the shift in $h_{\t z}^z$ is not a pure gauge from a purely bosonic point of view,
since it is not of the form $\partial_{\t z}w^z$ for any $w^z$.  
So this shift in $h_{\t z}^z$ changes the bosonic geometry in a nontrivial fashion and shifts
the even moduli $m_\alpha$ by bilinears in the $\eta$'s. 

In the context of supergeometry, this mixing of even and odd 
variables is quite natural and one should not expect to be able to define  bosonic moduli 
while ignoring the existence of odd variables.  So the result probably should not be a surprise.  It implies
that except in special cases in low
orders,  one should not expect to be able to avoid
the subtleties described
in sections \ref{harder} and \ref{yzzo}.

\subsection{Integrating Out  Odd Variables}\label{pcorp}

\subsubsection{General Procedure}\label{zico}

We will now make the considerations of section \ref{basodd} more concrete.  The worldsheet action of the heterotic
string is actually linear in the gravitino field $\chi_{\t z}^\theta$, so the dependence of the action on this field is entirely contained
in the linear formula (\ref{bitzo}).  (For Type II superstrings, matters are a little more complicated; the action has a term that involves
the product $\chi_{\t z}^\theta\chi_z^{\t \theta}$.)  For the gravitino field (\ref{zingo}), the gravitino coupling comes out to be
\begin{equation}\label{tyrog}I_\eta=\sum_{\sigma=1}^{2\sg-2}\frac{\eta_\sigma}{2\pi} \int_{\Sigma_\red} \d^2z\, \chi_{\t z}^{(\sigma)\theta} 
S_{z\theta}. \end{equation}   (This is an integral on the reduced space $\Sigma_\red$, with $\theta$ integrated out; $\theta$ is similarly integrated out in the following formulas.) 

There is also a corresponding coupling of $\d\eta_\sigma$ to the holomorphic antighost field $B_{z\theta}$.  We can deduce it from (\ref{extac}).
It actually is related to (\ref{tyrog}) by the BRST symmetry $\{Q_B,\eta_\sigma\}=\d\eta_\sigma$, $[Q_B,\beta_{z\theta}]=S_{z\theta}$:
\begin{equation}\label{bryrog}I_{\d\eta}=\sum_{\sigma=1}^{2\sg-2} \frac{\d\eta_\sigma}{2\pi}\int_{\Sigma_\red}\d^2z\,
 \chi_{\t z}^{(\sigma)\theta}\beta_{z\theta}.
 \end{equation} 
 
We have given the reader fair warning that in general it is not illuminating to integrate out the odd variables.  On the other hand, it is
also important to know what happens when we do integrate them out, both to make contact with the literature and more importantly
because integrating out the odd variables is useful for performing practical calculations in low orders.  

According to the above formulas, for each $\sigma$, the dependence of the integrand of the path integral on the pair $\eta_\sigma$, $\d\eta_\sigma$
is a simple factor
\begin{equation}\label{inzoop}\exp\left(-\frac{1}{2\pi}\int_{\Sigma_\red}\d^2z\,\chi_{\t z}^{(\sigma)\theta}\left(\eta_\sigma S_{z\theta}+
\d\eta_\sigma \beta_{z\theta} \right) \right).\end{equation}
If we integrate out the pair $\eta_\sigma$ and $\d\eta_\sigma$,
we get simply
\begin{equation}\label{pinzo}\delta\left(\int_{\Sigma_\red}\d^2 z\, \chi_{\t z}^{(\sigma)\theta} \beta_{z\theta}\right)\cdot\int_{\Sigma_\red}
\d^2 z\,\chi_{\t z}^{(\sigma)\theta} S_{z\theta}.\end{equation}
This is a standard formula; see \cite{superoperator,EHVerl} or eqn. (3.335) in \cite{DPh}.  Usually 
the differentials $\d\eta_\sigma$ are not introduced explicitly
in the formalism, and one says simply that (\ref{pinzo}) results from integration over $\eta_\sigma$.  Since an odd variable
is its own delta function, we can also write (\ref{pinzo}) with the two factors treated more symmetrically:
\begin{equation}\label{winzo}\delta\left(\int_{\Sigma_\red}\d^2 z\, \chi_{\t z}^{(\sigma)\theta} \beta_{z\theta}\right)\cdot\delta\left(\int_{\Sigma_\red}
\d^2 z\,\chi_{\t z}^{(\sigma)\theta} S_{z\theta}\right).\end{equation}

 The result  (\ref{pinzo}) or (\ref{winzo}) is invariant under a rescaling of the gravitino mode $\chi_{\t z}^{(\sigma)\theta}$,
 because the two factors transform oppositely.  A more fundamental explanation
of why such rescaling does not matter is  that the gravitino field (\ref{zingo}) is unchanged under
\begin{equation}\label{ydze}\chi_{\t z}^{(\sigma)\theta}\to \lambda \chi_{\t z}^{(\sigma)\theta} ,~~\eta_\sigma\to \lambda^{-1}\eta_\sigma,~~~\lambda\in\C^*,\end{equation}
and the extended action is invariant if we also rescale $\d\eta_\sigma$ by a factor of $\lambda^{-1}$.
So when we integrate out $\eta_\sigma$ and $\d\eta_\sigma$ -- an operation that is certainly invariant under a common
rescaling of $\eta_\sigma$ and $\d\eta_\sigma$ -- we get a result that is invariant under rescaling of the gravitino mode
$\chi_{\t z}^{(\sigma)\theta}$.

An important generalization of what we have just explained is that the operation of integrating out several odd moduli $\eta_1,\dots,\eta_s$
only depends on the linear span of the gravitino fields $\chi^{(1)\theta}_{\t z}\dots\chi^{(s)\theta}_{\t z}$.  The reason for this is
that in the definition of the gravitino field and in the extended action we can compensate for a linear transformation
\begin{equation}\label{underw}\begin{pmatrix}\chi_{\t z}^{(1)}\cr \vdots \cr \chi_{\t z}^{(s)\theta}\end{pmatrix}\to M\begin{pmatrix}\chi_{\t z}^{(1)}\cr \vdots \cr \chi_{\t z}^{(s)\theta}\end{pmatrix} \end{equation}  
for any invertible matrix $M$ by
\begin{equation}\label{bunder}\begin{pmatrix}\eta_1 & \dots & \eta_s\end{pmatrix} \to \begin{pmatrix}\eta_1 & \dots & \eta_s\end{pmatrix}M^{-1},~~
\begin{pmatrix}\d\eta_1 & \dots & \d\eta_s\end{pmatrix} \to \begin{pmatrix}\d\eta_1 & \dots & \d\eta_s\end{pmatrix}M^{-1}.\end{equation}
The combined operation leaves fixed the gravitino field and the extended action.
Integrating out the $\eta$'s and $\d\eta$'s is invariant under the transformation (\ref{bunder}).  So the factor
\begin{equation}\label{pinzor}\prod_{\sigma=1}^s\left(\delta\left(\int_{\Sigma_\red}\d^2 z\, \chi_{\t z}^{(\sigma)\theta} \beta_{z\theta}\right)\cdot\int_{\Sigma_\red}
\d^2 z\,\chi_{\t z}^{(\sigma)\theta} S_{z\theta}\right)\end{equation}
that comes by integrating out the $\eta$'s and $\d\eta$'s is invariant under (\ref{underw}), as one may readily verify.

\subsubsection{The Picture-Changing Operator}\label{pco}

The example of this construction that is most often considered is that each gravitino mode $\chi_{\t z}^{(\sigma)\theta}$ is a delta function
supported at some point $p_\sigma \in\Sigma_\red$:
\begin{equation}\label{infox}\chi_{\t z}^{(\sigma)\theta}=\delta_{p_\sigma}.\end{equation}
There is a subtlety here: $\chi_{\t z}^{(\sigma)\theta}$ is supposed to be a $(0,1)$-form valued in $T^{1/2}$.  The space of such forms
with delta function support at
 $p_\sigma$ is one-dimensional, but there is no natural way to pick a non-zero vector in this space.  So there is no natural
way to normalize the delta function in (\ref{infox}).  However, if we are planning to integrate out the odd variables, 
this does not matter very much, because of an observation in section \ref{zico}: 
the formalism is invariant under a rescaling of the gravitino modes 
(with a compensating rescaling of the odd modulus $\eta_\sigma$). 

For a delta-function gravitino supported at $p\in \Sigma_\red$, the factor  (\ref{pinzo}) associated with integrating out an odd  modulus becomes
\begin{equation}\label{nifox}\YY(p)=\delta(\beta(p))S_{z\theta}(p).  \end{equation}
This is called the picture-changing operator.  It was originally defined rather differently in \cite{FMS}.  The definition (\ref{nifox}) 
was first presented in \cite{EHVerl,EHV}, where the meaning of operators such as $\delta(\beta(p))$ was also
analyzed. (We explore this subject further in section \ref{betag} of the present paper.)   $\YY(p)$ is a  primary
field of conformal dimension 0, the product of the supercurrent
$S_{z\theta}$, which has dimension $3/2$, and $\delta(\beta)$, which has dimension $-3/2$.

 Actually, the definition of $\YY(p)$ needs some clarification, because there
is a singularity in the operator product of $S_{z\theta}(p)$ and $\delta(\beta(p))$.  The usual approach is to fix the ambiguity (which
involves an operator of the form $\delta'(\beta)\partial_z b_{zz}$, where $b_{zz}$ is the fermionic antighost)
by using BRST symmetry.  Equivalently, one may regularize the definition of $\YY(p)$ by using a smooth gravitino
wavefunction with very concentrated support and taking the limit as it approaches a delta function.

At least locally on the reduced space $(\M_L\times \M_R)_\red$, one can use picture-changing operators to integrate over all odd moduli.  For this, one picks points $p_\sigma$, $\sigma=1,\dots,2\g-2$,
and one lets 
\begin{equation}\label{dolme}\chi_{\t z}^{(\sigma)\theta}=\delta_{p_\sigma}.\end{equation}
Initially, we take the $p_\sigma$ to be distinct so that the $\chi_{\t z}^{(\sigma)\theta}$'s are linearly independent and have
a chance to provide a basis for $H^1(\Sigma_\red,T^{1/2})$.
With this choice of gravitino modes, integrating out the odd moduli will give a product of picture-changing operators
\begin{equation}\label{olme}\prod_{\sigma=1}^{2\sg-2}\YY(p_\sigma). \end{equation}

It is interesting to ask when the use of picture-changing operators to integrate out all the odd moduli
will lead to a spurious singularity. This will happen when there is 
a non-zero solution of the equation (\ref{humby}), which here reads
\begin{equation}\label{tomox}\partial_{\t z}y^\theta=\sum_{\sigma=1}^{2\sg-2} e_\sigma \delta_{p_\sigma}. \end{equation}
A solution $y^\theta$ of this equation is a section of $T^{1/2}$ that has simple poles at the points $p_1,\dots,p_{2\sg-2}$ (with residues
$e_1,\dots,e_{2\sg-2}$) and is holomorphic
elsewhere.  So $y^\theta$ is a holomorphic section of the line bundle $T^{1/2}(p_1+\dots+p_{2\sg-2})= 
T^{1/2}\otimes \O\left(\sum_{\sigma=1}^{2\sg-2} p_\sigma\right)$.  
This is, by definition, the line bundle whose holomorphic sections are sections of $T^{1/2}$ that may have simple poles at the $p_i$.
It has degree $\g-1$, since  $T^{1/2}$ has degree $-(\g-1)$ and allowing the poles increases the degree by
$2\g-2$.  A generic holomorphic line bundle of degree $\g-1$ has no holomorphic section; however, when one varies one complex parameter (a modulus of $\Sigma_\red$, or the choices of the $p_i$), it is generic for such a section to arise at special values of this
parameter.
In particular, if one  integrates out the odd moduli of $\Sigma$ via a product of picture-changing operators, one should expect
to encounter spurious singularities as the bosonic moduli -- the moduli of $\Sigma_\red$ -- 
are varied.  These are poles that appear where
\begin{equation}\label{dumo}H^0(\Sigma_\red,T^{1/2}(\SIgma_\sigma p_\sigma))\not=0.\end{equation}

So far, we have assumed that the points $p_\sigma$ are distinct.  However the criterion
(\ref{dumo})
for a spurious singularity makes perfect sense even if some of the $p_\sigma$ coincide.  In fact, the procedure
of integrating out odd moduli by using gravitino wavefunctions with delta function support has a generalization  (not a limit; the
distinction will become clear) when
some of the $p_\sigma$ coincide.  To explain this heuristically, let us choose a local coordinate $z$ such that $p_1$ and $p_2$
correspond, say, to $z=0$ and $z=\epsilon$.  So two of our gravitino wavefunctions are
\begin{equation}\label{timex}\chi_{\t z}^{(1)\theta}=\delta^2(z), ~~\chi_{\t z}^{(2)\theta}=\delta^2(z-\epsilon).\end{equation}
But the result of integrating out the odd moduli is invariant under a linear transformation (\ref{underw}) of the gravitino
modes, and as a special case of this, we can replace $\chi_{\t z}^{(2)\theta}$ with
\begin{equation}\label{limex}\h\chi_{\t z}^{(2)\theta}=-\frac{1}{\epsilon}\left(\chi^{(2)\theta}_{\t z}-\chi^{(1)\theta}_{\t z}\right)=
-\frac{1}{\epsilon}\left(\delta^2(z-\epsilon)-\delta^2(z)\right).\end{equation}
In this form, we can take the limit as $\epsilon\to 0$, getting
\begin{equation}\label{pimex}\h\chi_{\t z}^{(2)\theta}=\partial_z\delta^2(z).\end{equation}
Thus if one gravitino mode is a delta function at a given point, it is fairly natural to take a second one to be the derivative
of a delta function at the same point.  

By evaluating (\ref{pinzo}), we find that if $\chi_{\t z}^{(\sigma)\theta}=\partial_z \delta^2(z-z_0)$ for some $z_0$, then
integrating out the corresponding odd modulus $\eta_\sigma$ and its differential $\d\eta_\sigma$ gives a factor
\begin{equation}\label{himiny} \delta(\partial_z\beta(z_0))\partial_z S_{z\theta}(z_0). \end{equation}
Again some regularization of this singular product is required.  

Similarly, the picture-changing formalism has a generalization when any number of the points
$p_1,\dots,p_{2\sg-2}$ coincide.   If  $k$ of them are to coincide at some point in $\Sigma_\red$, say $z=0$,
we take $k$ of the gravitino wavefunctions to be  $\delta^2(z)$, $\partial_z\delta^2(z)$, $\dots,$ $\partial^{k-1}_z\delta^2(z)$.

The foregoing remarks must be treated with care.  It is true, as we have just seen,
 that the picture-changing formalism has a natural
analog when two or more of the points $p_\sigma$ coincide.   However, this analog is not simply 
the limit of the picture-changing formalism when,
say, $p_1\to p_2$. The reason is that there is a short-distance
singularity in the operator product $\YY(p_1)\YY(p_2)$ for $p_1\to p_2$.  
The above classical treatment amounted to normal-ordering this product and dropping the singular terms.
The singular terms are $Q_B$-exact, so that their contribution to $F(\J,\delta\J)$ is an exact form, but as usual such exact
forms must be treated carefully.  So the picture-changing formalism, defined initially for distinct points $p_\sigma$, has an analog
when some of the points are taken to coincide, but it does not in general have a limit as the points approach each other.

\subsubsection{Picture Number}\label{pnumber}

We already observed in discussing eqn. (\ref{ofus}) that on a genus $\g$ Riemann surface with no vertex operator
insertions, we need $2\g-2$ delta functions of the commuting ghost field $\beta_{z\theta}$ in order to get a sensible path integral. These insertions remove singularities that would otherwise arise from zero-modes of this field.
These delta functions may be in general delta functions of  arbitrary modes of the field $\beta_{z\theta}$, of the form
\begin{equation}\label{yder}\delta\left(\int_{\Sigma_\red}\d^2z \,\chi_{\t z}^{(\sigma)\theta} \beta_{z\theta}\right)  \end{equation} 
for any $c$-number gravitino field $\chi_{\t z}^{(\sigma)}$.
Any generic $2\g-2$ insertions of this kind will give a path integral that makes sense (in the absence of delta functions of $\gamma$,
as discussed shortly). 

We call an operator of the form (\ref{yder}) an operator of picture number 1.   Generically, the picture number 1 operators in
(\ref{yder}) are not local operators.  They become local operators -- although relatively unfamiliar ones -- if $\chi_{\t z}^{(\sigma)\theta}$
has delta function or derivative of delta function
support.  In that case, we get local operators of the form $\delta(\beta(z))$, $\delta(\partial_z\beta(z))$,
$\delta(\partial_z^2\beta(z))$, etc., all of picture number 1.  

According to generalities of quantum field theory, once we allow a local operator $\delta(\beta(z))$, we also
must allow its $z$-derivative $\partial_z\delta(\beta(z))$.  Formally
\begin{equation}\label{inko} \partial_z\delta(\beta(z))=\partial_z\beta(z)\cdot \delta'(\beta(z)).\end{equation}
This indicates that a reasonable formalism will have to include not just delta function operators $\delta(\beta(z))$
but more general distributional operators $\delta'(\beta(z))$, $\delta''(\beta(z))$, and so on.   In fact, an operator such
as $\delta'(\beta(z))$ can remove a $\beta$ zero-mode and help give a sensible path integral, just as $\delta(\beta)$ can.
We will describe this in more detail  in section \ref{betag}.  So we assign picture number 1 to
$\delta'(\beta(z))$, and similarly to $\delta''(\beta(z))$, $\delta'(\partial_z\beta(z))$, etc., just as to $\delta(\beta(z))$.  Similarly we consider
a nonlocal operator 
\begin{equation}\label{ydery}\delta^{(k)}\left(\int_{\Sigma_\red}\d^2z \,\chi_{\t z}^{(\sigma)\theta} \beta_{z\theta}\right)  \end{equation} 
(where $\delta^{(k)}$ is the $k^{th}$ derivative of a delta function) to have picture number 1.

All this matches integration theory on supermanifolds, as very briefly reviewed in section \ref{lightning}.
In that subject, one assigns picture number $-1$
to $\delta(\d\theta)$ (where $\theta$ is an odd coordinate so $\d\theta$ is an even differential) and also to $\delta'(\d\theta)$,
$\delta''(\d\theta)$, etc.  (See eqn. (\ref{condo}) for a concrete example using a form $\delta'(\d\theta)$.)  The role
of a picture number $-1$ operator is to reduce by 1 the number of $\d\theta$'s over which one has to integrate,
and any of these delta functions has that effect.  The reader who
wishes to understand more deeply 
the analogy between picture number in integration on supermanifolds and picture number in 
the $\beta\gamma$ system  is urged to consult \cite{Belthree}.

Now we will explain some important aspects of the $\beta\gamma$ system.  Some readers may want
to jump to section \ref{betag} (which does not depend on the intervening parts of this paper), where the following is explained more systematically.  Readers who
find what is stated in the next few paragraphs to be at least temporarily sufficient can simply proceed.  
In any event, the relevant facts are this.  The ghost field
$\gamma$ is a section of a line bundle $T^{1/2}$.  Suppose for simplicity that this line bundle has no holomorphic sections (which
is the case for $\g\geq 2$); then $\gamma$ has no zero-modes.
 The index of the $\partial_{\t z}$ operator acting on sections of $T^{1/2}$ is $-(2\g-2)$, so $\beta$ has  $2\g-2$ zero-modes and it takes $2\g-2$ insertions of $\delta(\beta)$ to get
a sensible path integral.  Now suppose that we include in the path integral a factor $\delta(\gamma(p))$ for some point $p$.
Since this forces $\gamma$ to vanish at $p$, it effectively replaces the line bundle $T^{1/2}$ by $T^{1/2}(-p)$ (whose sections are sections of $T^{1/2}$ that vanish
at $p$).  The index is now $-(2\g-1)$ and a sensible path integral requires $2\g-1$ operators of the general type $\delta(\beta)$.  
More generally, 
when delta functions of the commuting ghost $\gamma$ are considered, as well as delta functions of $\beta$,
one finds that  a $\beta\gamma$ path integral with $r$ delta functions of $\beta$ and $s$ delta functions of $\gamma$
is sensible only if
\begin{equation}\label{ruffo} r-s=2\g-2.\end{equation}
(This criterion is necessary but only generically sufficient, since exceptionally it can happen that the delta functions
do not remove all the zero-modes.)   This statement is valid for all $\g$, not just the case $\g\geq 2$ that we started with.

The condition (\ref{ruffo})  does not depend on exactly what sort of delta functions are considered; from this point of view,
we do not want to distinguish $\delta(\gamma)$ from $\delta(\partial_z\gamma)$, $\delta'(\gamma)$, etc., just
as we did not make such a distinction for delta functions of $\beta$.
So we assign picture number 1 to any delta function of $\beta$, regardless of details of its construction, and picture
number $-1$ to any delta function of $\gamma$.  Then the general statement is that 
that the net picture number of all operators must add up to $2\g-2$ in order for the path integral to be sensible.
Needless to say, superstring scattering amplitudes are always described by sensible path integrals.

The $\beta\gamma$ system also has an anomalous symmetry -- the holomorphic ghost number -- under which $\beta$ and
$\gamma$ have charges $-1$ and $1$, respectively.  This quantum number is also carried by $b$ and $c$ (the $\chi_{\t z}^\theta S_{z\theta}$
term in the action does not allow separate ghost number symmetries for $b,c$ and for $\beta,\gamma$, though there is a separate
antiholomorphic ghost number symmetry carried by $\t b,\t c$).  The $b,c$ system has a ghost number anomaly $-(3\g-3)$ and
the $\beta,\gamma$ system has an anomaly $2\g-2$, so the net ghost number anomaly is $-(\g-1)$.  So to get a non-zero path
integral, we need to insert a product of operators with a net ghost number $-(\g-1)$.  A typical example is the product of
delta functions of $b$ and $\beta$ that we generated in (\ref{doofus}). This was
\begin{equation}\label{ycco}\delta^{3\sg-3|2\sg-2}(B^{(\alpha)})=\prod_{\alpha'=1}^{3\sg-3}\delta(b^{(\alpha')}) \prod_{\alpha''=1}^{2\sg-2}
\delta(\beta^{(\alpha'')}).\end{equation}
Each $\delta(b)$ has ghost number $-1$ and each $\delta(\beta)$ has ghost number $1$.  (That is because under the
scaling $(b,\beta)\to \lambda^{-1}(b,\beta)$, which is the ghost number symmetry of the antighosts,
 $\delta(b)$ scales like $b$ but $\delta(\beta)$ scales oppositely to
$\beta$.)  So the product of delta functions in (\ref{ycco}) does have net ghost number $-(\g-1)$.

The assertion that the  ghost number is an anomalous symmetry 
means that a path integral with operator insertions of the wrong ghost number
is well-defined but vanishes.  Picture number is not really a symmetry in that sense, anomalous or not, since a path integral
with operator insertions of the wrong picture number is divergent or not well-defined, rather than being zero.

\section{The Neveu-Schwarz Sector}\label{vertex}

In this section, we begin our study of 
superstring vertex operators.  For the most part, it does not matter much which supersymmetric
string theory we consider, since  left- and right-moving worldsheet degrees of freedom can be treated
independently in many respects.  For definiteness, we concentrate on the heterotic string.   
In this case, there is a supersymmetric structure
only for right-movers, and there are two types of vertex operator -- Neveu-Schwarz (NS) and Ramond (R).  We begin
with the more straightforward NS case, deferring Ramond vertex operators to section \ref{ramond}.

\subsection{Neveu-Schwarz Vertex Operators}\label{nsop}

A Neveu-Schwarz vertex operator is simply inserted at a point $\t z;\neg z|\theta$ on a heterotic string worldsheet $\Sigma$.
This operation increases the holomorphic dimension of the moduli space
by $1|1$, the extra moduli being simply the values of $z$ and $\theta$
at which the vertex operator is inserted.  Similarly, the antiholomorphic dimension of the moduli space
is increased by $1|0$, the extra modulus being $\t z$.  

Let us try to understand, by analogy with section \ref{cvo}, what is the simplest type\footnote{Instead of the simplest
choice, one can ask for the most general possible choice. As discussed for the bosonic
string in section \ref{morge}, in general we could use arbitrary $Q_B$-invariant operators that
are annihilated by $b_0-\t b_0$.  To use this larger class of vertex operators, one needs a more elaborate formalism that
does not assume superconformal symmetry.} of vertex operator that we can use
to compute scattering amplitudes involving NS states.  

Just as in section \ref{vertop}, for it to be possible to express scattering amplitudes in terms of forms on a finite-dimensional
space (basically the moduli space of super Riemann surfaces with  punctures), we will want the vertex operators to depend
only on the ghost fields $\t c$, $c$, and $\gamma$, and not on their derivatives. 
 An equivalent condition is that 
\begin{equation}\label{igoz} b_n\V  = \t b_n\V =\beta_r\V=0, ~~n,r \geq 0,\end{equation}
where the $b$, $\t b$ modes were defined in eqn. (\ref{dobbo}) and
\begin{equation}\label{betamodes}\beta_r=\frac{1}{2\pi i}\oint \d z\, z^{r+1/2} \beta_{z\theta}. \end{equation}
(The need for the condition (\ref{igoz}) is explained in section \ref{scattering}.)

We certainly also want our vertex operators to be annihilated by the BRST charge $Q_B$.
As in our discussion of the bosonic string, if $Q_B\V=0$ and $\V$ also obeys (\ref{igoz}), 
then it follows that from a holomorphic point of view, $\V$ is a superconformal primary of dimension 0
\begin{align}\label{monkey}  L_n\V & =0, ~~n\geq 0 \cr  G_r\V& =0,~~r\geq 1/2, \end{align}
and similarly that from an antiholomorphic point of view, $\V$ is a conformal primary of dimension 0,
\begin{equation}\label{yonky}\t L_n\V=0, ~~n\geq 0.\end{equation}
This latter condition is familiar from the bosonic string.

The conditions (\ref{monkey}) have an intuitive meaning.  Looking back to (\ref{urv}), we see that the generators
$L_n$ and $G_r$ with $n,r\geq 0$ are precisely the ones that vanish at $z=\theta=0$.  
So  the point $z=\theta=0$ is invariant under the symmetries generated by those operators,
and if we want to insert a vertex operator at $z=\theta=0$ that preserves all the superconformal
symmetries that leave fixed this point, then that
 operator should  obey (\ref{monkey}).
Similarly, (\ref{yonky}) says that $\V$ is invariant under antiholomorphic changes of coordinate that leave
fixed the point at which it is inserted.  

  How should $\V$ depend on the ghost fields?  We can answer this as follows.
  Since inserting
an NS vertex operator increases the antiholomorphic dimension of moduli space by $1|0$ and the holomorphic dimension
by $1|1$, in a formalism similar to what we had for bosonic strings in section \ref{bosmeasure} or for the superstring vacuum
amplitude in section \ref{measure}, there will be an extra delta function insertion for each of $\t b$, $b$, and $\beta$.
We write this schematically as
\begin{equation}\label{monstro} \delta(\t b)\delta(b)\delta(\beta). \end{equation}
Let us consider first the holomorphic degrees of freedom.  The operator $\delta(b)\delta(\beta)$ has ghost number 0 and picture number
$1$.   We want $\V$ to be such that when we include $\V$ in the path integral and also include an operator with the quantum
numbers $\delta(b)\delta(\beta)$, the path integral remains sensible and non-zero.  
For this, $\V$ must have holomorphic ghost
number 0 and picture number $-1$.  The only operator with these quantum numbers that does not depend on derivatives
of $c$ or $\gamma$ is $c\delta(\gamma)$.  We need the factor $\delta(\gamma)$ to get picture number $-1$ (we cannot
use other delta functions such as $\delta(\partial_z\gamma)$ or 
$\partial_z\delta(\gamma)=\partial_z\gamma\delta'(\gamma)$, since
they  depend on derivatives of $\gamma$).  And given that $\delta(\gamma)$ has ghost number $-1$, we need a factor of $c$
to make the holomorphic ghost number of $\V$ vanish.

As a bonus, we can extend this argument to show that $\V$ cannot 
depend on the holomorphic antighosts $b$ and $\beta$.  Ordinary
functions of $b$ and $\beta$ would make the holomorphic ghost number negative, 
and delta functions of $\beta$ would make the
picture number positive.   

For the antiholomorphic degrees of freedom, we can make a shorter version of the same argument.  If when we insert $\V$
in the path integral, we can insert an extra factor of $\delta(\t b)$ with the path integral remaining non-zero, then $\V$ must
have antiholomorphic ghost number 1.  Given that $\V$ does not depend on the derivatives of $\t c$, it must be precisely
proportional to $\t c$, with no dependence on $\t b$.

The conclusion then is that an NS vertex operator that will lead to a simple superconformal formalism has the form
\begin{equation}\label{irdo}\V=\t c c\delta(\gamma) V, \end{equation}
where $V$ is  constructed from matter fields only.  In fact, as we will discuss, such an operator is $Q_B$-invariant if and only
if $V$ is a superconformal primary of antiholomorphic and holomorphic dimensions $(1,1/2)$.

The need for the factor  $\delta(\gamma)$ is perhaps daunting, though by now this is an old
story, going back to \cite{FMS,EHVerl}.  But we should not be faint of heart; we know from section \ref{measure} that operators
of the general form $\delta(\beta)$ are unavoidable, so we should be 
prepared to deal with $\delta(\gamma)$ as well.

Actually, the factor $c\delta(\gamma)$ has an intuitive explanation.  As in eqn. (\ref{moby}), we can write $c$ as $\delta(c)$ so
$c\delta(\gamma)$ becomes $\delta(c)\delta(\gamma)$.  The ghosts represent symmetry generators, but in the presence
of a vertex operator, only superconformal vector fields that vanish at the position of the vertex operator are symmetries.
So $c$ and $\gamma$ should both be constrained to vanish at the position of the vertex operator, and this is achieved
via the delta functions.\footnote{To elaborate on this slightly, let us go back to the
formula (\ref{durov}) for general superconformal vector fields.  The condition that $\nu_f$ and $V_g$ both leave fixed the
point $z=\theta=0$ is that $f(0)=g(0)=0$.  The fields $\gamma$ and $c$ are ghosts that correspond to the vector fields $f$ and
$g$, so we want $\gamma(0)=c(0)=0$.  This is enforced by the delta functions.}

Finally, we have to impose the condition of BRST invariance.  A vertex operator of the form $\V=\t c c \delta(\gamma)V$ is
$Q_B$-invariant if and only if $V$ is a conformal and superconformal primary (for antiholomorphic and holomorphic variables,
respectively) of dimension $(1,1/2)$.  Thus, $V$ must obey
\begin{align}\label{unsox} L^\XX_n V& =\frac{1}{2}\delta_{n,0}V,~~ n\geq 0\cr
                                               G^\XX_r V&=0 , ~~r>0 \cr
                                                \t L^\XX_n V&=\delta_{n,0}V,~~n\geq 0, \end{align}
                                                where the operators on the left are Virasoro and super Virasoro generators of the matter
                                                system.
The ghost factor $\t c c \delta(\gamma)$ is a primary of conformal dimension $(-1,-1/2)$, so $V$ being a primary of
dimension $(1,1/2)$ ensures that $\V$ is a primary of dimension $(0,0)$.    We refer to such a $\V$ as a superconformal
vertex operator.  According to the BRST version of the no-ghost theorem, 
every BRST cohomology class (at non-zero momentum) contains a representative that is a superconformal
vertex operator, so in particular such operators suffice for computing the $S$-matrix. 

Let us briefly explain the
role of the $\delta(\gamma)$ factor in $\V$, since this is the most unusual feature. In verifying that $Q_B\V=0$ for $\V$
a superconformal vertex operator, one needs
among other things to verify that $\V$ is annihilated by the following part of $Q_B$:
\begin{equation}\label{bosox} Q_B^*=\sum_{r\in\Z+1/2} \gamma_r G^\XX_{-r}.\end{equation}
Here   $\gamma_r$
are the  modes
\begin{align}\label{yodel} 
                      \gamma(z)=\sum_{r\in\Z+1/2}z^{-r+1/2}\gamma_r.
                                             \end{align}
                                             The conditions
(\ref{unsox}) ensure that $\gamma_r G_{-r}^\XX\V=0$ for $r<0$.  For $r>0$, we see from (\ref{yodel}) that
vanishing of $\gamma_r$ for $r$ positive means that $\gamma(z)$ vanishes at $z=0$.  In the presence of
the identity operator, $\gamma(z)$ would be regular but non vanishing at $z=0$; to make $\gamma$ vanish
at $z=0$, we need to insert an operator $\delta(\gamma(0))$.  That is why $\V(z)$ is proportional to $\delta(\gamma(z))$.

For our applications, we need to know that 
 the operator $\V=\t c c \delta(\gamma)V$ 
is $Q_B$-invariant not just in the ordinary sense but also in the extended sense in which $Q_B$ acts on $\J$.
This is true by the same sort of
 argument given at the end of section \ref{cvo}, using the antighost
equation of motion derived from the extended action (\ref{extac}).  The same goes for the superconformal
Ramond vertex operators that we will study in section \ref{ramond}.

\subsubsection{Other Picture Numbers?}\label{otherpictures}

In conventional language, we seem to have arrived at a unique choice for the picture numbers of the NS sector
vertex operators: because of the factor of $\delta(\gamma)$, they have picture number $-1$. 
(This is the picture number of the unintegrated form
of the vertex operator; passing to the integrated form of the vertex operator maps the picture number to
0, as we describe later.)  However, this resulted from a possibly innocent-looking assertion
that we made at the beginning: we claimed that adding an NS puncture increases the holomorphic dimension
of the moduli space of super Riemann surfaces by $1|1$.  This is the right answer for the conventional definition of
the moduli space of super Riemann surfaces with NS punctures.  Actually, as explained in section 4.3 of \cite{Wittentwo},
one can modify the definition of supermoduli space in a way that increases its odd dimension, and then one
can calculate using NS vertex operators of any desired integer picture number $k\leq -1$.  (One can choose the picture
number of each NS vertex operator independently, as long as they are all no greater than $-1$.)  
The main property of this more general formalism is that it can be reduced to the standard one by
integrating out the extra odd moduli.  We will not pursue the
generalization, which seems
 to have no benefit, except to answer the question, ``Is there a natural way to calculate scattering
amplitudes using NS vertex operators of picture number different from $-1$?'' 

There seems to be no natural way to compute scattering amplitudes, in general, using (unintegrated) NS
vertex operators
of picture number greater than $-1$.  To do this, one would want a version of supermoduli space with punctures
with an odd dimension less than the usual value, and this appears not to exist.  In genus 0, one
can integrate over odd moduli with the help of picture-changing operators, and then, after placing
the picture-changing operators at the positions of the external vertex operators, one can compute scattering
amplitudes with vertex operators of arbitrary picture number (with the right overall sum); see eqn. (\ref{ydex}).  
In higher
genus, such an approach will lead to the spurious singularities described in section \ref{measure},
and also to difficulties in understanding massless tadpoles, as we describe starting in section \ref{propagator}.

In section \ref{ramond}, we introduce Ramond vertex operators, and there a similar story holds.  The simplest
procedure is based on Ramond vertex operators of picture number $-1/2$, but if one wishes, one can
modify the definition of supermoduli space and compute with vertex operators of any picture number less than $-1/2$.

\subsection{Scattering Amplitudes For NS States}\label{scattering}

Now we would like to explain how to calculate amplitudes 
for the scattering of $\n$ string states, all in the NS sector, described by superconformal
vertex operators $\V_1,\dots,\V_\sn$.  The derivations in sections \ref{bosmeasure} and \ref{measure}
were phrased so as to generalize straightforwardly, so we can be relatively brief.

Combining the definitions of sections \ref{vertop} and \ref{exform}, we 
define a form $F_{\V_1,\dots,\V_n}(\J,\delta\J)$ on the space $\JJ$ of supercomplex structures:
\begin{equation}\label{forox}F_{\V_1,\dots,\V_n}(\J,\delta\J)=\int\D(\XX,B,C,\t B,\t C)\exp(-\h I)
\prod_{i=1}^n \V_i(p_i),\end{equation}
with $p_i$ being points in $\Sigma$.
This definition makes sense for vertex operators that are not necessarily superconformal, though it is
particularly useful in the superconformal case.

If $\V_1=\{Q_B,\W_1\}$ and the other $\V_i$ are all $Q_B$-invariant, then
\begin{equation}\label{norox}F_{\{Q_B,\W_1\},\V_2,\dots,\V_\ssn}+\d F_{\W_1,\V_2,\dots,\V_n}=0,\end{equation}
just as in eqn. (\ref{ondo}).
This relation is the basis for the proof of gauge-invariance, that is, the decoupling of BRST-trivial states
from the scattering amplitudes.  

As we will explain momentarily, superconformal symmetry can be used
in a simple way to analyze the scattering amplitudes only if the vertex operators $\V_i$ obey
\begin{equation}\label{nigoz}\t b_n\V=b_n\V=\beta_r\V=0,~~n,r\geq 0.\end{equation}
To prove gauge-invariance
in a superconformally invariant framework, one needs to know that if $\V$ obeys (\ref{nigoz}) and
can be written as $\{Q_B,\W\}$ for some $\W$, then $\W$ can be chosen to obey the same constraints:
\begin{equation}\label{imely}\t b_n\W=b_n\W=\beta_r\W=0,~~n,r\geq 0.\end{equation}
This  is shown  in appendix \ref{surreally}, by adapting the corresponding bosonic argument
of  appendix \ref{really}. 

To reduce to computations on a finite-dimensional moduli space, one would like to know that 
$F_{\V_1,\dots,\V_\ssn}(\J,\delta\J)$ is a pullback from the quotient $\JJ/\Diff_{p_1,\dots,p_\ssn}$, where
$\Diff_{p_1,\dots,p_\ssn}$ is the group of diffeomorphisms of $\Sigma$ that leaves fixed the punctures
at $p_1,\dots,p_\sn$.  For this, we need to know the usual two facts: (1) $F_{\V_1,\dots,\V_n}(\J,\delta\J)$
must be $\Diff_{p_1,\dots,p_\ssn}$-invariant, which is obvious from the definition; (2) $F_{\V_1,\dots,\V_n}(\J,
\delta\J)$ must vanish if contracted with a vector field on $\JJ$ that is a generator of 
$\Diff_{p_1,\dots,p_\ssn}$.  The explanation of this is very similar to what it was in section \ref{intmod}.  
Contraction with a vector field $q^z(\t z;\neg z|\theta)\partial_z+\frac{1}{2}D_\theta q^z\,D_\theta$ on $\Sigma$ will generate a shift
$\delta\J_{\t z}^z\to \delta\J_{\t z}^z+\partial_{\t z}q^z$.  This shifts the extended action (\ref{extac}) by
\begin{equation}\label{bextac}\h I\to \h I+\frac{1}{2\pi} \int_\Sigma\dzzt\,\partial_{\t z}q^z B_{z\theta}.
\end{equation}
Looking back to the ghost action (\ref{tomgor}), we see that we can restore the invariance
of the extended action by shifting $C^z$ by $C^z\to C^z-q^z$.  But we need to know that the
vertex operators $\V_1,\dots,\V_\sn$ are invariant under $C^z\to C^z-q^z$.  The only constraint on
$q^z$ is that $q^z=D_\theta q^z=0$ at the points $p_i$ (so that
the vector field $q^z\partial_z+\frac{1}{2}D_\theta q^zD_\theta$ leaves those points fixed).  There is
no constraint on the derivatives of $q^z$ with respect to $z$.
 Accordingly,
the condition for $\V_1(p_1),\dots,\V_n(p_n)$ to be invariant under $C^z\to C^z-q^z$ is that, although
$\V_1,\dots,\V_n$ may depend on the ghost fields $c$ and $\gamma$ at the points $p^i$, they do not depend 
on the derivatives
of those ghost fields with respect to $z$.  In other words, the condition that we need is precisely that $b_n\V=\beta_r\V=0$,
$n,r\geq 0$.  The remaining condition $\t b_n\V=0$, $n\geq 0$, arises by considering in a similar
way vector fields $q^{\t z}(\t z;\neg z|\theta)\partial_{\t z}$.  (As remarked in relation to eqn. (\ref{duffo}),
vector fields $q^\theta D_\theta$ have already been used to eliminate the possibility of deforming
the superconformal structure of $\SIgma$ independent of its complex structure, 
so we need not consider such vector fields now.)

The superdegree of the form $F(\J,\delta\J)$ can be determined by asking what generalization of
the delta function insertions of eqn. (\ref{doofus}) is needed to get a sensible and non-zero path integral
for the matter fields.  Given the way that we determined what class of vertex operators we wanted,
there is not much to say here: for each superconformal vertex operator $\V$, we need one
extra $\delta(\t B)$ insertion and $1|1$ additional $\delta(B)$ insertions.  So with $\n$ external
superconformal NS vertex operators, $F(\J,\delta\J)$ has antiholomorphic superdegree $3\g-3+\n|0$
and holomorphic superdegree $3\g-3+\n|2\g-2+\n$.  

So the superdegree of $F_{\V_1,\dots,\V_n}(\J,\delta\J)$
is that of a form that can be integrated over a supermanifold of dimension $6\g-6+2\n|2\g-2+\n$.
The right supermanifold is roughly speaking the quotient $\JJ/\Diff_{p_1,\dots,p_\ssn}$, from which 
$F_{\V_1,\dots,\V_\ssn}(\J,\delta\J)$ is a pullback; and roughly speaking this is the moduli space
of super Riemann surfaces of genus $\g$ with $\n$ NS punctures. 
Actually, we meet here the same subtlety as in  section \ref{harder}, 
and we will proceed as we did there.  

We use the real analyticity of the worldsheet path integral to regard the moduli $\t m_\beta$
of left-movers and $m_\alpha$, $\eta_\sigma$ of right-movers as independent complex variables.
Then $F_{\V_1\dots,\V_\ssn}(\J,\delta\J)$ can be interpreted as a form\footnote{To be
more precise, $F_{\V_1,\dots,\V_n}(\J,\delta\J)$ is the pullback of a   section of $\BBer(\M_L\times \M_R)$ that is
defined and holomorphic on a neighborhood of the ``diagonal'' $\Gamma$ that is introduced
shortly.} on a product $\M_L\times \M_R$, where $\M_L$ and $\M_R$ parametrize
respectively the antiholomorphic structure and the holomorphic structure of $\Sigma$.   This structure holds for all of the oriented closed
string theories; for the heterotic
string, $\M_L$ is a copy of $\M_{\sg,\sn}$, the moduli space of Riemann surfaces of genus $\g$ with
$\n$ punctures, while $\M_R$ is a copy of $\MM_{\sg,\sn}$, the moduli space of super Riemann surfaces
of genus $\g$ with $\n$ NS punctures.   Inside the reduced space $(\M_L\times\M_R)_\red$,
there is a ``diagonal'' $\Gamma$ characterized by saying that, when one reduces modulo the odd variables,
$\Sigma $ is real -- its holomorphic and antiholomorphic structures  are complex conjugates.  With a suitable
choice of coordinates, this means that $\t m_\alpha=\bar m_\alpha$.  Letting the odd moduli vary,
we ``thicken'' $\Gamma$ slightly to a cycle $\varGamma\subset \M_L\times \M_R$ of the same codimension.
Then $\varGamma$ is the integration cycle of the heterotic string with external NS vertex operators.
The genus $\g$ contribution to the scattering amplitude is
\begin{equation}\label{oroto} \langle \V_1\dots\V_\sn\rangle=\int_{\varGamma}F_{\V_1,\dots,\V_\ssn}(\J,\delta\J).
\end{equation}
We can make this formula concrete by the same methods of integrating over a slice and/or integrating 
out odd moduli as in section \ref{measure}. 

The formula (\ref{norox}) for gauge invariance becomes a statement about integration
over $\varGamma$:
\begin{equation}\label{noroxic}\int_{\varGamma}
F_{\{Q_B,\W_1\},\V_2,\dots,\V_\ssn}=-\int_{\varGamma}\d F_{\W_1,\V_2,\dots,\V_n}=0.\end{equation} 
The supermanifold version of Stokes's theorem can be applied to the right hand side,
so as for bosonic strings,  the only potential obstruction to gauge-invariance is the possible occurrence of
a surface term at infinity.  We return to this after analyzing the basic structure at infinity in section
\ref{propagator}.

The topic of this kind that perhaps requires more detail is the superstring analog of an integrated
vertex operator.  We turn to this next.  

\subsection{Integrated NS Vertex Operators}\label{intns}

For Neveu-Schwarz vertex operators  (but not for Ramond vertex operators),
there is a superconformally invariant notion of an integrated vertex operator, analogous to what we explained in
 section \ref{integrated} for the bosonic string.
Let $\V=\t c c\delta(\gamma)V$, where $V$ is a superconformal primary of dimension $(1,1/2)$ constructed
from the matter system only.  This is the right dimension so that  $V$ can be integrated over $\Sigma$
in a natural way.  Thus we can define
\begin{equation}\label{yozo}K_\V=\int_\Sigma \dzzt\,V(\t z;\neg z|\theta), \end{equation}
and instead of inserting $\V$ at a point in $\Sigma$, we can insert a factor of $K_\V$ in the worldsheet
path integral.  For many purposes, it is equivalent to insert an NS vertex operator on 
$\Sigma$ in its integrated
or unintegrated form.  

We can relate the two types of vertex operator by adapting the derivation of section \ref{horsy}.  We compare two
different descriptions {\it (i)} and {\it (ii)}.  In description {\it (i)}, we use local coordinates $\t w;\neg w|\zeta$ on $\Sigma$.
The vertex operator $\V=\t c c\delta(\gamma)V$ is inserted at $\t w=w=\zeta=0$.  The complex structure $\J$ depends
on moduli $\t \z;\neg\z|\btheta$ (as well as other moduli that are not relevant for this discussion).  In description {\it (ii)}, $\Sigma$
is described by local holomorphic coordinates $\t z;\neg z|\theta$.  The complex structure $\J$
does not depend on the moduli $\t\z;\neg\z|\btheta$, but the vertex operator $\V$ is inserted at $\t z;\neg z|\theta=\t\z;\neg\z|\btheta$.

In section \ref{horsy}, the key point in the derivation was the identity
\begin{equation}\label{uglos}\left(\frac{1}{2\pi} \int_{\Sigma}\d^2z g^z\partial_{\t z}b_{zz}\right) \cdot c^z(0)=g^z(0).\end{equation}
This identity -- and the corresponding one for antiholomorphic variables -- was used to remove the ghost factors $\t c c$ from the vertex
operator and replace them by an integration measure $\d \t \z\,\d \z$.  

For the superstring case, certainly we are going to need a superanalog of eqn. (\ref{uglos}).  We can
rewrite (\ref{uglos}) in terms of delta functions:
\begin{equation}\label{tuglos}\delta\left(\frac{1}{2\pi} \int_{\Sigma}\d^2z g^z\partial_{\t z}b_{zz}\right) \cdot \delta(c^z(0))=g^z(0).\end{equation}
In this form, it is easy to guess the superanalog:
\begin{equation}\label{zuglos}   \delta\left(\frac{1}{2\pi} \int_{\Sigma}\d^2z f^\theta\partial_{\t z}\beta_{z\theta}\right) \cdot \delta(\gamma^\theta(0))
=\frac{1}{f^\theta(0)}.\end{equation}                
This formula will be justified in section \ref{betag} (see eqn. (\ref{numbly})).

With the help of this identity, one proceeds just as in section \ref{horsy}.  We start in description {\it (i)}, which fits the framework 
of section \ref{scattering}.  In this framework, the integral over $\t \z,\z,$ and $\btheta$ is made by inserting suitable delta functions
defined in eqns. (\ref{happy}) and (\ref{doofus}).  We write schematically
\begin{equation}\label{pofus}\delta(\t B^{(\t \z)})\delta(B^{(\z)})\delta(B^{(\btheta)}) \end{equation}
for these delta functions.  In (\ref{happy}), we have explicit
formulas for them in terms of the derivatives of $\J$ with respect to $\t\z$, $\z$, and $\btheta$. These derivatives are very simple;
since the dependence of  $\J$ on $\t \z,\z,$ and $\btheta$
is induced from a diffeomorphism, we get formulas such as $\partial_\z \J_{\t w}^w=\partial_{\t w}v^w$, for some vector field $v^w\partial_w$, and
similarly for $\t \z$ and $\btheta$.
Inserting $\partial_\z\delta\J_{\t w}^w=\partial_{\t w}v^w$ in (\ref{happy})  and integrating by parts, the delta functions (\ref{pofus}) have arguments that naively
vanish by the equations of motion $\partial_{\t w}B_{w\zeta}=D_\zeta \t B_{\t w\t w}=0$.
So we get the opportunity to use the identities 
in (\ref{tuglos}) and (\ref{zuglos}) (and the antiholomorphic counterpart of (\ref{tuglos})).  These identities can be used to remove
the delta functions while also   removing the factors of $\t c c \delta(\gamma)$ from the
vertex operator.  The net effect of this and the change of coordinates to description {\it (ii)} is to transform the insertion
\begin{equation}\label{yofus}[\d\t\z;\neg\d\z|\d\btheta]\,\delta(\t B^{(\t \z)})\delta(B^{(\z)})\delta(B^{(\btheta)})\t c c\delta(\gamma)V,\end{equation}
which represents the  general procedure
for integration over $\t\z;\neg\z|\btheta$ with an insertion of $\V=\t c c\delta(\gamma)V$,  into simply
\begin{equation}\label{yoffus}[\d\t\z;\neg\d\z|\d\btheta]\, V(\t\z;\neg\z|\btheta).\end{equation}
This is the description in terms of an integrated vertex operator.    The steps that we have omitted are quite similar to those
in section \ref{horsy}.  

Just as for bosonic strings, the equivalence between integrated and unintegrated vertex operators is not quite valid at infinity in moduli
space (see fig. \ref{failint}
of section \ref{persp}).  The description by unintegrated vertex operators is always valid, while the description by integrated vertex operators
can lead to difficulty.

\subsubsection{More On Gauge Invariance}\label{morgauge}

The remarks of section \ref{moregauge} have a close analog for Neveu-Schwarz vertex operators of superstring theory.
The analogy is so close that we will be brief.

If a superconformal primary field $V$ of the matter system
of dimension $(1,1/2)$ is also a descendant (in which case it is a null vector), then
$\V=\t cc\delta(\gamma)V$ is BRST-trivial and should decouple.  In the case of massless string states, this has a natural
explanation by integration by parts on $\Sigma$.  If (in the heterotic string) $V$ is a  massless vertex operator that is also null, then $V=G_{-1/2}W$
where $W$ is a primary of dimension $(1,0)$, or  $V=\t L_{-1}W'$, where $W'$ is of conformal dimension $(0,1/2)$.
In these cases, the integrated vertex operator insertion is
\begin{equation}\label{yerf}\int_\Sigma [\d\t z;\neg\d z|\d\theta]\,V=\begin{cases} \int_\Sigma [\d\t z;\neg\d z|\d\theta]D_\theta W\\
\int_\Sigma [\d\t z;\neg\d z|\d\theta]\,\partial_{\t z}W'                                                ,  \end{cases}\end{equation} 
and  vanishes by integration by parts.   (To understand how to integrate by parts in the case of $D_\theta W$, see section 2.4.1 of
\cite{Wittentwo}.)

If instead $V$ is a null vector corresponding to a massive string state, then just as in section \ref{moregauge}, $\V$ is not a total
derivative on $\Sigma$.  Its decoupling has to be proved by using (\ref{noroxic}) and integrating by parts on $\varGamma$, 
not just by integrating by parts on $\Sigma$.

The first few examples of Neveu-Schwarz gauge parameters are described in appendix \ref{surreally}.

\subsection{Tree-Level Scattering}\label{treem}

Having understood the integrated version of an NS vertex operator, it is straightforward to compute tree amplitudes.

Let us first recall how this is done for the bosonic string.  We begin with $\n$ vertex 
operators $\V_i=\t c c V_i$, inserted at 
prescribed 
points  in a genus 0 Riemann surface $\Sigma_0$.  We can identify $\Sigma_0$ as the complex $z$-plane together
with a point at infinity.  We use the $SL(2,\C)$ symmetry of $\Sigma_0$ to map three of the $\V_i$, say the first three,
to arbitrary points $z_i\in \C$.  For the moduli of $\Sigma_0$, we can take the positions 
$z_4,\dots,z_n$ of the other vertex operators.
As in section \ref{integrated}, to integrate over those positions, we simply replace the unintegrated vertex 
operators $\V_i$
by their integrated counterparts $\int \d^2z_i \,V_i(\t z_i,z_i)$.  The scattering amplitude is then
\begin{equation}\label{mitrox}\left\langle
\prod_{i=1}^3 \t c c V_i(\t z_i,z_i)\prod_{j=4}^\sn \int \d^2 z_j\, V(\t z_j,z_j)\right\rangle.\end{equation}

To extend this to superstring theory with NS vertex operators, the main subtlety arises for the three-point function.
We start with three unintegrated vertex operators.  In the heterotic string, they take the form 
$\V_i=\t c c \delta(\gamma) V_i$.  
The moduli space $\M_{0,3}$ of an ordinary Riemann surface of genus 0 with three punctures is a 
point, but the supermoduli
space $\MM_{0,3}$ of a super Riemann surface of genus 0 with three NS punctures has dimension 
$0|1$.  So in computing
the heterotic string NS three-point function in genus zero, there are no antiholomorphic moduli and 
$0|1$ holomorphic ones.
So there are no $\delta(\t b)$ or $\delta(b)$ insertions, and one $\delta(\beta)$ insertion.  Via eqn. 
(\ref{zuglos}), we use the
$\delta(\beta)$ insertion to remove the $\delta(\gamma)$ from any one of the three vertex operators, say $\V_3$.
We still have to integrate over one odd modulus, which we take to be the odd coordinate $\theta_3$ of $\V_3$.
The integral is easily evaluated:
 \begin{equation}\label{tolbo} \int \d \theta_3 \t c c V_3(\t z_3;\neg z_3|\theta_3)=\t c c D_\theta V_3(\t z_3;\neg
 z_3|0).\end{equation}
The NS three-point
function in genus 0 is therefore
\begin{equation}\label{freddy}\biggl\langle \t c c\delta(\gamma) 
V_1(\t z_1,z_1|0) \,\,\t c c\delta(\gamma) V_2(\t z_2,z_2|0)\,\,
\t c c D_\theta V_3(\t z_3,z_3|0)\biggr\rangle.\end{equation} 
Here the $z_i$ and $\t z_i$ may be chosen arbitrarily.

When we add an additional NS vertex insertion, we gain $1|0$ antiholomorphic moduli and $1|1$ holomorphic moduli.
The extra moduli are simply the coordinates $\t z;\neg z|\theta$ at which an additional NS vertex operator is inserted.
For each added vertex operator $\V_j=\t c c \delta(\gamma)V_j$, 
we gain a full complement of delta functions $\delta(\t b)\delta (b)\delta(\beta)$.
As in section \ref{intns}, we can use the delta functions to convert the unintegrated vertex operator $\V_j$
into its integrated form $\int [\d\t z_j;\neg\d z_j|\d\theta_j]V_j(\t z_j;\neg z_j|\theta_j)$, whereupon the
scattering amplitude becomes
\begin{equation}\label{breddy}\left\langle \t c c\delta(\gamma) V_1(\t z_1,z_1|0)\,\, \t c c\delta(\gamma) V_2(\t z_2,z_2|0)
\,\,\t c c D_\theta V_3(\t z_3,z_3|0)\prod_{j=4}^\sn \int [\d\t z_j;\neg\d z_j|\d\theta_j]V_j(\t z_j;\neg z_j|\theta_j)
\right\rangle.\end{equation}
In this procedure for calculating the tree-level scattering, two vertex operators, namely $\V_1$ and $\V_2$,
have picture number $-1$, while the others have picture number 0.  In section \ref{intodd}, we describe a more
general formula for the tree-level $S$-matrix.

\begin{figure}
 \begin{center}
   \includegraphics[width=3in]{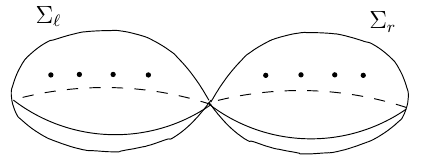}
 \end{center}
\caption{\small At infinity in moduli space, a genus 0 surface $\Sigma$ splits into two genus zero components
$\Sigma_\ell$ and $\Sigma_r$, joined at a common singularity.  The external vertex operators are distributed
between the two sides in an arbitrary fashion. In the case of external
NS vertex operators, to treat the compactification
of moduli space correctly, both $\SIgma_\ell$ and $\Sigma_r$  should
have two unintegrated vertex operators -- counting the singularity as one. 
Regardless of which two of the original  vertex operators we take in unintegrated form, this is not always the
case, since they might be both on $\Sigma_\ell$ or both on $\Sigma_r$.  
So the formalism based on integrated vertex operators does not treat correctly the compactification
of the moduli space.  In the more general formalism of section \ref{intodd} based on vertex operators
of general picture numbers, one has the same problem.  One would like the condition (\ref{zolot}) to hold on each side; but no choice of picture numbers
ensures this. }
 \label{common}
\end{figure}
To arrive at (\ref{breddy}), we have followed the usual procedure of integrating over some moduli by the use of
integrated vertex operators.  (Since we are in genus 0, we were able to integrate over all moduli this way.)
This procedure treats the interior of $\MM_{0,\sn}$ correctly, but actually does not
treat correctly the compactification.  One reason for this has nothing to do with worldsheet supersymmetry and
was described in fig. \ref{failint} of section \ref{persp}.  With super Riemann surfaces, there is actually a second,
somewhat  analogous problem.  At infinity in 
$\MM_{0,\sn}$, $\Sigma$ splits up into the union of two genus 0 components $\Sigma_\ell$ and $\Sigma_r$
joined at a common singularity (fig. \ref{common}).  
When this happens, a correct procedure requires that each of $\Sigma_\ell$ and
$\Sigma_r$ should have two unintegrated vertex operators -- counting the singularity as an unintegrated
vertex operator.  (We are here imposing the condition (\ref{ruffo})  on both $\SIgma_\ell$ and $\Sigma_r$.
If the path integrals on $\SIgma_\ell$ and $\Sigma_r$ are not separately sensible, we cannot expect the path
integral on $\Sigma$ to factor naturally when $\Sigma$ degenerates.  
For more on such factorization, and also for  the interpretation of the singularity as an unintegrated
vertex operator, see section \ref{propagator}.)  In eqn. (\ref{breddy}), this condition is satisfied if some of the integrated vertex operators
collide with one of the unintegrated ones, but not if they collide only with each other.

Even though (\ref{breddy}) does not treat the compactification of $\MM_{0,\sn}$ correctly, it does give
the right tree-level $S$-matrix.  In general, a method of integration that treats the compactification incorrectly
will differ from a correct treatment by surface terms at infinity.\footnote{This follows from the analysis in section \ref{yzzo}.  Any two
methods of integrating over odd variables differ by a total derivative, so in particular a treatment that is correct except at infinity
differs from a treatment that is correct everywhere by a surface term at infinity.}  In the case of a genus 0 scattering amplitude,
as long as the external momenta are sufficiently generic (so that the momentum flowing through the singularity
is off-shell), the relevant integral is  highly convergent at infinity
(or more precisely the calculation is made by analytic continuation from a region of external momenta
where the integral is convergent)
and there is no surface term.   In higher genus, a formalism that treats the region at infinity incorrectly
will lead to difficulty in understanding massless tadpoles and mass renormalization, but these are not problems in genus 0.

\subsection{Integration Over Odd Moduli}\label{intodd}

The procedure introduced in sections \ref{basodd} and \ref{pcorp} for integrating over odd moduli 
can straightforwardly be extended to include NS vertex operators.
We start again with a reduced Riemann surface $\Sigma_\red$ with spin bundle $K^{1/2}$; by 
adding an odd coordinate $\theta$, we build
a super Riemann surface $\Sigma$.  We insert NS vertex operators at $\n$ points $q_1,\dots,q_\sn\in\Sigma_\red$ 
(and thus at $\theta=0$).

In this framework, odd moduli are incorporated by turning on a gravitino field.
We expand the gravitino field as in (\ref{zingo}),
\begin{equation}\label{zingoy}\chi_{\t z}^\theta=\sum_{\sigma=1}^{2\sg-2+\sn}\eta_\sigma 
\chi_{\t z}^{(\sigma)\theta}.\end{equation}
We can still make a gauge transformation $\chi_{\t z}^\theta\to\chi_{\t z}^\theta+\partial_{\t z}y^\theta$. But now $y^\theta$
should vanish at $q_1,\dots,q_\sn$, so the gravitino modes should be understood as elements of 
$H^1(\Sigma_\red,T^{1/2}\otimes \O(-\sum_{i=1}^\sn q_i))$.
This space has dimension $2\g-2+\sn$ (which therefore is the odd dimension of the supermoduli space 
$\MM_{\sg,\sn}$ of genus $\g$ surfaces with $\n$ NS punctures),
and that is the number of gravitino modes that we need.  A spurious singularity will arise if the 
$\chi_{\t z}^{(\sigma)\theta}$ are not
linearly independent as elements of $H^1(\Sigma_\red,T^{1/2}\otimes \O(-\sum_i q_i))$.  

The general procedure for integrating out the odd moduli is precisely as described in section \ref{zico}, 
and leads to the insertions 
described in eqns. (\ref{pinzo}) or (\ref{winzo}).  If we wish, then as in section \ref{pco}, we can take 
the gravitino modes to have delta
function support at points $p_\sigma\in\Sigma_\red$, $\sigma=1,\dots,2\g-2+\n$.    The necessary insertions for
integrating out odd moduli are then simply a product of picture-changing operators, $\prod_{\sigma=1}^{2\sg-2+\sn}\YY(p_\sigma)$, where
$\YY(p_\sigma)$ was defined in (\ref{nifox}).  The condition (\ref{dumo}) for a spurious singularity becomes
\begin{equation}\label{yrfed} H^0\left(\Sigma_\red,T^{1/2}\otimes 
\O\left(\sum_{\sigma=1}^{2\sg-2+\sn} p_\sigma-\sum_{i=1}^\sn q_i\right)\right)\not=0.\end{equation}

The line bundle $\L= T^{1/2}\otimes \O(\sum_{\sigma=1}^{2\sg-2+\sn} p_\sigma-\sum_{i=1}^\sn q_i)$ has 
degree $\g-1$, and just as in the discussion
of eqns. (\ref{tomox}) and (\ref{dumo}), this implies that $H^0(\Sigma_\red,\L)$  generically vanishes, but can generically
become non-zero as one varies one complex parameter.   For example, the parameter in question 
could be the choice of one of the points
$q_i$ at which a vertex operator is inserted, or one of the points $p_\sigma$ at which a picture-changing
operator is inserted.  When $H^0(\SIgma_\red,\L)\not=0$, the gauge-fixing
procedure is incorrect, and the integration measure computed with this procedure acquires a pole.

There is one important situation in which spurious poles do not arise.  This is the case that 
$\Sigma$ has genus zero.  For $\g=0$,
the line bundle $\L$ has degree $-1$; 
a line bundle of this degree is unique up to isomorphism, and its sheaf cohomology vanishes.  

Even for $\g=0$, we still have to decide where to insert the picture-changing operators.  If the $p_\sigma$ do not
vary holomorphically with the $q_i$, the resulting formulas will be inelegant, though free of 
spurious singularities (and correct as long as the 
$p_\sigma$ do not vary too wildly near infinity in $\MM_{0,\sn}$).  The 
usual procedure \cite{FMS} to insure holomorphy and 
minimize extraneous choices
is to take the $p_\sigma$ to coincide with some of the $q_i$.  Concretely, one 
picks nonnegative integers $r_i$, $i=1,\dots,\n$
with $\sum_i r_i=\n-2$, and for each $i$ one takes $r_i$ of the $p_\sigma$ to 
coincide\footnote{We have understood in 
section \ref{pco} that there is no trouble letting several of the $p_\sigma$ coincide.} with $q_i$.
The vertex operators $\V_i=\t c c\delta(\gamma)V_i$
are thereby transformed into vertex operators $\V_i^{(s_i)}=\t c c \hat V^{(s_i)}_i$ 
of picture number $s_i= r_i-1$.  (One can show that $\hat V^{(0)}_i=D_\theta V_i$. Explicit formulas for the $\hat V^{(s_i)}$ 
quickly become complicated for $s_i>0$.)  One still has
to integrate over the $q_i$.  This can be done just as in bosonic string theory
by removing the factors $\t c c$ from $n-3$ of the vertex operators and integrating over
their positions. The scattering amplitude is thus
\begin{equation}\label{ydex}\left\langle\prod_{i=1}^3\t c c\V^{(s_i)}(\t z_i,z_i)
\prod_{j=4}^\sn\int \d^2z_j \hat V_j^{(s_j)} (\t z_j,z_j)\right\rangle.\end{equation}
We have fixed in an arbitrary way the positions of the first three vertex operators and integrated over the others.

In this fashion, we may compute tree amplitudes using NS vertex operators of any 
pictures\footnote{One may also remove
the restriction to $s_i\geq -1$ by modifying the definition of $\MM_{0,\ssn}$ 
so that more than one odd modulus is associated
to each puncture, as in  section 4.3 of \cite{Wittentwo}.}  
$s_i\geq -1$, constrained only
by
\begin{equation}\label{zolot}\sum_i s_i=-2.\end{equation}
  What we have just described is a generalization of
the procedure of section \ref{treem}, in which  two unintegrated vertex operators had picture $-1$ and all others
had picture 0.

An important detail  is the following.  The absence of spurious singularities means that the procedure for
gauge-fixing and integration over fermionic moduli is correct in the interior of $\MM_{0,\sn}$.  
It does not guarantee that this procedure
is correct at infinity.  In fact, regardless of what one chooses for the $s_i$, the formula (\ref{ydex}) does not 
treat correctly the compactification
of $\MM_{0,\sn}$.  This assertion is a generalization of what was explained at the end of section \ref{treem}.
To treat properly the compactification, when  $\Sigma$  breaks into a union of two components 
$\Sigma_\ell$ and $\Sigma_r$, joined
at a singularity (as in fig. \ref{common} of section \ref{treem}), one needs a condition analogous to (\ref{zolot}) on each of
$\Sigma_\ell$ and $\SIgma_r$ separately, with the singularity considered to contribute $s=-1$ on each
branch. (This statement amounts to imposing eqn. (\ref{ruffo}) on $\SIgma_\ell$
and $\Sigma_r$ separately, and  is a special case of what we will explain in section
\ref{pnpco}.)  Moreover, this should be the case irrespective of how the points
$q_1,\dots,q_\sn$ are distributed between $\Sigma_\ell$ and $\Sigma_r$.   No choice of the $s_i$ obeys these conditions,
so regardless of what pictures one chooses,
the picture-changing formalism always treats incorrectly the compactification of the supermoduli space.  

Nevertheless, for the same reason as in section \ref{treem}, the picture-changing formalism
computes tree amplitudes correctly.

\section{The Ramond Sector}\label{ramond}

\subsection{Ramond Insertion Points And Superconformal Algebra}\label{ramins}

A Neveu-Schwarz vertex operator is inserted at a generic point on a 
super Riemann surface $\Sigma$. By contrast, a Ramond vertex operator 
is inserted at a singularity of the superconformal structure of $\Sigma$.
(See for example section 4.1 of \cite{Wittentwo}.)  In the presence of a 
Ramond vertex operator, $\Sigma$ is still from a holomorphic
point of view a complex supermanifold of dimension $1|1$, with local complex 
coordinates $z|\theta$.  Moreover, the holomorphic
tangent bundle $T_R\Sigma$ of $\Sigma$ still has a distinguished subbundle $\D$.  But it is no longer true that it is possible
everywhere to pick local superconformal coordinates $\tau|\zeta$ so that $\D$ has a section 
of the form $D_\zeta=\partial_\zeta+\zeta\partial_\tau$.
Rather the local form near a Ramond puncture at $z=0$ is that $\D$ is generated by
\begin{equation}\label{imonkey}D^*_\theta=\partial_\theta+\theta z\partial_z.\end{equation}
For $z\not=0$, we can reduce to the superconformal form by setting
\begin{equation}\label{donkey} \tau|\zeta=(\log z)|\theta,\end{equation}
whence $D^*_\theta=D_\zeta$.  But at $z=0$, no change of coordinates will put $D^*_\theta$ 
(or any other section $f(z|\theta)D^*_\theta$ of
$\D$) in the superconformal form $\partial_\zeta+\zeta\partial_\tau$.  The obstruction is that
\begin{equation}\label{ketty}(D^*_\theta)^2=z\partial_z,\end{equation}
which vanishes at $z=0$.  By contrast, if a change of coordinates could put $D^*_\theta$ in the form 
$f(\tau|\zeta)(\partial_\zeta+\zeta\partial_\tau)$ where inevitably the function $f(\tau|\zeta)$ would be everywhere non-zero
(since $D^*_\theta$ is), then $D^*_\theta$ and $(D^*_\theta)^2$ would 
everywhere be linearly independent.  

Where $D^*_\theta$ and $(D^*_\theta)^2$ are linearly dependent, we say that the 
superconformal structure of $\Sigma$ is singular.
In our example, this occurs precisely for $z=0$.  The locus $z=0$ defines not a point but a divisor 
$\FF\subset \SIgma$, of dimension $0|1$. We call a divisor of this kind a Ramond divisor or Ramond puncture.

Now let us discuss superconformal symmetry in the presence of a Ramond puncture.  A superconformal
vector field $W$ is a vector field that preserves the subbundle $\D\subset T_R\Sigma$; concretely, this means
that the commutator of $W$ with $D^*_\theta$ is a ($z|\theta$-dependent) multiple of $D^*_\theta$.  A little
calculation reveals that odd and even superconformal vector fields take the form
\begin{align}\label{odeven}\nu_f& = f(z)(\partial_\theta-\theta z\partial_z) \cr
                                             V_g&=z\left(g(z)\partial_z+\frac{g'(z)}{2}\theta\partial_\theta\right).\end{align}
In particular, a basis of superconformal vector fields that are holomorphic except possibly for a pole at $z=0$ is given
by
\begin{align}\label{nodeven}G_r& =z^r\left(\partial_\theta-\theta z\partial_z\right)\cr
                                             L_n&= -z^{n+1}\partial_z-\frac{n z^n}{2}\theta\partial_\theta,\end{align}
with $n,r\in\Z$.  A short calculation reveals that    
\begin{align}\label{rovo} [L_m,L_n]& =(m-n)L_{m+n}\cr \{G_r,G_s\}&=2L_{r+s} \cr [L_m,G_r]&=
\left(\frac{m}{2}-r\right) G_{m+r}.\end{align}
This is the super Virasoro algebra (\ref{rov}), but now in the Ramond sector,
since the odd generators $G_r$ have integral grading.  

The subalgebra of the super Virasoro algebra that leaves fixed the divisor $\FF$ at $z=0$ is generated by $G_r,$ $L_n$,
$r,n\geq 0$. In contrast to the other vector fields just mentioned, $G_0$, when  restricted to $\FF$, is not zero:
\begin{equation}\label{usurf} G_0|_{z=0}=\partial_\theta. \end{equation}
Rather than vanishing when restricted to $\FF$,  $G_0$ generates a nontrivial automorphism of $\FF$:
\begin{equation}\label{surfo} \theta\to\theta+\alpha,\end{equation}
with $\alpha$ an odd constant.   Obviously, this symmetry does not leave $\FF$ fixed pointwise.
                                      
What are the simplest possible vertex operators?   For the bosonic string and the NS sector of superstrings,
the simplest vertex operators are what we have called conformal or superconformal vertex operators.  They
obey three conditions, beyond BRST-invariance:
\begin{enumerate}\item
 They are invariant under conformal or superconformal transformations that leave fixed the point at
which the vertex operator is inserted.  
\item
 They also obey a more subtle condition of invariance under shifting
the ghost fields by a conformal or superconformal vector field that leaves fixed the vertex operator insertions.  
(For NS vertex operators in superstring theory, 
this condition is stated in eqn. (\ref{nigoz}).)  \item They have the right ghost and (in the superstring case) picture numbers so
that their insertion in the path integral gives a sensible and non-zero result.
\end{enumerate}
 Condition 1  lets us deduce from the worldsheet  path
integral with insertions of BRST-invariant vertex operators $\V_1,\dots,\V_\sn$ a closed form
$F_{\V_1,\dots,\V_\ssn}(\J,\delta\J)$  in a superconformally invariant fashion. Condition 2 enables
one to prove that $F_{\V_1,\dots,\V_n}$ is a pullback from the appropriate finite-dimensional moduli space.
Condition 3 then suffices to single out a very special class of vertex operators.  (Condition 1 is a consequence
of Condition 2 plus BRST symmetry, so one can consider Condition 2 to be more fundamental.)

For the Ramond sector, we want to follow the same pattern. But first we have to decide if a Ramond vertex
operator should be inserted at a point in the divisor $\FF$ or should be associated to the whole divisor.
Both choices are viable (see section 4.3 of \cite{Wittentwo}), but the simpler formalism -- and the one
that is  compatible with using the standard definition of the moduli space of super Riemann surfaces with punctures --
associates a Ramond vertex operator to the whole divisor $\FF$, not to a particular point in it.  

In this case, the constraints that we want are that the vertex operator $\V$ should be invariant under all
the super Virasoro generators that leave fixed the divisor $\FF$:
\begin{equation}G_r\V=L_n\V=0,~~ r,n\geq 0.\end{equation}
Since $G_0$ does not leave fixed any point on  $\FF$, the $G_0$ constraint implies
that  $\V$ is associated to the whole divisor $\FF$, and not to a point in it.

As in our previous experience, 
the simplest vertex operators that can be used to compute the $S$-matrix
are given by a standard ghost vertex operator times a matter
vertex operator that will be a superconformal primary of the appropriate dimension.     More precisely, we will have
\begin{equation}\label{zorp}\V=\t c c \SSigma V,\end{equation}
where $\t c $ and $c$ are the usual anticommuting ghosts,
 $\SSigma$ is a special vertex operator of the $\beta\gamma$ system (first constructed in \cite{FMS} and
described  below), and
$V$ is a vertex operator of the matter system.
The matter vertex operator $V$ will be a superconformal primary:
\begin{align}\label{dolit} L^\XX_n V& = \frac{5}{8}\delta_{n,0}V,~~n\geq 0\cr
                                       G^\XX_rV& = 0,~~r\geq 0.\end{align}
 Here $L^\XX_n$ and $G^\XX_r$ are super-Virasoro generators of the matter system.
In particular, $V$ has holomorphic dimension $5/8$, which   is the appropriate value because $\SSigma$ has
dimension $3/8$.  (The value $5/8$ is the analog of dimension 1 for bosonic strings or dimension $1/2$
for NS vertex operators of superstring theory. Antiholomorphically, $V$ will be a primary of 
dimension 1 and $\SSigma$ will be the  identity operator, of dimension 0.)
The no-ghost theorem ensures that vertex operators of this kind, with $\SSigma$ as described shortly, suffice for computing the $S$-matrix.

To obtain from BRST invariance the conditions (\ref{dolit}), including the $G_0$ constraint, $\V$ will have to be
a vertex operator of picture number $-1/2$.
Indeed, the mode expansion of the $\beta\gamma$ fields reads    
\begin{equation}\label{helomb} \beta(z)=\sum_r z^{-r-3/2}\beta_r,~~~\gamma(z)=\sum_r z^{-r+1/2}\gamma_r, \end{equation}                                   
where $r$ takes integer values in the Ramond sector. (The meaning of these formulas for the Ramond sector
in the context of super Riemann surfaces is described in section \ref{trvo}.)  The BRST operator $Q_B$ contains a term
\begin{equation}\label{elomb}Q_B^*=\sum_{r\in\Z}\gamma_{-r} G^\XX_r.\end{equation}
If we want $Q_B$-invariance of $\V$ to give precisely the $G^\XX_r$ constraints in (\ref{dolit}), then $\SSigma$
should be annihilated by $\gamma_r$ precisely if $r>0$.  We will thus have 
\begin{equation}\label{melomb} \gamma_r\SSigma  = 0,~~r>0.\end{equation} 
On the other hand, Condition 2 above -- stating that $\SSigma$ should be invariant under shifts of $\gamma$ by an odd superconformal
vector field -- means for the Ramond sector that
\begin{equation}\label{telb} \beta_r\SSigma=0,~~ r\geq 0.\end{equation}
The constraints (\ref{melomb}) and (\ref{telb}) together uniquely uniquely determine $\SSigma$.  In
conventional language, they mean that $\SSigma$ corresponds to the $\beta\gamma$ vacuum of
picture number $-1/2$.  When we want to emphasize this, we denote it as $\SSigma_{-1/2}$.

As in our study of bosonic strings or the NS sector of superstrings, one can organize this discussion more systematically as follows.
Condition 2 requires eqn. (\ref{telb}).  States of the $\beta\gamma$ system that obey (\ref{telb}) but not (\ref{melomb})
can be obtained from $\SSigma_{-1/2}$ by acting with operators $\beta_{-m}$ or $\delta(\beta_{-m})$, $m\geq 1$;  As we explain in the next paragraph, the part of $\V$ constructed
from the $\beta\gamma$ system should have ghost number and picture number $-1/2$; $\SSigma_{-1/2}$ has these properties
and acting with $\beta_{-m}$ or $\delta(\beta_{-m})$ will either reduce its ghost number or increase its picture number.
(That the ghost number of $\SSigma_{-1/2}$ is  $-1/2$ is a standard result
 \cite{FMS},  which we will  explain in section \ref{trvo}.)  So the $\beta\gamma$ part of the vertex operator must be $\SSigma_{-1/2}$.

Now let  $\MM_{\sg,\sn_\NS,\sn_\Ra}$ be the  moduli space
of super Riemann surfaces $\Sigma$ of genus $\g$ with $\n_\NS$ punctures
and $\n_\Ra$ Ramond punctures (that is, $\n_\Ra$ divisors $\FF_i$, $i=1,\dots,\n_\Ra$ along which
the superconformal structure has the sort of singularity described in eqn. (\ref{ketty})).  For topological
reasons, $\n_\Ra$ is always even.   As explained,
for example, in \cite{Wittentwo}, the dimension of $\MM_{\sg,\sn_\NS,\sn_\Ra}$ is
\begin{equation}\label{occo}\dim \,\eusmm =3\g-3+\n_\NS+\n_\Ra|2\g-2+\n_\NS+\frac{1}{2}\n_\Ra.\end{equation}
In particular, this dimension receives a contribution $1|\frac{1}{2}$ for each Ramond puncture.  
That is why the simplest formalism uses Ramond vertex operators of picture number $-1/2$.
Adding $\n_\Ra$ Ramond punctures increases the odd dimension of supermoduli space by $\n_\Ra/2$,
so in the computation of a scattering amplitude, there will be an additional $\n_\Ra/2$ insertions of the general type
$\delta(\beta)$, each of picture number 1.  But as just sketched,
a natural superconformal formalism (with Ramond insertions associated to divisors rather than points)
 requires each Ramond vertex operator to have picture
number $-1/2$.  The balance is preserved nicely: adding $\n_\Ra$ Ramond punctures adds
$\n_\Ra/2$ insertions $\delta(\beta)$, each of picture number 1, and $\n_\Ra$ Ramond vertex operators, each of
picture number $-1/2$.  Since the operator $\delta(\beta)$ has ghost number 1, it follows also that $\SSigma$ should
have ghost number $-1/2$ and hence must obey (\ref{melomb}) as well as (\ref{telb}).

\subsubsection{The Alternative}\label{alternative}

The alternative to what we have described is to define vertex operators that are inserted at a point on
a Ramond divisor, as explained in section 4.3 of \cite{Wittentwo}. In this case, the definition of supermoduli
space and the dimension formula (\ref{occo})
are modified so that the contribution of a Ramond puncture to the odd dimension is $3/2$.
Moreover, if a Ramond vertex operator is supposed to be inserted at a point, one
must drop the $G_0$ constraint in (\ref{dolit}), and
accordingly one replaces (\ref{melomb}) with
\begin{align}\label{zelomb} \gamma_r\SSigma & = 0,~~r\geq 0\cr \beta_r\SSigma&=0,~~ r> 0.\end{align}
In this case, in conventional language, $\SSigma$ has picture number $-3/2$, again matching the contribution 
of the Ramond puncture to the
odd dimension of moduli space.  (Its ghost number is $-3/2$, also balancing that of the relevant
$\delta(\beta)$ insertions.)

More generally, as explained in \cite{Wittentwo}, one can calculate with Ramond vertex operators with any
picture number that is no greater than $-1/2$, if one suitably modifies the definition of supermoduli space.
We will stick with the simplest procedure, based on the standard definition of the supermoduli space and Ramond
vertex operators of picture number $-1/2$.   There seems to be no systematic procedure to compute
with Ramond vertex operators of picture number greater than $-1/2$, though this can certainly be
done in genus 0, as explained in \cite{FMS} and in section \ref{tramp} below. 

\subsubsection{A Pair Of Ramond Punctures}\label{refo}

Since it is an important point, we want to look in another way at the claim that superconformal symmetry
in the neighborhood of a Ramond puncture leads naturally to the use of Ramond vertex operators of picture number $-1/2$. 

First let us describe a general example with an arbitrary number of Ramond divisors.
We take $\C^{1|1}$ with coordinates $z|\theta$ and superconformal structure defined by
\begin{equation}\label{zorop}D^*_\theta=\partial_\theta+\theta w(z)\partial_z,\end{equation}
with
\begin{equation}\label{orp} w(z)=\prod_{i=1}^{\sn_\Ra}(z-z_i). \end{equation}
Thus there are Ramond punctures at $z=z_i$, $i=1,\dots,\n_\Ra$.
Superconformal vector fields are now 
\begin{align}\label{bideven}\nu_f& = f(z)(\partial_\theta-\theta w(z)\partial_z) \cr
                                             V_g&=w(z)\left(g(z)\partial_z+\frac{g'(z)}{2}\theta\partial_\theta\right) \end{align}

Let us specialize to the case $\n_\Ra=2$ and place the Ramond punctures at $z=0$ and $z=a$.
So the superconformal structure is defined by
\begin{equation}\label{omorp}D^*_\theta=\partial_\theta+\theta z(z-a)\partial_z.\end{equation}
Setting $f(z)=z^{s}$, $g(z)=z^{n}$, some interesting examples of superconformal vector fields that
are everywhere holomorphic are
\begin{align}\label{morp}  \nu_s &= z^{s}(\partial_\theta-\theta z (z-a)\partial_z)\cr
                       V_n&= z(z-a)\left(z^{n}\partial_z+\frac{nz^{n-1}}{2}\theta\partial_\theta\right),\end{align}
                       with $s,n\geq 0$. 

We want to see what this Lie algebra of superconformal symmetries looks like when viewed from near $\infty$
on the complex $z$-plane.  As a shortcut, since we only going to be considering the behavior for large $z$,
let us just set $a=0$.  The superconformal structure can be put in a standard form  by
setting $\theta=\theta^*/z$; indeed, $z|\theta^*$ are superconformal coordinates, as $D^*_\theta$ is a multiple
of $D_{\theta^*}=\partial_{\theta^*}+\theta^*\partial_z$.   Transforming to the coordinates $z|\theta^*$ and 
comparing to eqn. (\ref{urv}) for the superconformal
symmetries of $\C^{1|1}$, we see that $V_n$ corresponds to $L_{n+1}$
and $\nu_s$ corresponds to $G_{s+1/2}$.

Accordingly, the unbroken symmetries in the presence of two Ramond punctures, as seen from large $z$, are 
generated by $L_n$, $n\geq 1$ and $G_{r}$, with  $r$ half-integral and $\geq 1/2$.  (Happily, these form
a super Lie algebra!)  Missing are the following symmetry generators of $\C^{1|1}$:  $L_{-1}$, $L_0$, and $G_{-1/2}$.
Each insertion of a Ramond vertex operator has removed one bosonic symmetry; a pair of Ramond
vertex operator insertions has removed one fermionic symmetry.  

In terms of ghost fields $c$ and $\gamma$ that correspond to superconformal symmetries, a standard
NS vertex operator is proportional to $c\delta(\gamma)$ (or equivalently $\delta(c)\delta(\gamma)$).  By setting
$c$ and $\gamma$ to zero at $z=0$, an insertion of this operator removes the symmetries $L_{-1}$ and $G_{-1/2}$.
In a similar spirit,
the operator
that removes $L_{-1}$, $L_0$, and $G_{-1/2}$ is $c\partial c \delta(\gamma)$ (or $\delta(c)\delta(\partial c)\delta(\gamma)$).  
So if a Ramond sector vertex operator depends on the ghosts as $c\SSigma$, where $\SSigma$ is some
vertex operator constructed from the $\beta\gamma$ system, we would like the OPE's to read
\begin{equation}\label{thermo}c\SSigma(a)\, c\SSigma(0)\sim c\partial c \delta(\gamma),~~a\to 0. \end{equation}
(The symbol $\sim$ means that the operator on the right is the leading operator appearing in the OPE; there is no claim about the
$a$-dependence. In the present example, the higher order terms are BRST-trivial.)   The non-trivial part of the OPE is
\begin{equation}\label{wermo} \SSigma(a)\,\SSigma(0)\sim \delta(\gamma),~~a\to 0. \end{equation}

Since $\delta(\gamma)$ has picture number $-1$, $\SSigma$ must have picture number $-1/2$.
Thus we see in a slightly different way from before that if the vertex operators are supposed to naturally reflect
the superconformal symmetry of the worldsheet, then the Ramond vertex operators
should have picture number $-1/2$.

\subsection{Matter Vertex Operators In The Ramond Sector}\label{rvo}

In general, what can be integrated naturally\footnote{This and related facts cited below are explained much more fully in \cite{Wittentwo}.} 
over a heterotic string world sheet  $\SIgma$ without a choice of coordinates
is a $(0,1)$-form with values in $\D^{-1}$. 
For example, if $\t \tau;\neg\tau|\zeta$ are standard coordinates (meaning that $\tau|\zeta$ are holomorphic superconformal coordinates and $\t \tau$ is
antiholomorphic and close to the complex conjugate of $\tau$) and  $X(\t \tau;\neg \tau|\zeta)$ is a scalar superfield,
then $\partial_{\t \tau}X$ is a $(0,1)$-form and $D_\zeta X$ is a section of $\D^{-1}$. So $\partial_{\t \tau}XD_\zeta X$ is a $(0,1)$-form with
values in $\D^{-1}$.  This
makes it possible to write an action for  a collection of scalar fields $X^I$, $I=1\dots 10, $
describing a map from $\Sigma$ to spacetime, that is to the target space of the string theory.  If 
 $G_{IJ}$ is the metric of spacetime, then we can write the supersymmetric action
\begin{equation}\label{memo}I_X=\frac{1}{2\pi i}\int_\Sigma
[\d\t\tau;\neg\d\tau|\d\zeta]\,G_{IJ}(X^K)\partial_{\t \tau}X^I D_\zeta X^J,\end{equation}
which is globally-defined, independent of the choice of standard coordinates.

In the presence of a Ramond puncture supported on a divisor $\FF$, matters are a little different. Let $\O(-\FF)$ be the 
holomorphic line bundle whose
sections are holomorphic functions that vanish along $\FF$.  What can be integrated in a natural way
is a $(0,1)$-form valued in $\D\otimes \O(-\FF)$.  The Lagrangian density associated to the action $I_X$ is a $(0,1)$-form valued in 
$\D$, not $\D\otimes \O(-\FF)$, and the geometrical interpretation of the fields $X^I$ in terms of maps to spacetime does not
allow us to ``twist'' them and change this statement.   So the integrand of $I_X$ is a section of the wrong line bundle.  It will
have a pole along Ramond divisors, since
a section of $\D$ can be regarded as a section of $\D\otimes \O(-\FF)$ that has a pole along 
$\FF$.   

Let us go back to our local model of a Ramond divisor with coordinates $\t z;\neg z|\theta$ and superconformal structure
defined by $D^*_\theta=\partial_\theta+\theta z\partial_z$.  We cannot choose superconformal coordinates in a neighborhood of 
the Ramond divisor at $z=0$, precisely because the superconformal structure is degenerate there.  But we can certainly find
superconformal coordinates away from $z=0$:
\begin{align}\label{zop} \t\tau& =\log \t z\cr  \tau&=\log z \cr \zeta& = \theta.\end{align}
Using the relation  $D^*_\theta=D_\zeta=\partial_\zeta+\zeta\partial_\tau$, we can transform the action $I_X$  of eqn. (\ref{memo})
from the standard coordinate system
$\t\tau;\neg\tau|\zeta$ to the coordinates $\t z;\neg z|\theta$ that behave well along $\FF$.  The action acquires the expected pole from
the change of coordinates:
\begin{equation}\label{omibo}I_X=\frac{1}{2\pi i}\int_\Sigma[\d\t z;\neg\d z|\d\theta]\frac{1}{z}G_{IJ}(X^K)\partial_{\t z}X^ID^*_\theta X^J.\end{equation}

To understand the implications, it is convenient to expand $X^I=x^I+\theta\psi^I$, and perform the $\theta$ integral:\footnote{As usual, the connection that enters the covariant derivative $D/D\t z$ is the pullback to the worldsheet of
the Riemannian connection in spacetime.}
\begin{equation}\label{plombo}I_X=\frac{1}{2\pi}\int_{\Sigma_\red} \d^2z\, \left(G_{IJ}\partial_{\t z}x^I\partial_z x^J
+\frac{1}{z}G_{IJ}\psi^I \frac{D}{D\t z}\psi^J\right).
\end{equation} 
We see that the pole has disappeared from the bosonic part of the action, and affects only the fermions.  

To understand the fermion action, it actually helps to return to the description by $\tau$ and $\t\tau$.  (For present purposes,
one may identify $\t z$ and $\t \tau$ with $\bar z$ and $\bar \tau$.)  Since $\tau=\log z$ where $z$ is single-valued,
$\tau$ is subject to the identification 
\begin{equation}\label{imboxip}\tau\cong \tau+2\pi i.\end{equation}
So in terms of $\tau$, the reduced worldsheet $\Sigma_\red$ is a cylinder of circumference $2\pi$.  The point $z=0$ has
been projected to $\mathrm{Re}\,\tau=-\infty$.  

In terms of $\tau$, the fermion action is
\begin{equation}\label{zonus}I_\psi=\frac{1}{2\pi}\int_{\Sigma_\red}
\d^2\tau G_{IJ}(x^K)\psi^I \frac{D}{D\t \tau}\psi^J,\end{equation}
so it is just the standard Dirac action for a fermion of positive chirality, with no unusual factors.
Since they are single-valued functions of $z$, the 
$\psi^I$ are periodic under $\tau\to \tau+2\pi i$.  Thus we have landed in what is
usually called the Ramond
sector \cite{Ramond}.  From here, the quantization is standard.  Each of the $\psi^I$, $I=1,\dots,10$ has
a zero-mode along the circle that is parametrized by $\mathrm{Im}\,\tau$.  
Quantizing these zero-modes gives a multiplet of states
in the spinor representation of $SO(10)$.  
(In constructions with spacetime supersymmetry, the GSO projection \cite{GSO}
projects out one of the two chiralities of this spinor.)  
The corresponding vertex operators have dimension $5/8$; this
number is $10\cdot 1/16$ where 10 is the number of components 
of the field $\psi^I$ (in ten-dimensional superstring theory)
and $1/16$ is the dimension of the spin field for a single fermion.  
The relation to the spin field of the Ising model was elucidated in
\cite{FMS}.

We have arrived at a standard answer, but not quite in the most familiar way.  
There were no square roots or double-valued fields.
We associated the insertion of a Ramond vertex operator (in the matter sector, so far) to 
a divisor $\FF$ along which the heterotic string worldsheet $\Sigma$ 
was itself perfectly smooth and all fields and coordinates
were single-valued,
but the superconformal structure was degenerate, because $(D_\theta^*)^2$ vanishes along that divisor.
The relation of our derivation to the standard one is that 
locally one can put the superconformal structure of $\Sigma$ in a standard
form by setting $\theta=\theta'/z^{1/2}$, so that 
$D^*_\theta =z^{1/2}(\partial_{\theta'}+\theta'\partial_z)$.  The coordinates $z|\theta'$
are superconformal, but of course $\theta'$ is double-valued.  In the new coordinates,
 we expand $X=x+\theta'\psi'$, with $\psi'=\psi/z^{1/2}$.  
 Now there is no pole in the action, but $\psi'$ has a square root
 branch point at $z=0$.

\subsection{Ghost Vertex Operators In The Ramond Sector}\label{trvo}

With or without Ramond punctures, superconformal vector fields are 
sections of the line bundle $\D^2$.  In general, in quantization
of gauge theory, the ghosts transform as symmetry generators with reversed statistics.  So the holomorphic 
ghost field $C$ is an odd section of $\D^2$,
whether or not Ramond punctures are present.

The action density for the holomorphic ghosts is $\L_{BC}=B\partial_{\t z}C$, where 
$B$ is the antighost field.  In the absence of Ramond punctures, the Berezinian of $\SIgma$ in the
holomorphic sense is
$\D^{-1}$, so for it to be possible to integrate $\L_{BC}$, $B$ must be a section of $\D^{-3}$.
The action is then
\begin{equation}\label{ozotto}I_{BC}=\frac{1}{2\pi i}\int_\Sigma [\d \t z;\neg\d z|\d\theta]\,B\partial_{\t z}C. \end{equation}
This formula is actually valid in an arbitrary coordinate system, not necessarily superconformal.
Invariance under a change of the antiholomorphic coordinate $\t z$ is manifest.  Under an arbitrary
change of the holomorphic coordinates from $z|\theta$ to, say,  $z^*|\theta^*$, the holomorphic
measure $[\d z|\d\theta]$ transforms as
\begin{equation}\label{bizoott}[d z|\d\theta]=[\d z^*|\d\theta^*]\,  \Ber^{-1}\begin{pmatrix}\partial_{ z}z^*&
\partial_{ z}\theta^*\cr \partial_{\theta}z^*& \partial_{\theta}\theta^* \end{pmatrix}.\end{equation}
And similarly $C$ and $B$ transform as appropriate powers of the same Berezinian:
\begin{align}\label{metty} C & =  C^* \cdot \Ber^{-2}\begin{pmatrix}\partial_{ z}z^*&
\partial_{ z}\theta^*\cr \partial_{\theta}z^*& \partial_{\theta}\theta^* \end{pmatrix} \cr
B & =  B^* \cdot  \Ber^{3}\begin{pmatrix}\partial_{ z}z^*&
\partial_{ z}\theta^*\cr \partial_{\theta}z^*& \partial_{\theta}\theta^* \end{pmatrix}.\end{align}
So the holomorphic ghost action in the absence of Ramond
punctures can be written in the form (\ref{ozotto}) in terms of arbitrary holomorphic local
coordinates $z|\theta$ and antiholomorphic local coordinate $\t z$. 

There is one immediate change in the presence of Ramond punctures.  The Berezinian of
$\Sigma$ in the holomorphic sense is now not $\D^{-1}$ but $\D^{-1}\otimes \O(-\FF)$, where $\FF$
is the divisor of Ramond punctures (so $\FF=\sum_{i=1}^{\sn_\Ra}\FF_i$ 
if there are Ramond punctures supported on divisors
$\FF_1,\dots,\FF_{\sn_\Ra}$). So now, in order for the action $I_{BC}$ to make sense,
we must interpret $B$ as a section of $\D^{-3}\otimes \O(-\FF)$.  The fact that we can twist
$B$ in this way to accommodate the Ramond punctures means that, in contrast to the matter action $I_X$,
the ghost action $I_{BC}$ will have no pole along $\FF$.  However, the definition of the ghost fields
will be modified along $\FF$.

To compare to standard conformal field theory formulas, we want to reduce $I_{BC}$ to an action defined
on an ordinary Riemann surface.  For this, we take $\Sigma$ to be a split super Riemann surface
with Ramond punctures at points $z_i\in \Sigma_\red$.  For $\Sigma_\red$ of genus 0, the relevant
superconformal structure was described explicitly in eqn. (\ref{zorop}).  
To make the reduction, we need to know the restriction of $\D$ to $\Sigma_\red$.  
In the absence of Ramond punctures, the restriction
of $\D$ to $\Sigma_\red$ is $T^{1/2}$, a square root of the holomorphic tangent 
bundle $T\to\Sigma_\red$.  (A choice of square root
involves a choice of spin structure on $\Sigma_\red$, and this is therefore built into the construction of a super Riemann surface
with reduced space $\Sigma_\red$.)
In the presence of Ramond punctures at points
$z_1,\dots,z_{\sn_\Ra}$,
however, the restriction of $\D$ to $\Sigma_\red$ is not a square root of $T$ but 
a square root of $T\otimes \O(-q_1-\dots-q_{\sn_\Ra})$.
See for example section 4.2.4 of \cite{Wittentwo}.  Thus, constructing a super Riemann surface $\Sigma$ with Ramond punctures
at the points $q_1,\dots,q_{\sn_\Ra}$  requires a choice of a line bundle $\RR\to \Sigma_\red$ with an isomorphism
\begin{equation}\label{zombotto} \RR^2\cong T\otimes \O(-q_1-\dots-q_{\sn_\Ra}). \end{equation}
For our purposes here, we are interested in the local behavior near a single 
Ramond puncture, which we may call $q$, at $z=0$.  So
\begin{equation}\label{yombo}\RR^2\cong T\otimes \O(-q).\end{equation}
The dual relation reads
\begin{equation}\label{moob}\RR^{-2}\cong K\otimes \O(q),\end{equation}
where $K$ -- the dual of $T$ -- is the canonical bundle of $\Sigma_\red$.
$\RR$ is the restriction of $\D$ to $\Sigma_\red$.

We expand the ghost field $C$ in powers of $\theta$:
\begin{equation}\label{medzo}C(z|\theta)=\h c(z)+\theta \h\gamma(z).\end{equation}
The reason for the hats is that, as will become clear, $\h c$ and $\h\gamma$ do not quite
coincide with the ghost fields $c$ and $\gamma$ as conventionally defined.
$\h c$ is a section of $\RR^2\cong T\otimes\O(-q)$.  In other words, we can view $\h c$ as a
section of $T$ that vanishes at $z=0$.  Since $\theta$ is a section of $\RR$, we can understand
$\h\gamma$ as a section of $\RR$.

We make a similar expansion for $B$:
\begin{equation}\label{lixo}B=\hat \beta+\theta\hat b.\end{equation}
Here, as $B$ is a section  of $\RR^{-3}\otimes \O(-q)$, it follows that
$\hat\beta$ is a section of $\RR^{-3}\otimes\O(-q)$, and that 
$\hat b$ is a  section of
$\RR^{-4}\otimes \O(-q)$.   In view of (\ref{moob}), it is equivalent to say that $\h b$ is a section of
$K^2\otimes \O(q)$ -- in other words $\h b$ is a quadratic differential that may have a pole
at $z=0$ -- and $\h\beta$ is a section of $K\otimes \RR^{-1}$.

In terms of these variables, the ghost action becomes
\begin{equation}\label{ghob}I_{BC}=\frac{1}{2\pi}\int_{\SIgma_\red} \d^2z\,\left(\h b
\partial_{\t z}\h c+\h \beta\partial_{\t z}
\h\gamma\right).\end{equation}

This description of  the fermionic ghosts  $\h b$ and $\h c$ and the action 
describing them is almost standard. The only novelty is that  $\h c$ is constrained to
vanish at $z=0$ while $\h b$ is allowed to have a pole there.  We discuss the significance of this in section \ref{imox}.

More unusual is the 
description that we have reached for the commuting ghosts $\h\beta$ and $\h\gamma$.  There
are no square root branch points in sight.  $\h\gamma$ 
is a section of a line bundle $\RR$, but perfectly 
single-valued, and
similarly $\h\beta$ is a perfectly single-valued section of $K\otimes \RR^{-1}$.  The $\h\beta\h\gamma$ system is just
a system of chiral bosons coupled to a line bundle -- or loosely speaking, to an abelian gauge field.  
What then is the relation to the usual description \cite{FMS} by spin fields?

\subsubsection{The Usual Description Of The $\beta\gamma$ System}\label{usbeta}

As in section \ref{rvo}, we
 can transform to the usual description of the $\beta\gamma$ system by introducing a
 double-valued coordinate $\theta'=\theta\sqrt z$, so that $z|\theta'$ are superconformal
 coordinates but $\theta'$ has a branch point at $z=0$.    
 To understand how to transform $\h\gamma$ and $\h\beta$ under the change of coordinates,
 we recall that $\h\gamma$ is a section of $\RR\cong \D|_{\SIgma_\red}$, so that 
 $\h\gamma\partial_\theta$
 should be invariant under the redefinition of $\theta$.   So we transform to the new coordinates by
  $\h\gamma\partial_\theta =\gamma\partial_{\theta'}$,
 or $\h\gamma\sqrt z=\gamma$.
 In transforming $\h\beta$, we view it as a section of $\RR^{-1}\otimes K$, so we set 
  $\h\beta/\sqrt z=\beta$.  The conventional ghost fields $\gamma$ and $\beta$ in superconformal
  coordinates $z|\theta'$ are often denoted
  as $\gamma^{\theta'}$ and $\beta_{z\theta'}$; this sort of notation makes the transformation
  we just made more obvious.
  
  The fields
 $\h\gamma$ and $\h\beta$ are regular at $z=0$ -- as they are simply chiral bosons
 valued in a line bundle.  So the behavior of $\gamma$ and $\beta$ near $z=0$ is 
 \begin{equation}\label{rondo}\gamma\sim z^{1/2},~~\beta\sim z^{-1/2},~~z\to 0.\end{equation}   

In conventional language, eqn. (\ref{rondo}) means that $\gamma$ and $\beta$ are coupled
to a vertex operator $\SSigma_{-1/2}$ at $z=0$ that is a spin operator of picture number $-1/2$.  
This follows from the Ramond sector mode expansions
\begin{equation}\label{ducho} \gamma(z)=\sum_{r\in\Z} z^{-r+1/2}\gamma_r,~~\beta(z)=\sum_{r\in\Z}
z^{-r-3/2}\beta_r. \end{equation}
To get the behavior (\ref{rondo}), we want
\begin{align}\label{ormox} \gamma_r\SSigma_{-1/2}&=0, ~~r>0\cr 
\beta_r\SSigma_{-1/2}&=0,~~r\geq 0,\end{align}
and these are the picture number $-1/2$ conditions that we found in another 
way in eqns. (\ref{melomb}) and (\ref{telb}).

For completeness, we will explain why the operator $\SSigma_{-1/2}$ has 
ghost number $-1/2$ and dimension $3/8$.
The original explanation \cite{FMS} involved a transformation from fields 
$\beta_{z\theta'},$ $\gamma^{\theta'}$ to a new set
of fields $\phi, \eta,\xi$.  As these variables do not have a very transparent 
interpretation in terms of the geometry of super Riemann
surfaces,\footnote{The reader may object that also the double-valued coordinate 
$\theta'$ is not entirely natural in super Riemann surface theory.  Indeed, one may prefer to rephrase the 
computation we describe here in the coordinates
$z|\theta$ that behave well at $z=0$.} 
we will instead follow a method \cite{DFMS} that has been used in the theory of orbifolds. The short distance behavior of the
$\beta(z)\cdot \gamma(w)$ operator product for $z\to w$ is
\begin{equation}\label{ocon}\beta_{z\theta'}(z)\gamma^{\theta'}(w)\sim -\frac {1}{z-w}. \end{equation}
The $\beta\cdot\gamma$ two-point function in the presence of the operator $\SSigma_{-1/2}$ inserted at $z=0$ with no other operator
insertions\footnote{To be more exact, we assume that there are no more operator insertions except a second spin field at $z=\infty$.  This second
 spin field has picture number $-3/2$, as one may learn
by transforming $z\to z'=1/z$ and examining the behavior of $\beta_{z\theta'}$ and $\gamma^{\theta'}$ for $z'\to 0$.  More generally, as
explained in section \ref{pnumber},
a sensible $\beta\gamma$ path integral in genus 0 always involves a product of operator insertions of total picture number $-2$.}  is
\begin{equation}\label{bocon}\langle\beta_{z\theta'}(z)\gamma^{\theta'}(w)\rangle_{\SSigma_{-1/2}}=-\frac{1}{z-w}\sqrt{\frac{w}{z}}.\end{equation}
This formula is determined by the fact that it is has the right behavior for $z\to w$, for $z\to 0$, and for $w\to 0$, and also
has no other singularities and the
slowest possible growth at infinity (the last condition reflects the fact that we take the spin field at infinity to be a primary).  
The ghost number current of the $\beta\gamma$ system is $J_{\beta\gamma}(w)=-:\beta\gamma(w):$ or more explicitly
\begin{equation}\label{notty}J_{\beta\gamma}(w)=\lim_{z\to w}\left(-\beta(z)\gamma(w)-\frac{1}{z-w}\right), \end{equation}
where here normal-ordering is carried out by subtracting the vacuum expectation value $\langle-\beta(z)\gamma(w)\rangle_{\mathrm{vac}}=1/(z-w)$.
A short calculation reveals that in the presence of the operator $\SSigma_{-1/2}$, $J_{\beta\gamma}(w)$ has a simple pole at $w=0$
with residue $-1/2$,
\begin{equation}\label{otty}\langle J_{\beta\gamma}(w)\rangle_{\SSigma_{-1/2}} =-\frac{1}{2 w},\end{equation}
so $\SSigma_{-1/2}$ has ghost number $-1/2$.   Similarly, the stress tensor of the $\beta\gamma$ system is \begin{equation}\label{luster}T_{\beta\gamma}=:\partial_z\beta_{z\theta'}
\cdot \gamma^{\theta'}:-\frac{3}{2}\partial_z(:\beta_{z\theta'}\gamma^{\theta'}:).\end{equation}  A short calculation gives
\begin{equation}\label{zinotty}\langle T_{\beta\gamma}(w)\rangle_{\SSigma_{-1/2}}=\frac{3}{8w^2},\end{equation}
so that $\SSigma_{-1/2}$ has dimension $3/8$. 

It is straightforward to generalize this analysis to determine the conformal dimension and
the ghost number of the operator $\SSigma_{-t}$ that represents the $\beta\gamma$ ground state
with picture number $-t$, for any integer or half-integer 
$t$.   In this case, we want $\gamma(w)$ to behave as $w^t$ and
$\beta(z)$ to behave as $z^{-t}$ near $z=0$. 
   The analog of (\ref{bocon}) is 
\begin{equation}\label{boccon}\bigl\langle\beta_{z\theta'}(z)\gamma^{\theta'}(w)\bigr\rangle_{\SSigma_{-t}}=-\frac{1}{z-w}\left({\frac{w}{z}}\right)^t.\end{equation}
A calculation along the above lines gives
\begin{align}\label{prott}\bigl\langle J_{\beta\gamma}(w)\bigl\rangle_{\SSigma_{-t}}& =-\frac{t}{ w} \cr
\bigl\langle T_{\beta\gamma}(w)\bigr\rangle_{\SSigma_{-t}}& =-\frac{t(t-2)}{2w^2}.\end{align}
Thus $\SSigma_{-t}$ has ghost number $-t$ and dimension $-t(t-2)/2$.   For integer $t$, this
operator can be constructed from a product of delta functions, as in eqn. (\ref{zinnof}) for $t\geq 0$, and its
ghost number and dimension can be computed classically.

\subsubsection{The Definition Of The  Ghosts}\label{imox}

The description of  the anticommuting ghosts $\h b$ and $\h c$ that we arrived at  in eqn. (\ref{ghob})
 differs in precisely one way from the usual description.   In the conventional formulation, 
$c$ is a vector field -- a section of $T$ --
and $b$ is a quadratic differential -- a section of $K^2$.
Instead, in our derivation,   $\h c$ was a section of $T\otimes \O(-q)$, that is a vector field that is constrained to 
vanish at the Ramond puncture,
while $\h b$ was a section of $K^2\otimes \O(q)$, that is a quadratic differential that is allowed to have a 
pole at the Ramond puncture.

The relation between the two descriptions is that in the standard approach, vanishing of $c$ 
at the insertion point of a vertex operator is not part of the definition of $c$, 
but the vertex operator contains a factor of $c$, or equivalently of $\delta(c)$,
that enforces this vanishing.  Once this factor is included, $b$ can have a pole at the location of the vertex operator.  

Instead of including in a bosonic string conformal vertex operator a factor $\t c c$, as one usually does,
 we could declare $\t c$ and $c$ to be fields
(sections of $T_L\Sigma$ and $T_R\Sigma$, respectively) that vanish at the positions of all vertex operator insertions.  
Then we would say that the simplest conformal vertex operators are constructed from matter fields only.\footnote{
A constraint that
$\t c$ and $c$ should  vanish at a given  point $p$ could be enforced by incorporating additional variables
-- Lagrange multipliers -- in the path integral, as in eqn. (\ref{omito}).  If we do this, the path integral measure,
 instead of being conformally invariant as in the usual description,  transforms under
conformal transformations like the operator $\t c c(p)$, in other words like a field of dimension $(-1,-1)$ at    the point $p$; this reflects the scaling behavior of  the measure of the Lagrange
multipliers.
Hence  insertion of a $(1,1)$ matter vertex operator $V$ at the point $p$ is necessary
to restore conformal invariance of the path integral. Analogous statements hold for the NS and R sectors of superstrings.}
Defining $\t c$ and $c$ to vanish at vertex operator insertions is natural because $\t c$ and $c$ are associated to gauge
symmetries, and the symmetries of bosonic string theory  are worldsheet 
diffeomorphisms that leave fixed the vertex operator insertions.

Similarly, for the NS sector of superstrings, instead of including a factor $\t c c \delta(\gamma)$ in the vertex operator,
we could equivalently say that $\t c$, $c$, and $\gamma$ are all constrained to vanish at points at which vertex operators
are inserted.  This is natural for the same reason as in the last paragraph.

We did not adopt this viewpoint in the present paper, in part because it might have made the formulas look unfamiliar.
However, for the Ramond sector, this viewpoint has been more or less forced upon us.  What is usually regarded
as a spin operator $\SSigma$ of the $\beta\gamma$ system really appeared in the derivation of eqn. (\ref{ghob})
as a modification of the
definitions of the fields $\beta$ and $\gamma$.  $\gamma$, rather than being a section of $T^{1/2}$, was a section
of a more general line bundle $\RR$, and similarly $\beta$ was a section of $K\otimes \RR^{-1}$.  The vertex operator
was just this instruction about how to modify the definition of the fields.  The reason that
the Ramond sector gives a sharper message about how the formalism should be developed is that a Ramond puncture
is an intrinsic part of the geometry of a super Riemann surface, in contrast to an NS puncture (or a puncture on a purely
bosonic worldsheet), which can be viewed if one wishes as something extra that is tacked on to a preexisting super Riemann
surface.  At any rate, part of the message of the Ramond sector seems to be that we should view the traditional
factors $\t c c$, $\t c c \delta(\gamma)$, and $\t c c \SSigma$ in the vertex operators as shorthand ways of saying
how the definition of the ghosts is modified by the presence of the vertex operators.  

For more on some matters considered in this section, see section \ref{reparam}.

\subsection{Ramond Amplitudes}\label{tramp}

Now we will discuss scattering amplitudes including Ramond states.

Once one constructs BRST-invariant Ramond vertex operators, many steps follow in the familiar fashion.  Given any assortment
of Ramond and/or Neveu-Schwarz BRST-invariant
vertex operators $\V_1,\dots,\V_s$, the worldsheet integral computes for us a closed form $F_{\V_1,\dots,\V_s}(\J,\delta\J)$ on the
space of supercomplex structures.  If the vertex operators obey the usual conditions 
\begin{equation}\label{undoc}b_n\V_1=\beta_r\V_1=0,~~ n,r\geq 0,\end{equation} 
that make possible a superconformally invariant formalism, then  $F_{\V_1,\dots,\V_s}(\J,\delta\J)$ is
 as usual a pullback from an appropriate product  $\M_L\times\M_R$ of left and right moduli spaces. For example, for the heterotic
string, $\M_L$ is the moduli space of genus $\g$ Riemann surfaces with $s$ punctures, and $\M_R$ is the moduli space of genus $\g$ super Riemann
surfaces with an appropriate number of NS and Ramond punctures.  In the familiar way, one 
defines an integration cycle $\varGamma\subset \M_L\times \M_R$ and one defines the genus $\g$ contribution to the scattering
amplitude as $\int_{\varGamma}F_{\V_1,\dots,\V_s}(\J,\delta\J)$.  

There are also no surprises concerning gauge-invariance.
If for example $\V_1=\{Q_B,\W_1\}$, then one has the standard relation 
\begin{equation}\label{morz}F_{\{Q_B,\W_1\},\V_2,\dots,\V_s}+\d F_{\W_1,\V_2,\dots,\V_s}=0. \end{equation}
Here it is important to know that if $\V_1$ obeys the 
conditions 
(\ref{undoc}), and $\V_1=\{Q_B,\W_1\}$ for some $\W_1$, then we can choose
$\W_1$ to also obey the same conditions (see appendix \ref{gp}).  This enables one to pull $F_{\W_1,\V_2,\dots,\V_s}$  back
to the finite-dimensional space $\M_L\times \M_R$, and to deduce from (\ref{morz}) a relation between finite-dimensional integrals:
\begin{equation}\label{orz}\int_{\varGamma}F_{\{Q_B,\W_1\},\V_2,\dots,\V_s}+\int_{\varGamma}\d F_{\W_1,\V_2,\dots,\V_s}=0. \end{equation}
This relation shows as usual that gauge-invariance holds if there are no surface terms when one integrates by parts over $\varGamma$.

One important difference between Ramond vertex operators and the other cases is that there does {\it not}
exist, in a natural sense, an integrated version of a Ramond vertex operator.  The fundamental reason for this is that
the forgetful map (\ref{hobbo}) does not have an analog for Ramond punctures.  For NS punctures there is such a map:
\begin{equation}\label{plobbo}\begin{matrix} \Sigma& \to & \eusmm \cr && 
\downarrow \pi\cr & & \MM_{\sg,\sn_\NS-1,\sn_\Ra}.\end{matrix}\end{equation}
Here $\eusmm$ is the moduli space of genus $\g$ super Riemann surfaces with $\n_\NS$ NS punctures and $\n_\Ra$ Ramond
punctures.  Since a NS puncture is simply a chosen point in a pre-existing super Riemann surface, we can forget such a puncture
if we wish, and this gives the fibration (\ref{plobbo}).   An integral over $\eusmm$ can be reduced to an integral over  
$ \MM_{\sg,\sn_\NS-1,\sn_\Ra}$ by integrating first over the fibers of this fibration, with the help of the integrated NS vertex operator. There is no analogous forgetful map for Ramond punctures, since a Ramond 
puncture is part of the superconformal structure of $\Sigma$; there is no way to forget such a puncture while
keeping  fixed the rest of a super Riemann surface, and  there is no sensible
notion of two super Riemann surfaces being the same except with Ramond punctures in different places.  So there is
no superconformal notion of an integrated Ramond vertex operator.   Any definition of an integrated Ramond vertex operator depends on
a method of integrating over odd moduli, and has the limitations of such methods.

However, more or less everything else we have said, for instance in sections \ref{yttro}-\ref{pcorp}, about how
to compute $F(\J,\delta\J)$ is substantially unaffected by Ramond punctures.  A few minor differences are as follows.
We originally introduced the gravitino field $\chi_{\t z}^\theta$ in section \ref{basodd} as a section of $T^{1/2}$, a square root
of the holomorphic tangent bundle $T$ of the reduced space $\SIgma_\red$.  A choice of $T^{1/2}$ is equivalent to a choice of
spin structure on $\SIgma_\red$.  The reason that $\chi_{\t z}^\theta$ takes values in $T^{1/2}$ is that the restriction of $\D$
to $\Sigma_\red$ is isomorphic to $T^{1/2}$, in the absence of Ramond punctures.  However, in the presence of Ramond punctures at points
$q_1,\dots,q_{\sn_\Ra}$,
the restriction of $\D$ to $\Sigma_\red$  is not a square root of $T$ but a square root of $T\otimes \O(-q_1-\dots-q_{\sn_\Ra})$.
(This fact played a prominent role in section \ref{trvo}.)
In other words, the restriction of $\D$, which we denote $\RR$, possesses an isomorphism
\begin{equation}\label{zombox} \RR^2\cong T\otimes \O(-q_1-\dots-q_{\sn_\Ra}). \end{equation}
The degree of $\RR$  is
\begin{equation}\label{yomibo}\mathrm{deg}\,\RR=1-\g -\frac{1}{2}\n_\Ra.\end{equation}
With Ramond punctures, the gravitino field $\chi_{\t z}^\theta$  is a $(0,1)$-form with values in $\RR$.  If we include also NS punctures
at points $p_1,\dots,p_\NS$, then $\chi_{\t z}^\theta$ becomes a $(0,1)$-form valued in $\h\RR=\RR\otimes \O(-\sum_{i=1}^{\sn_\NS}p_i)$.
The degree of this line bundle is 
\begin{equation}\label{otombo}\mathrm{deg}\,\h\RR=1-\g-\n_\NS-\frac{1}{2}\n_\Ra.\end{equation}
A gauge transformation of the gravitino is $\chi_{\t z}^\theta\to \chi_{\t z}^\theta+\partial_{\t z}y^\theta$, where now $y^\theta$
is valued in $\h\RR$.  Modulo gauge transformations, $\chi_{\t z}^\theta$ is an element of $H^1(\Sigma_\red,\h\RR)$,
whose dimension is $\Delta=2\g-2+\n_\NS+\frac{1}{2}\n_\Ra$, which accordingly is the odd dimension of $\eusmm$.

In the picture-changing formalism, we select points $r_1,\dots, r_\Delta$ and write
\begin{equation}\label{picor}\chi_{\t z}^\theta=\sum_{\sigma=1}^{\Delta}\eta_\sigma\delta_{r_\sigma} \end{equation}
with anticommuting parameters $\eta_\sigma$.
A spurious singularity occurs where the gravitino modes in this expression do not furnish a basis of $H^1(\Sigma_\red,\h\RR)$.
The condition for this is now that
\begin{equation}\label{icor} H^0(\Sigma,\h\RR\otimes \O(\sum_{\sigma=1}^\Delta r_\sigma))\not=0.\end{equation}
The line bundle $\h\RR\otimes \O(\sum_{\sigma=1}^\Delta r_\sigma)$ has degree $\g-1$, as always.  Generically, the condition (\ref{icor})
will be satisfied as one varies one parameter, leading to a spurious pole.  However, in genus 0, a line bundle of degree $-1$ has
no holomorphic section, so the gauge-fixing prescription associated to the picture-changing formalism is always correct in the interior
of moduli space. The subtleties that occur at infinity in moduli space are not important for tree-level amplitudes at reasonably
generic external momenta, so there is no problem to compute scattering amplitudes of Ramond and Neveu-Schwarz states at
tree-level in the picture-changing formalism.   

\subsubsection{The Simplest Examples}\label{simplex}

The simplest examples of scattering amplitudes are those for which the odd dimension $\Delta$ of the moduli space vanishes, 
so that no picture-changing insertions are necessary.  This happens only for $\g=0$, for which
  $\Delta=-2+\n_\NS+\frac{1}{2}\n_\Ra$.   So $\Delta=0$ for $\n_\NS=1,$ $\n_\Ra=2$, and for $\n_\NS=0,$ $\n_\Ra=4$.

For $\n_\NS=1$, $\n_\Ra=2$, the even and odd dimensions of the moduli space both vanish.   The scattering amplitude is simply computed
as a three-point function on a genus 0 super Riemann surface of one NS vertex operator and two Ramond vertex operators:
\begin{equation}\label{onz}\biggl\langle \t c c \delta(\gamma)
V_1(\t z_1;\neg z_1|0) \, \t c c \SSigma_{-1/2}  V'_2(\t z_2;\neg z_2) \t c c\SSigma_{-1/2}  V'_3(\t z_3;\neg z_3) \biggr\rangle.\end{equation}
Here $V_1$ is a matter vertex operator in the NS sector, while $V'_2$ and $V'_3$ are Ramond sector matter vertex operators.
The points $z_1,z_2,$ and $z_3$ are arbitrary and there are no moduli to integrate over.
The formula is a shorthand for the following recipe.  Pick points $z_1,z_2,z_3$ in a purely bosonic $\CP^1$ and use the formula
(\ref{zorop}) to construct a super Riemann surface $\Sigma$ with Ramond divisors 
at $z_2$ and $z_3$, and no others.  Pick a matter
vertex operator $V_1(\t z;\neg z|\theta)$ in the NS sector and evaluate it 
at\footnote{The automorphism group of a genus 0 super Riemann surface with
two Ramond punctures has dimension $1|1$ and enables one to map a given 
NS puncture to a chosen point such as $\t z_1; \neg z_1|0$ in a unique way.
See for example section 5.1.4 of \cite{Wittentwo}.}  $\t z_1;\neg z_1|0$.  Insert matter vertex operators $V'_2$
and $V'_3$ at the Ramond divisors at $z=z_2$ and $z=z_3$, 
respectively.\footnote{These vertex operators are associated to Ramond divisors, so there
is no choice in where they are to be inserted once the geometry of $\Sigma$ is fixed.  
When we denote them as $V_i'(\t z_i;\neg z_i)$, $i=2,3$, this is
just meant as a reminder of which Ramond operator is inserted at which Ramond divisor.  
In any case, as these operators are associated to divisors,
not points, they do not depend on $\theta$, only on $\t z$ and $z$. There is no misprint in 
the fact that a $\theta$-dependence of $V_2'$ and $V_3'$
is not indicated in (\ref{onz}).}  Compute the path integral with these insertions to get the amplitude (\ref{onz}).

For $\n_\NS=0$, $\n_\Ra=4$, since there are no odd moduli, the moduli space is just 
the bosonic moduli space that parametrizes
four points in $\CP^1$, up to the action of $SL(2,\C)$.  We can use $SL(2,\C)$ to 
specify three points in an arbitrary way; then we integrate
over the fourth point.  The scattering amplitude is
\begin{equation}\label{pylot}\biggl\langle\prod_{i=1}^3\t c c 
\SSigma_{-1/2} V'_i(\t z_i;\neg z_i)\int \d^2z_4 \SSigma_{-1/2} V'_4(\t z_4;\neg z_4)\biggr\rangle,\end{equation}
where $V'_i$, $i=1,\dots,4$ are Ramond sector matter vertex operators.
This formula is again a shorthand for a recipe that involves constructing a super 
Riemann surface with the specified Ramond divisors and
inserting the indicated matter vertex operators at those divisors.  

Still in $\g=0$, let us consider amplitudes with  $\Delta>0$.  If $\n_\Ra\leq 4$, one has $\Delta\leq \n_\NS$.  
If one were to forget the NS insertions, there would be no odd moduli at all, so
one can integrate over all odd moduli by replacing
some of the NS vertex operators with their integrated versions.  If $\n_\Ra>4$, one additionally 
needs $\frac{1}{2}\n_\Ra-2$ 
picture-changing operators.    To minimize arbitrary choices and preserve holomorphy,
a relatively simple procedure is to pick $\frac{1}{2}\n_\Ra-2$ of the Ramond vertex operators 
and attach a single picture-changing operator
to each one.  Thus one calculates tree amplitudes using Ramond vertex operators of picture 
number $+1/2$ as well as $-1/2$,
as originally described in \cite{FMS}.  One may also compute tree amplitudes 
with more general choices of the pictures.

\subsection{Duality}\label{duality}

Let $\H$ be the space of all string states in a given string theory model
or -- by the state-operator correspondence -- the space of all vertex operators.
The two-point function in genus 0 gives a nondegenerate pairing $\omega:\H\times \H\to \C$ or equivalently an identification -- which
we also call $\omega$ --
between $\H$ and its dual space $\H^*$.  For brevity, we will describe this for open strings. As usual, the discussion for
a chiral sector of closed strings is almost the same. 
We take the string worldsheet (or its reduced space, in the case of superstrings) to be the upper half of
the complex $z$-plane.

The pairing $\omega$ is defined by a two-point function of vertex operators inserted at, say, $z=0$ and $z=1$. For vertex
operators $\V,\,\W$, we set
\begin{equation}\label{gormo}\omega(\V,\W)=\langle\V(0)\,\W(1)\rangle. \end{equation}
This formula defines a nondegenerate bilinear form on the space of all vertex operators.
This pairing is BRST-invariant in the sense that
\begin{equation}\label{torry}\omega(Q_B\,\U,\V)+(-1)^{|\U|}\omega(\U,Q_B\V)=0.\end{equation}
 (Here $|\U|$ is 0 or 1 depending on whether the state $\U$ is bosonic or fermionic.)  As a result,
 there is an induced pairing on the BRST cohomology: if $\V$ and $\W$ are in the kernel of $Q_B$, then
 $\omega(\V,\W)$ is invariant under $\V\to \V+Q_B\U$, or under  a similar transformation of $\W$.  This
 pairing on the BRST cohomology is  nondegenerate.  This assertion follows from the BRST version of the no-ghost
 theorem, which gives convenient lightcone representatives of the BRST cohomology classes, making the nondegeneracy of the pairing
 manifest.  An instructive example of passing from the pairing among all vertex operators to the pairing between physical states
 can be found in section \ref{massten}.

In the case of bosonic strings, $\omega$ has ghost number $-3$; this means that
if $\U$ and $\V$ have definite ghost numbers, the sum of their ghost numbers must
be 3 in order to have $\omega(\U,\V)\not=0$.  For example, if $U$ and $V$ are primary fields in the matter
sector, we can have
\begin{equation}\label{zoromo}\langle c\partial c U(0) \,c V(1)\rangle\not=0.\end{equation}
The insertions of $c\partial c(0)$ and $c(1)$ have the following intuitive interpretation.  To get a non-zero path
integral, we must remove the zero-modes of the ghost field $c$, by eliminating the $SL(2,\R)$ symmetry of $\Sigma$.  
We can do this  by restricting the diffeomorphism group
of $\Sigma$ to its subgroup generated by vector fields that vanish, together with their first derivative, at $z=0$ and that
also vanish at $z=1$.  To achieve this, we must place these constraints on the ghost field $c$, so we insert $\delta(c)\delta(\partial c)
=c\partial c$ at $z=0$, and $\delta(c)=c$ at $z=1$.  

It is convenient to write  $\H_n$ for the subspace of $\H$ consisting of states of ghost number $n$.  Then $\omega$ is a
nondegenerate pairing $\H_n\times \H_{3-n}\to \C$, or equivalently an isomorphism between $\H_{3-n}$ and the dual of $\H_n$:
\begin{equation}\label{orzot}\omega:\H_n^*\cong \H_{3-n}. \end{equation}

For superstring theory, one should specify the picture numbers as well as the ghost numbers of the states.  We write $\H_{n;k}$
for the space of vertex operators of ghost number $n$ and picture number $k$.  The inner product $\omega(\U,\V)$ is again defined
as in (\ref{gormo}); for it to be non-zero, the ghost numbers must add to 1 and the picture numbers to $-2$.  So $\omega$ is
a nondegenerate pairing $\H_{n;k}\otimes \H_{1-n;-2-k}\to \C$, or equivalently  an isomorphism
\begin{equation}\label{oior} \omega:\H_{n;k}^*\cong \H_{1-n;-2-k}. \end{equation}

In this paper, we almost always restrict ourselves to the canonical picture numbers, namely $-1$ for NS states and $-1/2$
for Ramond states.  For the NS sector, there is no problem with this restriction: we get a nondegenerate pairing (\ref{gormo}) or
an isomorphism (\ref{oior}) with $k=-2-k=-1$. Given matter primaries $U$ and $V$ with a non-zero two-point function in the matter sector, a  typical non-zero pairing in the full theory with the ghosts is
\begin{equation}\label{gotty}\langle c\partial c \delta(\gamma) U(0)\,c\delta(\gamma)V(1)\rangle\not=0. \end{equation}

For the Ramond sector, the picture number is a half-integer, and if one state has the canonical picture number $-1/2$, then
the second will have to have picture number $-3/2$.  This corresponds to an isomorphism
\begin{equation}\label{rotty}\omega:\H_{n;-3/2}^*\cong \H_{1-n;-1/2}.  \end{equation}

The fact that the dual of a state in the canonical picture is a state of picture number $-3/2$ can be given the following
interpretation.  The superconformal structure of a worldsheet whose reduced space is the upper half-plane and that
has Ramond punctures at $z=0$ and $z=1$ is generated by the odd vector field
\begin{equation}\label{orzo}D_\theta^*=\partial_\theta+z(z-1)\theta\partial_z. \end{equation}
This worldsheet has an automorphism group of dimension $0|1$, generated by the odd superconformal vector field
\begin{equation}\label{prozo}\nu=\partial_\theta-z(z-1)\theta\partial_z, \end{equation}
which is regular (and nonvanishing) at $z=\infty$.  To get a sensible worldsheet path 
integral (without a zero-mode of the
commuting ghost field $\gamma$), we have to remove this automorphism group.  
For this, we have to insert at one of the
two Ramond divisors -- either at $z=0$ or at $z=1$ -- a vertex operator 
that is associated to a point on the divisor in question,
not to the whole divisor.  In other words, as explained in section \ref{alternative},
one of the two vertex operators has to have picture number $-3/2$ rather than the
canonical value $-1/2$.

In section \ref{stad}, we will find that this occurrence in the 
duality of an operator of non-canonical picture number is related to
the fact that the Ramond sector propagator is proportional to $G_0/L_0$, in 
constrast to $1/L_0$ in the NS sector (or for
bosonic strings).  The field theory limit of this is that fermions obey  
first order wave equations (such as the Dirac equation for spin
1/2) in contrast to the second order wave equations obeyed by bosons.

\subsubsection{Massless Fermions In Ten Dimensions}\label{massten}

We will make the Ramond sector duality more explicit for the basic case of massless fermion vertex operators in ten-dimensional
Minkowski spacetime $\R^{1,9}$.  The basic fermion vertex operator constructed from the matter fields is \cite{FMS}  a spin field $\Stigma$   that
transforms in the spinor representation of $SO(1,9)$.  $\Stigma$ can have either $SO(1,9)$ chirality; for its components of positive
or negative chirality, we write $\Stigma_\alpha$ and $\Stigma^\beta$, respectively, with $\alpha,\beta=1,\dots, 16$.  $\Stigma_\alpha$
and $\Stigma^\beta$ transform oppositely under the GSO projection.
We also write $\Stigma_\alpha(p)$, $\Stigma^\beta(p)$ for the corresponding operators at spacetime momentum $p$ (for example,
$\Stigma_\alpha(p)=\Stigma_\alpha \exp(ip\cdot X)$, where $X^I$, $I=0\dots 9$  are  bosonic fields describing the motion of the string
in $\R^{1,9}$).  

Let $\SSigma_{-1/2}$ and $\SSigma_{-3/2}$ be the ground states of the $\beta\gamma$ system at picture number $-1/2$ and $-3/2$, respectively.
These operators transform oppositely under the GSO projection.  (For example, this follows from the fact that the part 
of the picture-changing operator
that acts on the $\beta\gamma$ system, namely $\delta(\beta)$, is GSO-odd, since\footnote{In treating bosonic integration
as an algebraic operation in the sense introduced in section  \ref{bosal}, one uses $\delta(\lambda\beta)=\lambda^{-1}\delta(\beta)$,
not $\delta(\lambda\beta)=|\lambda|^{-1}\delta(\beta)$.  See for example eqn. (3.37) of
\cite{Wittenone}.} it is odd under $\beta\to-\beta$.)    So if $\SSigma_{-1/2}\Stigma_\alpha(p)$ is GSO-even, then $\SSigma_{-3/2}\Stigma^\beta(p)$ is likewise GSO-even.  

Let $u^\alpha$ and $v_\beta$ be commuting $c$-number spinors of the indicated chirality.
The pairing in the Ramond sector at the lowest mass level is
\begin{equation}\label{monzo}\bigl\langle c\partial c\SSigma_{-1/2}u^\alpha\Stigma_\alpha(p)\cdot c\SSigma_{-3/2}v_\beta \Stigma^\beta(q)\bigr\rangle
=u^\alpha v_\alpha (2\pi)^{10}\delta^{10}(p+q). \end{equation}

Clearly, this pairing is nondegenerate in the space of all vertex operators of this type.  It remains nondegenerate if we pass to the BRST cohomology.
To do this, we must restrict $p$ and $q$ by $p^2=q^2=0$, which follows from $Q_B$-invariance.  
$Q_B$-invariance also imposes on $u^\alpha$ the constraint
\begin{equation}\label{bonzo} p^I\Gamma_{I\alpha\beta}u^\beta =0, \end{equation}
where $\Gamma_I$, $I=0,\dots,9$ are the gamma matrices.  This constraint is simply the Dirac equation written in momentum space, and
arises because the condition $Q_B(c\SSigma_{-1/2}u^\alpha\Stigma_\alpha(p))=0$ implies $G_0(u^\alpha\Stigma_\alpha(p))=0$, where $G_0$ acts on the massless level as the Dirac operator of field theory.
On $v_\beta$, $Q_B$-invariance imposes no such constraint,\footnote{The $G_0$ constraint on $u^\alpha\Stigma_\alpha(p)$ arises because $Q_B=\gamma_0G_0+\dots$ and $\gamma_0\SSigma_{-1/2}
\not=0$.   Because  $\gamma_0\SSigma_{-3/2}=0$, 
the condition $Q_B(c\SSigma_{-3/2}v_\beta
\Stigma^\beta(q))=0$ does not lead to a $G_0$ constraint on $v_\beta\Stigma^\beta(q)$.  Instead, there is a gauge-equivalence
on $v_\beta\Stigma^\beta(q)$, generated by a gauge transformation with gauge parameter 
$c\beta_0\SSigma_{-3/2}w^\gamma\Stigma_\gamma(q)$.}
but instead when we pass to the $Q_B$ cohomology, there is an equivalence
relation
\begin{equation}\label{wonzo}v_\beta\cong v_\beta+p^I\Gamma_{I\beta\gamma}w^\gamma,\end{equation}
for any $w^\gamma$.  Evidently, with the constraint (\ref{bonzo}) and the equivalence relation (\ref{wonzo}), the pairing (\ref{monzo}) remains
nondegenerate.

\section{The Propagator}\label{propagator}

\begin{figure}
 \begin{center}
   \includegraphics[width=5.5in]{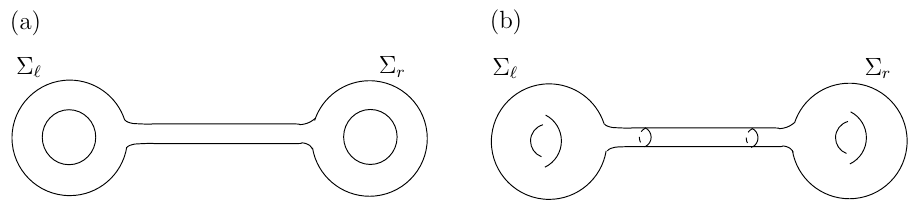}
 \end{center}
\caption{\small (a) A long strip connecting two parts of an open-string worldsheet.  This is meant to be a ``flat,'' purely two-dimensional
picture.
(b) A long tube connecting two parts of a closed-string worldsheet.}
 \label{longsurface}
\end{figure}

In field theory, the Feynman propagator can be represented as an integral over elapsed proper time.  
For example, for a scalar particle of mass $m$, the propagator in Euclidean signature is
\begin{equation}\label{ritto}\frac{1}{p^2+m^2}=\int_0^\infty\d s \exp\left(-s(p^2+m^2)\right), \end{equation}
where $s$ is the Schwinger proper time parameter. The two limits of the $s$ integral are both very important.
The ultraviolet region is $s\cong 0$.  The ultraviolet divergences of Feynman diagrams arise when the proper
time parameters associated to the propagators in a loop are all simultaneously small.  On the other hand, $s\to \infty$
is the infrared region.  The pole of the propagator for $p^2+m^2=0$ comes from the behavior of the integral for $s\to\infty$;
the pole results from an almost on-shell particle propagating in spacetime for a long proper time.  (The almost on-shell
particle also propagates a long distance in spacetime along the lightcone,
as one sees from the position space form of the propagator.)    
More generally, when one evaluates a Feynman
diagram, the singularities that are associated to unitarity all arise from poles of propagators, so they all 
arise when some or all proper time parameters  become large.  So we can think of large $s$ as the infrared
or on-shell region. 

In string theory, there is an immediate analog of the large $s$ region.  An open or closed string
propagating for a long proper time is described by a long strip or a long tube, respectively.  For examples of string
worldsheets containing such a  long strip or tube, see fig. \ref{longsurface}.  But one of the most fundamental
facts about perturbative string theory is that there is no ultraviolet region.  
   The roots of this statement go back nearly
 forty years to the original study of modular invariance in string theory \cite{Shapiro}.   In one description of
 a string worldsheet, it may appear that a proper time parameter becomes small in some limit.  But the same worldsheet
 always has an alternative description in which the proper time parameters are all bounded safely away from zero.
 
 The most precise statement that there is no ultraviolet region in string perturbation theory is the existence of
 the Deligne-Mumford compactification of the moduli space of Riemann surfaces (or of super Riemann surfaces).
 Moduli space can be compactified by adding only limit points that correspond to $s\to\infty$; there is no need
 for additional limit points that would correspond to $s\to 0$ or anything else.  This means that there is no potential
 for ultraviolet divergences in string theory; the delicate questions all involve the behavior in the infrared.  It also means
 that in any integration by parts on moduli space -- such as we have contemplated at many points in this paper,
 beginning with our discussion of eqn. (\ref{tarmob})
  -- the subtle issues involve the region $s\to\infty$ where one of the string
 states goes on-shell.
 
\begin{figure}
 \begin{center}
   \includegraphics[width=4.5in]{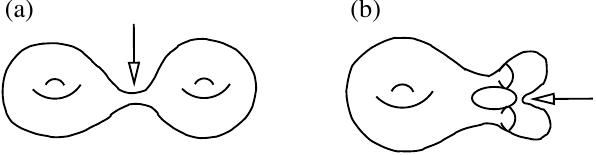}
 \end{center}
\caption{\small As explained in the text,  
a narrow neck in a Riemann surface -- marked here by an arrow -- is conformally equivalent
to a long tube.  The singularity or ``degeneration'' that occurs when the neck collapses
is said to be separating in (a), and non-separating in (b).}
 \label{collapse}
\end{figure}

 In the Deligne-Mumford compactification, one assigns a limit to a sequence of string worldsheets in which the length
 or a tube or strip diverges.   In doing this, one exploits worldsheet conformal invariance.  A string worldsheet
 with a long tube
 is conformally equivalent to a worldsheet (fig. \ref{collapse}) with a very narrow neck.  In the case of an ordinary Riemann surface,
 such a worldsheet is described locally by an equation\footnote{For the analogous case of a long strip, roughly speaking one takes $q$ real and identifies
 $x,y$ with $\bar x,\bar y$, whereupon the gluing formula of eqn. (\ref{itry}) is applicable
 to gluing of open-string worldsheets through a narrow neck.   For more detail, see  section 7.4 of \cite{Wittentwo}.}  
 \begin{equation}\label{itry}xy=q,\end{equation}
 where $x$ is a local complex parameter on one side of the narrow neck, $y$ is a local complex parameter on the other
 side, and $q$ is a complex modulus that controls the width of the neck.  The relation of this description to
 the ``long tube'' description is made simply by the  change of variables or conformal mapping
 \begin{equation}\label{oncd}x=e^\varrho,~~~y=qe^{-\varrho},\end{equation}
 with
 \begin{equation}\label{bocd}\varrho=\uu+i\varphi,~~~\uu,\varphi\in \R.\end{equation}
 Here $\varphi $ is an angular variable of period $2\pi$.  If the coordinates $x,y$ are valid for $|x|,|y|<1$, then the description by $\varrho$ is 
 good for $0>\uu>-\ln(1/ |q|)$.  So for $q\to 0$,
 the $\varrho$
 coordinate describes a tube of circumference $2\pi$ and length $\ln (1/|q|)$.   In the ``long tube'' description, the Riemann
 surface seems to diverge for $q\to 0$, but the description by $x$ and $y$ has a limit for $q=0$.  
 The equation simply becomes 
 $xy=0$, which describes two branches, one characterized by $x=0$ for any $y$,
 and one characterized by $y=0$ for any $x$, and meeting
 at a singularity at $x=y=0$.  This is a rather special singularity, called a node or ordinary double point, where two
 branches meet.\footnote{We have already discussed the singular limiting configurations for certain  cases
 in which one branch has genus 0; see fig. \ref{failint} of section \ref{persp}
 and fig. \ref{common} of section \ref{treem}.}  The Deligne-Mumford compactification of $\M_{\sg,\sn}$ is 
 achieved by allowing this type of singularity
 and no other.   The existence of this compactification is 
  a precise statement of the fact that in string perturbation theory there is an infrared region and
 no ultraviolet region.  Similarly, the existence of an analogous
 Deligne-Mumford compactification of the moduli space of super Riemann surfaces means
 that there is no ultraviolet region in superstring perturbation theory.
 For an introduction to the Deligne-Mumford compactification, including its extensions for super Riemann
 surfaces and for open strings, see sections 6 and 7.4 of \cite{Wittentwo}.
 
 We will proceed as follows.  In sections \ref{mopen} and \ref{mclosed}, we compute the
 behavior of the string measure  when a Riemann surface develops a long strip or tube.   This entails calculating what one may call
 the string propagator; in fact, with a standard gauge-fixing, string field theory leads precisely to the propagators that
 we will calculate
 (for example, see \cite{WittenSFT,GM}).  The goal is to show that in all
 cases (open and closed bosonic strings or superstrings, including all sectors of superstrings), the string propagator
 has the same singular behavior -- the same on-shell poles -- that one would expect in a field theory with the
 same particles and couplings.
 Apart from giving a nice
 illustration of the machinery that we have developed up to this point, 
 this is an essential step in proving that perturbative string scattering amplitudes have the same infrared 
 singularities (and more generally the same singularities due to on-shell intermediate particles)
 that one would
 expect in a field theory with the same particles and couplings. The main additional step required is an analysis of the on-shell
 factorization of the string amplitudes, which we discuss in sections \ref{ondang} and \ref{stad}.  Understanding what produces this
 factorization is also a good starting point to complete our description of the integration cycle of superstring perturbation
 theory; this is the topic of section \ref{delcycle}.    All this will prepare the ground for section \ref{massren}, in which we discuss the troublesome exceptional cases of on-shell behavior in 
 string perturbation theory: mass renormalization and massless tadpoles.  These are the only possible sources of BRST anomalies
 for closed oriented strings, though for open and/or unoriented strings, there are additional anomalies, which we will
 discuss in section \ref{anomalies}.
 
\subsection{Open-String Propagators}\label{mopen}

\subsubsection{Bosonic Open Strings}\label{bos}

We begin with the propagator of the bosonic open string.

We describe a strip $S$ of length $s$ and width $\pi$ by real coordinates $\varphi, \,\uu$ with
\begin{equation}\label{omzi} 0\leq \uu\leq s,~~~0\leq \varphi\leq \pi. \end{equation}
We take the metric on the strip to be 
\begin{equation}\label{omongo}\ds^2=\d\uu^2+\d\varphi^2.\end{equation}
Viewing $\uu$ as the ``time'' direction, a string propagating on the strip has a conserved Hamiltonian
\begin{equation}\label{pomzi} L_0=-\frac{1}{2\pi}\int_0^\pi\d\varphi \,T_{\uu\uu},\end{equation}
where $T_{ij}$ is the worldsheet stress tensor.  The odd-looking minus sign comes from writing
the formula in Euclidean signature.  Similarly the antighost field $b_{ij}$ has a zero-mode
\begin{equation}\label{gomzi}b_0=-\frac{1}{2\pi}\int_0^\pi\d\varphi\, b_{\uu\uu}. \end{equation}
(The ghost field $c^i$ also has a zero-mode on the strip, but this will be less important.)
It is useful to define
\begin{equation}\label{dez}z=-i\varrho=\varphi-i\uu.\end{equation}

In quantization of an open string, the  stress tensor and the antighost field are both free fields with left- and right-moving
components.   As usual, a free field on the strip is very similar to a chiral free field (purely left-moving or right-moving) on a tube
obtained by gluing together two copies of the strip along their boundaries.  A quick way to exhibit this is to extend $\varphi$
to a real variable and extend the definition of the fields by
\begin{align}\label{pez}T_{zz}(\uu,\varphi)&=T_{\t z\t z}(\uu,-\varphi)=T_{zz}(\uu,\varphi+2\pi),\cr
b_{zz}(\uu,\varphi)&=b_{\t z\t z}(\uu,-\varphi)=b_{zz}(\uu,\varphi+2\pi). \end{align}
The zero-modes are then
\begin{align}\label{olez}L_0& =\frac{1}{2\pi}\int_0^{2\pi}\d\varphi \,T_{zz}\cr
                                      b_0&=\frac{1}{2\pi}\int_0^{2\pi}\d\varphi\, b_{zz} \end{align}

The strip $S$ has a single real modulus, the parameter $s\geq 0$.  We want to learn how to integrate over $s$,
using the general recipe of section \ref{slice}.  In this procedure, we are supposed to work on a fixed worldsheet
with a metric that depends on the modulus, in contrast to the description we have given above, with a metric (\ref{omongo})
that does not depend on $s$ while the definition (\ref{omzi}) of the strip $S$ does depend on $s$.  To convert
to a description of the desired sort, we simply introduce a new ``time'' coordinate $t=u/s$, so that the definition of the strip
becomes
\begin{equation}\label{moco}  0\leq t\leq 1,~~~0\leq \varphi\leq \pi, \end{equation}  
and the metric is
\begin{equation}\label{ocop} \ds^2={s^2}\,\d t^2+\d\varphi^2.\end{equation}
According to eqn. (\ref{gor}), in integrating over $s$, we must insert in the path integral a factor of
\begin{align}\label{zocop}\Psi_s&=\frac{1}{4\pi}\int_S\d t\,\d\varphi \frac{\partial(\sqrt g g_{ij})}{\d s} b^{ij}
=\frac{1}{2\pi}\int_0^1\d t\int_0^{\pi}\d\varphi \,{s^2}b^{tt}\cr &=\frac{1}{2\pi}\int_0^1\d t\int_0^{\pi}\d\varphi \,b_{\uu\uu}
=\int_0^1\d t \,b_0=b_0.\end{align}
In the last step, we used the fact that $b_0$ is time-independent, so that $\int_0^1\d t\,b_0=b_0$.

We conclude that integration over the strip means integration over $s$ with an insertion of $b_0$.
The path integral on the strip without any insertion is described by the operator $\exp(-sL_0)$.  
So integrating over $s$ and including an insertion of $b_0$, we learn that the open-string propagator is
\begin{equation}\label{omino}\int_0^\infty \d s\, b_0\exp(-sL_0)=\frac{b_0}{L_0}.\end{equation}
It is also convenient to express the propagator as an integral over $q=e^{-s}$.  Here $q$, which clearly is real and ranges from $0$ to $1$
for $0\leq s<\infty$, is the open-string analog of the gluing parameter that we introduced for closed bosonic strings in (\ref{itry}).
In terms of $q$, the open-string propagator is
\begin{equation}\label{zomino}b_0\int_0^1 \frac{\d q}{q}q^{L_0}.\end{equation}

The operator $L_0$ is
\begin{equation}\label{inox}L_0=\frac{\alpha'}{4}p^2+N,\end{equation}
where $p$ is the momentum and
$N$ contains the contributions of ghost and matter oscillators.  For uncompactified bosonic strings, $N$ has  eigenvalues $-1,0,1,2,\dots$.
The condition for an open string to be on-shell is $L_0=0$, so the mass squared operator of the string is $M^2=4N/\alpha'$.
Thus we can write the propagator as
\begin{equation}\label{zinox}\frac{4}{\alpha'}\frac{b_0}{p^2+M^2}.\end{equation}
Since $b_0^2=0$, the operator $b_0$ projects the string propagation onto states annihilated by $b_0$, and means
(in a sense that we will state more precisely in section \ref{stad}) that only states annihilated by $b_0$ propagate.
 The factor $4/\alpha'$
is a normalization factor that could be eliminated, if we wish, by rescaling the string coupling constant and the external
vertex operators.\footnote{In ordinary field theory, one usually normalizes the fields so that the Feynman propagator is
$1/(p^2+m^2)$, but if one wishes, one can multiply one's fields by an arbitrary constant $\kappa$ and then the propagator
becomes $\kappa^2/(p^2+m^2)$.  Changing the normalization of the fields will rescale the coupling parameters and
the wavefunctions of external particles. The situation is precisely the same in string theory; multiplying the propagator
by a constant is equivalent to changing the string coupling constant and the normalization of vertex operators.}
Modulo the normalization factor and the projection on states annihilated by $b_0$, the open-string propagator
is just what one would guess from field theory.

The precise use of this propagator is as follows.  Suppose that as in fig. \ref{longsurface}(a), the strip is attached
to Riemann surfaces $\Sigma_\ell$ and $\Sigma_r$ at its left and right ends.   We want to keep $\Sigma_\ell$
and $\Sigma_r$ fixed and integrate over the length $s$ of the strip.\footnote{To be more fastidious, we include antighost
insertions on $\Sigma_\ell $ and $\Sigma_r$, to make the path integral non-zero, but do not integrate over their moduli.}
 Keeping $\Sigma_\ell$ and $\Sigma_r$ fixed,
the path integral on those surfaces generates quantum states $\psi_\ell$ and $\psi_r$.  The path integral on the
full surface $\Sigma$ is obtained by taking $\psi_\ell$ and $\psi_r$ as initial and final states for the propagation on the
strip.  The propagation on the strip is described by the propagator (\ref{zinox}), so the path integral  on $\Sigma$ after
integrating over $s$ is
\begin{equation}\label{pobo} Z_{\Sigma;s}=\biggl\langle \psi_\ell \biggl|\,\,\frac{b_0}{L_0} \,\,\biggr|\psi_r\biggr\rangle.\end{equation}

A complete calculation will of course include
also integrating over the moduli of $\Sigma_\ell$ and $\Sigma_r$; in fact 
integrating just over $s$ is only sensible in the region where $s$ is large.  What the computation that we have just done is good for
is to isolate the singularities that arise when an open string in a particular channel goes on shell.  As we have just seen,
these singularities are precisely the simple poles at $p^2+m^2=0$ that one would expect in field theory.

One can analyze in a similar fashion the contribution of a region in which any number of 
open-string states simultaneously
go on-shell.  We get a propagator $b_0/L_0$ in each channel, leading to 
a standard $1/(p^2+m^2)$ pole in every channel. Presently, we will obtain similar results
 for open superstrings and for all the closed-string theories.   Since the usual infrared and on-shell
singularities of Feynman amplitudes come from the poles of propagators, this makes it more or 
less obvious that string theory will have the same
infrared and on-shell singularities as field theory.    

To be more precise, this is more or less obvious once one incorporates
an important refinement, the Feynman $i\epsilon$, which will be described in section \ref{iepsilon}.   We go in more detail about the residue of the pole of the propagator in section  \ref{ondang}.

\subsubsection{The NS Propagator}\label{nspropagator}

Now we consider the NS sector of open superstrings. 

On the worldsheet of an open superstring, in addition to bosonic coordinates $\uu$ and $\varphi$,
there are holomorphic and antiholomorphic odd coordinates $\theta$ and $\t\theta$.   At the endpoints $\varphi=0,\pi$
of a strip $S$,
they are glued together, with some choices of signs.  In the NS sector, the signs are opposite at the two ends.
\begin{equation}\label{tomo}\t\theta=\begin{cases}\theta &\mbox {if} ~~\varphi=0\\
                                    -\theta &\mbox{if} ~~\varphi=\pi.\end{cases}\end{equation}
The minus sign can be moved from one end of the strip to the other by a change of variables $\t\theta\to -\t\theta$.    

Right-moving massless free fields  along the string depend on $\theta$ (as well as $z$) and left-moving ones depend on 
$\t\theta$ (and $\t z$).
By extending the range of $\varphi$ beyond the interval $0\leq \varphi\leq \pi$, one can combine these two
types of modes to a chiral field that depends only on $\varphi$ and $\theta,$ subject
to
\begin{equation}\label{yomo}\varphi\to\varphi+2\pi,~~\theta\to -\theta.\end{equation}
The minus sign, which comes about because the signs in (\ref{tomo}) are opposite, is the reason that quantization of
open strings with the boundary conditions (\ref{tomo}) leads to what is usually called the NS sector.

Now let us discuss the propagator in the NS sector.  In fact,  the minus sign in the boundary conditions makes
things simple.  The strip in the NS sector has no odd moduli.  An odd modulus would arise as always
from a mode of the gravitino field that cannot be gauged away.  To be more exact, for open superstrings we have left
and right gravitino fields $\chi_{\t z}^\theta$ and $\chi_z^{\t\theta}$ that are glued together on the boundary
by analogy with (\ref{tomo}):
\begin{equation}\label{tomolx}\chi_z^{\t\theta}=\begin{cases}\chi_{\t z}^\theta &\mbox {if} ~~\varphi=0\\
                                    -\chi_{\t z}^\theta &\mbox{if} ~~\varphi=\pi.\end{cases}\end{equation}
They fit together to a single gravitino field $\chi_{\t z}^\theta$ that is defined for all $\varphi$ and obeys
\begin{equation}\label{romox}\chi_{\t z}^\theta(\uu,\varphi+2\pi)=-\chi_{\t z}^\theta(\uu,\varphi).\end{equation} 
The minus sign ensures that $\chi_{\t z}^\theta$ can be gauged away by
\begin{equation}\label{womox}\chi_{\t z}^\theta\to \chi_{\t z}^\theta+\partial_{\t z}y^\theta,\end{equation}
as one may prove via a Fourier expansion with respect to $\varphi$, using the absence of a zero-mode.
Thus the strip has no odd moduli.  Similarly, the minus sign in (\ref{yomo}) means that the mode expansion
of the commuting ghost fields $\beta$ and $\gamma$ involves modes $\beta_r$, $\gamma_r$ with $r\in \Z+1/2$;
in particular, there are no zero-modes.

Accordingly, the derivation of the propagator is almost straightforward.  We only have to integrate over one
even modulus $s$, the ``length'' of the strip.  The same calculation as in section \ref{bos} shows
that the integral gives $\int_0^\infty \d s\, b_0\exp(-sL_0)=b_0/L_0$.  

To this we only have to add one detail and answer one question.
What we need to add is the GSO projection  \cite{GSO}.    The GSO projection, which removes
the tachyon from superstring theory, reflects the fact that the worldsheet path integral of superstring theory includes
a sum over worldsheet spin structures.  In general, this sum cannot be carried out independently of the integral
 over supermoduli space, because the spin structure is part of the structure of a super Riemann
surface and there is no way to sum over the spin structures of a super Riemann surface except as part of
the integral over all of the even and odd moduli.  However, in the region $s\to\infty$ that gives rise to the on-shell
singularities of a superstring amplitude, a very partial sum over spin structures makes sense: this is the sum
over pairs of spin structures that differ only by a minus sign ``twist'' in the $\uu$ direction.  
For a detailed explanation of this, see section 6.2.3 of \cite{Wittentwo}; the basic idea is 
also indicated in section \ref{supan} below. The sum over pairs of spin structures that differ only by this particular
twist
gives the GSO projection $\piGSO$ that removes half the states of the NS sector, including the tachyon.  With
this included,   the open superstring propagator in the NS sector is 
\begin{equation}\label{nsprop} \frac{b_0}{L_0}\piGSO.\end{equation}

We also need to answer a question: What are the states that are propagated by the
propagator (\ref{nsprop})?  To be more precise, what is the picture number of these states?

\begin{figure}
 \begin{center}
   \includegraphics[width=2.5in]{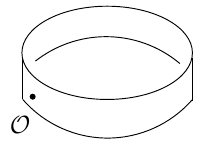}
 \end{center}
\caption{\small An annulus, built by gluing together the ends of a long strip, with an operator insertion $\O$.}
 \label{annulus}
\end{figure}

\subsubsection{Picture Number And The Propagator}\label{pcprop}
Before getting to picture number, let us remember that for an ordinary anomaly-free or anomalous symmetry
such as ghost number, the answer to a question like the last one
 is that states of all possible quantum numbers can
propagate down the strip.  To illustrate this in a precise context, let us glue together the two ends of the strip at $\uu=0$
and $s$ 
to make an annulus (fig. \ref{annulus}).  In the gluing, we insert
some operator $\O$.  The path integral on the annulus with this insertion is
\begin{equation}\label{ombic}\int_0^\infty \d s\,\Tr\,\O\,b_0\exp(-sL_0)\piGSO.\end{equation}
The trace is a sum over all states, including states with all possible values of the 
ghost number and other quantum
numbers.  

We have already explained at the end of section \ref{pnumber} one reason that picture number cannot
be interpreted as a symmetry, anomalous or not.  Here we are about to come upon another such reason.  
In explaining this, we first consider the case that $\O$ is  independent of $\beta$ and $\gamma$, 
so that $\beta$ and $\gamma$ are decoupled from the other fields  in (\ref{ombic}) and the $\beta\gamma$
path integral that we want
is simply the partition function.
We claim that the $\beta\gamma$ partition function on an annulus
in the NS sector is a sum only over states of picture number $-1$, with
no sum over the picture number of the states propagating around the annulus.

We will explain this in two ways.  First, this is the only answer that makes any sense; otherwise the annulus
path integral would diverge, since $L_0$ would be unbounded below.  
The ghost  vacuum with picture number $|\q\rangle $ is defined by the conditions
\begin{align}\label{drondo}\beta_r|\q\rangle&=0,~~~r>-\q-3/2\cr
                                          \gamma_r|\q\rangle&=0, ~~~r\geq \q+3/2.\end{align}
For any value of $\q$ other than $-1$, the state $|\q\rangle$ is annihilated by some $\beta_r$ or $\gamma_r$ with
$r<0$, and then the states $\beta_{-r}^k|\q\rangle$ or $\gamma_{-r}^k|\q\rangle$, $k=0,1,2,\dots$ are linearly
independent and have $L_0$ unbounded below.  So for $\q\not=-1$,  the trace in (\ref{ombic}) that is supposed
to give the annulus path integral does not make any sense, at least for generic $\O$.  For $\q=-1$, there is no
problem, since (\ref{drondo}) says that $|\q\rangle$ is annihilated by what one would want to call the annihilation operators,
\begin{equation}\label{rondop}\beta_r|\q\rangle=\gamma_r|\q\rangle=0,~~r>0.\end{equation}

For a deeper explanation, we should ask how we define the 
$\beta\gamma$ path integral so that it does make
sense.  We explore this question more fully in section \ref{betag}.  However, the 
basic idea is that the $\beta\gamma$
path integral without operator insertions is understood as a bosonic Gaussian integral that equals 
$1/\det \tilde\partial_{\beta\gamma}$, where 
$\t\partial_{\beta\gamma}$ is the kinetic operator of the $\beta\gamma$ system.   
Moreover $\det\,\t\partial_{\beta\gamma}$ (defined with any standard type of regularization)
 is a definite function of $s$ and therefore the question, ``What is the picture number
of the states that propagate around the annulus?'' is going to have a definite answer; 
we are not free to make
any choice.  As a shortcut to determine the answer, let us note that if we replace $\beta$ and $\gamma$
with fields $\beta^*$ and $\gamma^*$ of opposite statistics (and thus fermionic) 
but otherwise with the same
Lagrangian and boundary conditions, this simply replaces the path integral of the 
$\beta\gamma$ system by
its inverse.  So the $\beta^*\gamma^*$ path integral is $\det\tilde\partial_{\beta\gamma}$.
On the other hand, we certainly know how to interpret the $\beta^*\gamma^*$ path 
integral on an annulus as a sum over
quantum states.  On an annulus of circumference $s$ in the time direction, taking the 
fields $\beta^*$ and $\gamma^*$
to be periodic in the $s$ direction (this gives an insertion of $(-1)^F$, the operator 
that counts fermions mod 2), and
setting $q=e^{-s}$, the
$\beta^*\gamma^*$ path integral can be evaluated in a standard fashion as
a trace in  the fermion Fock space:
\begin{equation}\label{bongo}\det\,\t\partial_{\beta\gamma}= 
\Tr\,(-1)^F\exp(-s L_{0;\beta^*\gamma^*})=q^f\prod_{r=\frac{1}{2},\frac
{3}{2},\dots }(1-q^r).\end{equation}
(There is no analog of picture number for fermions, so there is no choice to be made in writing this formula.)
  Here $L_{0;\beta^*\gamma^*}$ is the Hamiltonian of the $\beta^*\gamma^*$ 
  system and $f$ is the ground state energy of the $\beta^*\gamma^*$ system in the NS sector.
The $\beta\gamma$ path integral on the annulus is therefore the inverse of this, or
\begin{equation}\label{wrongo}\frac{1}{\det\,\t\partial_{\beta\gamma}}=
q^{-f}\prod_{r=\frac{1}{2},\frac{3}{2},\dots}\frac{1}{1-q^r}.\end{equation}
As expected, we see that the energy levels of the $\beta\gamma$ system are bounded below; the $\beta\gamma$
path integral on the annulus is $q^{-f}$ times a series in positive powers of $q$.   Moreover, we see a standard
partition function for a Fock space of bosons, 
confirming that it does make sense to define a bosonic Gaussian integral as the inverse
of a corresponding fermionic Gaussian integral.  
Finally, we see that the ground state energy of the $\beta\gamma$ system \
is precisely minus that of the $\beta^*\gamma^*$ system.  According to \cite{FMS}, 
this is true precisely at picture number $-1$.

Now let us restore the operator $\O$ in eqn. (\ref{ombic}).  Part of what we have said is valid
in general.  The $\beta\gamma$ path integral defined as a generalized Gaussian integral 
 will always lead to a definite answer, 
with no freedom in the choice of picture number.
Moreover, if $\O$ is simply a polynomial in $\beta$ and $\gamma$ and their derivatives, its inclusion
does not really affect the above reasoning.  (The $\beta\gamma$ path integral with insertions
of elementary fields is analyzed in section \ref{elf}.)  

However, in general, it is possible to
find an $\O$ such that the states that propagate around the annulus have non-canonical
picture number.  A simple example with this property is $\O=\delta(\beta_{-1/2})\delta(\gamma_{1/2})$.
We have taken a product of picture-raising and picture-lowering operators, so that the path
integral with insertion of $\O$ remains sensible.  
The operator $\O$ projects onto states that are annihilated by $\beta_{-1/2}$.  The Fock vacuum
 $|\q\rangle$    with picture number $\q=-1$ is not annihilated by $\beta_{-1/2}$, but the corresponding
 Fock vacuum with $\q=0$ does have this property.  
 
 It can be shown using methods of section
 \ref{betag} that the $\beta\gamma$ path integral on an annulus with insertion of $\O$  computes
\begin{equation}\label{tugly}\Tr\,\O \,q^{L_0}\end{equation}
where the trace is taken in the $\beta\gamma$ Fock space with $\q=0$ (we denote the Fock
vacuum in this Fock space as $|0\rangle$).
This claim may seem to present a paradox:
 for $\q=0$, $L_0$ is not bounded
below, so how can the trace converge?  In fact, $L_0$ is unbounded below for $\q=0$
because the oscillator $\gamma_{1/2}$ lowers $L_0$ by $1/2$. But this causes no problem in
 the trace $\Tr\,\O q^{L_0}$, since the potentially dangerous states
 $\gamma_{1/2}^p|0\rangle$ with $p>0$ (and any other states constructed with the negative
 energy  creation operator $\gamma_{1/2}$)
 are annihilated by $\O=\delta(\beta_{-1/2})\delta(\gamma_{1/2})$, because of the second factor.  
 Such an apparently ``lucky'' rescue occurs in any sensible $\beta\gamma$ path
 integral that receives contributions from NS or R states whose picture number is such that $L_0$ is unbounded below.
 
So it is not true in general that in an arbitrary sensible $\beta\gamma$ path integral,
the $\beta\gamma$ propagator in the NS sector only propagates
states of picture number $-1$.   But this is true in the context of superstring perturbation theory. See
section \ref{pnpco} for further discussion.
                                    
\subsubsection{The Ramond Propagator}\label{ramondpropagator}
                                  
The Ramond sector of open superstrings is obtained by dropping some minus signs in the gluing relations.                                   
Thus, eqn. (\ref{tomo}) for the identification of the fermionic coordinates at the end of the strip becomes
\begin{equation}\label{tomotz}\t\theta=\begin{cases}\theta &\mbox {if} ~~\varphi=0\\
                                    \theta &\mbox{if} ~~\varphi=\pi,\end{cases}\end{equation}
with the same sign at both ends. The left- and right-moving supercurrents can accordingly be combined
to a holomorphic supercurrent $S_{z\theta}(\uu,\varphi)$ that is invariant under $\varphi\to\varphi+2\pi$. 
As a result, the odd generators $G_r$ of the super-Virasoro algebra
are graded by integers, and in particular, the algebra contains a zero-mode that commutes with the Hamiltonian
$L_0$:
\begin{equation}\label{omotz} G_0 =\frac{1}{2\pi}\int_0^{2\pi}\d\varphi \,S_{z\theta}(0,\varphi)  .\end{equation}
Similarly we can combine left- and right-moving antighost fields to a single holomorphic field $\beta_{z\theta}(\uu,\varphi)$ that
is also invariant under $\varphi\to\varphi+2\pi$, so that it has a zero-mode:
\begin{equation}\label{domotz}\beta_0=\frac{1}{2\pi}\int_0^{2\pi} \d\varphi\,\beta_{z\theta}(\uu,\varphi).\end{equation}
Finally, the gravitino field $\chi_{\t z}^\theta(\uu,\varphi)$ is now a periodic variable,
\begin{equation}\label{icoc}\chi_{\t z}^\theta(\uu,\varphi+2\pi)=\chi_{\t z}^\theta(\uu,\varphi). \end{equation}

The consequence of this last statement is that the gravitino field in the strip has a zero-mode that should be treated
as a modulus.  This can be represented by the constant gravitino field
\begin{equation}\label{nicoc}\chi_{\t z}^\theta=\eta,\end{equation}
where $\eta$ is a constant anticommuting parameter.  To gauge this mode away
by
\begin{equation}\label{picoc}\chi_{\t z}^\theta\to \chi_{\t z}^\theta+\partial_{\t z}y^\theta,\end{equation}
where $y^\theta$ should be invariant under $\varphi\to\varphi+2\pi$, we would have to let $y^\theta$ grow in the 
$\uu$ direction.  Since $\uu$ ranges from $0$ to $s$, a gauge transformation that grows
 with $\uu$ does not behave well for $s\to\infty$, where we are trying to extract on-shell poles.  
This explains at least heuristically why one should treat the constant mode (\ref{nicoc}) as a modulus
and not try to gauge it away. A precise explanation involves the Deligne-Mumford
compactification; see the discussion of eqn. (\ref{pirat}) below.

It is straightforward\footnote{Since there is
only one odd modulus associated to the strip, none of the subtleties of integrating over odd moduli come into play.
As explained in sections \ref{harder}, \ref{yzzo}, and \ref{basodd}, these subtleties become relevant when there are two or more odd
moduli.}
 to integrate over $\eta$ using the procedure of eqn. (\ref{pinzo}) or (\ref{winzo}). In fact, we can do this using the bosonic coordinates $\uu,\varphi$ without even transforming to $t=\uu/s$.
Replacing $\chi_{\t z}^{(\sigma)\theta}$ by 1, the factor (\ref{inzoop}) in the path integral becomes
\begin{equation}\label{middo}\exp\left(-\frac{1}{2\pi}\int_0^{2\pi}\d\varphi\int_0^s\d\uu\,\left(\eta S_{z\theta}+\d\eta \,\beta_{z\theta}\right)\right)=\exp\left(-s(\eta G_0+\d\eta \beta_0)\right).\end{equation}
The integral over $\eta$ and $\d\eta$ gives 
\begin{equation}\label{pikl}G_0\delta(\beta_0).\end{equation}
This is the only part of the evaluation of the propagator that is special to the Ramond sector.
The integral over $s$ gives the usual result $b_0/L_0$, and the sum over the possible twists of the fermions in the $\uu$
direction gives the GSO projection $\piGSO$.  So the Ramond sector propagator for open strings is
\begin{equation}\label{tyrk}\frac{b_0\delta(\beta_0)\piGSO G_0}{L_0}.\end{equation}
 
Clearly, only states that are annihilated by $b_0$ and $\beta_0$ and are invariant under $\piGSO$ propagate down
the strip.  To understand the result (\ref{tyrk}) more fully, we recall that in the Ramond sector, $G_0^2=L_0$,
so we can write the propagator as
\begin{equation}\label{tyrok}\frac{b_0\delta(\beta_0)\piGSO}{G_0}.\end{equation}
The operator $G_0$ is the string theory analog of the Dirac operator of field theory, and so 
$1/G_0$ (with some states projected out by the numerator) is the natural string theory analog of the usual Dirac propagator.
For example, for massless Ramond
states, $G_0$ reduces to $\frac{1}{2} (\alpha')^{1/2}\Gamma\cdot p$ (where $\Gamma^I$ are spacetime gamma matrices
and $p_I$ is the momentum), which apart from the factor $\frac{1}{2}(\alpha')^{1/2}$ is the massless Dirac operator
in momentum space as usually normalized.  
In general, $G_0=\frac{1}{2} (\alpha')^{1/2}\Gamma\cdot p + N_\Ra$, where $N_\Ra$ is the oscillator contribution to $G_0$ and
$2(\alpha')^{-1/2}N_\Ra$ is 
the mass operator for open strings in the Ramond sector. 

As in section \ref{nspropagator}, we should now ask what is the picture number of the Ramond states whose propagation
is described by this propagator.    For a simple example, let us glue together the two ends of the strip with
insertion of some operator $\O$  to make an annulus.  Thus the $\beta\gamma$ path integral, if
it can be interpreted in terms of a sum over states, computes $\Tr\,\O\delta(\beta_0) q^{L_0}$.
We cannot simply take $\O=1$, since the criterion of eqn. (\ref{ruffo}) for a sensible $\beta\gamma$
path integral would not be satisfied.  We need an operator of picture number $-1$, such as $\O=\delta(\gamma_0)$.    So we want to calculate a $\beta\gamma$ path integral on the strip
with insertion of $\delta(\gamma_0)\delta(\beta_0)$.    What is the picture number of the
states that contribute to this path integral?   
 The obvious candidates are $\q=-1/2$ and $-3/2$, since
those are the values of $\q$ at which $L_0$ is bounded below; this is clear from
    the definition (\ref{drondo}) of the ghost vacuum $|\q\rangle$ with picture number 
$\q$.   If we treat any of the $\beta_n$ or $\gamma_n$ with $n>0$ as a creation operator,
the trace $\Tr\,\delta(\gamma_0)\delta(\beta_0)q^{L_0}$ will certainly diverge.  And the explicit factor of $\delta(\beta_0)$ in
the propagator 
shows that the states that are propagated are annihilated by $\beta_0$.  Putting these
facts together, we find that in this minimal example, the states whose propagation is described
by the Ramond propagator have canonical picture number $\q=-1/2$.  As in the NS case, one
can construct sensible $\beta\gamma$ path integrals that receive contributions from states
of noncanonical picture number, but these are not the path integrals that arise in superstring
perturbation theory.

\subsection{Closed-String Propagators}\label{mclosed}

Now we turn to closed-string propagators.

\subsubsection{Closed Bosonic Strings}\label{bosclosed}

A long strip describing an almost on-shell open string has a single real modulus $s$, also 
usefully parametrized by $q=\exp(-s)$.  For closed strings, $s$ and
$q$ become complexified.  For closed strings, the gluing parameter $q$ appears  in formula
(\ref{itry}) that describes the gluing of two branches of a Riemann surface.  
But $q$ is naturally a complex parameter, one of the moduli
of the Riemann surface.  So for closed strings, $s$ must combine with a second real parameter to make a complex modulus.

Concretely, this second real modulus is obtained by ``cutting'' the long tube $T$  in fig. \ref{longsurface}(b) to separate the two ends,
and then rotating one piece relative to the other by an angle $\aalpha$ before gluing them back together.  Assuming that this rotation
cannot be extended as a symmetry of either $\Sigma_\ell$ or $\Sigma_r$, $\aalpha$ is a modulus of $\Sigma$.  In 
the Deligne-Mumford
compactification of $\M_{\sg,\sn}$, one only considers decompositions such that $\Sigma_\ell$ and $\Sigma_r$ (with their punctures
deleted) both have negative Euler
characteristic, and then it is automatically true that the symmetry does not extend and $\aalpha$ is a modulus.   This is also almost always true
for open and/or unoriented string worldsheets, though the exceptions turn out to be important.\footnote{\label{exc} The exceptional cases
are that $\Sigma_r$ (or $\Sigma_\ell$) is a disc or a copy of $\Bbb{RP}^2$ with only one puncture (the node).     These examples will be important in studying anomalies in section \ref{anomalies}.}

In one description of this situation, we can describe the tube $T$  by a flat metric that does not depend on $\alpha$
\begin{equation}\label{dimox}\ds^2=\d\uu^2+\d\varphi^2,~~   0\leq \uu\leq s,~~ 0\leq \varphi\leq 2\pi.\end{equation}
$\varphi$ is now an angular variable.  For closed strings, we have separate holomorphic and antiholomorphic Virasoro algebras,
each with its own zero-mode:
\begin{align}\label{foss}L_0&=-\frac{1}{2\pi}\int_0^{2\pi}\d\varphi\,T_{zz} \cr
                                   \t L_0 & =-\frac{1}{2\pi}\int_0^{2\pi}\d\varphi\,T_{\t z \t z}.\end{align}
Similarly, there are separate holomorphic and antiholomorphic antighost zero-modes.  For closed bosonic strings, these are
\begin{align}\label{boss}b_0&=-\frac{1}{2\pi}\int_0^{2\pi}\d\varphi\,b_{zz} \cr
                                   \t b_0 & =-\frac{1}{2\pi}\int_0^{2\pi}\d\varphi\,b_{\t z \t z}.\end{align}

To the tube $T$, we want to associate the real modulus $s$ that already appears in the metric (\ref{dimox}), and another real modulus
$\aalpha$.                                   
If $\Sigma$  is obtained by gluing $T$ at its ends onto
surfaces $\Sigma_\ell$ and $\Sigma_r$, then the extra modulus can be obtained by gluing 
$T$ onto $\Sigma_\ell$ in a way that is independent of $\aalpha$, but rotating it by an angle $\aalpha$ before
gluing it onto $\Sigma_r$.

This is a simple description, but it has two drawbacks.  It does not explain in what sense $\aalpha$ should be associated to $T$, rather
than to the whole surface $\Sigma$.  And it does not lend itself to computing the measure for the string path integral via
formulas such as eqn. (\ref{gor}).  For that purpose, we want to describe $\Sigma$
as a fixed two-manifold, independent of $\alpha$, but with a metric that depends on $\alpha$.  We can easily get such
a description by replacing $\varphi$ with a new angular coordinate
\begin{equation}\label{omonk}\h\varphi=\varphi-\aalpha f(\uu), \end{equation}
where $f(\uu)$ is any smooth function with $f(0)=0$, $f(s)=1$.  If $T$ is described by the coordinates $u,\h\varphi$,
then its metric becomes
\begin{equation}\label{zormonk}\ds^2=\d\uu^2+\d(\h\varphi+\aalpha f(\uu))^2.\end{equation}
Now the definition of the space $T$ and the gluing recipe are both independent of $\aalpha$, which appears only in the metric of $T$.

It is now straightforward to apply the recipe of eqn. (\ref{gor}).  The integral over $\alpha$ must be accompanied by an insertion of 
\begin{equation}\label{zormo}\Psi_\alpha=\frac{1}{4\pi}\int_0^{2\pi}\d\varphi \int_0^s\d\uu \,\frac{\partial(\sqrt g g_{ij})}{\partial\aalpha}b^{ij}.\end{equation}
We can evaluate $\Psi_\alpha$ in a simple way, since the $\aalpha$ dependence of the metric is equivalent to a change of coordinates.  
It follows from this that
\begin{equation}\label{zorki}\frac{\partial g_{ij}}{\partial\aalpha}=D_iv_j+D_jv_i \end{equation}
where $v$ is a vector field.  In fact,
\begin{equation}\label{orki}v=f(\uu)\frac{\partial}{\partial\varphi},\end{equation}
or in other words $v^\uu=0$, $v^\varphi=f(\uu)$.  (Thus, $\exp(\alpha v)$ is the transformation from coordinates $s,\h\varphi$ back to
$s,\varphi$.)  It follows that
\begin{equation}\label{lorik}\Psi_\alpha=\frac{1}{4\pi}\int_0^{2\pi}\d\varphi\int_0^s\d\uu \sqrt g b^{ij}(D_iv_j+D_jv_i).\end{equation}
Integrating by parts and using the equation of motion $D_ib^{ij}=0$, we find that we can evaluate $\Psi_\alpha$ as a surface term at $\uu=s$:
\begin{equation}\label{loriks}\Psi_\alpha=-\frac{1}{2\pi}\int_0^{2\pi}\d\varphi \,b_{\uu \varphi}(\varphi,s)=b_0-\t b_0.\end{equation}

The computation of $\Psi_s$, the ghost insertion that accompanies the integration over $s$, proceeds just as in eqn. (\ref{zocop}),
except that for closed strings, in the final step, we get $-\frac{1}{2\pi}\int_0^{2\pi}\d\varphi b_{\uu\uu}=b_0+\t b_0$.  So $\Psi_s\Psi_\alpha=
2\t b_0b_0$.  The operator that propagates a closed string through an imaginary time $s$ and rotates it by an angle $\alpha$ is
\begin{equation}\label{zerombo}\exp\left(-s(L_0+\t L_0)\right)\exp\left(-i\alpha(L_0-\t L_0)\right).\end{equation}
The closed bosonic string propagator is then
\begin{align}\label{merom}2\Psi_s\Psi_\alpha&\int_0^\infty\d s\int_0^{2\pi}\d\alpha\,\exp\left(-s(L_0+\t L_0)\right)\exp\left(-i\alpha(L_0-\t L_0)\right)
\cr &=4\pi\t b_0b_0 \delta_{L_0-\t L_0}\int_0^\infty\d s\exp(-s(L_0+\t L_0))=\frac{2\pi \t b_0b_0\delta_{L_0-\t L_0}}{L_0}.\end{align}
Here the integral over $\alpha$ has given a factor $\delta_{L_0-\t L_0}$ that ensures that only states annihilated by $L_0-\t L_0$ propagate through the tube.  The integral over $s$ gives a factor $1/L_0$ (or equivalently $1/\t L_0$), which
contains the expected closed-string poles. 

The closed-string gluing parameter is $q=\exp(-(s+i\alpha))$.  Its antiholomorphic counterpart is $\t q=\exp(-(s-i\alpha))$.  We can write the
closed bosonic string propagator as
\begin{equation}\label{erom}\t b_0b_0\int_{|q|\leq 1}\frac{\d^2q}{|q|^2}\,q^{L_0}\t q^{\t L_0}.\end{equation} 
 
\subsubsection{Closed Superstrings}\label{primo}

Closed superstring propagators can be obtained by combining the constructions that we have explained so far.

For the heterotic string, we have to consider the NS and Ramond sectors.  In the NS sector, the only moduli of a long tube
are the parameters $s$ and $\alpha$ that we already have considered.  The derivation of the propagator is precisely the same 
as in section \ref{bosclosed}, except that as in section \ref{nspropagator}, we have to include the GSO projection, which comes from
summing over fermionic twists in the $\uu$ direction.  So the NS sector propagator of the heterotic string is
\begin{equation}\label{omix}\frac{2\pi\t b_0 b_0 \delta_{L_0-\t L_0}\piGSO}{L_0}.\end{equation}
For the Ramond sector of the heterotic string, there is also a gravitino mode.  Just  as in section \ref{ramondpropagator},
integrating over the corresponding odd modulus gives a factor of $\delta(\beta_0)G_0$.  So the Ramond sector propagator is
\begin{equation}\label{bomix} \frac{2\pi\t b_0 b_0\delta(\beta_0) \delta_{L_0-\t L_0}G_0\piGSO}{L_0}= \frac{2\pi\t b_0 b_0\delta(\beta_0) \delta_{L_0-\t L_0}\piGSO}{G_0}.\end{equation}
Just as for open superstrings, the projection operators in the numerator place some restrictions on what classes of states can propagate,
and the factors $1/L_0$ and $1/G_0$ reproduce the poles that one would expect in field theory.

The propagator is one subject for which the generalization from the heterotic string to Type II superstrings merits some comment.  The left- and right-movers can independently be placed in the NS or R
sector, so overall there are four sectors, namely NS-NS, NS-R, R-NS, and R-R.  In each case, the  sum over fermionic twists in the $\uu$
direction
can be carried out separately for holomorphic and antiholomorphic degrees of freedom, giving separate GSO projections $\piGSO $ and $\tpiGSO$
for the two types of mode.  Apart from this, in the NS-NS sector, the derivation is the same as for closed bosonic strings and
the  propagator is
\begin{equation}\label{komix}\frac{2\pi\t b_0b_0 \delta_{L_0-\t L_0}\piGSO\tpiGSO}{L_0}.\end{equation}
In the R-NS and NS-R sectors, one also has a holomorphic or antiholomorphic gravitino mode.  Integration over the corresponding
odd modulus gives the familiar factors $\delta(\beta_0)G_0$ or $\delta(\t \beta_0)\t G_0$, so the propagator is
\begin{equation}\label{komixa}\frac{2\pi\t b_0b_0 \delta_{L_0-\t L_0}\delta(\beta_0)G_0\piGSO\tpiGSO}{L_0}\end{equation}
or
\begin{equation}\label{komixb}\frac{2\pi\t b_0b_0 \delta_{L_0-\t L_0}\delta(\t\beta_0)\t G_0\piGSO\tpiGSO}{L_0}.\end{equation}
Finally, in the R-R sector, there is both a holomorphic gravitino mode and an antiholomorphic one.  Integrating over the corresponding
odd moduli gives a factor of $\delta(\beta_0)G_0$ and also a factor of $\delta(\t\beta_0)\t G_0$, and the propagator is
\begin{equation}\label{komixc}\frac{2\pi\t b_0b_0 \delta(\beta_0)G_0\delta(\t\beta_0)\t G_0\delta_{L_0-\t L_0}\piGSO\tpiGSO}{L_0}.\end{equation} 

The only case that really requires discussion is the R-R  propagator; the meaning of the factor $G_0\t G_0$ in the numerator
may be unclear.  There is no problem of principle.  The factor of $1/L_0$ gives the expected poles for on-shell bosons. The factors $G_0$ and
$\t G_0$ in the numerator are nonsingular on-shell, so in matching to field theory expectations, those factors could be absorbed in the couplings
of the R-R fields rather than regarded as part of the propagator.  However, it is possible to learn more by considering
the massless states in the R-R sector.  For simplicity, we do this for uncompactified superstrings in $\R^{10}$.  The gauge-invariant field strength 
of a massless
R-R field can be regarded as a bispinor $\phi_{\alpha\beta}$, where $\alpha$ and $\beta$ are spinor indices in $\R^{10}$ (one of which
comes by quantizing  holomorphic degrees of freedom of the string, and one by quantizing antiholomorphic degrees of freedom).  The
field $\phi_{\alpha\beta}$ is subject to a chirality projection on each index (coming from the GSO projections), but this will not affect our
remarks here.  From a field theory point of view, treating the R-R fields as free fields, the two-point function of $\phi_{\alpha\beta}$ at momentum $p$ 
is
\begin{equation}\label{xenfox}\langle \phi_{\alpha\beta}(p)\phi_{\alpha'\beta'}(-p)\rangle=
\frac{(\Gamma\cdot p)_{\alpha\alpha'}(\Gamma\cdot p)_{\beta\beta'}}{p^2}+{\rm constant}.\end{equation}
Here $\Gamma\cdot p$ is the Dirac operator in momentum space.  (Eqn. (\ref{xenfox}) has been used in computing
the gravitational anomalies of R-R fields; see eqn. (48)  of \cite{AlWitten}.)  This agrees, up to an inessential constant factor,
 with the low energy limit of the massless propagator (\ref{komixc}), 
since for massless R-R states,
 $G_0=\frac{1}{2}\alpha'^{1/2}(\Gamma\cdot p)_{\alpha\alpha'}$ and $\t G_0=\frac{1}{2}\alpha'^{1/2}(\Gamma\cdot p)_{\beta\beta'}$.

We conclude that for massless  states,
the R-R propagator for Type II superstrings matches the field theory two-point function of the R-R field strength,
not the two-point function of the R-R gauge field. Hence in perturbation theory, for oriented closed superstrings, R-R fields couple
only via their field strength. In particular they
decouple at zero momentum, ensuring that there is no R-R analog of the subtleties involving massless NS tadpoles that we will explore beginning
in section \ref{massren}.  For open and/or unoriented superstrings, there is a more complicated story that we will study in section \ref{anomalies}.

\subsection{The Feynman $i\epsilon$}\label{iepsilon}

Our discussion of the propagators is so far missing a crucial detail: the Feynman $i\epsilon$.

\subsubsection{The Propagator In Lorentz Signature}\label{lorsig}

In Euclidean signature, $p^2$ is positive-definite, and the propagator $1/(p^2+m^2)$ is singular only at $p=0$.
In Lorentz signature -- where we ultimately must work in order to compute scattering amplitudes -- $p^2$ is no longer
positive-definite and the propagator has a pole on-shell at $p^2+m^2=0$.  The appropriate treatment of this pole
was explained long ago by Feynman; it is essential in Lorentz signature to include the Feynman $i\epsilon$, replacing
$1/(p^2+m^2)$ by $1/(p^2+m^2-i\epsilon)$, where $\epsilon$ is an infinitesimal positive quantity, and one takes the limit $\epsilon\to 0$
at the end of any computation.  The Feynman $i\epsilon$, in other words, is a recipe to avoid the pole at $p^2+m^2=0$.

The Feynman $i\epsilon$ similarly must be incorporated in any string theory computation.  How to do this has been explained in
\cite{Berera,OtherWitten}. The latter reference contains a detailed explanation in the spirit of the present paper; we will be much more
brief here.

We start with bosonic open strings.
Upon setting $q=e^{-\tt}$, the integral in (\ref{zomino}) becomes
\begin{equation}\label{helbo} \frac{1}{p^2+m^2} =\int_0^\infty \d\tt \exp(-\tt (p^2+m^2)). \end{equation}
Here $\tt$ is a Euclidean Schwinger parameter.   We can think of it as a proper time parameter in Euclidean signature.
To get a Lorentz signature propagator, we should integrate over a proper time parameter in Lorentz signature.
One way to do this is to set $\tt=i\tau$ and integrate over real positive $\tau$.  The integral becomes oscillatory and needs
a convergence factor $\exp(-\epsilon\tau)$, where $\epsilon$ is taken to zero at the end.  We get the Feynman propagator
\begin{equation}\label{fleob} \frac{1}{p^2+m^2-i\epsilon}=i\int_0^\infty \d \tau\exp(-i\tau(p^2+m^2)-\epsilon\tau). \end{equation}

What can be the analog of this in string theory?  The open string modulus $q$ is naturally real, as is  $\tt=-\log q$.
What is worse, $q$ is only defined when it is small; equivalently, in the context of string theory, $\tt$ is only defined
when it is large.   What can it mean to make $\tt$ imaginary?

\begin{figure}
 \begin{center}
   \includegraphics[width=3in]{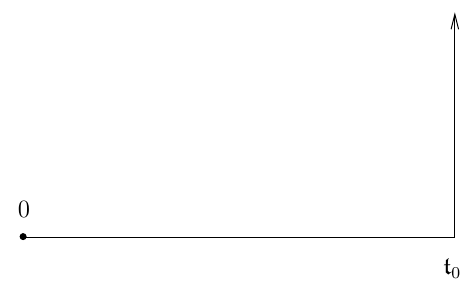}
 \end{center}
\caption{\small An integration contour that is useful in generalizing the Feynman $i\epsilon$ to string theory.}
 \label{usefulone}
\end{figure}

A procedure that is valid in field theory and is closer to what we can do in string theory is to  integrate over real $\tt$ from
0 up to some very large value $\tt_0$ and then to continue in the positive imaginary direction (fig. \ref{usefulone}).  This procedure,
which gives the right answer in field theory, can be generalized to string theory if we interpret it correctly.   We will give just an outline of how
this goes, referring to the above-cited papers for more detail.

Let $\Sigma_0$ be an open-string worldsheet, that is, a  Riemann surface with boundary, and let $\Gamma$ be the moduli 
space of conformal structures on $\Sigma$.  $\Sigma_0$ has a closed oriented double cover $\Sigma$.  Let $\M$ be the moduli 
space of complex structures on $\Sigma$.  

Then $\Gamma$ is a middle-dimensional real cycle in $\M$.  More specifically, $\Gamma$ is a component of the fixed point
set of an antiholomorphic involution of $\M$.  (Such matters are discussed in section 7 of \cite{Wittentwo}.)  This implies
in particular that we can regard $\M$ as a complexification of $\Gamma$.  Likewise any covering space $\mathcal T$ of $\M$ is a complexification
of $\Gamma$.  Saying that $\M$ or $\T$ is a complexification of $\Gamma$ means that, locally, a real analytic function on $\Gamma$ can
be analytically continued to a holomorphic function on $\M$ or $\T$.

In particular, near an open-string degeneration, one can define the real, positive function $q$ on $\Gamma$; likewise
we can define $\tt=-\log q$.  To be more precise, these functions can be defined when $q$ is sufficiently small or when $\tt$ is
sufficiently large.  Moreover, $\tt$ can be defined as  a real-analytic function on $\Gamma$, so  it can be analytically continued to a holomorphic
function on $\M$.   This function is well-defined for sufficiently large $\mathrm{Re}\,\tt$.  As is explained in \cite{OtherWitten},
one must replace $\M$ by a cover thereof to make $\mathrm{Im}\,\tt$ single-valued.  Once this is done, the integration contour
in fig. \ref{usefulone} makes sense in string theory.

Let us explain concretely what this means for the special case that $\Sigma_0$ is an annulus without punctures.  
The moduli space of conformal
structures on $\Sigma_0$ is a half-line parametrized by a positive parameter $\tt$ (one can think of $\Sigma_0$ as an annulus
of width $\pi$ and circumference $\tt$).   Thus in this special case, $\tt$ is naturally-defined for all positive values, not just when it is large.  
The oriented double cover of $\Sigma_0$ is a genus 1 Riemann surface $\Sigma$.  The Teichmuller space $\mathcal T$
of $\Sigma$ (which is a cover
of its moduli space $\M$) is a copy of the upper half-plane, parametrized by a complex variable $\tau$.  If $\Sigma$ is the oriented
double cover of an annulus $\Sigma_0$, then $\tau$ is imaginary and the relation between $\tau$ and $\tt$ is $\tau=i\tt$.  Writing
this relation in the form $\tt=-i\tau$, we see that $-i\tau$ is a holomorphic function on $\mathcal T$ that coincides with $\tt$
on $\Gamma$ and thus represents the analytic continuation of $\tt$ from a function on $\Gamma$ to a holomorphic function on $\mathcal T$.

At this point, we can explain what the Feynman $i\epsilon$ means for the special case of an annulus.  Instead of integrating over
real $\tt$, we integrate $\tt$ over the contour that was sketched in fig. \ref{usefulone}, or equivalently, we integrate $\tau=i\tt$ over the contour
in the upper half-plane that corresponds to this.\footnote{ To show that what we need to integrate is holomorphic in $\tt$,
observe that the partition function on the annulus is $\Tr\,\exp(-\tt H)$, where $H$ is the
Hamiltonian, and this is manifestly holomorphic in $\tt$ for ${\mathrm{Re}}\,\tt>0$.  In general, an open-string worldsheet 
that is close to an open-string degeneration is built by gluing two surfaces $\Sigma_{\ell}$ and $\Sigma_{r}$ via a long strip of
length $\tt$, as in fig.   \ref{longsurface}(a).  The path integral on such a worldsheet is
  the matrix element of $e^{-\tt H}$ between initial and final states
that are determined by the path integrals on $\Sigma_{\ell}$ and $\Sigma_{r}$.  Keeping $\mathrm{Re}\,\tt$ large (where this
description  is meaningful)
and giving $\tt$ an imaginary part, such a matrix element is holomorphic in $\tt$.}     We include the usual convergence factor $\exp(-\epsilon\,\mathrm{Im}\,\tt)$.
  Of course, we do not need to integrate precisely over the stated contour.  Any contour homologous to this one and with the same behavior at infinity will do.

The generalization for other topologies proceeds in the same way, treating each open-string degeneration as we have
just described.   One defines the scattering amplitudes by integrating over an integration cycle $\varGamma$ that coincides with the
naive one $\Gamma$ except near a degeneration, where $\tt$ is integrated over a contour like that of fig. \ref{usefulone}.
One can find more detail in
\cite{OtherWitten}, but further details are not really
essential for the remainder of the present paper.   However, one point is really worth spelling out. 

\subsubsection{An Interesting Analogy}\label{intan}

 Naively open-string
scattering amplitudes are computed by integrating over the moduli space $\Gamma$ of conformal structures on the open-string worldsheet.
However, as we have seen, the Feynman $i\epsilon$ means that in reality, we must integrate (with a convergence factor)
over a more general cycle $\varGamma$ in a complexification
of $\Gamma$.  The appropriate
complexification of $\Gamma$ is a cover $\mathcal T$ of the complex moduli space $\M$. 

In the supersymmetric case,  given an open-string worldsheet $\Sigma_0$,
 it seems that, for technical reasons involving supermoduli, there is no natural definition of a moduli space $\Gamma$ of superconformal structures on $\Sigma_0$.  
(See section 5 of \cite{Wittenone} and section 7 of \cite{Wittentwo}.)    Because of this fact, the naive idea that open-superstring scattering amplitudes are computed
by integrating over a moduli space $\Gamma$ of open-superstring worldsheets in not really correct.  Instead, one must integrate
over a cycle $\varGamma$ in a complex supermanifold $\MM$ that parametrizes superconformal structures on a closed super
Riemann surface $\Sigma$ that corresponds to the oriented double cover of $\Sigma_0$.   

Thus, whether we consider the Feynman $i\epsilon$ or the details of supermoduli, the conclusion is similar.  Open-string scattering
amplitudes must be defined by integrating not over the naive moduli space $\Gamma$, but over a more general cycle $\varGamma$
in a suitable complexification of $\Gamma$, either because the naive $\Gamma$ does not exist (open superstrings) or because
it exists but does not incorporate the Feynman $i\epsilon$ 
(open bosonic strings).   For open superstrings, we have both problems and we must
define $\varGamma$ in a suitable cover of $\MM$.

In each case, there is no natural choice of $\varGamma$.  Any homologous cycle with the same behavior at infinity is equally good.

The two issues of the Feynman $i\epsilon$ and the Deligne-Mumford compactification
 are actually complementary in the following sense. At a generic degeneration
not associated to mass renormalization or massless tadpoles  (see section \ref{brstanom}), the Feynman $i\epsilon$ is
important and we integrate over a contour such as that of fig. \ref{usefulone}.  In particular, we do not integrate up to $\mathrm{Re}\,\tt=
\infty$, where the compactification occurs, so we do not see the details of the Deligne-Mumford compactification.\footnote{This is
the reason for something that was explained in a slightly different way
in sections \ref{treem} and \ref{intodd}: tree-level scattering amplitudes can be computed
correctly with a procedure that treats correctly the interior of moduli space, but does not treat the compactification correctly.
At tree level, one does not encounter tadpoles or mass renormalization.}  (We can still use the Deligne-Mumford compactification
to analyze the residue of the pole in the propagator, as discussed later.)
On the other hand, at special degenerations (associated to mass renormalization or massless tadpoles)
 at which $p^2+m^2$ is identically 0, the Feynman propagator $1/(p^2+m^2-i\epsilon)=i/\epsilon$ has a pole
at $\epsilon=0 $ and the Feynman $i\epsilon$ is not useful.  That case, as we will discuss starting in section \ref{brstanom}, calls for a different approach that
does require a knowledge of the compactification.

\subsubsection{Closed Strings}\label{closedan}

All this persists if we consider the Feynman $i\epsilon$ for closed strings.  Let $\Sigma$ be the worldsheet of a closed
bosonic string, and $\M$ the moduli space of conformal structures on $\Sigma$.  To exploit the fact that the functions and measures
usually encountered in string perturbation theory are real-analytic, we can view $\M$ as the diagonal in a product $\M_L\times \M_R$,
where $\M_R$ and $\M_L$ parametrize respectively holomorphic and antiholomorphic structures on $\Sigma$.   (In the bosonic case,
$\M_L$ and $\M_R$ are isomorphic spaces with opposite complex structures; each is naturally isomorphic to $\M$.)   A real-analytic
function on $\M$ can be analytically continued to a holomorphic function on $\M_L\times \M_R$, or on a covering space thereof.

A closed-string degeneration is described by vanishing of a complex parameter $q$; the Euclidean proper time variable is for closed
strings
$\tt=-\log |q|$.  (This coincides with the integration variable $s$ that appears in eqn.  (\ref{merom}) above.)   As $\tt$ is a real-analytic function
on $\M$, it can be continued locally to a holomorphic function on $\M_L\times \M_R$.   One wants to replace $\M_L\times \M_R$ by
a cover that is large enough so that ${\mathrm{Im}}\,\tt$ is single-valued at every degeneration, but small enough that one can find
in it an integration cycle that is topologically the same as the diagonal $\M\subset \M_L\times \M_R$.  For this, we observe that the
universal cover of $\M$ is a Teichmuller  space $\T$, with $\M=\T/F$ where $F$ is a discrete group (the
mapping class group).   The universal cover of $\M_L\times \M_R$ is therefore $\T_L\times \T_R$, the product of two
Teichmuller spaces.  A suitable cover of $\M_L\times \T_R$ is $(\T_L\times \T_R)/F$, with the diagonal
action of $F$ on the two factors.  This contains $\M$, embedded as the diagonal, but is ``large'' enough so that $\mathrm{Im}\,\tt$
is single-valued at each degeneration.  Then the contour of fig. \ref{usefulone} makes sense in this situation: 
it is part of the definition of an integration cycle $\varGamma\subset (\M_L\times \M_R)/F$ that coincides with the diagonal 
$\M\subset (\M_L\times \M_R)/F$
except near a degeneration.
To incorporate for bosonic strings the Feynman $i\epsilon$, we replace $\M$ with $\varGamma$ and include  in the integral a convergence
factor $\exp(-\epsilon\,{\mathrm{Im}}\,\tt)$.

Everything is the same for closed superstrings, except that in this case there is no natural moduli space of closed superstring
worldsheets to begin with.  Even without the Feynman $i\epsilon$,
all that one can naturally define in any case is an integration cycle $\varGamma\subset \M_L\times \M_R$
(where $\M_R$ and $\M_L$ parametrize holomorphic and antiholomorphic structures on the worldsheet).   This framework,
which is forced on us by properties of supermoduli, is in any event what we need to incorporate the Feynman $i\epsilon$.   For this,
we have to replace $\M_L\times \M_R$ by its cover $(\T_L\times \T_R)/F$
 and define $\varGamma$ in this cover.

\subsection{The On-Shell And Infrared Behavior Of String Theory}\label{ondang}
\subsubsection{Overview}\label{lover}

\begin{figure}
 \begin{center}
   \includegraphics[width=5in]{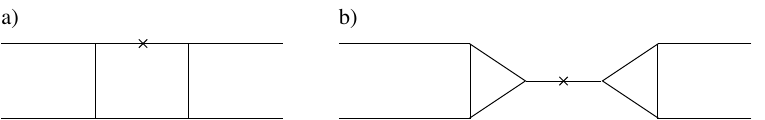}
 \end{center}
\caption{\small Cutting an internal line (marked $\times$) in a Feynman diagram will reduce the number of loops by 1, as
in (a), or disconnect the diagram, as in (b).}
 \label{cutxx}
\end{figure} 
At this stage, it is almost clear that string theory amplitudes will match the on-shell and infrared singularities that one would
expect in a field theory with the same particles and interactions.  In field theory,  the  singularities  come from the poles
of propagators.  In string theory, the singularities can only come from integration over the length of a long tube or strip,
and we have seen that the singularities generated in this way precisely match the poles in a corresponding field theory propagator.

To pursue this further, we need a certain factorization property of the worldsheet path integral of string theory.  In field theory,
one can always imagine ``cutting'' an arbitrary line in a Feynman diagram to make a diagram with one less propagator
and two more external lines.  Cutting a line either causes a Feynman diagram to become disconnected or else reduces the 
number of loops by 1  (fig. \ref{cutxx}) .

The cutting procedure is most
useful when the line that is cut is on-shell or almost on-shell.  Of course, one can simultaneously
cut several lines, taking the corresponding lines to be on-shell.    The main purpose of cutting is to understand the singularity
that a Feynman diagram develops when one or more internal lines go on-shell.    A fundamental fact about  Feynman
diagrams is the following ``factorization'' property: evaluating  a diagram with the
momentum flowing through a given line held fixed is equivalent to cutting that line, treating its ends as external lines,
 and evaluating what remains of the diagram.  

\begin{figure}
 \begin{center}
   \includegraphics[width=4.5in]{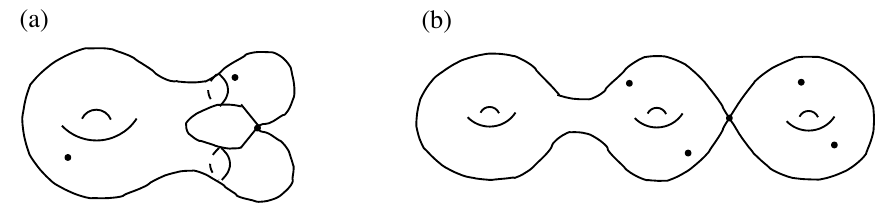}
 \end{center}
\caption{\small (a) A nonseparating degeneration reduces the genus of $\Sigma$ by 1 while adding two punctures -- one
on each side.  So this divisor in $\h\M_{\sg,\sn}$ is a copy of $\h\M_{\sg-1,\sn+2}$.  (b) A separating degeneration splits
$\Sigma$ in two components, whose genera $\g_1$ and $\g_2$ sum to $\g$.  The punctures are divided between the two
components and an extra puncture is added on each branch. So this divisor in $\h\M_{\sg,\sn}$ is a copy of $\h\M_{\sg_1,\sn_1+1}\times
\h\M_{\sg_2,\sn_1+1}$, where $\g_1+\g_2=\g$, $\n_1+\n_2=\n$. }
 \label{longofig}
\end{figure}

We need an analogous fact in
string theory.  It can be understood from the Deligne-Mumford compactification of moduli space
(for background to much of what follows, see section 6 of \cite{Wittentwo}).
For illustration, until section \ref{supan} we consider bosonic closed strings only.
We write $\h\M_{\sg,\sn}$ for the Deligne-Mumford compactification of the
moduli space $\M_{\sg,\sn}$ of Riemann surfaces $\Sigma$ of genus 
$\g$ and $\n$ punctures.  $\h\M_{\sg,\sn}$ has a number of
``divisors at infinity'' corresponding to the possible ways that $\Sigma$ 
can degenerate.  A nonseparating degeneration
reduces the genus of $\Sigma$ by 1 while adding two punctures 
(fig. \ref{longofig}(a)).  The corresponding divisor $\fD_{\mathrm{nonsep}}\subset
\h\M_{\sg,\sn}$ is  simply the moduli space of Riemann surfaces with 
the appropriate genus and number of punctures:
\begin{equation}\label{zitto}\fD_{\mathrm{nonsep}}\cong\h\M_{\sg-1,\sn+2}. \end{equation}
A separating degeneration divides $\Sigma$ into two components while dividing 
the punctures in some way between the two
sides (fig. \ref{longofig}(b)); in addition, the singularity counts as an extra puncture on 
both sides.  So the corresponding divisor $\fD_{\mathrm{sep}}
\subset \h\M_{\sg,\sn}$
factorizes:
\begin{equation}\label{bitto}\fD_{\mathrm{sep}}\cong\h\M_{\sg_1,\sn_1+1}\times 
\h\M_{\sg_2,n_2+1}, \end{equation}
where
\begin{equation}\label{itto}\g_1+\g_2=\g,~~\n_1+\n_2=\n.  \end{equation}
Moreover, as we will now explain, the measure on moduli space that comes 
from the worldsheet path integral factorizes at these degenerations
in a way that is very similar to the behavior of a Feynman diagram when a line is cut. 

\subsubsection{The Separating Case}\label{sepcase}

\begin{figure}
 \begin{center}
   \includegraphics[width=4.5in]{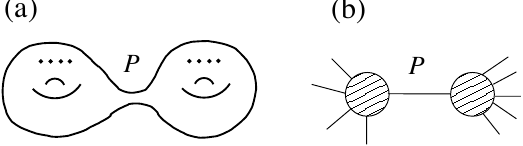}
 \end{center}
\caption{\small (a) A separating degeneration of a Riemann surface with momentum $P$ flowing between the left
and right.  (b) The field theory analog, in which a total momentum $P$ enters on the left and exits on the right.}
 \label{longfigo}
\end{figure} 

In either field theory or string theory, separating degenerations are simpler to analyze.
The total momentum $P$ flowing through the separating line from the left to the right of the graph or worldsheet
of fig. \ref{longfigo}
 is equal to the sum of all momenta flowing in from the left
or out to the right.  The amplitude has a pole $1/(P^2+m^2)$  whenever
a particle $\sigma$ that propagates between the two parts of the diagram 
is  on-shell at momentum $P$.  In field theory, the residue of the pole is
simply obtained by evaluating the rest of the diagram, with the separating line cut and $\sigma$ 
and its conjugate (its antiparticle) attached at its two ends.  We need a similar result in string theory.

The key assertion is that the worldsheet path integral of string theory behaves in the way
that is suggested by the description (\ref{bitto}) of the compactification divisor.  Let us consider a Riemann surface $\Sigma$
that is degenerating to a pair of surfaces $\Sigma_\ell$, $\Sigma_r$.    We pick local coordinates $x$ and $y$ on $\Sigma_\ell$
and $\Sigma_r$.  Picking points $x=a$ in $\Sigma_\ell$ and $y=b$ in $\Sigma_r$, we glue the two surfaces together by
\begin{equation}\label{yrof}(x-a)(y-b)=q. \end{equation}
Thus, at $q=0$, the point $x=a$ in $\Sigma_\ell$ is glued to the point $y=b$ in $\Sigma_r$.  For $q\not=0$, the two branches
join together smoothly.

Near the degeneration, the moduli of $\Sigma$ consist of the following: the gluing parameter $q$; the positions $a$ and $b$
of the extra punctures in $\Sigma_\ell$ and $\Sigma_r$ at which the gluing is made; and the other moduli of $\Sigma_\ell$ and
$\Sigma_r$.  The worldsheet path integral can be evaluated as follows.    
The integral over $q$ gives the closed bosonic string propagator
\begin{equation}\label{zeromb}\t b_0b_0\int_{|q|\leq 1}\frac{\d^2q}{|q|^2}\,q^{L_0}\bar q^{\t L_0}.\end{equation} 
An important detail here is that the precise upper limit on the $q$ integral is not important.  The pole comes entirely
from an arbitrarily small neighborhood of $q=0$.    Indeed, the pole arose in section \ref{bosclosed} as the contribution of
a string state that flows through the narrow neck between $\Sigma_\ell$ and $\Sigma_r$.  Let $\sigma$ be one of the string
states that contributes to the pole; let $\V^\sigma$ be the corresponding vertex operator,
and let $ \V_\sigma$ be the vertex operator for the antiparticle of $\sigma$. 
 By the state-operator
correspondence of conformal field theory, propagation of $\sigma$ through the neck
 can be represented in an effective path integral on $\Sigma_r$
by the insertion of  $\V^\sigma$ at the point $b$; similarly it is represented in an effective path integral on $\SIgma_\ell$
by insertion of $\V_\sigma$.
(For more on this, see \cite{OldPolch} and also section \ref{stad} below.)  The residue of the pole due to the string state $\sigma$
is computed by integration over $\fD_{\mathrm{sep}}=\h\M_{\sg_1,\sn_1+1}\times \h\M_{\sg_2,\sn_2+1}$ (that is, over $a,$ $b$,
and all the other moduli of $\SIgma_\ell $ and $\SIgma_r$) with the insertions  of $V_\sigma$ and $\V^\sigma$ just described, along with
any other vertex operators that may be present.
This gives the same sort of description of the residue of a pole that we had in field theory.

We have implicitly assumed that the momentum $P$ flowing between $\Sigma_\ell$ and $\Sigma_r$
is generically off-shell.  This is true
as long as there are two or more external states attached to both $\Sigma_\ell$ and $\Sigma_r$ (for example, in fig.
\ref{longfigo}(a), there are four external states on each side).   In this case, 
the $1/(P^2+m^2)$ singularity of the propagator  gives
a pole as a function of the external momenta; these poles are important in the physical
interpretation of scattering amplitudes.  If the number of external particles on
 $\Sigma_\ell$ and/or $\SIgma_r$ is either 1 or 0,  then $P$ is automatically on-shell, independent of the external momenta,
and the propagator has a $1/0$ singularity.
We will return to these cases in section \ref{massren}.

\subsubsection{The Nonseparating Case}\label{nonsepcase}

\begin{figure}
 \begin{center}
   \includegraphics[width=2.5in]{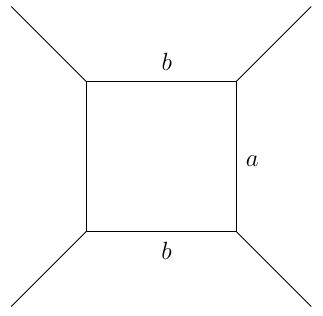}
 \end{center}
\caption{\small In this contribution to an $S$-matrix element, when
the line labeled $a$ goes to zero momentum, the lines labeled $b$ are forced to be on-shell.  For
spacetime dimension $d\leq 4$,
this leads to infrared divergences in gravity and in gauge theory with massless gauge fields.
These divergences follow from very general properties of phase space and momentum conservation,
along with the low energy limits of the gauge and gravitational couplings.  So they are common to field theory and string theory. 
}
 \label{lofig}
\end{figure} 
In field theory, when a nonseparating line goes on-shell, one has to integrate over the momentum
 flowing through the relevant line.  In $D$-dimensional Minkowski spacetime with $d>2$ (possibly multiplied by a compact
 manifold of some sort), there is no infrared singularity associated to the momentum integral for a single generic propagator, even
 for a massless field,  since the integral
\begin{equation}\label{boggo}\int \frac{\d^dp}{(2\pi)^d}\frac{1}{p^2}\end{equation}
is convergent at $p=0$.  The divergence at $d\leq 2$ is common to field theory and string theory, since it only depends
on the pole of the propagator and on $d$-dimensional phase space.  The usual on-shell infrared divergences for theories
with massless fields in $d\leq 4$ dimensions arise because in some cases (fig. \ref{lofig}) taking the momentum of one line to zero
forces adjacent lines to be on-shell, leading to an infrared behavior that is less convergent than that of (\ref{boggo}). Again,
this depends only on momentum conservation, the poles of propagators, and $d$-dimensional phase space, as well as the
low energy limits of couplings, so it does
not distinguish field theory and string theory in any way.\footnote{In four-dimensional field theories with massless particles,
infrared divergences in the perturbative $S$-matrix are often dealt with via dimensional regularization.  One works in $4-\varepsilon$
dimensions and one takes the limit $\varepsilon\to 0$ only after imposing a lower bound on the energy of an observable
soft particle.  One presumably could do something similar in string theory, by compactifying from $\R^4$ to $\R^{4-\varepsilon}
\times T^\varepsilon$, where $T^\varepsilon$ is a torus of dimension $\varepsilon$.} 

In field theory, the  most interesting singularities that come from cutting of nonseparating lines are the singularities associated
to unitarity.  Nothing happens in a generic Feynman diagram when a single generic nonseparating line goes on-shell,
but if enough nonseparating lines go on-shell that their removal would ``cut'' the diagram in two disjoint pieces, then
one gets singularities associated to unitarity. These singularities come ultimately from the poles of propagators and the way
the Feynman rules factorize when a line is cut.   The contributions of these singularities depend crucially on the Feynman $i\epsilon$.

In string theory, one has the same poles and -- via  (\ref{zitto}) and reasoning
that we have just sketched above -- the same sort of factorization of the worldsheet path integral. Moreover,
the Feynman $i\epsilon$ can be incorporated along lines explained in section \ref{iepsilon}. With this in place, it is reasonable 
to expect that perturbative string theory is unitary.    A proof of unitarity that proceeds roughly along these
lines, by adapting to string theory what is known about field theory with the Feynman propagator, has been given recently
\cite{Sen4}.  

An older approach to unitarity in string theory was based on light cone gauge, in which unitarity is manifest.   Light cone
string diagrams give a triangulation  of the moduli space of ordinary Riemann surfaces
 \cite{GidWo}, and this 
 has been used in bosonic string theory to establish the equivalence of light cone string 
 perturbation theory to the covariant description
 \cite{Giddings}.  Analogous arguments have also been developed in superstring theory
 \cite{Mandelstam,ADP}.  
Finally, we mention that a rare example in which analytic properties of a string theory loop amplitude have been
analyzed explicitly with manifest unitarity can be found in \cite{DPhtwo}.

\subsubsection{The Leading Singularity}\label{leadcase}

Much easier than describing unitarity, or giving a full description of all singularities of scattering amplitudes,
 is to describe the analog in string theory of what in field theory is called the ``leading
singularity'' of an amplitude.  (For a modern explanation and application of this notion, see \cite{bcf}.)

\begin{figure}
 \begin{center}
   \includegraphics[width=4.5in]{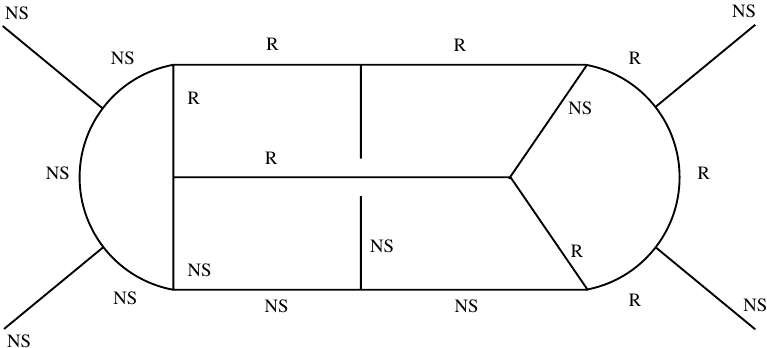}
 \end{center}
\caption{\small A trivalent graph in which the lines represent NS or R states and the
vertices represent genus 0 three-point functions of type $\NS^3$
or $\NS\cdot \Ra^2$.  Leading singularities of string theory amplitudes are associated to such trivalent graphs.  The example
considered here is a five-loop amplitude with four external NS lines.  Internal lines and vertices are of both types.}
 \label{longfig}
\end{figure}  

Any string worldsheet $\Sigma$ with $\g>0$ or with $\g=0$ and more than three punctures has a strictly positive number
of bosonic moduli and can undergo a separating
or nonseparating degeneration.  In the nonseparating case, this reduces the genus of $\Sigma$; in the separating case,
$\Sigma$ is decomposed into two components each of which has a smaller genus and/or fewer punctures than $\Sigma$.
We can continue this process, degenerating $\Sigma$ again or degenerating its components, until
finally $\Sigma$ is built out of a collection of three-punctured spheres, glued together pairwise along their punctures.
This gives the string theory analog of a Feynman diagram in which all possible
internal lines have been placed on-shell.   The simultaneous
poles in all possible channels give the leading singularity.

From a completely degenerated Riemann surface, we can build a trivalent graph in which a three-punctured sphere is represented
by a trivalent vertex, while the gluing of two punctures is represented by a line between two vertices.  
Leading singularities come from these trivalent graphs. Though
we have not yet described degenerations of super Riemann surfaces,
the generalization of these trivalent graphs to   superstrings is so straightforward that we will describe it without further ado.  In -- for example -- the heterotic
string, there are two types of string states -- NS and R -- and there are two types of three-punctured spheres, which we will call
$\NS^3$ (three punctures of NS type) and $\NS\cdot \Ra^2$ (one NS puncture and two R punctures).  
Leading singularities
of heterotic string amplitudes are associated to trivalent graphs labeled as in fig. \ref{longfig}, with two types of internal
line and two types of vertex.  The vertices represent the $\NS^3$ genus 0 three-point function described in eqn. (\ref{freddy}), and
the $\NS\cdot \Ra^2$ genus 0 three-point function described in eqn. (\ref{onz}).   These trivalent
graphs have an obvious analogy with Feynman diagrams of field theory.  This is part of the match between
the on-shell singularities of field theory and string theory.

\subsubsection{Factorization Of Superstring Amplitudes}\label{supan}

In our explanation of how to compare the singularities of string theory and field theory amplitudes, there were several key
ingredients, which we may roughly summarize as follows: 
\begin{enumerate}
\item{A singularity develops when a certain modulus $q$ vanishes and $\Sigma$ degenerates.
Integration over $q$ produces poles that are analogous
to the poles of a Feynman propagator.}
\item{The locus $q=0$ is a divisor $\fD$ in the moduli space; it can be factored as in 
(\ref{zitto}) or (\ref{bitto}).}
\item{The integral over $\fD$ that gives the residue of a pole due to a given string state $\sigma$ is itself a string
theory scattering amplitude, now with extra insertions of the vertex operator $\V_\sigma$ and its conjugate on the two
sides.}
\end{enumerate}

All of these ingredients have analogs in superstring theory, with just a few added wrinkles.
The bosonic gluing formula $xy=q$ has two different superanalogs, one for the NS sector and one for the R sector (these
were originally constructed by Deligne \cite{Deligne,DeligneLectures}; for more detail on the following, see section 6 of \cite{Wittentwo}).
In the NS sector, we glue together two copies of $\C^{1|1}$, one with 
superconformal coordinates $x|\theta$ and one with superconformal coordinates $y|\psi$, via
\begin{align}\label{wondo}  xy & = -\varepsilon^2 \cr
                                           y\theta & = \varepsilon\psi \cr
                                            x\psi & =-\varepsilon\theta\cr
                                             \theta\psi&=0. \end{align}
It is convenient to define
\begin{equation}\label{dolf}q_\NS=-\varepsilon^2. \end{equation}
$q_\NS$ is the closest analog of the bosonic gluing parameter $q$.   For given $q_\NS$, $\varepsilon$ is determined only up to
sign.  The sum over the possible signs of $\varepsilon$ gives the GSO projection in the NS sector, as explained in section 6.2.3 of
\cite{Wittentwo}.   

To describe a Ramond degeneration, we again glue    two copies of $\C^{1|1}$ with local coordinates $x|\theta$ and $y|\psi$.  
But now     we endow each copy with a superconformal structure defined                               
 by the odd vector fields
\begin{equation}\label{omy}D^*_\theta =\frac{\partial}{\partial\theta}+\theta x\frac{\partial}{\partial x},~~D^*_\psi =\frac{\partial}{\partial\psi}+\psi y\frac{\partial}{\partial y}.\end{equation}
This is chosen to describe Ramond punctures at $x=0$ and $y=0$, respectively.
The gluing formulas are now simply
\begin{align}\label{tormo} xy& = q_\Ra \cr
                        \theta& =\pm \sqrt{-1}\psi.\end{align}
The GSO projection in the Ramond sector comes from the sum over the sign that is explicitly written in eqn. (\ref{tormo}).
The parameter $q_\Ra$ plays the role of the bosonic gluing parameter $q$ and the NS gluing parameter $q_\NS$.

Now let us move on to the second item in the above list.  In the NS sector, the factorizations (\ref{zitto}) and (\ref{bitto}) have
immediate analogs.  For example, let $\h{\MM}_{\sg,\sn_\NS,\sn_\Ra}$ be the Deligne-Mumford compactification of $\MM_{\sg,
\sn_\NS,\sn_\Ra}$, the moduli space of super Riemann surfaces of  genus $\g$ with  $\n_\NS$ NS punctures and $\n_\Ra$ Ramond
punctures.  The nonseparating divisor $\fD_{\mathrm{nonsep}}$ has the same sort of description as for bosonic
Riemann surfaces.  At a nonseparating degeneration, the genus is reduced by 1 and $\n_\NS$ increases by 2; an extra
NS puncture appears on each branch:
\begin{equation}\label{midon}\fD_{\mathrm{nonsep}}\cong\h{\MM}_{\sg,\sn_\NS+2,\sn_\Ra}. \end{equation}
This is the obvious analog of eqn. (\ref{zitto}). Similarly the obvious analog of eqn (\ref{bitto})  holds for separating
degenerations.  Any separating divisor has the form
\begin{equation}\label{idon}\fD_{\mathrm{sep}}\cong \h{\MM}_{\sg_1,\sn_{\NS,1}+1,\sn_{\Ra,1}}\times
 \h{\MM}_{\sg_2,\sn_{\NS,2}+1,\sn_{\Ra,2}},
\end{equation}
with $\g_1+\g_2=\g$, $\n_{\NS,1}+\n_{\NS,2}=\n_\NS$, and $\n_{\Ra,1}+\n_{\Ra,2}=\n_\Ra$.

Given this, the analog of the third item is clear.  
We can describe the residue of the pole due to an on-shell NS state in the same way that we did for bosonic strings in 
sections \ref{sepcase} and \ref{nonsepcase}.    Integration over $\q_\NS$ gives poles due to on-shell string states; 
if a string state $\sigma$ contributes such a pole, then according to the
state-operator correspondence of superconformal field theory, vertex operators $\V_\sigma$ and 
$\V^\sigma$ appear on the two
branches. 
Integration over the remaining moduli gives a description of the residue of the pole due to $\sigma$ in terms of
a scattering amplitude with the extra insertions of $\V_\sigma$ and $\V^\sigma$.

By contrast, the most obvious analog of eqns. (\ref{midon}) and (\ref{idon}) does {\it not} hold for Ramond degenerations.  
The reason is that
Ramond punctures are really divisors.  
For brevity, we will describe the situation for separating degenerations; the same
idea holds for nonseparating ones.  Suppose that we are given two super 
Riemann surfaces $\Sigma_\ell$ and $\Sigma_r$
each with a distinguished Ramond divisor.  Suppose further that near each of these divisors we are given local
coordinates $x|\theta$ or $y|\psi$, with the superconformal structure being as in (\ref{omy}).  The distinguished
divisors are given respectively by $x=0$ and by $y=0$, and are parametrized by $\theta$ and by $\psi$.
 We want to glue together
these two divisors to make a super Riemann surface $\Sigma$.  We can do this by gluing $\theta=\pm \sqrt{-1}\psi$
as in (\ref{tormo}).  But we can also introduce an odd parameter $\eta$ and 
make the gluing $\theta=\pm\sqrt{-1}(\psi+\eta)$.
Here $\eta$ is an odd modulus that we should for our present purposes associate 
to the gluing and not to $\Sigma_\ell$ or $\Sigma_r$.
We call $\eta$ the fermionic gluing parameter.

One can generalize (\ref{midon}) or (\ref{idon}) for a Ramond degeneration if one takes proper acount of the fermionic
gluing parameter.  For example, instead of a separating Ramond divisor $\fD_{\mathrm{sep}}$
being a product  $\h{\MM}_{\sg_1,\sn_{\NS,1},\sn_{\Ra,1}+1}\times
 \h{\MM}_{\sg_2,\sn_{\NS,2},\sn_{\Ra,2}+1}$, it is a fiber bundle over such a product. 
 We denote the fibration as
  \begin{equation}\label{pirat}\bm\Pi: \fD_{\mathrm{sep}}\to \h{\MM}_{\sg_1,\sn_{\NS,1},\sn_{\Ra,1}+1}\times
 \h{\MM}_{\sg_2,\sn_{\NS,2},\sn_{\Ra,2}+1}.\end{equation}
  The fibers have dimension $0|1$
 and are parametrized by the fermionic gluing parameter.  There is a precisely analogous fibration for nonseparating 
 degenerations. 
 
Now let us consider the pole due to an on-shell
Ramond string state.  What is the residue of such a pole?
In the separating case, imitating what we have done for bosonic strings and for the NS sector of superstrings,
we would like to express this residue in terms of an integral over 
$\h{\MM}_{\sg_1,\sn_{\NS,1},\sn_{\Ra,1}+1}\times
 \h{\MM}_{\sg_2,\sn_{\NS,2},\sn_{\Ra,2}+1}$.  
 If we simply integrate over  $q_\Ra$, we will extract in the usual way a pole $1/L_0$, and we learn that its residue can be computed
 as an integral over $\fD_{\mathrm{sep}}$.  To further reduce to an integral over $\h{\MM}_{\sg_1,\sn_{\NS,1},\sn_{\Ra,1}+1}\times
 \h{\MM}_{\sg_2,\sn_{\NS,2},\sn_{\Ra,2}+1}$, we have to integrate over the fibers of the fibration $\bm\Pi$.
In other words, we have to integrate over the fermionic gluing parameter $\eta$.
 
 This parameter was introduced in another way in section \ref{ramondpropagator}.  As we explained there, integration over $\eta$
 gives a factor of $G_0$.  So the contribution of an on-shell Ramond state to a scattering amplitude is $G_0/L_0=1/G_0$
 times an integral over $\h{\MM}_{\sg_1,\sn_{\NS,1},\sn_{\Ra,1}+1}\times
 \h{\MM}_{\sg_2,\sn_{\NS,2},\sn_{\Ra,2}+1}$.   Just as in the case of an NS degeneration, the operator-state correspondence
 of conformal field theory tells us how to compute the residue of the pole associated to an on-shell Ramond sector
 string state $\sigma$: we integrate over $\h{\MM}_{\sg_1,\sn_{\NS,1},\sn_{\Ra,1}+1}\times
 \h{\MM}_{\sg_2,\sn_{\NS,2},\sn_{\Ra,2}+1}$ with insertions of the appropriate vertex operators $\V_\sigma$ and
 $\V^\sigma$ on the two sides. 
 
 To compare the
 description of the gluing parameter that we have given here with that of section \ref{ramondpropagator}, one may observe
 the following.  In the local model (\ref{tormo}), including the fermionic gluing parameter is equivalent to acting on $y|\psi$
 with a superconformal transformation $\exp(\eta G_0)$, where $G_0$ is the superconformal vector field
 \begin{equation}\label{primox}G_0=\frac{\partial}{\partial\psi}-\psi y\frac{\partial}{\partial y} \end{equation}
 that acts nontrivially on the Ramond  divisor at $y=0$. (In particular, at $y=0$, $\exp(\eta G_0)$ transforms $\psi$ to $\psi+\eta$.)
  Because of this geometrical fact, a shift in the fermionic gluing parameter
 acts on the string state flowing between the two branches of $\Sigma$ by $\exp(\eta G_0)$, which is the result
 that we obtained in another but related way in
 section \ref{ramondpropagator}.  
 
 \vskip .5cm
 \noindent{\bf Remark.} A further remark will be important when we study anomalies in section \ref{anomalies}.
 There is an important special case in which the fermionic gluing parameter does {\it not} represent a modulus.   
 If $\Sigma_r$ has a superconformal symmetry
 that acts on the divisor at $y=0$ by $\psi\to \psi+\eta$ (or if $\Sigma_\ell$ has a superconformal symmetry
 that acts as such a shift at $x=0$) then the fermionic gluing parameter is not a modulus of the super Riemann surface
 $\Sigma$ made by gluing of $\Sigma_\ell$ and $\Sigma_r$.  
 In practice, this happens
 only for the Ramond-Ramond sector of open and/or unoriented superstring theory, and only for very particular cases of $\Sigma_r$
 or $\Sigma_\ell$.  

\subsubsection{Picture Numbers And Picture-Changing Operators}\label{pnpco}

\begin{figure}
 \begin{center}
   \includegraphics[width=3in]{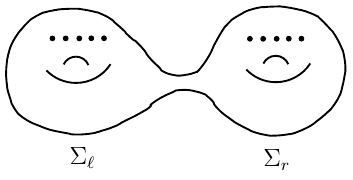}
 \end{center}
\caption{\small A super Riemann surface $\SIgma$ decomposes to a union of two components
$\Sigma_\ell$ and $\Sigma_r$, meeting at a point (or at a Ramond divisor).}
 \label{ready}
\end{figure} 
As we have just explained, the physical interpretation of superstring theory depends on the fact
that the residue of a pole at $L_0=0$ can be interpreted as a sum of contributions of on-shell states.
Thus (fig. \ref{ready}), if a super Riemann surface $\Sigma$ splits locally into a product of two
branches $\Sigma_\ell $ and $\Sigma_r$ meeting at a point (or at a Ramond divisor), 
then the contribution of an on-shell
 string state $\sigma$ to the pole can be evaluated by inserting vertex operators $\V_\sigma$
and $\V^\sigma$ on $\Sigma_\ell$ and $\SIgma_r$, respectively.  

This statement reflects the factorization of the divisor at infinity, which for a separating NS degeneration
 takes the form (\ref{idon}):
\begin{equation}\label{lidon}\fD_{\mathrm{sep}}\cong \h{\MM}_{\sg_1,\sn_{\NS,1}+1,\sn_{\Ra,1}}\times
 \h{\MM}_{\sg_2,\sn_{\NS,2}+1,\sn_{\Ra,2}},
\end{equation}
where $\SIgma_\ell$ has genus $\g_1$ with $\n_{\NS,1}+1$ NS punctures and $\n_{\Ra,1}$ Ramond
punctures, while the corresponding values for $\Sigma_2$ are $\g_2, \,\n_{\NS,2}+1,$ and $\n_{\Ra,2}$.
At a separating
Ramond degeneration, the divisor at infinity has an analogous but slightly more complicated structure described in
eqn.  (\ref{pirat}).  

There is one very simple but important observation to add to what we have already said.  Interpreting the
residue of the pole by using the factorization of the divisor at infinity requires that the vertex operators
$\V_\sigma$ and $\V^\sigma$ should have the canonical picture numbers, namely $-1$ in the NS sector and
$-1/2$ in the R sector.  Indeed, as we know from sections \ref{nsop} and \ref{ramins}, in 
a superconformal formalism, only vertex operators of  canonical
picture can be inserted at a puncture of NS or R type, assuming that those punctures are defined
in the conventional way. One can modify the
definition of a puncture by endowing it with more structure
in a way that increases the odd dimension of the moduli space (see
section 4.3 of \cite{Wittentwo}) and this makes it possible to compute in a superconformal
formalism using  vertex operators of picture number more negative than $-1$ or $-1/2$.
If we wish, though there is no apparent benefit in doing so,
we can do this for the punctures at which external vertex operators in fig. \ref{ready} are inserted.
   But we do not have any such freedom for the ``internal'' punctures
that appear when $\Sigma$ degenerates; they are the standard punctures described in sections \ref{nsop} and
\ref{ramins}, and in a superconformally-invariant formalism, the vertex operators inserted at those punctures
will have canonical picture number.

Another type of comment about picture number may be helpful.
In this paper, we never make any fundamental statements in terms of positions of picture-changing operators (PCO's);
we view the PCO's as a possible method of integration over odd moduli, but not a basic definition.
Still, it may be helpful to spell out in terms of PCO's what it means to treat correctly 
the compactification of the moduli space.
To the extent that the decomposition (\ref{idon}) of the divisor $\fD_\mathrm{sep}$ is important, one would like
the procedure for integrating over odd moduli using the PCO's to respect this factorization.  This
means that  when $\Sigma$ decomposes to two intersecting components $\Sigma_\ell$ and $\Sigma_r$, the location of the PCO's should have a limit as $q_\NS\to 0$,
and $\Sigma_\ell$ and $\Sigma_r$ should each have the right number of PCO's to describe its odd moduli.
Moreover, the PCO's should be placed to avoid spurious singularities on $\Sigma_\ell $  and on $\Sigma_r$.
(To compute correctly with PCO's, they
 must be used piecewise, avoiding spurious singularities, as described in section \ref{basodd}. One follows
 this same procedure along both $\Sigma_\ell$ and $\Sigma_r$.)
With notation as above, 
the odd dimension of the moduli space of $\SIgma_\ell$ is $(2\g_1-2)+(\n_{\NS,1}+1)+\n_{\Ra,1}/2$,
and that is the number of PCO's that should be placed on $\Sigma_\ell$.  
The corresponding number of PCO's on $\SIgma_r$
is then $(2\g_2-2)+(\n_{\NS,2}+1)+\n_{\Ra,2}/2$.   And if $\SIgma_\ell$ and $\Sigma_r$ 
undergo further separating degenerations,
the conditions just stated should be satisfied again on each of the resulting components.  

All this has a close analog for Ramond degenerations, with one correction: for a Ramond degeneration,
 there is one odd modulus associated to the 
gluing rather than to $\Sigma_\ell$
or $\Sigma_r$.  So as $\Sigma$ degenerates via formation of a long neck, one of the PCO's should be placed
in the neck.

It is a slightly tricky question to what extent one will actually get wrong answers in the picture-changing formalism
if one distributes the PCO's incorrectly when $\Sigma$ degenerates.  In genus 0, as long as the momentum
flowing between $\SIgma_\ell$ and $\Sigma_r$ is generically off-shell, one can distribute the PCO's between
$\Sigma_\ell$ and $\Sigma_r$ in an arbitrary fashion, treating the compactification of the moduli space
incorrectly, and still get the correct tree-level $S$-matrix.  This was explained in section \ref{treem}.  
It is not entirely clear if this remains true in higher genus; the spurious singularities may cause trouble.
(The recipe explained in section \ref{basodd} to compute piecewise and avoid spurious singularities certainly
cannot be extended over the compactification if one treats the compactification incorrectly.)
At any rate,  in higher genus, one will run into degenerations (associated to massless tadpoles and mass renormalization) for which
the momentum flowing between $\Sigma_\ell$ and $\Sigma_r$ is generically on-shell.  In our treatment of 
such questions, starting in section \ref{massren}, it will be clear that one should expect trouble if one does
not treat the compactification of moduli space correctly.  

\subsection{The States That Contribute To The Pole}\label{stad}

Now we want to explain that only physical states contribute to the singularities described in section \ref{ondang}.
(For a slightly different view with some further details, see \cite{OldPolch}.)

\subsubsection{Bosonic Open Strings}\label{bosop}

To illustrate the idea, we consider bosonic open strings.  The propagator is $b_0/L_0$, and the residue of the pole at
$L_0=0$ is
\begin{equation}\label{monkeys}\Res=b_0\Pi_0,\end{equation}
where $\Pi_0$ is the projector onto states with $L_0=0$.  To define $\Pi_0$, one expands any state $\V\in\H$
in eigenstates of $L_0$ and defines $\Pi_0\V$ as the component of $\V$ with $L_0=0$. This can be done in  a sector
of fixed spacetime momentum; in such a sector, the spectrum of $L_0$ is discrete.   The residue $\Res$
is BRST-invariant:
\begin{equation}\label{onkey}\{Q_B,\Res\}=\{Q_B,b_0\}\Pi_0=L_0\Pi_0=0.\end{equation}
Of course, $\Res$ has ghost number $N_\gh=-1$.

Let $\H$ be the space of all string states.  We can view $\Res$ as a linear map $\Res:\H\to \H$, or equivalently as
an element of $\H\otimes \H^*$, where $\H^*$ is the dual space to $\H$.    Since $\Res$ has ghost number $-1$, it maps
states of ghost number $n+1$ to states of ghost number $n$, so 
it is, more specifically, an element of
\begin{equation}\label{dolbo} \oplus_{n\in \Z}\H_n\otimes \H_{n+1}^*,\end{equation}
where $\H_n$ is the subspace of $\H$ with $N_\gh=n$, and $\H_n^*$ is the dual of $\H_n$.  

So far, we have viewed the propagator and its residue as maps from one string state to another.  But for what follows,
 it will be more convenient to view $\Res$ more symmetrically as an element of $\H\otimes \H$, representing
a pair of string states that are to be inserted at the two ends of the long strip.  For this, we simply use the duality $\H_n^*\cong \H_{3-n}$ of eqn. 
(\ref{orzot}) , via which we can identify 
$\Res$ with an element
\begin{equation}\label{olbo}\Res' \in\oplus_{n\in\Z} \H_n\otimes \H_{2-n},\end{equation}
or more simply as an element 
\begin{equation}\label{zunk}\Res'\in (\H\otimes\H)_2,\end{equation}
where $(\H\otimes\H)_2$ is the subspace of $\H\otimes \H$ consisting of pairs of states with total ghost number 2.

Since $\Res$ and the duality map are both  BRST-invariant,
it follows that $\Res'\in (\H\otimes \H)_2$ is BRST-invariant.  Hence, it helps to know the cohomology of the BRST operator $Q_B$
acting on $\H\otimes \H$.  This is simply the tensor product of two copies of the cohomology of $Q_B$ acting on $\H$.
The non-zero cohomology groups of $Q_B$ acting on $\H$ are as follows\footnote{Here and also in the corresponding discussion
below of open superstrings, we omit possible Chan-Paton factors.  If present, these must be included in an obvious way, taking
the tensor product of what is described momentarily with an algebra of matrices acting on the Chan-Paton factors.}:
\begin{enumerate}
\item{In $N_\gh=0$, there is the identity operator 1, with zero spacetime momentum.}
\item{In $N_\gh=1$, the cohomology can be identified with the space of states $\V=c\,\UU$, where $\UU$ is a dimension 1 conformal
primary of the matter system.}
\item{In $N_\gh=2$, the cohomology can be identified with the space of states $\V=c\partial c\,\UU$, with $\UU$ as before.}
\item{In $N_\gh=3$, there is the operator $c\partial c\partial^2c$, with zero spacetime momentum.}
\end{enumerate}

It follows immediately that the $N_\gh=2$ cohomology of $Q_B$ acting on $\H\otimes \H$ is, at any non-zero momentum,
the tensor product of two copies of the $N_\gh=1$ cohomology of $Q_B$ acting on $\H$.  It is convenient now to let $\UU_i$
be a basis of matter primaries of $L_0=1$, and $\UU^i$ a dual basis (so that the genus zero two-point functions in the matter
theory are $\langle \UU_i(0)\UU^j(1)\rangle=\delta^j_i$).  Then we have 
\begin{equation}\label{zorom}\Res'=\sum_i c\,\UU_i\otimes c\,\UU^i+Q_B\X,\end{equation}
for some $\X\in (\H\otimes\H)_1$, of the general form
\begin{equation}\label{omboc}\X=\sum_j \S_j\otimes \T_j,~~~\S_j,\T_j\in\H.\end{equation}

The contribution $\sum_i c\,\UU_i\otimes c\,\UU^i$ in (\ref{zorom}) is important, so we will explain it in detail.
Let us go back to the original definition of $\Res$ as the map $b_0\Pi_0$ from $\H_{n+1}$ to
$\H_n$.  As such, $\Res$ maps $c\partial c\,\UU_i$ to $c\,\UU_i$; indeed, $c\partial c\UU_i$ is invariant under $\Pi_0$, and $b_0$
removes the $\partial c$ factor.  On the other hand, the dual of $c\partial c\,\UU_i$
is $c\,\UU^i$.  Combining these statements, the cohomologically nontrivial part of $\Res$ is $\sum_i c\,\UU_i\otimes c\,\UU^i$, as claimed
in (\ref{zorom}).  

Eqn. (\ref{zorom}) implies that the residue of a scattering amplitude at $L_0=0$,
 with all other vertex
operators being BRST-invariant, can be computed from  a sum over physical states
where we insert $c\,\UU_i$ on one side of the degeneration, $c\,\UU^i$ at the other side, and sum over $i$.
The residue receives no contribution from the BRST-trivial term $Q_B\X$.   
That $Q_B\X$ decouples is possibly most obvious for a separating degeneration, at which the worldsheet path
integral factorizes as a product of path integrals on the two components of the worldsheet.  In this case, the path integral
with an insertion of $Q_B(\S_j\otimes \T_j)=Q_B\S_j\otimes \T_j+(-1)^{|\S_j|}\S_j\otimes Q_B\T_j$ vanishes, since a BRST-trivial
state $Q_B\S_j$ or $Q_B\T_j$ is inserted on one branch or the other.  To show the decoupling of $Q_B\X$ at a nonseparating
degeneration, we have to use in a slightly
more general way than we have done before
the fundamental identity (\ref{ool}) for the decoupling of BRST-trivial states:
\begin{equation}\label{omxo}\d F_\Omega+ F_{Q_B\Omega}=0. \end{equation}
 Given any BRST-invariant vertex operators $\V_1,\dots,\V_\sn$,
we set $\Omega=\V_1\V_2\cdots\V_\sn {\X}$ (where the $\V_i$ and the two factors of ${\X}$ are inserted at $\n+2$ distinct punctures) and learn that
\begin{equation}\label{pomxo}\d F_{\V_1\cdots \V_\sn {\X}}+F_{\V_1\cdots \V_\sn Q_B{\X}}=0. \end{equation}
The contribution of $Q_B{\X}$ to a genus $\g$ non-separating residue is therefore
\begin{equation}\label{lomxo}\int_{\M_{\ssg-1,\ssn+2}}F_{\V_1\cdots\V_\ssn Q_B{\X}}=-\int_{\M_{\ssg-1,\ssn+2}}\d F_{\V_1\cdots \V_\ssn {\X}}.\end{equation}   
Thus this contribution vanishes, modulo possible anomalous contributions at infinity in moduli space.  The contributions at infinity in arguments
such as this will be  analyzed in section \ref{massren} and do not affect our present discussion.

For future reference, we make a small aside. 
In studying gauge-invariance and BRST anomalies, it will be useful to replace the residue $\Res=b_0\Pi_0$ with
$\t\Res=\Pi_0$.  $\t\Res$ is BRST-invariant but of course not annihilated by $b_0$.  Duality maps $\t\Res$ to a $Q_B$-invariant element 
$\t \Res'\in(\H\otimes \H)_3$. At non-zero spacetime momentum, by arguments similar to those above,
\begin{equation}\label{yrfe}\t\Res'=\sum_i\left( c\partial c\, \UU_i\otimes c\,\UU^i+ c\,\UU_i\otimes c\partial c\,\UU^i\right)+Q_B\Y,\end{equation}
with $\Y\in (\H\otimes \H)_2.$ 
Note that $\t\Res'$ is invariant under exchanging the two factors of $\H\otimes \H$.  
At zero momentum, there is an additional term
\begin{equation}\label{zyrfex}\left(c\partial c\partial^2c\otimes 1+1\otimes c\partial c\partial^2c\right). \end{equation}

Returning to our main theme, we now have to discuss the role of conformal invariance in the above analysis.  
A careful reader may have noted a small sleight of hand in our explanation of the decoupling of $Q_B\X$.  We have based the present paper
on a conformally-invariant formalism in which the only vertex operators considered are conformal or superconformal vertex operators
(primary fields that do not depend on derivatives of the ghost fields).  
But as we briefly described in section \ref{morge}, 
it is also possible to develop a more general formalism in which conformal invariance is not assumed and vertex operators are more 
general $Q_B$-invariant
local
operators (annihilated by $b_0-\t b_0$  in the case of closed strings).  In such a formalism, the argument for decoupling
of $Q_B\X$ proceeds exactly as stated above.  However, in our conformally-invariant formalism, we cannot define a correlation function
with insertion of $Q_B\X$ unless $Q_B\X$ is a conformal vertex operator. 
More precisely, $Q_B\X\in \H\otimes \H$ must be annihilated by $L_n$ and $b_n$, 
$n\geq 0$, acting on either of the two factors of $\H$ in $\H\otimes \H$.  On the other hand, in a conformally-invariant formalism, the residue
of a pole at $L_0=0$ must be conformally invariant, so the BRST-trivial contribution to the residue must come from an operator
$Q_B\X\in \H\otimes \H$ that obeys these conditions.   So at least what we have to show
to vanish is well-defined.
Given that $Q_B\X$ is annihilated by $L_n$ and $b_n$, $n\geq 0$, in each factor,
the argument of appendix \ref{really}, applied to $\H\otimes \H$ rather than $\H$, 
shows that we can assume that $\X$ obeys the same conditions, and then the argument for decoupling of $Q_B\X$ proceeds as above.

\subsubsection{Open Superstrings}\label{zopen}

In open superstring theory, we let $\H_{n,m}$ be the space of string states of ghost number $n$ and picture number $m$.

In the NS sector, there is not much new to say.  As explained in section \ref{pnpco}, we set the picture number to the canonical value $m=-1$. 
The residue of a scattering amplitude at $L_0=0$ is still $\Res=b_0\Pi_0$, which we can understand
as an element of  $\oplus_n \H_{n,-1}\otimes \H^*_{n+1,-1}$.  Via the duality $\H^*_{n,-1}\cong \H_{1-n,-1}$ (eqn. (\ref{oior})), we can identify
$\Res$ with a $Q_B$-invariant element $\Res'\in (\H\otimes \H)_{0,-1\otimes-1}$, where the notation means that $\Res'\in \H\otimes \H$
has ghost number 0 and has picture number $-1$ in each factor.  

At this stage, we can now give a slightly different explanation of the fact that we should use vertex operators of picture number
$-1$ in each factor, relying on the form of the duality map rather than on a knowledge of the factorization of the divisor at infinity
in the moduli space.  If we take $\Res$ to act on states of  picture number $m$,
then after applying the duality map,
$\Res'$ will take values in $(\H\times \H)_{0,m\otimes (-2-m)}$, with picture numbers $m$ and $-2-m$
in the two factors.  There appears to be no moduli space suitable for  computing superstring scattering amplitudes
using vertex operators of positive or zero picture number; on the other hand, the only way to make both $m$ and $-2-m$ negative
is to set $m=-2-m=-1$.  So to understand the pole at $L_0=0$, we will have to use NS vertex operators with
picture number $-1$.  A similar argument can be applied later in the Ramond sector to show that the residue of a pole
should be computed with vertex operators of the canonical picture number $-1/2$.

The cohomology of $Q_B$ acting on $(\H\otimes \H)_{-1\otimes -1}$ 
(where we allow states of any ghost number
but set the picture number to $-1$ in each factor) is as follows:
\begin{enumerate}
\item{In $N_\gh=-1$, there is the operator $c\delta'(\gamma)$, 
with zero spacetime momentum. (In conventional language,
this operator is related to the identity
operator by picture-changing.)}
\item{In $N_\gh=0$, the cohomology can be identified with the 
space of states $\V=c\delta(\gamma)\UU$, where $\UU$ is a dimension 1/2 superconformal
primary of the matter system.}
\item{In $N_\gh=1$, the cohomology can be identified with the 
space of states $\V=c\partial c\delta(\gamma)\UU$, with $\UU$ as before.}
\item{In $N_\gh=2$, there is the operator 
$c\partial c\delta(\gamma)\partial\gamma$, with zero spacetime momentum.}
\end{enumerate}

Given this, the results of section \ref{bosop} have straightforward analogs. 
The analog of eqn. (\ref{zorom}), at non-zero spacetime momentum, is
\begin{equation}\label{melfo}\Res'=\sum_i c\delta(\gamma)\UU_i\otimes c\delta(\gamma)\UU^i+Q_B\X.\end{equation}
Eqn. (\ref{yrfe}) similarly has an obvious analog constructed by including factors of $\delta(\gamma)$, and eqn. (\ref{zyrfex})
has an analog with $c\delta'(\gamma)$ and $c\partial c\delta(\gamma)\partial \gamma$ replacing 1 and $c\partial c\partial^2c$.   

The Ramond sector involves some new elements.  The residue of the pole at $L_0=0$ is
\begin{equation}\label{moop}\Res=b_0\delta(\beta_0)G_0 \Pi_0. \end{equation}
This operator commutes with the ghost number, and increases the picture number by 1.  We are interested in the case
that the image of $\Res$ -- which is the operator that will be inserted on one of the two branches of a degenerating string worldsheet -- has
the canonical picture number $-1/2$.  So we can view $\Res$
as a linear transformation from $\H_{n;-3/2}$ to $\H_{n;-1/2}$, or in other words as an element of $\oplus_n \H_{n;-1/2}\otimes \H^*_{n;-3/2}$.
The duality (\ref{rotty}) lets us identify $\Res$ with an element
\begin{equation}\label{komo}\Res'\in (\H\otimes \H)_{1;-1/2\otimes-1/2},\end{equation}
where the notation means that $\Res'$ has overall ghost number 1 and has picture number $-1/2$ in each factor.

At picture number $-1/2$, the cohomology of $Q_B$ acting on $\H$ is nonvanishing only for ghost number $1/2$ or $3/2$.
At ghost number $1/2$, the cohomology 
 is generated by states $c\SSigma_{-1/2}\UUU$, where $\SSigma_{-1/2}$ is the $\beta\gamma$ ground state of picture number $-1/2$, and  
$\UUU$ is a matter primary of dimension $5/8$.  Similarly, at ghost number $3/2$, the cohomology is generated by states $c\partial c\SSigma_{-1/2}\UUU$.   

Given the facts stated in the last paragraph, the general form of $\Res'$ must be
\begin{equation}\label{lomo}\Res'=\sum_i c\SSigma_{-1/2}\UUU_i\otimes c\SSigma_{-1/2}\UUU'_i +Q_B\X,\end{equation}
where $\UUU_i$ and $\UUU'_i$ are matter primaries of dimension $5/8$. The exact term $Q_B\X$ will decouple as usual,
so the pole can be evaluated by insertions on the two branches of the physical states
$c\SSigma_{-1/2}\UUU_i$ and $ c\SSigma_{-1/2}\UUU'_i $.
However, because of the factor of $G_0$ in the original formula (\ref{moop}) for
$\Res$, the relation between $\UUU_i$ and $\UUU'_i$ is not as simple as  in the cases that we have considered so far. 

We will explain in detail what happens for the important case of  massless Ramond states of uncompactified 
ten-dimensional superstring theory.  The residue $\Res$ maps the picture number $-3/2$ state 
$c\partial c\SSigma_{-3/2}\Stigma^\alpha(p)$
to $c\SSigma_{-1/2}p^I\Gamma_I^{\alpha\beta}\Stigma_\beta(p)$.  On the other hand, the dual of 
$c\partial c\SSigma_{-3/2}\Stigma^\alpha(p)$ is $c\SSigma_{-1/2}\Stigma_\alpha(-p)$.  So the contribution 
of massless Ramond states of momentum $p$ to the
residue is
\begin{equation}\label{incob}\sum_{\alpha\beta}(p\cdot \Gamma)^{\alpha\beta} \,c\SSigma_{-1/2}\Stigma_\alpha(p)\otimes
c\SSigma_{-1/2}\Stigma_\beta(-p). \end{equation}
To match this to what one would expect from field theory, we simply observe that $(p\cdot \Gamma)^{\alpha\beta}$ is
the numerator of the Dirac propagator (describing the propagation of a massless fermion field of momentum $p$ and
specified  $SO(1,9)$  chirality),
while $c\SSigma_{-1/2}\Stigma_\alpha(p)$ is the string theory operator that describes the coupling of such a fermion to other fields.

\subsubsection{Analog For Closed Strings}\label{bclosed}

For the most part, the analog of all this for closed strings is fairly clear, so we will be brief.  But a few points are worthy of note.

The residue of the pole in the closed bosonic string propagator is $\Res=\t b_0b_0\t\Pi_0\Pi_0$, where $\t\Pi_0$ and $\Pi_0$ are respectively
the projectors to $\t L_0=0$ and to $L_0=0$.  By analogy to what we did  for open strings, 
this can be converted to an element $\Res'\in(\H\otimes \H)_4$, with $\H$
being the space of physical states and the subscript indicating the total ghost number.  Reasoning as for open strings gives
at non-zero momentum
\begin{equation}\label{zobo}\Res'=\sum_i\t c c\,\UU_i\otimes \t c c \,\UU^i+\{Q_B,\X\}.\end{equation}
The sum runs over a complete set of conformal vertex operators.
The conditions for dropping the $\{Q_B,\X\}$ term are the same as they were for open strings.

In studying anomalies, it will be important to know what happens if the residue $\Res$ is replaced by
 $\t\Res=(b_0-\t b_0)\t\Pi_0\Pi_0$.  This can  be
mapped to an element $\t\Res'\in(\H\times\H)_5$. At non-zero 
momentum, the analog of (\ref{yrfe}) reads
\begin{equation}\label{dilonkey}\t\Res'=\sum_i\bigl(\t c c(\t\partial \t c+\partial c)\UU_i\otimes \t c c\, \UU^i+\t c c\,
 \UU_i\otimes \t c c (\t\partial\t c+\partial c)\UU^i\bigr)+\{Q_B,\X\}.\end{equation} 
At zero momentum, there are exceptional contributions analogous to (\ref{zyrfex}).  The reason for this is that at zero momentum,
the closed bosonic string has cohomology at ghost number 1, generated (for bosonic strings in $\R^{26}$) by the operators $c\partial X^I$ and 
$\t c \t\partial X^I$, $I=0,\dots,25$,
that are related to momentum and winding number symmetries (see section \ref{mansym}), 
as well as cohomology at ghost number 4, obtained by multiplying those
operators by, respectively,  $\t c\t\partial\t c\t\partial^2\t c$ and $c\partial c\partial^2c$.  We can make an element of $(\H\otimes \H)_5$
by multiplying exceptional classes of ghost number 1 and 4. At zero momentum, these classes appear in
an exceptional contribution to $\t\Res'$:
\begin{equation}\label{opal}\sum_I\left( \t c\,\t\partial\t c\,\t\partial^2 \t c \cdot c\partial X^I\otimes c\partial X_I+c\partial X_I\otimes
  \t c\,\t\partial\t c\,\t\partial^2 \t c \cdot c\partial X^I\right)+z\leftrightarrow\t z.\end{equation}
  
The ingredients needed to generalize this to closed superstrings should be fairly clear from what we said about open superstrings
in section \ref{zopen}.  Perhaps the only real subtlety concerns the generalization of the exceptional zero-momentum contribution
(\ref{opal}).  In the NS sector, the analog of $c\partial c  \partial^2 c$ is $c\partial c\partial^2c \delta(\gamma)\delta(\partial\gamma)$,
and the analog of $c\partial X^I$ is $c\delta(\gamma)D X^I$.  In the Ramond sector, there is no analog of the identity operator
or of $c\partial c\partial^2c$, but  $c\partial X^I$ has an analog, namely  the supersymmetry generator $\S_\alpha = c\SSigma_{-1/2}\Stigma_\alpha$ 
studied in section \ref{superc}.  This is a $p=0$ limit of the chiral Ramond vertex operator $c\SSigma_{-1/2}\Stigma_\alpha e^{ip\cdot X}$. Because
of the factor of $p$ in the residue formula (\ref{incob}), which arose from integration over the fermionic gluing parameter, it appears that
there is no exceptional zero-momentum contribution to $\t\Res'$ in Ramond sectors.  Under certain conditions, there is a loophole in
this reasoning, but we defer this to section \ref{anomalies}.

\subsubsection{Another Look At The Pole Of The Propagator}\label{anlook}

Since the pole of the string theory propagator is so important, we will explain another way to look at it.
For brevity, we consider only open strings or a chiral sector of closed strings.
We start with bosonic string theory.

We consider gluing of two Riemann surfaces $\Sigma_\ell$ and $\Sigma_r$ with local parameters $x$ and $y$.
We slightly generalize the usual gluing relation $xy=q$ so that the gluing is centered at $x=a$ and $y=b$:
\begin{equation}\label{zonxo}(x-a)(y-b)=q.\end{equation}
  We want to determine how the worldsheet path integral depends on $a,b,$ and $q$,
keeping other moduli fixed.  Focusing on these three moduli makes sense when (but only when) $q$ is small.  The analysis
we will give is suitable for understanding the singular behavior for $q\to 0$.

In a first pass,
we will ignore insertions of operators other than the identity operator and represent the dependence of the path integral
on the moduli $a,b,$ and $q$ by a three-form $\Omega=f(a,b,q)\d a\,\d b\, \d q$. 
 By translation symmetry, we  assume that $f$ is independent of $a$ and $b$. 
 The dependence of $f$ on $q$ can be determined from the scaling symmetry
\begin{equation}\label{tonzo}(x,a)\to (\lambda  x,\lambda a),~~(y,b)\to (\t\lambda y,\t\lambda b),~~q\to\lambda\t\lambda q.\end{equation}
of the gluing equation (\ref{zonxo}).  This implies that $f$ must be a constant multiple of $1/q^2$, so the
contribution to the worldsheet path integral in which the operator inserted on each branch is the identity operator is
\begin{equation}\label{mondo}\Omega\sim \d a\,\d b \,\frac{\d q}{q^2}. \end{equation}

The formula in the last paragraph involves integrated vertex operators in the sense of section \ref{integrated}. 
An analogous formulas with unintegrated vertex operators is simply
\begin{equation}\label{pondo}\Omega\sim c(a)\otimes c(b) \,\frac{\d q}{q^2}.\end{equation}
The meaning of this formula is that to evaluate this contribution to the scattering amplitude, we are supposed to insert
$c(a)$ on $\Sigma_\ell$, and $c(b)$ on $\Sigma_r$, and integrate over $q$,  as well as over the moduli of $\Sigma_\ell$ and
$\Sigma_r$.   Those latter moduli include $a$ and $b$. The passage from unintegrated to integrated vertex operators replaces $c(a)$ and $c(b)$ by one-forms
$\d a$ and $\d b$.  
Notice that $c(a)$ and $\d a$ transform the same way under scaling; indeed, $c(a)$ has mass dimension $-1$
(since it has $L_0=-1$) and hence length dimension 1, just like $\d a$.  The factor $\d q/q^2$ in (\ref{pondo}) is $\d q \,q^{L_0-1}$
where $L_0=-1$ for the operator $c$.  

This discussion is rather formal, since the operator $c$ is not $Q_B$-invariant and there is
no way to isolate or define its contribution to a scattering amplitude.  In fact, that is clear from eqn. (\ref{pondo}):
to evaluate the contribution of the operator $c$ to a scattering amplitude, we would have to integrate the  form $\d q/q^2$ near $q=0$, where that form is unintegrable.   
The operator
$c$ can be viewed as the tachyon vertex operator $c\exp(ik\cdot X)$ at $k=0$, so  the $\d q/q^2$ singularity reflects the
existence of the tachyon; there is no such singularity in tachyon-free theories.  Let us consider a more general contribution
in which (unintegrated) vertex operators $\V_\ell$ and $\V_r$ are inserted on $\Sigma_\ell$ and $\Sigma_r$, 
respectively, and suppose that they have conformal dimension $h_\ell$ and $h_r$.
This means that they scale under (\ref{tonzo}) as $\V_\ell\to \lambda^{-h_\ell}\V_\ell$, $\V_r\to \t\lambda^{-h_r}\V_r$.
A scale-invariant generalization of (\ref{pondo}) exists only if $h_\ell=h_r$, and takes the form
\begin{equation}\label{londo}\V_\ell(a)\otimes \V_r(b)\,\d q\,q^{L_0-1}\end{equation}
where we consider $L_0$ to act on $\V_\ell$ or $\V_r$, so that it can be set to $h_\ell$ or $h_r$.  The integral
over $q$ near $q=0$  now has a pole at $L_0=0$ and the residue of the pole can be computed by inserting the operator
$\V_\ell(a)\otimes \V_r(b)$ on the two sides.

In generalizing this to superstring theory, for brevity we will consider only the NS sector.  
We start with super Riemann surfaces $\Sigma_\ell$
and $\Sigma_r$ with local coordinates $x|\theta$ and $y|\psi$, respectively.  We glue them by slightly generalizing
(\ref{wondo}) so that the gluing is centered at 
$a|\alpha\in\Sigma_\ell$ and $b|\beta\in\Sigma_r$:
\begin{align}\label{iblondo} (x-a-\alpha\theta)(y-b-\beta\psi)& =-\varepsilon^2 \cr
                                         (y-b-\beta\psi)(\theta-\alpha) & = \varepsilon(\psi-\beta) \cr
                                          (x-a-\alpha\theta)(\psi-\beta)&= -\varepsilon(\theta-\alpha)\cr
                                                              (\theta-\alpha)(\psi-\beta)& = 0.\end{align}
For the moment, all we really need to know of this formula is the scaling symmetry
\begin{align}\label{ipooz} (x,a,\alpha)\to &(\lambda x,\lambda a,\lambda^{1/2}\alpha) \cr
                                     (y,b,\beta)\to & (\t\lambda y,\t\lambda b,\t\lambda^{1/2}\beta)\cr
                                      \varepsilon\to& (\lambda\t\lambda)^{1/2}\varepsilon. \end{align}
This is enough to determine that the analog of (\ref{mondo}) is, in an obvious notation,
\begin{equation}\label{nipz}\Omega\sim  [\d a|\d\alpha]\otimes [\d b|\d\beta]\,\, \frac{\d\varepsilon}{\varepsilon^2}.\end{equation}
The analog of (\ref{pondo}) is similarly
\begin{equation}\label{zipondo}\Omega\sim c\delta(\gamma)\otimes c\delta(\gamma) \,\,\frac{\d\varepsilon}{\varepsilon^2}.\end{equation}  Finally with general operator insertions, the analog of (\ref{londo}) is
\begin{equation}\label{imondox} \V_\ell(a|\alpha)\otimes \V_r(b|\beta)\,\d\varepsilon\,\varepsilon^{2L_0-1}.\end{equation}
We can also write this in terms of $q_\NS=-\varepsilon^2$ using
\begin{equation}\label{plintto}\d \varepsilon \,\varepsilon^{2L_0-1}\sim \d q_\NS \,q_\NS^{L_0-1}. \end{equation}
Again the integral over $\varepsilon$ or $q_\NS$ has a pole at $L_0=0$, whose residue can be computed
by inserting $\V_\ell(a|\alpha)\otimes \V_r(b|\beta)$ on the two sides. 

Several well-known but
relatively subtle facets of superstring theory follow from these simple formulas.  Since the GSO projection
comes from the sum over the sign of $\varepsilon$, and the  form $\d\varepsilon/\varepsilon^2$
 in (\ref{nipz}) is odd under $\varepsilon\to-\varepsilon$,
we see that the identity operator -- or equivalently the RNS tachyon -- is removed by the GSO projection. 
Also, the behavior $\d\varepsilon/\varepsilon^2$ in (\ref{nipz}), compared to $\d q/q^2$ in (\ref{mondo}), shows that the tachyon
mass squared in superstring theory is one-half what it is in bosonic string theory.   

Returning to the bosonic string (the superanalog of what we are about to explain is fairly evident),
the only transformations of the local coordinates $x$ and $y$ that we have considered so far are the affine
linear transformations $x\to \lambda(x-a)$, $y\to \t\lambda(y-b)$.    Suppose that we replace
$x$ and $y$ with  general local coordinates $\h x(x)$ and $\h y(y)$, and that the points $x=a$, $y=b$
correspond in the new coordinate system to $\h x=\h a$ and $\h y=\h b$.  Then to first order in $q$, the gluing is described by
\begin{equation}\label{uttz}(\h x-\h a)(\h y-\h b)=\h q,\end{equation}
where
\begin{equation}\label{yuttz}\h q=q\frac{\partial \h a}{\partial a}\frac{\partial \h  b}{\partial b}.\end{equation}
The fact that $q$ is rescaled under a change of local parameters means that it is really not best understood as a
complex number but as a section of a complex
line bundle over the divisor $\fD$ at infinity; the fact that this rescaling is by the product
 of a function ${\partial \h a}/{\partial a}$ of the local parameters on $\Sigma_\ell$
  times a function $ {\partial \h  b}/{\partial b}$ of the
local parameters on $\Sigma_r$  means that  the line bundle in question  is the tensor product of a line bundle over $\h\M_\ell$ with
one over $\h\M_r$.  For more on this, see, for instance, section 6 of  \cite{Wittentwo}.

Beyond first order in $q$, the relation between $\h q$ and $q$ becomes nonlinear.    $\Omega$ can be expanded
in powers of $q$ with contributions of the form written in (\ref{londo}) that come  
from both primary fields and descendants.  The descendant
contributions can be related to the primary contributions using  invariance under reparametrizations of $\Sigma_\ell$
and $\Sigma_r$, but in doing so, one has to use
the nonlinear transformation of the gluing parameter under reparametrizations.

\subsection{The Deligne-Mumford Compactification And The Integration Cycle}\label{delcycle}

Having introduced the Deligne-Mumford compactification in section \ref{ondang},
we can now complete the description given in section \ref{harder} of the integration cycle of superstring perturbation theory.
In section \ref{hets}, we consider the heterotic string.  Since the idea of choosing an integration cycle as a step in
formulating superstring perturbation theory may be unfamiliar, in section \ref{lanex} we consider an example.
Finally, in section \ref{banex}, we consider the other superstring theories.  The description we give here of the
integration cycle resolves what in the 1980's was described as an ambiguity in superstring perturbation theory
\cite{ARS}.

\subsubsection{The Heterotic String}\label{hets}

We first recall the problem as presented in section \ref{harder}.
The worldsheet $\Sigma$ of a heterotic string has holomorphic moduli $m_1\dots m_p|\eta_1\dots\eta_s$
and antiholomorphic moduli $\t m_1\dots \t m _p$.   Roughly speaking, one wants to integrate over the
cycle $\varGamma$ defined by taking the $\t m_i$ to be the complex conjugates of the $m_i$.  But this is not well-defined
(unless supermoduli space is projected) because of the possibility of shifting the $m_i$ by expressions of quadratic
and higher order in the $\eta_j$.  In the interior of moduli space, we simply choose any integration cycle $\varGamma$ such that 
$\bar{\t m}_i=m_i$, modulo fermion bilinears.  There is no natural choice, but 
any two choices are homologous.
If, therefore, $\varGamma$ were compact, the choice of $\varGamma$ would not matter at all; as it is, we have
to specify how $\varGamma$ should behave at infinity in moduli space.

We are now in a position to answer this question.  The inspiration for the answer comes from the factorization described in section \ref{ondang}.
 For simplicity in terminology,
let us consider a separating degeneration in the NS sector.
The idea behind the factorization was  roughly that the integral over the bosonic gluing parameter $q_\NS$
gives a pole $1/L_0$.  At $q_\NS=0$, the other moduli of $\Sigma$ are  the moduli of $\Sigma_\ell$ and $\SIgma_r$
(including the moduli of the points on $\Sigma_\ell$ and $\Sigma_r$ 
at which these surfaces are glued together to make $\Sigma$), and the integral over those other moduli will then
give the residue of the pole.  

 In more detail, the heterotic string worldsheet $\Sigma$ has an
antiholomorphic gluing parameter $\t q$ as well as the holomorphic gluing parameter $q_\NS$.  The $1/L_0$ pole
is supposed to come from an integration over $\t q$ and $q_\NS$ keeping fixed the moduli of $\Sigma_\ell$ and $\Sigma_r$, and the pole specifically arises at $q_\NS=\t q=0$. 
This suggests that the definition of $\varGamma$ should let us set $\t q=q_\NS=0$ without affecting the other moduli.  To this
end, the definition of $\varGamma$ should include a condition\begin{equation}\label{muro}\bar{\t q}=q_\NS \end{equation}
or
\begin{equation}\label{zuro}\bar{\t q}=q_\NS(1+\O(\eta_i\eta_j)),\end{equation}
but not
\begin{equation}\label{yuro}\bar{\t q}=q_\NS+\O(\eta_i\eta_j).\end{equation}
If as in (\ref{yuro}), we were to add to the relation between $\t q$ and $q_\NS$ a fermion bilinear with a coefficient
that does not vanish at $q_\NS=0$, then naively speaking, we could not set $\t q=q_\NS=0$ without disturbing the other moduli.

It turns out that the right procedure in defining $\varGamma$
is to impose (\ref{muro}) or (\ref{zuro}), as opposed to (\ref{yuro}), but the reason
that this is necessary is a little subtle.   The most critical case to understand
is the case that  the
momentum flowing between $\Sigma_\ell$ and $\Sigma_r$ is on-shell regardless of the external momenta.  
As usual, the problem first arises when there are two odd moduli.  It is possible for the measure that
must be integrated to compute a heterotic string amplitude to behave near $q_\NS=\t q=0$ as
\begin{equation}\label{nolfo}\Xi=[\d\t q;\d q_\NS|\d\eta_1,\d\eta_2]\,\t q^{-1}.\end{equation}
When this is the case, $\int\Xi$ is invariant under $q_\NS\to q_\NS(1+\eta_1\eta_2)$ but not under $q_\NS\to q_\NS+\eta_1\eta_2$.
Under any infinitesimal change of coordinates, $\int\Xi$ changes by a total derivative. For the particular case $q_\NS\to q_\NS+\eta_1\eta_2$, the
shift in $\int\Xi$ is
\begin{equation}\label{teflon}\int\Xi\to \int\Xi+\int [\d\t q;\d q_\NS|\d\eta_1,\d\eta_2]\eta_1\eta_2\frac{\partial}{\partial q_\NS}
\frac{1}{\t q}. \end{equation}
But this is non-zero, because once we interpret $q_\NS$ as $\bar{\t q}$, the expression $\partial_{q_\NS}(1/\t q)$ has a delta
function contribution at $q_\NS=\t q=0$.  
For more detail on how this phenomenon arises in heterotic string computations, see 
section \ref{lanex} and appendix \ref{leisurely}.  

Once one knows that a condition is required on the behavior of the integration cycle at infinity, the right condition must
be  (\ref{muro}) or more precisely (\ref{zuro}), since this is the only condition that can be stated using the data at hand.

Indeed, because of facts explored in section 6.3 of \cite{Wittentwo} and also at the end of section
\ref{anlook} above, only the more general condition (\ref{zuro}) really makes
sense, and not the more restrictive (\ref{muro}).  In brief, $\t q$ and $q_\NS$ are really not 
complex numbers
but  sections of complex line bundles
that we will call $\t \L$ and $\L$, respectively. 
($\t\L$ and $\L$ are line bundles over the left and right moduli spaces $\M_L$ and $\M_R$, respectively.)
 By $\t q$ or $q_\NS$ we mean a holomorphic section of $\t\L$ or of $\L$
that has a simple zero on the compactification divisor.  This defines them  only up to
\begin{equation}\label{zomboo} \t q\to e^{\t\varphi}\t q,~~q_\NS\to e^\varphi q_\NS, \end{equation}
for some functions $\t\varphi$ and $\varphi$.  Because of this, a condition as precise as (\ref{muro}) would not be well-defined.
However, once one reduces modulo the odd moduli $\eta_1\dots\eta_s$ and identifies $\M_{L,\red}$ as the complex
conjugate of $\M_{R,\red}$,
$\t \L$ is the complex conjugate of $\L$.  Therefore, a well-defined
condition is that $\bar{\t q}=q_\NS e^\phi$ where $\phi$ vanishes modulo the odd moduli.  This is what we have written in (\ref{zuro}).  

At this point, we can complete the description of the heterotic string integration cycle $\varGamma$.  An inductive procedure
is involved.  When $\Sigma$ degenerates, we impose the condition (\ref{zuro}).  Beyond this, as long as 
$\Sigma_\ell$ and $\Sigma_r$ are smooth, $\varGamma$ may be any
cycle of the right dimension
whose reduced space is the diagonal in $\M_{L,\red}\times \M_{R,\red}$.  When $\Sigma_\ell$ or $\Sigma_r$
degenerates, one needs a further condition with the same form as (\ref{zuro}).  This continues until we finally reach a maximal
degeneration, such as the one depicted schematically in fig.  \ref{longfig}.

It is noteworthy that the condition (\ref{zuro}) that we actually need makes sense, and a more precise condition (\ref{muro})
that we do not need does not make sense.  This is the story of superstring perturbation theory: precisely what one needs
is true, and in general no more.  

The condition (\ref{zuro}) is formulated in terms of  $q_\NS=-\varepsilon^2$ rather than
in terms of $\varepsilon$, so for given $\t q$, the sign of $\varepsilon$ is not fixed; the sum over this sign leads to the GSO
projection.  (Changing the sign of $\varepsilon$ while keeping the other variables fixed only makes sense for $\varepsilon\to 0$,
but the poles at $L_0=0$ come from the behavior for $\varepsilon\to 0$, so
 the sum over the two signs of $\varepsilon$ does make sense
in analyzing which states contribute a pole at $L_0=0$.)

For a Ramond degeneration of the heterotic string, we simply replace $q_\NS$ by $q_\Ra$ in the foregoing.  For the other
superstring theories, just a few minor modifications are needed; see section \ref{banex}.  For a treatment of some of these
issues in terms of conditions on  the placement of picture-changing operators, see \cite{AMS}.

\subsubsection{An Example}\label{lanex}

An example explored in the literature \cite{DIS,ADS,GrSe}  gives a good illustration of these ideas.  
In this case, the degeneration involves a collision of two NS vertex operators.  On a heterotic string worldsheet with
local coordinates $\t z;\neg z|\theta$, we consider two NS punctures, at say $\t z;\neg z|\theta =\t u_1;\neg u_1|\zizeta_1$ and
$\t u_2;\neg u_2|\zizeta_2$.  The degeneration occurs at $\t u_1\to \t u_2$, $u_1\to u_2$.  The gluing parameters are
\begin{align}\label{onorm}\t q&=\t u_1-\t u_2 \cr
                                                  q_\NS&=u_1-u_2-\zizeta_1\zizeta_2.\end{align}
The only nontrivial point is the term $-\zizeta_1\zizeta_2$ in $q_\NS$.  This term is easily motivated from global supersymmetry
(it is determined by invariance under the global supersymmetry generator $G_{-1/2}=\partial_\theta-\theta\partial_z$)   and is
derived\footnote{The factor of $1/2$ in eqn. (6.28) of that paper is inessential, since $q_\NS=-\varepsilon^2$ is
only defined up to $q_\NS\to q_\NS e^\phi$.}  in section 6.3.2 of \cite{Wittentwo}.  

At this stage, we need to know one fact that will be more systematically developed in section \ref{inttad}.  Infrared singularities
in string theory are regularized by placing a lower bound on the magnitude of the gluing parameters.  In the present
example, the cutoff can be $|\t q|\geq \epsilon$, where we take $\epsilon\to 0$ at the end of the computation.  Since the integration
cycle is defined by a condition such as $\bar{\t q}=q_\NS$, the infrared cutoff also bounds $q_\NS$ away from zero.

Now let $u_{12}=u_1-u_2$ and $\t u_{12}=\t q=\t u_1-\t u_2$.  The reasoning of section \ref{hets}, specialized to this situation,
together with the infrared cutoff procedure stated in the last paragraph,
amounts to saying  
that near $u_{12}=\t u_{12}=0$, instead of integrating over $\t u_{12}, u_{12}, \zizeta_1$, and $\zizeta_2$ with 
$\bar{\t u}_{12}=u_{12}$, we want
to integrate over $\t q$, $q_\NS$, $\zizeta_1,$ and $\zizeta_2$ with $\bar{\t q}=q_\NS$.  This is precisely the conclusion of
the above-cited papers, where it is shown that the alternative procedure leads to the wrong answer.

In appendix \ref{leisurely}, we give a more detailed and precise account of this example.    See also \cite{More} for much more
detail.

\subsubsection{Other Superstring Theories}\label{banex}

Here we extend the description of the integration cycle to Type II and Type I superstring theories.  There are only a few details to explain. 

Let $\Sigma$ be a Type II superstring worldsheet.  It has holomorphic and antiholomorphic moduli $m_1\dots m_p|\eta_1\dots \eta_s$
and $\t m_1\dots \t m_p|\t \eta_1\dots\t\eta_{\t s}$.  (The numbers of holomorphic and antiholomorphic even moduli are always equal,
but this is not so in general for the odd moduli; $s$ and $\t s$ can be unequal if there are operator insertions of types NS-R and/or R-NS.)

Let us consider for definiteness in the notation a degeneration of NS-NS type.  There are then holomorphic and antiholomorphic
gluing parameters $q_\NS$ and $\t q_\NS$. (For other degenerations, simply replace $\t q_\NS$ by $\t q_\Ra$ and/or $q_\NS$ by
$q_\Ra$ in the following.)  As in the case of the heterotic string, roughly speaking,
we want to constrain the behavior at infinity of the integration cycle $\varGamma$ by setting $\bar{\t q}_\NS=q_\NS$.
However, there is a problem here: $\t q_\NS$ is a function on a complex supermanifold (the moduli space of super Riemann
surfaces with punctures) and there is no natural notion of complex-conjugating a function on a complex supermanifold.
The problem becomes obvious if we recall that $\t q_\NS$ is only well-defined up to multiplication by a function $\exp(\t\phi)$ where
$\t\phi$ may depend on the $\t\eta_j$.   So it does not make much sense to complex conjugate $\t q_\NS$ without being able
to complex conjugate the $\t\eta_j$, but the complex conjugates of the $\t\eta_j$ are certainly not part of the formalism.

Writing the relation between $\t q_\NS$ and $q_\NS$ in the form $\t q_\NS=\bar q_\NS$ would merely move the problem from
antiholomorphic to holomorphic degrees of freedom.  
For the heterotic string, we were able to avoid this issue, because $\t q$ was a function on an ordinary complex manifold, where
it makes sense to take the complex conjugate of a function. 

The appropriate procedure (see \cite{Wittenone}, especially section 5, for background) 
is to describe the relation between $\t q_\NS$ and $q_\NS$
in parametric form.  $\varGamma$ is a smooth cs supermanifold, meaning that it is parametrized by even and odd coordinates
$t^1\dots t^{2p}|\zeta^1\dots \zeta^{s+\t s}$, where the $t^i$ can be considered real modulo the 
odd variables $\zeta^j$.  The locus at infinity in $\varGamma$ is of codimension $2|0$ and we pick coordinates so that
this locus is defined by $t^1=t^2=0$.  Then $q_\NS$ and $\t q_\NS$ are defined in terms of functions on $\varGamma$ by any condition of the form
\begin{align}\label{potro} q_\NS(1+\O(\eta^2))& =t^1+it^2\cr
                                            \t q_\NS(1+\O(\t \eta^2))&=t^1-it^2. \end{align}
In this way, one formulates a relation that roughly corresponds to $\bar{\t q}_\NS=q_\NS$ or $\t q_\NS=\bar q_\NS$ without
complex-conjugating either $\t q_\NS$ or $q_\NS$.

Since the asymptotic behavior of $\varGamma$ has been defined via conditions on $q_\NS$ and $\t q_\NS$, the signs of
the gluing parameters $\varepsilon$ and $\t\varepsilon$ are unspecified.  The sum over these signs leads to the chiral
GSO projection -- separate GSO projections for holomorphic and antiholomorphic modes.  As usual, reversing the signs of
$\varepsilon$ and $\t\varepsilon$ while keeping fixed the other variables only makes sense for $\varepsilon,\t\varepsilon\to 0$,
but this suffices for studying the pole at $L_0=0$.

For open superstrings, a basic fact to start with (see section 7.4 of \cite{Wittentwo} as well as section \ref{geometry}
below for more detail on
much that follows) is that on an ordinary Riemann surface with boundary, the gluing parameter $q$
associated to an open-string degeneration is real and positive (modulo the odd variables).  
In particular, an open-string degeneration is associated
to a boundary of moduli space at $q=0$.   The same is true of certain closed-string degenerations of
open and/or unoriented strings.  Because of this, the integration cycle $\varGamma$ of open and/or unoriented superstring theory
 is going to be a supermanifold with boundary.  This notion is a little delicate; see for 
example section 3.5 of \cite{Wittenone}.  The essential point is that to make sense of supermanifolds with boundary and especially to make sense of integration 
on them,
one needs an equivalence class of nonnegative functions $\rho$ with a first order zero along the boundary, modulo rescaling $\rho\to \rho e^f$
for a function $f$. Roughly speaking, 
$\rho$ will be 
the gluing parameter $q_\NS$ or $q_\Ra$.  A fuller explanation involves the following.

For bosonic strings, there is a natural moduli space of open and/or unoriented Riemann surfaces. But this does not
appear to have a superstring analog.\footnote{It does have an analog in the nonsupersymmetric
and tachyonic Type 0 string theory, in which holomorphic and antiholomorphic odd coordinates are complex
conjugates.}
Just as we explained for closed oriented superstrings in section \ref{harder}, it seems that open and/or unoriented
superstring world sheets  have a natural integration cycle (up to homology) but not a natural 
moduli space.
To construct the integration cycle, we begin by associating to 
an open and/or unoriented superstring worldsheet $\Sigma$ a certain double cover that is a closed oriented
super Riemann surface $\h\Sigma$.  Let $\h{\MM}$ be the moduli space that parametrizes deformations of $\h\SIgma$. It
is a complex supermanifold, say of complex dimension $p|s$, with even and odd moduli $m_1\dots m_p|\eta_1\dots\eta_s$.
The integration cycle for open and/or unoriented super Riemann surfaces is a (real) codimension $p|0$ cycle
$\varGamma\subset \h{\MM}$ that is defined, roughly speaking, by taking the $m_i$ to be real, modulo the
odd variables.  To be more exact, inside the reduced space $\h{\MM}$, there is a natural real cycle $\Gamma$
of codimension $p$ that parametrizes the deformations of the reduced space of $\Sigma$, 
and $\varGamma$ is defined by thickening this in the fermionic directions. 

Just as for closed oriented superstrings,
$\varGamma$ is not uniquely determined, but the possible choices are homologous.  So as usual, all we need to do
is to put a condition on how $\varGamma$ should behave at infinity.
For this, we require first a few generalities.  $\varGamma$ will be again a smooth cs supermanifold
with coordinates $t^1\dots t^p|\zeta^1\dots\zeta^s$, where the $t^i$ can be considered real modulo the $\zeta^j$.
But now, as observed above, $\varGamma$ will be  a supermanifold with boundary. This
 means that we can pick the coordinates, locally, so that $t^1$ plays
a special role; $t^1\geq 0$ on $\varGamma$ and the boundary of $\varGamma$ is defined by $t^1=0$.  Of course,
the function $t^1$ with these properties is not uniquely determined; it is determined modulo $t^1\to e^f t^1$, where $f$ is
a function on $\varGamma$. (Importantly, this does not allow a substitution such as $t^1\to t^1+\zeta^1\zeta^2$, though
it does allow $t^1\to t^1(1+\zeta^1\zeta^2)$.)   The integration cycle $\varGamma$ for open superstrings 
near, say, an NS degeneration is
constrained to obey
\begin{equation}\label{otro} t^1=q_\NS(1+\O(\eta^2)).\end{equation}
Here $q_\NS$ is the complex gluing parameter of the double cover $\h\Sigma$, defined in the usual way.
(For a Ramond degeneration, just replace $q_\NS$ by $q_\Ra$.) 
What we have done, roughly speaking, is to put the assertion that $q_\NS$ is real and nonnegative 
near the boundary of $\varGamma$
in parametric form, without mentioning the complex conjugate of $q_\NS$ or  making any other claim that does not make
sense on a complex supermanifold. 

Another way to express the idea of the last paragraph is to say that the boundary of the integration cycle $\varGamma$
is contained in the divisor  ${\fD}\subset{ \MM}$ defined by $q_\NS=0$.  

For Type II superstrings as well as for open and/or unoriented superstrings, what we have explained is the condition
that should be imposed
on the integration cycle when a single degeneration occurs. For multiple degenerations, one places such conditions
on the gluing parameters at each degeneration. 
                                                                                      
\section{BRST Anomalies, Massless Tadpoles, And All That}\label{massren}

\subsection{BRST Anomalies}\label{brstanom}

At many points in the present paper, beginning in section \ref{ginv}, we have seen that the proof of gauge-invariance
and related properties relies on integration by parts on moduli space.
One simply uses eqn. (\ref{tarmob})  together with Stokes's theorem (or the superanalogs of these formulas) to express a scattering amplitude with
insertion of a BRST-trivial state such as $Q_B\W_1$ in terms of a boundary integral:
\begin{equation}\label{tarmib}\int_{\varGamma}F_{\{Q_B,\W_1\},\V_2,\dots,\V_\ssn} 
=-\int_{\partial\varGamma} F_{\W_1,\V_2,\dots,\V_\ssn}. \end{equation}
Here $\varGamma$ is the integration cycle of the worldsheet path integral, as described in section \ref{delcycle}.
(For bosonic string theory, it is simply the moduli space $\M_{\sg,\sn}$.) 
The scattering amplitude with insertion of $Q_B\W_1$ is zero 
if the boundary integral on the right hand side of (\ref{tarmib}) vanishes.

Actually, this phrasing of the problem is a little too schematic.
In closed oriented string theories,
$\varGamma$ has no boundary, but the forms that must be integrated over $\varGamma$ are singular
along a divisor $\fD$ (which has multiple components, corresponding to different ways that $\Sigma$ may
degenerate).  We place a cutoff on the integral on the left of (\ref{tarmib})
by removing a small neighborhood of $\fD$; details of this will be described.  By $\partial\varGamma$, we mean the
boundary of the resulting cutoff version of $\varGamma$. Then we consider the limit in which the cutoff is removed;
there is no anomaly if the right hand side of (\ref{tarmib}) vanishes in this limit.

By now we have assembled the ingredients to understand when a problem may actually arise. Infinity in moduli
space is the region in which one of the gluing parameters vanishes.  The gluing parameters
are  $q=e^{-s}$ for open strings or $q=e^{-(s+i\alpha)}$ for
closed strings, and vanish for $s\to\infty$. We cut off $\varGamma$ by placing an upper bound on $s$,
so to get a component of $\partial\varGamma$, we simply set $s$ to a large value\footnote{More precisely, we set $s$ to be the sum of a 
large constant and an arbitrary, but fixed, function of the
remaining worldsheet moduli. 
This more careful statement is needed because $q$ is a section of a complex line bundle rather than a complex
number, as explained at the end of section \ref{anlook}. We cannot simply set $q$ to 0 or $s$ to $\infty$,
since $F_{\W_1,\V_2,\dots,\V_\ssn}$ is typically singular at $q=0$.  See section \ref{lazy} for more on these points.} 
rather than integrating over it.  

The basic idea of what will happen is visible for open bosonic strings.  Propagation of an open string through
a proper time $s$ is described by a factor $\exp(-sL_0)$.  To construct the open bosonic string propagator in section \ref{bos},
we multiplied by $b_0$ and integrated over $s$, giving 
\begin{equation}\label{limely} b_0\int_0^\infty \d s\,\exp(-sL_0). \end{equation}
In looking for a possible boundary contribution on the right hand side of a formula like eqn. (\ref{tarmib}), we do not
want to integrate over $s$, but rather to set it to a large value.  So we drop the $b_0$ factor and the integration in
(\ref{limely}).  We are left simply with a factor $\exp(-sL_0)$, so a boundary term can only arise if $\exp(-sL_0)$
is nonvanishing for large $s$.  

An immediate consequence is that only certain types of degeneration can
generate anomalies. Let us consider first
a typical separating degeneration, such as that of fig. \ref{longfigo} in section \ref{sepcase}, with more than one external particle on each side.
The momentum $P$ flowing through the separating line is a sum of several external momenta and is generically 
not on-shell.  Moreover, for suitable  external momenta, $P^2$ can have any real or complex value.  In this situation, there is no possible anomaly.  This follows from
the relation $L_0=(\alpha'/4)P^2+N$, where $N$ is the mass squared operator of the string.  In a region of external
momenta in which the real part of $P^2$ is sufficiently positive, $\exp(-sL_0)$ vanishes for large $s$.  So for such external
momenta, there is no surface term at large $s$.  The general result for the scattering amplitude can be obtained
by analytic continuation from the region just indicated, and so possesses no anomaly associated to
a degeneration of this type.

A somewhat similar argument shows that there are never anomalies associated to nonseparating degenerations.
Nonseparating degenerations only arise in loop amplitudes (such as the one-loop diagram of fig. \ref{collapse}(b) in section \ref{bos}), 
and loop amplitudes do not make sense in theories with
tachyons.  So to discuss anomalies associated to nonseparating degenerations, we should consider only tachyon-free
theories, that is we should assume that $N\geq 0$ (possibly after projecting to the GSO-invariant part of the spectrum
in the case of superstring theory).  In the case of a nonseparating degeneration, the momentum $P$ flowing through
the separating line is an integration variable (the loop momentum).  The loop integration can be performed over the cycle
on which $P$ is real in Euclidean signature,
so that $L_0$ is positive semidefinite and vanishes only when $N$ and $P$ both vanish.  Away from 
$N=P=0$, $\exp(-sL_0)$ vanishes exponentially. 
Since $P=0$ has measure zero, there is no anomaly associated to the fact that $L_0=0$ just for $P=0$.
(Even for $N=0$, the integral $\int \d^DP\exp(-sL_0)=\int \d^DP \exp(-s(\alpha'/4)P^2)$ vanishes for large $s$,
though only as a power of $s$.)

\begin{figure}
 \begin{center}
   \includegraphics[width=4.5in]{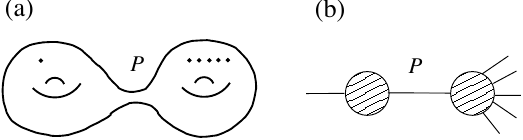}
 \end{center}
\caption{\small With only one external particle to the left (or right) of a separating degeneration, the momentum $P$
flowing through the separating line is constrained to be on-shell.  This is the configuration associated with mass renormalization. The configuration is sketched in string theory and in field theory in (a) and (b), respectively.}
 \label{hoopla}
\end{figure}

\begin{figure}
 \begin{center}
   \includegraphics[width=4.5in]{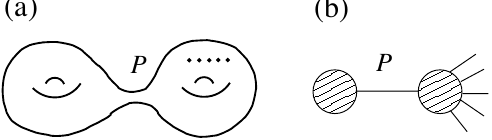}
 \end{center}
\caption{\small With no external particles at all to the left (or right) of a separating degeneration, the momentum $P$
flowing throught the separating line automatically vanishes.  Hence a massless particle flowing through this line is
on-shell.  This is the most troublesome configuration for superstring perturbation theory.}
 \label{zoopla}
\end{figure}

Going back to separating degenerations, there are two critical cases that actually do cause trouble.
These are the cases, sketched 
in  figs. \ref{hoopla} and \ref{zoopla},
 in which the momentum $P$ flowing
through the separating line is automatically on-shell. (The figures are drawn for closed strings, though
this involves jumping slightly ahead of our story.) In fig. \ref{hoopla},  there is just one external particle to the left
 of the separating line, so $P$ is equal to
the momentum of that external particle and in particular is constrained to be on-shell.  In fig. \ref{zoopla},
there are no external particles at all on the left, and momentum conservation forces $P=0$, which is on-shell
in the case of a massless particle.  

These then are the troublesome cases for superstring perturbation theory.  We will review the analogous
issues in field theory in section \ref{fieldrev}, after which we explain
some basics of what happens in string theory.  The issues that arise here
will  occupy us for most of the rest of this paper.

What we have said so far generalizes straightforwardly to the other string theories. For example, consider
closed bosonic string theory.  To extract a boundary term involving a closed-string degeneration, we modify eqn.
(\ref{merom}) for the closed-string propagator in the following way.  We keep the integration over $\alpha$ and
the associated factor of $\Psi_\alpha=b_0-\t b_0$.  But we set $s$ to a large constant and omit the factor of $\Psi_s$
that is associated to the $s$ integral.  We are left with
\begin{equation}\label{yzes}2\pi (b_0-\t b_0) \delta_{L_0-\t L_0}\exp(-s(L_0+\t L_0)). \end{equation}
By essentially the same arguments that we have given for open strings, an 
anomaly can only arise from states of $L_0=\t L_0=0$, and only for separating degenerations of the types sketched
in figs. \ref{hoopla} and \ref{zoopla}.  In passing from bosonic string theory to superstring theory, the only modifications
of the above formulas are the obvious factors of $\delta(\beta_0)G_0$ in the Ramond sector.  These factors do not affect the
discussion of which types of degeneration can contribute anomalies.  

\subsection{Review Of Field Theory}\label{fieldrev}

Before trying to understand what is happening in string theory, it helps to review how the troublesome cases are understood
in field theory.

The two cases are completely different.  In fig. \ref{hoopla}, let the external particle flowing in from the left have momentum
$p$ and mass $m$.  Momentum conservation implies that the momentum $P$ of the separating internal line is simply
equal to $p$. 
In an $S$-matrix element, the external particle is always on-shell, so $p^2+m^2=0$.  Consequently,
if we directly evaluate the contribution to the $S$-matrix from a Feynman diagram such as that of fig. \ref{hoopla}(b),
we are sitting on top of the pole $1/(P^2+m^2)$ of the propagator.   We cannot simply calculate the $S$-matrix by
summing Feynman diagrams with on-shell external particles.

The solution to this problem is well-known.
Instead of computing an $S$-matrix element directly,
one introduces a local field $\O$ that can create the particle in question from the vacuum, and one computes
matrix elements of $\O$.  This enables one to vary the momentum $p$ 
(and therefore also $P$) away from its mass-shell, and then by searching for a pole
in the matrix element of $\O$, one recovers
the appropriate $S$-matrix element.  The effect of diagrams such as that of fig. \ref{hoopla}(b) is to shift the position of
the pole away from its tree-level value; this is called mass renormalization.  
Mass renormalization generally\footnote{\label{noggin} There is a potential exception 
in the case of a theory that at tree level has a massless
scalar; perturbative effects of mass renormalization might make the scalar tachyonic, in which case the theory has no perturbative
expansion near the originally considered classical vacuum.  In practice, if there is a good reason -- such as supersymmetry
or a spontaneously broken continuous
bosonic symmetry -- to have a massless scalar at tree level, this usually prevents mass renormalization
for the scalar in question. However, if (as in three-dimensional theories with only two supercharges, for example), supersymmetry does allow perturbative mass renormalization for massless scalars, then perturbation theory must
be developed around a minimum of an appropriate effective potential, not around an arbitrary classical vacuum. } does not affect the  existence of a sensible perturbation expansion for a given theory.
A physically sensible perturbation expansion remains physically sensible, whatever the mass renormalization effects may be.

Quite different is the configuration of
fig. \ref{zoopla}. There are  no particle insertions at all to the left of the separating line, so the momentum flowing through
that line is $P=0$.  Thus the separating line is on-shell if and only if it is associated to a field  of mass 0. Let us call this field 
$w$.   The left
half of  fig. \ref{zoopla}(b)  represents a contribution to the ``tadpole'' of $w$ -- the amplitude for a single $w$ quantum to disappear into
the vacuum.   By Lorentz invariance, this tadpole vanishes unless $w$ is Lorentz-invariant.  In practice, $w$ is
either a massless scalar field or else the trace of the metric tensor.

\begin{figure}
 \begin{center}
   \includegraphics[width=2.5in]{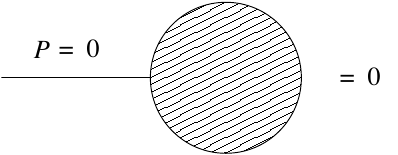}
 \end{center}
\caption{\small The shaded region represents symbolically the ``tadpole'' of a massless scalar field -- its amplitude
to disappear into the vacuum.  
In general in quantum field theory, to make possible
 a sensible perturbation expansion around a chosen classical vacuum, the tadpoles of massless scalar fields must vanish.}
 \label{oopla}
\end{figure} 
If such a massless tadpole is non-zero, then the associated contributions to the scattering amplitude
are proportional to $1/0$ -- that
is, they are proportional to the scalar propagator $1/(P^2+m^2)$ at $P=m=0$.  
Accordingly,
for a theory to have a perturbation expansion around an initially assumed classical vacuum, the tadpoles for all massless
scalars, computed in that vacuum, must vanish (fig. \ref{oopla}).
Otherwise, one has to search for an extremum of an appropriate quantum effective potential.
Nonvanishing of the tadpole means that this extremum will not coincide with the initially assumed value.  

What can happen in detail depends on the underlying classical potential for the field $w$.  Rather than trying to be general,
let us discuss the situation that prevails in ten-dimensional supergravity and in  other theories that 
arise as low energy limits of superstring compactifications
 to $d>2$ Lorentzian dimensions.  
Whenever there is a perturbative string theory, there is always at least one massless scalar field at the classical
level -- the dilaton field $\phi$.  Perturbation theory is an expansion in $g_\st=e^\phi$, which is called the string coupling constant.
In particular, the effective action at string tree level is (in the closed-string sector) proportional to $1/g_\st^2=e^{-2\phi}$,
and so the classical potential for $\phi$, if not identically zero, is a multiple of $e^{-2\phi}$.
Since this function has no stationary point as a function of $\phi$, a classical vacuum that could represent the
starting point of superstring perturbation theory can only exist if the potential is identically zero as a function of $\phi$ at the
classical level.
By familiar arguments \cite{DS}, the relationship $g_\st=e^\phi$  implies that a stable vacuum accessible to perturbation theory
exists if and only if the effective potential for the dilaton field 
is identically zero in perturbation theory, not just classically (in particular, the dilaton mass renormalization must vanish
as well as the dilaton tadpole).  The basic idea here is that if a non-zero effective potential is generated in loops, then
for weak coupling or $\phi<<0$, it is dominated by the non-zero contribution of lowest order and has no critical point.\footnote{The
genus $\sg$ contribution to the effective potential is a multiple of $g_\st^{2\ssg-2}=\exp((2\sg-2)\phi)$.  When one adds
this to the kinetic energy of the metric and the gravitational field, which at the classical level is a multiple of
$e^{-2\phi}(R+4\partial_I\phi \partial^I\phi)$, one finds that to get a classical solution in a homogeneous spacetime
-- even with a cosmological constant -- one needs a cancellation between contributions to the effective potential associated to different values of $\sg$.  
Such cancellations may occur but only if $g_\st$ is sufficiently large, so they
are not accessible to perturbation theory.}
If other moduli are present -- in addition to the expectation value of $\phi$ -- this argument holds for any values they may have.
So for perturbation theory to work, the effective potential must vanish identically 
as a function of the moduli of the compactification -- or at least all moduli that  parametrize vacua
that are accessible to perturbation theory.

\subsubsection{The Importance Of Spacetime Supersymmetry}\label{superimp}

Such a result will not hold without a special reason.  The reason usually considered is spacetime supersymmetry.  Typically, in the field theory
that arises as the low energy limit of a supersymmetric
string compactification to $d>3$ dimensions (or a  compactification with
more than 2 unbroken supercharges to $d=3$), it is straightforward to show that supersymmetry remains unbroken
to all orders of perturbation theory.  Conversely, in the relatively rare cases that supersymmetry
is unbroken at tree level but is spontaneously broken by loop effects, it is typically possible to predict this using 
the low energy effective field theory.

For example, the arguments that are relevant for analyzing the behavior in perturbation theory of a compactification
to 4 dimensions with $\N=1$ supersymmetry
are well-known.  If absent at tree level, a tadpole for the dilaton and other classical moduli can be triggered only by loop
corrections to the superpotential for chiral superfields, or a $D$-term for a $U(1)$ gauge field.   
A simple argument using holomorphy\footnote{The superpotential depends holomorphically on a complex field whose
real part is the dilaton and whose imaginary part is an axion-like field that in perturbation theory decouples at zero 
momentum.  This decoupling, plus the known dependence of loop corrections on the dilaton field, implies \cite{DStwo}
that a superpotential cannot be generated in loops.} shows that a superpotential 
is not generated in loops.  As for a $D$-term, this can be generated
in perturbation theory for a $U(1)$ gauge field that lacks a $D$-term at tree level,
but only at the one-loop level.\footnote{The Fayet-Iliopoulos $D$-term is a constant, independent of all fields.
The known dependence of loop corrections on the dilaton field implies \cite{DSW} that such a constant can only be generated
at one-loop order.}  These are the necessary facts to understand the fate in perturbation theory
of a classical supersymmetric vacuum of  closed oriented strings. For open and/or unoriented strings, one also
has to analyze one-loop anomalies,
which we defer to section \ref{anomalies}.

Supersymmetry breaking by loop effects is only harder to come by in models with more unbroken supersymmetry at tree level,
including all supersymmetric compactifications with $d>4$.  Rather than reviewing all the possible cases here, we will
just jump to $d=10$.
 A rather general reasoning shows that if a  supersymmetric action is going to have
a non-vanishing potential energy function $V(\phi)$ (or possibly $V(\phi,a)$ for Type IIB supergravity, which has a second scalar
field $a$ as well as the dilaton $\phi$), then the supersymmetrization of this interaction will require a non-derivative
Yukawa-like coupling
$W_{ij}(\phi)\psi^i\psi^j$ that is bilinear in fermion fields $\psi^i$.  In general, these fermion fields might
have spin 1/2 and/or spin 3/2.  In most
of the ten-dimensional supergravity theories, Lorentz invariance makes it impossible to write such a Yukawa-like coupling.
(For example, in the supergravity limit of the heterotic string, the massless neutral fermions are a spin $3/2$ 
field $\psi_I$ of one chirality
and a spin $1/2$ field $\lambda$ of opposite chirality; no Lorentz-invariant bilinear $\psi^2$, $\psi\lambda$, 
or $\lambda^2$ exists. There
are also massless chiral fermions in the adjoint representation of the gauge group, and again there is no 
Lorentz-invariant and gauge-invariant fermion bilinear.)
The only ten-dimensional supergravity theory in which a non-zero effective potential (including a constant potential -- a cosmological
constant) is not excluded by this simple argument is the Type IIA supergravity.  This theory actually does admit a supersymmetric
deformation with a nontrivial potential \cite{Romans}.  
However, this deformation violates the symmetry $(-1)^{F_L}$,
which assigns the value $-1$ (or $+1$) to states in the left-moving Ramond (or NS) sector, and which is conserved in perturbative
Type IIA superstring theory.  When supplemented by that last statement, low energy field theory predicts that a nontrivial potential
cannot be generated in any of the ten-dimensional superstring theories in perturbation theory.

We have here emphasized tadpoles for massless scalars, rather than for the trace of the metric tensor.  From a field
theory point of view, in the absence of massless scalars (or at least in the absence of massless scalar tadpoles),
a tadpole for the trace of the metric tensor simply represents the generation of a cosmological constant.  Assuming that this
tadpole is finite and small, a perturbative
expansion might exist in de Sitter or Anti de Sitter spacetime.  This situation is not realized in perturbative string theory (at least not for
$d>2$), because the structure of string perturbation theory is such
that if the dilaton tadpole vanishes, the vacuum energy also vanishes
and there is no tadpole for the trace of the metric.  It follows that in string theory compactifications, it is not necessary to
study a tadpole for the trace of the metric separately from the  dilaton tadpole.

\subsubsection{A Subtlety At Zero Momentum}\label{remark}

It probably is no coincidence that in addition to it being unnecessary to analyze a tadpole for the trace of the
metric separately from that for the dilaton, it is also difficult to do so.  The reason for the last statement is that although
at non-zero momentum, there are separate conformal or superconformal vertex operators for the dilaton and the graviton, at zero momentum
there is only a single Lorentz-invariant conformal or superconformal
vertex operator, which represents the coupling of a linear combination of the dilaton and the
trace of the metric.

For example, for bosonic strings, the operator in question is $\t c c \eta_{IJ}\t\partial X^I\partial X^J$, while for
the heterotic string, it is $\t c c \delta(\gamma)\eta_{IJ}\t\partial X^I D_\theta X^J$.  (Here $\eta_{IJ}$ is the Lorentz 
metric.) These formulas have simple
modifications for the other superstring theories.
In each case, there is no other conformal or superconformal primary with the same quantum numbers. This operator
couples to a linear combination of the dilaton and the trace of the metric, as analyzed in  \cite{BZ}.

It is also true from the point of view of the low energy effective
 field theory that there is only one tadpole condition at $\g$-loop order.  That is because the $\g$-loop contribution
to the 
effective potential has a known dependence on the dilaton $\phi$ and the spacetime metric $G$, namely it is proportional
to $\sqrt{\det{G}} \exp((2\g-2)\phi)$.

It is possible to construct separate $Q_B$-invariant
vertex operators for the dilaton and the trace of the metric, but the ``second'' vertex
operator is not conformal or superconformal.  As we remarked in section \ref{morge}, methods to compute with such
more general vertex operators are known \cite{OldPolch,PN}, but involve some extra machinery beyond what we have
described in the present paper.  For our purposes, we do not need to develop this machinery, since the tadpole
of an operator that is not superconformal cannot appear in a superconformal formalism.

\subsection{Back To String Theory}\label{zumico}

The reader may have noticed that although we began in section \ref{brstanom} with
a discussion of BRST anomalies in string theory, the field theory discussion of section \ref{fieldrev} focused
on infrared singularities -- poles associated to mass renormalization and massless tadpoles -- rather than anomalies.
The two effects are different, but closely related.  Both involve the effects of on-shell intermediate particles.

It is possible to understand heuristically why, in string theory,
mass renormalization and massless tadpoles are associated to BRST anomalies.  BRST symmetry in string theory
incorporates the  mass shell condition for string states, so if the mass shell 
condition is modified by quantum mass renormalization, this will show up as an
anomaly in the classical BRST symmetry.  Likewise, BRST symmetry incorporates the equations of motion
for background fields.  A non-zero massless tadpole means that the equations of motion have not been satisfied,
so BRST symmetry will fail.

\begin{figure}
 \begin{center}
   \includegraphics[width=5in]{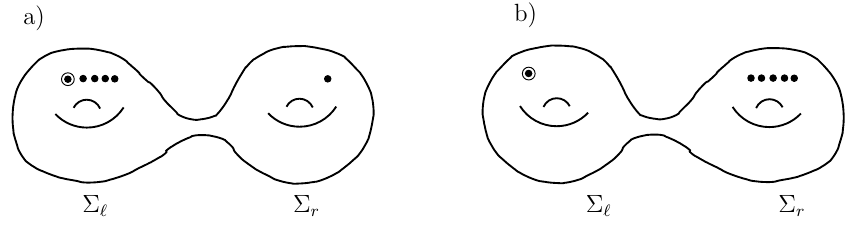}
 \end{center}
\caption{\small A string worldsheet $\Sigma$ degenerates to a union of two branches $\Sigma_\ell$ and $\Sigma_r$
in a manner  related to mass renormalization and anomalies.  
A dot represents a generic conformal vertex
operator which may be any of $\V_2,\dots,\V_\sn$, while a dot surrounded by a circle represents the $Q_B$-trivial
vertex operator $\V_1=\{Q_B,\W_1\}$.  There are two interesting configurations -- a separating degeneration (a) on which
all but 1 of $\V_2,\dots,\V_\sn$ are on the same branch as $\V_1$, and a separating degeneration (b) in which they are
all on the opposite branch. }
 \label{kopla}
\end{figure}

 \begin{figure}
 \begin{center}
   \includegraphics[width=2.5in]{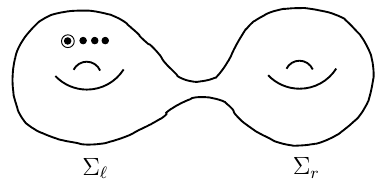}
 \end{center}
\caption{\small A separating degeneration related to tadpoles and the associated anomalies.  All vertex operators,
including the BRST-trivial vertex operator labeled by a circle, are on the same branch.}
 \label{nopla}
\end{figure} 
In figs. \ref{kopla} and \ref{nopla}, we have sketched the separating degenerations that can lead to BRST anomalies.  
In each case, we consider a product
of $\n$ conformal or superconformal vertex operators $\V_1,\dots,\V_\sn$ on a string worldsheet $\Sigma$, of genus
$\g$.  These operators
are all $Q_B$-invariant, have canonical ghost number and (in the superstring case) picture number, and are annihilated
by $b_n,\beta_m$, $n,m\geq 0$.  We single out one of these operators, say $\V_1$, to be BRST-trivial, obeying
$\V_1=\{Q_B,\W_1\}$,
where $\W_1$ obeys the same conditions as the $\V_i$ except that its ghost number is 1 less (and, of course, $\W_1$ is
not $Q_B$-invariant).  $\Sigma$ degenerates to a union of two components $\Sigma_\ell$ and $\SIgma_r$, where we can
assume that $\V_1$ is inserted on, say, $\Sigma_\ell$.  There are two interesting cases related to mass renormalization
-- the operators $\V_2,\dots,\V_n$ may be all but one on $\Sigma_\ell$ or all on $\Sigma_r$, as in figs. \ref{kopla}(a) and
(b), respectively.   There is essentially only one configuration associated to massless tadpoles, with all vertex operators
on $\Sigma_\ell$, as in fig. \ref{nopla}.

To properly understand any of these three cases, we need a sharper result than the one that we obtained in section
\ref{brstanom}.  Let us first recall the description in section \ref{stad} of the residue of the pole at $L_0=0$.  In that
analysis, we showed that the residue of the pole can be computed by inserting on each side conformal vertex
operators associated to physical states.  For example, for bosonic open strings, according to eqn. (\ref{zorom}),
the residue is computed by inserting $\sum_i c\UU_i\otimes c\UU^i$, where the sum runs over a complete set of matter
primaries of $L_0=1$ (modulo null vectors), 
and the two factors are to be inserted on $\Sigma_\ell$ and $\Sigma_r$, respectively.
An essential point here is that the residue of the pole can be computed by inserting only the vertex operators
associated to physical states, not to more general states of $L_0=0$.  (In fact, the conformally-invariant formalism
assumed in this paper does not enable us to define insertions of such more general vertex operators.)

By contrast, in section \ref{brstanom}, we associated the anomaly to states of $L_0=0$ propagating between $\Sigma_\ell$
and $\SIgma_r$, but we did not restrict these to physical states.  To understand the BRST anomalies associated to the
processes sketched in the above figures, we need a sharper result, more similar to that of section \ref{stad}, relating
the anomaly to couplings of physical states.

There is an obvious and almost correct guess for how to compute the anomaly.   
One step in the derivation in section \ref{brstanom} was to omit
the usual factor of $b_0$ that appears in the residue of the $1/L_0$ pole.  According to (\ref{yrfe}), this means that
we need to replace $\sum_i c\UU_i\otimes c\UU^i$ by \begin{equation}\label{numbody}
\sum_i (c\partial c\UU_i \otimes c\UU^i + c\UU_i\otimes c\partial c\UU^i).\end{equation}
So on either $\Sigma_\ell$ or $\Sigma_r$ we must insert an operator $c\partial c \UU_i$ rather than $c\UU_i$, with
ghost number greater by 1 than the canonical value.  On the other hand, when we calculate the anomaly
by evaluating the right hand side of eqn. (\ref{tarmib}), one of the operators inserted on $\Sigma_\ell$ is
the operator $\W_1$, whose ghost number is less by 1 than the canonical value.  So to get a non-zero
path integral, we should insert $c\partial c\UU_i$ on $\Sigma_\ell$ and the conformal vertex operator 
$c\UU^i$ on $\Sigma_r$.

Both the derivation and the conclusion in the last paragraph are oversimplified.  This must be so for the following
simple reason: our formalism
does not allow an insertion of the operator $c\partial c\UU_i$, since it is not annihilated by $b_0$.   A more precise
explanation is given in section \ref{lazy}.  However, it is true that, as the heuristic argument in the last paragraph
suggests, the anomaly can be evaluated as a sum of contributions in each of which a conformal or superconformal
vertex operator  is inserted on $\Sigma_r$.  Thus, the anomaly
can be evaluated as a path integral on $\Sigma_r$ with an insertion of
\begin{equation}\label{pilb} \O=\sum_i a_i \Y^i, \end{equation} 
where the sum runs over all conformal or superconformal vertex operators $\Y^i$ of the appropriate momentum
and the coefficients $a_i$ are computed by path integrals
on $\Sigma_\ell$.  For open bosonic strings, the $\Y^i$ are $c\UU^i$, where $\UU^i$ are matter primaries of $L_0=1$.
The analogs for the other string theories are familiar.

We can now get a clearer understanding of the various BRST anomalies.
Let us begin with the tadpole-like diagram of fig. \ref{nopla}. 
The genus $\g_r$ of $\Sigma_r$  is strictly positive (or the indicated 
degeneration would not arise in the Deligne-Mumford
compactification).\footnote{\label{ygr} With the right definition of the genus, this statement holds in all string theories,
not just closed oriented string theory.
For any possibly open or unorientable string worldsheet $\Sigma$ of Euler characteristic $\chi$,
we define the genus $\sg$ by $\chi=2-2\sg$.  For example, by this definition, a disc has $\sg=1/2$.  The general topological
constraint
for fig. \ref{nopla} to be relevant in string perturbation theory is then indeed simply $\sg_r>0$.}  For such a degeneration, the anomaly can be computed
by inserting $\O=\sum_i a_i \Y^i$ on $\Sigma_r$, where the $\Y^i$ are conformal or superconformal primaries of 
spacetime momentum
$P=0$.  Thus the anomaly is $\sum_i a_i \langle \Y^i\rangle_{\sg_r}$, where $\langle\Y^i\rangle_{\sg_r}$ is the tadpole of
$\Y^i$ in genus $\g_r$.  In particular, if the massless tadpoles vanish, then the anomalies associated to the tadpole
diagrams also vanish.   The reader would probably guess this result based on sections \ref{brstanom} and \ref{fieldrev}.

The basic fact about fig. \ref{nopla} is really the tadpole (the perturbative vacuum does not satisfy
the loop-corrected equations of motion) rather than the BRST anomaly (which reflects the fact that in string theory
the equations of motion
are built into the BRST symmetry).  For oriented closed strings, there is not really a BRST anomaly independent of the
equations of motion.  Matters are different for open and/or unoriented strings, as we analyze in section \ref{tadanomalies}.
In that case, anomalies have a life of their own, independent of the equations of motion.

The next most obvious case is perhaps fig. \ref{kopla}(a) in which there is just one vertex operator, say $\V_\sn$,
supported on $\Sigma_r$.  The anomaly is
\begin{equation}\label{ticox}\sum_i a_i \langle \Y^i \,\V_\sn\rangle_{\sg_r}. \end{equation}
By momentum conservation, we can assume that the spacetime 
momentum carried by $\Y^i$ is minus that of $\V_\sn$.
The expression $\langle \Y^i\,\V_\sn\rangle_{\sg_r}$ represents a genus $\g_r$ coupling of two physical states; it is
a contribution to mass renormalization for the physical states in question.  Thus  if the state
corresponding to $\V_\sn$ undergoes no 
mass renormalization in genus $\g_r$, then there is no anomaly of this type.  
This again is probably the result
that the reader would guess.  

We are left with the anomaly of fig. \ref{kopla}(b).    This 
is best-understood as a loop correction to classical BRST symmetry.  At tree-level, the
state $\{Q_B,\W_1\}$ decouples from the $S$-matrix.  The genus $\g_\ell$ of $\Sigma_\ell$ is strictly
positive,\footnote{See footnote \ref{ygr} for the most general version of this statement.} 
or the configuration of fig. \ref{kopla}(b) does not arise in the Deligne-Mumford compactification.
The anomaly of fig. \ref{kopla}(b) means that in genus $\g\geq \g_\ell$, 
the coupling of $\{Q_B,\W_1\}$ is not zero but equals  the coupling of $\O=\sum_ia_i\Y^i$ in 
genus $\g_r$.  In other words, what decouples
is not $\{Q_B,\W_1\}$ but 
\begin{equation}\label{monok} \QQ_B(\W_1)=\{Q_B,\W_1\}-g_\st^{2\sg_\ell}\O(\W_1),  \end{equation}
 where we write $\O(\W_1)$ (and not just $\O$) to emphasize
that $\O$ is a linear function of $\W_1$. Since $\O(\W_1)$ is a physical vertex operator, 
obeying $\{Q_B,\O(\W_1)\}=0$, the deformation
preserves the fact that  $\QQ_B^2=0$, to this order.  Thus, BRST
symmetry is retained in this order, but deformed.   We discuss the physical interpretation 
of this phenomenon in section \ref{burgo}.

Going back to fig. \ref{kopla}(a),
one might be slightly puzzled about why mass renormalization of the string state associated to $\V_\sn$ causes
an anomaly in the BRST symmetry $\V_1\to \V_1+\{Q_B,\W_1\}$ associated to particle 1. An intuitive reason for
this has already been explained.  To show vanishing of 
$\langle\{Q_B,\W_1\}\,\V_2\dots\V_\sn\rangle$, one needs BRST invariance of
$\V_2,\dots,\V_\sn$.  But since classical BRST symmetry incorporates the mass shell conditions, mass renormalization
of the string states associated to $\V_2,\dots,\V_\sn$ implies a correction to the condition of BRST invariance for
those vertex operators, leading to a failure of BRST-invariance for $\V_1$.

We can write a formula for the correction to BRST symmetry of physical states by representing  the anomaly
by the operator insertion in (\ref{numbody}).  Given this and defining $y^i=\langle c\UU^i \,\V_\sn\rangle_{\Sigma_r}$, 
the anomaly
amounts to replacing $\{Q_B,\V_\sn\}$ by $\sum_i y^i c\partial c\UU_i$.  The corrected BRST symmetry is 
\begin{equation}\label{pilox}\QQ_B(\V_\sn)=\{Q_B,\V_\sn\}+\sum_i y^i c\partial c\UU_i, \end{equation}
a formula described in \cite{Sen}.  Eqns. (\ref{monok}) and (\ref{pilox}) give the leading corrections to the BRST symmetry
for two adjacent values of the ghost number, corresponding to gauge parameters and physical states, respectively.  A systematic
framework to study such deformations to all orders has been developed \cite{Sen4}, though this has not been expressed in the
super Riemann surface language.

\subsection{Gauge Symmetry Breaking}\label{burgo}

Our next task is to explain the physical meaning of 
the correction to  BRST symmetry 
described in eqn. (\ref{monok}).  At least for gauge symmetries of massless string states,
this correction represents spontaneous breaking of gauge symmetry. (For massive string states, the gauge symmetries are already
spontaneously broken at tree level, so the corrections to the BRST transformations do not seem to have an interpretation
in terms of spontaneous symmetry breakdown.)

For example, $\V_1$ might be the vertex operator of a gauge field $A$ that is massless at tree level.  In bosonic open string
theory, we might have $\V_1=c\,\varepsilon\cdot \partial X\exp(ip\cdot X)$, where $p$ and $\varepsilon$ are momentum
and polarization vectors, obeying $p^2=\varepsilon\cdot p=0$.  In this example, the gauge parameter is $\W_1=\exp(ip\cdot X)$;
it generates the gauge transformation $\varepsilon\to \varepsilon-i\lambda p$ (where $\lambda$ is a constant).
We could consider in a similar way any other string theory that has a massless gauge field at tree level, coming from an open-
or closed-string state; the specific formulas
for $\V_1$ and $\W_1$ will not be important for now.  The operator $\W_1$ is not the vertex operator of a physical
state, since it is not $Q_B$-invariant, but
 $\O(\W_1)$, if not zero,
 is the vertex operator of a physical state with the same Poincar\'e quantum numbers as $\W_1$. So it carries spacetime momentum $p$ and is invariant under the little group (the subgroup of Lorentz transformations that
keep $p$ fixed).  In other words, $\O(\W_1)$ is the vertex operator of a scalar field $\sigma$ that is massless at tree level.  
That is the key point.
The formula $\QQ_B(\W_1)=Q_B(\W_1)-g_\st^{2\sg_\ell}\O(\W_1)$ tells us that  a gauge transformation $\varepsilon\to \varepsilon-i\lambda\,
p$
generated by $\W_1$ must be accompanied by an insertion of the $\O(\W_1)$ vertex operator with a coefficient
$-g_\st^{2\sg_\ell}\lambda$ or in other words by the shift
 \begin{equation}\label{wobbly}\sigma\to \sigma-g_\st^{2\sg_\ell}\lambda.\end{equation}   This leads to what is usually called
gauge symmetry breaking.  The essential phenomenon (but without the factor of $g_\st^{2\sg_\ell}$) is exhibited by the basic
Stueckelberg model of $U(1)$ gauge symmetry breaking,
\begin{equation}\label{incvo}\int\left(F_{IJ}F^{IJ}+(\partial_I\sigma+A_I)^2\right), \end{equation}
which is invariant under $A\to A+\partial\lambda$, $\sigma\to \sigma-\lambda$.  In vacuum, $\sigma=0$ up to a gauge transformation; in expanding around this vacuum, $A$ is a massive spin 1 particle.

The power of $g_\st$ given in eqn. (\ref{wobbly}) is correct if $A$ and $\sigma$ are both closed-string states or both 
open-string states; it requires some modification otherwise, as we discuss presently.  This power of $g_\st$, which is not
a constant but equals $e^\phi$ with $\phi$ the dilaton field, is unnatural in the context of gauge symmetry breaking
and this gives one reason to suspect that the phenomenon just described will be hard to realize in string theory.  
This is true and will be explained below.

For another example, $\V_1$ might be the vertex operator for the massless spin $3/2$ gravitino field $\psi_I$ in a supersymmetric
string compactification.  In that case, $\W_1$ would have the Poincar\'e quantum numbers of a spin 1/2 field obeying the massless
Dirac equation, and
$\O(\W_1)$ is the vertex operator for a physical fermion field $\lambda$ that is massless at tree level. (The fermion fields corresponding
to $\V_1$ and to $\O(\W_1)$ have the same spacetime chirality.)  The anomaly means that a gravitino gauge transformation,
which at the linearized level is $\psi_I\to \psi_I+\partial_I\zeta$, where $\zeta$ is a $c$-number
spinor field with the same chirality
as $\psi$, must be accompanied by 
\begin{equation}\label{ofra}\lambda\to \lambda-g_\st^{2\sg_\ell}\zeta.\end{equation}
  $\lambda$ is usually called the Goldstino,
and what we have just described is spontaneous breaking of supersymmetry.

For a further example, one might be tempted to consider $p$-form gauge fields arising in the Ramond-Ramond
sector of superstrings.  However, these are not a useful example because in perturbative string
theory there are no Ramond sector gauge parameters\footnote{This is related to the fact that the vertex operator of a Ramond-Ramond field, in the canonical picture which we use to
compute the $S$-matrix, describes the $p+1$-form field strength rather than the $p$-form gauge field. See
section \ref{primo} for one explanation of that fact.} at the massless level; the first Ramond sector gauge parameter arises at the first massive level (see eqn. \ref{tolver}).  
Instead, the last example relevant to our discussion arises for the $B$-field of closed oriented strings.   If $\V_1$ 
is the $B$-field vertex operator, then the corresponding
gauge parameter \begin{equation}\label{mibz}\W_1=\varepsilon_I \cdot (c^z\partial_z X^I-\t c^{\t z}\partial_{\t z}X^I) \exp(ip\cdot X),~~~p^2=\varepsilon\cdot p=0,\end{equation} has the Poincar\'e quantum numbers of a $U(1)$ gauge field, and $\O(\W_1)$, if not zero,
actually is the vertex operator for a $U(1)$ gauge field $A$ that is massless at tree level.  The anomaly means that
a $B$-field gauge transformation $B_{IJ}\to B_{IJ}+\partial_I\lambda_J-\partial_J\lambda_I$ 
must be accompanied by $A_I\to A_I -g_\st^{2\sg_\ell}\lambda_I$.  This results in the breakdown of $B$-field gauge-invariance;
$A$ can be gauged away and $B$ becomes massive \cite{KR}.  

Based on field theory intuition, one might surmise -- correctly -- that the phenomena we have described here are actually
rather rare among string theory compactifications.  We have already explained in section \ref{superimp} that spontaneous
supersymmetry breaking by loops -- in models in which supersymmetry is unbroken at tree level -- can occur only under
rather special circumstances.  The only known models are those described in \cite{DSW}, involving the generation of a Fayet-Iliopoulos
D-term at one-loop order.
For ordinary bosonic gauge symmetries, the difficulty in triggering symmetry breaking by loops
is even more obvious.  If all charged scalar fields have positive masses
at tree level, then perturbative quantum corrections cannot trigger gauge symmetry breaking; if there is a charged scalar
field with a negative mass squared at tree level, then the gauge symmetry was spontaneously broken already classically.
For a gauge symmetry unbroken at tree level to be spontaneously broken by weak quantum corrections, there must be a charged
scalar field that is massless at tree level and acquires a negative mass squared from perturbative quantum corrections.

This is obviously rather special, but it can be natural if supersymmetry accounted for existence of massless 
charged scalars at tree level and is spontaneously
broken by loop effects, as in the models of \cite{DSW}. 
But even then, this mechanism for gauge symmetry breaking as a result of loop effects does not correspond to the BRST anomaly
described in eqn. (\ref{monok}).  That anomaly is a loop correction to the gauge transformation laws of the fields, not to
the masses.

Field theory offers only less guidance about how spontaneous breakdown of $B$-field gauge symmetry might be triggered by small
quantum corrections.  

\subsubsection{Examples}\label{examples}

Given all this, one might despair of finding an example in which the perturbative BRST anomaly of eqn. (\ref{monok})
occurs for bosonic gauge symmetries of massless modes of string theory.  In fact, this happens only in a very restricted way.
To understand where to look, it helps to start with something that was explained in sections \ref{moregauge} and \ref{morgauge} 
for closed bosonic strings and the NS sector of closed superstrings.  (The phenomenon under discussion does not
occur for the R-R sector, as explained above.)  

The key point is that in those cases, the gauge invariances of massless fields can be proved by integration by parts on
the worldsheet $\Sigma$, as opposed to the moduli space of Riemann surfaces. 
We consider a closed-string vertex operator $\V$ whose integrated form is $V$.
$V$ is a $(1,1)$-form constructed from the matter fields only, and couples via the integral
$\int_\Sigma V$.   A gauge transformation
acts by $V\to V+\d W$, where $W$ is a 1-form constructed from the matter fields.  For example, if $V$ is the vertex operator
of the $B$-field of the closed oriented bosonic string, then
 \begin{equation}\label{roxt}W=\varepsilon_I \,\d X^I\exp(ip\cdot X),~~~p^2=\varepsilon\cdot p=0.\end{equation}
The gauge symmetry is simply 
 the vanishing of  $\int_\Sigma \d W$.
In closed-string theory,
there is no problem with the vanishing of that integral.
The only singularities in the integral occur when $W$ meets another vertex operator
$\X$, 
in which case the worldsheet $\Sigma$ splits off a genus zero component containing $W$ and $\X$.  This is not
a degeneration of the form 
of fig. \ref{kopla}(b) that can lead to quantum corrections to gauge-invariance; the total momentum carried by $W$ and $\X$
is generically off-shell, so a collision between them does not lead to an anomaly.

The situation for open-string vertex operators is no different.  
If $\W$ is an open-string vertex operator, its integrated form is a 1-form
$W$, constructed from the matter fields, that couples via $\oint_{\partial\SIgma}W$.  A gauge transformation acts
by $W\to W+\d P$, where now $P$ is a scalar or 0-form operator constructed from the matter fields.  But $\d P$ decouples,
since $\oint_{\partial\Sigma}\d P=0$.  Again, there may be singularities where $P$ meets another open-string vertex
operator, but they do not affect the vanishing of the integral. 

But something new does happen in the combined theory of open and closed strings, specifically for the case of a 
closed-string gauge parameter. Let us go back to the closed-string vertex operator $\V$ with integrated form $V$
and gauge-invariance $V\to V+\d W$.
 Clearly, if $\Sigma$ has a nonempty boundary, then $\int_\Sigma\d W$ need not vanish.  Rather,
by Stokes's theorem, it equals $\int_{\partial\Sigma}W$.  Here we can regard $W$ as an open-string vertex operator,
in integrated form.  For example, in the case that $\V$ was the vertex operator for the BRST-invariant $B$-field,
the gauge parameter $W$ was presented in (\ref{roxt}), and we recognize it as the classic formula for the vertex
operator of a massless open-string gauge field.  When $W$ is understood as an open-string vertex operator, rather
than a closed-string gauge parameter, 
we will denote it as  $W_{\mathrm{open}}$. The corresponding unintegrated vertex operator is  $\W_{\mathrm{open}}=cW_{\mathrm{open}}$.

Clearly, in this situation,
the closed-string mode that is represented (in integrated form) by $V=\d W$ does not decouple from the $S$-matrix;
rather its coupling is that of the open-string mode with integrated
vertex operator $W_{\mathrm{open}}$, as originally explained long ago \cite{KR}.  
(The original construction was for breakdown of $B$-field gauge invariance; to get a similar model with breaking of
a $U(1)$ gauge symmetry, it suffices to compactify the target space on a circle, whereupon part of the $B$-field
becomes a $U(1)$ gauge field which undergoes the same mechanism.)
In our terminology, the fact that the closed-string
pure gauge mode $\V=\{Q,\W\}$ (or in integrated form $V=\d W$)  does not decouple,
but rather couples via the open-string
vertex operator $W_{\mathrm{open}}$, is expressed in the form
\begin{equation}\label{melno}\O(\W)=\W_{\mathrm{open}},~~\W_{\mathrm{open}}=cW_{\mathrm{open}}. \end{equation}

Indeed, this type of
 example  perfectly illustrates eqn. (\ref{monok}).  The boundary term that prevents decoupling of $\d W$ occurs
when $W$ approaches the boundary of $\Sigma$.  From the point of view of the Deligne-Mumford compactification,
this means that $\Sigma$ splits off an additional component $\Sigma_\ell$
 which is a disc that contains $\W$, but contains none of the other
vertex operators that may originally have been present.  (See section \ref{geometry} for more on such open-string
degenerations.)  In other words, this example fits the framework of fig. \ref{kopla}(b),
with $\Sigma_\ell$ being a disc, $\V_1=\{Q,\W_1\}$ being a closed-string vertex
operator, $\O(\W_1)$ an open-string vertex operator, and no restriction on $\Sigma_r$ or $\V_2,\dots,\V_\sn$.

Though it perfectly fits the framework of fig. \ref{kopla}(b), the example challenges the terminology that we used
in describing that figure.   A worldsheet with disc topology generates the classical action of open strings, but 
is a correction of some sort -- it is not clear that one should call it a quantum correction -- for the closed strings.  With the
most natural way to normalize the vertex operators, there is no factor of $g_\st$ in the correction to BRST symmetry
in this example; instead the transformation $\W\to \O(\W)$ maps closed-string gauge parameters to open
string vertex operators (whose natural scaling with $g_\st$ is different).  
For example, for the breakdown of $B$-field gauge symmetry, the action is naturally written
\begin{equation}\label{xix}\int\d^Dx\left(\frac{1}{g_\st^2}H_{IJK}^2+\frac{1}{g_\st}(\partial_I A_J-\partial_J A_I
+B_{IJ})^2\right),~~~H=\d B. \end{equation}
The gauge symmetry is $\delta B_{IJ}=\partial_I \lambda_J-\partial_J\lambda_I$, $A_I\to A_I-\lambda_I$,
with no factor of $g_\st$.  

Symmetry breaking by mixing of open and closed strings can also occur for spacetime
supersymmetry; this is discussed in section \ref{strapple}.  But, even though there are severe constraints,
 there are additional possibilities for spontaneous breaking of spacetime supersymmetry by loop effects
that do not have analogs for gauge symmetries associated to massless bosonic gauge fields.
That is because (section \ref{tramp}) there is no forgetful map for Ramond punctures.  
Decoupling of a pure gauge mode of the gravitino field can only be proved by a full 
integral over the moduli space of super Riemann
surfaces, not simply by an integral over $\Sigma$.  As a result, some of the 
arguments that apply for bosonic gauge symmetries cannot be used to constrain supersymmetry breaking by
loops.  In particular, in the models
described and studied in \cite{DSW,DIS,ADS}, the phenomenon of fig. \ref{kopla}(b) does 
occur, with $\Sigma_\ell$ having
genus 1.  From the point of view of low energy effective field theory, what happens is the following.  
At tree level, the theory
has a massless neutral spin 1/2 field $\lambda$ with a supersymmetry transformation law 
$\delta\lambda = D+\dots$,
where $D$ is the auxiliary field in the vector multiplet associated to a $U(1)$ gauge symmetry.
At tree level, $D$ vanishes in a vacuum with unbroken $U(1)$ gauge symmetry. At one-loop order,
 the value of $D$ in
such a vacuum
is shifted by a constant $\zeta$ (called the  Fayet-Iliopoulos term) and so the transformation law becomes
$\delta\lambda=\zeta+\dots$, where the $\dots$ terms are bilinear and higher order in charged fields.  The constant term means that, 
at least if we keep the $U(1)$ unbroken, supersymmetry is spontaneously broken.  (For a framework, not expressed in terms
of super Riemann surfaces, in which one can study the restoration of supersymmetry with the $U(1)$ symmetry spontaneously
broken, see \cite{Sen5}.)

From the vantage point  of the present paper, the main importance of 
the phenomenon depicted in fig. \ref{kopla}(b) is
that it is a failure mode for superstring perturbation theory.  We have to know 
that loop corrections do not trigger the breaking
of spacetime supersymmetry in order to have any hope of using supersymmetry to prove the vanishing of massless tadpoles.  
Happily, there is no difficulty, as supersymmetry breaking by loops is highly constrained and it is possible to effectively determine
when it occurs. 

\subsubsection{Some Miscellaneous Remarks}\label{misc}

We conclude this discussion with some miscellaneous remarks.

First of all, spontaneous gauge symmetry breaking always leads to mass renormalization, since the gauge field gains a non-zero
mass.  When this happens, as in any situation with mass renormalization,
the $S$-matrix diverges if one attempts to calculate
on the classical mass shell. 

Although spontaneous gauge symmetry breaking always leads to mass renormalization for a massless gauge field,
the converse is not true. (In contrast to gauge symmetry breaking, the phenomenon that we are about to describe
is limited to abelian gauge fields.)  For example, consider a closed-string compactification to four dimensions
with a $U(1)$ gauge field $A$ and also a two-form
gauge field $B$.   At tree level, the action is schematically
\begin{equation}\label{dombo}\int_{\R^4}\d^4x \, e^{-2\phi}\left((\d A)^2+(\d B)^2\right). \end{equation}
At one-loop order, one may generate a gauge-invariant interaction
\begin{equation}\label{zombo}\int_{\R^4}B\wedge \d A. \end{equation}
This interaction can only be generated at one-loop order, since the presence of a non-zero power of $g_\st=e^\phi$
would spoil the  invariance under the $B$-field gauge transformations
$B\to B+\d\Lambda$; the proof of this invariance requires integration by parts.  
If generated, the interaction (\ref{zombo}) leads to mass renormalization for $A$ and $B$, without
modifying the classical gauge-invariance.  The fact that this interaction can only be generated at one-loop order means
that it is possible, in a given string theory compactification, to effectively determine whether it is generated or not.  The main known
case in which such a one-loop interaction is generated is the class of models studied in \cite{DSW,DIS,ADS}.  The mechanism
for generating this interaction in these models is as follows: in certain heterotic string compactifications to four dimensions, the ten-dimensional 
Green-Schwarz interaction $B\wedge \Tr \,F^4+\dots$  (which arises in heterotic string theory at one-loop order)
generates the interaction (\ref{zombo}). When this interaction is generated at one-loop order, the one-loop $S$-matrix
elements of $A$ or $B$ are divergent, because of the mass renormalization or more precisely
the on-shell pole that results from the $A$-$B$ mixing.\footnote{This example is rather tricky.  Even when
the interaction (\ref{zombo}) is generated at one-loop level, the on-shell one-loop
 two-point function $\langle A\,B\rangle$
vanishes.  (This follows just from conservation of angular momentum.)  Accordingly there is no one-loop BRST anomaly.  However, as stated in the text, the mixing
and the resulting pole
do cause a divergence in one-loop $S$-matrix elements of $A$ or $B$.}

From a field theory point of view, in such a four-dimensional model,
one can dualize $B$  to a scalar field $\sigma$.  In the dual description, $A$ gets a mass from gauge symmetry
breaking.  Outside of perturbation theory, there is no distinction between massless spin-zero particles that are
associated to spin-zero fields and those that are associated to two-form fields; nonperturbative dualities can exchange
different descriptions of the same massless modes.  
In string perturbation theory, however, there are two kinds of massless spin-zero mode
in four dimensions
-- those whose vertex operator is derived from a zero-form and those whose vertex operator is derived from a two-form.
Therefore, in perturbation theory, the two mechanisms for an abelian gauge field to gain a mass, by mixing with a spin-zero
mode of one of the two types, are distinct.  Both mechanisms are severely restricted, in ways that we have
explained.

\subsubsection{Restricting To The Massless $S$-Matrix}\label{massmat}

Although perturbative mass renormalization is heavily constrained for massless gauge fields, it is  almost ubiquitous for massive string 
states.
The only obvious exception is that in some models, massive BPS states are not subject to mass renormalization.
 
This leads us to a problem that will  place a significant restriction on what we can accomplish in the rest of this
paper.   As soon as there is mass renormalization, the usual conformal or superconformal 
framework of superstring perturbation theory does not work.   It is necessary to go slightly off-shell in order to proceed.
This can be done by endowing all Riemann surfaces or super Riemann surfaces with local parameters at punctures
(in the super Riemann surface case, the analog of a local parameter is a local system of superconformal coordinates).
This approach goes back to \cite{OldPolch,PN} and has been much developed recently for superstrings (in the language of 
picture-changing operators, not super Riemann surfaces); for example see  \cite{Sen1,Sen2,Sen3,Sen4,
Sen5,Sen6}.

However, in the present paper, to avoid an extra layer of complication, we prefer to avoid these issues.
Therefore, we will limit ourselves in the rest of this paper to studying the $S$-matrix of massless particles.  Moreover
we will consider only supersymmetric theories and more specifically only theories in which spacetime supersymmetry
and general considerations of low energy field theory suffice to show that in perturbation theory
there is no spontaneous supersymmetry breaking and no mass renormalization
for any of the particles that are massless at tree level.   (This includes most supersymmetric theories, as we have explained.)
 
The restriction to the massless $S$-matrix is unfortunate.  However, in one sense the restriction
is less severe than it might appear.  Massive particles that decay to stable ones are most precisely understood as resonances
in the $S$-matrix of massless particles, so it is really only 
the existence of {\it stable} massive particles that obstructs our ability to compute
the complete $S$-matrix in perturbation theory.  For example, in 
four of the five ten-dimensional superstring theories, the massive string states are all unstable against decay to massless
ones, so the massless $S$-matrix is the complete $S$-matrix in perturbation theory.  The exception is the 
$\mathrm{Spin}(32)/\Z_2$ heterotic string, whose perturbative spectrum includes stable massive particles in the spinor representation of the gauge group.

\subsection{Massless Tadpoles In String Theory}\label{turgo}

Ultimately, the most critical question for superstring perturbation theory is to make sure that the vacuum state that one attempts
to construct in perturbation theory is not destabilized by massless tadpoles.  The show this, we will use spacetime supersymmetry.
As just explained, we consider only the massless $S$-matrix, and only string compactifications in which there is no perturbative
mass renormalization for massless particles.  

Additionally, to begin with, we consider only closed oriented strings.  This leads to several simplifications.
One immediate simplification is that for closed, oriented strings, we only  have to consider tadpoles in the NS sector (or the NS-NS sector
in the case of Type II superstrings).  R-R tadpoles are prevented both by the $(-1)^{F_L}$ symmetry\footnote{The superconformal
formalism only works when the expectation values of R-R fields vanishes.  This ensures the existence in perturbation theory
of a
symmetry $(-1)^{F_L}$ that acts as $+1$ or $-1$ on states from the left-moving NS or R sector.} of closed oriented Type II
string theory, and by the decoupling of R-R fields at zero momentum.  In open and/or unoriented superstring
theory, there are some new ingredients  (there is no $(-1)^{F_L}$ symmetry,  and the decoupling of R-R fields
at zero momentum can fail under certain circumstances), and R-R tadpoles become relevant. We discuss these matters
 in  section \ref{anomalies}.

The proof that the amplitudes are spacetime supersymmetric and free of massless tadpoles will proceed by induction in the genus
$\g$ of the string worldsheet.  Assuming that there are no massless tadpoles and that spacetime supersymmetry holds up to genus
$\g-1$, we will show that this is also true in genus $\g$.  

The arguments are deferred to section \ref{tadpoles}.   However, one important detail about the interpretation
of the result will be explained here.

Whenever one has a perturbative superstring $S$-matrix, it depends on at least one modulus, namely the string coupling
constant $g_\st=e^\phi$.  There may possibly be other moduli.  Let $\ZZ$ be the moduli space that is parametrized by $g_\st$ and
any  other moduli.  The output of our discussion will be to show that the perturbative $S$-matrix is well-defined as a function
on $\ZZ$.  But in general there is no reason to expect $\ZZ$ to have a natural parametrization in terms of fields such as $\phi$.  

We will find that in general there is no such natural parametrization.  This will be the conclusion of an analysis
in section \ref{inttad} of infrared divergences and their regularization.   We will find that when massless tadpoles
vanish, the integrals that have to be evaluated to compute
the $S$-matrix are convergent, but only conditionally so.
Different infrared regulators lead to results for the $S$-matrix that differ by reparametrizations of $\ZZ$.  (Some of the issues have
been treated in \cite{LaN}.)

What we have just explained has an analog for wavefunction renormalization (section \ref{wavefunction}).  When
mass renormalization vanishes, this causes the cancellation of certain logarithmically divergent contributions
to the scattering amplitudes.  One is left with integrals that are conditionally convergent.   The result of regulating
these conditionally convergent integrals is unique up to a coupling-dependent renormalization of the vertex operators
that are used to compute the scattering amplitudes.  So the $S$-matrix is naturally determined, but the vertex operators
that are used to evaluate it are not.  

Though this will not be explored in the present paper,
one anticipates that in compactifications with much unbroken supersymmetry, $\ZZ$ will often have
 a natural parametrization or at least
a small class of natural parametrizations.   This would correspond in our analysis in section \ref{inttad} to using
supersymmetry to find a distinguished infrared regulator.

\subsection{What Happens When The  Tadpole Vanishes}\label{inttad}

\subsubsection{Conditionally Convergent Integrals}\label{concon}

In describing how to proceed when tadpoles vanish, we will use the language of bosonic string theory  since
the subtleties of worldsheet and spacetime supersymmetry  play no essential role.  We ignore the tachyon of bosonic
string theory as we are really interested in applying the reasoning that follows to tachyon-free superstring theories.   (For
a systematic approach to these issues, not expressed in the super Riemann surface language, see \cite{Sen4}.)

\begin{figure}
 \begin{center}
   \includegraphics[width=2.5in]{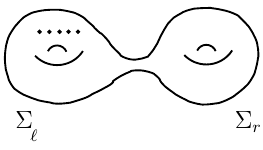}
 \end{center}
\caption{\small  A Riemann surface $\Sigma$ splits into two components $\Sigma_\ell$ and $\Sigma_r$.  All external vertex operators
are on $\Sigma_\ell$.}
 \label{qoneck}
\end{figure} 
We consider the separating degeneration of fig. \ref{qoneck} in which a Riemann or super Riemann surface
surface $\Sigma$ splits into two components $\Sigma_\ell$ and $\Sigma_r$; the delicate case is that all external vertex
operators are on one branch, say $\Sigma_\ell$.  
In bosonic string theory, the splitting is described by an equation
\begin{equation}\label{opcite}xy=q, \end{equation}
where $x$ and $y$ are local parameters on $\Sigma_\ell$ and $\SIgma_r$ respectively, and $q$ is a complex parameter
that controls the degeneration.  
In superstring theory, $q$ is replaced by an analogous parameter $q_\NS$ or $q_\Ra$, defined in section \ref{supan},
and the analysis then proceeds in the same way.

 In general, $\Sigma$ is a Riemann surface of genus $\g$ with $\n$ punctures, and
$\Sigma_l$ and $\Sigma_r$ are of genus $\g_\ell$ and $\g_r$ with $\g_\ell+\g_r=\g$.    The  complex structures of $\SIgma$
is parametrized by 
$\M_{\sg,\sn}$,  the moduli space of Riemann surfaces of genus
$\g$ with $\n$ punctures.  We write $\h\M$ for the Deligne-Mumford compactification of $\M_{\sg,\sn}$ that arises if one
allows degenerations such as the one at $q=0$.

If we set $q=0$ in (\ref{opcite}), then $\Sigma$ decomposes as the union of $\Sigma_\ell$
and $\Sigma_r$, joined together at the single point $x=y=0$.  This defines a divisor $\fD\subset\h\M$, and this divisor
has a very simple structure, as already explained in eqn. (\ref{itto}):
\begin{equation} \label{divst}\fD=\h\M_\ell\times \h\M_r.\end{equation}
Here $\h\M_\ell$ and $\h\M_r$ are the Deligne-Mumford compactions of $\M_\ell$ and $\M_r$, which parametrize
complex structures on $\Sigma_\ell$ and $\Sigma_r$, respectively.
This just says that at $q=0$, the moduli of $\Sigma_\ell$ and $\SIgma_r$ (including the choices of the points at which they
are  glued together to make $\Sigma$) can be varied independently, and a choice
of those moduli determines $\Sigma$.

if all external vertex operators are on $\Sigma_\ell$, then
$\h\M_\ell$ is a copy of $\h\M_{\sg_\ell,\sn+1}$, which parametrizes a genus $\g_\ell$ surface with $\sn+1$ punctures.
The 
 $\n+1^{th}$ puncture is the point at which
$\Sigma_\ell$ is glued to $\Sigma_r$.   Similarly in this situation, $\h\M_r$ is a copy of $\h\M_{\sg_r,1}$.   

Near $\fD$, $\h\M$ is parametrized by $q$ as well as $\fD$ and can be approximated
as  a small neighborhood of $\fD$ embedded as the zero section of a  complex line bundle $\N\to \fD$.  $\N$ is the normal bundle
to $\fD$ in $\h\M$.  The parameter $q$ in (\ref{opcite}) is not well-defined as a complex number, because its definition  depends
on the choices of local  parameters $x$ and $y$.  Rather, $q$ is a linear function on the line bundle $\N$ (or equivalently a section of
$\N^{-1}$).    The transformation of $q$ under a change in the local parameter
was described explicitly in eqn. (\ref{yuttz}).  
The only aspect of that formula that we will use is that $q$ is well-defined up to a transformation
\begin{equation}\label{zottoz} q\to e^{f_\ell +f_r}q.\end{equation}
Here $f_\ell$ is a function on $\h\M_\ell$ and reflects the dependence of $q$ on the choice of $x$; similarly, $f_r$
is a function on $\h\M_r$ and reflects the dependence of $q$ on the choice of $y$. The factorized form of (\ref{zottoz})
 reflects the fact that $\N$ is the tensor
product of a line bundle over $\h\M_\ell$ and one over $\h\M_r$:
\begin{equation}\label{timo}\N=\L_\ell\otimes \L_r,\end{equation}
as described in more detail in section 6.1.3 of \cite{Wittentwo}.

The $\g$-loop contribution to a scattering amplitude is  obtained by integrating over  $\h\M$  a differential form $F(g|\delta g)$, which
of course depends on the momenta $p_i$ and the other quantum numbers $\zeta_i$ of the external string states:
\begin{equation}\label{ombo}\A_g(p_1,\zeta_1;\dots;p_n,\zeta_n)=\int_{\h\M} F_{p_1,\zeta_1;\dots;p_n,\zeta_n}(g|\delta g) \end{equation}
We usually write just $F(g|\delta g)$ or simply $F$  rather than $F_{p_1,\zeta_1;\dots;p_n,\zeta_n}(g|\delta g)$.
The contribution to $F$ of a massless
scalar of zero momentum that propagates between $\SIgma_\ell$ and $\Sigma_r$ was analyzed in section \ref{mclosed}.  It
factors as the product of $\d^2q /\bar q q$ multiplied by a form $\G_\ell$ on $\h\M_\ell$ and a form $\G_r$ on $\h\M_r$. (We use here the language of bosonic string theory and
do not distinguish $\t q$ from $\bar q$; $\d^2 q$ is short for $-i \d\bar q\,\d q$.)  The forms $\G_\ell$ and $\G_r$ describe
the coupling of a single massless scalar of zero momentum to the surfaces $\Sigma_\ell$ and $\SIgma_r$, respectively, in
addition to the $\n$ external strings  that are already coupled to $\Sigma_\ell$.
The singular part
of the integral at $q=0$ is thus
\begin{equation}\label{zoffo}\A_{g,\mathrm{sing}}=\sum_{\alpha=1}^s \int \frac{\d^2 q }{\bar q q}
\int_{\h\M_\ell}\G_{\ell,\alpha}\int_{\h\M_r}\G_{r,\alpha}. \end{equation}
The sum runs over all massless scalars $\phi_\alpha$, $\alpha=1,\dots,s$,  that might contribute tadpoles.  The factor
$1/\bar q q$ is $\bar q^{\t L_0-1}q^{L_0-1}$ with $\t L_0=L_0=0$.  

The integral (\ref{zoffo}) diverges at $q=0$.  On the other hand, our hypothesis that the $\g$-loop tadpoles vanish means that
\begin{equation}\label{ofto}\int_{\h\M_r}\G_{r,\alpha}=0,~~\alpha=1,\dots, s. \end{equation}
So the divergence at $q=0$ is multiplied by 0, rather like the $0/0$ that one gets in field theory if the tadople
vanishes.

Suppose that we cut off the integral by placing a lower bound $|q|\geq \epsilon$ for some small $\epsilon>0$.  
The integral\footnote{$\Lambda$ is an irrelevant
cutoff at large $|q|$.  The approximation (\ref{zoffo}) to the integral over $\h\M$ is only valid for $q$ small.}
\begin{equation}\label{fto}\int_{\epsilon<|q|<\Lambda}\frac{\d^2q}{\bar q  q} \end{equation}
diverges for $\epsilon\to 0$ as $-4\pi \log \epsilon$, but the coefficient of $\log \epsilon$ vanishes after integration over 
$\h\M_\ell\times\h\M_r$, because of
(\ref{ofto}).    So the cutoff integral 
\begin{equation}\label{oytro}\A_{\sg,\epsilon}=\int_{\h\M_\epsilon}F(g|\delta g),\end{equation}
where $\h\M_\epsilon$ is the region in $\h\M$ defined by $|q|\geq\epsilon$, has a limit as $\epsilon\to 0$.  Vanishing of the tadpoles
has made the scattering amplitude convergent.

This is not the whole story, however, since $q$ is really not a complex number, as assumed in the above derivation, 
but a section of the line bundle $\N^{-1}$.  
This bundle is topologically nontrivial, so there is no way to trivialize it (even if we do not ask to do so holomorphically). 
However, what we need to make sense of the cutoff that we used is not a trivalization of $\N$ but merely 
a hermitian metric on $\N$, or equivalently on $\N^{-1}$.
The condition $|q|>\epsilon$ means that the norm of $q$, computed using some chosen hermitian metric on $\N^{-1}$, is greater than $\epsilon$.
A change in the hermitian metric on $\N^{-1}$ would be equivalent to replacing the cutoff condition $|q|>\epsilon$ 
by $|q|>e^h\epsilon$, for some real-valued function $h$
on $\h\M_\ell\times \h\M_r$.  This has the
effect of shifting $\log \epsilon$ to $\log \epsilon + h$.  This does not affect the fact that $\A_{g}$ has a limit for $\epsilon\to 0$, but it shifts the value of
the limit by
\begin{equation}\label{imbo}\A_{\sg}\to\A_{\sg}-4\pi  \sum_\alpha\int_{\h\M_\ell\times\h\M_r}h\,\G_{\ell,\alpha}\G_{r,\alpha}.\end{equation}
This is certainly not zero in general, so we seem to be in trouble.  

What saves the day is the following.\footnote{Note added in November 2023:  Ashoke Sen and Barton Zwiebach have pointed out that in the following,
I should have specified that the metric on $\L_r$ is chosen once and for all, independent of $\h\M_\ell$.   A similar comment applies at several later points in this
section.}
Since $\N^{-1}$ is the tensor product of a line bundle on $\L_\ell^{-1}\to \h\M_\ell$ and a line bundle   $\L_r^{-1}\to \h\M_r$, 
it is natural to choose the metric on $\N^{-1}$ to be the tensor product of a metric on $\L_\ell^{-1}$ and a metric on $\L_r^{-1}$.
This means that we can naturally restrict to the case
that $h=h_\ell+h_r$, where $h_\ell$ and $h_r$ are functions on $\h\M_\ell$ and $\h\M_r$, respectively.  Using the vanishing tadpole
condition (\ref{ofto}), we see that $h_\ell$ does not contribute to the integral in (\ref{imbo}).
Defining
\begin{equation}\label{zorky}\Delta_{\sg_r}\phi_\alpha=-\frac{1}{4\pi }\int_{\h\M_r}h_r \G_{r,\alpha},\end{equation}
the metric dependence of $\A_{\sg}$ is
\begin{equation}\label{porky}\A_{\sg}\to \A_\sg+\sum_\alpha \Delta_{\sg_r}\phi_\alpha\int_{\h\M_\ell} \G_{\ell,\alpha}.\end{equation}
Here $\int_{\h\M_\ell}\G_{\ell,\alpha}$ is the genus $\g_\ell$ contribution to a scattering amplitude with the $\n$ vertex operators that
we started with plus one more 
vertex operator that represents a $\phi_\alpha$ field at zero momentum.  This insertion gives the derivative of the scattering
amplitude with respect to $\phi_\alpha$. So we can interpret (\ref{porky})
as
\begin{equation}\label{zorkyl}\A_\sg\to \A_\sg+\sum_\alpha\Delta_{\sg_r}\phi_\alpha\frac{\partial}{\partial\phi_\alpha}\A_{\sg_\ell},\end{equation}
where $\A_{\sg_\ell}$ is the genus $\g_\ell$ contribution to the scattering amplitude under study.   

However, what we have analyzed here is a particular degeneration corresponding to the decomposition $\g=\g_\ell+\g_r$.  For
the full story, we have to sum over all such degenerations, analyzing each one in the same way.  This gives
\begin{equation}\label{morky} \A_\sg\to \A_\sg+\sum_{\sg_\ell+\sg_r=\sg}\sum_\alpha\Delta_{\sg_r}\phi_\alpha\frac{\partial}{\partial\phi_\alpha}\A_{\sg_\ell}.
\end{equation}

The full perturbative scattering amplitude is $\A=\sum_{\sg=0}^\infty \A_\sg$. We include factors of the 
string coupling constant  $g_\st$ in the definition of
$\A_\sg$ (so explicitly $\A_\sg$ is proportional to $g_\st^{2\sg-2}$), so this infinite sum is really an expansion 
in powers of $g_\st$. In perturbation theory, one views this sum over genus as a formal power series in powers
of $g_\st$.  The reason to not write explicitly  the powers of $g_\st$ is that
we do not want to treat the dilaton, whose expectation value determines $g_\st$, differently from the rest of the $\phi_\alpha$.
Similarly, we define $\Delta\phi_\alpha=\sum_{\sg_r=0}^\infty \Delta_{\sg_r}\phi_\alpha $.  Now, summing (\ref{morky}) over $\g$,
we find that the dependence of $\A$ on the choice of infrared cutoff used in calculating it is
\begin{equation}\label{torky}\A\to \A+\sum_\alpha \Delta\phi_\alpha \frac{\partial}{\partial\phi_\alpha}\A.\end{equation}

\begin{figure}
 \begin{center}
   \includegraphics[width=2.5in]{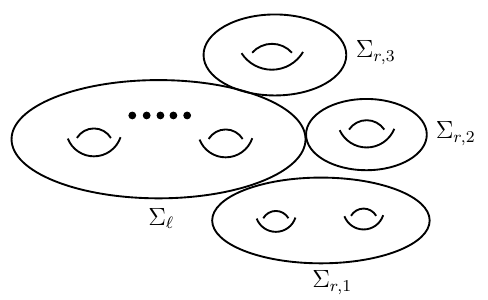}
 \end{center}
\caption{\small  A Riemann surface that degenerates to a union of $\Sigma_\ell$ with several other branches $\Sigma_{r,i}$, $i=1,2,3$.}
 \label{zumbox}
\end{figure} 
This is still not the full answer because in arriving at this formula, we have only included contributions in which 
$\Sigma_\ell$, to which the external vertex operators are attached, couples through a massless scalar of zero momentum to a single
additional component $\SIgma_r$.  In reality, $\Sigma_\ell$ may couple to any number $k\geq 0$ of 
such components (fig. \ref{zumbox}).   Associated to each
such component is its own gluing parameter $q$ with its logarithmically divergent measure 
$\d^2q/\bar qq$.  We regulate each such
integral as before by a suitable choice of  hermitian metric.  The dependence on the choice of infrared regulator always gives the
same factor $\sum_\alpha\Delta\phi_\alpha\cdot \V_\alpha$ found above, where by $\V_\alpha$ we mean 
the insertion in the scattering amplitude of a zero
momentum $\phi_\alpha$ vertex operator.  As usual, $\V_\alpha$ can be replaced by a derivative of the 
scattering amplitude with respect
to $\phi_\alpha$.  Summing over $k$ and remembering to include a factor of $1/k!$ since the 
components disappearing into the vacuum
are equivalent, we exponentiate the result in (\ref{torky}).  At this level, the full dependence of the scattering amplitude on the choice of
infrared regulator takes the form
\begin{equation}\label{lorky}\A\to \exp\left(\sum_\alpha\Delta\phi_\alpha \frac{\partial}{\partial\phi_\alpha}\right)\A.\end{equation}

The operator
\begin{equation}\label{toft}\mathcal K= \sum_\alpha \Delta\phi_\alpha\frac{\partial}{\partial\phi_\alpha}\end{equation}
is a vector field on the parameter space of the string compactifications under consideration, or in other words on the moduli space
$\ZZ$ of string theory vacua.  So $\exp(\mathcal K)$ is a diffeomorphism of that parameter space 
(in the sense of formal power series in $g_\st$).  What we learn from this analysis is that when massless tadpoles vanish,
string perturbation theory constructs a natural family of perturbative $S$-matrices
parametrized by $\ZZ$, but -- at this level of generality -- without a natural choice 
of parametrization of $\ZZ$.  Different infrared regulators
will give results that differ by a reparametrization of $\ZZ$.

\begin{figure}
 \begin{center}
   \includegraphics[width=2.5in]{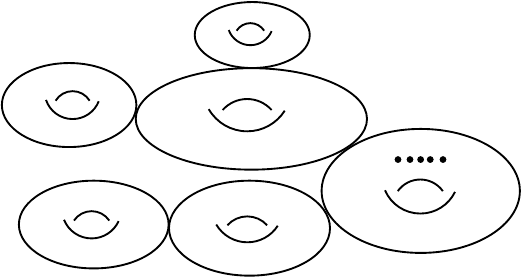}
 \end{center}
\caption{\small  A ``tree'' of Riemann surfaces, with all vertex operators inserted on one component.}
 \label{lubbo}
\end{figure} 

What we have just described is really only part of an inductive procedure.  
The moduli spaces $\M_\ell$ and $\M_r$
are themselves not compact, and in evaluating the integrals $\int_{\M_\ell}\A_\ell$ and $\int_{\M_r}\A_r$, one may have met tadpoles
in lower order.  Inductively, once  all integrals up to genus $\g-1$ have been defined, one applies the above reasoning
 to  the genus $\g$ integrals.  The full analysis amounts to regularizing the contributions of ``trees''
of Riemann surfaces, with all external vertex operators on the same component,
as indicated in fig. \ref{lubbo}.  Consider any component $\Sigma^*$ in such a tree other than the component that contains the external vertex operators; let $\g^*$ be its genus.  In general, $\Sigma^*$ is joined at double points to some number $s>0$ of other components,
and the analysis of the simultaneous $\d^2q/\bar q q$ singularities associated to all the double points  
leads to a sum of terms, each of which is proportional to a correlation function $\langle \V_{\alpha_1}\dots\V_{\alpha_s}\rangle_{\sg^*}$,
where the $\V_{\alpha_i}$ are vertex operators of moduli fields at zero momentum.  
If all such correlation functions vanish (after integration over the moduli of $\Sigma^*$)
 then the analysis of general trees is similar to what we have already explained and the result is the same, 
except that $\K$ becomes a nonlinear function of the $\Delta\phi_\alpha$. 
Vanishing of the zero-momentum correlation functions amounts
to the condition that the effective potential is identically zero as a function of the $\phi_\alpha$.  It is desireable but more difficult
to develop the theory without this assumption, so as to include scalar fields that are massless without being moduli.
 
One can consider in a similar way infrared divergences associated to massless open strings.
 An open-string degeneration is still described by a gluing formula $xy=q$
 (or its superanalog), but now $q$ is real.  The 
 $\d^2q/\bar q q$ singularity that we have analyzed is replaced by a slightly simpler $\d q/q$ singularity, which has the
 same logarithmic divergence and can be treated the same
 way.  Different regularizations of the massless open-string degenerations differ by shifts in the expectation 
 values of massless open-string
 fields of spin zero.

 \subsubsection{Another  Infrared Regulator}\label{anop}

 Here, we will explain another type of infrared regularization that can be used in analyzing these questions.  
 The results will be equivalent to what we had before.  
 One reason to describe this alternative regulator is to emphasize that different types of infrared regulator are
 possible and do lead to equivalent results. 
 
We return to the basic question of studying the integral (\ref{ombo}) that defines a scattering amplitude
near the separating degeneration at $q=0$.
The form $F$ can be approximated near $q=0$ as
\begin{equation}\label{zingort}F_0=\sum_\alpha\frac{\d^2q}{\bar q q}\wedge\G_{\ell,\alpha}\wedge\G_{r,\alpha}.\end{equation}
An important point, which applies also to some expressions written below, is that the form $F_0$ is invariant 
under $q\to e^f q$, where $f$ is a function on $\h\M_\ell\times\h\M_r$,
because $\G_{\ell,\alpha}\wedge \G_{r,\alpha}$ is a form of top degree on $\h\M_\ell\times \h\M_r$.    So $F_0$ is well-defined, 
independent of precisely how we define the $q$ parameter,
as a differential form on what we will call $\h\M_0$,  
the total space of the line bundle $\N^{-1}\to \h\M_\ell\times \h\M_r$.    Suppose we can  write $F_0=\d\Lambda$,
where $\Lambda$ is a form of real codimension 1.  Then we modify the cutoff definition (\ref{oytro}) by adding a boundary term:
\begin{equation}\label{ogglo}\A_{g,\epsilon}=\int_{\h\M_{\epsilon}}F-\int_{\partial\h\M_{\epsilon}}\Lambda.\end{equation}
Here $\h\M_{\epsilon}$ is defined by $|q|\geq \epsilon$, and $\partial\h\M_{\epsilon}$ its  its boundary, defined by  $|q|=\epsilon$.
If $F=\d\Lambda$, Stokes's theorem implies that the right hand side of (\ref{ogglo}) is 
invariant under $\epsilon\to e^h\epsilon$.
If instead we actually have $\d\Lambda=F_0$, where $F_0$ coincides with $F$ only near $q=0$,
then the correct conclusion from Stokes's theorem is
that the limit of $\A_{g,\epsilon}$ for $\epsilon\to 0$ is invariant under $\epsilon\to e^h\epsilon$.

We have eliminated the need to choose a hermitian metric on $\N$, but now we need to find a  
way to write $F_0=\d\Lambda$.  For this,
let us restrict to a point in $\h\M_\ell$ and view $q$ as a section of the line bundle $\L_{r}^{-1}\to \h\M_r$. 
Consider the differential form
\begin{equation}\label{melf}\chi_{r,\alpha}= \frac{\d^2 q}{\bar q q}\G_{r,\alpha}.\end{equation}
We view this as a top form on what we will call $\X_r$ -- the total space of the line bundle $\L_{r}^{-1}\to \h\M_r$, 
with the zero section removed.  
The reason that we remove the zero section is that $\chi_{r,\alpha}$ is singular at $q=0$.  
We note that $\chi_{r,\alpha}$ is invariant under $q\to e^f q$
for any function $f$, so in particular it is well-defined.   We want to find a  form $\lambda_{r,\alpha}$ with 
\begin{equation}\label{morfo}\chi_{r,\alpha}=\d\lambda_{r,\alpha}.\end{equation}
Abstractly, $\lambda_{r,\alpha}$ exists since (because of the noncompactness of $\X_r$) the 
top-dimensional cohomology of $\X_r$ vanishes.
We will explain shortly how to make a fairly nice choice of $\lambda_{r,\alpha}$, but for the moment suppose this has been done.
Then in the cutoff formula (\ref{ogglo}), we take
\begin{equation}\label{abc}\Lambda=\sum_\alpha\G_{\ell,\alpha}\wedge \lambda_{r,\alpha}.\end{equation}  
Clearly $\d\Lambda=F_0$, as desired.

Let us now investigate the extent to which (\ref{ogglo}) depends on the choice of $\lambda_{r,\alpha}$.  
Since we are supposed to obey $\d\lambda_{r,\alpha}
=\chi_{r,\alpha}$, we can only add to $\lambda_{r,\alpha}$ a closed form.  On the other hand, 
if we add an exact form to $\lambda_{r,\alpha}$,
then $\Lambda$ will change by an exact form and (\ref{ogglo}) is invariant.  So we are only interested
 in shifting $\lambda_{r,\alpha}$ by a form
$\Delta\lambda_{r,\alpha}$ that is closed but not exact.  Let us write $\X_\epsilon$ 
for the subspace of $\X$ defined by $|q|=\epsilon$.  $\X$ is contractible onto $\X_\epsilon$
(and topologically, $\X_\epsilon$ does not depend on the metric used in the condition $|q|=\epsilon$).   The 
only invariant information in $\Delta\lambda_{r,\alpha}$,
modulo exact forms, is the ``period''
\begin{equation}\label{zobob}\Delta\phi_\alpha=-\int_{\X_\epsilon}\Delta\lambda_{r,\alpha}.\end{equation}
 If we do shift $\lambda_{r,\alpha}$ in this fashion, then $\Lambda$ is shifted by
\begin{equation}\label{obob}\Lambda\to\Lambda+\sum_\alpha \G_{\ell,\alpha}\wedge \Delta\lambda_{r,\alpha},\end{equation}
and when we insert this in (\ref{ogglo}), and take the limit $\epsilon\to 0$, we find that  $\A_{\sg}$ shifts by
\begin{equation}\label{bobo}\A_{\sg}\to \A_{\sg}+\sum_\alpha\Delta\phi_\alpha\int_{\h\M_\ell}\G_{\ell,\alpha}.\end{equation}
This is the familiar result of eqn. (\ref{porky}), and the rest of the analysis proceeds from there.

To conclude, we will describe a nice class of choices for $\lambda_{r,\alpha}$.  
A naive way to proceed is as follows.  The vanishing 
tadpole condition (\ref{ofto}) implies that the form $\G_{r,\alpha}$ on $\M_r$ is exact,
$\G_{r,\alpha}=\d \beta_{r,\alpha}$, for some $\beta_{r,\alpha}$.  So one might try
\begin{equation}\label{welf}\lambda^{(0)}_{r,\alpha}= \frac{\d^2q}{\bar q q}\beta_{r,\alpha}.  \end{equation}
But this expression is not invariant under $q\to e^f q$, so it only makes sense once 
one is given a trivialization of $\L_r^{-1}$.  Under $q\to e^f q$,
we have $\lambda^{(0)}_{r,\alpha}\to \lambda^{(0)}_{r,\alpha}-i
\left(\d \bar f (\d q/q)+\d\bar q/\bar q\cdot \d f\right)\beta_{r,\alpha}$. 
Note that $\Delta\lambda^{(0)}_{r,\alpha}=-i\left(\d \bar f (\d q/q)+\d\bar q/\bar q\cdot \d f\right)
\beta_{r,\alpha}$ is closed (since $d\bar q/\bar q$, $\d q/q$,
$\d f\wedge \beta_{r,\alpha}$, and $\d\bar f\wedge\beta_{r,\alpha}$ are all closed; the last two are closed
because they are top forms on $\h\M_r$), so this shift in $\lambda^{(0)}_{r,\alpha}$ 
does not affect the condition $\d\lambda^{(0)}_{r,\alpha}=
\chi_{r,\alpha}$.   The upshot of this is that instead of (\ref{welf}), we should try
\begin{equation}\label{gelf}\lambda_{r,\alpha}= \frac{\d^2q}{\bar q q}\beta_{r,\alpha}
+\frac{\d q}{q}\gamma_{r,\alpha}+\frac{\d\bar q}{\bar q}\tilde\gamma_{r,\alpha}, \end{equation}
where $\gamma_{r,\alpha}$ and $\tilde\gamma_{r,\alpha}$ are top forms on $\h\M_r$.  There is no problem in globally
solving $\d\lambda_{r,\alpha}=\chi_{r,\alpha}$
with $\lambda_{r,\alpha}$ of this form, though the way of writing $\lambda_{r,\alpha}$ as a sum of 
the three indicated terms depends on a choice
of local trivialization of $\L_r^{-1}$.  To prove that a global choice of $\lambda_{r,\alpha}$ of the claimed
form does exist, one covers $\M_r$ by small open sets on which one can pick a trivialization of $\L_r^{-1}$, so
that one can choose $\lambda_{r,\alpha}$ in the form (\ref{welf}).  Since $\beta_{r,\alpha}$ is globally defined,
two such local solutions differ by terms of the form $\Delta\lambda=(\d q/q) \Delta\gamma_{r,\alpha}+(\d\bar q/\bar q)\Delta\tilde\gamma_{r,\alpha}$.
The obstruction to modifying
 the local choices of $\lambda_{r,\alpha}$ by adding terms of the form $\Delta\lambda$
 so that they fit together into a global 
$\lambda_{r,\alpha}$ is given by a one-dimensional cohomology class of a smooth manifold, namely $\h\M_r$, with values
in a coherent sheaf.  Such cohomology always vanishes above dimension zero, so there is no obstruction to finding 
$\lambda_{r,\alpha}$.

\subsubsection{Wavefunction Renormalization}\label{wavefunction}

Starting in section \ref{massmat}, we restricted attention to the case that mass renormalization vanishes.
To understand mass renormalization in superstring perturbation theory requires a more general formalism than 
the one developed in the present paper.   But there is something to say even if mass renormalization
vanishes.

When mass renormalization vanishes, we meet a question that is exactly analogous to the question that we have
been studying in the context of tadpoles.  Mass renormalization is associated to a degeneration of the type
sketched in fig. \ref{kopla}(b), with a single external vertex operator $\V$ on one side, say on $\SIgma_\ell$,
and arbitrary insertions on $\Sigma_r$.  Near such a degeneration, the genus $\g$ contribution to the scattering
amplitude $\A_\sg$ has a singular behavior that is just
like that of the integral (\ref{zoffo}) that  we have studied in the tadpole case.  The behavior near the degeneration is
\begin{equation}\label{zoffort}\A_{\sg,\mathrm{sing}}=\sum_{\alpha=1}^s \int \frac{\d^2 q }{\bar q q}
\int_{\h\M_\ell}\G_{\ell,\alpha}\int_{\h\M_r}\G_{r,\alpha}. \end{equation}
The label $\alpha$ now runs over all physical string states that at tree level are degenerate in mass 
with the external particle associated to the vertex operator $\V$.

The integral has a potential logarithmic divergence near $q=0$, but absence of mass renormalization means that
\begin{equation}\label{offort}\int_{\h\M_\ell}\G_{\ell,\alpha}=0\end{equation}
for all $\alpha$.   This ensures  the cancellation of the logarithmic divergence in $\A_{\sg,\mathrm{sing}}$.  We are now
in  a familiar situation.  Vanishing of mass renormalization ensures
that the integral for $\A_\sg$ converges, but only conditionally so.  If one introduces an infrared cutoff by restricting
to $|q|\geq \epsilon$, then $\A_\sg$ has a limit for $\epsilon\to 0$, but this limit depends on the hermitian metric that was
used to define what we mean by $|q|$.  

The interpretation is quite similar to what it was in the tadpole case.  By imitating the previous arguments,
one shows that the change in $\A_\sg$ resulting from a change
in the hermitian metric is equivalent to a scattering amplitude computed on $\Sigma_r$ with an insertion of 
one more vertex operator, which reflects the output of the path integral on $\Sigma_\ell$.  This vertex operator is 
a linear
combination of the vertex operators $\V_\alpha$ that represent physical string states that have the same momentum
and therefore the same mass as the one
corresponding to $\V$.  Thus the dependence on the choice of cutoff is equivalent to a wavefunction renormalization
$\V\to \V+\sum_\alpha c_\alpha \V_\alpha$.  The constants $c_\alpha$ are of order $g_\st^{2\sg_\ell}$.

Note that $\V$ itself is a linear combination of the $\V_\alpha$.
It is helpful to generalize the problem slightly by inserting on $\SIgma_\ell$ an arbitrary linear combination of the $\V_\alpha$,
rather than making a  particular choice as we did in the above presentation.  Then the dependence on the choice of cutoff
would give us a general $s\times s$ wavefunction renormalization matrix, as one might expect in field theory. 
The $S$-matrix is natural, but there is no
natural notion in general of computing it using vertex operators that are independent of the string coupling constant
and the other moduli.

\subsection{More Detail On The Anomaly}\label{lazy}

In section \ref{zumico}, we gave a somewhat heuristic explanation of an important result.  The result concerned
a separating degeneration in string theory, in which a worldsheet $\Sigma$ decomposes to two components
$\Sigma_\ell$ and $\Sigma_r$, joined at a double point, with a BRST-trivial vertex operator $\{Q_B,\W_1\}$ inserted on $\Sigma_\ell$.
The claim is that any BRST anomaly arising in this situation is always proportional to the amplitude obtained by inserting
at the double point on $\Sigma_r$ a physical state vertex operator $\O(\W_1)$  that depends linearly on $\W_1$.  
To be more precise, this is supposed to be true in an inductive sense: assuming that there are no BRST anomalies in genus 
less than $\g$, any BRST anomaly in genus $\g$ should have the property just stated.

The relevant ideas can be explained somewhat more simply for the case of an open-string degeneration.   Moreover,
supersymmetry will not be important and we will use the language of bosonic strings.   Gluing for open strings is described
by the usual gluing formula
\begin{equation}\label{zildo} xy = q, \end{equation}
where $q$ is real and positive and $x$ and $y$ are local parameters on $\Sigma_\ell$ and $\Sigma_r$ that are real along
the boundary.  For $q\to 0$, we glue the boundary point $x=0$ in $\Sigma_\ell$ to the boundary point $y=0$ in $\Sigma_r$.
In section \ref{zumico}, we used the parameter $s$ defined by $q=e^{-s}$, which parametrizes the length of a long
strip joining the two branches, but here it will be more convenient to work with $q$,
since this makes it easier to describe what is happening at $q=0$ or $s=\infty$.
The condition $q=0$ defines a component $\B$
of the boundary of $\h\M_{\sg,\sn}$.  The last statement, which is explained in more detail in section 7.4 of \cite{Wittentwo} (and
also in section \ref{geometry} below),
makes open-string degenerations slightly simpler than closed-string degenerations for our present purposes, and that
is why we consider this case.  $\B$ is a product of moduli spaces:
\begin{equation}\label{doof}\B\cong \h\M_\ell\times \h\M_r.\end{equation}
This statement is an open-string analog of the closed-string statement (\ref{bitto}).  For open superstrings, $q$ is replaced
by the analogous parameter $q_\NS$ or $q_\Ra$.

In section \ref{zumico}, motivated by eqn. (\ref{yrfe}), we claimed that the anomaly can be evaluated by inserting at the double
point
\begin{equation}\label{zyrfe}\sum_i\left( c\partial c\, \UU_i\otimes c\,\UU^i+ c\,\UU_i\otimes c\partial c\,\UU^i\right).\end{equation}
The sum runs over a complete set of physical open-string states with momentum and spin that match those of  $\W_1$.
We have omitted the BRST-trivial term in (\ref{yrfe}), which  can be dropped because we assume that there are no BRST anomalies
in lower orders.   The formula (\ref{zyrfe}) suggests that, with $\V_i=c\UU_i$, the anomaly can be computed
by a path integral on $\Sigma_r$ with insertion of
\begin{equation}\label{dradful} \O(\W_1)=\sum_ia_i \V_i, \end{equation} where the coefficient $a_i$ is to be computed by a path integral on 
$\Sigma_\ell$ with an insertion of $c\partial c\UU_i$ (along with $\W_1$ and possibly other vertex operators).

Although it is true that the anomaly can be computed from an insertion of an operator $\O(\W_1)$ 
of the form indicated in (\ref{dradful}),
the proposed formula for the coefficients $a_i$ is oversimplified.  
This must be the case, as the insertion of $c\partial c\UU_i$ cannot arise in our formalism, for this operator is not annihilated by $b_0$.

To 
analyze the problem more systematically, we return to the basic eqn.
(\ref{tarmib}) for the anomaly, which we repeat for convenience:
\begin{equation}\label{zarmib}\int_{\h\M_{\sg,\sn}}F_{\{Q_B,\W_1\},\V_2,\dots,\V_\ssn} 
=-\int_{\partial\h\M_{\sg,\sn}} F_{\W_1,\V_2,\dots,\V_\ssn}. \end{equation}
Since $\W_1$ and $\V_2,\dots,\V_\sn$ are all  conformal vertex operators --  conformal 
primary fields of dimension 0
annihilated by the antighost
modes $b_n$, $n\geq  0$ (and by $\beta_r$, $r\geq 0$, in the superstring case) -- the forms $F_{\{Q_B,\W_1\},\V_2,\dots,\V_\ssn}$ and $F_{\W_1,\V_2,\dots,\V_\ssn}$ are
both pullbacks from $\h\M_{\sg,\sn}$.   The former is a top form that can be integrated over $\h\M_{\sg,\sn}$, and the latter is a form
of codimension 1 that can be integrated over a codimension 1 submanifold, such as the boundary of $\h\M_{\sg,\sn}$.  
In particular, we will study the behavior near the
component $\B$ of the boundary.

If $F_{\W_1,\V_2,\dots,\V_\ssn}$ were non-singular along $\B$, then we could literally restrict it to $\B\cong \h\M_\ell\times \h\M_r$.
The restriction could be expressed in terms of insertion at the double point of some bilinear expression in local operators,
schematically $\S=\sum_\alpha \RR_\alpha\otimes \RR'_\alpha$, where $\RR_\alpha$ and $\RR'_\alpha$ are local operators on
$\Sigma_\ell$ and $\Sigma_r$, respectively.  But $\RR_\alpha$ and $\RR'_\alpha$ must be conformal vertex operators
 because of the conformal invariance of our formalism, and $\S$ must have total ghost number 3.
These conditions are incompatible, as the maximum possible ghost number of a conformal vertex operator is 1 (corresponding
to conformal vertex operators $\V=c\UU$, where $\UU$ is a matter primary of dimension 1), and $1+1<3$.  We conclude that
if $F_{\W_1,\V_2,\dots,\V_\ssn}$ is nonsingular along $\B$, so that its restriction to $\B$ can be defined, then this restriction is
actually 0, and in particular there is no anomaly.  (Essentially the same argument applies
to open superstring theory, using the constraints on superconformal vertex operators explained in sections \ref{vertex} and 
\ref{ramond}.  The extension to closed-string degenerations is also straightforward.)  

In general,  $F_{\W_1,\V_2,\dots,\V_\ssn}$ is singular at $q=0$.  In fact, the singularity can be computed via the insertion
\begin{equation}\label{ozark}\frac{\d q}{q}\sum_i c\,\UU_i\otimes c\,\UU^i \end{equation} 
that is familiar from section \ref{propagator} and especially from
 eqn. (\ref{zorom}).  (The operators $c\,\UU_i$ that appear here are the most general conformal vertex operators of ghost number 1.
 The conformal invariance of the formalism ensures that other operators cannot arise.)
At first sight, it may seem that we are not interested in this contribution to $F_{\W_1,\V_2,\dots,\V_\ssn}$, since we are interested
in setting $q$ to a small ``constant,'' rather than integrating over $q$.  However, this is not correct because $q$ is really a section of a real
line bundle, rather than a real number, so setting $q$ to a constant value is not a natural operation.  (The dependence of $q$ on the
choice of local parameters was described explicitly in eqn. (\ref{yuttz}).)  It is meaningful to multiply $q$ by a small positive constant
or equivalently to add a large constant to $s$, but there is no natural notion of setting $q$ to a constant value.  We will see that the 
$\d q/q$ term has to be included to write a conformally invariant formula for the coefficients $a_i$ of eqn.
(\ref{dradful}).

In writing eqn. (\ref{ozark}), we are using a sort of hybrid formalism.  The expression $\d q/q$ is an explicit one-form on the normal
direction to $\B$ in $\h\M$.  In addition, the path integral on $\Sigma=\Sigma_\ell\cup\Sigma_r$ with
insertion of $\sum_i cU_i\otimes c U^i$ will generate a codimension 1 form on $\M_\ell\times\M_r$.  (This
form has codimension 1 because one of the operators inserted on $\Sigma_\ell$, namely $\W_1$,
 is a gauge parameter rather than the vertex
operator for a physical state.)  So eqn. (\ref{ozark})
describes the singularity along $\B$ of the codimension 1 form $F_{\W_1,\V_2,\dots,\V_\ssn}$ on $\h\M$.

In addition to the singular terms (\ref{ozark}), $F_{\W_1,\V_2,\dots,\V_\ssn}$ has the  contributions of
eqn. (\ref{zyrfe}) that are nonsingular for $q\to 0$.  Combining them gives
\begin{equation}\label{bozark}\frac{\d q}{q}\sum \,c\,\UU_i\otimes c\,\UU^i+\sum_i \left(c\partial c\,\UU_i\otimes c\,\UU^i+ c\,\UU_i\otimes c\partial c\,\UU^i\right),\end{equation}
a formula that suffices for evaluating $F_{\W_1,\V_2,\dots,\V_\ssn}$ modulo terms that vanish along $\B$.
The fact that the same matter primary fields $\UU_i$ appear in these singular and nonsingular contributions to $F_{\W_1,\V_2,\dots,\V_\ssn}$ 
can be explained as follows.  To evaluate the codimension 1 differential form $F_{\W_1,\V_2,\dots,\V_\ssn}$, we are supposed to make all
possible insertions of antighost modes.  
To evaluate a term in $F_{\W_1,\V_2,\dots,\V_\ssn}$ that is proportional to $\d s=\d q/q$, one of the insertions should be a $b_0$
insertion in the narrow neck. This removes $\partial c$ from the operators.  To evaluate contributions to $F_{\W_1,\V_2,\dots,\V_\ssn}$
with no $\d q/q$, we omit the $b_0$ insertion (making instead an extra antighost insertion somewhere else), 
in which case we are left with $\partial c$ in the operators.  But the matter
primaries that appear are the same.

Now let us show that the combined formula (\ref{bozark}) behaves correctly with respect to reparametrizations of $\Sigma_\ell$ and $\Sigma_r$.
It suffices to consider reparametrizations of $\Sigma_\ell$.  Consider a reparametrization $x\to \h x(x)$ that leaves fixed the point at which
the gluing occurs, so that $\h x=0$ at $x=0$.  The operators $c\,\UU_i$ and $c\partial c\UU_i$ are primary fields of dimension 0, so they are unaffected
by this reparametrization.  The reparametrization multiplies $q$ by the constant $\left.\partial \h x/\partial x\right|_{x=0}$
(see eqn. (\ref{yuttz})).  The differential form $\d q/q$ is invariant under this rescaling.  So eqn. (\ref{bozark}) is reparametrization-invariant.

Just as in section \ref{intmod}, we need an additional condition beyond reparametrization invariance
to ensure that $F_{\W_1,\V_2,\dots,\V_\ssn}$ is a pullback from moduli space.   The additional condition says that $F_{\W_1,\V_2,\dots,\V_\ssn}$ should
vanish if contracted with a vector field induced from a diffeomorphism of $\Sigma$.  In the present context, such a vector field is induced
from a vector field that generates a reparametrization of $\Sigma_\ell$ that leaves fixed the point $x=0$ along with the corresponding
rescaling of $q$:
\begin{equation}\label{mozark}v(x)\partial_x +\left.\frac{\partial v(x)}{\partial x}\right|_{x=0}q\partial_q,~~~~v(0)=0.\end{equation}
(Of course, we also consider in a similar way reparametrizations of $\Sigma_r$.)  As we know from section \ref{intmod},
contraction with the vector field on the space of metrics that is induced from the vector field $v(x)\partial_x$ on $\Sigma_\ell$
has the effect of $c\to c+v$.
 On an operator supported at $x=0$, since $v(0)=0$, this leaves $c(0)$ unchanged and        this acts by
$\partial c(0)\to  v'(0)$.  On the other hand, contraction with $ v'(0) q\partial_q$ maps $\d q/q $ to $ v'(0)$.
The two contributions cancel (a minus sign appears because the contraction operation $\partial c(0)\to v'(0)$
 anticommutes with the fermionic field $c$), so the
expression in (\ref{bozark}) does have the desired property to make  $F_{\W_1,\V_2,\dots,\V_\ssn}$  a pullback.

To evaluate the anomaly, we observe the following.  First of all, an insertion of $c\partial c\,\UU^i$ on $\Sigma_r$ will vanish because the
ghost number of this operator is too large by 1.  
So the anomaly will come from an insertion of $\sum_i a_ic\,\UU^i$ on $\Sigma_r$, where the coefficients $a_i$
can be computed by integration on $\Sigma_\ell$. To compute the $a_i$, we cannot set $q=0$ and integrate over $\M_\ell$, since the form we want to integrate is
singular at $q=0$.  Rather, we define an integration cycle $\M^*_\ell$ that is isomorphic to $\M_\ell$ by taking $q$ to be small
(for example, we could fix any metric on the real line bundle where $q$ takes values and define $\M_\ell^*$ by
 $|q|=\epsilon$, with $\epsilon $ a small positive
constant).  
If we set
\begin{equation}\label{telgoc}\X_i=\frac{\d q}{q}c\,\UU_i+c\partial c\,\UU_i, \end{equation}
then the coefficients $a_i$ are 
\begin{equation}\label{belgo}a_i=\int_{\M_\ell^*} F_{\W_1,\dots,\X_i}. \end{equation}
In other words, they are obtained by a path integral in which we insert on $\Sigma_\ell$
the expression $\X_i$
as well as $\W_1$ and any other operators (indicated by the ellipses in (\ref{belgo}))  that were present 
on $\Sigma_\ell$ at the beginning, 
and then integrate over $\M_\ell^*$.
It is hopefully now clear that the $\d q/q$ term in $\X_i$ is necessary  here.

For open-string degenerations in superstring theory, everything is almost the same, with $q$ replaced by $q_\NS$ or $q_\Ra$, and
with $c\,\UU_i$ replaced by its superconformal analog.  
For Ramond degenerations, one has to include the integration over the fermionic
gluing parameter.   For closed bosonic strings, the gluing parameter is a complex variable $q=\exp(-(s+i\alpha))$. 
To evaluate the anomaly, we want to integrate over $\alpha$ while fixing $|q|$. 
As in the derivation of the propagator in section \ref{bosclosed}, the integration over $\alpha$ is associated to an insertion of $b_0-\t b_0$
and leads to a factor or $2\pi \delta_{L_0-\t L_0}$.  
Just as in the open-string case, there is no natural operation
of setting $|q|$ to a constant and instead one must develop a formalism with terms proportional to $\d |q|/|q|$.  The anomaly
form $F_{\W_1,\V_2,\dots,\V_\ssn}$ has terms proportional to $\d|q|/|q|$ that can be computed by making a $b_0+\t b_0$ insertion
and regular terms that can be computed by instead making a different antighost  insertion.  The analog of eqn. (\ref{bozark}) is
\begin{equation}\label{powark}\frac{\d(|q|^2)}{|q|^2}\sum_i\t c c \UU_i \otimes \t c c \UU^i+\sum_i\left(\t c c (\t\partial \t c+\partial c)\UU_i\otimes
\t c c \UU^i+\t c c \UU_i\otimes \t c c (\t\partial \t c+\partial c)\UU^i\right).\end{equation}
This leads to an obvious analog of the formula (\ref{belgo}) for the anomaly coefficients.  For closed superstrings, one has left
and right gluing parameters $q_\ell$ and $q_r$ (as in section \ref{delcycle}, one can define the integration cycle so that they are 
complex conjugates  modulo the odd variables).  Eqn.
(\ref{powark}) has an immediate analog, with $|q|^2$ replaced by $q_\ell q_r$, and the vertex operators $\t c c \,\UU_i$ replaced
by their superconformal analogs. In Ramond sectors, one also integrates over the fermionic gluing parameters.

The main conclusions of this analysis -- such as eqns. (\ref{bozark}) and (\ref{powark}) -- must be supplemented at zero momentum
with exceptional terms whose origin we first saw in eqns. (\ref{zyrfex}) and (\ref{opal}).

\section{Spacetime Supersymmetry And Its Consequences}\label{tadpoles}

We focus here on the spacetime supersymmetry of the $S$-matrix.  In fact, in any supersymmetric compactification of string theory,
there is at tree level a massless field of spin $3/2$, known as the gravitino.  Its tree-level couplings are 
constrained by spacetime supersymmetry
and in particular at zero momentum these couplings are proportional to the matrix elements of the supercurrent. 
Perturbative corrections cannot alter the fact that the gravitino couplings are non-zero at zero momentum.

Assuming that the perturbative $S$-matrix exists and the gravitino remains massless in perturbation theory,
spacetime supersymmetry of the perturbative $S$-matrix follows just from these facts.
Indeed, just as the
existence of a massless spin 1 particle with nonvanishing couplings at zero momentum implies conservation of electric
charge, and a massless spin 2 particle with nonvanishing couplings at zero momentum must interact like a graviton \cite{SW},
the existence of a massless spin 3/2 particle with nonvanishing couplings at zero momentum implies spacetime
supersymmetry \cite{GP}.  (Such arguments were originally applied to string theory in \cite{Yoneya,SS}.)

We elaborate on this point in section \ref{revarg}.  However, the information one can gain from such arguments
appears to be not quite adequate for our purposes. 
To address the tadpole problem, we seem to need an argument
that can be formulated directly at zero momentum in spacetime, not by taking a limit from non-zero momentum.  Also, we should
not assume {\it a priori} that the perturbative $S$-matrix exists; spacetime supersymmetry is supposed to be an ingredient in proving this.

So in section \ref{superc}, we describe a more precise and stringy proof of spacetime supersymmetry.  The main tool is what
 one might call the spacetime supercurrent.  In the standard conformal field theory language, this 
is simply the holomorphic (or antiholomorphic) fermion vertex operator of \cite{FMS},
at zero spacetime momentum.  We then go on to show in what sense this operator generates spacetime supersymmetry, and 
how it can be used to show -- under appropriate conditions -- that massless tadpoles vanish and hence the $S$-matrix exists.  

\subsection{Massless Particles and Conserved Charges}\label{revarg}

Here we will review the arguments \cite{SW,GP} showing that massless particles of spin $\geq 1$ couple at zero 
momentum to conserved
charges.  The couplings of a given massless field might vanish at zero momentum (this is the case in 
perturbative superstring theory for
massless Ramond-Ramond gauge fields in the absence of D-branes), but if a massless field of spin $\geq 1$ has nonvanishing 
couplings at zero momentum,
then it couples to a conserved charge.  We will be brief since 
the considerations here are not novel, and in any event we will introduce a more explicit approach to conserved charges
in string theory
in section \ref{superc}.  The purpose of this section is to explain what can be understood based on general arguments
that do not involve the details of string theory.

\subsubsection{Gauge Theory}\label{gt}

\begin{figure}
 \begin{center}
   \includegraphics[width=2in]{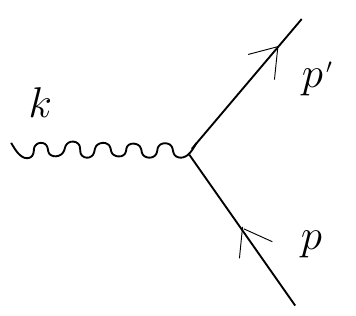}
 \end{center}
\caption{\small  The vertex coupling a massless ``photon'' to a massive charged scalar.}
 \label{lubbox}
\end{figure} 
Let us consider an $S$-matrix element describing the coupling of a massless spin 1 ``photon'' 
of momentum $\kk$ and polarization vector $\varepsilon$ (with $\kk^2=\kk\cdot \varepsilon=0$)
to $\n$ additional particles of masses $m_\ii$, momenta $p_\ii$, and 
charges $e_\ii$, with $\ii=1,\dots,\n$.  Gauge invariance means
that $S$-matrix elements must vanish if $\varepsilon=\kk$.  We consider the photon to be soft, 
meaning that we will study the limit $\kk\to 0$,
and we consider the other particles to be non-soft, meaning that their momenta will have non-zero limits for $\kk\to 0$.
For simplicity, we take the non-soft particles to have spin 0, in which case, their propagators (in Lorentz signature) are
\begin{equation}\label{dooksee} \frac{i}{p_\ii^2-m_\ii^2}.\end{equation}
For small $\kk$, the vertex by which the photon couples to the $\ii^{th}$  non-soft particle is (fig. \ref{lubbox})
\begin{equation}\label{cornyx}-ie_\ii \,\varepsilon\cdot (p_\ii+p_\ii'),~~p_\ii'=p_\ii+\kk.\end{equation}
  The specific
form of the trilinear vertex assumed in (\ref{cornyx}) follows 
from\footnote{This statement holds above four dimensions.  In four dimensions, 
there is a more general possibility involving magnetic charge.
This is not relevant to superstring perturbation theory, as there are no magnetic monopoles in superstring perturbation theory.}
 gauge invariance and Lorentz invariance applied to the on-shell three-point function.

 \begin{figure}
 \begin{center}
   \includegraphics[width=2.5in]{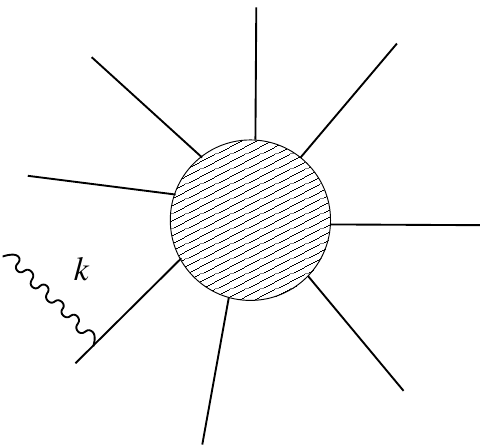}
 \end{center}
\caption{\small The most singular contributions to scattering of a soft photon come from processes in which the photon
is attached to an external charged line, as sketched here. }
 \label{nubbox}
\end{figure}  
The condition of gauge invariance can be usefully analyzed 
 for small $\kk$.  For $\kk\to 0$, the most singular contributions to the scattering amplitude $\A$ come
from processes in which the soft photon is attached to one of the external lines (fig. \ref{nubbox}).  
 In perturbative field theory, one can understand this picture in terms of Feynman diagrams,
but more generally one can understand it simply as a pictorial description of a certain singularity of the $S$-matrix, associated
to an on-shell particle in a particular channel.
The contribution to $\A$ 
with the soft photon attached
to the $\ii^{th}$ external line factors as
\begin{equation}\label{zorny} -ie_\ii\, \varepsilon\cdot(p_\ii+p_\ii')\frac{i}{(p_\ii')^2-m_i^2}=\frac{e_\ii\, \varepsilon\cdot p_\ii}{\kk\cdot p_\ii},\end{equation} 
 times an amplitude $\A'$ with no external photon and with the $\ii^{th}$ external momentum shifted slightly from $p_\ii$ to $p'_\ii$.
(We used $p_i^2=m_i^2$, $k^2=0$, and $p'_i=p_i+k$; we also dropped a term proportional to $k$ in the numerator.)
The shift in  $p_i$ is unimportant in the soft limit, by which we mean the limit  $\kk\to 0$ with the other momenta fixed.  
Adding all  contributions with a soft photon attached to an external line,  the scattering amplitude $\A$ with the soft photon
behaves for $\kk\to 0$ as
\begin{equation}\label{lorny}\A\sim\sum_{\ii=1}^\sn \frac{e_\ii\,\varepsilon \cdot p_\ii}{\kk\cdot p_\ii}\,\A'.\end{equation}
Therefore,  in the soft limit, the condition for $\A$ to 
vanish if $\varepsilon=\kk$ is 
\begin{equation}\label{orny}\sum_\ii e_\ii = 0, \end{equation}
or in other words conservation of electric charge.  We simplified the reasoning slightly by assuming that the non-soft particles were all of spin 0; for the general case, see \cite{SW}.

\begin{figure}
 \begin{center}
   \includegraphics[width=6.5in]{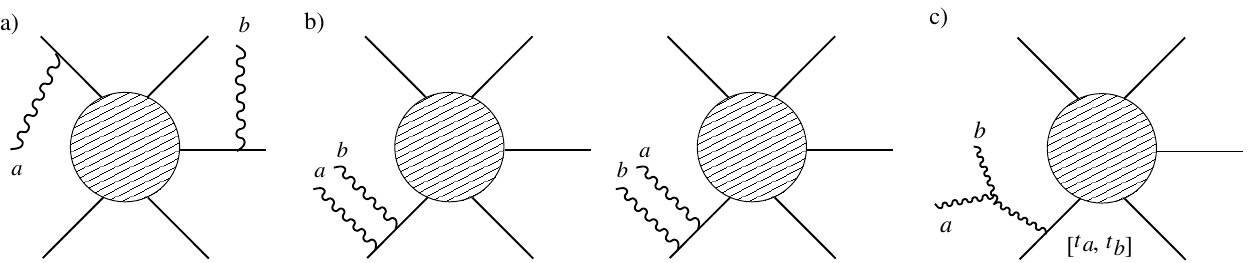}
 \end{center}
\caption{\small Couplings of two soft photons of types $a$ and $b$ to non-soft particles.  The most singular contributions
to the $S$-matrix arise from {\it (a)} processes in which the photons attach to distinct external lines, {\it (b)} processes
in which they couple to the same external line in one of the two possible orderings, and {\it (c)} processes involving
a trilinear coupling of massless fields of spin 1. The label $[a,b]$ in (c) indicates a massless spin 1 field coupling to $[t_a,t_b]$.}
 \label{ubbo}
\end{figure} 
It is instructive to generalize this to the case of several massless fields of spin 1.  
In that case, it is important to consider the possibility that there are several non-soft particles of the same mass,
so rather than speaking of the $i^{th}$ non-soft particle in a given amplitude, we speak 
of the $i^{th}$ mass level of non-soft particle in that amplitude.  A massless
field close to zero momentum can only couple a non-soft particle to another particle of the same mass.
If there are $s$ massless ``photons,''
then in general the  coupling of the $a^{th}$ one
 to the  non-soft particles 
will take the form  \begin{equation}\label{tongo}-i t_{a,i}\, \varepsilon\cdot (p+p'), \end{equation}
where $t_{a,i}$ is a matrix that acts on the states at the $i^{th}$ mass level.\footnote{We absorb the 
coupling constants in the definition of the 
$t_{a,i}$, and
again we assume for simplicity that the non-soft fields have spin 0.}
First let us consider processes in which a single soft ``photon'' of type $a$ is coupled to non-soft particles of momenta $p_1,\dots,p_\sn$.
As in (\ref{lorny}), the amplitude behaves for small photon momentum $\kk$ as
\begin{equation}\label{lornoy}\A\sim\sum_{\ii=1}^\sn \frac{t_{a,\ii}\,\varepsilon\cdot p_\ii}{\kk\cdot p_\ii}\,\A',\end{equation}
and gauge-invariance implies that
\begin{equation}\label{borny}\sum_i t_{a,i}\A'=0. \end{equation}
This is usually described by saying that the spin-one field of type $a$ couples to the conserved charge $t_a$; one interprets $t_{a,i}$
as the matrix by which the symmetry $t_a$ acts on the $i^{th}$ mass level.  
So far, this is not
really a significant generalization of (\ref{orny}); as long as we consider only one massless field of spin 1, we can diagonalize the
charge matrices $t_{a,i}$ and reduce to the previous result.
Now consider an amplitude with two soft photons, say of types $a$ and $b$, with momenta $\kk_a$, $\kk_b$ and polarization vectors 
$\varepsilon_a$,
$\varepsilon_b$,  coupling
to non-soft particles.  The most singular contributions are found by attaching the soft particles on external lines.  If we attach the two soft
photons to two different external lines (fig. \ref{ubbo}(a)), we get a contribution
\begin{equation}\label{worny}\sum_{\ii\not=\ii'}  \frac{t_{a,\ii}\,\varepsilon_a \cdot p_\ii}{\kk_a\cdot p_\ii}  \frac{t_{b,\ii'}\,
\varepsilon_b \cdot p_{\ii'}}{\kk_b\cdot p_{\ii'}}  \,\A' \end{equation}
to the scattering amplitude.
We test  gauge invariance for particle $a$ by setting $\varepsilon_a=\kk_a$, whereupon this contribution  becomes
\begin{equation}\label{corny}\sum_{\ii\not=\ii'}  t_{a,\ii} \frac{t_{b,\ii'}\,
\varepsilon_b \cdot p_{\ii'}}{\kk_b\cdot p_{\ii'}}  \,\A'=-\sum_i  \frac{t_{b,\ii}\,
\varepsilon_b \cdot p_\ii}{\kk_b\cdot p_{\ii}}t_{a,\ii}\A', \end{equation}
where in the last step, we use (\ref{borny}) (and the fact that $t_{a,i}$ commutes with $t_{b,\ii'}$ for $\ii\not=\ii'$).  
Evidently, these contributions to the scattering amplitude are not gauge-invariant by themselves.
Another contribution comes by attaching the two soft photons to the same external line.
This can be done  in two possible orderings (fig. \ref{ubbo}(b)).   The resulting contribution to the amplitude is, in the soft limit $\kk_a,\kk_b\to 0$,
\begin{equation}\label{monzoo}\left(\sum_\ii\frac{t_{a,\ii} \,\varepsilon_a\cdot p_\ii}{\kk_a\cdot p_\ii}
\frac{t_{b,\ii}\,\varepsilon_b\cdot p_\ii}{ (\kk_a+\kk_b)\cdot p_\ii}
+a\leftrightarrow b\right)\A'.\end{equation} 
Upon setting $\varepsilon_a=\kk_a$, this becomes
\begin{equation}\label{surmog}\left(\sum_\ii \frac{t_{b,\ii}\varepsilon_b\cdot p_\ii}{\kk_b\cdot p_\ii}t_{a,\ii}
+\sum_\ii [t_{a,\ii},t_{b,\ii}]\frac{\varepsilon_b\cdot p_\ii}
{(\kk_a+\kk_b)\cdot p_\ii}\right)\A'. \end{equation}
The first term in (\ref{surmog}) cancels the contribution (\ref{corny}) from insertions on distinct lines.  To cancel the second term, we need
a new singular contribution (fig. \ref{ubbo}(c)) 
involving a massless spin 1 particle that couples to $[t_a,t_b]$; this field must participate in a trilinear vertex with
the external photons of types $a$ and $b$. 
We are beginning to uncover here  the basic structure of Yang-Mills theory: the conserved charges form a Lie algebra, and 
the structure constants of this Lie algebra
determine trilinear couplings of massless spin 1 fields.

All of these considerations are valid only if the spin 1 fields under discussion are truly massless.  
The soft limit does not make sense for massive spin 1 particles.

\subsubsection{Gravity And Supergravity}\label{gs}

What we have summarized in section \ref{gt} has close analogs for theories with massless particles of spin 3/2 or 2.
We refer the reader to the references  and merely state the conclusions.   In the context of a Poincar\'e invariant
$S$-matrix, a massless particle
of spin 2 that does not decouple in the zero-momentum limit must couple at zero momentum to the
stress tensor, with a universal coefficient \cite{SW}.  This statement can be expressed as the equality of 
gravitational and inertial mass, something that is usually deduced
from the Principle of Equivalence. The proof is rather similar to the proof of charge conservation
in the spin 1 case.  Going farther in this vein, one can deduce the Einstein equations (with possible
higher derivative corrections), starting simply with a Poincar\'e-invariant theory of a 
massless spin 2 particle -- the ``graviton'' -- that does not decouple at zero momentum \cite{SWtwo}.
A derivation of this along the lines of what we explained for Yang-Mills theory would proceed by considering
a scattering amplitude with two or more soft gravitons coupled to non-soft particles; as we saw for Yang-Mills theory,
gauge-invariance would require nonlinear interactions among the gravitons.

There is a similar story for massless fields of spin 3/2.  In the context of a Poincar\'e invariant $S$-matrix, a
 massless spin 3/2 field -- the ``gravitino'' -- that does not decouple
at zero momentum must couple at zero momentum to a conserved supersymmetry current \cite{GP},
and the $S$-matrix must be supersymmetric.
The proof is again rather similar to the proof of charge conservation in the spin 1 case.  
To show that the conserved supercharges obey the standard supersymmetry algebra, one considers $S$-matrix
elements with two soft gravitinos.  From gauge-invariance of such $S$-matrix elements, one can deduce \cite{GP}
that the supersymmetry algebra must take a standard form and also that in addition to the gravitino, there must
be a massless spin 2 particle, the graviton.  (If there are massless fields of spin 1, then in the right dimensions and
with the right amount of unbroken supersymmetry,  it is possible for the supersymmetry
algebra to have central charges.)  The low
energy structure is that of supergravity.
The reasoning involved is similar to the reasoning by which we showed in section \ref{gt} that if there are massless
spin 1 fields coupled to $t_a$ and to $t_b$, then there must be one coupled to $[t_a,t_b]$.  

The output of this reasoning is a supersymmetric identity obeyed by the $S$-matrix, which can be stated
as follows.  Let $\A'$ be an amplitude for scattering of particles with masses $m_1,\dots,m_\sn$.  Let
$Q_\alpha$ be the supercharges of the theory and let $Q_{\alpha,i}$ be the matrices by which these act
at the $i^{th}$ mass level.  Then
\begin{equation}\label{orob}\sum_{i=1}^\sn Q_{\alpha,i}\A'=0. \end{equation}
The resemblance to eqn. (\ref{borny}) is hopefully obvious, and the derivation by taking a $\kk\to 0$ limit
of a scattering amplitude with a soft gravitino is similar.  Of course, the relation (\ref{orob}) only holds
if the gravitino is truly massless; otherwise one cannot take the limit $\kk\to 0$.  In perturbation theory
around a supersymmetric classical background, the relation holds if the gravitino remains massless in perturbation
theory.

\subsection{Gauge Symmetries In String Theory And Conserved Charges}\label{uperc}

We will now examine the relation between gauge symmetries and conserved charges in string theory.
We begin in section \ref{mansym}
with symmetries of bosonic string theory and symmetries coming from the NS (or NS-NS)
sector of superstring theory.   These are easy to understand because the conserved charges in question
are manifest symmetries of string perturbation theory.  Then starting in section \ref{superc}, we investigate spacetime supersymmetry.

\subsubsection{Manifest Symmetries Of The Worldsheet Theory}\label{mansym}

Let us begin with the  closed bosonic string theory in $\R^{26}$, and consider for illustration the gauge parameter
$\W=c\varepsilon_I \partial X^I \,e^{i\kk\cdot X}$.  For $\kk^2=\varepsilon\cdot \kk=0$, this is a conformal primary
of dimension 0, and of course it obeys our usual condition of not depending on $\partial c$ or $\t\partial\t c$.
The corresponding null state is
\begin{equation}\label{bumox}\V=\{Q_B,\W\}=
\t c ci\kk_J \varepsilon_I \t\partial X^J\partial 
X^I e^{i\kk\cdot X}\end{equation}
and is a linear combination of longitudinal graviton and $B$-field vertex operators.  
(The orientation-reversed gauge parameter
$\t\W=\t c\varepsilon_I\t\partial X^I e^{i\kk\cdot X}$
generates a different linear combination of graviton and $B$-field gauge transformations; symmetric and
antisymmetric combinations are associated to graviton or $B$-field gauge transformations only.)  

The decoupling of the pure gauge  mode $\V$ can be proved using the BRST machinery, but as we have described
in sections \ref{moregauge} and \ref{examples}, for gauge transformations of massless states of the bosonic string, a more explicit
approach is available. The vertex operator $\V$ can be written
\begin{equation}\V=\t c c \t\partial(\varepsilon_I\partial X^I e^{i\kk\cdot X}),\end{equation}
so its integrated version is the total derivative $V=\t\partial(\varepsilon_I\partial X^I e^{i\kk\cdot X})$.  As $V$ is
a total derivative, its integral $\int_\Sigma V$ vanishes, and this ensures the decoupling of the null state represented
by $\V$, establishing gauge-invariance for the fields in question.

The arguments reviewed in  section \ref{revarg} relating  gauge-invariance and conserved charges  make perfect sense in string theory as well as in field theory.  These arguments
involve taking a limit of $S$-matrix elements
for $\kk\to 0$.  However, it will become clear that one can
learn more by setting $\kk=0$ at the beginning, whereupon $\V=V=0$, as is evident in eqn. (\ref{bumox}).
The relation $\V=\{Q_B,\W\}$ becomes 
\begin{equation}\label{ortz} 0=\{Q_B,\W\} \end{equation}
and the relation $V=\t\partial(\varepsilon_I\partial X^I e^{i\kk\cdot X})$ becomes
\begin{equation}\label{rotz}0=\t\partial J,~~J=\varepsilon_I\partial X^I.\end{equation}

So $J$ is a conserved current and therefore generates a manifest symmetry of the worldsheet theory.  
Actually, once we set $\kk$ to zero, the constraint $\varepsilon\cdot \kk=0$ becomes vacuous and any $\varepsilon$ is allowed.  
So we get a whole
family  of conserved currents
\begin{equation}\label{bonko}J^I=\partial X^I. \end{equation}
For the bosonic string in $\R^{26}$, 
the symmetry
associated to these conserved currents is simply translation invariance $X^I\to X^I+a^I$ (with constant $a^I$) and the associated
conserved quantity is
 the energy-momentum (or more briefly the momentum).  
 Conservation of $J^I$ can be applied to closed-string amplitudes on a worldsheet $\Sigma$ of any
 genus and therefore
 the associated conservation law is valid to all orders of bosonic closed-string perturbation theory.
  
 Of course, we can also define
 a second conserved current $\t J^I=\t\partial X^I$.  In $\R^{26}$, the currents $J^I$ and $\t J^I$ are associated
 to the same conserved quantities -- the energy-momentum.  But if some of the $X^I$ are circle-valued, as is appropriate to
describe strings in $\R^{26-m}\times  T^m$, where $ T^m$
is an $m$-torus, then the conserved charges generated by $J^I$ are linear combinations of
momentum and winding numbers, while   $\t J^I$ are associated
to different linear combinations of the same conserved quantities.  The linear combinations $J^I_+= \star \d X^I$
and $J^I_-= \d X^I$ are associated to momentum and winding, respectively. 

In the case of a theory with open as well as closed bosonic strings, the statements of
 the last two paragraphs  assume that
 the boundary conditions along $\partial\SIgma$ are invariant under the symmetry generated by the current
 under consideration (this
 is so precisely if the normal component of the current vanishes
 along $\partial\Sigma$).  Otherwise, mixing of closed and open strings can spontaneously break
 a closed-string gauge symmetry, as explained in section \ref{examples}.  Some examples of boundary conditions
 that preserve or do not preserve a symmetry are as follows \cite{DLP,Horava}.  
   Neumann boundary conditions for a scalar field $X^I$  preserve $J_+^I$  but not $J_-^I$, while Dirichlet
boundary conditions preserve $J_-^I$ but not $J_+^I$.

\begin{figure}
 \begin{center}
   \includegraphics[width=3.5in]{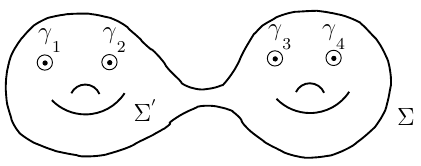}
 \end{center}
\caption{\small $\Sigma'$ is made by omitting small open balls around each of the vertex operator
insertions $\V_1,\dots,\V_\sn$ in the string worldsheet $\Sigma$.  The boundary of $\Sigma'$ is a union of circles $\gamma_1,\dots,\gamma_\sn$ that
enclose the $\V_i$. In the example sketched here, $\n=4$.}
 \label{bunzo}
\end{figure}
Symmetries associated to conserved currents on the string worldsheet are usually so obvious in perturbation theory
that a sophisticated formalism is not really necessary. That is certainly so in the case of translation invariance. 
However, for comparison with what we will say about
spacetime supersymmetry, it is instructive to derive a Ward identity, as follows.  
A vertex operator $\V$ is said to have momentum $p^I$ if
\begin{equation}\label{oink}\frac{1}{2\pi\alpha'}\oint_\gamma J^I \cdot \V=p^I\V. \end{equation}
The integral is taken over a contour $\gamma$ that encloses the operator $\V$ once in the counterclockwise direction. 
Now consider $\n$ vertex operators $\V_1,\dots,\V_\sn$ of momenta $p_1,\dots,p_\sn$, inserted at points $u_1,\dots,u_\sn
\in\SIgma$.  Let $\Sigma'$ be obtained from $\Sigma$ by omitting a small  open ball around each of the $u_i$
(fig. \ref{bunzo}).       
The boundary of $\Sigma'$ is a union of circles $\gamma_1,\dots,\gamma_\sn$, with $\gamma_i$ wrapping
once around $u_i$ in the counterclockwise direction.  Since $\d J=0$, we have
\begin{equation}\label{zoink}0=\left\langle \int_{\Sigma'} \d J \,\cdot\,\V_1\dots\V_n\right\rangle=\sum_{i=1}^\sn\left\langle
\V_1\dots\V_{i-1}\left(\oint_{\gamma_i}J\cdot \V_i\right)\,\V_{i+1}\dots\V_\sn\right\rangle.\end{equation}
No integration over moduli  is relevant here; we make the appropriate antighost insertions
so that the correlation function in question is not trivially zero, but we do not integrate over any moduli.
The correlation function $\left\langle J\,\V_1\dots\V_\sn\right\rangle$ is singular when $J$ approaches any of the $\V_i$.
It is only because of this singularity that the integral over a small circle $\gamma_i$ surrounding one of the operators
is non-zero.  According to (\ref{oink}), the contribution of the singularity can be evaluated by replacing $\oint_{\gamma_i}J^I\cdot \V_i$
with $2\pi \alpha' p_i^I\,\V_i$, so that (\ref{zoink}) becomes
\begin{align}\label{plook}0
 =\left(\sum_{i=1}^n p_i\right)
\bigl\langle\V_1\dots\V_\sn\bigr\rangle.\end{align}
Thus the correlation function $\left\langle\V_1\dots\V_\sn\right\rangle$ vanishes unless the charge associated to $J$
is conserved,
\begin{equation}\label{yelf}\sum_{i=1}^\sn p_i=0.\end{equation}
This is analogous to the conservation laws that we derived in section \ref{gt}.

 There have been two main steps in our reasoning:
\begin{enumerate}\label{zorg}
\item{At $\kk=0$, the gauge parameter $\W$ becomes BRST-invariant; the relation $\V=\{Q_B,\W\}$ reduces to $\{Q_B,\W\}=0$.
}
\item{The integrated version of the vertex operator is a total derivative, $V=\d J$, so the fact that $V=\V=0$
at $\kk=0$ means that $J$ becomes a conserved current on the string worldsheet.}
\end{enumerate}

The first of these two steps is completely general and applies to gauge symmetries of all massless states in all of the bosonic
and supersymmetric string theories.  It just reflects the fact that gauge transformations of massless
fields are proportional to the derivative of the gauge parameter, so that the gauge transformations act trivially at $\kk=0$.

The second step is more special.
It  applies  to gauge symmetries of arbitrary massless states of the bosonic string.
It also applies to gauge symmetries of massless superstring states that come from the NS (or  NS-NS)  sector,
 because the integrated vertex operator corresponding to a massless NS null vector  is 
a total derivative on the worldsheet, as explained in section \ref{morgauge}.  So momentum and winding
symmetry of superstring theory can be treated exactly as we have described for the bosonic string.  
For example, for the heterotic string, the relevant conserved currents are $\t\partial X^I$ and $D_\theta X^I$. 

However, the second step listed above does not hold for spacetime supersymmetry, to which we turn next.

\subsubsection{Spacetime Supersymmetry}\label{superc}

In discussing spacetime supersymmetry in closed, oriented string theory, we will take the heterotic string as the basic
example.  As we explain in section \ref{anomalies},
the analog for Type II superstring theory involves no essential novelty, but some new things
do happen for open and/or unoriented superstrings.

We consider  a supersymmetric compactification of the heterotic string to $\R^d$.  The value of $d$, the number of unbroken
supersymmetries, and the chirality (if $d$ is of the form $4k+2$, $k\in\Z$) of the supersymmetry generators will not play a major role.  So we will not specify these  and we will attempt to keep the notation generic.

The gravitino gauge parameter, at momentum $\kk$, is the operator
\begin{equation}\label{morgo}\W(\kk,u)= c\,\SSigma_{-1/2}u^\alpha\Stigma_\alpha \,e^{i\kk\cdot X},\end{equation}
inserted at a Ramond divisor. Here $\kk^2=0$ and $u^\alpha$ is a $c$-number spinor that obeys the massless Dirac
equation in momentum space,
\begin{equation}\label{torgo}(\gamma\cdot \kk)_{\alpha\beta}u^\beta=0.\end{equation}
Also, as in section \ref{ramond}, $\SSigma_{-1/2}$ represents the $\beta\gamma$ ground state at picture number $-1/2$
(or in other words in the presence of the Ramond divisor) and $\Stigma_\alpha$, which transforms as a spinor
under rotations of $\R^d$, is the fermion vertex operator of \cite{FMS}.  (We need not be concerned
with the details of how $\Sigma_\alpha$ depends
on the variables that describe the compact dimensions, if there are any.)  

The null state 
\begin{equation}\label{yotz}\V(\kk,u)=\{Q_B,\W(\kk,u)\} =i\t c c \kk\cdot \t\partial X\, \SSigma_{-1/2}
u^\alpha\Stigma_\alpha e^{i\kk\cdot X} \end{equation} 
is the vertex operator for a longitudinal gravitino.  Because this vertex operator is BRST-trivial, the corresponding state
will decouple from the $S$-matrix.   As in \cite{GP}, and as summarized in section \ref{gs},
decoupling of this state implies 
spacetime supersymmetry of the $S$-matrix, provided the $S$-matrix exists and the gravitino is exactly massless.  

To prove the vanishing of massless tadpoles and the existence of the $S$-matrix, it seems best to use a formalism
in which $\kk$ is set to 0 from the outset.  Precisely at $\kk=0$, we have $\V(0,u)=0$ and thus $\{Q_B,\W(0,u)\}=0$. 
This is the first main step in section \ref{mansym}.  However, there is no analog of the second step.
We can think of $\W(0,u)$ as being holomorphic, in the sense that it varies holomorphically with the moduli of $\Sigma$.  
(What this means is explained more precisely in the next paragraph.)
But we cannot view $\W(0,u)$ as a holomorphic function (or form) on $\Sigma$ because the only place
that it can be inserted on $\Sigma$ is at a Ramond divisor.  The only way to ``move'' the Ramond divisor at which $\W(0,u)$ is inserted
is to vary the moduli of $\Sigma$, and among these moduli, there is no distinguished one that controls only the position of a given Ramond
divisor.  
Hence, the  only type of
``integration'' involving $\W(0,u)$ that is possible is integration over the moduli of $\Sigma$.
Accordingly, we will have to prove spacetime supersymmetry by integration over the moduli space, not just by integration over $\Sigma$.
It is because of this that it is possible \cite{DSW,DIS,ADS}, though somewhat unusual, for spacetime supersymmetry to be
spontaneously broken in loops though unbroken at tree level.  This is impossible for those gauge symmetries for which the second step
of section \ref{mansym} goes through.

Since $\V(0,u)=0$, we have $\t\partial \W(0,u)=0$.  Does not this mean that $\W(0,u)$ 
can be viewed as a conserved current?  Here it helps
to remember (see section 5 of \cite{Wittenone})
 that the precise interpretation of a heterotic string worldsheet $\Sigma$ is that it is a smooth 
cs supermanifold of dimension $2|1$ embedded
in a product $\Sigma_L\times \Sigma_R$, where $\Sigma_L$ is an ordinary Riemann surface and 
$\Sigma_R$ is a super Riemann surface.
The symbol $\t\partial$ is $\partial_{\t z}$, where  $\t z$ is a local holomorphic parameter on $\Sigma_L$.  
The equation $\t\partial \W(0,u)=0$ reflects the fact that $\W(0,u)$ is the product of the identity
operator on $\Sigma_L$ (which is annihilated by $\t\partial$) times an object on $\Sigma_R$.  That object, which we may as
well just call $\W(0,u)$, can only be inserted at a Ramond puncture and there is no way to ``move'' it 
except by varying the moduli of $\Sigma_R$.
The statement that $\W(0,u)$ varies holomorphically with the moduli of $\Sigma$ just means that it 
varies holomorphically with the moduli of $\Sigma_R$
and is independent of those of $\Sigma_L$.

Once we set $\kk=0$, the condition $\kk^2=0$ is satisfied and the Dirac equation (\ref{torgo}) holds for all $u$.  
So, as in the case
of translation symmetry, we can drop
$u$ from the definition and take the basic object to be
\begin{equation}\label{olgo}\S_\alpha =c\,\SSigma_{-1/2}\Stigma_\alpha. \end{equation}
This is the worldsheet operator that, in a sense that we will explore, generates spacetime supersymmetry.

On a heterotic string worldsheet $\Sigma$ of genus $\g$, 
we will consider a
correlation function of $\S_\alpha$ together with $\n_\NS$ physical state vertex operators from the 
NS sector and $\n_\Ra$ such
operators from the Ramond sector.
(The total number of Ramond punctures is $\n_\Ra+1$, so $\n_\Ra$ must be odd.)  It is
convenient to set $\n=\n_\NS+\n_\Ra$ and denote the physical state
vertex operators simply as $\V_1,\dots,\V_\sn$, without specifying which are of which type.

By the machinery described in sections \ref{measure}-\ref{ramond}, we can define a form 
$F_{S_\alpha\V_1\dots\V_\ssn}(\J,\delta\J)$ on the appropriate
moduli space.  We can view it as a holomorphic form on $\M_L\times \M_R$ (defined in a suitable 
neighborhood of the ``diagonal'' in the reduced
space), where $\M_L$ and $\M_R$ are the moduli spaces of $\Sigma_L$ and $\SIgma_R$, respectively.  
However, $F_{\S_\alpha \V_1\dots\V_\ssn}$
is not an integral form of top degree that could be integrated over the usual integration 
cycle\footnote{We define this cycle
in the usual way for a heterotic string worldsheet of genus $\sg$ with $\n_\NS$ NS punctures 
and $\n_\Ra+1$ Ramond punctures,
just as in section \ref{delcycle}.} $\varGamma\subset \M_L\times \M_R$ to
compute a scattering amplitude.  The reason for this is that the ghost number of $\S_\alpha$ is 
lower by 1 than that of a physical state
vertex operator; accordingly $F_{\S_\alpha\V_1\dots\V_\ssn}$ is a form of codimension 1.  After all, 
$\S_\alpha$ is a symmetry generator,
not the vertex operator of a physical state.  The exterior derivative operator $\d$ increases the degree by 1, so it would
map $F_{\S_\alpha\V_1\dots\V_\ssn}$ to a form of top degree.
  But since $\S_\alpha$ and $\V_1,\dots,\V_\sn$ are all $Q_B$-invariant, the usual relation 
  $\d F_\X=-F_{\{Q_B,\X\}}$ tells us in
this case that
\begin{equation}\label{zolgo}\d F_{\S_\alpha\V_1\dots\V_\ssn}=0.\end{equation}
Since this is the case, we have via the supermanifold version of Stokes's theorem
\begin{equation}\label{nolgo}0=\int_\varGamma \d F_{\S_\alpha\V_1\dots\V_\ssn}=
\int_{\partial\varGamma}F_{\S_\alpha\V_1\dots\V_\ssn}.\end{equation}
This formula is the Ramond sector analog of (\ref{zoink}); it involves the use of Stokes's theorem on  
$\varGamma$, not just on $\Sigma$.  
We will study this formula exactly the way that we studied the general formula (\ref{tarmib}) 
for decoupling of pure gauge modes.

In fact, there are only two differences from that rather general case.  First, the left hand 
side is now 0, since $\V(0,u)=0$.  This means that, rather than a relation saying that a pure gauge mode decouples,
we will get, under appropriate conditions, a conservation law, saying that a certain linear combination of scattering amplitudes vanishes. 
Second, since $\S_\alpha$ carries zero momentum in spacetime, some considerations of section \ref{massren} will be modified.

\subsubsection{The Supersymmetric Ward Identity}\label{bt}

\begin{figure}
 \begin{center}
   \includegraphics[width=4.5in]{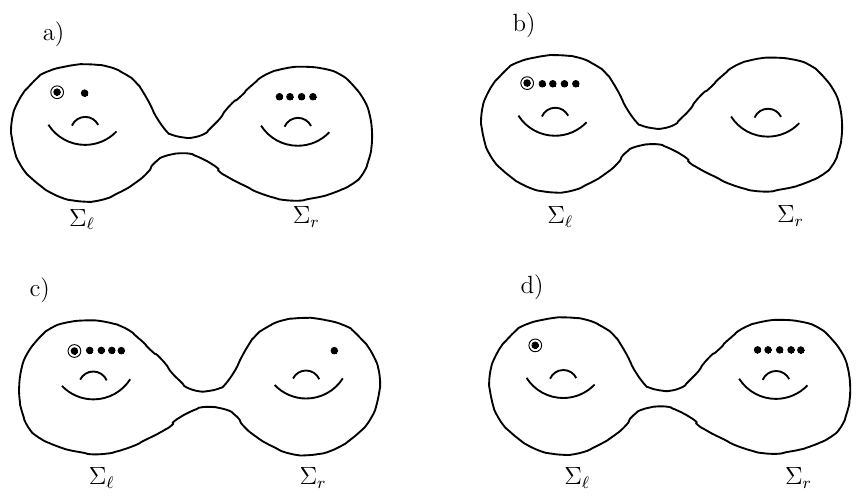}
 \end{center}
\caption{\small Four types of separating degeneration that potentially can contribute to the supersymmetric Ward identity. The
dot enclosed by a circle represents the supercurrent $\S_\alpha$; the other dots represent vertex operators of physical
states.}
 \label{rufus}
\end{figure}
Which components of $\partial\varGamma$ can contribute on the right hand side
of eqn. (\ref{nolgo})? The analysis of this question is very similar to the analysis in section \ref{brstanom} of which boundary components
can contribute BRST anomalies.  In fact, the only relevant difference between the right hand sides of (\ref{nolgo}) and of the general
BRST anomaly formula (\ref{tarmib}) is that in (\ref{nolgo}), we have set $\kk=0$.

Just as in section \ref{brstanom}, we only need to consider separating degenerations, at which
$\Sigma$ splits into two components $\Sigma_\ell$ and $\Sigma_r$, say of genera $\g_\ell$ and $\g_r$,
joined at a double point; we  label these so that $\S_\alpha$ is contained in $\Sigma_\ell$.  And a separating degeneration
can only contribute if the momentum-carrying vertex operators $\V_1,\dots,\V_\sn$ are divided between 
$\Sigma_\ell$ and $\Sigma_r$ in a way that forces the total momentum flowing between $\SIgma_\ell$ and $\Sigma_r$ to be on-shell.

In our previous analysis, BRST anomalies came from three types of separating degenerations, pictured in
figs. \ref{kopla} and \ref{nopla} of section \ref{zumico}.  Setting $\kk=0$ adds only one novelty.  A fourth
type of degeneration  can contribute; if $\Sigma_\ell$ contains $\S_\alpha$ and precisely one additional
vertex operator $\V_i$, then the total momentum flowing between the two branches is precisely that of $\V_i$
and hence is on-shell.

For convenience, we display the four potentially relevant degenerations in fig. \ref{rufus}.  They can each be
analyzed as described in sections \ref{zumico} and \ref{lazy}.  In each case, $\Sigma_\ell$ can be replaced
by a conformal or superconformal vertex operator $\O$ that is inserted on $\Sigma_r$.   $\O$ can be determined
by a path integral on $\Sigma_\ell$, as described most precisely in section \ref{lazy}.

The physical interpretation of the four interesting types of degeneration is as follows. After replacing $\Sigma_\ell$ by an insertion
of $\O$ on $\Sigma_r$,  fig. \ref{rufus}(c) is
equal to an on-shell two-point function $\langle \O\V_i\rangle_{\sg_r}$ (where $\V_i$ is the vertex operator
in $\Sigma_R$).  This is a matrix element of mass renormalization.  For reasons explained in section \ref{massmat},
we restrict ourselves to the $S$-matrix of massless particles, and we only consider  compactifications in which 
supersymmetry prevents mass
renormalization for massless particles.  (This condition is satisfied in the ten-dimensional superstring
theories and, because of nonrenormalization theorems for the superpotential and the Fayet-Iliopoulos terms,
 in most supersymmetric compactifications to 4 or more dimensions.) Thus, we will not have
to analyze fig. \ref{rufus}(c).

In fig. \ref{rufus}(b), $\O$ carries zero momentum and  is the vertex operator of a massless boson
of zero momentum.  Thus the boundary contribution from fig. \ref{rufus}(b) is proportional to the massless
tadpole $\langle\O\rangle_{\sg_r}$.  We will prove in section \ref{vantad} that massless tadpoles vanish in 
models in which supersymmetry is unbroken in perturbation theory.  For the moment, let us simply assume that this
is true and also that we are studying a model in which supersymmetry is unbroken in perturbation theory.  Given this, we need not worry about fig. \ref{rufus}(b).

Fig. \ref{rufus}(d) was analyzed in section \ref{burgo} (without specializing to $\kk=0$). If $\O\not=0$, this degeneration
 will lead to an extra
contribution to the supersymmetric Ward identity, and 
supersymmetry will be spontaneously broken, in the vacuum under study.  Since $\O$ in this degeneration is completely determined
by $\S_\alpha$, it is natural to denote it as $\O_\alpha$.  $\O_\alpha$ is the vertex operator for a physical
fermion field of spin 1/2 -- the Goldstino or Goldstone fermion of spontaneous supersymmetry breaking -- at zero momentum.
In practice, in any given model, it is possible to explicitly show that $\O_\alpha\not=0$ for some $\g_\ell$ (for example, $\O_\alpha\not=0$
for $\g_\ell=1$ in the models studied in \cite{DIS,ADS,DSW}), or else to argue that $\O_\alpha=0$ on general grounds.

For example,\footnote{Arguments similar to the following were used in section \ref{superimp} in analyzing the implications of supersymmetry
for massless tadpoles.}  in the ten-dimensional superstring theories, $\O_\alpha=0$ simply because there is no massless
fermion with the same chirality as $\S_\alpha$ and the same transformation under the global symmetries
of perturbation theory. To spell this out in more detail, consider first 
the heterotic string.  This theory has a massless
neutral fermion of spin 1/2, but its chirality is opposite from that of the spacetime supersymmetry generator
$\S_\alpha$, so it cannot arise as a contribution to $\O_\alpha$.  In Type I and Type IIB superstring theories, there
again is no massless neutral fermion with the same chirality as $\S_\alpha$.  In Type IIA superstring theory, there is
no such fermion with the same chirality and the same transformation under $(-1)^{F_L}$ as $\S_\alpha$.\footnote{
In Type IIA and Type IIB, one must consider both holomorphic and antiholomorphic supersymmetry generators,
but this does not affect what we have just stated.  
In Type I, there is more to the story when open and/or unoriented string worldsheets are considered; see section \ref{anomalies}.}

Below ten dimensions, we need more detailed arguments.  We can use low energy effective field theory to analyze the problem
because we have seen in section \ref{burgo} that in the lowest order of perturbation theory in which it occurs, spontaneous 
supersymmetry breaking means that the effective action is supersymmetric with a correction (\ref{ofra}) to the transformation
law of the Goldstino. Thus if the supersymmetric Ward identity has a contribution in $\g$-loop order from fig. \ref{rufus}(b), this means 
that in $\g$-loop order, the low energy effective action is spacetime supersymmetric but describes spontaneous breakdown of
supersymmetry.   One expects that in this case, massless tadpoles will arise in $2\g$-loop order and perturbation theory
will break down, but in $\g$-loop order one will only see an effective action that describes spontaneous supersymmetry breaking.
Accordingly, in compactifications to four-dimensions with $\N=1$ supersymmetry, nonrenormalization
of the superpotential by loops \cite{DStwo} and nonrenormalization of Fayet-Iliopoulos $D$-terms beyond one-loop
\cite{DSW} means that spontaneous supersymmetry breaking by loops is limited to the models in which it is known
to occur  (and perhaps some of their close relatives in string theories other than the $\mathrm{Spin}(32)/\Z_2$ heterotic
string).  In higher dimensions or with more unbroken supersymmetry,
the constraints are only more severe.

At any rate, we limit ourselves here to models in which general considerations show that spacetime supersymmetry
cannot be spontaneously broken by loop effects.  This being so, the only boundary contributions that remain
are those of fig. \ref{rufus}(a).  These are the ``new'' contributions in which the vertex operators on $\Sigma_\ell$
are $\S_\alpha$ and just one of the $\V_i$.  In this case, $\O$ is bilinear in $\S_\alpha$ 
and in $\V_i$, so we will denote it
as $Q_\alpha(\V_i)$, where $Q_\alpha$ is a transformation between superconformal vertex 
operators of the NS and R
sectors.  From its definition, $Q_\alpha$ manifestly commutes with the spacetime momentum 
so in particular it acts within the space of physical string states at a given mass level.
It also clearly has the same spinor
quantum numbers as $\S_\alpha$.  We define $Q_\alpha$ to be the spacetime supersymmetry charge.

Obviously, there are precisely $\n$ degenerations of the type of fig. \ref{rufus}(a), corresponding to $\n$ possible
choices for which of the $\V_i$ is on $\Sigma_\ell$.  Let us write $\fD_i$ for the $i^{th}$ 
such boundary component.  The contribution of $\fD_i$ to the right hand
side of eqn. (\ref{nolgo}) is an $\n$-particle scattering amplitude with $\V_i$ replaced by $ Q_\alpha(\V_i)$.  
So the vanishing of
eqn. (\ref{nolgo}) becomes
\begin{equation}\label{ibob}\sum_{i=1}^\sn\langle \V_1\dots \V_{i-1}Q_\alpha(\V_i)\V_{i+1}\dots\V_n\rangle=0.\end{equation}
This is our supersymmetric Ward identity, the precise analog of the field theory relation (\ref{orob}). 

In general, $\partial\varGamma$ has many different algebraic components, corresponding to different ways that the string worldsheet
$\Sigma$ may degenerate.  But these components can intersect each other, since $\Sigma$ can undergo multiple degenerations.
Because of this, it is not clear {\it a priori} that the contributions to the right hand side of (\ref{nolgo}) of individual components of $\varGamma$
are well-defined.  However, the $\fD_i$ do not intersect each other (see fig. 6 in section 6.1.2 of \cite{Wittentwo} for an explanation),
so that if they are the only relevant divisors (in other words, if the other configurations sketched in fig. \ref{rufus} do not
contribute), there is no problem in defining their individual contributions.

\subsubsection{Comparison To The Standard Description}\label{modox}

We would now like to compare our definition of the spacetime supersymmetry generators to the standard one \cite{FMS}.

\begin{figure}
\begin{center}
\includegraphics[width=4.5in]{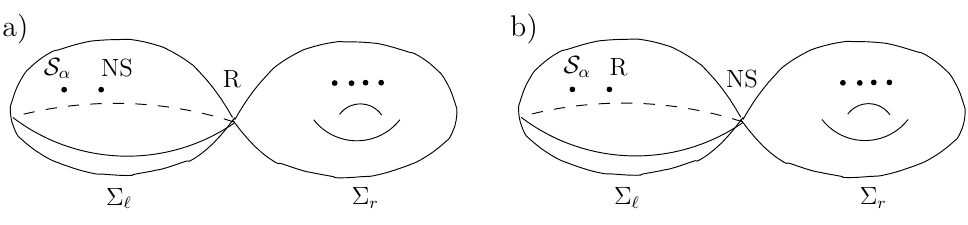}
\end{center}
\caption{\small The $\g_\ell=0$ contribution to the action of $Q_\alpha$ on a vertex operator $\V$.  Whether $\V$ is
of NS or R type, because of the $\S_\alpha$ insertion in $\Sigma_\ell$, the vertex operator inserted at the double point is 
of the opposite type, as sketched in (a) and (b).  So
in either case, $\Sigma_\ell$ is a super Riemann surface of genus 0 with
one NS puncture and two R punctures, and therefore has no even or odd moduli.  However, there is an important difference between
the two cases; if $\V$ is of NS type as in (a), 
the vertex operator at the double point is of R type, so there is a fermionic gluing parameter.
If $\V$ is of R type as in (b), the vertex operator at the double point is of NS type and there is no fermionic gluing parameter.}
\label{diffgen}
\end{figure}

In fact, we can recover the standard definition by just setting $\g_\ell=0$.  In this case, $\Sigma_\ell$ is simply a three-punctured
sphere.  The subtleties of section
\ref{lazy} are irrelevant, since a three-punctured sphere has no bosonic moduli, and  we can omit the 
$\d |q|^2/|q|^2$ terms from eqn.
(\ref{powark}).   To evaluate $Q_\alpha(\V)$, for a superconformal vertex operator $\V$, we need
to compute a three-point function on $\Sigma_\ell$ (fig. \ref{diffgen}).
  A genus zero three-point function is determined by the operator product expansion, which was also the basis
for the definition of the supercharges given in \cite{FMS}, so it should come as no surprise that the two definitions agree.

When we evaluate $Q_\alpha(\V)$,   because of the $\S_\alpha$ insertion in
$\Sigma_\ell$, the string state that flows through the double point connecting $\Sigma_\ell$ to $\Sigma_r$ is of opposite NS or R
type from $\V$, as indicated in the figure, so regardless of whether $\V$ is of NS or R type, $\Sigma_\ell$ contains one NS insertion
and two R insertions.  Accordingly, $\Sigma_\ell$ has no even or odd moduli (a super Riemann surface of genus 0 with one NS
insertion and two R insertions has no moduli).  However, there is an interesting difference between the two cases.
If $\V$ is of NS type, then the vertex operator at the double point is of R type and hence in the gluing of  $\Sigma_\ell$ to $\Sigma_r$,
there is a fermionic gluing parameter.  If $\V$ is of R type, there is no fermionic gluing parameter. 

Eqn. (\ref{powark}), or its superconformal analog, applies uniformly for all $\g_\ell$, but to calculate just for $\g_\ell=0$, it is easier
to remember where eqn. (\ref{powark}) came from.  A degeneration in which $\Sigma$ splits off a genus 0 component $\Sigma_\ell$ that
contains two vertex operators (in our case $\S_\alpha$ and $\V$) 
is the way that in the Deligne-Mumford compactification one describes
a collision between two vertex operators.  So let us consider an operator product $\S_\alpha(z)\V(0)$.  For the degeneration
that occurs as $z\to 0$, the gluing parameter $q_\NS$
or $q_\Ra$ is simply equal to $z$.  
So our usual procedure for cutting off the moduli space amounts, in this example, to requiring $|z|\geq \epsilon$
for some small $\epsilon$, and the boundary of the cutoff moduli space is the circle $|z|=\epsilon$.
We wish to integrate over this circle.    It is convenient to define
\begin{equation}\label{seemly}\S_\alpha = c\Stt_\alpha,~~ \Stt_\alpha=\SSigma_{-1/2}\Sigma_\alpha. \end{equation}
By essentially the usual procedure of passing from a vertex operator to its integrated form, as described in sections \ref{horsy} and
\ref{intns},
the integral  can be accomplished by replacing $c$ with a one-form $\d z$, so that $\S_\alpha$ is
replaced by its ``integrated form'' $\d z\,\Stt_\alpha$.  Then we simply integrate over the circle $|z|=\epsilon$:
\begin{equation}\label{picolo}\frac{1}{2\pi i}\oint_{|z|=\epsilon}\d z\, \Stt_\alpha \cdot \V.\end{equation}
(We have taken the liberty of dividing the integral by  $2\pi i$ to agree with standard normalizations.)
The integral picks out the residue of the pole in the operator product $\Stt_\alpha(z)\cdot \V(0)$.
If $\V$ is in the Ramond sector, this residue is the genus 0 approximation to $Q_\alpha(\V)$; if $\V$ is in the NS sector, we still
have to integrate over the fermionic gluing parameter, which can be accomplished by multiplying by a picture-changing operator.\footnote{One
integrates over the fermionic gluing parameter by acting with $\delta(\beta_0)G_0$, while the usual picture-changing recipe is to multiply
with the picture-changing operator $\YY(z')$ (defined in eqn. (\ref{nifox})), with $z'\to 0$.  
In general, these operations coincide modulo $Q_B(\dots)$; on superconformal
vertex operators with the usual simple dependence on the ghosts, they coincide precisely. A useful reference on such matters is
\cite{Belthree}.}  
What we have arrived at 
is the definition of the spacetime supercharges $Q_\alpha$ given in \cite{FMS}.

It is interesting that we have a framework to compute higher genus corrections to the supersymmetry generators $Q_\alpha$, namely
the contributions with $\g_\ell>0$.   But to make this transparent, it really needs to be combined with an understanding of mass
renormalization, presumably via an off-shell approach such as that of \cite{Sen1,Sen3}.
From the point of view of low energy effective field theory, the representations of the supersymmetry algebra are completely determined by
the particle masses, spins, and possibly the central charges (in string theory compactifications in which these are present in the
supersymmetry algebra).  One expects therefore that  renormalizations of the masses and possibly the central charges
completely determine the corrections to the $Q_\alpha$.

Our definition of the $Q_\alpha$ is actually a very close cousin of what is briefly explained in \cite{Beltwo}.  The main difference is that
we use a superconformally invariant formalism, while  \cite{Beltwo} is based on a more general formalism along the lines of section
\ref{morge}.

\subsubsection{More On Momentum Conservation}\label{morecon}

The definition of the spacetime supercharge in section \ref{superc} was notably more sophisticated
than the treatment of bosonic symmetries such as momentum and winding in section \ref{uperc}.
It does not seem that the definition of spacetime supersymmetry can be reduced to the more elementary ideas
used in section \ref{uperc}, but there is no problem to go in the opposite direction and describe bosonic gauge symmetries
in the same language that we used for supersymmetry.

For example, let us consider the momentum and winding symmetries of the bosonic string, either in $\R^{26}$ or
toroidally compactified.  The gauge parameters
for the sum and difference of the metric and $B$-field are $\W=c\varepsilon\cdot \partial X \exp(ip\cdot X)$,
$\t\W=\t c\varepsilon\cdot \t\partial X\exp(ip\cdot X)$, with $p^2=p\cdot \varepsilon=0$.  Taking $p=0$, $\varepsilon$
becomes arbitrary and we define $\P^I=c\partial X^I$, $\t\P^I=\t c\t\partial X^I$.  These $Q_B$-invariant operators
can be used to define
conservation laws in the same way that we did for supersymmetry.  In the case of  $\P^I$, for example,
we use the correlation function $\langle \P^I\V_1,\dots\V_\ssn\rangle$
to define a differential form $F_{\P^I,\V_1,\dots,\V_\ssn}$ on $\h\M_{\sg,\sn+1}$.
It is closed and of codimension 1, so as usual
\begin{equation}\label{luggy}0=\int_{\h\M_{\sg,\sn+1}} 
\d F_{\P^I\V_1\dots\V_\ssn}=\int_{\partial\h\M_{\sg,\sn+1}} F_{\P^I\V_1\dots\V_\ssn}.  \end{equation}
Analysis of the surface terms here leads to a conservation law, which is the same one that we deduced in section \ref{uperc} using
the conserved worldsheet current $J^I=\partial X^I$.  

For the heterotic string, one defines similarly $\P^I=c\delta(\gamma)D_\theta X^I$, $\t \P^I=\t c\t\partial X^I$.
The Type II analog is evident.  Again, the same formalism can be used to deduce momentum and winding conservation.  
In the language of \cite{FMS}, the operator $\P^I$ is related by picture-changing to the  conserved worldsheet current $D_\theta X^I$
that is used in the more elementary explanation. (To be more precise, $\P^I$ is a picture-changed version of
$\partial_zX^I=\int \d\theta D_\theta X^I$.)
The advantage of the more abstract approach to momentum and winding symmetry using
the operator $\P^I$ rather than the current $D_\theta X^I$
is that it enables one to treat momentum conservation and spacetime supersymmetry in the same
framework.  When we study the
spacetime supersymmetry algebra in section \ref{superalg}, it will be hard to avoid treating the different spacetime symmetries
in the same framework.

Though it is not clear that this is useful,
one can also place in the same framework the conservation laws associated to massless spin 1 gauge fields.
We will just mention a few illustrative examples.  In the fermionic description of the $\mathrm{Spin}(32)/\Z_2$ heterotic string,
one describes the left-moving current algebra via fermionic primary fields $\Lambda_a$, $a=1,\dots,32$ of dimension $(1/2,0)$.
The associated current operators are $\RR_{ab}=\t c \Lambda_a\Lambda_b$ and can be used in
an argument along the lines of eqn. (\ref{luggy}) to establish the relevant global symmetries. In Type I superstring theory,
the usual massless gauge fields are open-string modes.  Including the Chan-Paton factors (we usually do not
make them explicit  in this paper), the corresponding vertex operators are $\V_T=c\delta(\gamma)\varepsilon\cdot \partial X \exp(ip\cdot X) T$,
where $T$ is a group generator acting on the Chan-Paton factors.  The corresponding gauge parameter (see eqn. (\ref{wonzop}))
is $\W_T=c\delta'(\gamma)\exp(ip\cdot X)T$.  Setting $p=0$, we define the current operator $\RR_T=c\delta'(\gamma)T$,
and one can use this operator, inserted on $\partial\SIgma$, 
to prove the global symmetries associated to Chan-Paton gauge-invariance by
following the procedure of eqn. (\ref{luggy}).  Of course, what we have just described are rather long routes to establish
 symmetries that actually are manifest in the perturbative formalism.

The  fields of perturbative string theory that are usually called gauge fields but
that cannot be put in this framework are the $r$-form gauge-fields of the R-R sector of Type II superstring
theory (possibly enriched with D-branes and/or orientifold planes).   
They are not really gauge fields in the relevant sense.
There are no R sector gauge parameters at the massless level of perturbative string theory
(the first R sector gauge parameter is described in eqn.
\ref{tolver}).  Moreover,  the elementary string states 
are all neutral under the R-R gauge symmetries,
so R-R gauge symmetry does not lead to a  non-trivial conservation law  in scattering of these states. R-R gauge
invariance comes into play when one constrains the D-branes and orientifold planes via R-R tadpole cancellation;
see section \ref{tadanomalies}.

\begin{figure}
\begin{center}
\includegraphics[width=2.5in]{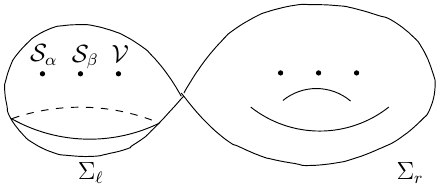}
\end{center}
\caption{\small As a preliminary to evaluating the anticommutator $\{Q_\alpha,Q_\beta\}$, we restrict the closed integral form
 $F_{\S_\alpha\S_\beta\V_1\dots\V_\ssn}(\J,\delta\J)$ to a locus $|q|=\epsilon$, where the divisor indicated here is defined by $q=0$.
   After this restriction,  $F_{\S_\alpha\S_\beta\V_1\dots\V_\ssn}(\J,\delta\J)$ still has codimension 1. }
\label{itz}
\end{figure}

\subsubsection{The Supersymmetry Algebra}\label{superalg}
Since we have obtained the standard supersymmetry generators, at least in the $\g_\ell=0$ approximation, it is fairly obvious
that we must get the standard supersymmetry algebra, at least in that approximation.  It is interesting to see how to compute
this algebra, in the present formulation.   First let us recall how we would do this at non-zero
momentum.  For this, we can follow a procedure described in section \ref{gs}.  Given any $\n$ non-soft
string states represented by vertex operators $\V_1,\dots,\V_\sn$, we would add two vertex operators  for
soft gravitinos, compute a $\g$-loop contribution by integrating a suitable integral form on the appropriate 
integration cycle $\varGamma$,
and deduce the supersymmetry algebra by studying the limit of this integral as the gravitino momentum goes to zero.

Instead here we want to describe an analogous procedure with the soft gravitino vertex operators replaced by supersymmetry
generators $\S_\alpha$ and $\S_\beta$, which carry zero momentum.  The usual procedure will extract from the worldsheet path integral a closed integral form
 $F_{\S_\alpha\S_\beta\V_1\dots\V_\ssn}(\J,\delta\J)$
on the same cycle $\varGamma$ as before, but now, as two physical state 
vertex operators have been replaced by $\S_\alpha$ or $\S_\beta$
(whose ghost number is smaller by 1), $F(\J,\delta \J)$ is a form of codimension 2.  So it certainly cannot be integrated over 
$\varGamma$.  

We can reduce the gap in dimension from 2 to 1 by restricting 
$F(\J,\delta\J)$ from $\varGamma$ to one of its boundary components $\B$.
(This is done in the usual way by a relation such as $|q|=\epsilon$, where $q$ is a complex gluing parameter; such a relation
reduces the real dimension by 1.)  We pick $\B$ to correspond (fig. \ref{itz}) to a separating
degeneration in which $\Sigma$ splits into two components $\Sigma_\ell$ and $\Sigma_r$, where $\Sigma_\ell$ contains
precisely $\S_\alpha$, $\S_\beta$, and one more vertex operator that we will call 
simply $\V$; all other vertex operators are in $\Sigma_r$.

We still cannot integrate $F(\J,\delta\J)$ over $\B$ since even after restriction to $\B$, it is of codimension 1.  
However, $\B$ itself has boundaries associated to further degenerations. 
Since $\d F(\J,\delta\J)=0$, we have a relation just analogous to (\ref{nolgo}) but with $\varGamma$ replaced by $\B$:
\begin{equation}\label{inco}0=\int_\B\d F(\J,\delta\J)=\int_{\partial\B}F(\J,\delta\J).\end{equation}
Among the boundary component of $\B$, those that arise by degeneration of $\Sigma_r$ are not relevant, since all vertex operators
on $\Sigma_r$ are those of physical states; the path integral on $\Sigma_r$ is already generating a form of top degree. We only care about degenerations of $\Sigma_\ell$.

\begin{figure}
\begin{center}
\includegraphics[width=6.5in]{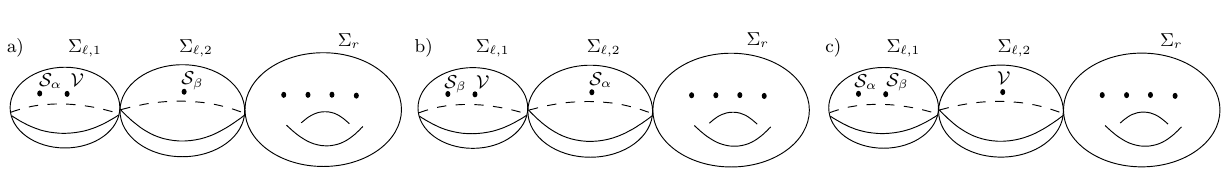}
\end{center}
\caption{\small Stokes's theorem applied to the configuration of  fig. \ref{itz} says that the sum of these three contributions
vanishes.  This yields the spacetime supersymmetry algebra.}
\label{witz}
\end{figure}

\begin{figure}
\begin{center}
\includegraphics[width=6.5in]{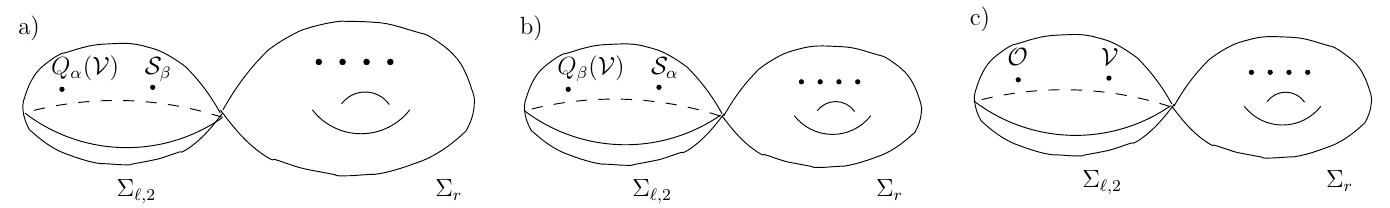}
\end{center}
\caption{\small As described in the text, these three pictures represent a step in evaluating the contributions
from the three pictures in fig. \ref{witz}.  The operator labeled $\O$ in (c) is bilinear in $\S_\alpha$ and $\S_\beta$.}
\label{nitz}
\end{figure}
To keep things simple,\footnote{And also because of not fully understanding the exceptional
zero-momentum contributions that will enter momentarily.  One expects no higher loop corrections
to the supersymmetry algebra, probably because there are no contributions in which the genus of
$\Sigma_{\ell, 1}$ in fig. \ref{witz}(c) is positive.  To show this, one needs to analyze the exceptional
contributions.}
we will only analyze the genus 0 contribution to the supersymmetry algebra.
If $\Sigma_\ell$ has genus 0, it can degenerate in precisely three ways, sketched in fig. \ref{witz}.  
In each of these cases, $\Sigma_\ell$ splits into two components $\Sigma_{\ell,1}$ and $\Sigma_{\ell,2}$ where
only the second intersects $\Sigma_r$.  We can compute the corresponding amplitudes by iterating ideas we have already
explained.  In each case, we can replace $\Sigma_{\ell,1}$ by some operator insertion on $\Sigma_{\ell,2}$.  In fig. \ref{witz}(a),
this operator is by definition $Q_\alpha(\V)$. Replacing $\Sigma_{\ell,1}$ by an insertion of $Q_\alpha(\V)$ on $\Sigma_{\ell,2}$,
we arrive at fig. \ref{nitz}(a), where now we can replace $\Sigma_{\ell,2}$ by an insertion of $Q_\beta(Q_\alpha(\V))$ on $\Sigma_r$.
A similar analysis of fig. \ref{witz}(b), leads via  fig. \ref{nitz}(b)  to an insertion of $Q_\alpha(Q_\beta(\V))$ on $\Sigma_r$.
Finally, in fig. \ref{witz}(c), we can replace $\Sigma_{\ell,1}$ by an insertion of an operator that is bilinear in  $\S_\alpha$ and $\S_\beta$;
we call this operator $\O_{\S_\alpha,\S_\beta}$ or simply $\O$.  And then from fig. \ref{nitz}(c), we can replace $\Sigma_{\ell,2}$ by an insertion on $\Sigma_r$ of an operator
bilinear in $\O$ and $\V$, which we call $\O(\V)$.   The identity (\ref{inco}) becomes
\begin{equation}\label{texo}Q_\alpha(Q_\beta(\V))+Q_\beta(Q_\alpha(\V))+\O(\V)=0.\end{equation}
Hence 
\begin{equation}\label{mexo}\O=-\{Q_\alpha,Q_\beta\},\end{equation}
and we can determine the supersymmetry algebra by a path integral on $\Sigma_{\ell,1}$ in fig. \ref{witz}(c).

In this figure, the momentum flowing between $\Sigma_{\ell,1}$ and $\Sigma_{\ell,2}$ is zero, so to understand the path integral
on $\Sigma_{\ell,1}$,  we need the exceptional
zero-momentum contribution whose origin we have first seen in formulas such as (\ref{zyrfex}) and (\ref{opal}) for bosonic strings.
The relevant contribution is 
\begin{equation}\label{tornz}\t c \,\t\partial\t c\,\t\partial^2\t c \,c\delta(\gamma)D_\theta X^I\otimes c\delta(\gamma) D_\theta X_I+\dots
\end{equation}
where we omit similar terms (with the two factors of the tensor product exchanged, or holomorphic factors exchanged with analogous
antiholomorphic ones) that do not contribute to evaluating the supersymmetry algebra.   In eqn. (\ref{tornz}), we see
the operator $\P^I=c\delta(\gamma)D_\theta X^I$ whose relation to energy-momentum conservation was described in section
\ref{morecon}.

  However,
as in section \ref{modox}, it is easier, for $\Sigma_{\ell,1}$ of genus 0, to recognize that what we are trying to calculate
is simply a term in the operator product expansion.  The same reasoning that led to (\ref{picolo}) gives
\begin{equation}\label{ironz}\O_{\S_\alpha,\S_\beta}=\frac{1}{2\pi i}\oint_{|z|=\epsilon}\d z\,\Stt_\alpha(z)\S_\beta(0).\end{equation}
This is the standard answer, in the sense that the supersymmetry algebra is computed in 
\cite{FMS,Knizhnik} from just this operator product.   For superstrings in $\R^{10}$, the
 right hand side of (\ref{ironz}) can be evaluated to give 
$\gamma_{I\alpha\beta}\P^I= c\delta(\gamma)\ggamma_{I\alpha\beta} D_\theta X^I$.
(Not coincidentally, the same operator $c\delta(\gamma)D_\theta X^I$  appears in (\ref{tornz}), giving another route to the same result 
for $\{Q_\alpha,Q_\beta\}$.) Here, as we explained in section \ref{morecon}, the operator $\P^I$ is related to energy-momentum
conservation in precisely the same way that $\S_\alpha$ is related to spacetime supersymmetry.  So we have arrived
at the usual spacetime supersymmetry algebra.

In the language of \cite{FMS}, the operator $\P^I= c\delta(\gamma)D_\theta X^I$ is 
related by picture-changing to  the holomorphic current
 that generates translation symmetry (or after toroidal compactification, a linear combination of translation
and winding symmetry).  The picture-changing is
a way to describe the integration over the fermionic gluing parameter in fig. \ref{nitz}(c).

A brief explanation of a calculation of the supersymmetry algebra somewhat similar to what we have explained
but in a rather different language can be found in section 6.3 of \cite{Beltwo}.

For superstring theory compactified to $\R^d\times Z$ with $d<10$ for some space $Z$,  the operator product in eqn. (\ref{ironz}) may involve,
in addition to the $\P^I$, also operators associated to  symmetries of $Z$.  In that case, generators of these symmetries
appear as central charges in the supersymmetry algebra.  If not prevented by arguments of holomorphy or other
considerations of low energy effective
field theory, there may be loop corrections to the central charges, which would appear as 
loop corrections to the supersymmetry algebra.  This may be most likely for $d\leq 3$, where constraints of low energy
field theory can be less powerful.

\subsection{Vanishing Of Massless Tadpoles}\label{vantad}

Finally we come to the question of proving that perturbative massless tadpoles vanish in those supersymmetric compactifications
in which supersymmetry is not spontaneously broken in perturbation theory.

The one fact that we need to know is that if $\V_\phi$ is the vertex operator of a massless neutral spin-zero field $\phi$ at zero momentum, 
then there is
always a fermion vertex operator $\V_{\psi_\alpha}$ of a zero-momentum neutral fermion field $\psi_\alpha$ 
that satisfies a relation 
\begin{equation}\label{boxco}\V_\phi=\sum_\alpha Q_\alpha(\V_{\psi_\alpha}).\end{equation}
One can show this by explicitly exhibiting $\V_{\psi_\alpha}$ in each case.
For example,
 in ten-dimensional heterotic string theory, the only $\V_\phi$ is 
 \begin{equation}\label{colder}\V_\phi=\t c c\delta(\gamma)\t\partial X_I D X^I,\end{equation}
 and the corresponding $\V_{\psi_\alpha}$ is 
 \begin{equation}\label{oxco}\V_{\psi_\alpha}=\ggamma_I^{\alpha\beta} 
 \t c c \t\partial X^I \SSigma_{-1/2}\varSigma_\beta.
 \end{equation}
(The $\ggamma_I$ are spacetime gamma matrices and $\varSigma_\beta$ is a spin field of the matter system.)

Rather than exhibit such formulas in all the cases, we prefer to observe that the result actually
follows from the general form of the low energy supersymmetry transformations.
Under spacetime supersymmetry, $\phi$ always transforms into a fermion field, $Q_\alpha(\phi)=\psi_\alpha$,
and $\psi_\alpha$ 
is non-zero at zero momentum.  On the other hand, at zero momentum 
we have $Q_\alpha(\psi_\beta)=0$, assuming that
supersymmetry is not spontaneously broken.\footnote{In supersymmetric 
field theories with unbroken supersymmetry (which
excludes a constant term in the transformation law of a fermion field), massless 
neutral spin 1/2 fields transform under supersymmetry
into derivatives of scalars and also into 
the field strengths of abelian gauge fields. These vanish at zero momentum. Note
as well that 
 the derivative of a scalar and the field strength of a gauge field are not Lorentz scalars,
 so for that reason alone, they could not
 contribute to $\sum_\alpha Q_\alpha(\V_{\psi_\alpha})$.}  
Now consider turning on $\phi$ and $\psi_\alpha$ at zero momentum, perturbing the worldsheet
action by $\phi \V_\phi+\psi_\alpha \V_{\psi_\alpha}$.  For this class of perturbations 
to be closed under spacetime supersymmetry, since
$\phi$ transforms into $\psi_\alpha$ while $\psi_\alpha$ is invariant, 
$\V_{\psi_\alpha}$ must transform into $\V_\phi$, giving
a relation (\ref{boxco}).  

Supersymmetric invariance of $\phi \V_\phi+\psi_\alpha \V_{\psi_\alpha}$ gives one more interesting condition:
at zero momentum,
$\V_\phi$ must be invariant under supersymmetry.   This simply means that the perturbation 
by $\V_\phi$ preserves spacetime supersymmetry,
so the expectation value of $\phi$ is a modulus of the supersymmetric theory, at least to first order.

Going back to eqns. (\ref{boxco}) and (\ref{oxco}),
there is actually an interesting detail here, which echoes a comment in section \ref{remark}.  
Just as $\V_\phi$ is the vertex
operator for a linear combination of the dilaton and the trace of the metric, so $\V_{\psi_\alpha}$ is the vertex operator
for a linear combination of gravitino and dilatino fields.  
(By the dilatino, we mean the spin 1/2 field in the ten-dimensional $\N=1$ supergravity
multiplet.)  At non-zero momentum, there are separate superconformal vertex operators for 
dilatons and gravitons and likewise
for dilatinos and gravitinos.
At zero momentum, in a superconformal formalism, there is only one $\V_\phi$, so only 
one linear combination of the tadpoles
of the dilaton and the trace of the metric is a potential obstruction to the validity of 
perturbation theory, and there is only one $\V_{\psi_\alpha}$
that we can use to prove the vanishing of this one obstruction.  In a more general 
formalism as summarized in section \ref{morge}, we could
potentially encounter separate tadpoles for the dilaton and the trace of the graviton, and 
we would have two separate $\V_{\psi_\alpha}$'s to
deal with these two tadpoles.

\begin{figure}
 \begin{center}
   \includegraphics[width=5.5in]{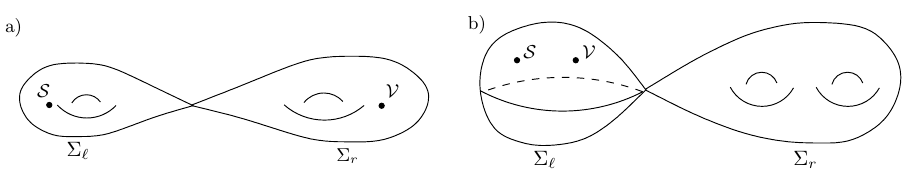}
 \end{center}
\caption{\small With only two vertex operators, only two types of separating 
degeneration are relevant to the supersymmetric
Ward identity.  They are related to (a) spontaneous supersymmetry breaking or (b) the massless tadpole.}
 \label{curry}
\end{figure}
Given the relation (\ref{boxco}), the proof that massless tadpoles vanish in $\g$-loop 
order is very short.  We simply consider
the two-point function 
\begin{equation}\label{jokkop}\sum_\alpha\langle\S_\alpha \,\V_{\psi_\alpha}\rangle\end{equation} in genus $\g$.  
As usual, we use this correlator to define a closed integral form $F_{\S,\V}$ on the relevant integration
cycle $\varGamma$, leading to the usual relation
\begin{equation}\label{pkil}0=\int_\varGamma\d F_{\S\,\V}=\int_{\partial\varGamma}F_{\S\,\V}.\end{equation}
With only two vertex operators,
there are only two
types of boundary contribution that may contribute to the relation (\ref{pkil}); the two 
vertex operators are contained in opposite 
components or the same component of $\Sigma$ (fig. \ref{curry}).   The two cases 
correspond respectively to spontaneous supersymmetry
breaking (fig. \ref{curry}(a)) and massless tadpoles (fig. \ref{curry}(b)).

In fig. \ref{curry}(a), the two branches $\Sigma_\ell$ and $\Sigma_r$ both have positive 
genus, or else this degeneration does not
arise in the Deligne-Mumford compactification.  Since $\g_\ell+\g_r=\g$, it 
follows that $\g_\ell, \,\g_r\leq \g-1$.  In fig. \ref{curry}(b),
$\g_r>0$ but $\g_\ell$ may vanish, so $\g_r$ can equal $\g$.

A very short argument now shows the vanishing of massless tadpoles.
Inductively, suppose that we know that massless tadpoles vanish and spacetime 
supersymmetry is unbroken up to and including genus $\g-1$.
Then almost all boundary contributions to the relation (\ref{pkil}) vanish.  The only  contribution that is not ruled
out by the inductive hypothesis comes from the degeneration of
fig. \ref{curry}(b) with $\g_\ell=0$, $\g_r=\g$.  Using (\ref{boxco}), this is simply the genus $\g$ tadpole 
$\langle\V_\phi\rangle_\sg$.
So that tadpole vanishes.

Once we know that the massless tadpoles vanish in genus $\g$, the $\g$-loop amplitudes 
make sense and the arguments of 
section \ref{bt} show that these amplitudes have spacetime supersymmetry, in general  possibly
with spontaneous supersymmetry breaking in $\g$-loop order.  If  (as in most cases) general 
arguments show that  spacetime supersymmetry
is not spontaneously broken at $\g$-loop order, then the inductive step is complete and we 
can repeat the argument for genus $\g+1$. 

Just one more comment is perhaps called for here.  It was crucial for the inductive argument 
that if spacetime supersymmetry is unbroken
up to and including
$\g$-loop order, there are no massless tadpoles below $\g+1$-loop order.  We deduced this 
from general properties of the Deligne-Mumford
compactification.  But in fact, low energy effective field theory 
implies that if supersymmetry is unbroken up to $\g$-loop order, then
massless tadpoles appear only in order $2\g$.

\section{Open And/Or Unoriented Strings}\label{anomalies}

\subsection{Overview}\label{anomover}

In  section \ref{tadpoles}, we took the heterotic string as the basic example of a string theory with spacetime supersymmetry.
In that string theory, the supersymmetry generator $\S_\alpha$ is a right-moving or holomorphic field, and the distinction between NS and R sectors
only exists for right-moving and not for left-moving  degrees of freedom on the string worldsheet.

The analysis of section \ref{tadpoles} extends almost at once to oriented closed Type II superstring theory.  In this theory, 
there are two types of spin 0
field, arising either in the NS-NS sector or the R-R sector.  In the superconformal framework, 
expectation values of  R-R fields cannot be conveniently incorporated, so they are assumed
to vanish; as a result, there is a global
symmetry $(-1)^{F_L}$ that distinguishes between  the left-moving NS and R sectors and
ensures the vanishing of R-R tadpoles.  So only NS-NS tadpoles need to be considered.

In oriented closed Type II superstring theory, there are separate spaces $\S_L$ and $\S_R$ of left- and right-moving
supersymmetry generators. (In ten dimensions and in the obvious compactifications, $\S_L$ and $\S_R$ have the same dimension, but this
is not true in general; for example \cite{FK}, it is not always true for asymmetric orbifolds.)
The left- and right-moving symmetries can each be
treated exactly as we treated the supersymmetry generators of the heterotic string.
For example, every left- or right-moving supersymmetry generator $\S_\alpha'\in \S_L$ or $\S_\beta''\in\S_R$
is associated to a spacetime supersymmetry generator $Q_{\alpha}'$ or $Q_\beta''$, each 
defined exactly by the procedure of section \ref{bt}.
  $Q_\alpha'$ exchanges the left-moving NS and R sectors, and $Q_\beta''$ does the same
for right-moving ones. $Q_\alpha'$ anticommutes with $Q_\beta''$, since there is no short distance singularity
between $\S_\alpha'$ and $\S_\beta''$. 
If $\V_{\NS,\NS}$ is a spin-zero and momentum zero
 superconformal vertex operator from the NS-NS sector, then eqn. (\ref{boxco}) bifurcates into two
separate formulas, one involving  $Q_\alpha'$ and one involving $Q_\beta''$:
\begin{equation}\label{zaxco}\V_{\NS,\NS}=\sum_\alpha\{Q_\alpha',\V^\alpha_{\Ra,\NS}\}=\sum_\beta\{Q_\beta'',
\V^\beta_{\NS,\Ra}\},\end{equation}
where $\V^\alpha_{\Ra,\NS}$ and $\V^\beta_{\NS,\Ra}$ are vertex operators of the indicated types.
Each of these relations holds for the same reasons as (\ref{boxco}), and 
either one of them can be used, independent of the other,
as input for the proof of vanishing of massless tadpoles given in section \ref{vantad}.

There is much more to say if we generalize Type II superstring theory to include  
open and/or unoriented string worldsheets,
by including either orientifold planes or D-branes.  In this paper,\footnote{More general cases can be studied
similarly.  In all cases, one aims to prove that the low energy behavior in string theory is consistent
with what one would expect from the appropriate low energy effective field theory.} orientifold planes and D-branes are assumed
to preserve the full $d$-dimensional Poincar\'e symmetry of an underlying 
oriented closed-string compactification to $\R^d\times Z$, for some $Z$ (here $d\geq 2$). In the presence of D-branes and/or orientifolds, the symmetry $(-1)^{F_L}$ is lost  and no longer prevents
 R-R tadpoles. Such tadpoles are possible
and play an important role.  Also, separate conservation 
of $Q_{\alpha}'$ and $Q_{\beta}''$ is not possible when the worldsheet
can be open and/or unoriented.  One must form linear combinations, for reasons we explain momentarily.
This ends up leading to a subtle interplay between
NS-NS and R-R fields.   

\subsubsection{Orientifolds And D-Branes}\label{ordb}

In practice, the string worldsheet $\Sigma$ can be unorientable in an orientifold theory.   
Such a theory is constructed
starting from an underlying Type II superstring theory (possibly with D-branes included, as discussed shortly),
by projecting the string states onto states that are invariant
 under an orientifold projection $\Omega$.  Here $\Omega$ is defined 
 by combining a diffeomorphism of the string that reverses its orientation with some symmetry of the target spacetime
 (or more generally some $\sigma$-model symmetry).
 In the orientifold theory, in traversing a loop in $\Sigma$ around which its
orientation is  reversed, every left-moving field and in particular every left-moving 
supersymmetry generator  is exchanged with some right-moving
one.  This exchange corresponds to an invertible map $\phi:\S_L\to \S_R$; in particular, in this
situation, $\S_L$ and $\S_R$ always have the same dimension.  Unorientability means that
we cannot distinguish the spacetime supersymmetry associated to $\S_\alpha'\in\S_L$ from that associated
to $\phi(\S_\alpha')\in \S_R$.  So for unoriented superstrings, the unbroken supersymmetries are associated
to the sums of left- and right-moving supercharges, $\S_\alpha'+\phi(\S_\alpha')$. A standard way to express this
reasoning is to say that these sums are $\Omega$-invariant and so make sense after projecting  to the orientifold theory.

If $\Sigma$ can have a boundary, which is the case when D-branes are present,
then the purely left- and/or right-moving bulk supersymmetries are broken;
any unbroken supersymmetries 
are linear combinations of left- and right-moving ones.   This is usually
proved by considering the conditions for a worldsheet supercurrent to be conserved in the
presence of a boundary; we give a slightly more precise explanation in section \ref{strapple}.
We are mainly interested in the case that there are some unbroken supersymmetries, since unbroken
supersymmetry will be an ingredient in proving the vanishing of massless tadpoles. 

A prototype is 
Type I superstring theory.  To construct this theory, one starts with Type IIB superstring theory, in which 
the spaces $\S_L$ and $\S_R$ are isomorphic; 
they are 16-dimensional, and transform under the ten-dimensional
Lorentz group $SO(1,9)$ as a spinor
of, say, positive chirality. A spacetime-filling orientifold plane and associated 
D-branes are then introduced in a way
that preserves half of the supersymmetry.  We can consider $\phi$ to be 1.
The unbroken supersymmetries are $Q_\alpha=Q_\alpha'+Q_\alpha''$, 
where $\alpha$ is a positive chirality spinor
index.  They correspond to the supersymmetry
generators $\S_\alpha=\S_\alpha'+\S_\alpha''$.   

\subsubsection{What We Will Learn}\label{zorog}

After establishing some fundamentals in section \ref{zoogo}, we analyze
tadpoles
in open and/or unoriented superstring theory in  section \ref{moretad}. What emerges is more complicated
than for closed oriented superstring theory: instead of a proof that NS-NS tadpoles vanish, we 
get a formula that in a sense relates
NS-NS tadpoles to R-R tadpoles.  The precise statement is rather delicate and 
leads to a severe constraint on R-R tadpoles:
they arise only when the worldsheet $\Sigma$ has the topology of a disc or of $\Bbb{RP}^2$.  Since 
only finitely many topologies
can contribute to R-R tadpoles, it is possible to completely evaluate those tadpoles in a given superstring theory.
Moreover, the disc and $\Bbb{RP}^2$ both 
 have Euler characteristic 1 and hence contribute in the same order of superstring
perturbation theory -- the lowest order in which open and/or unoriented worldsheets
appear at all.  Accordingly it is possible for their contributions to cancel.  A celebrated calculation \cite{GS} that triggered what is sometimes
called the first superstring revolution showed, in effect,
that this cancellation occurs in Type I superstring theory precisely if
the Chan-Paton gauge group of the open strings is $SO(32)$.  More generally,
R-R tadpole cancellation is a standard and important
ingredient in constructing superstring models with orientifolds and/or D-branes.

From general 
considerations such as those
of section \ref{fieldrev}, one might expect that R-R tadpoles would be associated to 
infrared divergences, just like other massless tadpoles.
There is a certain important sense in which this is not true: R-R tadpoles lead to gauge and 
gravitational anomalies in spacetime,
not  directly to infrared divergences.   It is true \cite{GStwo}
that infrared divergences cancel in (for example) Type I superstring theory
precisely when R-R tadpoles cancel, but the reason for this is that spacetime supersymmetry
relates R-R tadpoles to NS-NS tadpoles, which in turn are associated to infrared divergences.
Our explanation of these facts relies on the fermionic gluing parameter associated to a Ramond degeneration; for 
an earlier approach, see \cite{CaiPol}.   
The relation of anomaly cancellation (cancellation of R-R tadpoles) to the cancellation of infrared
divergences (cancellation of NS-NS tadpoles)
is important in superstring model-building.

\subsection{Fundamentals}\label{zoogo}

\subsubsection{Geometry}\label{geometry}

The Deligne-Mumford compactification of the moduli space of oriented closed Riemann surfaces $\Sigma$ 
is a manifold without boundary.\footnote{To be more precise, it is an orbifold or in fancier language a 
stack  without boundary, rather than a manifold. This refinement will not be important and we will loosely
refer to the moduli space as a manifold.}  Compactification
is achieved by adjoining to the ordinary moduli space $\M$ a divisor $\fD$ at infinity; $\fD$ is a union of irreducible
components $\fD_\alpha$, each defined by the vanishing of an appropriate complex gluing parameter $q_\alpha$.
As usual, we write $\h\M$ for the compactification.

When we speak loosely of the ``boundary'' of $\h\M$, we really mean the following.  Because the integration 
measures of interest
are typically 
singular at $q_\alpha=0$, we introduce an infrared cutoff by requiring $|q_\alpha|\geq \epsilon$ for some small positive $\epsilon$.
(As usual, the definition of  $|q_\alpha|$ depends on a suitable choice of hermitian metric.)  
This inequality gives a cutoff version of $\h\M$ that we might call $\h\M_\epsilon$.  It is really 
$\h\M_\epsilon$ that has a boundary,
namely the boundary at $|q_\alpha|=\epsilon$.  The ``boundary'' contributions to Ward identities of 
closed oriented string theories,
as studied in sections \ref{massren} and \ref{tadpoles},
are always boundary contributions in this sense.  They are more precisely contributions from the divisors
at infinity in $\h\M$. For short, we call them the contributions of virtual boundaries.


\begin{figure}
 \begin{center}
   \includegraphics[width=4.5in]{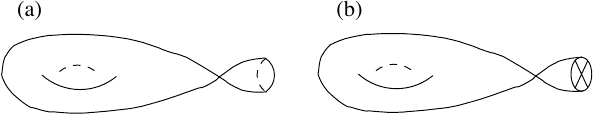}
 \end{center}
\caption{\small  
These closed-string degenerations, in which the right-hand component $\SIgma_r$ is a disc, as in (a), 
or a copy of $\Bbb{RP}^2$, as in (b), are the exceptional cases in which the gluing parameter $q$ is real and positive.
Typically, for closed-string degenerations, gluing depends on a complex parameter.  In these particular cases,
the argument of the gluing parameter can be absorbed in a rotation of 
$\Sigma_r$ around its intersection with $\Sigma_\ell$.}
 \label{xandy}
\end{figure}
\begin{figure}
 \begin{center}
   \includegraphics[width=5.5in]{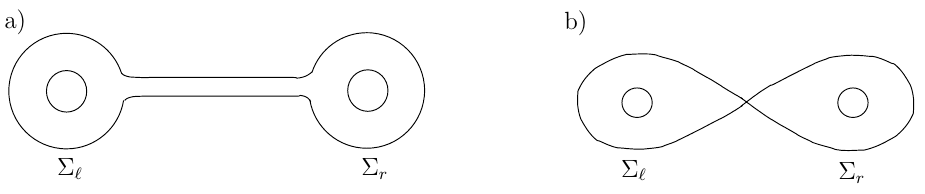}
 \end{center}
\caption{\small  
(a) An open-string worldsheet with a long strip connecting two branches $\Sigma_\ell$ and $\Sigma_r$.  
(b) Compactification is achieved by allowing an open-string
degeneration, shown here, to represent the limit that the length of the strip becomes infinite.}
 \label{blandy}
\end{figure}

For open and/or unoriented Riemann surfaces, things are different;  the relevant moduli spaces have boundaries
in the naive sense.
Technically, the details are a little easier to describe for ordinary Riemann surfaces, so we consider that case first.
(What follows is a  summary of matters explained more fully in section 7.4 of \cite{Wittentwo}.)  
For open and/or unoriented Riemann surfaces, $\h\M$ has three types of boundary component.  Two of them arise   
from closed-string degenerations in which, exceptionally, the gluing parameter $q$ is real and nonnegative rather than
being, as usual, a complex parameter.
  This    
happens (fig. \ref{xandy}) when $\Sigma$ decomposes to a union of components $\Sigma_\ell$
and $\Sigma_r$, with one of them, say $\Sigma_r$,   
being a disc ${\sD}$ with one puncture or a copy of ${\Bbb{RP}}^2$ with  
one puncture.  Precisely in those two cases, $\Sigma_r$ has a $U(1)$ 
symmetry group (consisting of rotations around the puncture)
that can be used to eliminate the argument of the gluing parameter; accordingly, one can take $q$  to be real and
nonnegative.  
The condition $q=0$ then defines a  boundary component 
of the compactified moduli space.
 The third type of boundary component 
arises from what one might call an open-string degeneration, in which (fig. \ref{blandy})   
$\Sigma$ is a union of two components joined by a long strip  rather than a long
tube. In other words, this degeneration is associated to an on-shell open string rather than an on-shell closed string.
 If $s$ is the length of the strip, one defines the gluing parameter $q=e^{-s}$; it is real and positive,   
and compactification of the moduli space is achieved by including the degeneration 
at $q=0$.  Again, the condition $q=0$
defines a boundary component of the compactified moduli space.


All this is for open and/or unoriented bosonic strings.  The supersymmetric case involves some further details.
For open and/or unoriented superstring worldsheets $\Sigma$, it seems that there is no natural definition
of a moduli space, but instead one can define an integration cycle $\varGamma$ suitable for superstring perturbation theory.   (This has been discussed in sections \ref{harder} and \ref{banex}, as well as section 7.4
of \cite{Wittentwo}.) The compactification $\h\varGamma$ is a smooth supermanifold
(or more precisely superorbifold) with boundary.  Its boundaries again correspond\footnote{In addition to
these ordinary boundaries, $\h\varGamma$ has virtual boundaries associated to generic closed-string
degenerations, just as in the case of oriented closed strings.}  to the degenerations of
figs. \ref{xandy} and \ref{blandy}.  In superstring theory, each of these degenerations is of two types.   In fig. \ref{xandy},
the closed-string state that propagates through the double point may be of either NS-NS or R-R type.  In fig. \ref{blandy},
the open-string state that propagates through the double point may be of either NS or R type.

In each of these cases, the gluing parameter $q$ is real and nonnegative modulo the odd variables.  Moreover,
it is defined up to
\begin{equation}\label{zolt}q\to e^\phif q,\end{equation}
where $\phif$ is real modulo the odd variables.  The existence of such a function $q$, vanishing on the boundary, is part of
the definition of a smooth supermanifold with boundary; see for instance section 3.5 of \cite{Wittenone}.  With this
definition, integration of a smooth measure is
a well-defined operation on a supermanifold with boundary; there is no integration ambiguity.  

In general, of course, in superstring
perturbation theory, one is not dealing with smooth measures.  The thorniest singularity that can arise at $q=0$ is the $\d q/q$
singularity associated to a massless  tadpole (this will be a closed-string tadpole in fig. \ref{xandy}, or an
open-string tadpole in fig. \ref{blandy}).  When the integrated tadpole vanishes, there is a natural
procedure  described in section \ref{inttad} to regularize the singularity and evaluate the resulting integral.

As usual, we can be a little more precise than (\ref{zolt}); we can restrict to $\phif=\phif_\ell+\phif_r$, where $\phif_\ell$
depends only on the moduli of $\Sigma_\ell$ and $\phif_r$ depends only on the moduli of $\Sigma_r$.  The indeterminacy
of $q$ is really only of the form
\begin{equation}\label{intox}q\to e^{\phif_\ell+\phif_r}q.\end{equation}  
This has an important
implication for the exceptional closed-string degenerations of fig. \ref{xandy}.  In those cases, $\Sigma_r$, being a disc
or a copy of $\Bbb{RP}^2$ with only one puncture, has no moduli at all.  So $\phif_r$ is a function on a point, in other words
a constant.  Now suppose that $\Sigma_\ell$ is the same in figs. \ref{xandy}(a) and (b), and denote
the gluing parameter as  $q_{\Sigma_\ell,\sD}$ or $q_{\Sigma_\ell,\Bbb{RP}^2}$ 
depending on whether $\Sigma_r$ is a disc $\sD$ or a copy of $\Bbb{RP}^2$.
Since $\Sigma_\ell$ is the same in the two cases, we can use the same local parameters on $\Sigma_\ell$ in defining
$q_{\Sigma_\ell,\sD}$ and $q_{\Sigma_\ell,\Bbb{RP}^2}$.  If we do this, a change in 
those local parameters multiplies $q_{\Sigma_\ell,\sD}$ and $q_{\Sigma_\ell,\Bbb{RP}^2}$
by the same factor $e^{\phif_\ell}$, and this cancels out of the ratio $q_{\Sigma_\ell,\sD}/q_{\Sigma_\ell,\Bbb{RP}^2}$.  
What about $\phif_r$?
 If one has what one regards
as a preferred way to fix a local coordinate at the puncture of a once-punctured disc or $\Bbb{RP}^2$, then one can
also eliminate the indeterminacy in this ratio due to $\phif_r$.  But even if one does not wish to select preferred local
coordinates in these two special cases, the ambiguity due to $\phif_r$ only affects the ratio 
by a multiplicative constant, since $\phif_r$ is itself a constant in each of the two cases.  So the ratio 
of gluing parameters is well-defined
up to a positive multiplicative constant:
\begin{equation}\label{oozer}\frac{q_{\Sigma_\ell,\sD}}{q_{\Sigma_\ell,\Bbb{RP}^2}}\to e^\kappa\frac{q_{\Sigma_\ell,\sD}}
{q_{\Sigma_\ell,\Bbb{RP}^2}},~~\kappa\in\R.\end{equation}  Here $\kappa$ is entirely independent of $\Sigma_\ell$.
This is ultimately important in canceling infrared divergences.

\subsubsection{BRST Anomalies}\label{trapple}

Obviously, once we consider superstring theories with open as well as closed strings, we have to consider
gauge transformations $\V\to \V+\{Q_B,\W\}$, where now $\V$ may be an open-string vertex operator. 
We also have to allow for the fact that in open and/or unoriented string theory, the integration cycle $\h\varGamma$
may have actual boundaries, as well as virtual boundaries associated to closed-string degenerations.

Still,  our analysis in section \ref{massren} carries over, with only a few changes.  For example,
just as before, BRST anomalies in a scattering amplitude can come only from 
degenerations of a worldsheet $\Sigma$ to two branches $\Sigma_\ell$ and $\Sigma_r$, of such a type that the momentum
flowing between the two branches is automatically on-shell. The anomalies can therefore only come from the 
obvious generalizations of  figs. \ref{kopla} and \ref{nopla} of section \ref{zumico}, as follows:  in general, 
 $\Sigma_\ell$ and $\Sigma_r$ may be  open and/or unoriented surfaces,
they may be joined at an open-string degeneration rather than a closed-string degeneration, and some or all of the
external vertex operators may be open-string vertex operators.  These generalizations do not affect much that we said
previously, and the  conclusion relating anomalies to 
 mass renormalization and  tadpoles is almost unaffected.  
 
 Clearly, mass renormalization for open-string states
 is  now relevant. We avoid this as usual by restricting to the massless $S$-matrix in models in which supersymmetry
 prevents mass renormalization for massless particles. 
 
 Also,  tadpoles may now arise for either open- or closed-string states, and closed-string tadpoles
 may be of either NS-NS or R-R type.   The open-string tadpoles do not
 lead to much novelty, as will be clear in section   \ref{strapple}. 
The R-R tadpoles do involve novelties, related to the
peculiar fermionic gluing parameter that appears at a Ramond degeneration; see  section \ref{moretad}.
In particular, hexagon anomalies that appear at one-loop order in Type I superstring theory if the
gauge group is not $SO(32)$ \cite{GS} have their origin in R-R tadpoles.  
 
As we will see, the real importance of the exceptional closed-string degenerations
of fig. \ref{xandy} is their relation to R-R tadpoles.
They also represent potential new contributions to NS-NS tadpoles, but in this
they are not unique. Open and/or unoriented superstring theory has more options for the worldsheet
topology than closed oriented superstring theory, and all of them are equally relevant in analyzing NS-NS
tadpoles. 

\subsubsection{Spacetime Supersymmetry For Open And/Or Unoriented Superstrings}\label{strapple}

The analysis of spacetime supersymmetry from section \ref{bt} must likewise 
be generalized to open and/or unoriented superstring theory.   
The basic idea as before is to analyze the identity
\begin{equation}\label{mursaka}0=\int_{\partial\h\varGamma}F_{S_\alpha \V_1\dots\V_\ssn} \end{equation}
where now $\V_1\dots\V_\sn$
may be open- or closed-string vertex operators, and we  must consider real as well as virtual boundary components
of the compactified integration cycle $\h\varGamma$.  (For the moment, $\S_\alpha$ may be a left- or right-moving supersymmetry
generator or a linear combination.)
For the same reasons as before, contributions to this formula
can only come from the four types of degeneration sketched in  fig. \ref{rufus} of section \ref{bt}, generalized in the obvious way to allow for
open and/or unoriented string worldsheets and open-string vertex operators.

\begin{figure}
 \begin{center}
   \includegraphics[width=5.5in]{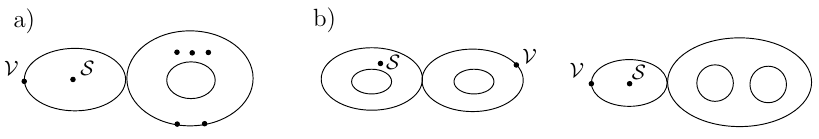}
 \end{center}
\caption{\small  
(a) The action of a supercharge $Q_\alpha$ associated to a supercurrent $\S_\alpha$ on an open-string vertex operator
$\V$
is obtained by evaluating the contribution to the Ward identity (\ref{mursaka})  of degenerations of this type, with only
$\S_\alpha$ and $\V$ on the left, and any collection of open and/or closed-string vertex operators
on the right. (b) The proof that massless open-string tadpoles vanish comes by considering
the correlator $\langle \S_\alpha \,\V\rangle$, where $\V$ is an open-string vertex operator.  The Ward identity has
only the two contributions sketched here; they are associated respectively to 
 spontaneous breaking of the supersymmetry $Q_\alpha$ and to the tadpole of the vertex operator $\{Q_\alpha,\V\}$.  
 If  there is unbroken supersymmetry in perturbation theory, then the vanishing of 
massless open-string tadpoles follows.}
 \label{tufus}
\end{figure}
 In particular,
 we use the open-string analog of the degeneration of fig. \ref{rufus}(a) to define  
  how the supercharge
 $Q_\alpha$ associated to $\S_\alpha$ acts on an open-string vertex operator $\V_i$.
 The contribution for the case that $\Sigma_\ell$ is a disc is sketched in fig. \ref{tufus}(a).
 (Just as for closed strings,
 there may also be loop corrections to $Q_\alpha(\V_i)$, presumably associated
 to mass renormalization.)  As before,
 as long as spontaneous supersymmetry breaking, mass
renormalization, and tadpoles do not come into play, the analogs of the other degenerations in
fig. \ref{rufus} do not contribute.  Under this hypothesis, one gets a linear supersymmetric
 Ward identity among scattering amplitudes, with the standard form of eqn. (\ref{ibob}).  
 In particular, under these conditions, massless open-string tadpoles vanish
 (fig. \ref{tufus}(b)).

However, the question of spontaneous 
supersymmetry breaking does have some special features in the presence of open
strings.  What we will now describe is the analog for spacetime supersymmetry of spontaneous 
breaking of $B$-field gauge-invariance
by mixing of open and closed strings, as described in \cite{KR} and 
in section \ref{examples}.

 Matters are particularly simple for a theory of oriented open and closed superstrings,
in other words an extension of Type II superstring theory with D-branes but no orientifold plane.
It is instructive to examine this case even though in higher orders such theories are frequently anomalous.
(The anomaly, which we study in section \ref{tadanomalies}, arises in a higher order of perturbation
theory than the effect we will describe now, since it comes from an annulus diagram rather than a disc.)

\begin{figure}
 \begin{center}
   \includegraphics[width=4.5in]{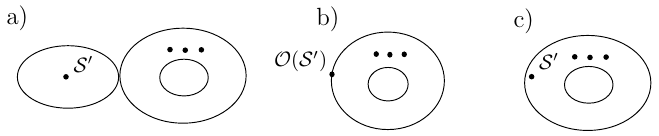}
 \end{center}
\caption{\small  
 (a) Supersymmetry breaking by mixing of open and closed strings is reflected in a non-zero contribution
to the Ward identity from the degeneration shown here, in which the left branch is a disc and contains
 only the supercurrent $\S'$.  The right branch in the example shown is an annulus with several closed-string vertex operators.
 (b) As usual,
this contribution can be evaluated via an insertion on the right branch of an operator $\O(\S')$ -- the vertex
operator for the Goldstone fermion, which in this mechanism for supersymmetry breaking is an open-string mode.
(c)  At tree level (that is, when the left branch is a disc), the operator $\O(\S')$ 
is simply the limit of the supercurrent $\S'$ as the point at which it is inserted approaches the boundary;
this limit is conformally equivalent to the degeneration in (a).
 Hence, $\O(\S')$ at tree level is a matrix element of the bulk-to-boundary analog of the operator product expansion.}
 \label{zotto}
\end{figure}
Let us consider the supersymmetry-breaking degeneration of fig. \ref{rufus}(d), in which a surface $\Sigma$
splits into two components $\Sigma_\ell$ and $\SIgma_r$, with only a 
supercurrent  inserted on
$\Sigma_\ell$.
But now we focus on the special case that $\Sigma_\ell$
is a disc  (fig. \ref{zotto}(a)); and for definiteness, we consider the supercurrent to
be a left-moving one $\S'_\alpha$.  By familiar reasoning, the contribution to the path integral from the degeneration
indicated in the figure can be computed by replacing $\Sigma_\ell$ by an insertion of an operator $\O(\S'_\alpha)$ on
$\Sigma_r$ (fig. \ref{zotto}(b)).  $\O(\S'_\alpha)$ is the vertex operator of a physical open-string state of spin 1/2 (and momentum zero) -- a Goldstone
fermion.  
Specifically for the case that $\SIgma_\ell$ is a disc, the operator $\O(\S'_\alpha)$ is always non-zero.  Indeed,
for $\Sigma_\ell$ a disc, the degeneration of fig. \ref{zotto}(a) is conformally equivalent to a limit in which
the operator $\S'_\alpha$ approaches the boundary of $\SIgma$ from the interior (as in fig. \ref{zotto}(c)).   Thus,
for $\Sigma_\ell$ a disc, $\O(\S'_\alpha)$ is simply a matrix element of the bulk-to-boundary analog of the OPE \cite{DD}. 
 A holomorphic field such as $\S'_\alpha$ always has a non-zero limit as it approaches the boundary.  
It can in general be oversimplified to think of a Ramond vertex operator as a conformal field, since 
in this viewpoint it is difficult to incorporate
the odd moduli of a super Riemann surface.  However,
the behavior as $\S'_\alpha$ approaches the boundary can be computed by a local bulk-to-boundary computation
in which odd moduli, which are global in nature,
 play no important role.  So for this purpose, one may view $\S'_\alpha$ as just
another type of holomorphic conformal field, and it is inevitably non-zero along the boundary of $\Sigma$.

The fact that $\O(\S'_\alpha)\not=0$ means that the presence of a boundary definitely breaks the supersymmetry
generated by a left-moving supercurrent $\S'_\alpha$.  Similarly, if 
$\S''_{\t\alpha}$ is a right-moving supercurrent, then
$\O(\S''_{\t\alpha})$ is non-zero and the supersymmetry generated by $\S''_{\t\alpha}$ 
is definitely broken by the boundary condition.
The only way to find a spacetime  supersymmetry that is not broken by the boundary condition is to find
$\S'_\alpha$ and $\S''_{\t\alpha}$ such that
\begin{equation}\label{dunkly}\O(\S'_\alpha)+\O(\S''_{\t\alpha})=0. \end{equation}
We write $\S_L$ for the space of left-moving supercurrents $\S'_\alpha$ and $\S_L^*$ for the subspace of $\S_L$ such
that this eqn. (\ref{dunkly}) is satisfied for some right-moving supercurrent $\S''_{\t\alpha}$. 
 For $\S'_\alpha\in\S_L^*$, the solution $\S''_{\t\alpha}$ is unique (since $\O(\S'')$ is non-zero
 for all non-zero right-moving supercurrents $\S''$) and of course it is linear in $\S'_\alpha$.  We denote 
 $\S''_{\t\alpha}$  as $\phi_{\text D}(\S'_\alpha)$, where the notation is meant to 
 remind us that $\phi_{\text D}$ reflects the effects of D-branes.
 Thus a generator of unbroken supersymmetry is a linear combination 
 \begin{equation}\label{pikof}\S_\alpha=\S'_\alpha+\phi_{\text D}(\S'_{\alpha})\end{equation} of
 left and right-moving supercurrents.
 
 A simple example  \cite{SchW} that illustrates these ideas is Type IIB superstring theory in $\R^{10}$, with
D9-branes. Half of the bulk supersymmetry is spontaneously broken 
by the coupling to the branes.  The Goldstone fermion is an open-string mode.

In unoriented superstring theory, the starting point is simpler. Even in bulk, the only 
supercurrents that can be defined without a choice
of orientation are mixtures of left- and right-moving
supercurrents of an underlying Type II theory.  In the notation of section \ref{anomover}, such mixtures take the form
$\S_\alpha=\S'_\alpha+\phi(\S'_\alpha)$, where $\phi$ is the orientifold projection viewed as a map from left-moving
to right-moving supercurrents.  In  a theory with D-branes as well as an orientifold 
projection, the condition that the mixing
of open and closed strings at the level of the disc amplitude does not spontaneously break supersymmetry is that
$\O(\S_\alpha)=0$.  Supersymmetries that are unbroken in the presence of both 
the D-branes and the orientifold projection
are derived from left-moving supercurrents $\S'_\alpha$ such that 
$\phi(\S'_\alpha)=\phi_{\text D}(\S'_\alpha)$.  Henceforth
we restrict to such supercurrents and drop the distinction between $\phi$ and $\phi_{\text D}$.

After analyzing the disc amplitudes, one can ask whether higher order perturbative corrections 
trigger further spontaneous
breaking of supersymmetry.   Just as for closed oriented strings, simple arguments can usually 
be given to show that this does
not occur (or, more exceptionally, that it does).  For example, in section \ref{bt}, we 
gave an elementary argument of spacetime chirality
to show that in ten-dimensional Type I superstring theory, the supersymmetry that is allowed by the orientifold projection 
is not spontaneously broken in perturbation theory.
Given such an argument, one has the
right ingredients for the considerations that we will present next.

\subsection{NS-NS and R-R Tadpoles}\label{moretad}

\subsubsection{Preview}\label{preview}

We now come to a  point at which open and/or unoriented superstring theories are really different.  
For closed oriented superstrings, we were able in section \ref{vantad} to 
use spacetime supersymmetry
to prove the vanishing of all massless tadpoles.   But for  
open and/or unoriented superstrings, spacetime supersymmetry
does not lead to a result as strong as this.  Given our usual assumptions, supersymmetry leads
 to a relation that expresses
NS-NS tadpoles in terms of R-R tadpoles, schematically
\begin{equation}\label{tuboff}\langle \V_{\text{NS-NS}}\rangle+\langle\V_{\text{R-R}}\rangle =0, \end{equation}
but not to a general proof that the tadpoles vanish.

There is a surprise here.  Physical R-R fields decouple at zero momentum, even in
open and/or unoriented superstring theory.  This is a consequence of the fact that R-R fields are $r$-form
gauge fields (for various values of $r$, depending on the model), which couple only via their $r+1$-form 
field strengths.  For example, the field strength of
a physical R-R scalar field is a closed one-form, which vanishes at zero spacetime momentum.  In section
\ref{primo}, we essentially deduced this vanishing as a consequence of integration over fermionic gluing parameters.
That being so, one might not expect R-R tadpoles.

It is indeed true that there are no tadpoles for physical R-R fields.    
What we will call an R-R tadpole is the expectation
value of a certain R-R vertex operator, but this is {\it not} the superconformal
vertex operator of a physical R-R field.  Why do we have to worry about R-R fields outside of the usual pantheon
of superconformal vertex operators?  The answer is subtle and involves the fact that at certain exceptional
R-R degenerations, the usual fermionic gluing parameter is absent.  Explaining this is our main goal.

The net effect is that R-R tadpoles arise 
from string worldsheets of only two possible topologies: a disc or $\Bbb{RP}^2$.  
As already noted in section \ref{zorog}, it follows that in a given superstring compactification, R-R tadpoles
can be completely calculated.  

To convince oneself without any technicalities that the R-R operators that lead to tadpoles
cannot be vertex operators of physical states, it suffices to consider the most classic example
of a superstring theory in which R-R tadpoles are important.  This is Type I superstring theory in ten
dimensions, where R-R tadpoles underlie the classic computations \cite{GS,GStwo} of anomalies and NS-NS
tadpoles.  (This is explained and the key points are analyzed in \cite{CaiPol}, in a language somewhat different from ours.)  This theory has
no physical R-R field of spin zero, so the tadpole must be associated to some R-R vertex operator
that is not the vertex operator of a physical field.

\subsubsection{Ward Identity With R-R Vertex Operators}\label{zoward}

We start with a simplified explanation of why R-R vertex operators appear in studying NS-NS tadpoles for
open and/or unoriented superstrings.   First
we consider the illustrative case of a Type II superstring compactification modified
by adding D-branes but not orientifold planes.

In a Type II model, let $\V_{\text{NS-NS}}$ be the vertex operator for a spin 0 field of zero momentum whose
tadpole we wish to study.  As already observed in the discussion of eqn. (\ref{zaxco}), in the absence of D-branes
and orientifold planes, we could establish the vanishing of the tadpole using spacetime supersymmetries associated
to either a left- or right-moving supercurrent.  For example, we can take a left-moving supercurrent, and find a relation
at zero momentum
\begin{equation}\label{dilbo}\V_{\text{NS-NS}}=\sum_\alpha\{Q'_\alpha,\V^\alpha_{\text{R-NS}} \},  \end{equation}
where $\V^\alpha_{\text{R-NS}}$ is a fermion vertex operator of R-NS type.
From this we deduce the vanishing of the NS-NS tadpole as in section \ref{vantad}, assuming that supersymmetry is
not spontaneously broken.

When we add D-branes, the supersymmetry generated by the 
left-moving supercurrent $\S'_\alpha$ is always broken by the boundary conditions and instead we must
use a linear combination $\S'_\alpha+\S''_\alpha$ (or $\S'_\alpha+\phi_{\text D}(\S'_\alpha)$, in the notation of 
eqn. (\ref{pikof})) that generates an unbroken supersymmetry.  What happens now, roughly speaking,
 is that the supersymmetry
$Q''_\alpha$ generated by $\S''_\alpha$ acts on the right-moving part of the vertex operator $\V^\alpha_{\text{R-NS}}$,
transforming it into a vertex operator of R-R type that we will schematically call $\V_{\text{R-R}}$. So in the presence
of D-branes, eqn. (\ref{dilbo}) must be replaced by something more like
\begin{equation}\label{zilbo}\V_{\text{NS-NS}}+\V_{\text{R-R}}=\sum_\alpha\{Q'_\alpha+Q''_\alpha,\V^\alpha_{\text{R-NS}}\}.
\end{equation}
This formula is very schematic and we will see that it needs some corrections.

If we consider an orientifold plane as well as (or instead of) D-branes, everything is much the same except that we must use
operators invariant under the orientifold projection.  So in the starting point, the operator $\V^\alpha_{\text{R,NS}}$ must be replaced
by a linear combination of operators from the R-NS and NS-R sectors.  With orientifold
planes, a better (though still schematic) statement of the identity that constrains R-R tadpoles is
\begin{equation}\label{ilbo}\V_{\text{NS-NS}}+\V_{\text{R-R}}=\sum_\alpha\{Q'_\alpha+Q''_\alpha, \V^\alpha_{\text{R-NS}}
+\V^\alpha_{\text{NS-R}}\}. \end{equation}

These formulas suggest that spacetime supersymmetry for open and/or unoriented superstring
theory will only tell us that NS-NS tadpoles can be expressed in terms of R-R tadpoles, not that they each vanish
separately.  In a sense, that is the right answer, but we are still several steps removed from a correct explanation.

A more careful derivation, as we explain shortly, shows that if eqn. (\ref{ilbo}) is understood as a relation
between physical state vertex operators, then $\V_{\text{R-R}}$ must be multiplied by a factor of the spacetime
momentum $k_I$. This factor is related to the  
spacetime supersymmetry algebra $\{Q_\alpha,Q_\beta\}=\Gamma^I_{\alpha\beta}P_I$; by virtue of this
formula, when we act with $Q_\alpha$ on the vertex operators,  a factor of momentum has to appear somewhere.
The purpose of section \ref{supac} is to explain that this factor actually
appears multiplying $\V_{\text{R-R}}$.
From this it seems that upon setting $k_I=0$ to study the tadpoles, the R-R contribution will
disappear and we will learn, just as for Type II, that the NS-NS tadpoles vanish.  That conclusion is mistaken,
but only for the case that the superstring worldsheet $\Sigma$ is a disc or a copy of $\Bbb{RP}^2$.  It seems difficult
to understand the exceptions through a corrected version of a formula such as (\ref{ilbo}); we will really have to go
back to the identity (\ref{mursaka}) that was the starting point in defining the spacetime supercharges $Q_\alpha$.

\subsubsection{A Factor Of Momentum}\label{supac}

To proceed, instead of being abstract, we consider the illustrative example of Type I superstring theory in
ten dimensions.   For the more general case of a compactification to $d<10$ dimensions, one essentially
repeats the analysis once for every relevant zero-mode wavefunction in the internal space \cite{Polother}.  
These wavefunctions
behave as constants in the following analysis, so keeping track of them would mostly modify only our notation. 

The action of spacetime supersymmetry on vertex operators can be computed \cite{FMS} using the
operator product expansion of superconformal field theory.  In doing so, one can treat the relevant
vertex operators as products of holomorphic and antiholomorphic factors.\footnote{The vertex operator of a string state of momentum $k$ contains
a factor $\exp(ik\cdot X)$ that only factorizes locally,
but this is good enough for computing OPE's.  The factors $\exp(ik\cdot X)$  in eqns. (\ref{lotty}) and (\ref{yotty}) below are understand
as functions of
the right-moving and left-moving parts of $X$, respectively. In compactifications, the zero-mode wavefunctions
mentioned in the last paragraph do not factorize, but they behave as constants in the specific OPE's we will need
and do not affect the analysis.}  
The relevant building blocks of physical state vertex operators
are the right-moving bosonic and fermionic massless vertex operators
\begin{align}\label{lotty}\Y^I & = c\delta(\gamma) D_\theta X^I \exp(ik\cdot X) \cr
                                      \ZZ_\alpha&= c\SSigma_{-1/2}\Stigma_\alpha \exp(ik\cdot X) ,\end{align}
and their left-moving counterparts                                       
\begin{align}\label{yotty}\t\Y^I & = \t c\delta(\t\gamma)D_{\t\theta} X^I \exp(ik\cdot X) \cr
                                      \t\ZZ_\alpha&= \t c\t\SSigma_{-1/2}\t\Stigma_\alpha \exp(ik\cdot  X) .\end{align}
The expressions we have written here
have the canonical picture numbers $-1$ for bosons and $-1/2$ for fermions;
 as we know from sections \ref{nsop}
and \ref{ramins},  superconformal vertex operators appropriate for evaluating the scattering amplitudes
by integration over the usual moduli spaces exist only at those picture numbers.   
$\SSigma_{-1/2}$  represents the $\beta\gamma$ ground state at picture number $-1/2$,
and $\Stigma_\alpha$ is the spin field of the matter system, projected to the positive chirality part so that $\ZZ_\alpha$
is invariant under the GSO projection $\piGSO$.  For $k\not=0$, the physical state conditions require us to take certain
linear combinations of the above operators (such as $\varepsilon_I \Y^I$ with $\varepsilon\cdot k=0$ or
$u^\alpha\ZZ_\alpha$, with $(\Gamma\cdot k)_{\alpha\beta}u^\alpha=0$).   At $k=0$, this is unnecessary.
Physical state vertex
operators are constructed by multiplying left- and right-moving factors of this type.  For example, the NS-NS vertex operator
whose tadpole we need to analyze is the operator $\V^\phi=\t\Y_I \Y^I$ at zero momentum.    

Finally, we will also need the fermion vertex operators at picture number $-3/2$:
\begin{align}\label{wittle} \ZZ_*^\alpha & = c\SSigma_{-3/2}\Stigma^\alpha \exp(ik\cdot X) \cr
                                        \t \ZZ_*^\alpha & = \t c\t\SSigma_{-3/2}\t\SIgma^\alpha \exp(ik\cdot \t X). \end{align}
Here $\SSigma_{-3/2}$ represents the $\beta\gamma$ ground state now at picture number $-3/2$, and $\Stigma^\alpha$
is the chirality $-1$ part of the spin field of the matter system.  
$\SSigma_{-3/2}$ and $\Stigma^\alpha$ are both GSO-odd, so their product is even (see section \ref{massten}).
Thus $\ZZ_*^\alpha$ is GSO-invariant, and so similarly is  $\t\ZZ_*^\alpha$.   Importantly, they have opposite spacetime chirality
from $\ZZ_\alpha$ and $\t \ZZ_\alpha$.
The right-moving picture-changing operator $\YY$ maps $\ZZ_*^\alpha$ to       $\ZZ_\beta$, modulo a BRST-exact operator,
 with an important factor of the
momentum:
\begin{equation}\label{ilzog} \YY\cdot \ZZ_*^\alpha=(\Gamma\cdot k)^{\alpha\beta}\ZZ_\beta+\{Q_B,\cdot\}.\end{equation}
There is of course a similar formula for left-movers.  

Since $\ZZ_*^\alpha$ and $\t\ZZ_*^\alpha$ are not 
superconformal vertex operators,                                  
they cannot be used in computing scattering amplitudes by integration over the usual moduli space of super Riemann surfaces.
  But it will turn out that it is difficult to discuss R-R tadpoles without considering these
operators.
 
As we have already discussed in section \ref{modox} in the context of oriented closed superstrings,
the usual procedure \cite{FMS}   to construct the spacetime supersymmetry generators is as follows.
The basic ingredient is  the spin 1 holomorphic field
$\h\S_\alpha =\SSigma_{-1/2}\Stigma_\alpha$, which we have written at the canonical and most convenient picture number $-1/2$.  
This field may loosely speaking be regarded
as a conserved current on the string worldsheet, and as such it generates a symmetry --  
spacetime supersymmetry.  
This interpretation is valid locally, though globally it somewhat obscures the super Riemann surface
geometry.
(What we call the supercurrent in the present paper  is not $\h\S_\alpha$ but $\S_\alpha = c\h\S_\alpha$.)  
The action of spacetime supersymmetry on a vertex operator $\V$ is usually defined by extracting
a pole in the OPE $\h\S(z)\V(w)$, and if necessary applying a picture-changing operation so as to return to a canonical value of the
picture number.  
For example
\begin{equation}\label{ivorz}\h\S_\alpha(z) \ZZ_\beta(w)\sim \frac{1}{z-w}\Gamma^I_{\alpha\beta}\Y_I.  \end{equation}
In this case, picture-changing is unnecessary, and one interprets the formula to mean that
\begin{equation}\label{nivorz}\{Q_\alpha,\ZZ_\beta\}=\Gamma^I_{\alpha\beta}\Y_I. \end{equation}
On the other hand, since $\h\S$ has picture number $-1/2$ and $\Y_I$ has picture number $-1$, the product
$\h\S\cdot \Y_I$ has picture number $-3/2$, so in this case we will want to 
apply a picture-changing operation to
map back to the canonical picture number, which for Ramond vertex operators is $-1/2$.   
The pole in the OPE is
\begin{equation}\label{bivorz}\h\S_\alpha(z)\Y_I(w)\sim \frac{1}{z-w}
\Gamma_{I\alpha\beta}\ZZ_*^\beta.\end{equation}
After picture-changing via eqn.  (\ref{ilzog}), one therefore defines
\begin{equation}\label{wivorz}\{Q_\alpha,\Y_I\}
=(\Gamma_I \Gamma\cdot k)_\alpha{}^\beta \ZZ_\beta. \end{equation}
Eqns. (\ref{nivorz}) and (\ref{wivorz}) are compatible\footnote{To demonstrate this is a little tricky.
One must use the fact that $\Y^I$ only appears in a combination $\varepsilon_I\Y^I$
where $\varepsilon\cdot k=0$, and one must also use the fact that $k_I\Y^I$ is $Q_B$-trivial. And similarly,
one must use the fact that $\ZZ_\alpha$ only appears in a combination $u^\alpha\ZZ_\alpha$ where $k\cdot\Gamma u=0$.}
with the expected 
supersymmetry algebra $\{Q_\alpha,Q_\beta\}
=\Gamma^I_{\alpha\beta}P_I$.  We will also want later a variant of (\ref{bivorz});
with $\S_\alpha=c\hat\S_\alpha$, we have
\begin{equation}\label{bivorzo}\S_\alpha(z)\Y_I(w)\sim
\Gamma_{I\alpha\beta}\partial c\ZZ_*^\beta.\end{equation}

The purpose of this explanation has been to show that the action of $Q_\alpha$ 
on a massless NS vertex operator to produce
a Ramond vertex operator has no factor of the momentum $k_I$ (eqn. (\ref{nivorz})), while its 
action  on a massless
Ramond vertex operator to produce a vertex operator in the NS sector is proportional to 
$k_I$ (eqn. (\ref{wivorz})).   
Accordingly, in a more careful
derivation, the R-R vertex operator $\V_{\text{R-R}}$ on the left hand side of eqns. (\ref{zilbo}) or (\ref{ilbo}) would
be multiplied by a factor of $k_I$.

Setting $k=0$ to study tadpoles, this contribution would therefore disappear.  
This result is consistent with the intuition that
R-R fields decouple at zero momentum, but it leaves one wondering: How can  R-R tadpoles can possibly appear? 

\subsubsection{When The Fermionic Gluing Parameter Disappears}\label{fermglue}

The answer to the last question is that under certain conditions, there is no picture-changing in the derivation
of the supersymmetric Ward identity.

\begin{figure}
 \begin{center}
   \includegraphics[width=2.5in]{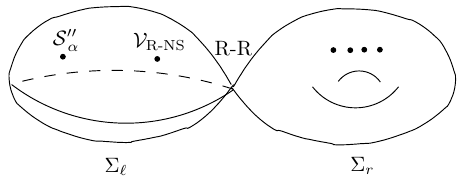}
 \end{center}
\caption{\small  
 A right-moving supercurrent acting on a vertex operator of R-NS type to produce
 an R-R vertex operator. }
 \label{loner}
\end{figure}
Let us return to our definition of the supercharge, based on the process sketched in fig. \ref{diffgen} of section 
\ref{bt}.   
What in the usual description is called picture-changing in the action of the supercharge $Q_\alpha$
on an NS vertex operator arises in our point of view from integration over the fermionic gluing parameter in
fig. \ref{diffgen}(a).   If we want the usual picture-changing not to appear, we need to find a reason that
the usual fermionic gluing parameter is missing.  
The way to eliminate it is the same as the way to eliminate the argument of
the closed-string gluing parameter $q$, as described in section \ref{geometry} and fig. \ref{xandy}.    If either
$\Sigma_\ell$ or $\Sigma_r$ has a fermionic symmetry that acts by shifting 
the fermionic divisor at which $\Sigma_\ell$
and $\Sigma_r$ are glued together, then the fermionic gluing parameter can be 
absorbed in an automorphism of $\Sigma_\ell$ or $\Sigma_r$.

For convenience, we repeat the relevant picture here (fig. \ref{loner}) for the case of a right-moving
supercurrent $\S_\alpha''$ acting on a vertex operator $\V_{\text{R-NS}}$ from the R-NS sector.  An R-R state
is propagating between $\Sigma_\ell$ and $\Sigma_r$.  If $\Sigma_\ell$ and $\Sigma_r$ were 
completely generic, 
then to evaluate the matrix element represented by this picture, we would have to integrate 
over both a left-moving and right-moving
(antiholomorphic and holomorphic) fermionic gluing parameter.  

However, we will take $\Sigma_\ell$ to have 
genus 0, and in this case something special happens.
From an antiholomorphic point of view, the operator $\S_\alpha''$ is the identity
and hence $\Sigma_\ell$ is a genus 0 surface with only two punctures, associated to the insertion of
$\V_{\text{R-NS}}$ and the intersection of $\Sigma_\ell$ and $\Sigma_r$.  These are both Ramond
punctures.  A genus 0 super Riemann surface with only two Ramond punctures has a fermionic automorphism
which can be used to remove the antiholomorphic gluing parameter.  
One can describe such a surface $\Sigma_\ell$ by coordinates $\t z|\t \theta$ (including a divisor at $\t z=\infty$), 
with the superconformal structure being defined
by $D^*_{\t\theta}=\partial_{\t\theta}+\t\theta\t z\partial_{\t z}$.  The Ramond divisors are  defined by $\t z=0$ and $\t z=\infty$; we call
them $\t\FF_0$ and $\t\FF_\infty$, respectively.
$\Sigma_\ell$ admits the odd superconformal vector field
\begin{equation}\label{zonik}\t\nu=\partial_{\t \theta}-\t\theta\t z\partial_{\t z}.\end{equation}
For future reference, we note that
\begin{equation}\label{onk}\t\nu^2=-\t z\partial_{\t z}.\end{equation}
When restricted to either $\t\FF_0$ or $\t\FF_\infty$, $\t\nu$ reduces to $\partial_{\t \theta}$ (to see this for $\t\FF_\infty$,
it helps to transform $\t z\to 1/\t z$), and generates the symmetry
\begin{equation}\label{wonok} \t\theta\to \t\theta+\alpha,\end{equation}
with an anticommuting parameter $\alpha$.  For more on this, see sections 5.1.4 and 7.4.4 of \cite{Wittentwo}.
 The symmetry of $\Sigma_\ell$ that we have just described, since it shifts $\t\theta$ 
 by an arbitrary constant, can be used to transform away the gluing parameter that arises
when $\t{\FF}_0$ or $\t\FF_\infty$ is glued to a Ramond divisor in $\Sigma_r$.

From a holomorphic point of view, $\Sigma_\ell$ has a third puncture and no fermionic automorphism.  According, if $\Sigma_r$ is
generic, then
to evaluate the action of the supercharge, we will have to integrate over one fermionic gluing
parameter, the holomorphic one.  This will reproduce what in the conventional approach comes from picture-changing.

To eliminate the holomorphic fermionic gluing parameter from fig. \ref{loner}, $\Sigma_r$ will
have to have a fermionic automorphism.  A closed, oriented super Riemann surface  $\Sigma_r$ with at least
one puncture (the point at which $\SIgma_r$ intersects $\Sigma_\ell$) appearing in the Deligne-Mumford
compactification\footnote{We exclude
the case that $\Sigma_r$ is a genus 0 surface with at most two punctures, since this case does not
arise in the Deligne-Mumford compactification.} never has such an automorphism, so
for closed oriented superstrings, it is never possible to get rid of the fermionic gluing parameter.
If $\Sigma_r$ is open and/or unoriented, it can have a fermionic automorphism, but only in two special
cases: $\Sigma_r$ must be a disc or a copy of $\Bbb{RP}^2$, with precisely one puncture which must
be of R-R type.  In either of these cases, the closed oriented double cover $\hat\Sigma_r$  of 
$\Sigma_r$ is a genus 0 super Riemann
surface with two punctures of R type, and its fermionic symmetry can again be described as in eqn. 
(\ref{zonik}).  Moreover, just as before, this symmetry generates a shift of the Ramond divisors at which
gluing occurs and hence can be used to transform away the fermionic gluing parameter.

It is no coincidence that the two cases in which $\Sigma_r$ has a fermionic automorphism are the
two cases in which $\Sigma_r$ has a continuous bosonic symmetry group, whose role was discussed in section 
\ref{geometry}.  If $\Sigma_r$ has an odd superconformal
vector field $\nu$, then (as illustrated in eqn. (\ref{onk})), $\nu^2$ is an even 
superconformal vector field that generates a one-parameter
bosonic symmetry group of $\Sigma_r$.

\begin{figure}
 \begin{center}
   \includegraphics[width=4in]{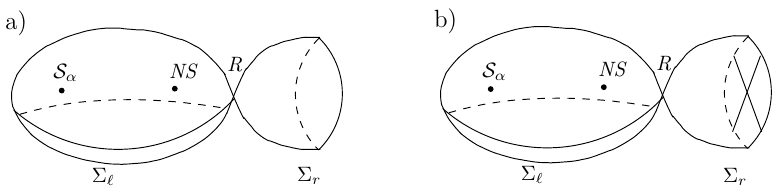}
 \end{center}
\caption{\small  
  Contributions to the supersymmetric Ward identity that give an R-R operator without a fermionic
  gluing parameter.   $\Sigma_r$ in fig. \ref{loner} must be either  (a) 
   a disc or (b) a copy of $\Bbb{RP}^2$, in either case with no vertex operator
  insertions.}
 \label{zumbotz}
\end{figure}

\begin{figure}
 \begin{center}
   \includegraphics[width=2.25in]{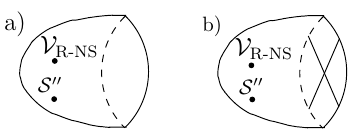}
 \end{center}
\caption{\small  
   Correlation function $\langle \S''\,\V_{\text{R-NS}}\rangle$ on (a) a disc or (b) a copy of $\Bbb{RP}^2$.        }
 \label{umbotz}
\end{figure}

In short, the cases in which there is no fermionic gluing parameter are the cases that $\Sigma_r$ is a disc
or a copy of $\Bbb{RP}^2$ (fig. \ref{zumbotz}).   Since the absence of a fermionic gluing parameter
in those two cases is an important result, we will explain it in another way. The worldsheets in fig. 
\ref{zumbotz} arise by degeneration (as the two vertex operators approach each other) from
the  smooth  worldsheets $\Sigma$ sketched in fig. \ref{umbotz},
consisting of a disc or a copy of $\Bbb{RP}^2$ with insertions of the two vertex operators $\S''_\alpha$
and $\V_{\text{R-NS}}$.  In each of these cases, the closed oriented double cover of $\Sigma$
is a genus zero surface with one NS puncture and two R punctures.  
(Since the supercurrent $\S_\alpha''$ is the identity operator from an antiholomorphic point of
view, its insertion point lifts on the double cover to a single R puncture.  But the insertion point
of $\V_{\text{R-NS}}$ lifts on the closed oriented double cover to a pair of punctures, one of NS type
and one of R type.)
A surface of genus 0 with one NS puncture and two R punctures
 has no fermionic moduli, and there are still none when one degenerates from the smooth
worldsheets of fig. \ref{umbotz} to the singular ones of fig. \ref{zumbotz}.  
So in particular, there are no fermionic gluing parameters in that figure.

Since picture-changing does not come into play, the R-R vertex operator $\V_{\text{R-R}}$ that propagates
between $\Sigma_\ell$ and $\Sigma_r$ in fig. \ref{zumbotz} has picture numbers $(-1/2,-3/2)$.
Actually, in open and/or unoriented string theory, we should use a mixture of left- and right-moving
supercurrents, such as $\S_\alpha=\S'_\alpha+\phi(\S'_\alpha)$, rather than the right-moving one that we
considered for brevity in explaining how the fermionic gluing parameter can disappear.  
Also, in the unoriented case, a closed-string fermion vertex operator cannot simply come from the R-NS
sector; it must be a linear combination of R-NS and NS-R vertex operators.  We denote such a linear
combination by $\V_{\text{NS/R}}$.  These refinements do not affect the essence of what we have said.

\subsubsection{The Tadpoles}\label{manytad}

We are finally ready to analyze the tadpoles of open and/or unoriented superstring theory.  
We consider on a string worldsheet $\Sigma$ of genus $\g$
the Ward identity associated to a two-point function $\langle \S_\alpha\,\V^\alpha_{\text{NS/R}}\rangle$,
with $\S_\alpha$ chosen as the generator of a supersymmetry that is unbroken in perturbation theory, and
a suitable NS/R vertex operator $\V^\alpha_{\text{NS/R}}$.

Since supersymmetry is unbroken, the Ward identity
\begin{equation}\label{dolto}0=\int_{\partial\hat\varGamma}F_{\S_\alpha \V^\alpha_{\text{NS/R}}}\end{equation}
will receive contributions only from one-point functions or in other words tadpoles.  Just as explained
in section \ref{zoward}, the commutator of a supercharge with a vertex operator of mixed NS/R type will
in general be a linear combination of NS-NS and R-R vertex operators.  However, if $\Sigma$ (or more precisely
its reduced space) is anything
other than a disc or $\Bbb{RP}^2$, the R-R contribution vanishes at zero momentum after integrating
over the fermionic gluing parameter in fig. \ref{loner}.  So for generic $\Sigma$, the Ward identity just
tells us that the contribution of $\Sigma$ to an NS-NS tadpole vanishes:
\begin{equation}\label{bunky}\langle \V_{\text{NS-NS}}\rangle_\Sigma = 0 .\end{equation}
Hence in supersymmetric compactifications of open and/or unoriented superstrings that obey the usual
mild conditions ensuring that supersymmetry is not spontaneously broken in perturbation theory, 
there are no NS-NS tadpoles except possibly for the case that $\Sigma$ is a disc or a copy of $\Bbb{RP}^2$.

Precisely in those two cases, there is no fermionic gluing parameter that would kill the R-R contribution.
Such a contribution may appear and the identity becomes schematically
\begin{equation}\label{wunky}\langle \V_{\text{NS-NS}}\rangle_\Sigma+\langle V_{\text{R-R}}\rangle_\Sigma
=0. \end{equation}
Accordingly, supersymmetry tells us not that NS-NS tadpoles vanish, but that they vanish
if and only if the R-R contribution also vanishes.  But simplicity persists because the 
R-R contributions arise only from a disc or
$\Bbb{RP}^2$.  Since those Riemann surfaces both have Euler characteristic 1, they arise in the
same order of perturbation theory and their contributions should be added together. The tadpoles of any given
NS vertex operator $\V_{\text{NS-NS}}$ vanish to all orders of perturbation theory
if and only if precisely one number vanishes,
namely the sum of the disc and $\Bbb{RP}^2$ contributions to the corresponding R-R tadpole.  

Let us make all this more concrete for the important example of Type I superstring theory in ten dimensions.  
In this case, 
in the notation of section \ref{supac}, the only NS-NS vertex operator that might have a tadpole is
$\V_{\text{NS-NS}}=\t\Y_I \Y^I$.  The supercurrent of interest is the usual Type I supercurrent
$\S_\alpha=\S'_\alpha+\S''_\alpha$.  The NS/R vertex operator that is related by supersymmetry to $\V_{\text{NS-NS}}$
is $\V_{\text{NS/R}}^\alpha=\Gamma_I^{\alpha\beta}(\t \Y^I\ZZ_\beta+\t\ZZ_\beta \Y^I)$.  
In deriving the Ward identity, since there is no fermionic gluing parameter, we simply use (\ref{bivorzo}) to
evaluate the R-R contribution.
 So the R-R vertex operator that contributes
in the Ward identity is the operator
\begin{equation}\label{icorn} \V_{\text{R-R}}=\t\partial\t c\t\ZZ_*^\alpha\,\ZZ_\alpha+
\t\ZZ_\alpha \,\partial c\ZZ_*^\alpha. \end{equation}

Although this operator is not a superconformal vertex operator that we could use to calculate scattering
amplitudes in the usual way, it does have a well-defined one-point function on a disc or $\Bbb{RP}^2$.
A naive way to explain why is that  a disc or $\Bbb{RP}^2$ with only one puncture has no even or odd
moduli.  Therefore, in computing the one-point function in question, we do not need to know how
to integrate over moduli.   We give a better explanation momentarily.  

The one-point function of $\V_{\text{R-R}}$ on a worldsheet $\Sigma$ whose reduced
space is a disc or $\Bbb{RP}^2$ is indeed non-zero.
This can actually be demonstrated by a simple computation on the closed oriented double cover $\hat\Sigma$
of $\Sigma$. $\hat\Sigma$ is a genus 0 super Riemann surface in the purely holomorphic sense,
with two Ramond punctures.  Passing to $\hat\Sigma$ separates the holomorphic and
antiholomorphic factors in $\V_{\text{R-R}}$, which lift to holomorphic operators at the two distinct Ramond punctures of $\h\Sigma$.
So the one-point function $\langle \V_{\text{R-R}}\rangle_\Sigma$
is proportional to the purely holomorphic two-point function $\langle \partial 
c\ZZ_*^\alpha(z)\, \ZZ_\alpha(z')\rangle_{\hat\Sigma}$.  This two-point function is non-zero (see eqn. (\ref{monzo})).  

It is tricky to correctly normalize the contributions of a disc and $\Bbb{RP}^2$ and thereby show (for example)
that the R-R tadpole cancels in Type I superstring theory precisely for gauge group $SO(32)$.
Factorization from an annulus and a Mobius strip is a convenient way to do this \cite{GS}; annulus and
Mobius strip amplitudes can be conveniently normalized using their Hamiltonian interpretation in the open
string channel.  The boundary state formalism for superstrings \cite{CLNY} is also useful in constraining and understanding
R-R tadpoles.  For instance, the operator
$\V_{\text{R-R}}$ was related to anomalies  using this formalism \cite{Yost}.

The geometrical meaning of the one-point function of $\V_{\text{R-R}}$ on 
a disc or $\Bbb{RP}^2$ 
 is as follows.  As explained in sections \ref{otherpictures} and \ref{alternative}
of this paper and in more detail in section 4.3 and Appendix C 
of \cite{Wittentwo}, in superstring perturbation theory,
one can calculate systematically
with vertex operators of picture number more negative than the canonical values, provided that one
suitably modifies the definition of supermoduli space in a way that increases its odd dimension.  Usually
this does not add much, since one can immediately integrate over the extra
odd moduli and reduce to  vertex operators of canonical picture number.  We are discussing
here the one case in which that step is not possible.  As explained in section \ref{alternative} above and 
in \cite{Wittentwo}, a Ramond operator of picture
number $-1/2$ is associated to a Ramond divisor, while one of picture number $-3/2$ is associated
to a Ramond divisor together with a choice of a point on the divisor; the choice of point usually
adds an odd modulus.
On a purely holomorphic super Riemann surface, a Ramond divisor is a subvariety $\FF$ of dimension
$0|1$.  For Type II superstrings (possibly enriched with D-branes or  orientifold planes), an R-R vertex
operator of canonical picture 
numbers $(-1/2,-1/2)$ is associated to a product $\t{\FF}\times \FF$ of antiholomorphic and holomorphic
submanifolds of dimension $0|1$.  An R-R vertex operator of picture numbers $(-3/2,-1/2)$ or $(-1/2,-3/2)$ is
associated to $\t{\FF}\times \FF$ with a choice of a point on $\t{\FF}$ or on $\FF$, respectively.
For a generic choice of the string worldsheet $\Sigma$, the choice of 
a point on $\t\FF$ or ${\FF}$ adds one odd modulus, and by integrating over this odd modulus, 
we can reduce to the case of a vertex operator of picture number $(-1/2,-1/2)$.   However, by now we know that a disc or
$\Bbb{RP}^2$ with only one R-R puncture has a fermionic automorphism
that acts by shifting $\t{\FF}$  and ${\FF}$ and hence can be used 
to gauge away the choice of a point on $\t{\FF}$ or $\FF$ (just as earlier we used it to gauge away a gluing parameter).  Thus the choice of this point does not
constitute an odd modulus that one could integrate over to reduce to the canonical picture numbers $(-1/2,-1/2)$.
So the R-R one-point function on
a disc or $\Bbb{RP}^2$ is the only computation that one can perform with vertex operators of picture
$(-3/2,-1/2)$ and $(-1/2,-3/2)$ but which one cannot reduce to a computation with vertex operators
of canonical picture numbers. 

We are actually here
running into precisely the exceptional case \cite{BeZw} in which there is no isomorphism between vertex
operators of different picture numbers.  At non-zero momentum, there are always picture-changing isomorphisms
between vertex operators of different picture number, but this fails at zero momentum in precisely
the case that we have just encountered.  

\subsubsection{What Happens When R-R Tadpoles Cancel?}\label{whatcancel}

It remains to discuss the following two questions.  What happens when R-R tadpoles cancel?  And
what happens when they do not cancel?  
The second question is tricky and is reserved to section \ref{tadanomalies}.
The first question is more straightforward; all we have to do is to restate what we have said in section \ref{inttad} in
a context in which vanishing of tadpoles depends on a cancellation between worldsheets of different topologies.

Let us consider a supersymmetric
model (for example Type I superstring theory in ten dimensions with gauge group $SO(32)$)
in which the R-R tadpoles on a disc $\sD$ and on $\Bbb{RP}^2$ are separately non-zero, but add up to zero.
Then likewise, the same is true of the NS-NS tadpoles on $\sD$  and on $\Bbb{RP}^2$.
We write $\V_{\text{NS-NS}}$ for the zero-momentum NS-NS vertex operator that has a tadpole (we can
always pick a basis of operators so that there is just one such operator),
and let $\langle \V_{\text{NS-NS}}\rangle_\sD$ and $\langle \V_{\text{NS-NS}}\rangle_{\Bbb{RP}^2}$
be its tadpoles.  Tadpole cancellation means that 
\begin{equation}\label{tadcan}\langle \V_{\text{NS-NS}}\rangle_\sD+\langle \V_{\text{NS-NS}}\rangle_{\Bbb{RP}^2}=0.
\end{equation}

\begin{figure}
 \begin{center}
   \includegraphics[width=4.5in]{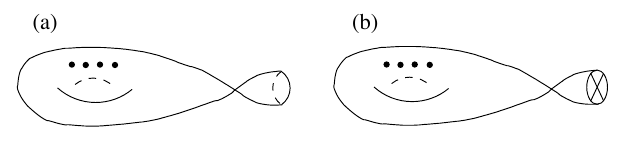}
 \end{center}
\caption{\small  
Infrared divergences in scattering amplitudes due to the NS-NS tadpole on (a) a disc, and (b) $\Bbb{RP}^2$.
When R-R tadpoles cancel, these infrared divergences also cancel in a natural way.}
 \label{xilandy}
\end{figure}
Consider a scattering amplitude $\A_{\V_1\dots\V_\ssn}$  with  vertex operators
$\V_1\dots\V_\sn$.
An NS-NS  degeneration in which one component is
a disc or $\Bbb{RP}^2$ with no vertex operator insertions  (fig. \ref{xilandy}) will
make an infrared-divergent contribution to this scattering amplitude.
(There are no infrared divergences associated
to R-R tadpoles.  This will be explained in section \ref{tadanomalies}.)  
To get physically sensible scattering amplitudes, we will need to cancel these infrared divergences in a natural way.

We let $\Sigma_1$ be a  string worldsheet that can degenerate as in  fig. 
\ref{xilandy}(a), to an intersection of an arbitrary worldsheet $\Sigma_\ell$ with a disc $\sD$.
  Similarly, let
$\Sigma_2$ be a string worldsheet that has the degeneration of fig. \ref{xilandy}(b), with
branches $\Sigma_\ell $ and $\Bbb{RP}^2$.  The contributions of $\Sigma_1$ and $\Sigma_2$ to 
the scattering amplitude $\A_{\V_1\dots\V_\ssn}$ are infrared-divergent; we will give a natural
recipe to cancel the divergence and get a finite sum. 
As in section \ref{geometry}, we write $q_{\Sigma_\ell,\sD}$
and $q_{\Sigma_\ell,\Bbb{RP}^2}$ for the gluing parameters in the two cases. Let $m_1\dots|\dots\eta_s$ 
be the even and odd moduli of $\Sigma_\ell$ (including the point at which it meets $\Sigma_r$) and let $\M_\ell$ be
the corresponding moduli space.  Near $q_{\Sigma_\ell,\sD}=0$,
the integral that gives the contribution of $\Sigma_1$ to the scattering amplitude looks like
\begin{equation}\label{dono}I_1\sim \int [\d m_1\dots|\dots \d\eta_s] \,\G_1(m_1\dots;q_{\Sigma_\ell,\sD}|\dots\eta_s)\frac{\d q_{\Sigma_\ell,\sD}}{q_{\Sigma_\ell,\sD}},\end{equation}
where the  $\d q_{\Sigma_\ell,\sD}/q_{\Sigma_\ell,\sD}$ singularity reflects the 
tadpole, and the function $\G_1(m_1\dots;q_{\Sigma_\ell,\sD}|\dots\eta_s)$ in general depends on all even and odd moduli,
including $q_{\Sigma_\ell,\sD}$.
  Similarly, the contribution of $\Sigma_2$ to the same scattering amplitude looks near
$q_{\Sigma_\ell,\Bbb{RP}^2}=0$ like 
\begin{equation}\label{donox}I_2\sim\int [\d m_1\dots|\dots \d\eta_s] \,\G_2(m_1\dots;q_{\Sigma_\ell,\Bbb{RP}^2}|\dots\eta_s)\frac{\d q_{\Sigma_\ell,\Bbb{RP}^2}}{q_{\Sigma_\ell,\Bbb{RP}^2}},\end{equation}
with some function $\G_2$ that depends on all moduli, including $q_{\Sigma_\ell,\Bbb{RP}^2}$.
Both integrals have logarithmic divergences whose coefficients can be extracted by setting 
$q_{\Sigma,\sD}$ to zero in $\G_1$ and setting
$q_{\Sigma_\ell,\Bbb{RP}^2}$ to zero in $\G_2$.  
The coefficients of the divergences are
\begin{align}\label{bondox}I_{1,\log}&= \int_{\M_\ell} [\d m_1\dots|\dots \d\eta_s] \,\G_1(m_1\dots;0|\dots\eta_s) \cr
I_{2,\log}&= \int_{\M_\ell} [\d m_1\dots|\dots \d\eta_s] \,\G_2(m_1\dots;0|\dots\eta_s).\end{align}

 As usual, $\G_1(m_1\dots;q_{\Sigma_\ell;\sD} |\dots\eta_s)$ and $\G_s(m_1\dots;q_{\Sigma_\ell;\Bbb{RP}^2}|\dots\eta_s)$ factor at $q_{\Sigma_\ell;\sD}=q_{\Sigma_\ell;\Bbb{RP}^2}=0$:
\begin{align}\label{pncon}\G_1(m_1\dots;0|\dots\eta_s)&= \G_0(m_1\dots|\dots\eta_s) \,\langle \V_{\text{NS-NS}}\rangle_\sD  \cr
\G_2(m_1\dots;0|\dots\eta_s)&= \G_0(m_1\dots|\dots\eta_s) \,\langle \V_{\text{NS-NS}}\rangle_{\Bbb{RP}^2},\end{align}
where $\G_0(m_1\dots|\dots\eta_s)$ depends only on $\Sigma_\ell$.  In fact, integration of 
$[\d m_1\dots|\dots\d \eta_s]\,\G_0$ over $\M_\ell$ gives the contribution of $\Sigma_\ell$ to a scattering
amplitude $\A_{\V_1\dots\V_\ssn;\V_{\text{NS-NS}}}$ with an insertion of $\V_{\text{NS-NS}}$ at zero momentum, as well as insertions of $\V_1,\dots,\V_\sn$:
\begin{equation}\label{blingo}\A_{\V_1\dots\V_\ssn;\V_{\text{NS-NS}}}=\int_{\M_\ell}[\d m_1\dots|\dots\d \eta_s]\,\G_0.
\end{equation}   The reasoning here should be familiar from section \ref{inttad}.

In view of (\ref{pncon}),  tadpole cancellation
(\ref{tadcan})
ensures that logarithmic divergences cancel
\begin{equation}\label{moto} I_{1,\log}+I_{2,\log}=0. \end{equation}

This condition is enough to eliminate the divergent part of $I_1+I_2$, but not enough by itself to give meaning to the
finite remainder.  We might introduce infrared cutoffs in $I_1$ and $I_2$
by restricting to $|q_{\Sigma_\ell,\sD}|,\,|q_{\Sigma_\ell,\Bbb{RP}^2}|\geq \epsilon$, where $\epsilon$ is a small
positive constant and the definitions of $|q_{\Sigma_\ell,\sD}|$ and $|q_{\Sigma_\ell,\Bbb{RP}^2}|$ depend on
arbitrary choices of metric on the relevant line bundles.  Tadpole cancellation  is enough to
ensure that the sum $I_1+I_2$ has a limit for $\epsilon\to 0$, but in general this limit will depend on the choices
of metric.  

This problem is very similar to the one that we grappled with in section \ref{inttad}, and its resolution is similar.  
There is a natural ratio $q_{\Sigma_\ell,\sD}/q_{\Sigma_\ell,\Bbb{RP}^2}$ up to a multiplicative constant, as
we asserted in eqn. (\ref{oozer}).  So once we make an arbitrary choice of infrared cutoff $|q_{\Sigma_\ell,\sD}|\geq \epsilon$
in $I_1$, the corresponding choice $|q_{\Sigma_\ell,\Bbb{RP}^2}|\geq \epsilon$ in $I_2$ is naturally determined,
up to a multiplicative constant $q_{\Sigma_\ell,\Bbb{RP}^2}\to e^{-\kappa}q_{\Sigma_\ell,\Bbb{RP}^2}$.

If we transform  $q_{\Sigma_\ell,\sD}$ and $q_{\Sigma_\ell,\Bbb{RP}^2}$
by common factors 
of $e^h$, for some function $h$,
then the limits of $I_1$ and $I_2$ for $\epsilon\to 0$ are both shifted. 
For example, similarly to the derivation of (\ref{imbo}), the limit of $I_1$ for $\epsilon\to 0$ is shifted by
\begin{align}\label{molox}
I_1\to &I_1- \int_{\M_\ell} [\d m_1\dots|\dots \d\eta_s] \,h
\,\G_1(m_1\dots;0|\dots\eta_s)\cr
 =&I_1- \langle \V_{\text{NS-NS}}\rangle_{\sD}\int_{\M_\ell}[\d m_1\dots|\dots\d\eta_s]\,h\,\G_0(m_1,\dots|
\dots\eta_s),
\end{align}
where we used the factorization condition (\ref{pncon}).
Similarly
\begin{equation}\label{motlox}
I_2\to
 I_2- \langle \V_{\text{NS-NS}}\rangle_{\Bbb{RP}^2}\int_{\M_\ell}[\d m_1\dots|\dots\d\eta_s]\,h\,\G_0(m_1,\dots|
\dots\eta_s).
\end{equation}
So the tadpole cancellation condition (\ref{tadcan}) ensures that the sum $I_1+I_2$ does not depend on $h$.
On the other hand, a transformation $q_{\Sigma_\ell,\Bbb{RP}^2}\to e^{-\kappa}q_{\Sigma_\ell,\Bbb{RP}^2}$ with
constant $\kappa$ (and with no such change in $q_{\Sigma_\ell,\sD}$)
transforms $I_2$, in the limit $\epsilon\to 0$, by 
\begin{align}\label{bto}I_2\to I_2+{\kappa} \langle \V_{\text{NS-NS}}\rangle_{\Bbb{RP}^2}\int_{\M_\ell}[\d m_1\dots|\dots\d\eta_s]\,\G_0(m_1,\dots|
\dots\eta_s), 
\end{align} which is just the special case of (\ref{motlox}) with $h=-\kappa$.  So in view of (\ref{blingo}),
 the effect of the rescaling by  $e^{-\kappa}$ on the scattering amplitude $\A_{\V_1\dots\V_\ssn}$ is
\begin{equation}\label{homero}\A_{\V_1\dots\V_\ssn}\to \A_{\V_1\dots\V_\ssn}
+{\kappa} \langle\V_{\text{NS-NS}}\rangle_{\Bbb{RP}^2}
\A_{\V_1\dots\V_\ssn;\V_{\text{NS-NS}}}.\end{equation}
This shift in the scattering amplitude can be interpreted as the result of shifting the scalar field that
couples to $\V_{\text{NS-NS}}$ by the constant ${\kappa}\langle\V_{\text{NS-NS}}\rangle_{\Bbb{RP}^2}.$
So the change in the $S$-matrix under the shift by $\kappa$ can be absorbed in a field redefinition.
This is the answer one should expect from section \ref{inttad}.

\subsection{What Happens When R-R Tadpoles Do Not Cancel?}\label{tadanomalies}

Now let us ask what happens if the R-R tadpoles on a disc and $\Bbb{RP}^2$ do not add to zero.

In this case, in  a supersymmetric model,
there will also be a nonvanishing sum of the NS-NS tadpoles on a disc and $\Bbb{RP}^2$,
leading to an infrared divergence and spoiling the validity of perturbation theory.

But what is the significance of the  R-R tadpoles themselves?
At first, one might assume that R-R tadpoles will lead to an infrared divergence in the R-R channel, but a little reflection might
make one skeptical.  After all, infrared divergences should be associated with propagation
of on-shell physical fields, and the R-R vertex operator $\V_{\text{R-R}}$ whose one-point function is 
the tadpole is not the vertex operator
of a physical field.

What happens is rather \cite{GS,CaiPol} that an R-R tadpole leads to an anomaly -- a failure of gauge invariance.
This is a more serious deficiency than an infrared divergence.  An infrared divergence in perturbation theory
can possibly be eliminated by expanding around a different -- possibly time-dependent -- classical solution.
But an anomaly not related to any infrared divergence is an overall inconsistency of a theory, independent
of the choice of a particular classical state or quantum solution.  To be more exact, it is clear in field theory that anomalies
are not affected by the choice of a quantum state, or of a classical solution around which one expands in order
to do perturbation theory.  One expects the same in string theory, though string theory is not well enough understood
to make it possible to make this statement completely clear.

\begin{figure}
 \begin{center}
   \includegraphics[width=4in]{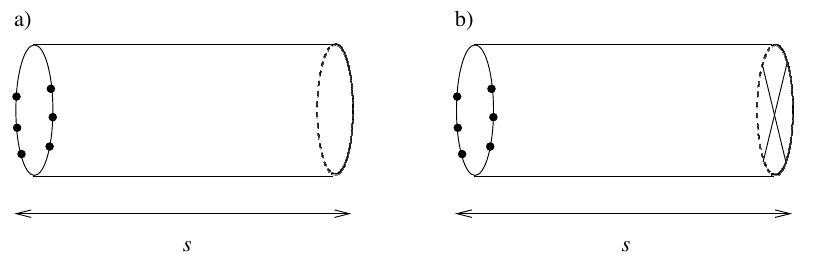}
 \end{center}
\caption{\small  
(a) A cylinder $\Sigma$ of width $s$ with $\n$ open-string vertex operators attached to its left boundary $\Sigma_\ell$ (pictured
for $\n=6$).
(b) A Mobius strip of width $s$ with $\n$ open-string vertex operators attached to its boundary.}
 \label{clandy}
\end{figure}

\begin{figure}
 \begin{center}
   \includegraphics[width=4in]{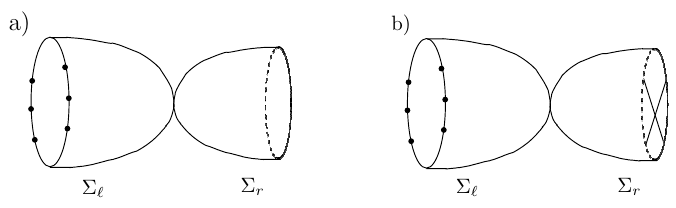}
 \end{center}
\caption{\small  
For $s\to\infty$, the string worldsheets in fig. \ref{clandy} undergo a closed-string degeneration, as pictured here.
$\Sigma_\ell$ is a disc with vertex operator insertions, while $\Sigma_r$ is a disc in (a), or a copy of $\Bbb{RP}^2$ in (b).  These
particular closed-string degenerations are the ones that produces R-R tadpoles and other exceptional behavior.}
 \label{cilandy}
\end{figure}
The classic example \cite{GS,GStwo} of a scattering amplitude that is affected by R-R tadpoles is the one-loop  
contribution to scattering of massless bosonic  open-string states.  We aim here to study this example in the framework
of the present paper.
The  worldsheets $\Sigma$ that contribute to open-string scattering at one-loop order
are  a cylinder or a Mobius strip.  We focus on the ``single trace'' contribution
to the scattering amplitude.  This means that if $\Sigma$ is a cylinder,
we take all open-string vertex operator insertions $\V_1,\dots,\V_\sn$ on its left boundary component, which
we call $\partial_\ell\Sigma$ (fig. \ref{clandy}). It will become clear that the anomaly arises precisely in this case.
 If $\Sigma$ is a Mobius strip, it has only one boundary
component, which we again call $\partial_\ell\Sigma$.  As the width $s$ of $\Sigma$ goes to infinity, $\Sigma$
undergoes a closed-string degeneration (fig. \ref{cilandy}) with $\Sigma_\ell$ a disc with $\n$ open-string insertions
and  $\Sigma_r$ a disc  or a copy of $\Bbb{RP}^2$.  For the case that the closed string propagating from left
to right in fig. \ref{clandy} is in the R-R sector, we want to investigate whether this type of closed-string degeneration,
which is associated to R-R tadpoles, leads to infrared divergences and/or anomalies.

Whether $\Sigma$ is a cylinder or a Mobius strip, its  closed oriented double cover $\hat\Sigma$  has genus 1.  
A spin structure on $\hat\Sigma$ is specified by a pair of binary choices.  We fix
one of these choices by asking that the closed-string state propagating from left to right in fig. \ref{clandy} should
be in the R-R sector.  This still leaves another binary choice to determine
the spin structure of $\hat\Sigma$.  This second binary choice
will determine whether the spin structure of $\hat\Sigma$ is either odd or even.  From a spacetime point of view,
the resulting scattering amplitudes are parity-violating (odd spin structure on $\hat\Sigma$) or parity-conserving (even
spin structure).   We will consider both cases.

If the spin structure of $\hat\Sigma$ is odd, then both $\beta$ and $\gamma$ have a single zero-mode.  This
being so, the vacuum path integral $\langle 1\rangle$ of the $\beta\gamma$ system is not well-defined.  The simplest
well-defined and non-zero path integral is obtained by inserting a single factor $\delta(\beta)$  and a single factor
$\delta(\gamma)$ to remove the zero-modes:
\begin{equation}\label{orony}\bigl\langle \delta(\gamma(z_1))\,\delta(\beta(z_2))\bigr\rangle\not=0. \end{equation}
Instead of local operators $\delta(\gamma(z_1))$ and $\delta(\beta(z_2))$, we could similarly
consider delta functions of integrated modes of $\gamma$ or $\beta$.  Any generic
delta functions will remove the zero-modes and give a sensible and non-zero
path integral.  The same goes
for other formulas below.
If we replace $\delta(\gamma)$ in (\ref{orony}) by $\delta'(\gamma)$, the path integral vanishes:
\begin{equation}\label{borony}\bigl\langle \delta'(\gamma(z_1))\,\delta(\beta(z_2))\bigr\rangle=0. \end{equation}
One way to explain why is that the operator $\delta'(\gamma)$ has ghost number 1 less than that of $\delta(\gamma)$
(it transforms as $\lambda^{-2}$ rather than $\lambda^{-1}$ under $\gamma\to\lambda\gamma$) 
so (\ref{borony}) vanishes because the integrand has the wrong $\beta\gamma$ ghost number.  We can 
restore the ghost number balance by adding an
insertion of $\gamma$; this gives another non-zero path integral:
\begin{equation}\label{torny}\bigl\langle\delta'(\gamma(z_1))\,\delta(\beta(z_2))\,\gamma(z_3)\bigr\rangle\not=0.\end{equation}
If the assertions (\ref{borony}) and (\ref{torny}) are not already clear, they will hopefully become so in section \ref{betag}.

If the spin structure of $\hat\Sigma$ is even, then neither $\beta$ nor $\gamma$ has a zero-mode.  This being so, the
$\beta\gamma$ vacuum amplitude $\langle 1\rangle$ is well-defined and non-zero.  But that will not really be relevant as we will
always have $\delta(\beta)$ and $\delta(\gamma)$ insertions. The other statements in the last paragraph concerning
which correlation functions are nonvanishing remain valid
with an even spin structure.  

The anticommuting ghost and antighost fields $b$ and $c$ are not sensitive to 
the spin structure.  They both have one zero-mode,
so the vacuum correlation function $\langle 1\rangle$ vanishes, but the $bc$ path integral with one $b$ insertion and
one $c$ insertion is non-zero:
\begin{equation}\label{balon}\langle b(z_1)c(z_2)\rangle\not=0.\end{equation}
Overall, then, typical non-zero correlation functions for the $bc\beta\gamma$ ghost system are
\begin{align}\label{along}\bigl\langle b c\,\delta(\beta)\,\delta(\gamma)\bigr\rangle & \not= 0 \cr
                                       \bigl \langle b c\, \delta(\beta)\,\delta'(\gamma)\,\gamma\bigr\rangle & \not=0.\end{align}
We have written the formulas of the last few paragraphs in terms of holomorphic ghost fields $bc\beta\gamma$,
but in the context of open and/or unoriented superstrings, one should
remember that upon lifting to $\h\Sigma$, there is no difference
between $bc\beta\gamma$ and their antiholomorphic counterparts 
$\t b\t c\t\beta\t\gamma$. Both holomorphic and antiholomorphic ghosts on $\Sigma$ descend
from holomorphic variables on $\hat\Sigma$.
 So in (\ref{along}), to write the formulas on $\Sigma$ rather
than $\h\Sigma$, one can for example replace $b$ with $\t b$ or $\beta$
with $\beta+\t \beta$.
                                        
We will look for an infrared divergence (and later for an anomaly) in the scattering of $\n$ 
massless bosonic open-string states, with momenta $k^{(i)}$ and polarization vectors $\varepsilon^{(i)}$.  
It is convenient to represent $\n-1$ of the external vertex operators, say $\V_2,\dots,\V_\sn$, in their integrated form.\footnote{In general,
the use of integrated vertex operators has a drawback that we have explained  in sections 
\ref{persp} and \ref{treem}: it makes it difficult to describe correctly the compactification of the moduli space.  For our purposes
here, there is no problem, because the use of integrated open-string vertex operators does not cause any trouble
in understanding the closed-string R-R degeneration of interest.  As long as the momenta $k^{(i)}$ are generic,
the only problem that might be difficult to properly understand in terms of integrated open-string vertex operators
is the open-string tadpole that may arise (in some models) when the insertion points of all $\sn$ open-string vertex
operators become coincident.} For example, in Type I superstring theory, the 
corresponding factor in the worldsheet path integral is
\begin{equation}\label{zalong}\prod_{i=2}^\sn \oint_{\partial_\ell\Sigma} [\d z|\d\theta]\varepsilon^{(i)}_I D_\theta X^I
\exp(ik^{(i)}\cdot X).\end{equation}  (More generally, one should include Chan-Paton factors and restrict
the integral to a region in which the $\V_i$ are inserted on $\partial_\ell \Sigma$ in a definite cyclic order.)
The only important property of this formula for what follows is that it does not depend on the ghosts.  
It is not convenient to try to represent the last vertex operator $\V_1$ in integrated form; because of the symmetries
of $\Sigma$, the position at which $\V_1$ is inserted is not a modulus.  
So we use the unintegrated form of the vertex operator,
which for Type I is
\begin{equation}\label{palong}\V_1=c\delta(\gamma)\varepsilon^{(1)}\cdot 
D_\theta X \exp(ik^{(1)}\cdot X). \end{equation} 

Now we are ready to integrate over the moduli of $\Sigma$, to look 
for an infrared divergence and/or an anomaly.
We have to integrate over the proper time $s$ that the closed string 
propagates from left to right in fig. \ref{clandy}. 
The closed-string gluing parameter is $q=e^{-s}$.  We also have to 
integrate over the sum of holomorphic
and antiholomorphic gluing parameters.  We do not have to 
integrate over either the argument of $q$ or the difference
of holomorphic and antiholomorphic gluing parameters, since these 
can be absorbed in automorphisms of $\Sigma_r$, as we have learned 
in sections \ref{geometry} and \ref{fermglue}.  So we modify the usual construction of the closed
string propagator by integrating over only $s$ and the sum of 
fermionic gluing parameters.  This gives\footnote{In the following,
$s_0$ is an arbitrary lower cutoff in the integral over $s$.  We are 
really only interested in a possible infrared divergence
for $s\to\infty$.}
\begin{equation}\label{orzoot} (b_0+\t b_0)\int_{s_0}^\infty \d s 
\,\exp(-s(L_0+\t L_0)) \,\delta(\beta_0+\t \beta_0)(G_0+\t G_0).
\end{equation}
Including also (\ref{palong}), the factors in the worldsheet 
path integral that depend on the ghosts are 
\begin{equation}\label{orzoto} c\delta(\gamma)\varepsilon^{(1)}\cdot 
D_\theta X \exp(ik^{(1)}\cdot X)(b_0+\t b_0)\int_{s_0}^\infty \d s 
\,\exp(-s(L_0+\t L_0)) \,\delta(\beta_0+\t \beta_0)(G_0+\t G_0).
\end{equation}

Explicitly written in eqn. (\ref{orzoto}), though somewhat 
scattered,  are the factors $b c \delta(\beta)\delta(\gamma)$ that we need to get
a non-zero path integral (eqn. (\ref{along})).  If we could 
drop the $G_0+\t G_0$ factor in eqn. (\ref{orzoto}), 
the integral over $s$ would give the infrared divergence 
that one would naively expect from an R-R tadpole.
However the factor of $G_0+\t G_0$ eliminates the apparent infrared divergence.  We write
\begin{equation}\label{halong}G_0=G_0^\XX+G_0^\gh,\end{equation}
where $G_0^\XX$ and $G_0^\gh$ are respectively the 
$G_0$ operators of the matter and ghost systems.  
Similarly
\begin{equation}\label{malong} \t G_0=\t G_0^\XX+\t G_0^\gh. \end{equation}
The massless on-shell R-R state that contributes the 
tadpole\footnote{For Type I superstring theory, it was described
explicitly in eqn. (\ref{icorn}). It is annihilated by $G_0^\XX$ and by $\t G_0^\XX$ because
these are some of the defining conditions of a superconformal vertex operator.} is annihilated by $G_0^\XX$ 
and $\t G_0^\XX$.  Nor do $G_0^\gh$ and $\t G_0^\gh$
help.  We may as well consider $G_0^\gh$, which has terms 
$b\gamma$ and $\beta\partial c$. Insertion of a single factor of either of these
in (\ref{orzoto}) causes the path integral to vanish.  For instance, 
an insertion of $b\gamma$ lowers the $bc$ ghost number by 1
and raises the $\beta\gamma$ ghost number by 1; either of these 
shifts causes the path integral to vanish.  

So there is no infrared divergence from a closed-string 
degeneration of R-R type.  Now let us instead look
for an anomaly.  For this, we should replace one of the 
open-string vertex operators, say $\V_1$, by $\{Q,\W_1\}$,
where the gauge parameter $\W_1$ obeys the usual superconformal conditions.
There is an anomaly if the resulting amplitude does not vanish.
For massless open-string states from the NS sector,
the appropriate gauge parameter\footnote{This comes from eqn. 
(\ref{wonzop}), with $\Phi_0=\exp(ik^{(1)}\cdot X)$.} is
\begin{equation}\label{omigos}\W_1=c\delta'(\gamma)\exp(ik^{(1)}\cdot X)       .\end{equation}
To evaluate the anomaly, according to section \ref{brstanom}, we consider an amplitude
with insertion of $\W_1$ instead of $\{Q,\W_1\}$, and we dispense with the integration over
$s$.  
Instead of integrating over $s$, we set $s$ to a large value, and omit
the factor $(b_0+\t b_0)$ that is associated to the integral over $s$.  
So for computing anomalies rather than
infrared divergences, eqn. (\ref{orzoto}) must be replaced by 
\begin{equation}\label{norzoto}c\delta'(\gamma)\exp(ik^{(1)}\cdot X) 
\exp(-s(L_0+\t L_0)) \delta(\beta_0+\t\beta_0)
(G_0+\t G_0). \end{equation}  The anomaly is computed from a path integral 
with this insertion (multiplied by
the factor (\ref{zalong}) that does not involve the ghosts) in the limit of large $s$.

Now we are in the opposite situation from before.  If we drop the $G_0+\t G_0$ factor, 
the path integral will
vanish since the $bc$ and $\beta\gamma$ ghost numbers are both wrong.  
Comparing to the second line of (\ref{along}),
we see that to get a non-zero path integral, we need a $b$ (or $\t b$) insertion and a 
$\gamma$ (or $\t \gamma$) insertion.
But $G_0^\gh$ contains a term $b\gamma$, and similarly $\t G_0^\gh$ contains a term $\t b\t\gamma$.
So it is possible to get an anomaly.

Whether we actually get an anomaly depends on whether the contributions from 
$G_0^\gh$ and $\t G_0^\gh$
add or cancel.  This depends on whether the spin structure of $\Sigma$ is even or 
odd (which determines the sign
in the mixing between $\t\gamma$ and $\gamma$).  A more careful study of the 
boundary states \cite{Yost}, \cite{CaiPol}
shows that the anomaly comes from the odd spin structure. Hence, as one would expect 
from field theory, at one-loop order, 
the anomaly affects only the parity-violating amplitudes. The contribution from the odd spin
structure vanishes  if $\n$ is too small because
of the zero-modes of the matter fermions $\psi^I=D_\theta X^I$, $I=1,\dots,d$.
Each of the $\n-1$ integrated vertex operators in (\ref{zalong}) can absorb two of these $d$ zero-modes,
so an anomaly can only occur if $\n-1\geq d/2$.  In $d=10$, the condition is $\n\geq 6$ and the anomaly
is traditionally called a hexagon anomaly.
 
\section{More On The $\beta\gamma$ System}\label{betag}

\subsection{Preliminaries}\label{startpt}

The goal of the present section 
is to increase our comfort level with the commuting $\beta\gamma$ ghosts
of superstring theory.  As in \cite{EHVerl,EHV,OL} and as explained in section 3.2.2 above, we treat the
$\beta\gamma$ path integral algebraically, using the rules of Gaussian integration, rather
than literally trying to interpret it as an integral.  

We particularly want to gain more experience with the delta function operators that
are ubiquitous in superstring perturbation theory.   Super Riemann surfaces will play no role; we work simply
on an ordinary  compact Riemann surface $\Sigma$ of genus $\g$.  We assume that
$\Sigma$ is closed and oriented. The generalization
to  open and/or unoriented surfaces is immediate, 
since  the combined $\t\beta\t\gamma\beta\gamma$ system on 
an open and/or unoriented surface $\Sigma$
is equivalent to a holomorphic $\beta\gamma$ system on the closed oriented double cover $\h\Sigma$
of $\Sigma$.

Our basic approach is to compare the $\beta\gamma$ system to a system of fields $\beta^*$ and 
$\gamma^*$ that are identical in all respects except that they have the opposite statistics.
So $\beta$ and $\gamma$ are governed by the action
\begin{equation}\label{ilmotok}I_{\beta\gamma}
=\frac{1}{2\pi}\int_\Sigma\d^2z\,\beta\partial_{\t z}\gamma \end{equation}
and $\beta^*$ and $\gamma^*$ are governed by an identical-looking action
\begin{equation}\label{zilmotok}I_{\beta^*\gamma^*}
=\frac{1}{2\pi}\int_\Sigma\d^2z\,\beta^*\partial_{\t z}\gamma^*. \end{equation}
We define an anomalous ghost number symmetry that assigns 
the values $1$ and $-1$ to $\gamma$ and $\beta$,
respectively, and similarly for $\gamma^*$ and $\beta^*$.

In superstring theory, to begin with $\beta$ is a section of $K^{3/2}$ and $\gamma$ a section of $K^{-1/2}$,
where $K^{1/2}$ is a square root of the canonical line bundle
$K$ of $\Sigma$.  However, the $\beta\gamma$ system makes sense more generally with $\gamma$ a section of
an arbitrary holomorphic line bundle $\L$, and $\beta$ a section of $K\otimes\L^{-1}$.   For example, if $\L=K^s$,
then $\gamma$ is a primary field of spin or 
dimension $s$ and $\beta$ is a primary of dimension $1-s$.  In superstring
perturbation theory, once one includes Ramond punctures, one has to study the $\beta\gamma$ system
with a variety of choices of $\L$, so we do not want to restrict to the case $\L=K^{-1/2}$.

The simplest case is actually $\L=K^{1/2}$, so that $\beta$ and $\gamma$ have spin 1/2.  This being so,
if we chose $K^{1/2}$ to define an even spin structure, then for generic complex structure of $\Sigma$,
neither $\beta$ nor $\gamma$ has zero-modes.  This allows a minor simplification in the presentation,
since it means that we do not need to introduce delta function operators (which can remove the zero-modes)
at the very start.  Ultimately, however, we certainly want to introduce and study them.

Let $M=\partial_{\t z}/2\pi$ be the kinetic operator of the $\beta\gamma$ or $\beta^*\gamma^*$ system.
(One often writes $\bar\partial$ for what we call $\partial_{\t z}$.)
The path integral of the $\beta^*\gamma^*$ system is a determinant:
\begin{equation}\label{irtops}\int \D\beta^*\,\D\gamma^*\,\exp(-I_{\beta*\gamma*})=\det\,M. \end{equation}
Similarly, the path integral of the $\beta\gamma$ system is an inverse determinant:
\begin{equation}\label{birto}\int \D\beta\,\D\gamma\,\exp(-I_{\beta\gamma}) =\frac{1}{\det\,M}. \end{equation}
The theory of  determinants of differential operators, and their anomalies, is rather subtle, but we will assume that
it is known, so we will not discuss the properties of the holomorphic object $\det\,M$.
  We focus here only on understanding the correlation functions of the $\beta\gamma$ system.  

For $\L=K^{1/2}$, there generically are no $\beta$ or $\gamma$ zero-modes, but such modes may occur (for
$\g\geq 3$)
as the complex structure of $\Sigma$ is varied.   When this happens, $\det\,M=0$, so the partition function of the
$\beta^*\gamma^*$ system vanishes.  On the other hand, the partition function of the $\beta\gamma$ system
acquires a pole.  In superstring perturbation theory, such poles are called spurious singularities;  they result from an incorrect gauge-fixing procedure.  Superstring perturbation 
theory with a correct
gauge-fixing procedure always leads to a sensible and finite $\beta\gamma$ path integral; when there are zero
modes, there always are delta function operators that remove them.  

It will be convenient to write $\langle 1\rangle$ for the vacuum amplitude, or in other words the path integral
with no operator insertion.   We write $\langle \O\rangle$ for an unnormalized path integral
with insertion of an operator (or product of operators) $\O$.
The normalized expectation value $\langle\O\rangle_N$ is defined as the ratio
\begin{equation}\label{ytre}\langle\O\rangle_N=\frac{\langle\O\rangle}{\langle 1\rangle}. \end{equation}

\subsubsection{The Importance Of Ghost Number}\label{ghcon}

The algebraic treatment that we will give for the $\beta\gamma$ system would not work well for a bosonic
system without a conserved ghost number symmetry.  To see why, let us consider a bosonic Gaussian integral with finitely
many variables $x_1,\dots,x_n$ and a quadratic form $(x,Nx)=\sum_{a,b}N_{ab}x_ax_b$.  With a suitable normalization
of the measure $\d^n x$, the Gaussian integral gives
\begin{equation}\label{torox}\int\d^nx \,\exp\left(-\frac{1}{2}(x,Nx)\right)=\frac{1}{\sqrt{\det N}}.  \end{equation}
The square root poses an immediate problem.  To pick the correct sign of the square root requires some use of calculus,
not just algebra.

A bigger problem is that our main idea is of comparing a bosonic system to a fermionic one  does not work
well for an abstract Gaussian integral with no ghost number symmetry.  Suppose that we replace the bosonic variables
$x_1,\dots,x_n$ with fermionic ones $x_1^*,\dots,x_n^*$.  A quadratic action would have to be 
$(x^*,\t Nx^*)=\sum_{a,b}\t N_{ab}x^*_ax^*_b$ where now $\t N$ is an antisymmetric, rather than 
symmetric, bilinear form.  But there is no natural
way to convert a symmetric bilinear form to an antisymmetric one, so there is no natural map in general from a 
bosonic Gaussian
integral to a fermionic one.

Both problems are removed if we assume the existence of a ghost number symmetry, with the $x_a$ split into
variables $\beta_i$ of $N_\gh=-1$ and variables $\gamma_i$ of $N_\gh=+1$.  This corresponds to taking
\begin{equation}\label{orotz}N=\begin{pmatrix} 0 & M \cr M^t & 0 \end{pmatrix}, \end{equation}
whereupon
\begin{equation}\label{porotz}\sqrt{\det N}=\det M, \end{equation}
and there is no problem with the square root.  Moreover, there is now a natural map from a symmetric form $N$ to
an antisymmetric one $\t N$, namely
\begin{equation}\label{orotzic}\t N=\begin{pmatrix} 0 & M \cr -M^t & 0 \end{pmatrix}.\end{equation}

Finally, there is no difficulty in this general context to understand the case that the ghost number symmetry is
anomalous.  This means that the number of variables with $N_\gh=-1$ differs from the number of variables with $N_\gh=1$,
so that the measure $\d \beta_1\dots \d \gamma_s$ (or $\d\beta^*_1\dots\d\gamma_s^*$) transforms nontrivially under
ghost number.  That being so, only operator insertions that transform in an appropriate fashion under ghost number
can have non-zero expectation values.  In the bosonic case, as well as balancing the ghost number, the operator
insertions will also have to remove the zero-modes of $M$ that are inescapably present if the number of $\beta$'s and
$\gamma$'s are unequal.   Within the class of operators that we will consider in our algebraic treatment -- polynomials
and/or delta functions of the $\beta$'s and $\gamma$'s -- the ones that can remove the zero-modes are the delta
function operators.  That is one reason that they are  a crucial part of the formalism.  Removing the correct number of zero-modes amounts
to balancing the picture number as well as the ghost number.

\subsection{Correlation Functions For Spin $1/2$}\label{corf}

\subsubsection{Elementary Fields}\label{elf}

As explained in section \ref{startpt}, we begin with the case that all fields have spin 1/2.
The rules of Gaussian integration tell us how to compute the expectation value of a product of
elementary fields $\beta^*$ and $\gamma^*$ or $\beta$ and $\gamma$.  
The normalized two-point function is 
\begin{equation}\label{bikely}\langle \gamma(u)\beta(w)\rangle_N=\langle\gamma^*(u)\beta^*(w)\rangle_N
=S(u,w),  \end{equation}
where $S(z,z')$, which is called the propagator, is the inverse 
of the kinetic operator $M=\partial_{\t z}/2\pi$.  $S(z,z')$ can be understood as a section of the line
bundle $K^{1/2}\boxtimes K^{1/2}\to \Sigma\times \Sigma$ (that is, $S(z,z')$ is a section of $K^{1/2}$ in each
variable) with a pole of unit residue
on the diagonal.  It is antisymmetric under $z\leftrightarrow z'$ and is holomorphic away from the diagonal:
\begin{equation}\label{zoto}\frac{\partial_{\t z}}{2\pi}S(z,z')=\delta^2(z,z').\end{equation}
Our assumption that $\beta$ and $\gamma$ have no zero-modes ensures that $S(z,z')$ exists and is unique.

The expectation value of any product of elementary fields is directly constructed in the usual way
from the propagator.
For fermions
\begin{equation}\label{nikely}\bigl\langle \gamma^*(u_1)\gamma^*(u_2)\dots\gamma^*(u_s)\beta^*(w_s)\beta^*(w_{s-1})
\dots \beta^*(w_1)\bigr\rangle_N=\sum_\pi (-1)^\pi \prod_{i=1}^sS(u_i,w_{\pi(i)}),\end{equation}
where the sum runs over all permutations $\pi$ of $s$ objects, and $(-1)^\pi$ is 1 or $-1$ for even or
odd permutations.  More succinctly, the right hand side of (\ref{nikely}) is $
\det P_{(s)}$,
where $P_{(s)}$ is the $s\times s$ matrix whose $ij$ matrix element is $P_{(s)ij}=S(u_i,w_j)$.
For bosons, we need not worry about the ordering of the factors.
The correlation function is given by a formula like (\ref{nikely}), but without the factor $(-1)^\pi$:
\begin{equation}\label{hikely}\biggl\langle\prod_{i=1}^s\gamma(u_i)\,\prod_{j=1}^s\beta(w_j)\biggr\rangle_N=
\sum_\pi  \prod_{i=1}^sS(u_i,w_{\pi(i)}).
\end{equation}

\subsubsection{Delta Function Operators}\label{delfn}

Now we consider delta function operators.  For fermions this is straightforward, since $\delta(\gamma^*)=\gamma^*$
and $\delta(\beta^*)=\beta^*.$   Hence $\langle\delta(\gamma^*(u))\delta(\beta^*(w))\rangle_N=S(u,w)$.  It will be more
convenient to express this in terms of the unnormalized path integral:
\begin{equation}\label{memot}\langle \delta(\gamma^*(u))\delta(\beta^*(w))\rangle = (\det M) \,S(u,w). \end{equation}

How can we go from here to a corresponding formula for 
$\langle\delta(\beta)\delta(\gamma)\rangle$ in the bosonic
case?   The idea is to express the unnormalized path integral 
$\langle \delta(\gamma^*(u))\delta(\beta^*(w))\rangle$ as a Gaussian
integral.  We introduce an anticommuting variable $\sigma^*$ that we 
consider to have ghost number $-1$, like $\beta^*$,
and a variable $\tau^*$ that has ghost number $1$ like $\gamma^*$.  
Then we have\footnote{Signs are most simple
if in writing the exponent, we place fields of ghost number $-1$ to the left of 
fields of ghost number $+1$, just
as we did in defining the classical action $I_{\beta^*\gamma^*}$.}
\begin{equation}\label{pottery}\int \d \tau^*\,\d\sigma^*\,\,\exp(-\sigma^*\gamma^*(u)-\beta^*(w)\tau^*)
=\gamma^*(u)\beta^*(w)=\delta(\gamma^*(u))\,\delta(\beta^*(w)). \end{equation}  
Given this, we see that the unnormalized path integral (\ref{memot}) can 
actually be represented as a Gaussian integral
\begin{align}\label{zemo}\bigl\langle \delta(\gamma^*(u))\delta(\beta^*(w))\bigr\rangle
=\int \D\beta^*\,\D\gamma^* \,\d\tau^*\,\d\sigma^*\exp\left(-I_{\beta^*\gamma^*}
-\sigma^*\gamma^*(u)-\beta^*(w)\tau^*\right). \end{align}
It is convenient to combine all variables of ghost number 1 to 
$\hat\gamma^*=(\gamma^*(z),\tau^*)$ and all variables
of ghost number $-1$ to $\hat\beta^*=(\beta^*(z),\sigma^*)$.  We also define the extended action
\begin{equation}\label{miroto}\h I_{\h\beta^*\h\gamma^*}=I_{\beta^*\gamma^*}+
\sigma^*\gamma^*(u)+\beta^*(w)\tau^*.
\end{equation}
This function is a homogeneous and quadratic function of the full set of variables.  
It is convenient to view $\h\beta^*$ and $\h\gamma^*$ as row and column vectors
\begin{equation}\label{dombolo}\h\beta^*=(\beta^*(z)\,\,\sigma^*),~~~
\h\gamma^*=\begin{pmatrix}\gamma^*(z)\cr \tau^*
\end{pmatrix},\end{equation}
and to view the extended action as $(\h\beta^*,\h M\,
\h\gamma^*)$, where $\h M$ is an extended version of $M$.
The integral in  (\ref{zemo}) is simply a fermionic Gaussian integral with kinetic operator $\h M$, so
\begin{equation}\label{onoxt}\bigl\langle 
\delta(\gamma^*(u))\delta(\beta^*(w))\bigr\rangle=\det\,\h M.\end{equation}

Comparing to eqn. (\ref{memot}), we see that we must have
\begin{equation}\label{bonoxt}\det\h M = (\det M )\,S(u,w). \end{equation}
The enterprising reader can verify this directly, thinking of $\h M$ as a matrix that is obtained by adding one
row and one column to the $\infty\times \infty$ matrix $M$.

It is now fairly obvious how to imitate this for bosons.  
We just drop the $*$'s everywhere and repeat all the steps
with bosonic variables.  We introduce bosonic variables $\sigma,\tau$ of 
ghost numbers $-1$ and 1 and write
\begin{equation}\label{omito}\delta(\gamma(u))\delta(\beta(w)) 
=\int \d \tau\,\d\sigma \,\exp(-\sigma \gamma(u)-\beta(w)\tau).
\end{equation}
This representation of the delta function
in the context of an algebraic treatment of Gaussian integrals was explained in the discussion of
eqn. (\ref{tofo}).
So if as in the bosonic case we define the extended action 
\begin{equation}\label{mirotopic}\h I_{\h\beta\h\gamma}=I_{\beta\gamma}+\sigma\gamma(u)+\beta(w)\tau,
\end{equation}
then we can write $\bigl\langle\delta(\gamma(u))\delta(\beta(w))\rangle$ as a Gaussian integral:
\begin{equation}\label{ziroro}\bigl\langle\delta(\gamma(u))\delta(\beta(w))\bigr\rangle=\int \D \h\beta\,\D\h\gamma
\exp(-\h I_{\h\beta \h\gamma})=\frac{1}{\det\h M}.\end{equation}
In view of (\ref{bonoxt}), this is the same as $1/(\det M\cdot S(u,w))$.  The normalized two-point function
$\bigl\langle \delta(\gamma(u))\,\delta(\beta(w))\bigr\rangle_N$ is therefore
\begin{equation}\label{irororo}  \bigl\langle \delta(\gamma(u))\,\delta(\beta(w))\bigr\rangle_N= \frac{1}{S(u,w)}. \end{equation}

This formula has many interesting consequences and generalizations.  For one thing, we can immediately determine
the dimensions of the operators $\delta(\beta)$ and $\delta(\gamma)$.  The propagator $S(u,w)$ is the two-point
function of operators $\beta^*$ and $\gamma^*$ of dimension 1/2  (since $\beta^*$ and $\gamma^*$ are sections of 
$\L=K^{1/2}$); this is encoded in the way that it transforms under
 changes of coordinates.    The inverse function $1/S(u,w)$ transforms oppositely, so if it is a two-point function,
 then it is the two-point function of operators of dimension $-1/2$.  This  should not be a surprise.
 Our algebraic treatment of the delta functions ensures that under scaling, $\delta(\beta)$ and $\delta(\gamma)$
 transform oppositely to $\beta$ and $\gamma$.  But $\beta$ and $\gamma$ have conformal dimension 1/2
 so naturally $\delta(\beta)$ and $\delta(\gamma)$ have dimension $-1/2$.    Similarly,
 we will see in section \ref{zomix} that $\delta(\beta)$ and $\delta(\gamma)$ have opposite ghost numbers from $\beta$ and $\gamma$.
 
 The short distance behavior of $S(u,w)$ is $S(u,w)\sim 1/(u-w)$ for $u\to w$, corresponding to the OPE $\gamma^*(u)\beta^*(w)
 \sim 1/(u-w)$.  So $1/S(u,w)\sim u-w$, giving the OPE
 \begin{equation}\label{dolmus} \delta(\gamma(u))\delta(\beta(w)) \sim u-w. \end{equation}
 
In general, a fermionic Gaussian integral vanishes if and only if there is a fermion zero-mode.  Let us investigate this
condition in the context of the extended action $\h I_{\h \beta^*\h\gamma^*}$.  We will look for a zero-mode of
$\h\gamma^*$.  (With $\L=K^{1/2}$, the index of the operator $\h M$ vanishes, and there is a $\h\gamma^*$
zero-mode if and only if there is a $\h\beta^*$ zero-mode. We will write an explicit formula shortly.)  
The classical equations for $\h\gamma^*$ are
\begin{align}\label{delmo} \gamma^*(u) & = 0 \cr
    \frac{\partial_{\t z}\gamma^* }{2\pi}+\delta^2(z,w)\tau^* & =0. \end{align}
The second condition says that $\gamma^*(z)$ is holomorphic away from $z=w$, with at most a simple pole (of residue
$\tau^*$) at $z=w$.  Taking account also of the first condition, we see that $\gamma^*(z)$ is a section of $K^{1/2}$
that vanishes at $z=u$ and may
have a pole at $z=w$, but is otherwise holomorphic.

We can summarize all this by saying that $\gamma^*(z)$ is a holomorphic section of $K^{1/2}\otimes \O(w)\otimes \O(u)^{-1}$.
Under what conditions does such a section exist?  We have assumed that 
$K^{1/2}$ has no holomorphic sections,
so upon allowing a pole at $w$, there is at most a one-dimensional space of 
holomorphic sections.  Explicitly, such
a section is the propagator $S(z,w)$, regarded as a function of $z$ for fixed $w$.  
This section (with $\tau^*$
given by eqn. (\ref{delmo})) gives a $\h\gamma^*$ zero-mode 
 if and only if $\gamma^*(z)$  vanishes at $z=u$, that is if and only if $S(u,w)=0$.  
 So $S(u,w)=0$ is the condition
under which the Gaussian integral of the $\h\beta^*\h\gamma^*$ system should 
vanish, and that is indeed what we see
in eqn.  (\ref{bonoxt}).  If and only if there is a $\h\gamma^*$ zero-mode, 
there is also a $\h\beta^*$ zero-mode, with
$\beta^*(z)$ now given by $S(u,z)$ as a function of $z$ for fixed $u$.

By the same token, in the case of bosons, the $\h\beta\h\gamma$ system has a 
zero-mode if and only if $S(u,w)=0$,
and that should be the condition under which the $\h\beta\h\gamma$ path 
integral is not well-defined.  And this of course
is what we see in eqn. (\ref{irororo}).   The pole in that formula when there are zero-modes 
is a typical example of what in superstring perturbation
theory is called a spurious singularity.  The $\beta\gamma$ path 
integral with $\delta(\beta)$ and $\delta(\gamma)$ insertions
can in general develop a pole, but in superstring perturbation theory with
a correct gauge fixing, the $\delta(\beta)$ and $\delta(\gamma)$ insertions
are always such as to keep one away from the poles.  

\subsubsection{Multiple Delta Function Insertions}\label{iroz}

More generally, we would like to understand how to calculate a $\beta\gamma$ path integral with an arbitrary collection of
delta function insertions  $\bigl\langle \prod_{i=1}^s\delta (\gamma(u_i))\prod_{j=1}^s\delta(\beta(w_j))\bigr\rangle$.  
We can do this
by the same reasoning as before.  The corresponding fermionic correlation function
$\bigl\langle \delta(\gamma^*(u_1))\dots \delta(\gamma^*(u_s))\delta(\beta^*(w_s))\dots \delta(\beta^*(w_1))\bigr\rangle$
is completely equivalent to the correlator (\ref{nikely}) of a product of elementary fields.  On the other hand,
by adding auxiliary variables $\sigma_i^*$, $\tau_i^*$, $i=1,\dots,s$, this fermionic correlation function can
be expressed as a fermionic Gaussian integral:
\begin{equation}\label{ziddo}\int \D\beta^*\,\D\gamma^*\prod_{i=1}^s\d\tau_i^*\,\d\sigma_i^* \,\,\exp\left(-I_{\beta^*\gamma^*}
-\sum_i \sigma_i^*\gamma^*(u_i)-\sum_j\beta^*(w_j)\tau_j^*\right).\end{equation}
Dropping all the $*$'s, we get a bosonic Gaussian integral that computes 
$\bigl\langle \prod_{i=1}^s\delta (\gamma(u_i))\prod_{j=1}^s\delta(\beta(w_j)\bigr\rangle$.  From this, we deduce that
 the normalized correlation function of a product of delta function operators is simply the inverse of (\ref{nikely}):
\begin{equation}\label{domore} \biggl\langle \prod_{i=1}^s\delta (\gamma(u_i))\prod_{j=1}^s\delta(\beta(w_j))\biggr\rangle_N
=  \frac{1}{\sum_\pi (-1)^\pi \prod_{i=1}^sS(u_i,w_{\pi(i)})}.\end{equation}
Just as for $s=1$, the poles of this correlation function arise from zero-modes of the extended system in which
it can be computed as a Gaussian integral.   One will never encounter such poles in superstring perturbation theory with
a correct gauge-fixing.
  
Eqn. (\ref{domore}) has a number of interesting consequences.  First of all, this formula is antisymmetric in the $u_i$
and also antisymmetric in the $w_j$, so we see that the local operators $\delta(\gamma)$ and also $\delta(\beta)$ are fermionic.

We have already analyzed the $\delta(\gamma)\cdot \delta(\beta)$ operator product in eqn.  (\ref{dolmus}).
Now we can similarly study the $\delta(\gamma)\cdot \delta(\gamma)$ and $\delta(\beta)\cdot\delta(\beta)$ operator
products.  For fermions, 
we have $\gamma^*(u_1)\gamma^*(u_2)\sim -(u_1-u_2)\gamma^*\partial \gamma^*(u_2)$.
(This statement is shorthand for saying that the product $\gamma^*(u_1)\gamma^*(u_2)$ is for $u_1-u_2\to 0$
the product of $u_1-u_2$ with a dimension 2 primary that is conveniently denoted $\gamma^*\partial\gamma^*$.
That the leading operator that appears in the product $\gamma^*(u_1)\gamma^*(u_2)$ is of dimension 2
can be deduced from (\ref{nikely}).)
Equivalently, for fermions $\delta(\gamma^*(u_1))\cdot \delta(\gamma^*(u_2))\sim -(u_1-u_2)\delta(\gamma^*)
\delta(\partial\gamma^*).$   Since the bosonic correlation function is the inverse of the fermionic one,
it follows that the product $\delta(\gamma(u_1))\cdot \delta(\gamma(u_2))$ for $u_1\to u_2$ is the product
of $1/(u_1-u_2)$ times an operator
that is conveniently denoted $-\delta(\gamma)\delta(\partial\gamma)$:
\begin{equation}\label{moxo}\delta(\gamma(u_1))\delta(\gamma(u_2))\sim- \frac{1}{u_1-u_2}
\delta(\gamma)\delta(\partial\gamma)(u_2). \end{equation}
Since $\delta(\gamma)$ has dimension $-1/2$, the factor of $1/(u_1-u_2)$ implies
that the operator $\delta(\gamma)\delta(\partial\gamma)$ has dimension $-2$,
the opposite of the dimension of the corresponding fermionic operator.  Note that $\delta(\partial\gamma)$ is again
fermionic, and in particular $\delta(\gamma)\delta(\partial\gamma)=-\delta(\partial\gamma)\delta(\gamma)$.

More generally, the leading singularity as $t$ operators $\delta(\gamma)$ approach each other is a primary field
of dimension $-t^2/2$ that can conveniently be denoted
\begin{equation}\label{diffo}\varTheta_{-t}=\delta(\gamma)\delta(\partial\gamma)\dots\delta(\partial^{t-1}\gamma). \end{equation}
The justification for the notation is that near a point at which $\varTheta_{-t}$
 is inserted, the elementary field $\gamma$
has a zero of order $t$.  This can be established using formulas of section \ref{zomix}. Similarly, $\beta$ has a pole
of order $t$ at an insertion point of this operator. In conventional language,
the fact that $\gamma$ has a zero of order $t$ at a point at which 
$\varTheta_{-t}$ is inserted
 while $\beta$ has a pole of order $t$ at such a point
means that this operator  is associated to the ground state in picture number $-t$.  The ground state with positive picture
number is similarly constructed from delta functions of $\beta$.

Eqns. (\ref{moxo}) and (\ref{diffo}) have obvious analogs with $\beta$ replacing $\gamma$.
The interested reader can deduce additional OPE relations from eqn. (\ref{domore}), such as
\begin{equation}\label{omore}\delta(\gamma(u_1 ))\, \delta(\beta)\delta(\partial\beta)(u_2)\sim {(u_1-u_2)^2}\delta(\beta(u_2)).\end{equation}

\subsubsection{Mixed Correlation Functions}\label{zomix}

So far we have understood how to compute correlation functions of elementary fields $\beta$ and $\gamma$, and also
of delta function insertions $\delta(\beta)$ and $\delta(\gamma)$.  For superstring perturbation theory, in general one also
needs mixed correlation functions such as $\langle\delta(\gamma(u))\delta(\beta(w))\gamma(u') \beta(w')\rangle$.
These can be computed by slightly extending what we have described so far. 

First of all, for fermions, since there is no difference between a delta function and an elementary field,
this correlation function is just the correlation function (\ref{nikely}) of a product of elementary fields.  
However, it will be useful to write (\ref{nikely}) in terms of unnormalized path integrals, and also to break
the symmetry between the operator insertions by writing two operators as delta functions and the other two as
elementary fields:
\begin{equation}\label{hombbo}\bigl\langle\delta(\gamma^*(u))\,\delta(\beta^*(w))\,\gamma^*(u')\,\beta^*(w') \bigr\rangle
=\det M\,\left(S(u,w)S(u',w')-S(u,w')S(u',w)\right). \end{equation}
We will now proceed as in eqn. (\ref{pottery}),
introducing new variables $\tau^*$ and $\sigma^*$ and using the exponential
representation of the delta functions.  As before, we introduce extended variables
$\hat\gamma^*=(\gamma^*(z),\tau^*)$ and $\hat\beta^*=(\beta^*(z),\sigma^*)$, and define
the extended action $\h\I_{\h\beta^*\h\gamma^*}$.  Then we can write the correlation
function that we want as a two-point function in the extended Gaussian theory with this extended action:
\begin{equation}\label{hombibo}\bigl\langle\delta(\gamma^*(u))\,\delta(\beta^*(w))\,\gamma^*(u')\,\beta^*(w') \bigr\rangle
=\int \D\h\beta^*\,\D\h\gamma^* \,\exp(-\h I_{\h\beta^*\h\gamma^*})\,\gamma^*(u')\,\beta^*(w'). \end{equation}
This is simply a Gaussian integral with insertion of elementary fields $\gamma^*(u')$ and $\beta^*(w')$.  
So
\begin{equation}\label{ombicer}\bigl\langle\delta(\gamma^*(u))\,\delta(\beta^*(w))\,\gamma^*(u')\,\beta^*(w') \bigr\rangle
=(\det\h M)\, \h S(u',w'), \end{equation}
 where as before $\h M$ is the extended kinetic operator,
and $\h S(u',w')$ is a matrix element of the inverse operator to $\h M$, which we will call the extended propagator.

Using the formula (\ref{bonoxt}) for $\det\h M$, and eqn. (\ref{hombbo}) for the correlation function in (\ref{ombicer}),
we deduce a formula for the relevant matrix elements of the extended propagator:
\begin{equation}\label{murox} \h S(u',w')=\frac{1}{S(u,w)}\left(S(u,w)S(u',w')-S(u,w')S(u',w)\right). \end{equation}

Now it is straightforward to understand the analog of this for bosons.  To compute a mixed correlation function
$\bigl\langle \delta(\gamma(u))\,\delta(\beta(w))\,\gamma(u')\,\beta(w')\bigr\rangle$, we introduce the new variables
$\sigma$ and $\tau$ and use the integral representation of the delta functions.  In this way, we arrive at (\ref{hombibo})
without the $*$'s.  Performing the Gaussian integral gives the same result as before, except that in the case of a bosonic
Gaussian integral, the determinant appears in the denominator:
\begin{equation}\label{onoxoto}   \bigl\langle\delta(\gamma(u))\,\delta(\beta(w))\,\gamma(u')\,\beta(w') \bigr\rangle
=\frac{1}{\det\h M}\, \h S(u',w').\end{equation}
Upon using (\ref{bonoxt}) and (\ref{murox}) and also multiplying by $\det M$ to pass to normalized correlation functions,
we get finally
\begin{equation}\label{pnoxt} \bigl\langle\delta(\gamma(u))\,\delta(\beta(w))\,\gamma(u')\,\beta(w') \bigr\rangle_N
=  \frac{1}{S(u,w)^2} \left(S(u,w)S(u',w')-S(u,w')S(u',w)\right).\end{equation}

Given this, we can work out the OPE of an elementary field times a delta function operator.  For example,
eqn. (\ref{pnoxt}) has a simple zero for $u'\to u$.  This means that the leading contribution to the product of
$\gamma(u')$ and $\delta(\gamma(u))$ is an operator of dimension 1 
that is naturally understood as $\partial \gamma \cdot\delta(\gamma)$:
\begin{equation}\label{belview}\gamma(u')\delta(\gamma(u))\sim (u'-u)\,\,\partial\gamma\cdot\delta(\gamma)(u). \end{equation}
Similarly, eqn. (\ref{pnoxt}) has a simple pole for  $w'\to u$.  This means that the leading contribution to the product
$\beta(w')\cdot \delta(\gamma(u))$ involves an operator of dimension $-1$.  
It is natural to interpret this operator as $\delta'(\gamma(u))$:
\begin{equation}\label{elview}\beta(w')\delta(\gamma(u))\sim \frac{1}{w'-u}\delta'(\gamma(u)). \end{equation}
The justification for this notation comes from the OPE of $\gamma$ and $\delta'(\gamma)$, which one can compute
by letting $u'$ approach $u$ in (\ref{pnoxt}) after having let $w'$ approach $u$ to define the operator
that we are calling $\delta'(\gamma)$.  After we extract the pole
of (\ref{elview}) as $w'\to u$, there is no further singularity as $u'\to u$, but there is a minus sign because
of the sign of the second term in (\ref{pnoxt}).  Apart from this minus sign, the result of taking
$u'\to u$ after $w'\to u$ gives back the two-point function $\langle \delta(\gamma(u))\,\delta(\beta(w))\rangle_N$
with the operators $\gamma	(u')$ and $\beta(w')$ simply omitted, 
so\footnote{To fully verify that the operator on the right hand
side of this OPE is indeed $-\delta(\gamma(u))$ rather than a new operator that we have not seen before, 
we really
should extend this analysis to consider a limit $u'\to u$ after $w'\to u$ in an arbitrary correlation
function containing multiple insertions of $\delta(\gamma)$, $\delta(\beta)$, $\gamma$, and $\beta$, not just the example
considered in the text.  This is fairly straightforward given what we have explained, and is left to the reader.}
\begin{equation}\label{hinor}\gamma(u')\delta'(\gamma(u))\sim -\delta(\gamma(u)). \end{equation} 
The name $\delta'(\gamma)$ for the operator in question is an intuitive way to express this fact.  
Eqns. (\ref{belview}) and (\ref{elview}), taken together, mean in the terminology of \cite{FMS} that the operator
$\delta(\gamma)$ is the vertex operator of the $\beta\gamma$ vacuum of picture number $-1$.  Similarly, we have
\begin{align}\label{mippo}\gamma(u')\delta(\beta(w))&\sim \frac{1}{ u'-w}\delta'(\beta(w)) \cr
                                          \beta(w')\delta(\beta(w))&\sim (w'-w)\,\,\partial\beta\cdot\delta(\beta)(w). \end{align}
These formulas mean that, in conventional language, $\delta(\beta)$ is the vertex operator of the $\beta\gamma$
ground state at picture number $+1$.                            Similarly, the operator in  eqn. (\ref{diffo}) represents
the $\beta\gamma$ vacuum with picture number $-t$, and its cousin with $\gamma$ replaced by $\beta$ represents
the vacuum with picture number $+t$.               
                                   
To determine the ghost number of the operators $\delta(\gamma)$ and $\delta(\beta)$, we can proceed as follows.   
The ghost number current of the $\beta\gamma$ system is defined as the normal-ordered expression
$J_{\beta\gamma}(u')=-:\beta\gamma:(u')$.  Taking the limit as $w'\to u'$ in (\ref{pnoxt})
after subtracting out the pole term to define the normal-ordered operator $J$, we get the three-point function
$\bigl\langle \delta(\gamma(u))\delta(\beta(w)) J_{\beta\gamma}(u')\bigr\rangle$.  This three-point function has simple
poles as $u'\to u$ or $u'\to w$.  The poles come from the term $-S(u,w')S(u',w)$ in the numerator of (\ref{pnoxt}).  
The residues of these
poles determine the ghost numbers of the operators $\delta(\gamma) $ and $\delta(\beta)$.
A quick and largely error-proof way to proceed is to observe that we could compute the ghost
numbers of the elementary fields $\gamma$ and $\beta$ in the same way starting with the four-point function
\begin{equation}\label{milro}\bigl\langle\gamma(u)\,\beta(w)\,\gamma(u')\,\beta(w')\bigr\rangle
=S(u,w)S(u',w')+S(u,w')S(u',w). \end{equation}
Again from the behavior for $w'\to u'$, we extract the three-point function $\bigl\langle \gamma(u)\,\beta(w)\,
J_{\beta\gamma}(u')\bigr\rangle$, and then from the residues of the poles at $u'=u$ or $u'=w$, one finds
the ghost numbers of the elementary fields 
$\gamma$ and $\beta$.  The only material difference in the two computations is that
the crucial term $S(u,w')S(u',w)$ appears with opposite sign in (\ref{milro}) relative to (\ref{pnoxt}).  As a result,
the operators $\delta(\gamma)$ and $\delta(\beta)$ have opposite ghost numbers from $\gamma$ and $\beta$;
thus $\delta(\gamma)$ and $\delta(\beta)$ have ghost numbers $-1$ and $+1$, respectively.  

Once this is known, the OPE's given above suffice to determine the ghost numbers of the other operators that
we have encountered.  For example, (\ref{elview}) implies that $\delta'(\gamma)$ has ghost number $-2$
and (\ref{moxo}) implies that $\delta(\gamma)\delta(\partial\gamma)$ also has ghost number $-2$.

Now that we have encountered operators such as $\delta'(\gamma)$, it is 
instructive to analyze the operator product in (\ref{moxo}) in higher orders.   To shorten the formulas, we replace $u_1$ and $u_2$ by $u$ and $0$, and also write
$\gamma'$ and $\gamma''$ for $\partial\gamma$ and $\partial^2\gamma$.
For $u\to 0$, we have $\gamma(u)=\gamma(0)+u\gamma'(0)
+\frac{u^2}{2}\gamma''(0)+\dots$, so
\begin{align}\label{pomore}\delta(\gamma(u))\delta(\gamma(0))=\delta\left(\gamma(0)+u\gamma'(0)+\frac{u^2}{2}\gamma''(0)+\dots\right)\delta(\gamma(0)).
\end{align}
Since $\delta(a+b)\delta(a)=\delta(b)\delta(a)$ and $\delta(\lambda b)=\delta(b)/\lambda$, this is 
\begin{align}\label{pomorem}\delta\left(\gamma(0)+u\gamma'(0)+\frac{u^2}{2}\gamma''(0)+\dots\right)\delta(\gamma(0))&=
\delta\left(u\gamma'(0)+\frac{u^2}{2}\gamma''(0)+\dots\right)\delta(\gamma(0))\cr&=\frac{1}{u}\delta\left(\gamma'(0)+\frac{u}{2}\gamma''(0)+\dots\right)\delta(\gamma(0)).
\end{align}
Now we simply expand $\delta\left(\gamma'(0)+\frac{u}{2}\gamma''(0)+\dots\right)=\delta(\gamma'(0))+\frac{u}{2}\gamma''(0)\delta'(\gamma'(0))+\dots$ to get
\begin{equation}\label{milkob}\delta(\gamma(u))\delta(\gamma(0))\sim\frac{1}{u}\delta(\gamma')\delta(\gamma)(0)+\frac{1}{2}\gamma''\delta'(\gamma')\delta(\gamma)(0)
+\dots,\end{equation}
where the expansion can be straightforwardly carried out to any desired order.  All these manipulations are valid in the context of an algebraic treatment of Gaussian integrals.

\subsubsection{Extended Delta Functions}\label{exdelta}

In superstring perturbation theory, one encounters more general delta functions that are not necessarily local
operators. Let $f$ and $g$ be $(0,1)$-forms on $\Sigma$ with values in $K^{1/2}$ and define
\begin{equation}\label{gilg}\gamma_f =\int_\Sigma f(z)\gamma(z),~~\beta_g=\int_\SIgma g(z)\beta(z).\end{equation}
We would like to calculate corresponding correlation functions such as $\bigl\langle \delta(\gamma_f)\,\delta(\beta_g)\bigr
\rangle_N.$

For fermions, this would be completely straightforward.  With $\gamma^*_f$ and $\beta^*_g$ defined by the obvious
analogs of eqn. (\ref{gilg}), we have
\begin{equation}\label{hyter}\bigl\langle \delta(\gamma^*_f)\,\delta(\beta^*_g)\bigr\rangle_N=\bigl\langle \gamma^*_f\,\,\beta^*_g\rangle_N
=\int_{\SIgma\times\Sigma}f(z) S(z,z') g(z'). \end{equation}
Repeating the derivation in section \ref{delfn}, we find that the normalized bosonic correlation function
is the inverse of this:
\begin{equation}\label{numbly}\bigl\langle \delta(\gamma_f)\,\delta(\beta_g)\bigr
\rangle_N=\frac{1}{\int_{\SIgma\times\Sigma}f(z) S(z,z') g(z')}. \end{equation}

In superstring perturbation theory, one meets operators $\delta(\beta_g)$ for
arbitrary $g$, though usually one only encounters $\delta(\gamma_f)$ for the case that $f$ is a delta function.
All our statements about correlation functions with multiple delta function insertions or with insertions of elementary
fields as well as delta function operators have fairly immediate analogs in the presence of nonlocal
operators such as $\delta(\beta_g)$ and/or $\delta(\gamma_f)$.

If we specialize eqn. (\ref{numbly}) to the case that $f$ and $g$ are delta functions, we recover eqn. (\ref{irororo}).
More generally, if $f$ and $g$ are derivatives of delta functions, we get formulas such as
\begin{equation}\label{zumbil}\bigl\langle\delta(\gamma(u))\,\delta(\partial\beta(w))\bigr\rangle_N
=\frac{1}{\partial _w S(u,w)}.\end{equation}
Here $\delta(\partial\beta)$ is a local operator, though not a primary field.  (In fact, a Virasoro module containing
a state corresponding to $\delta(\partial\beta)$ is not a highest weight module and does not contain a primary state.
So $\delta(\partial\beta)$ is also not a descendant.  Still, it is possible to make sense of its correlation functions.
By contrast, the operator $\delta(\beta)\delta(\partial\beta)$
is primary.)

Hopefully the $\beta\gamma$ system with spin $1/2$ does not retain much mystery.  The generalization
that we need for superstring perturbation theory is only a short step, to which we turn next.

\subsection{The $\beta\gamma$ System With Zero-Modes}\label{melsp}

Now we generalize the $\beta\gamma$ system to the case that $\gamma$ is a 
section of a holomorphic line bundle $\L$
and $\beta$ is a section of $K\otimes \L^{-1}$.  The main novelty is that 
depending on the choice of $\L$, $\beta$ or
$\gamma$ may have zero-modes.  According to the Riemann-Roch theorem, the number of $\gamma$
zero-modes minus the number of $\beta$ zero-modes is $n_\gamma-n_\beta=1-\g+\mathrm{deg}\,\L$,
where $\mathrm{deg}\,\L$ is the degree of $\L$.  For a generic
choice of the moduli of $\Sigma$ or $\L$, the total number of zero-modes is the minimum required
by the Riemann-Roch theorem, which means that generically either $n_\gamma=0$ or $n_\beta=0$.  Correlation
functions have poles at  values of the moduli  at which the number of zero-modes
exceeds the minimum required by the Riemann-Roch theorem, so it is natural to begin with the generic
 case.
Ordinarily,\footnote{There are a few exceptional cases.  For $\sg=0$ and no more than 2 Ramond
punctures, one has $n_\gamma\not=0$, $n_\beta=0$.  The $\beta\gamma$ system can be described
exactly as we do below, with the roles of $\beta$ and $\gamma$ reversed. We actually do this
for $n_\Ra=2$ in section \ref{reparam} below.  For $\sg=1$ and no Ramond
punctures, in an odd spin structure, $n_\beta=n_\gamma=1$.   What we will say can be adapted to this case,
starting with the fact that for the $\beta^*\gamma^*$ system, the vacuum amplitude vanishes
in this example because of fermion zero modes, but $\langle \gamma^*(u)\beta^*(w)\rangle\not=0$.} 
in superstring perturbation theory, $\L$ is such that $n_\beta\not=0$, $n_\gamma=0$.
In the absence of Ramond punctures, one has $\L=K^{-1/2}$, with $\mathrm{deg}\,\L=1-\g$, $n_\gamma=0$,
$n_\beta=2\g-2$.  With Ramond punctures, $\mathrm{deg}\,\L$ becomes more negative, $n_\beta$ becomes
larger, and $n_\gamma$ remains zero. 

For these reasons, in developing the theory, we will assume that $n_\gamma=0$, $n_\beta>0$.
Since $\gamma$ and $\beta$ zero-modes are respectively holomorphic sections of $\L$ and of 
$K\otimes \L^{-1}$, our hypothesis means that $H^0(\Sigma,\L)=0$, $H^0(\Sigma,K\otimes \L^{-1})\not=0$.  
According to Serre duality, $H^0(\Sigma,\L)$ is dual to $H^1(\Sigma,K\otimes \L^{-1})$, and therefore
our hypothesis implies that
\begin{equation}\label{delbo} H^1(\Sigma,K\otimes\L^{-1})=0. \end{equation}

\subsubsection{The Minimal Delta Function Insertion}\label{mindel}

Setting $n_\beta=t$, let us suppose that $\beta$ has zero-modes $\y_1,\dots,\y_t$, obeying
\begin{equation}\label{poler}M^{\mathrm{t}}\y_i(z)=0, ~~i=1,\dots,t,\end{equation}
with\footnote{The transpose of the operator $M=\partial_{\t z}/2\pi$ acting on $\gamma$ is the operator
$M^{\mathrm{t}}=-\partial_{\t z}/2\pi$ acting on $\beta$.} $M^{\mathrm{t}}=-\partial_{\t z}/{2\pi}$.  This being so, 
the vacuum path integral of the $\beta\gamma$ system is not well-defined.
To make sense of it, we need delta function operators that will remove the zero-modes from the path integral.

As usual, it is easier to begin with fermions.  
So we consider a $\beta^*\gamma^*$ system
with $\gamma^*$ and $\beta^*$ being fermionic variables that 
are sections respectively of $\L$ and of $K\otimes \L^{-1}$.
The vacuum amplitude $\langle 1\rangle$ vanishes because of the zero 
modes $\y_i(z)$.  A minimal non-zero correlation
function is $\langle \beta^*(w_1)\dots \beta^*(w_t)\rangle$, with enough $\beta^*$ insertions to absorb the zero-modes.  

The correlation function $\langle \beta^*(w_1)\dots\beta^*(w_t)\rangle$ will be annihilated by 
$\partial_{\t w}$ in each of its
arguments.  This ensures that it is a linear combination of expressions $\prod_{i=1}^t \y_{j_i}(w_i)$ 
for some sequence
$j_1,\dots,j_t$.  But fermi statistics imply that  $\langle \beta^*(w_1)\dots\beta^*(w_t)\rangle$ 
must be antisymmetric in
$w_1,\dots,w_t$.  This implies that it must be proportional to $\det N_{(t)}$, where $N_{(t)}$ is the 
$t\times t$ matrix whose
$ij$ matrix element is $N_{(t) ij}=\y_i(w_j)$.  The constant of proportionality is by definition $\det'M$, 
the determinant of $M$
in the space of $\beta\gamma$ fields with the zero-modes divided out.  So for fermions
\begin{equation}\label{okok}\langle \beta^*(w_1)\dots\beta^*(w_t)\rangle={\det}'M\,\det\,N. \end{equation}

As usual, we can write this more suggestively as
\begin{equation}\label{blokok}\langle \delta(\beta^*(w_1))\dots\delta(\beta^*(w_t))\rangle=
{\det}'M\,\det\,N. \end{equation}
The reader will probably now not be surprised that the result for bosons is simply the inverse of this:
\begin{equation}\label{zokok}\langle \delta(\beta(w_1))\dots \delta(\beta(w_t))\rangle =
\frac{1}{{\det}'\,M}\frac{1}{\det N}. \end{equation}
To deduce this, we simply use the fact that the left-hand side of (\ref{blokok}) can be written as a Gaussian integral.
As in eqn. (\ref{ziddo}), we add variables $\tau^*_i$, $i=1,\dots,t$, of ghost number 1, and write
\begin{equation}\label{bokok} \delta(\beta^*(w_1))\dots\delta(\beta^*(w_t))=\int \d\tau_1^*\dots\d\tau_t^*
\exp\left(-\sum_i \beta(w_i)\tau_i^*\right). \end{equation}
Letting $\h\gamma^*=(\gamma^*(z),\tau_1^*,\dots,\tau_t^*)$, we define the extended action
\begin{equation}\label{dolfig}\h I_{\beta^*\h \gamma^*}=I_{\beta^*\gamma^*}+
\sum_{i=1}^t\beta^*(w_i)\tau^*_i, \end{equation}
and then 
\begin{equation}\label{ruolf}\langle \delta(\beta^*(w_1))\dots\delta(\beta^*(w_t))\rangle=
\int \D\beta^* \,\D\h\gamma^*
\exp(-\h I_{\beta^*\h\gamma^*}). \end{equation}
Now that we have expressed the correlation function of interest as a Gaussian integral, the generalization to
bosons is straightforward.  We simply drop the $*$'s everywhere, introducing bosonic 
variables $\tau_1,\dots,\tau_t$
and the extended action
\begin{equation}\label{dolif}\h I_{\beta\h \gamma}=I_{\beta\gamma}+\sum_{i=1}^t\beta(w_i)\tau_i, \end{equation}
where $\h\gamma=(\gamma(z),\tau_1,\dots,\tau_t)$.  Then 
\begin{equation}\label{rugolf}\langle \delta(\beta(w_1))\dots\delta(\beta(w_t))\rangle=\int \D\beta \,\D\h\gamma
\exp(-\h I_{\beta\h\gamma}). \end{equation}
The result of the bosonic Gaussian integral is inverse to the fermionic one, so we arrive at (\ref{zokok}).

This formula shows that $\langle \delta(\beta(w_1))\dots\delta(\beta(w_t))\rangle$ has a pole precisely when the
matrix $N$ has a non-zero kernel.  This occurs precisely if a non-zero linear combination 
of the zero-modes $\y_1,\dots,\y_t$
vanishes at all the points $w_1,\dots,w_t$, so that the insertion of the delta functions 
$\delta(\beta(w_1))\dots\delta(\beta(w_t))$
does not remove all of the zero-modes from the functional integral.   

\subsubsection{More General Correlation Functions}\label{moregen}

To understand more general correlation functions, we again begin with fermions. 

Consider a correlation function with $t+1$ 
insertions $\beta^*(w_i)$, $i=1,\dots,t+1$
and a single insertion $\gamma^*(u)$.  We would like to compute the correlation function 
\begin{equation}\label{inzo}\bigl\langle \gamma^*(u)
\,\beta^*(w_1)\dots \beta^*(w_{t+1})\bigr\rangle.\end{equation}   As a function of any one variable $w_i$, this correlation
function is a holomorphic section of $K\otimes \L^{-1}$ except for a simple pole at $w_i=u$.   Differently
put, the correlation function in its dependence on $w_i$ is a holomorphic section of $K\otimes \L^{-1}\otimes
\O(u)$.  

The space $H^0(\Sigma,K\otimes \L^{-1}\otimes \O(u))$ has dimension $t+1$.  This follows\footnote{The Riemann-Roch theorem says that $\mathrm{dim}\,H^0(\Sigma,K\otimes \L^{-1}\otimes \O(u))-
\mathrm{dim}\,H^1(\Sigma,K\otimes \L^{-1}\otimes \O(u))=t+1$.  But $H^1(\Sigma,K\otimes \L^{-1}\otimes \O(u))$
vanishes by virtue of (\ref{delbo}) and the long exact cohomology sequence derived from the
short exact sequence of sheaves $0\to K\otimes \L^{-1}\to
K\otimes \L^{-1}\otimes \O(u)\to \left.\L^{-1}\right|_{u}\to 0$.}
 from the
Riemann-Roch theorem plus the assumption (\ref{delbo}).  Of the $t+1$ linearly independent holomorphic
sections of $K\otimes \L^{-1}\otimes \O(u)$, $t$ are the modes $\y_1,\dots,\y_t$ that actually
come from holomorphic sections of $K\otimes \L^{-1}$ with no pole at $u$.  To get a basis
of $H^0(\SIgma,K\otimes \L^{-1}\otimes \O(u))$, we need one more section of $K\otimes \L^{-1}$
that actually does have a pole at $u$.  We can denote that last section as $f_u(w)$.  It is convenient to ask that in its dependence
on $u$, $f_u(w)$ should be a section of $\L$; we denote it as $S(u,w)$,
a section of $\L\boxtimes (K\otimes \L^{-1})\to\Sigma\times\Sigma$  (that is, $S(u,w)$ is a section of $\L$
in the first variable and a section of $K\otimes \L^{-1}$ in the second) with a pole on the diagonal.  We can constrain $S(u,w)$ by
requiring that the residue of the pole at $w=u$ is 1 (the fact that this condition makes sense is the reason to define $S(u,w)$ as
a section of $\L$ in its dependence on the first variable); once we do this, $S(u,w)$ is uniquely determined
modulo the possibility of adding a linear combination of the zero-modes $\y_i$ with $u$-dependent
coefficients
\begin{equation}\label{ipz}S(u,w)\to S(u,w)+\sum_i h_i(u)\y_i(w). \end{equation}
$S(u,w)$ is the closest analog of the propagator, given the presence of the zero-modes $\y_i$.

Now we can analyze the correlation function (\ref{inzo}) by the same reasoning that led to (\ref{okok}).
As a function of any of the $w_j$, this correlation function is a linear combination of the $t+1$ modes $\y_i(w_j)$,
$i=1,\dots,t$, and $S(u,w_i)$.  Fermi statistics implies that as a function of the $t+1$ variables $w_1,\dots,
w_{t+1}$, the correlation function must be, up to a multiplicative constant, the
``Slater determinant'' constructed from those $t+1$ modes.  
This Slater determinant is by definition the determinant of the $t+1\times t+1$ matrix
\begin{equation}\label{slater}N_{(t+1)} =\begin{pmatrix}\y_1(w_1) & \y_1(w_2) & \dots & \y_1(w_{t+1})\cr
\y_2(w_1) &\y_2(w_2)& \dots & \y_2(w_{t+1}) \cr
& & \ddots & \cr
\y_t(w_1) & \y_t(w_2) & \dots & \y_t(w_{t+1})\cr
S(u,w_1)& S(u,w_2) & \dots & S(u,w_{t+1})  \end{pmatrix}.   \end{equation}
Note that this determinant is not affected by the nonuniqueness (\ref{ipz}) of $S(u,w)$ (and therefore
we do not need to ask whether there is a global choice of $S(u,w)$).

The constant multiplying ${\det}\,N_{(t+1)}$ in the correlation function is precisely $\det'\,M$, the same factor
that appeared in   (\ref{okok}).  To show this, we simply take the limit that one of the $w_i$ approaches
$u$, and use the condition that the residue of $S(u,w)$ at $u=w$ is 1.  
So finally, we get a formula for the correlation function (\ref{inzo}):
\begin{equation}\label{binzo}\bigl\langle \gamma^*(u)
\,\beta^*(w_1)\dots \beta^*(w_{t+1})\bigr\rangle={\det}'\,M \,\,\det N_{(t+1)}. \end{equation}

There are two ways to extrapolate from this formula to a corresponding formula for bosons. 
In one approach, we replace all operators $\beta^*$ or $\gamma^*$ with $\delta(\beta^*)$ or $\delta(\gamma^*)$:
\begin{equation}\label{rinzo}\bigl\langle \delta(\gamma^*(u))
\,\delta(\beta^*(w_1))\dots \delta(\beta^*(w_{t+1}))\bigr\rangle={\det}'\,M \,\,\det N_{(t+1)}. \end{equation}
In the now familiar fashion, by using an exponential representation of the delta functions, one
can express the left hand side as a fermionic Gaussian integral.  The analogous correlation function
for bosons can be expressed as the corresponding bosonic Gaussian integral, so for the $\beta\gamma$
system, we have
\begin{equation}\label{linzo}\bigl\langle \delta(\gamma(u))
\,\delta(\beta(w_1))\dots \delta(\beta(w_{t+1}))\bigr\rangle=\frac{1}{{\det}'\,M }\frac{1}{\det N_{(t+1)}}. \end{equation}

The second thing we can do is to replace only the fields $\beta^*(w_1),\dots,\beta^*(w_t)$ in (\ref{binzo})
with delta functions, and leave  $\gamma^*(u)$ and $\beta^*(w_{t+1})$ as elementary fields.
In this case, it is convenient to consider a ratio of (\ref{binzo}) and the corresponding correlator (\ref{blokok})
with $\gamma^*(u)$ and $\beta^*(w_{t+1})$ omitted:
\begin{equation}\label{ozook}\frac{\bigl\langle \gamma^*(u)
\,\delta(\beta^*(w_1))\dots \delta(\beta^*(w_{t}))\,\,\beta^*(w_{t+1})\bigr\rangle}{\bigl\langle \delta(\beta^*(w_1))\dots \delta(\beta^*(w_{t}))\bigr\rangle}=\frac{\det\,N_{(t+1)} }{\det\, N_{(t)}}. \end{equation}
We can regard the left hand side as the propagator in the fermionic Gaussian theory of eqn. (\ref{dolfig}).  As
a propagator, this ratio is unchanged in replacing fermions by bosons.  So for the $\beta\gamma$ system
\begin{equation}\label{ozolk}\frac{\bigl\langle \gamma(u)
\,\delta(\beta(w_1))\dots \delta(\beta(w_{t}))\,\,\beta(w_{t+1})\bigr\rangle}
{\bigl\langle \delta(\beta(w_1))\dots \delta(\beta(w_{t}))\bigr\rangle}=\frac{\det\,N_{(t+1)} }{\det\, N_{(t)}}. \end{equation}
Using (\ref{zokok}) for the denominator, we find
\begin{equation}\label{tozolk}\bigl\langle \gamma(u)
\,\delta(\beta(w_1))\dots \delta(\beta(w_{t}))\,\,\beta(w_{t+1})\bigr\rangle=\frac{\det\,N_{(t+1)} }
{{\det}'M\,(\det\, N_{(t)})^2}. \end{equation}
But actually, for some purposes, the normalized two-point function (\ref{ozolk}) of the operators
$\gamma(u)$ and $\beta(w_{t+1})$ in the presence of the given delta function insertions is more useful.
Once one has this formula, the correlation function of an arbitrary collection of elementary fields
in the presence of the delta function insertions follows via the general free-field formula of eqn. (\ref{hikely}).

It is straightforward to extend all this to a correlation function with additional delta function insertions.
Given any $s$ points $u_1,\dots,u_s\in\Sigma$, and $t+s$ points $w_1,\dots,w_{t+s}$,
we define a $(t+s)\times (t+s)$ Slater determinant
$\det N_{(t+s)}$, where
\begin{equation}\label{ivob} N_{(t+s)}=\begin{pmatrix}\y_1(w_1) & \y_1(w_2) & \dots & \y_1(w_{t+s})\cr
\y_2(w_1) &\y_2(w_2)& \dots & \y_2(w_{t+s}) \cr
& & \ddots & \cr
\y_t(w_1) & \y_t(w_2) & \dots & \y_t(w_{t+s})\cr
S(u_1,w_1)& S(u_1,w_2) & \dots & S(u_1,w_{t+s}) \cr
&& \ddots& \cr
S(u_s,w_1)& S(u_s,w_2) & \dots & S(u_s,w_{t+s})
 \end{pmatrix}
.\end{equation}
Then
\begin{equation}\label{dynamics}\bigl\langle\gamma^*(u_1)\dots \gamma^*(u_s)\beta^*(u_1)\dots
\beta^*(u_{t+s})\bigr\rangle={\det}'M\,\det\, N_{(t+s)}.\end{equation}
For bosons, this extrapolates to a generalization of eqn. (\ref{linzo}) with $s>1$:
\begin{equation}\label{dynamicox}\bigl\langle\delta(\gamma(u_1))\dots \delta(\gamma^*(u_s))
\delta(\beta^*(u_1))\dots
\delta(\beta^*(u_{t+s}))\bigr\rangle=\frac{1}{\det'M}\,\frac{1}{\det\, N_{(t+s)}}.\end{equation}
After replacing $t$ and $s$ by $t+1$ and $s+1$, it also extrapolates to the following generalization of eqn. (\ref{tozolk}) for the two-point function
of elementary fields $\beta$ and $\gamma$ in the presence of any number of delta function insertions:
\begin{equation}\label{tozolko}\bigl\langle \delta(\gamma(u_1))\delta(\gamma(u_2))\dots\delta(\gamma(u_s))
\gamma(u_{s+1})
\,\delta(\beta(w_1))\dots \delta(\beta(w_{t+s}))\,\,\beta(w_{t+s+1})\bigr\rangle=\frac{\det\,N_{(t+s+1)} }
{{\det}'M\,(\det\, N_{(t+s)})^2}. \end{equation}
The normalized correlation function of $\beta$ and $\gamma$ in the presence of the delta function insertions is 
\begin{equation}\label{tozolkox}\frac{\bigl\langle \delta(\gamma(u_1))\delta(\gamma(u_2))\dots\delta(\gamma(u_s))
\gamma(u_{s+1})
\,\delta(\beta(w_1))\dots \delta(\beta(w_{t+s}))\,\,\beta(w_{t+s+1})\bigr\rangle}
{\bigl\langle \delta(\gamma(u_1))\delta(\gamma(u_2))\dots\delta(\gamma(u_s))
\,\delta(\beta(w_1))\dots \delta(\beta(w_{t+s}))\bigr\rangle}
=\frac{\det\,N_{(t+s+1)} }
{\det\, N_{(t+s)}} .\end{equation}
This formula again
has an immediate extension to the case of a larger number of elementary field insertions in the presence
of the same delta functions.  One just uses eqn. (\ref{hikely}) to express the $2n$-point functions of elementary
free fields in terms of the two-point functions.

From these general formulas for correlation functions, 
one can evaluate the operator products of the general $\beta\gamma$ system,
for an arbitrary collection of elementary fields and delta function operators. As in the spin 1/2 case, a variety of 
interesting additional operators  appear in this operator
product expansion. 
The operators built from products of elementary fields and delta functions  are all of the local operators
of the $\beta\gamma$ system that one needs in superstring perturbation theory.
(They correspond to the operators of the ``small Hilbert space'' in the language of \cite{FMS}.)
It is convenient in superstring perturbation theory to also include nonlocal delta function operators
described in section \ref{exdelta}; as we have seen, the extension of the theory to include such
operators is straightforward.

\subsubsection{Spin Fields?}\label{spinf}

At this point, the reader may well ask what has happened to
the usual spin fields \cite{FMS} of superstring perturbation theory.  
The answer to this question is that in fact, the spin fields are hidden in what we have already explained.
(See section \ref{imox} for related remarks.)

In superstring perturbation theory, in the absence of Ramond insertions, the line bundle $\L$ is $K^{-1/2}$,
or more precisely it is endowed with an isomorphism $\L^2\cong K^{-1}$.  In the presence of Ramond
punctures supported at points\footnote{We consider a split super Riemann surfaces
so that the positions of the Ramond divisors can be described by points in an ordinary Riemann surface
$\Sigma$.  This assumption does not affect the local structure, which is of interest here.}
 $p_1,\dots,p_{\sn_\Ra}\in\Sigma$, the isomorphism becomes
 \begin{equation}\label{plooz} \L^2\cong K^{-1}\otimes_{i=1}^{\sn_\Ra}\O(-p_i). \end{equation}
 
 In particular, with arbitrary Ramond insertions, the $\beta\gamma$ system is the system that we
 have already described, with a particular choice of $\L$.  There is no need to discuss separately
 a case in which spin fields are included.
 
 Still, the reader may feel that something is missing from what we have said so far.  In fact,
 what is missing is not a discussion of spin fields, but a discussion of reparametrization invariance.

\subsubsection{Reparametrization Invariance}\label{reparam}

  For general
$\L$, the question of reparametrization invariance does not arise.  Given an abstract holomorphic
line bundle $\L$ over a Riemann surface $\Sigma$, there is no natural way to lift local reparametrizations of
$\Sigma$ to symmetries of $\L$.  For this question to make sense, $\L$ must be defined in terms of the geometry
of $\Sigma$, such that we know how reparametrizations of $\Sigma$ should act on $\L$.

For example, this is the case if $\L=K^s$ for some $s$; reparametrizations of $\Sigma$ lift in a natural way to an action on
$K$ and also on $K^s$.  

The space of local operators of the $\beta\gamma$ system is independent of $\L$, in the following sense.
A local operator is a locally-defined notion, and since any $\L$ is locally trivial, the choice of $\L$ cannot
affect the space of local operators.  Likewise, operator product relations can be computed locally and thus
are independent of $\L$.

What depends on $\L$ is the way that reparametrizations act on the local operators.  Let us consider,
for example, the vertex operator (\ref{diffo}) that represents the ground state of the $\beta\gamma$ system at picture
number $-t$:
\begin{equation}\label{zinnof}\SSigma_{-t}=
\delta(\gamma)\delta(\partial\gamma)\dots\delta(\partial^{t-1}\gamma).\end{equation}
It can be defined for any $\L$. The question of its conformal dimension makes sense if $\L$ is
of the form  $\L=K^s$ for any $s$, but the answer
certainly depends on $s$.  Indeed, $\gamma$ has dimension $s$. Hence, reasoning
classically, 
$\delta(\gamma)$ has dimension $-s$, and $\delta(\partial^k\gamma)$ has dimension $-s-k$, 
so  $\SSigma_{-t}$ has dimension $-ts-t(t-1)/2$.  These
formulas are also valid quantum mechanically, for the following reasons. The fact that $\delta(\gamma)$ has
the opposite conformal dimension from $\gamma$ follows from the usual mapping from the $\beta^*\gamma^*$ system
to the $\beta\gamma$ system.  Given this, the conformal dimension of 
$\SSigma_{-t}$ can be deduced from operator product
relations such as eqn. (\ref{moxo}), leading to the classical result $-ts-t(t-1)/2$.  Ultimately, the reason that classical
formulas for the dimensions are valid is that we are looking at operators made only from $\gamma$
and there is no singularity in the $\gamma\cdot\gamma$ operator product.  In superstring perturbation
theory in the absence of Ramond punctures, this discussion applies with
$s=-1/2$ and hence the dimension of $\SSigma_{-t}$ is $-t(t-2)/2$.

Now let us discuss what happens in the presence of a Ramond puncture. 
In the presence of a Ramond puncture at a point $p$,
 $\L$ is endowed with an isomorphism
\begin{equation}\label{tino}\L^2\cong K^{-1}\otimes \O(-p).\end{equation}
(We do not specify the positions of other Ramond punctures, since we will be looking at the local behavior near
$p$.)  The line bundle $\O(-p)$ is canonically trivial away from $p$; it is trivialized by the section 1
(understood as a section of $\O(-p)$ with a simple pole at $p$).  So away from $p$, $\L$ is presented with an
isomorphism to $K^{-1/2}$, and therefore it makes sense to ask how reparametrizations of the punctured $z$-plane
act on the operators of the $\beta\gamma$ system.  The generators of such reparametrizations comprise the
Virasoro algebra, so in particular the Virasoro algebra acts on the operators that can be inserted at $p$.  
This action, however, is different from what it would be in the absence of the Ramond puncture at $p$.

To spell this out a little more fully, we will compare two descriptions of the $\beta\gamma$ system in the presence of a Ramond puncture -- the
same two descriptions that were considered in section \ref{trvo}. Pick a local holomorphic coordinate $z$ that vanishes at the point $p$.
Then the line bundle $K^{-1}\otimes \O(-p)\cong T\otimes \O(-p)$ is trivialized locally by the section $z^{-1}\partial_z$.
$\L$ can be trivialized locally by a section $s$ whose square maps to $z^{-1}\partial_z$ under the isomorphism (\ref{tino}).
This condition determines $s$ only up to sign.
Make a choice of sign and denote $s$ as  $(z^{-1}\partial_z)^{1/2}$.  In the notation of section \ref{trvo}, we write the ghost
field $\gamma$ near $z=0$ as
\begin{equation}\label{tomog} \h\gamma(z) (z^{-1}\partial_z)^{1/2},\end{equation}
where the function $\h\gamma(z)$ is single-valued and regular near $z=0$.  In this description, there is no spin operator
at $z=0$.  The more conventional approach is to take advantage of the fact that $\O(-p)$ is canonically trivial
when the point $z=0$ is omitted, so on the complement of that point, the line bundle $K^{-1}\otimes \O(-p)$ is isomorphic
to $K^{-1}\cong T$, and is trivialized by the section $\partial_z$.  So away from $z=0$, the isomorphism (\ref{tino}) means that
$\L$ can be trivialized locally by a section $s^\diamond$ whose square maps to $\partial_z$.  Again, $s^\diamond$ is locally determined
up to sign, but now there is no consistent way to define the sign; $s^\diamond$ has a monodromy around $z=0$, since
\begin{equation}\label{moomog}s=\frac{1}{z^{1/2}} s^\diamond, \end{equation}
and $s$ has no monodromy.   Let us write $(\partial_z)^{1/2}$ for $s^\diamond$.
The conventional description of the $\beta\gamma$ system is to write $\gamma$ near $z=0$ as 
\begin{equation}\label{otmog}\gamma^\diamond(z) (\partial_z)^{1/2}.\end{equation}
Clearly, the function $\gamma^\diamond(z)$ has a monodromy around $z=0$.  $\gamma^\diamond(z)$ is usually denoted
simply as $\gamma(z)$ (and we did so in section \ref{trvo}).  

The advantage of the description by $\gamma^\diamond(z)$ (and its conjugate $\beta^\diamond(z)$) is that the action of
reparametrizations is clear.  Away from $z=0$, the  $\beta^\diamond\gamma^\diamond$ system is a standard $\beta\gamma$ system
of spins $(3/2,-1/2)$; the local operators and operator product relations, the 
stress tensor, and the transformation of local operators under reparametrizations are all the same as if there were no Ramond puncture
at $z=0$.  Of course, the local operators at $z=0$ look exotic from this
point of view; they are spin fields.  However, the $\beta^\diamond\gamma^\diamond$ description gives a very effective way to
determine how those spin fields transform under reparametrizations;
for one approach relying on bosonization, see \cite{FMS}, and
for another approach, relying on a knowledge of how the stress tensor is defined away from $z=0$, 
see section \ref{usbeta} above.  In the $\h\beta\h\gamma$ description, there is nothing unusual
about the operators that are inserted at $z=0$, but it is less obvious how to compute their behavior under reparametrization. 

\subsubsection{Another Interpretation Of The Delta Function Insertions}\label{anint}

We will conclude by explaining another interpretation of the formula (\ref{tozolkox}) for the
two-point function of elementary fields $\gamma$ and $\beta$ in the presence of arbitrary
delta function insertions.  We  simplify notation by writing   $u$ and $w$ instead of $u_{s+1}$ and $w_{t+s+1}$,
and we write the left hand side of eqn. (\ref{tozolkox}) 
as $\bigl\langle \gamma(u)\beta(w)\bigr\rangle_{N,\delta}$,
where the notation is meant to indicate a normalized two-point function in the presence of
a collection of delta function insertions:
\begin{equation}\label{perfu}\bigl\langle\gamma(u)\beta(w)\bigr\rangle_{N,\delta}=\frac{\det\,N_{(t+s+1)}}
{\det\,N_{(t+s)}}. \end{equation}
We want to consider this as a function of $u$ and $w$ with the locations of the delta function
insertions held fixed.  This means that the denominator is a constant, and all zeroes and poles
come from the numerator $\det\,N_{(t+s+1)}$.

This Slater determinant has a simple pole at $u=w$, with a coefficient such that the residue of
 $\bigl\langle \gamma(u)\beta(w)\bigr\rangle_{N,\delta}$ at $u=w$ is 1.  It also has simple poles
at $u=w_i$, $i=1,\dots,t+s$, and simple zeroes at $u=u_i$, $i=1,\dots,s$, with no other zeroes
or poles.  These conditions
define the normalized two-point function $\bigl\langle\gamma(u)\beta(w)\bigr\rangle_{N,\t\L}$,
with no delta functions at all but with the line bundle $\L$ that enters the definition of the $\beta\gamma$ system
replaced by another line bundle
\begin{equation}\label{deggo}\t \L=\L\otimes_{i=1}^{s} \O(-u_i)\otimes _{j=1}^{t+s}\O(w_j).\end{equation}
Thus, the $\beta\gamma$ system defined with a line bundle $\L$ and an arbitrary collection
of delta function insertions is simply equivalent to a $\beta\gamma$ system with a different
line bundle $\t\L$ and no delta function insertions at all.

Hopefully the bosonic delta function operators
of superstring perturbation theory do not retain much mystery.

\appendix

\section{Pullback Of Differential Forms}\label{pullback}

Suppose that a group $G$ acts on a manifold $Y$ with quotient $M=Y/G$.   Let $y_i$, $i=1,\dots,n=\dim\,Y$ be local
coordinates on $Y$.  A differential form on $Y$ is a function $F(y_1\dots y_n|\d y_1\dots \d y_n)$.  

$F$ is said to be a pullback from $M$ if it can be written as $F(x_1\dots x_m|\d x_1\dots \d x_m)$, where $x_i$, $i=1,\dots,m=\dim\,M$
are local coordinates on $M$.  We can pick the local coordinates of $Y$ to be the $x_i$ and also local coordinates 
$f_i$, $i=1,\dots,n-m$
on the fibers of the fibration $Y\to M$.  For $F$ to be a pullback from $M$ means that it is independent of the $f_i$ and of the $\d f_i$.

This is equivalent to the following two conditions:

(1) $F$ must be $G$-invariant.

(2) $F$ must be annihilated by contraction with any of the vector fields that generate the action of $G$.

Concretely, $G$ is generated by vector fields of the general form
\begin{equation}\label{onx} V_a=\sum_{i=1}^{n-m} v_{a,i}\frac{\partial}{\partial f_i}\end{equation}
and the corresponding contraction operators are
\begin{equation}\label{tonx} \mathbf{i}_{V_a} =\sum_{i=1}^{n-m}v_{a,i}\frac {\partial}{\partial \d f_i}.\end{equation}
The condition that $\mathbf{i}_{V_a}F=0$, $a=1,\dots,n-m$ means precisely that $F$ is independent of the $\d f_i$,
and then the condition that $F$ is $G$-invariant means that it is also independent of the $f_i$.  (The matrix $v_{a,i}$ is invertible
if $G$ acts freely on $M$, but in any event the vector fields $V_a$ span the tangent space to the fibers of the projection $Y\to M$, so the condition
that $\mathbf{i}_{V_a}F=0$ for all $a$ means that $F$ is independent of all $\d f_i$.)

This argument is equally applicable if $n$ and/or $m$ is infinite.
In the application in section \ref{vacamp}, we have $Y=\JJ$, the space of all complex structures on a given surface $\Sigma$ of genus
$\g$;
$M=\mathcal M_\sg$ the moduli space of Riemann surfaces; and $G=\D$, the group of orientation-preserving diffeomorphisms of
$\Sigma$. So $n$ is infinite but $m$ is finite.
The argument also generalizes without difficulty to supermanifolds and supergroups, a fact we exploit in section \ref{measure}.
In the context of supermanifolds, the reasoning applies to forms of any picture number.  

\section{Bosonic String Gauge Parameters}\label{really}

\subsection{Examples}\label{exreally}

In bosonic string theory, in compactification to $\R^d$ with $d\geq 2$,
every physical state of non-zero momentum is associated to a primary field $V$ of dimension
$(1,1)$ constructed from matter fields only.\footnote{This statement and some others below are part of the BRST version of the
no-ghost theorem; see \cite{KO,FGZ,FO,Th}, or, for example, section 4.4 of \cite{Polch}.  For the original no-ghost theorem, see \cite{Brower,GT}.} 
 However, in general this representation is not unique; $V$ can be shifted by adding to it a null vector.
  
A primary field $V$ is said to be a null vector if it is a Virasoro descendant, meaning that it is of the form
\begin{equation}\label{zorbo}V=\sum_{n=1}^\infty L_{-n}U_n,\end{equation}
 with Virasoro generators $L_{-n}$ and some states $U_n$.  
(In practice, if $V$ is a state of definite momentum, the sum over $n$ is a finite
sum since there is a lower bound on the possible dimension of $U_n$.)  If $V$ is of this form, then according to the no-ghost
theorem, the corresponding
BRST-invariant vertex operator $\V= c V$ is a BRST commutator, $\V=\{Q_B,\W\}$ for some $\W$, and should
decouple from the $S$-matrix.\footnote{For brevity, in this appendix, we consider either open strings or a chiral sector of closed strings,
so we omit the antiholomorphic ghosts $\t c$.  Also, motivated by the operator-state correspondence
of conformal field theory,  we write just $Q_B\W$ instead of $\{Q_B,\W\}$.}

As we saw in section \ref{ginv}, to analyze this decoupling, it helps to know that we can assume that
\begin{equation}\label{odz}b_n\W=0,~~n\geq 0,\end{equation}
or equivalently that $\W$ is constructed using only $c$  and not its derivatives. 
Before presenting any general theory, we will describe the first few examples.  

The first case of a null vector is
\begin{equation}\label{zelm} V=L_{-1}\Phi_0,\end{equation}
where $\Phi_0$ is a matter primary of dimension 0.  In this case, $\V=cV$ obeys $\V=Q_B\W$, with
\begin{equation}\label{elm}\W=\Phi_0.\end{equation}
So $\W$ does not involve $c$ at all.  The next case is
\begin{equation}\label{welm}V=\left(L_{-2}+\frac{3}{2}L_{-1}^2\right)\Phi_{-1},\end{equation}
where $\Phi_{-1}$ is a matter primary of dimension $-1$.  In this case $\V=c V$ is of the form $Q_B\W$ with
\begin{equation}\label{pelm} \W=bc\Phi_{-1}+\frac{3}{2}L_{-1}\Phi_{-1}.\end{equation}
So $\W$ depends on $c$, though not on its derivatives.

 \subsection{General Proof}\label{genproof} 
  
 The BRST version of the no-ghost theorem says that 
 if $V$ is a Virasoro primary of dimension 1 constructed from matter fields that is also null, meaning
 that it is a Virasoro descendant in the sense of eqn. (\ref{zorbo}),
then $\V=cV$ is of the form
 $Q_B\W$ for some $\W$.  We want to show that in general, not just in the examples
 described above,  we can pick $\W$ so that $b_n\W=0$, $n\geq 0$.     
 
 There is actually a surprising shortcut to this conclusion, provided by the appendix to \cite{Th}.
 If $V$ is a dimension 1 primary that can be expanded as in eqn. (\ref{zorbo}), then actually
 this expansion can be drastically shortened and put in the very special form
 \begin{equation}\label{zizobo}V=L_{-1}\Phi_0+\left(L_{-2}+\frac{3}{2}L_{-1}^2\right)\Phi_{-1},\end{equation}
 where $\Phi_0$ and $\Phi_{-1}$ are primaries of the indicated dimension.
 In other words, a dimension 1 primary $V$ that is null is always a linear combination of null vectors of the sort that
 we analyzed in appendix \ref{exreally}.  So we can use equations (\ref{elm}) and (\ref{pelm}) to explicitly
 write $\V=cV$ in the desired form $\V=\{Q,\W\}$ where $\W$ is the primary
 \begin{equation}\label{loggo}\W=\Phi_0+\left(bc+\frac{3}{2}L_{-1}\right)\Phi_{-1}. \end{equation}
 We have actually learned more than we needed.  We aimed to prove that $\W$ can be chosen not to depend
 on the derivatives of $c$; we have learned that it can be chosen to depend on derivatives of neither $b$ nor $c$.
 The significance of this is unclear.
 
 The considerations in \cite{Th} are closely related to the no-ghost theorem, and here we will give
 an alternative argument that is closely related to the BRST version of the no-ghost theorem and especially
 to the proofs given in \cite{KO} and in section 4.4 of \cite{Polch}.
  This argument will have two steps: {\it (i)} a general argument to show 
 that we can assume that $\W$ does not involve $\partial^nc$, $n\geq 2$; {\it (ii)} a more special argument to
 show that we can assume that $\W$ also does not involve $\partial c$.
 
 Step {\it (ii)} goes as follows.  Associated to every matter primary $V$ of dimension 1, there are a dual pair of $Q_B$-invariant
vertex operators, namely $\V=cV$ and $\V^*=c\partial c V$.  The no-ghost theorem shows 
 that $\V$ is trivial in $Q_B$-cohomology if and only if $\V^*$ is (and if and only if $V$ is a descendant).  
 Now suppose that $\V^*=Q_B\W^*$ for some $\W^*$. After averaging over the compact group generated by $L_0$, we can assume that $L_0\W^*=0$;  the whole analysis that follows will be made
 in the subspace with $L_0=0$.  
Our reasoning in step {\it (i)} will apply
equally to $\V^*$ and $\V$ and show that, if $\V^*=Q_B\W^*$ for some $\W^*$, we can assume that 
$b_n\W^*=0$, $n>0$.  Then setting $\W=-b_0\W^*$, we see that $b_n\W=0,~n\geq0$.  Moreover
$Q_B\W=-Q_Bb_0\W^*=b_0Q_B\W^*=b_0\V^*=\V$.  So we have found an operator $\W$ with the desired properties
$Q_B\W=\V$ and $b_n\W=0$, $n\geq 0$.  

As for step {\it (i)},
we define a grading on the space of operators by assigning degree 0 to  matter fields as well as $b$, $c$,  $\partial c$, and derivatives
of $b$, but degree 1 to $\partial^nc$, $n\geq 2$.  We call this grading the $c''$-degree.  So for example, the operator $bc\,\partial X \partial^2c\,\partial^5c$
has $c''$-degree 2, with a contribution of 1 from  $\partial^2c$ and from $\partial^5c$ and no contribution from $b$, $c$, or $\partial X$.   
(The name $c''$-degree
is motivated by the fact that we are counting the number of times that the second or higher derivative of $c$ appears in an operator.)
The $c''$-degree of any operator is non-negative.

We can write the BRST operator $Q_B$ as
\begin{equation}\label{ombix}Q_B=Q_>+Q_0,\end{equation}
where $Q_>$ increases the $c''$-degree by 1, and $Q_0$ leaves it unchanged.
The equation $Q_B^2=0$ implies that $Q_>^2=0$, so we can define the cohomology of  $Q_>$.  

Moreover, since $Q_>$ increases the $c''$-degree by a definite amount (namely 1), we can define the cohomology of $Q_>$ for
states of any given $c''$-degree.  We will show below that this cohomology vanishes except for $c''$-degree 0.
Let us first explain how this implies what we want.

 We assume that there exists some $\W$ such that
\begin{equation}\label{omon}Q_B\W=\V,\end{equation}
where $\V$ has $c''$-degree 0. (So $\V$ can be either $cV$ or $c\partial cV$, where $V$ is constructed from matter fields only.)  We assume
that $\V$ has energy-momentum $k$, and we can require that $\W$ has the same energy-momentum.
Suppose that the expansion of $\W$ in operators of definite $c''$-degree is $\W=\sum_{r=0}^t \W_r$ and to
begin with suppose that $t$, which is the maximal $c''$-degree of any term in $\W$, is greater than 0. ($t$ is finite since the $c''$-degree
of states of fixed energy-momentum  and $L_0=0$ is bounded.)
Since the right hand side of (\ref{omon}) has $c''$-degree 0, (\ref{omon}) implies that $Q_>\W_t=0$.
If it is true that the cohomology of $Q_>$ vanishes for $c''$-degree greater than 0, it follows that $\W_t=Q_>\U_{t-1}$
for some $\U_{t-1}$ of $c''$-degree $t-1$.  This being so, we can replace $\W$ by $\W-Q_B\U_{t-1}$ without
disturbing eqn. (\ref{omon}); so we reduce to the case that the maximum $c''$-degree of any term in $\W$ is
$t-1$.  Continuing in this way, we reduce to the case that $\W$ has maximum $c''$-degree 0.  
At this point,
we have $b_n\W=0$, $n>0$ (since the operators $b_n$ with $n>0$ lower the $c''$-degree), and we have completed step {\it (i)}.

It remains  to show that the cohomology of $Q_>$ vanishes as claimed for positive $c''$-degree.   For this we
will imitate the proof of the no-ghost theorem, as presented in section
4.4 of \cite{Polch}, whose notation and reasoning we  follow as closely as possible. 
 
The argument assumes that the matter sector of the theory has at least two free fields $X^0$ and $X^1$,
with Lorentz metric $-(\d X^0)^2+(\d X^1)^2$.  We use a lightcone basis $X^\pm =(X^0\pm X^1)/\sqrt 2$.  The
corresponding oscillators $\alpha_m^\pm$, $m\in \Z$, obey $[\alpha_m^+,\alpha_n^-]=-m\delta_{m+n}$,
with other commutators vanishing.  We introduce the operator
\begin{equation}\label{zobof}N_*^\lc=\sum_{m\geq 1}\frac{1}{m}\alpha_{-m}^+\alpha_m^-, \end{equation}
which counts  minus the number of $+$ excitations and has eigenvalues $0,-1,-2,\dots$.  (Our $N_*^\lc$ is the relevant half of
the definition in eqn. (4.4.8) of \cite{Polch}.)   $Q_>$ has a decomposition
\begin{equation}\label{obog}Q_>=Q_{>,1}+Q_{>,0}+Q_{>,-1},\end{equation}
where $[N^\lc_*,Q_{>,j}]=jQ_{>,j}$, so that $Q_{>,j}$ shifts $N^\lc_*$ by $j$ units.  The fact that $Q_>^2=0$
implies that $Q_{>,1}^2=0$, so we can compute its cohomology.  

Assuming that the energy-momentum $k$ carried by $\V$ is non-zero, we can pick our
coordinates so that $k^+\not=0$.  (Because of this step, the analysis given here fails for $k=0$.
This case requires special treatment and has unusual properties; see section \ref{remark}.)
Explicitly
\begin{equation}\label{toff}Q_{>,1}=-(2\alpha')^{1/2}k^+\sum_{m\geq 1} \alpha^-_{m} c_{-m}.\end{equation}
(From $Q_1$ as defined in eqn. (4.4.13) of \cite{Polch}, we have omitted the terms that reduce the $c''$-degree.)
A convenient way\footnote{Since $Q_{>,1}$ is bilinear in oscillators, and commutes with $L_0$, 
one can also simply
compute its cohomology via a mode-by-mode analysis.  We will proceed that way in the superstring case, though it would
also be possible to imitate the argument involving $R$ and $S$.}  to compute the cohomology of $Q_{>,1}$ is to define
\begin{equation}\label{roff} R= \frac{1}{(2\alpha')^{1/2}k^+}\sum_{m\geq 1}\alpha_{-m}^+b_m, \end{equation}
which reduces $N^\lc_*$ by 1.
We also define
\begin{equation}\label{proff} S=\{Q_{>,1},R\}= \sum_{m=1}^\infty \left(m c_{-m}b_m-\alpha^+_{-m}\alpha^-_m \right) ,\end{equation}
which commutes with $N^\lc_*$.

The operator $S$ is positive semi-definite.  Its kernel consists of states annihilated  by $b_m$ and also by  $\alpha^-_m$, with 
$m>0$.  In particular, states annihilated by $S$ have  $c''$-degree 0.

Since $S$ commutes with $Q_{>,1}$, the cohomology of $Q_{>,1}$ can be decomposed in the eigenspaces of
$S$.  $Q_{>,1}$ annihilates the kernel of $S$, as is clear from the description of that kernel in the last paragraph.
On the other hand, the cohomology of $Q_{>,1}$ vanishes for $S\not=0$.  Indeed, if $S\Psi_0=s\Psi_0$ with $s\not=0$,
and $Q_{>,1}\Psi_0=0$, then $\Psi_0=s^{-1}\{Q_{>,1},R\}\Psi_0=Q_{>,1}(s^{-1}R\Psi_0)$, so $\Psi_0$ vanishes in the cohomology
of $Q_{>,1}$.

These statements imply that the cohomology of $Q_{>,1}$ is supported at $c''$-degree 0.  Now let us examine the cohomology of $Q_>$.

We will give an argument similar to one above.  Suppose that $\Y$ is an element of the cohomology of $Q_>$ at 
$c''$-degree $q>0$, and let $\Y=\sum_{r=-t}^{-t'}\Y_r$
be the expansion of $\Y$ in states $\Y_r$ of $N^\lc_*=r$.   This expansion is a finite sum since $N^\lc_*$ 
is bounded above and  below in the space of states of fixed
energy-momentum with $L_0=0$; in fact, $-t'\leq 0$ since $N^\lc_*$ is bounded above by 0.  If $Q_>\Y=0$, then $Q_{>,1}\Y_{-t'}=0$.  Since the cohomology
of $Q_{>,1}$ vanishes at positive $c''$-degree, there is a state $\U_{-t'-1}$ of $N^\lc_*=-t'-1$ (and with  $c''$-degree 1 less than that of $\Y$)
such that
$Q_{>,1}\U_{-t'-1}=\Y_{-t'}$.  Without changing the $Q_>$ cohomology class of $\Y$, we can replace $\Y$ by $\Y-Q_>\U_{-t'-1}$,
whose expansion in eigenstates of $N^\lc_*$ now runs over eigenvalues that are bounded above by $-t'-1$.  After repeating this process
finitely many times, we eventually reach the lower bound on $N^\lc_*$ in the chosen sector and reduce to the case $\Y=0$.
So the $Q_>$ cohomology vanishes for positive $c''$-degree.

\section{Superstring Gauge Parameters}\label{surreally}

\subsection{Examples}

The goal here is to generalize the results of appendix \ref{really} to superstring theory.
We begin by we describing 
the first few examples of null vectors and gauge parameters in superstring theory. 

In the Neveu-Schwarz sector, a 
physical state that should decouple from the $S$-matrix is associated to a primary
field $V$ of dimension $1/2$ constructed from the matter system
 that is also null.  In other words, $V$ is a superconformal descendant, that is it can be written
\begin{equation}\label{polver}V=\sum_{n>0} L^\XX_{-n}W_n+
\sum_{r>0}G^\XX_{-r} \Lambda_r, \end{equation} 
where $L_n^\XX$, $G_r^\XX$ are the superconformal generators of the matter system and
$W_n$, $\Lambda_r$, $n,r>0,$ are some states of the matter system. 
The first such null vector arises at the massless level and takes the form
\begin{equation}\label{olver}V=G^\XX_{-1/2}\Phi_0,\end{equation}
where $\Phi_0$ is a matter primary of dimension 0.  The corresponding superconformal 
vertex operator $\V=c\delta(\gamma)V$
can be written $\V=\{Q_B,\W\}$, with
\begin{equation}\label{wonzop}\W=c\delta'(\gamma)\Phi_0. \end{equation}
In terms of modes, if $|-1\rangle$ represents the $\beta\gamma$ ground state of picture number $-1$,
obeying 
\begin{equation}\label{yderly}\beta_r|-1\rangle=\gamma_r|-1\rangle=0, ~~r> 0,\end{equation}
then
\begin{equation}\label{muroxic}\W=-c_1\beta_{-1/2}|-1\rangle\otimes \Phi_0.\end{equation}
$\W$ is annihilated by $b_n$ and $\beta_r$, $n,r\geq 0$.

The first example of gauge-invariance for massive Neveu-Schwarz states  is associated to a level
1 null vector
\begin{equation}\label{gontor}V=\left(G^\XX_{-3/2}+2G^\XX_{-1/2}L^\XX_{-1}\right)\Phi_{-1}\end{equation}
of the matter system.  Here $\Phi_{-1}$ is a matter primary of dimension $-1$.
The corresponding superconformal vertex operator  $\V=c\delta(\gamma)V$ is $\V=\{Q_B,\W\}$, with
\begin{equation}\label{ontor}\W= \left(\delta(\gamma)G^{\XX}_{-1/2}
-c\beta\delta(\gamma)+c\delta'(\gamma)L_{-1}^\XX\right)\Phi_{-1}.   \end{equation}
In modes
\begin{equation}\label{turox}\W=\left( G_{-1/2}^\XX-c_1\beta_{-3/2}
-c_1\beta_{-1/2}L_{-1}^\XX   \right)|-1\rangle\otimes \Phi_{-1}.\end{equation}
$\W$ is annihilated by the modes $b_n,\beta_r$, $n,r\geq 0$. 

In the Ramond sector, a physical state that should decouple from the $S$-matrix is associated to a 
primary field $V$ of dimension $5/8$  constructed from the matter system only that is a descendant in the
sense that it can be written as in (\ref{polver}), again with $n,r>0$. (Now $r$ takes integer values.
We do not allow a term $G^\XX_0\Lambda_0$ in the sum in (\ref{polver}), or else 
we would be claiming that the
usual massless fermions should decouple.)  There are no massless string states of this form. 
The first non-trivial example arises at the first massive level of the Ramond sector and is
\begin{equation}\label{tolver} V=\left(L^\XX_{-1}-\frac{1}{2}G^\XX_{-1}
G^\XX_0\right)\Phi, \end{equation}
where  $\Phi$ is a matter
primary of dimension $-3/8$. The corresponding vertex operator
is $\V=c\SSigma_{-1/2}V$, where $\SSigma_{-1/2}$ is the $-1/2$ picture 
spin field of the $\beta\gamma$ system, obeying
\begin{align}\label{zecho} \beta_n\SSigma_{-1/2} & =0,~~n\geq 0,\cr \gamma_n\SSigma_{-1/2}&=0,~~n>0. \end{align} 
We have $\V=\{Q_B,\W\}$, with
\begin{equation}\label{irod}\W=\left(1-G_0^\gh G_0^\XX\right)\SSigma_{-1/2}\Phi, \end{equation}
where $G_0^\gh$ is the $G_0$ operator of the ghost system.  In terms of ghost oscillator modes, this can be written
\begin{equation}\label{trod}\W=\left(1+\frac{1}{2}c_1\beta_{-1}G_0^\XX\right)\SSigma_{-1/2}\Phi.\end{equation}
In particular $\W$ depends non-trivially on $c$, as is typical of gauge parameters for massive string states, but is
annihilated by $b_n$ and $\beta_r$, $n,r\geq  0$.

\subsection{General Proof}\label{gp}

Now we want to show that in general, gauge parameters in superstring theory
can be chosen to be annihilated by antighost modes $b_n$ and $\beta_r$, $n,r\geq 0$,
so that in analyzing gauge-invariance,  a superconformal formalism is possible.  

Let us consider first the NS sector.  Let $\V=c\delta(\gamma)V$ be a superconformal vertex
operator, where $V$ is a superconformal primary of the matter system of dimension 1/2.
The no-ghost theorem says that $\V=\{Q_B,\W\}$ for some $\W$ if and only if $V$ is a null vector in
the sense that it can be written as in eqn. (\ref{polver}).  To the matter primary $V$ we can also
associate the $Q_B$-invariant operator $\V^*=c\partial c \delta(\gamma)V$. The no-ghost theorem
says further that $\V^*$ is BRST-trivial, $\V^*=\{Q_B,\W^*\}$ for some $\W^*$, if and only if $V$ is a null
vector. In these statements, we can assume that $\W$ and $\W^*$ are annihilated by $L_0$.

We aim to show that if $\V=\{Q_B,\W\}$ for some $\W$, then $\W$ can be chosen to be independent
of $\partial^nc$ and $\partial^m\gamma$, with $n,m> 0$.  Equivalently, we want to show that
one can chose $\W$ to be annihilated by antighost modes $b_n$ and $\beta_r$, $n,r\geq 0$.

It is possible to generalize either
 of the two arguments described in section \ref{genproof}, but here we will present a generalization only
 of the second one.\footnote{For a generalization of the first one, see \cite{Thorntwo}.} 
As before, the argument consists of {\it (i)} a general proof to show that we can
assume that $\W$ and $\W^*$ do not involve $\partial^n c$, $n\geq 2$ or $\partial^m\gamma$, $m\geq 1$;
and {\it (ii)} a special argument to show that we can further assume that $\W$ does not involve $\partial c$.

Step {\it (ii)} is exactly as it was in section \ref{genproof}: once we find $\W^*$  obeying $\V^*=\{Q_B,\W^*\}$,
annihilated by $L_0$, 
and independent of $\partial^n c$, $n\geq 2$ and $\partial^m\gamma$, $m\geq 1$, we simply
set $\W=b_0\W^*$.  Then $\W$ obeys all the desired properties.  

For step {\it (i)}, we imitate the proof in section \ref{genproof}.  We define what we will call the
$c''\gamma'$ degree of an operator to be the number of times that $\partial^nc$, $n\geq 2$ or
$\partial^m\gamma,$ $m\geq 1$ appear in an operator.  For example, $c\partial c \partial^2c \delta'(\gamma)
\partial\gamma$ has $c''\gamma'$ degree 2.  We write the BRST operator as
\begin{equation}\label{meldox} Q_B=Q_>+Q_0,\end{equation}
where $Q_>$ increases the $c''\gamma'$ degree by 1, and $Q_0$ leaves it unchanged.  
The condition $Q_B^2=0$ implies that $Q_>^2=0$, so we can define the cohomology of $Q_>$.  
Just as in section \ref{genproof}, it suffices to show that
the cohomology of the operator $Q_>$ is supported at $c''\gamma'$ 
degree 0.

To prove this, just as in the bosonic case, we introduce lightcone coordinates 
$X^\pm =(X^0\pm X^1)/\sqrt 2$, and an operator $N^{\mathrm{lc}}_*$ equal to minus
the number of $+$ excitations.  Now, however, we must consider fermionic as well as bosonic $+$
excitations.  We expand the worldsheet fermions as $\psi^I(z)=\sum_{r\in\Z+1/2}\psi^I_r/z^{r+1/2},
~I=0,\dots,9$
with $\{\psi^I_r,\psi^J_s\}=\eta^{IJ}\delta_{rs}$, and define light cone operators
$\psi^\pm_r=(\psi^0_r\pm \psi^1_r)/\sqrt 2$, obeying $\{\psi^+_r,\psi^-_s\}=\delta_{rs}$.  Then we define
\begin{equation}\label{zobooot}N_*^\lc=\sum_{m\geq 1}\frac{1}{m}\alpha_{-m}^+\alpha_m^-
-\sum_{r\geq 1/2}\psi_{-r}^+\psi^-_r.  \end{equation}
As before, we expand $Q_>$ as a sum of terms $Q_{>,k}$ that shift $N^{\mathrm{lc}}_*$ by $k$:
\begin{equation}\label{onco}Q_>=Q_{>,1}+Q_{>,0}+Q_{>,-1}.\end{equation}
Exactly as in section \ref{genproof}, it suffices to show that the cohomology of $Q_{>,1}$ is
supported at $c''\gamma'$-degree 0.   Again, the explicit form of $Q_{>,1}$ makes this an easy result:
\begin{equation}\label{ytoff}Q_{>,1}=-(2\alpha')^{1/2}k^+\sum_{m\geq 1} \alpha^-_{m} c_{-m}+
(2\alpha')^{1/2}k^+\sum_{r\geq 1/2}\gamma_{-r}\psi^-_r.\end{equation}

To make a similar analysis for the Ramond sector, we mainly need to change our terminology slightly.
A basis of states  of the $\beta\gamma$ system at picture number $-1/2$ is
given by
\begin{equation}\label{ofrro}\prod_{s\geq 1}\beta_{-s}^{n_s}\prod_{r\geq 0}\gamma_{-r}^{m_r}\,\SSigma_{-1/2},
\end{equation} where all but finitely many $n_s$ and $m_r$ are zero.
We say that such a state has $\dot\gamma$ degree $\sum_{r\geq 0}m_r$.  
After including the $bc$
ghosts and the matter fields, we define the $c''\dot\gamma$-degree of a state to be the 
sum of the $c''$-degree and the $\dot\gamma$-degree of that state.  
(The $c''\dot\gamma$-degree is bounded in states of picture number $-1/2$ and 
fixed $L_0$ and ghost number.)  Consider a pair of $Q_B$-invariant states $\V=c\SSigma_{-1/2}\Phi$,
$\V^*=c\partial c\SSigma_{-1/2}$, where $\Phi$ is a Ramond-sector matter primary of dimension $5/8$.
According to the no-ghost theorem, $\V$ and $\V^*$ are $Q_B$-trivial
if and only if $\Phi$ is a null vector.  In this case, we want to prove that we can write $\V=\{Q_B,\W\}$,
with $\W$  not depending on derivatives of $c$ and constructed only from states of $m_r=0$ in
(\ref{ofrro}).  As in the other examples, it suffices to find $\W^*$ of $c''\dot\gamma$-degree 0 with
$\V^*=\{Q,\W^*\}$; then we set $\W=b_0\W^*$.

To find a suitable $\W^*$,
we make the same expansion
as in (\ref{meldox}), but now using the $c''\dot\gamma$-degree, and again it will suffice to show
that the cohomology of $Q_>$ is supported at $c''\dot\gamma$-degree 0.  
The worldsheet fermion fields still have an expansion $\psi^I(z)=\sum_{r\in\Z}\psi^I_r/z^{r+1/2}$,
but now with integer $r$.  
In defining $N^{\mathrm {lc}}_*$,
we now need to be careful with the treatment of modes with $r=0$.  We set
\begin{equation}\label{zoboolot}N_*^\lc=\sum_{m\geq 1}\frac{1}{m}\alpha_{-m}^+\alpha_m^-
-\sum_{r\geq 0}\psi_{-r}^+\psi^-_r.  \end{equation}  
Again we expand $Q_>$ as in (\ref{onco}) in terms of operators $Q_{>,k}$ that shift $N_*^{\mathrm{lc}}$
by $k$, and it suffices to show that the cohomology of $Q_{>,1}$ is supported at $c''\dot\gamma$ degree 0.
This follows directly from the explicit form:
\begin{equation}\label{yotoff}Q_{>,1}=-(2\alpha')^{1/2}k^+\sum_{m\geq 1} \alpha^-_{m} c_{-m}+
(2\alpha')^{1/2}k^+\sum_{r\geq 0}\gamma_{-r}\psi^-_r.\end{equation}

\section{An Example Of Fermion Integration}\label{leisurely}

Here we explore in more detail the example briefly cited in section \ref{lanex}.  This will also enable
us to show some of the subtleties of fermionic integration in a concrete example.   (The same example is explored in much
more detail in \cite{More}.)

The example, which was studied in \cite{DIS,ADS,GrSe},
involves a one-loop heterotic string amplitude with two NS vertex operators, with an even spin structure.  We will not describe here
the full string theory context for this computation.  We will simply describe what is involved in this example in constructing the right
integration cycle and integrating over it.

First we describe the string worldsheet.  From a holomorphic point of view, a heterotic string worldsheet $\Sigma$ is a super Riemann surface.
A genus one super Riemann surface  with an even spin structure can be described by superconformal coordinates $z|\theta$
with the equivalence relations
\begin{align}\label{mune} z\cong &z+1 \cr
                    \theta  \cong & -\theta\end{align}
and
\begin{align}\label{zune} z\cong & z+\tau \cr
                    \theta\cong & \theta.\end{align}
From an antiholomorphic point of view, $\Sigma$ is an ordinary genus 1 Riemann surface, described by a complex coordinate $\t z$
with the equivalence relations
\begin{equation}\label{une} \t z\cong \t z+1 \cong \t z+\t \tau.\end{equation}
We will just set $\t \tau=\bar\tau$, though the general formalism would let us relax this slightly.  $\tau$ is an even modulus of $\SIgma$,
and in superstring perturbation theory, one integrates over it.  However, the interesting subtleties do not involve
the integral over $\tau$, and for our purposes we will just set $\tau$ to a constant.

\begin{figure}
 \begin{center}
   \includegraphics[width=3.5in]{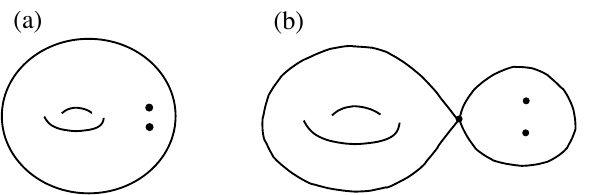}
 \end{center}
\caption{\small  (a) A Riemann surface $\Sigma$ of genus 1 with two marked points.  (b) As the two marked points approach each other,
$\Sigma$ splits into two components $\Sigma_\ell$ and $\Sigma_r$.  The external vertex operators
are on $\Sigma_r$, which has genus 0.}
 \label{broneck}
\end{figure} 
In general, the two NS vertex operators can be inserted at arbitrary points $z|\theta=u_1|\zeta_1$ 
and $z|\theta=u_2|\zeta_2$.  The interesting subtlety
will arise when the two vertex operators approach each other, or in other words when $u_1\to u_2$.    As depicted in fig. \ref{broneck},
this is actually a special case of a separating degeneration of a Riemann surface.   The natural gluing parameter at this degeneration is
\begin{equation}\label{satis}q_\NS=u_1-u_2-\zeta_1\zeta_2,\end{equation}
as already explained in section \ref{lanex} (and in 
 section 6.3.2 of \cite{Wittentwo}).  We should stress that what is
important in (\ref{satis}) is that the function whose vanishing defines the separating degeneration is $u_1-u_2-\zeta_1\zeta_2$,
rather than, say $u_1-u_2$ or $u_1-u_2-2\zeta_1\zeta_2$.  The precise coefficient multiplying $u_1-u_2-\zeta_1\zeta_2$ is not relevant
and has been chosen for convenience in (\ref{satis}).  (The normalization used in \cite{Wittentwo} actually differs from this by a factor of 2.)

A genus 1 surface has a translation symmetry $z\to z+\mathrm{constant}$.  Using this, we can set, say, $u_2=0$.  We write simply $u$ instead
of $u_1$. With this choice, the insertion points of the two NS vertex operators are $\t z;\neg  z|\theta=\t u;\neg u|\zeta_1$
and $0;\neg 0|\zeta_2$.   The equivalence relations (\ref{mune}), (\ref{zune}), and (\ref{une}) become
\begin{align}\label{tine} u\cong & \,u+1 \cr
                                      \zeta_1\cong &\, -\zeta_1\cr
                                       \zeta_2\cong & \,\zeta_2 \cr
                                        \t u\cong& \,\t u + 1 \end{align}
                                        and
 \begin{align}\label{stine} u\cong &\, u+\tau \cr
                                      \zeta_1\cong & \,\zeta_1\cr
                                       \zeta_2\cong & \,\zeta_2 \cr
                                        \t u\cong& \,\t u + \bar \tau \end{align}
The gluing parameters are
\begin{equation}\label{gline}\t q = \t u\end{equation}
and
\begin{equation}\label{ugline} q_\NS=u-\zeta_1\zeta_2.\end{equation}

The moduli space $\M_R$ is parameterized by $u|\zeta_1,\zeta_2$, and the
moduli space $\M_L$ is parameterized by $\t u$, in each case with equivalence relations stated in (\ref{tine}) and (\ref{stine}).
Now we want to define an integration cycle $\varGamma\subset \M_L\times \M_R$.  We will aim for the simplest choice of $\varGamma$,
meaning that its reduced space $\varGamma_\red$ will be the ``diagonal,''  by which we mean the subspace of $\M_{L,\red}\times \M_{R,\red}$ defined by $\bar{\t u}=u$.  Once we specify $\varGamma_\red$, the most general possible choice of
$\varGamma$ is made by generalizing the equation $\bar{\t u}=u$ to include the odd variables.  Since in this example, we only have two
odd variables $\zeta_1$ and $\zeta_2$, the most general choice of $\varGamma$ is defined by 
\begin{equation}\label{hobo}\bar{\t u}=u+\zeta_1\zeta_2 h(\t u, u) \end{equation}
for some function $h(u,\t u)$.  For $\varGamma$ to be invariant under the equivalences (\ref{tine}) and (\ref{stine}), we require
\begin{equation}\label{zomboy} h(\t u+1, u+1)=- h(\t u,u),~~~h(\t u+\bar \tau, u+\tau)=h(\t u, u).\end{equation}
These conditions would allow us to set $h=0$, but they would not allow us, for example, to set $h=-1$.  However, we need one more
condition for $h$.  At the degeneration $u=\t u=0$, $\varGamma$ is supposed to be defined by $\bar{\t q}=q_\NS$, and this
implies that we want
\begin{equation}\label{rombo} h(0,0)=-1.  \end{equation}

We cannot set $h$ to a constant, because no constant is compatible with both (\ref{rombo}) and (\ref{zomboy}).  Functions $h$ obeying
the conditions do exist, but there is no canonical choice.  And therefore, there is no canonical choice for the integration cycle $\varGamma$
of superstring perturbation theory in this situation.  However, any two choices give homologous integration cycles.  Indeed, if $h_1$
and $h_2$ are any two functions that obey the conditions, we can interpolate between them via
\begin{equation}\label{zrobo} h_\lambda = \lambda h_1+(1-\lambda)h_2,~~0\leq\lambda\leq 1,\end{equation}
and this gives an explicit homology between the two choices of $\varGamma$.  

Now let us discuss what sort of measure we want to integrate over $\varGamma$.  Perhaps the most obvious idea is to consider
a section of the Berezinian of $\M_L\times \M_R$ of the simple form $[\d \t u;\neg \d u|\d\zeta_1\,\d\zeta_2]$. This measure behaves
well at infinity, since it is invariant under the change of variables $u\to u-\zeta_1\zeta_2$, where according to (\ref{ugline}),
$u-\zeta_1\zeta_2$ is a good coordinate at infinity.    However, $[\d \t u;\neg \d u|\d\zeta_1\,\d\zeta_2]$ is
odd under (\ref{tine}) and so cannot be the form we want to integrate.  Instead we consider an integration form 
\begin{equation}\label{tryox}\Omega=[\d \t u;\neg \d u|\d\zeta_1\,\d\zeta_2]\,P(\t u),\end{equation}
where
\begin{equation}\label{ryox} P(\t u+1)=-P(\t u),~~P(\t u+\bar \tau)=P(\t u),\end{equation}
so that  $\Omega$ respects the necessary equivalences and is well-defined and holomorphic near $\Delta^\star\subset \M_L\times\M_R$.  
We further require that $P(\t u)$ is a meromorphic function of $\t u$
whose only singularity is a simple pole with residue 1 at $\t u=0$.  These conditions uniquely determine $P(\t u)$:
\begin{equation}\label{yox} P(\t u)=\sum_{n,m\in\Z}\frac{(-1)^n}{\t u+n+m\bar\tau}.\end{equation}

We now want to evaluate the integral
\begin{equation}\label{picoro}I=\int_{\varGamma}\Omega.\end{equation}
How such an integral arises in superstring perturbation theory is explained in \cite{DIS,ADS}.  The singularity of $\Omega$ at
$\t u = u = 0$ is $\d\t u\,\d u/\t u$, milder than the $\d\t u\,\d u/\t u u$ singularity that leads to subtleties with massless tadpoles, but
singular enough to lead to the interesting phenomenon that we are about to describe. 

To evaluate the integral, we simply use the condition (\ref{hobo}) that defines the integration cycle to solve for $u$.  Since
(\ref{hobo}) says that $u=\bar{\t u}\mod\,{\zeta_1\zeta_2}$, and since $(\zeta_1\zeta_2)^2=0$,   we can solve (\ref{hobo}) by 
\begin{equation}\label{ziro} u=\bar{\t u}-\zeta_1\zeta_2 h(\t u, \bar{\t u}).\end{equation}
This enables us to express the measure $[\d \t u;\neg \d u|\d\zeta_1\,\d\zeta_2]$ in terms of $\t u, \bar{\t u},\zeta_1,$ and $\zeta_2$ only:
\begin{equation}\label{miro}  [\d \t u;\neg \d u|\d\zeta_1\,\d\zeta_2]=
\left(1-\zeta_1\zeta_2\frac{\partial h}{\partial\bar{\t u}}\right)[\d\t u\,\d\bar{\t u}|\d\zeta_1\,\d\zeta_2] .\end{equation}
So our integral is
\begin{equation}\label{slort}I=\int_\varGamma[\d\t u\,\d\bar{\t u}|\d\zeta_1\d\zeta_2]
\left(1-\zeta_1\zeta_2 \frac{\partial h}{\partial\bar{\t  u}}\right)P(\t u).\end{equation}
Integrating first over the fermions, via the Berezin integral $\int[\d\zeta_1\,\d\zeta_2]\,1=0$,   
$\int[\d\zeta_1\,\d\zeta_2]\zeta_1\zeta_2=1$,  we reduce to an ordinary integral over
the torus $\Sigma_\red$:
\begin{equation}\label{lort}I=-\int_{\Sigma_\red}\d\t u\wedge \d\bar{\t u}\,\frac{\partial h}{\partial \bar{\t u}}P(\t u). \end{equation}
Integrating by parts and using the fact that $\partial_{\bar {\t u}}P(\t u)=2\pi \delta^2(\t u)$, because of the pole of $P(\t u)$ at $\t u=0$, 
and the fact that $h(0,0)=-1$, we get
\begin{equation}\label{ort}I=4\pi i.\end{equation}
As expected, the integral did not depend on the choice of $h$, as long as it behaves correctly at infinity.  

\subsection{What We Have Learned}\label{learned}

One lesson to learn from this example  
is that the evaluation of a superstring scattering amplitude depends crucially on knowing how the integration cycle $\varGamma$ is
supposed to behave near infinity.  Another lesson is that although the integral has a well-defined
value, there is no natural answer to the question of where on moduli space the
answer came from; this depends on the unnatural choice of $\varGamma$.  

It is true that after integration by parts,
the answer in this example seemed  to come from a delta function at $u=0$.  However, this appears to be
special to this low genus example, somewhat analogous to the fact that in genus 1, the vanishing of the dilaton
tadpole can be established by summing over spin structures.  

The basic reason that in this example it is possible to isolate a delta
function contribution  at infinity
is that the moduli spaces $\M_L$ and $\M_R$ are holomorphically split (this notion was defined in section \ref{harder}).
$\M_R$ is holomorphically split by the map that takes $z|\zeta_1\zeta_2$
to $z$, and $\M_L$ is trivially split since it is purely bosonic.  When $\M_L$ and $\M_R$ are holomorphically split, we have a natural
holomorphic map $\pi:\M_L\times \M_R\to \M_{L,\red}\times \M_{R,\red}$, and then there is a fairly natural choice of integration cycle,
namely $\varGamma_0=\pi^{-1}(\Delta)$, where $\Delta\subset \M_{L,\red}\times \M_{R,\red}$ is the ``diagonal.''  
The only thing that may be wrong with $\varGamma_0$ is that it may have the wrong
behavior at infinity.  In our example, this is the case; $\varGamma_0$ corresponds to taking $h$ to
be identically zero, while the desired behavior at infinity is $h(0,0)=-1$.  Still, we can pick $\varGamma$ to 
coincide with $\varGamma_0$ except
in a small neighborhood of infinity, so one can write the superstring amplitude as an integral over $\varGamma_0$ plus
a correction at infinity.  Our example has the further property that the bulk contribution -- 
the integral over $\varGamma_0$ --
vanishes.  In a more generic situation, there is no holomorphic splitting \cite{DonW}, and one should expect that there is no natural 
choice of integration cycle even
away from divisors at infinity, and no natural way to write a superstring amplitude as the sum of a ``bulk'' 
contribution and a contribution
at infinity.

\subsection{The Moduli Space As An Orbifold}\label{melnox}

Now let us look more closely at the moduli space $\MM_{1,2,0;+}$ that parametrizes a super Riemann surface $\Sigma$ with two NS punctures.
As above, it can be parametrized by $u|\zeta_1,\zeta_2$, with the equivalences in (\ref{tine}) and (\ref{stine}).  

However, there is one more symmetry that we should take into account.  The super Riemann surface $\Sigma$ that we started with in eqns. (\ref{mune}) and (\ref{zune})
has the additional symmetry
\begin{align}\label{plox} z& \to -z \cr \theta&\to \pm \sqrt{-1}\theta. \end{align}
The transformation of $\theta$ by $\pm \sqrt{-1}\theta$ ensures that $\varpi=\d z-\theta\d\theta$, which defines the superconformal structure, maps to a multiple
of itself.   With some choice of the sign, let us call this automorphism $\kappa$.  There is no natural choice of sign and indeed we must
allow both signs.  $\kappa^2$ is the universal symmetry
$z|\theta\to z|-\theta$ of any split super Riemann surface, and $\kappa^3$ is obtained from $\kappa$ by reversing the sign.  

Accordingly, we should impose on the variables $u|\zeta_1,\zeta_2$ the equivalence relation
\begin{align}\label{zlox} u & \to -u\cr
                                           \zeta_i & \to \pm \sqrt{-1} \zeta_i, ~i=1,2. \end{align}
generated by $\kappa$.    $\kappa^2$ is the universal symmetry 
\begin{align}\label{zilox} u & \to u \cr
                                          \zeta_i & \to -\zeta_i,~i=1,2\end{align}
that reverses the sign of all odd moduli, and again $\kappa^3$ is obtained from $\kappa$ by reversing the sign.                                         
                                         
So $\MM_{1,2,0;+}$ must be understand as an orbifold (or stack).  There is a $\Z_2$ automorphism group, generated by $\kappa^2$, whenever $\zeta_1=\zeta_2=0$. This
automorphism group is enhanced to $\Z_4$, generated by $\kappa$, at $u=\zeta_1=\zeta_2=0$.  In fact, taking account the equivalences (\ref{tine}) and (\ref{stine}), the automorphism
group is enhanced  to $\Z_4$ at the four   points $u=0,1/2,\tau/2$, and $(1+\tau)/2$, with $\zeta_1=\zeta_2=0$.   

$\kappa$ maps $q_\NS=u-\zeta_1\zeta_2$ to $-q_\NS$, consistent with the claim that the gluing parameters are well-defined up to multiplication
by an invertible function.

We did not take the automorphism $\kappa$ into account in  our discussion of integration.  However, the form (\ref{tryox}) that we integrated is $\kappa$-invariant,
with $\kappa$ acting on $\t u$ by $\t u \to -\t u$.

\vskip 1cm
\noindent {\bf Acknowledgments} Research partly supported by NSF grant  PHY-0969448. I thank 
N. Berkovits, L. Brink, P. Goddard,
 J. Maldacena,
 V. Pestun, J. Polchinski, M. Rangamani,  A. Sen, N. Seiberg,  H. Verlinde, and B. Zwiebach  and especially
K. Becker, P. Deligne, G. Moore,  D. Robbins, B. Safdi, and C. Thorn for helpful 
comments and discussions.  M. Turansick assisted with the figures.

\bibliographystyle{unsrt}

\end{document}